\documentclass[a4paper,11pt,twoside,draft,fleqn]{book}
\usepackage{longtable,url,mathptmx,fancyhdr,tocloft}

\usepackage{setspace}  
\usepackage{tocbibind}
\usepackage{natbib}
\usepackage{bibentry}

\usepackage{geometry}
\renewcommand{\subsectionmark}[1]{} 
\usepackage{mathptmx}
\usepackage{setspace}
\usepackage[dvips]{color}
\usepackage[dvips,final]{graphicx}
\usepackage{lscape}

\newcommand{\thesisdate}{September 2004}
\begin{document}

\begin{titlepage}
{                               

  \thispagestyle{empty}         

  \centering

  \vspace*{1cm}

  {\bfseries\Huge
   Progenitors of \\
    Core-Collapse Supernovae\\
    }

  \strut
  \vfill
\begin{figure}[!h]
\begin{center}
\includegraphics[height=90mm,angle=0]{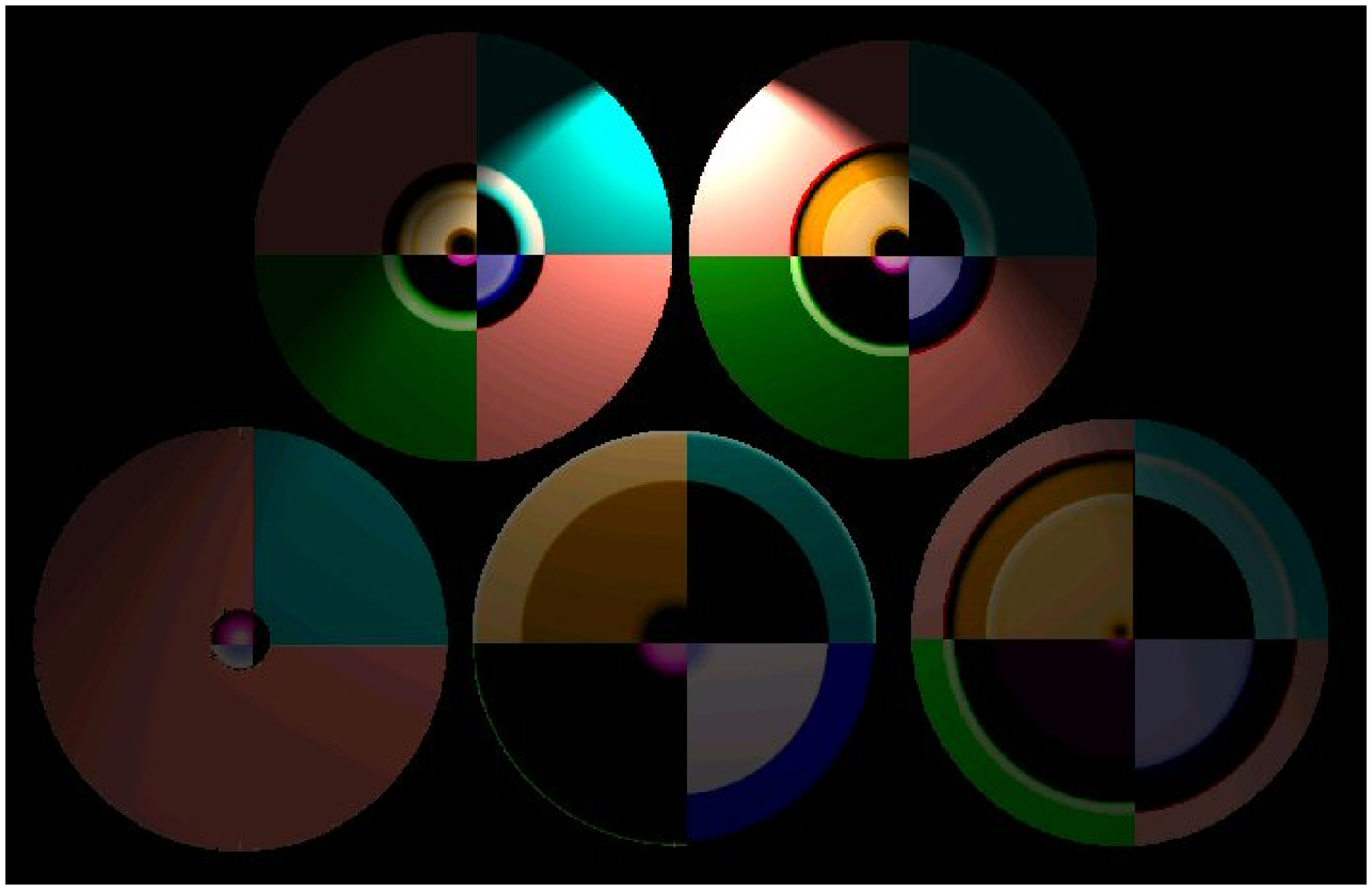}
\end{center}
\end{figure}
  {\large
    \textsc{This Dissertation is Submitted for\\
      the Degree of Doctor of Philosophy}\\
    \textit{\textbf{September 2004}}
  }
    
  \vfill
  \strut
  {\LARGE John James Eldridge}

  \vspace{1cm}
  {\Large \scshape
    Institute of Astronomy\\
    \&\\
    Fitzwilliam College\\
    \vspace{0.5\baselineskip}
    \centerline{University of Cambridge}
  }
}
\end{titlepage}

\cleardoublepage
\thispagestyle{empty}
\vspace*{5cm}
\begin{center}
  For Dad, Mum, Mark and Kate.
\end{center}

\cleardoublepage
\thispagestyle{empty}
\vspace*{5cm}
\begin{center}
``Uh-oh, is this going to be another crazy experiment that crosses a line man was not meant to cross?''\\
\textit{Fry, Futurama, I second that emotion.}\\
\end{center}

\chapter*{Declaration}
\addcontentsline{toc}{chapter}{Declaration}

I hereby declare that my thesis entitled \textit{Progenitors of Core-collapse Supernova} is not
substantially the same as any that I have submitted for a degree or
diploma or other qualification at any other University.

I further state that no part of my thesis has already been or is being
concurrently submitted for any such degree, diploma or other
qualification.

This dissertation is the result of my own work and includes nothing
which is the outcome of work done in collaboration except where
specifically indicated in the text. Those parts of this thesis which
have been published or accepted for publication are as follows.
\begin{itemize}
\item The construction of new Opacity tables in chapters 2 has been published in, \\Eldridge J.J. \& Tout C.A., 2004, MNRAS, 348, 201.
\item Parts of the work in chapters 3 and 4 is based on the work published in, \\Eldridge J.J. \& Tout C.A., 2004, MNRAS, 353, 87.
\end{itemize}

Various figures throughout the text are reproduced from the work of
other authors, for illustration or discussion. Such figures are always
credited in the associated caption.

This thesis contains fewer than 50,000 words.

\vspace{4cm}
\parbox{10cm}{%
  John James Eldridge\\
  Cambridge, \thesisdate
}

\chapter*{Summary}
\addcontentsline{toc}{chapter}{Summary}
\markboth{Summary}{Summary}

The progenitors of core-collapse supernovae are stars with an initial mass greater than about $8M_{\odot}$. Understanding the evolution of these stars is necessary to comprehend the evolution and differences between supernovae.

We have constructed new and unique opacity tables to increase model accuracy during the latest stages of stellar evolution. We have investigated how initial mass, initial composition and mass loss affects the progenitors and their populations. There are many prescriptions for mass loss. Different research groups use their preferred rates. We have compared 12 different prescriptions and determined which provides the best fit to observations. We use our preferred mass-loss scheme to make suggestions as to the source of the differences between supernova types from our progenitor models.

Binary evolution is considered in order to search for low luminosity SN progenitors and progenitor types not possible from single stars. Removal of the hydrogen envelope is more common and we find quite different hydrogen deficient SN progenitors. We discuss the implications of our binary models for ultra-luminous X-ray sources and gamma-ray bursts. We present an estimation of the mass distribution for black holes at various metallcities showing that massive black holes are not formed until very low metallicities. Finally we combine the single star and binary results to determine their relative populations and compare to observations. However it is not possible to draw many firm conclusions because of the uncertainty in observations to date.

\chapter*{Acknowledgements}
\addcontentsline{toc}{chapter}{Acknowledgements}
\markboth{Acknowledgements}{Acknowledgements}
		
\begin{center}
"I confess I am at somewhat a loss for words."  \\
\textit{Spock, The Alternative Factor.}\\
\end{center}

Well here I am. Almost finished, just need to print the entire thing out, and type a `few' words to say thank you to all of those who have aided and supported me during the production of this thesis. This goes beyond just the time writing this document, further than the long time taken to actually do the research, it goes to the time when I was growing as a person, learning about physics and the world.

First I must thank my supervisor Dr Christopher Adam Tout. He has guided me through my three years with care and patience. He has introduced me to the wonders of stellar evolution and also taught me a vast amount of English grammar! Chris has a deep and wide knowledge of astrophysics which he is always keen to share. He also has an in depth knowledge of alcoholic beverages and social etiquette in which he has also provided tuition. So thank you Chris!

Next I must thank the members of the Cambridge stellar evolution group. In my first year we were a bit of a rag-tag band of mercenaries but we have come together to form a productive group. Everyone's enthusiasm and knowledge was contagious. Thanks go to Dr Pierre Lesaffre, Dr Maria Lugaro and Dr Gijs Nelemens. Thank you to Dr Rob Izzard for persuading me to choose Chris as a supervisor because `he buys you beer' and for helpful advice over three years. Thanks to Richard Stancliffe for being down the corridor when I needed to talk to someone about stars, the code and life. Also thank you for proofreading this thesis. Thanks to Ross Church for also proofreading this thesis and discussing duplicitous thoughts. The entire lot of you have made the actual study part of my time in Cambridge brilliant and vibrant. I would also like to thank Dr Onno Pols, Dr Philipp Podsaidlowski and Dr Alexander Heger for helpful discussions and comments.

Thank you to all of the rest of the students, post-docs and staff at the IoA. My time in the department has been great. Special thanks go to some people including Will Thorne, Adrian Turner, Lisa Voigt, Dan Price, Adam Amara (a.k.a. Andy Azmara), Joanna Smith, Antonio Vale, Ali Basden and Shoko Jin. You've all been good friends and we've had some great times together so thank you all! Also knowing others were going through this hell of writing a thesis made it that little bit more survivable even if some of you did finish before me! Thanks to everyone else in the IoA, especially Justyn Maund, Maggie Hendry and all supernovae observers, for their feedback and input. Oh and thank you Muon!

While most of my work is done at the department, Fitzwilliam college has been a fantastic place to be based at during my studies. I would also like to thank the staff and fellows of the college for their help and support over my time at the college. Especially Dr Guy Pooley, Dr Bill Allison, Dr Gassan Yassin and Prof Brian Johnson.

Right well now with most of the serious work related stuff out of the way its time to get `crazy'. At my IoA interview I was asked why did I want to study astronomy and astrophysics? It caught me off guard and I didn't give the right reason. The reason is science fiction. From a young age I was exposed to Doctor Who, Star Trek and The Hitch-Hikers Guide to the Galaxy. Once at primary school my headteacher complained that all I read was science fiction! Watching Star Trek: The Next Generation, 5pm weekdays after school and then Doctor Who and Blake 7 on a Sunday morning has a lasting effect. Seeing science in such a positive light made me want to pursue knowledge and understanding to the highest degree. Now as I perform research, while I may not be on a starship, I am as close as possible to the career I've always wanted to follow. There are many other authors that have inspired and excited me: Issac Asimov, A.E. Van Vogt, Arthur C. Clarke, Douglas Adams, Larry Niven and Frank Hampson. Thank you to you and your writings, they have inspired me no end.

In most sci-fi adventures there is always action to make it a tad more exciting, this led to me wanting to learn to fight and be able to defend myself. When I got to university the opportunity presented itself with the Cambridge University Tae-Kwon Do club. After going to one lesson, and a certain incident with a French girl, I was hooked. It took me five years, but I finally passed my black belt grading, one of the things I'm most proud of. But along that path I had a lot of help and instruction. The members of the club are not only training partners but many of them are also my friends who have provided something to occupy me outside work. So thank you to you all who I traded kicks, punches and drinks with! Especially Dr Peter Smielewski, Dr Mareque Steele, Dr Phil Green, John Harvey, Graham Russell, Jen Hake, Rob Morgan-Warren, Andy Baldwin, Jules Jacobson, and everyone else otherwise this would go on for ages. Also I want to say thank you to Tina Atkin for fun times and being there at the start of my PhD. And thank you to Caroline Chupin for being a good friend and providing distractions outside of work and TKD.

Thanks to all the friends I've had as an undergraduate and postgrad at college. Especially Louise Benzimra, Baraba Harvey, everyone who was on the MCR committee with me and everyone else. Again too many people to mention! Also thanks to other friends, Ross, Laura, Richard, Maddie, Nick and anyone I haven't mentioned.

Well this is starting to get a bit long really, but I suppose there are a lot of people to thank! Don't worry we are getting towards the end. I now want to thank my extended family for your support and putting up with me forgetting anniversaries and birthdays! My Aunts and Uncles, Jim, Liz, Marion, Terry, Nina, Frank, Pam and Geoff. My cousins, Debbie, Harry, Stuart, Sarah, Justin, Vicky and Keeley. And the kids! Georgia, Lucas, Olivia and Charlotte.  And thanks to my grandparents. Thanks for being proud, fun times, producing an amazingly wonderful and large family, and raising those two people who are my parents.

The one person who has been there for me over the past year, put up with me being ultra-stressed, has taught me alot about myself and who I am and deserves special thanks is Julie Chih-Chun Wang. Thank you Julie for your love, understanding, patience, and other things I won't go into here!

I'm a lucky person to have some many friends and family to thank. There are however some friends, three guys and three girls, who I need to single out. They have helped me through my most difficult times, and together we have had some of the most fantastically fun times too. The guys are the best people to have a beer with in the world! Matt Hartley is the best scientist I know. Calum McFarlane is the best person to have with you when hiking for his witty remarks and true Scottish grit. Dave Ellis is the one guy I would want as my wingman in a dogfight. Thanks guys!

The three girls I must thank are the best friends anyone could want. They are all wonderful and have helped me to come to accept some of the things I always hated about myself and have put up with me moaning more than most. They are the first people I turn to when something bad or good happens. Amber Jenkins is caring, supportive and great fun to be around, thank you Amber. Sandie Bessis is the woman who made me realise there was so much more to do and experience in life. I would not be the person I am today if I hadn't met her, thank you Sandie. Elizabeth Stanway is a saint. There is no one I have probably been more annoying around than her and yet we are still friends. She cooks the most amazing roast dinner this side of the Thames and she shares my interest in sci-fi. Unfortunately she didn't have the sense of the other two to leave Cambridge after our undergraduate days so has had to put up with more than the others! Thank you Liz.

Well that's almost it. `Surely there can't be anyone else to thank?' I hear you cry. There is. My brother and sister are the best brother and sister anyone could want. I admire and love them for the people they are and for making being the middle child not so bad! Thanks you two.

Finally the people that brought me into this world and were the most fantastic parents anyone could wish for. For being understanding, generous, caring, supportive, and just letting me be me. They're always there at the end of the phone when I need to talk. Dad, thanks for introducing me to science fiction, especially Dan Dare. Mum, thanks for providing me with a lifelong love of good food by being the chef that cooks the best Roast dinner in the World! Thanks for getting together originally! Thank you both so very very much for everything.

Well now I've got tears in my eyes from all of that I'd better go blow my nose and get a beer. Its taken time to get here but thats it. Finished. Thanks again everyone. Think I'll let two quotes do the talking....
\vspace{-0.5cm}

\begin{center}
\begin{tabular}{cc}
"Live long and prosper."  								&"I'm going to drink until I reboot."\\
\textit{Spock, Amok Time, Stardate 3372.7.}	&\textit{Bender, Futurama, Hell is Other Robots.}\\
&\\
\end{tabular}
\\
\vspace{-0.1cm}
\noindent{JJ, Cambridge, 1/9/2004.}
\end{center}

\cleardoublepage
\tableofcontents

\listoftables
\addtocontents{lot}{\protect\thispagestyle{fancy}}

\listoffigures
\addtocontents{lof}{\protect\thispagestyle{fancy}}

\chapter{Introduction}
\section{Looking at stars}

\begin{center}
\begin{tabular}{lc}
{\textbf{Pumbaa:}}&{  ``Timon, ever wonder what those sparkly dots are up there?''}\\
{\textbf{Timon:}}&{  ``Pumbaa, I don't wonder; I know.''}\\ 
{\textbf{Pumbaa:}}& { ``Oh. What are they?''}\\
{\textbf{Timon:}}&{  ``They're fireflies that got stuck up in that bluish-black thing''.}\\
{\textbf{Pumbaa:}}&{ ``Oh, gee. I always thought they were balls of gas burning billions of miles away.''}\\
{ \textbf{Timon:}}&{  ``Pumbaa, with you, everything's gas.''}\\
& \textit{\small The Lion King.}\\
\end{tabular}
\end{center}

Everything starts with gas. Slowly over time diffuse interstellar gas collapses under its own gravity, forming a rough sphere that gradually contracts and heats up. Eventually the centre becomes hot enough to ignite nuclear reactions and forms a gravitationally confined nuclear fusion reactor, a new born star, the giver of life. The elements we need to exist were created in stars during their long lifetimes and sometimes their fiery ends. It is difficult to accept the concept that once the nuclei in our bodies were in such an extreme environment.

Stars are the most important objects in the Universe. Even though they make up less than 3\% of the matter they make up 100\% of the visible matter. They are a major apparatus for studying the universe and its evolution. This includes the remnants left from their deaths and the discarded material from stellar winds during their life. The main reason why we began studying the stars must have been their phenomenal beauty in the night sky. On a clear and dark night it is easy to become captivated if you do something very simple, look up. Mankind has, with little doubt, always been drawn to the sky since we started thinking and shaping our environment. We initially thought that the sky was fixed and any changes we observed would directly affect our fate and destiny. At least this allowed enterprising astronomers to secure funding by always providing a good spin on astronomical events for the rulers of their countries.

Once permanent written records of the sky were made we had to abandon the idea that the sky was unchanging forever. It was realised that the sky did evolve: stars move relative to each other, they are not as fixed as we thought. Planets would be the most obvious stars wandering across the sky. However things changed forever when the once in a lifetime spectacular event occurred, a supernova. Luckily this happened when two of the most famous astronomers in history, Tycho Brahe and Johannes Kepler, were alive. The supernovae of 1572 and 1604 captivated them both and they began to provide theories on what they were. Some of the ideas were wilder than others. The most interesting idea was that the supernova of 1604 was seen at the same time as a conjunction between a number of planets and it was thought that this cosmic ballet gave birth to the new star.

\begin{figure}
\begin{center}
\includegraphics[height=140mm,angle=270]{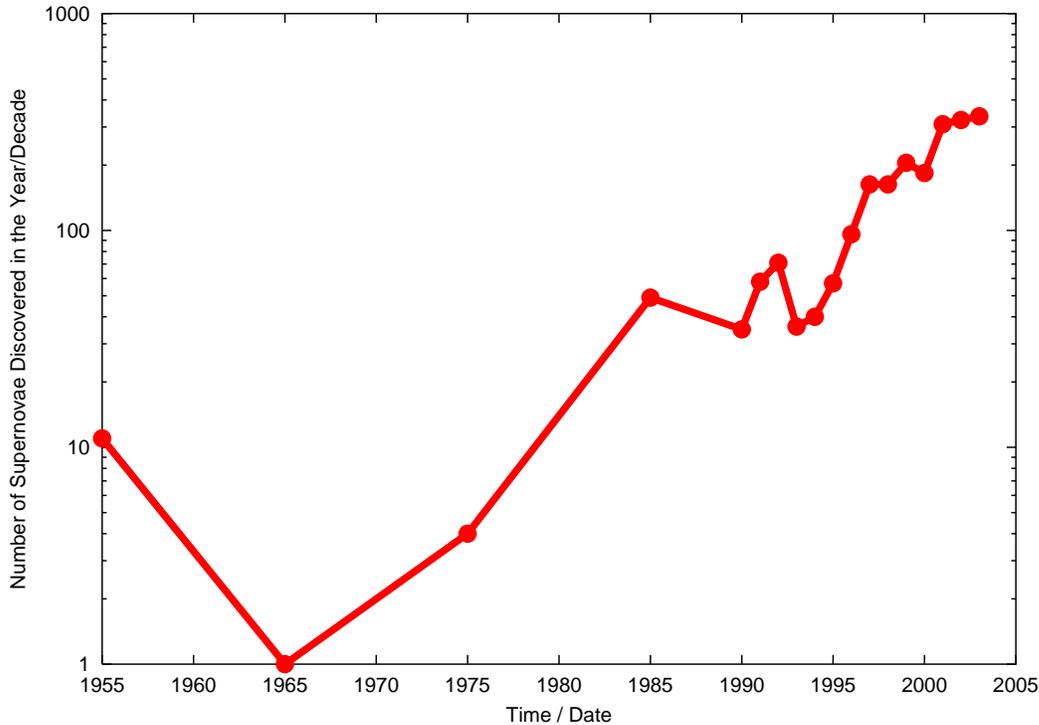}
\caption[Discovery rate of SNe up to 2003.]{Discovery rate of SNe up to 2003. Before 1990 supernovae are binned into decade totals, after 1990 binned into annual totals.}
\label{obsrates}
\end{center}
\end{figure}

We now know that supernovae are in fact the death of stars. The events have been observed in history as far back as the beginning of the second millennium according to current records. If we assume that Homo sapiens have been around for 300,000 years then there have been about 6,000 supernovae in our galaxy during our presence on the planet. Assuming we can only see nearby supernovae then maybe 2,000 of these would have been noticeable in the sky\footnote{Of these 588 would have been type Ia, 235 type Ibc and 1,177 would have been type II.}. We have now observed over 2000 supernovae in nearby galaxies by diligent observing schemes. Figure \ref{obsrates} shows the number of supernovae seen over the time we have been searching for them. The future of supernova studies looks bright. The growth in the area has primarily been driven by cosmologists using type Ia supernovae to determine the geometry of our Universe but the study of supernovae themselves is now starting to be seen as the new important output of the observations. The study of supernovae will filter down to the study of stellar evolution, limiting the possible outcomes forcing us to refine our models and approach a comprehension of stars and their fiery deaths.

\section{Stars and their Evolution}
\begin{center}
``Space is big. You just won't believe how vastly hugely, mind-bogglingly big it is. I mean, you may think it's a long way down the road to the chemists, but that's just peanuts to space.''\\
\textit{The Hitch-Hikers Guide to the Galaxy, by Douglas Adams.}\\
\end{center}
The scale of stars is beyond everyday experience. The temperature, radius, luminosity and density are all far beyond the environmental parameters humans can survive. What is most confounding is the large range of variation between stars themselves. Four factors determine how stars evolve, initial mass, initial composition, mass loss and duplicity.

Initial mass determines the pressures and temperatures the core of a star achieves and therefore how far the nucleosynthesis and evolution progress. Initial composition affects the evolution only slightly moving the mass ranges for different behaviour to occur. We take the composition of stars to be the amounts of hydrogen, helium and metals, metals being everything that isn't hydrogen or helium. When the Universe first formed the only constituents were hydrogen and helium and a small fraction of other light elements and isotopes. The first generation of stars produced carbon, oxygen and other elements. This pollution of the Universe with fusion waste caused the stars to change. Increasing contamination increases the opacity of stellar plasmas and makes stars less compact. The extra carbon, nitrogen and oxygen enable a star to burn hydrogen more effectively by catalysing the reaction.

\begin{figure}
\begin{center}
\includegraphics[height=140mm,angle=90]{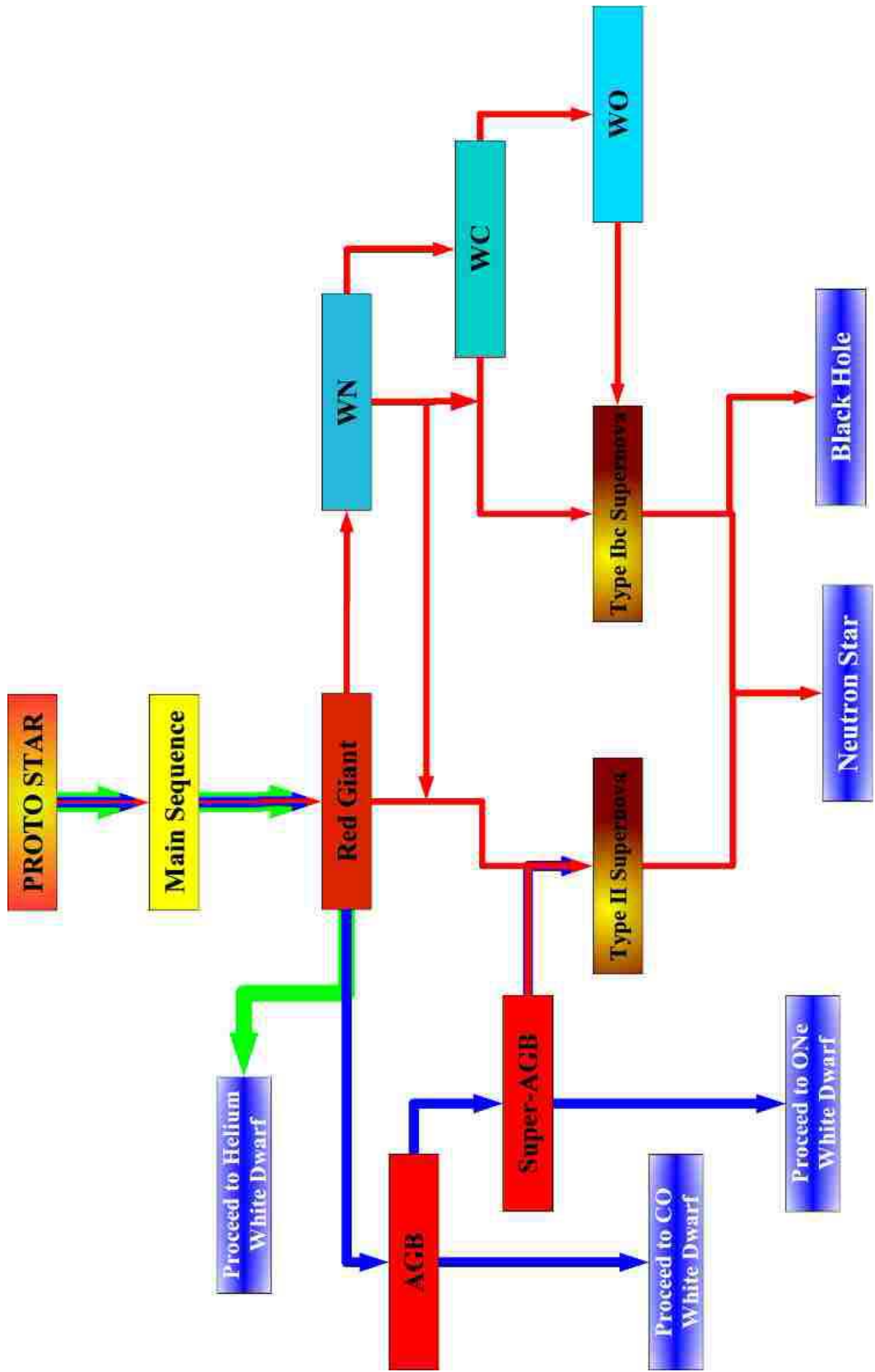}
\caption[Flow diagram for single star evolution.]{Flow diagram for single star evolution. Green - paths available for low mass stars, blue - paths available for intermediate mass stars and red - paths available for high mass stars.}
\label{singlestars}
\end{center}
\end{figure}

Mass loss reduces the mass of a star during its evolution. In the most extreme cases this can completely remove the hydrogen (from a star) leaving it as a naked helium star. Mass loss can be achieved by stellar winds or by a companion if the star is in a binary. The latter can also provide the opportunity for mass gain with mass transferred from one to another. We describe duplicity in more detail below. Figures \ref{hrplot1}, \ref{hrplot2} and \ref{hrplot} contain three Hertzsprung-Russel (HR) diagrams showing examples of the evolution of stars with different initial masses. We describe the main different classes of star below.

\subsection{Low-Mass Stars, $0.08 \leq M/M_{\odot} < 0.8$}
These are stars that burn hydrogen very slowly on the main sequence. Once all hydrogen is exhausted and the helium core is formed the burning moves into a shell around the core. This causes the star to become a red giant. The exact reason why a star becomes a red giant is unknown but it is thought to be due to nuclear burning in a shell. As the star becomes a red giant a deep convective envelope forms that mixes material from the core, mainly helium, to the surface. This is the event of first dredge-up. The core is supported by electron degeneracy pressure. Shell burning continues until the envelope is ejected by a wind during the giant phase and a helium white dwarf is formed. The evolution timescale of these stars is longer than that of the age of the universe so very few of these stars will have evolved all the way to helium white dwarfs unless they were in a binary system. In figure \ref{singlestars} we show the evolution paths of these stars, low-mass star evolution paths are shown in green. Figure \ref{hrplot1} shows the HR diagram for a $0.5M_{\odot}$ star. It slowly evolves on the main sequence and moves up the giant branch. Mass loss grows with ascension of the giant branch and slowly removes the envelope. Eventually removal of the envelope causes the giant to collapse to a white dwarf. This is when the star moves across the HR diagram to the white dwarf cooling track along which the star slowly loses heat to come into thermal equilibrium with the cosmic microwave background radiation.

\subsection{Intermediate Mass Stars, $0.8 \leq M/M_{\odot} < 8$}
These stars evolve as before but do ignite helium in the core. In the lowest mass stars this happens in a degenerate core and a helium flash occurs. In degenerate material the pressure only depends on density. Therefore as the temperature increases the reaction runs away. Energy output increases rapidly and the region continues to expand until the degeneracy is removed and the burning continues quiescently. For more massive stars helium ignites non-degenerately. The helium burning results in the star dropping down the red giant branch. The presence of a central energy source removes the cause for the star to be a giant.

\begin{figure}
\begin{center}
\includegraphics[height=129mm,angle=270]{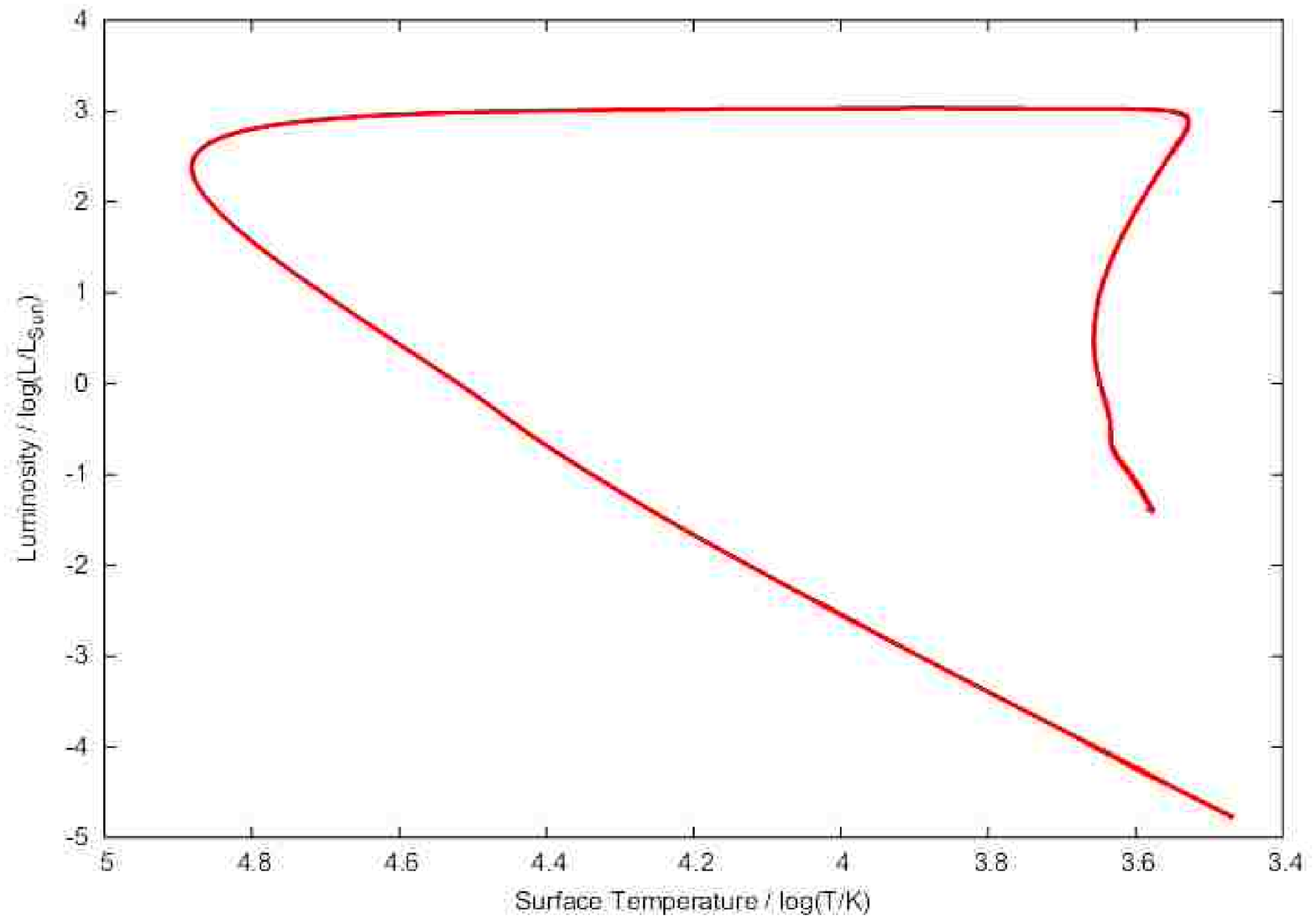}
\caption[HR diagram for a low mass star, $M_{\rm initial}=0.5M_{\odot}$.]{HR diagram for a low mass star, $M_{\rm initial}=0.5M_{\odot}$. The surface temperature changes little during the main sequence and giant branch however once the envelope is removed and the star becomes a white dwarf to move on to the cooling track.}
\label{hrplot1}
\includegraphics[height=129mm,angle=270]{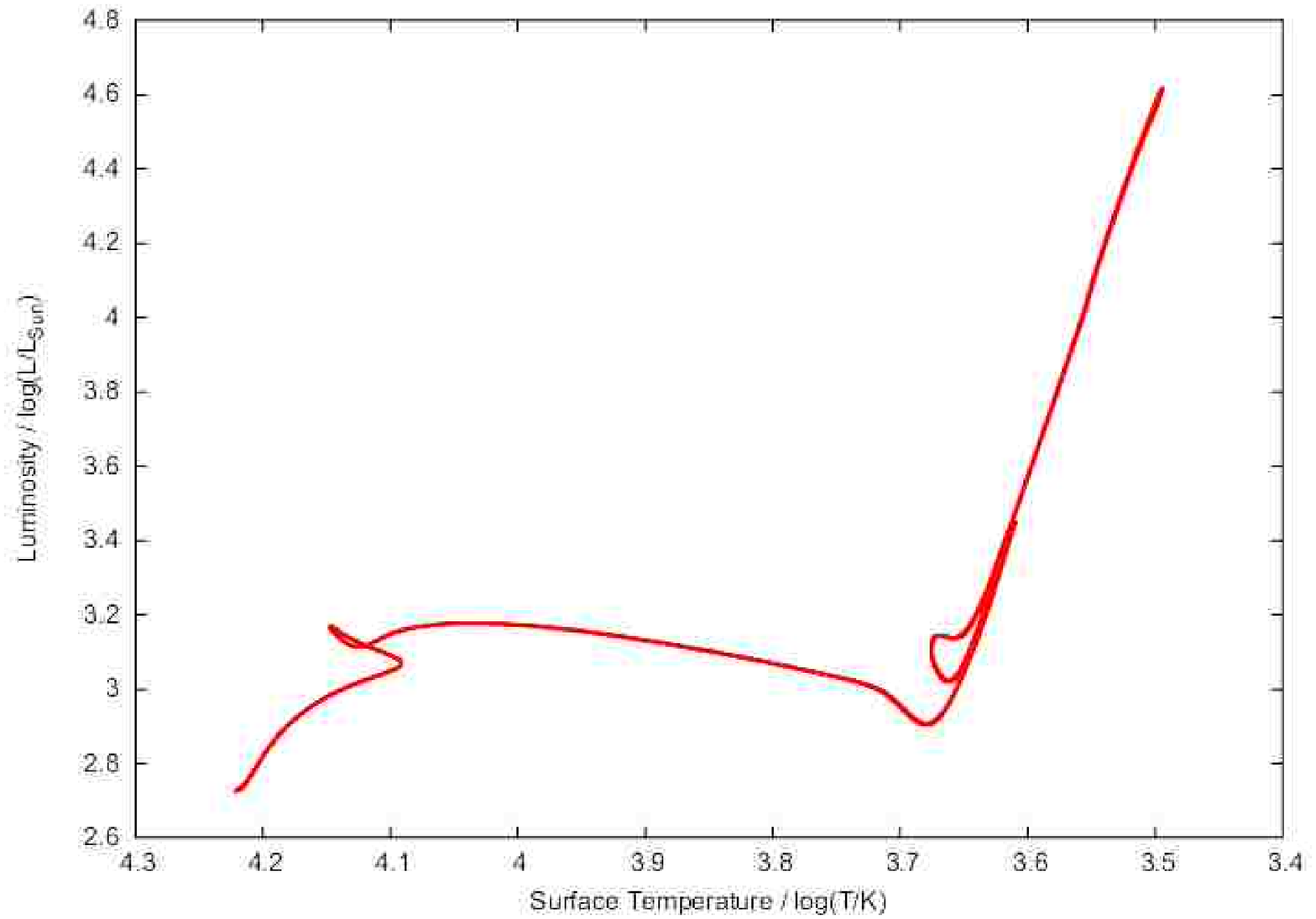}
\caption[HR diagram for an intermediate mass star, an AGB star, $M_{\rm initial}=5M_{\odot}$.]{HR diagram for an intermediate mass star, an AGB star, $M_{\rm initial}=5M_{\odot}$. The star moves up the main sequence along the Hertzspung gap before moving up the giant branch until helium ignites at which point it moves around the blue loop before reascending the giant branch and undergoing thermal pulses at the tip of the branch.}
\label{hrplot2}
\end{center}
\end{figure}

The burning continues until the production of a carbon oxygen (CO) core. The CO core grows as the helium burning shell burns outward. With the end of central helium burning the star re-ascends the giant branch on the asymptotic giant branch, AGB. Second dredge-up, similar to first dredge-up, now occurs and again mixes nuclear processed material to the surface. In the most massive stars the convective envelope can penetrate deeply into the helium core to assist the helium burning shell to catch up with the hydrogen burning shell. The star's luminosity increases as the hydrogen burning shell is now at a higher temperature. Eventually the two burning shells are in close proximity. The arrangement is unstable and thermal pulses occur. The cycle is that the helium burning shell ignites, quickly burning the helium in the thin region up to the hydrogen burning shell. This pushes the hydrogen burning shell into lower temperature regions and the luminosity of the star decreases. The helium burning only occurs for a few decades. Afterwards the hydrogen burning shell reignites and burns for around 10,000 years producing more helium until the helium burning shell reignites and the cycle begins again. The detail of these thermal pulses is extremely complex and uncertain and leads to third dredge-up that mixes nuclear burnt material to the surface at each thermal pulse. Because of this AGB stars are important sources of nucleosynthesis. Eventually they loose their envelopes forming CO white dwarfs.

Towards the maximum mass of this range carbon ignition is possible in the core. This occurs in a shell around the centre in the less degenerate region. Slowly the degeneracy of the inner regions is lifted and the burning continues inwards. Once this brief phase of burning is complete the star becomes a Super-AGB star. It is identical to an AGB star but has an oxygen neon (ONe) core. Then there is an overlap with high mass stars. In this work if the stars lose their envelope and form ONe white dwarfs they are intermediate-mass stars.

In figure \ref{hrplot2} the $5M_{\odot}$ star is an AGB star. We see it evolves on the main sequence. At the end of core hydrogen burning it moves across the Hertzspung gap as its surface cools and then ascends the giant branch. Once helium ignites the star drops down and goes round the blue loop before reascending the giant branch. Eventually at the top of the giant branch the star undergoes thermal pulses and its final position varies slightly. The core grows very slowly during thermal pulses and mass loss by the star's wind will remove the hydrogen envelope to leave a white dwarf as the star's remnant.

\subsection{High-Mass Stars, $M \geq 8M_{\odot}$}

These stars are the progenitors of supernovae and do not form white dwarfs. Their remnants are neutron stars or, in the most extreme cases, black holes. The bottom end of this range merges with intermediate mass stars and they are Super-AGB stars that go supernova. ONe cores can undergo core-collapse by electron capture on to $^{24}$Mg and other isotopes present. With increasing mass second dredge-up does not occur before the end of nuclear burning reactions and burning continues all the way to the formation of iron group elements at the centre of the star. Since iron is the most stable element no further reactions are exothermic and thus the core collapses forming a neutron star or black hole. The release of the energy from the core collapse initiates the supernovae.

\begin{figure}
\begin{center}
\includegraphics[height=129mm,angle=270]{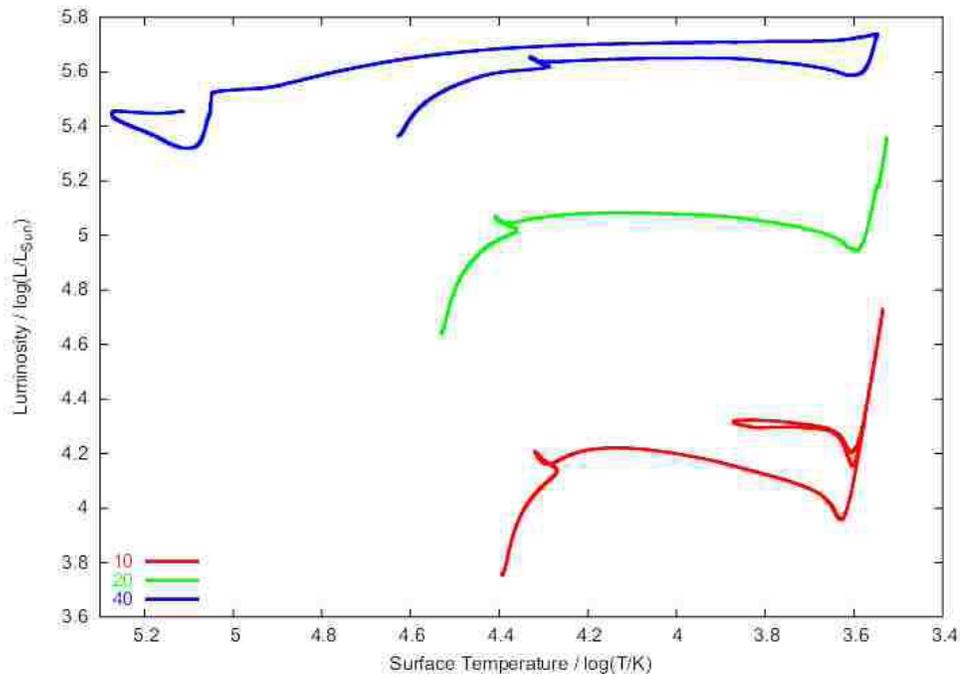}
\caption[HR diagram for three massive stars, $M_{\rm initial}=10,\, 20$ and $40\,M_{\odot}$.]{HR diagram for three massive stars, $M_{\rm initial}=10,\, 20$ and $40\,M_{\odot}$. Evolution is similar to the intermediate mass stars although only the $10M_{\odot}$ goes round the blue loop and they do not undergo second dredge-up to experience the large change in luminosity as they reascend the giant branch. The $40M_{\odot}$ loses most of its mass on the giant branch and moves back across the diagram to the blue region to become a Wolf-Rayet star.}
\label{hrplot}
\end{center}
\end{figure}

The evolution to this end point has many different paths. The essential factors are initial mass and mass loss. These are linked since to a first approximation more massive stars undergo more severe mass loss. The mass loss can be so severe that all hydrogen is removed from the star so that it becomes a naked helium star. If the initial mass is greater than around $20M_{\odot}$ the helium star is in fact a Wolf-Rayet star. These have strong mass loss from their surface and have extremely high surface temperatures, over an order of magnitude greater than the sun. There are a sequence of Wolf-Rayet stars, WN, WC and WO, where the second letter indicates the dominant element in the spectrum, nitrogen, carbon or oxygen. If the mass loss is only mild the final outcome of the star is a red giant.

In figure \ref{hrplot} we show the evolution of 10, 20 and $40M_{\odot}$ stars. We can see the $10M_{\odot}$ star has a similar final luminosity to the AGB star on the HR diagram. The $20M_{\odot}$ star evolves in a similar manner but the final motion up the giant branch is smaller. The $40M_{\odot}$ star goes up the red giant branch and experiences mass loss so severe that it moves back to the blue and becomes a Wolf-Rayet star.

\subsection{Stellar Winds and Mass Loss}
Mass loss modifies the evolution of a star by affecting the surface conditions and the total mass of the star. While theoretical mass-loss rates do exist \citep{KD2002,VKL2001,WRwinds2} most are empirical due to our limited understanding of the mechanism of stellar winds in most cases. Mass-loss rates must be divided into two main categories, those for Wolf-Rayet (WR) stars that have lost their hydrogen envelope, and pre-WR or hydrogen-rich evolution. The nature of mass loss changes once the hydrogen envelope has been removed and observations show quite different stellar winds. WR stars have greater mass loss than a main sequence star of the same luminosity. The mechanism for mass loss in OB stars is radiatively driven winds. The WR star mechanism is not known for certain. It could be driven by radiation, pulsations, magnetic fields or some other uncertain process \citep{WRwinds1,WRwinds2}.

The results of evolution are strongly dependent on the mass-loss prescription. Most observations of the mass loss from stars are at solar metallicity and there are few observations at different metallicities. The commonly adopted scaling with metallicity is that $\dot{M}(Z) = \dot{M}(Z_{\odot}) \times (Z/Z_{\odot})^{0.5}$ due to the work of theoretical models of \citet{K87,K91}. This scaling arises from the assumption that stellar winds are line driven and with lower surface opacity at lower metallicity there are weaker winds. However while there is agreement that mass loss scales in this form there is a range of suggested values for the exponent. It is assumed that this scaling applies at all stages of evolution even though there is some evidence that it varies with spectral type. For example \citet{VKL2001} suggest that for hot stars the exponent should be 0.69. This seems likely because the exponent assumes the winds are line driven while mass loss from red giants is driven by some quite different unestablished mechanism. At very low metallicites where no observations exist we must rely on theoretical mass-loss rates of \citet{KD2002} or extrapolate the empirical rates into this region.

The most recent rates for WR are from \citet{NL00} and these depend on luminosity and surface abundance. They find that the mass-loss rate depends strongly on composition as well as luminosity. This dependence gives lower mass-loss rates than the earlier and widely used rates of \citet{Langer} which link the mass-loss rate to the WR star's mass and the equivalent rates that \citet{NL00} also derive. However it is common to use the rates of Langer but lower them by a factor of two or three to bring them into line with the more recent observations of clumping in WR winds \citep{WRclump1}.

\subsection{Duplicitous Stars}
Single stars are somewhat unusual. Most stars are in binaries or higher order multiple systems and higher mass stars are more likely to have a companion. The presence of a companion gives the opportunity for new evolutionary paths. These are mass loss either by the star slowly expanding until the radius of the star reaches a point where the gravity of the secondary is greater than the primary and matter is transferred or the star could rapidly expand and a common envelope could occur, where the star's core and its companion orbiting in the envelope which is removed by some currently unknown process. The evidence that both occur is that we see systems which can only be explained if mass transfer has occurred. For example in Algol the less massive star is more evolved than the more massive star and this can only be explained if the mass ratio has changed at some point in the past. Planetary nebulae contain a large number of close binaries. We also see helium stars in tight orbits that can only be explained if a common envelope has occurred.

Binary systems also provide opportunities for mass gain. If one star is losing mass it is probable that the other star is gaining this discarded matter. This can increase the mass of the secondary star to the point where it will go SNe when it would not have done so without it. There are other binary effects and the physics is quite uncertain and much has still to be studied in this region even though we currently have some basic models that fit observations reasonably well. In any study of stars an understanding of binary stars is of fundamental importance.

\section{Numerical Astronomy}
\begin{center}
``Computers make efficient and practical servants, but I do not wish to serve under them.'' \\
\textit{Spock, Star Trek, The Ultimate Computer.}
\end{center}
Whenever there is something to understand it is important to make a model. By changing the parameters of a model we can see how the output (of our model) changes and thus gain insight into the physical object. The modelling of stars started with Lord Kelvin's calculations for the lifetime of the Sun if it was powered by gravitational collapse. However theoretical stellar evolution started with Schwarzschild, Henyey and others using physical laws to create increasingly sophisticated stellar models. The process invented was similar to that used now, the solution of differential equations by numerical rather than analytical processes.

The invention of the computer during the Second World War to break codes is probably the most important invention that has shaped modern society. Later on the invention of the silicon transistor and affordable home computers pioneered by Sir Clive Sinclair accelerated the impact and about twenty years later provided a source of research students with great experience with computers. The first to apply computers to help our study of stellar evolution were \citet{henyey}. While then computer time was limited and at a premium, we now discard it, wasting it to show pretty screensavers. There is therefore vast CPU time now available. We are now able to run multiple models with different parameters to gain further understanding.

This dissertation is the result of 70 years of CPU time, that is the equivalent of one computer calculating for 70 years. This has been achieved by use of multiple computers over the shorter time of 3 years. However it is important that even a gargantuan amount of data is not in itself useful. Computers are incredible idiots. They do exactly what we tell them to do.

\section{Supernovae}
\begin{center}
``I've never seen a supernova blow up but if it's anything like my old Chevey Nova\\ it'll light up the night sky!''\\
\textit{Fry, Futurama, Roswell that Ends Well.}
\end{center}
Supernovae (SNe) have most probably been observed since the dawn of astronomy. The earliest known surviving record of a supernova was made in the eleventh century by ancient Chinese astronomers (see Stephenson \& Green 2002 for the details of these early observations) but the modern study of SNe began with Baade and Zwicky in the 1930s when they realised SNe are more luminous and rarer than the more common novae. The high luminosities and broad spectral lines led them to conclude that SNe are extreme explosions, the result of a dying star collapsing to a neutron star. This basic model still holds. Sixty-five~years of work has vastly increased our understanding but many questions about SNe remain unanswered.

A core-collapse SN occurs once a star has a core (usually of iron group elements) that cannot be supported by any further nuclear fusion reactions or electron degeneracy pressure. This leads to the core collapsing to a neutron star or black hole. In the process it releases a large flux of neutrinos that interact weakly with the surrounding stellar envelope depositing a large amount of energy ($10^{44} {\rm J}$) that can drive the star to explode. In massive cores supported by electron degeneracy pressure core-collapse can be initiated by electron capture on to nuclei.

Important questions are, `Which stars give rise to which SNe and leave which type of compact remnant?' The answers provide a test of stellar evolution theory, give information for galactic chemical evolution and predict SNe birth rates and remnant populations. To solve these questions we must use our understanding of stellar evolution to model stars up to the point at which they become SNe. Three main factors affect the answer, the initial mass and metallicity of the star and mass-loss (or mass-gain in some binary systems) during its lifetime. \citet{H03}, using the models described in \citet{WHW02}, \citet{vanb03} and \citet{podsibin1} are example of studies of the effect of binaries. While their studies are informative each only investigates a single mass-loss scheme, does not investigate the lowest mass for a star to become a SN nor describes the SN progenitors and does not fully explore the factor of metallicity. These factors limit comparison with observations. We expand upon this work and add details on the progenitors of SNe. These details are important for surveys of progenitors now underway by various groups, e.g. \citet{S03a} and \citet{VD03}.

There are two main types of SNe, those without hydrogen in their spectra, type I, and those with hydrogen in their spectra, type II. Type I progenitors have lost their hydrogen envelopes and are white dwarfs or WR stars, while type II progenitors have retained their hydrogen envelopes. This gives us a basic method to discriminate between them.

\subsection{Type II}
A star gives rise to a type II SNe if there is any hydrogen in its envelope. If enough of the original hydrogen is retained then it gives rise to a type IIP supernova. The light curve plateau phase is powered by a moving hydrogen ionisation front. Type IIL SNe only retain a small fraction of their hydrogen and the light curve decays linearly powered by the decay of $^{56}$Ni. There are a few other subtypes, IIb are similar to IIL but are thought to have less hydrogen and begin looking like a type-II SN but eventually look like a type-Ib SN. Type IIn have narrow lines in their spectra indicating slow expansion speeds. They are thought to be the result of interaction with a dense circumstellar environment. Type IIpec are where all the other SNe go if they don't fit into one of the above categories.

\subsection{Type I}
Type I SNe are divided into types Ia, Ib and Ic. SNe Ia are extremely bright events thought to be the explosive carbon burning of a degenerate white dwarf that has reached the Chandrasekhar mass ($M_{\rm Ch}$) by accretion; this is not a core collapse event. However types Ib and Ic are core-collapse SNe in helium stars that have lost their hydrogen envelopes in a wind or by binary interaction \citep{EW88}. They are discriminated by the presence or absence of helium lines in the spectrum. However the differentiation between type Ib or Ic from limited observations can sometimes be difficult so it is common to label them together as Ibc. The largest progenitors are thought to give no display because the core is so massive that even with a large explosion energy nothing escapes the forming black hole. The exception would be if a jet driven SNe occurred that makes a black hole and produces an observable display \citep{MWH01}.

\subsection{Hypernovae and Gamma-Ray Bursts}
Some supernovae such as SN2003dh are extremely energetic with around ten times the energy of normal supernovae and are called hypernovae (HNe). Gamma-ray bursts (GRBs) are objects where a tremendous burst of gamma-rays is observed over a few seconds outputting a similar amount of energy. It is thought these events are supernovae and all similar events of the same process but in slightly different progenitors. This is confirmed by observations of GRBs associated with SNe such as SN1998bw/GRB980425 and SN2003dh/GRB030329. It is also possible that hypernovae are the results of failed GRBs according to the models of \citet{MW99}. The relationship between SNe, HNe, GRBs and their relative rates is quite uncertain and much remains to be confirmed. The current situation is discussed by \citet{podsi04}.

\section{Dead Stars}
\begin{center}
``I always wanna feel this way,\\
Just like a Phoenix from the Flames.''\\
\textit{Robbie Williams, Phoenix from the Flames.}
\end{center}

A star dies when it can no longer produce energy from nuclear fusion reactions. Low and intermediate-mass stars lose their envelopes leaving a cooling core while massive stars undergo supernovae. These events produce different stellar corpses. These dead stars are normally less luminous than stars that are producing energy by nuclear fusion. However the remnants are not necessarily dead objects but can be reborn to shine once more. If they have a binary companion that overfills its Roche Lobe it is possible for these remnants to accrete matter. The more compact the object the more extreme the conditions achieved in the accretion flow. As matter spirals towards the surface it can achieve high velocities and become heated to a high temperature so radiation is emitted. Neutrons stars and black holes when they accrete material from a companion are extreme X-ray emitting objects and many such binaries are observed. These spectacular objects were discovered by the first X-ray telescopes and provide direct evidence for the existence of these remnants.

\begin{figure}
\begin{center}
\includegraphics[height=130mm,angle=270]{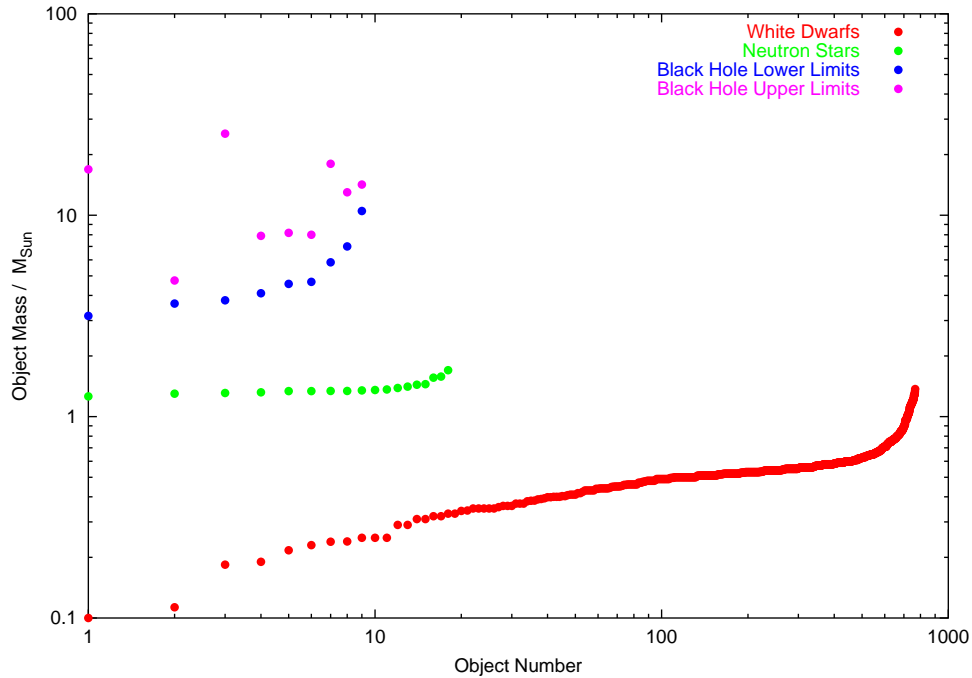}
\caption[Measured masses for compact remnants.]{Measured masses for compact remnants. Object number is the indentifier used for each object ranked in order of lowest to highest mass.}
\label{remnants}
\end{center}
\end{figure}

We show in figure \ref{remnants} the number of objects known in each of the remnant classes and their masses. We can see there is some overlap between these objects and the uncertainty is greatest in black hole masses. It is not surprising that most known remnants are white dwarfs since these should be more common as they come from low mass stars and they are more populous because the initial mass function makes low-mass stars more probable than higher-mass stars.

\subsection{White Dwarfs}

These are essentially the cores of low and intermediate mass stars once no further nuclear reactions are possible. They are supported by electron degeneracy pressure and calculations show that the maximum mass of a star supported by degeneracy pressure is the Chandrasekar mass, $M_{\rm Ch}$, which is around $1.4M_{\odot}$ although the exact value varies very slightly with composition. They usually have a thin layer of hydrogen on their surface, the remains of the envelope, and cool slowly over time. They are roughly the size of the Earth, however the more massive the star the smaller the radius so that as the mass approaches $M_{\rm Ch}$ the size shrinks to around 10km.

White dwarfs come in three flavours depending on how far the nuclear reactions in the progenitor's core have progressed. They can be made of helium (a HeWD), a mixture of carbon and oxygen (a COWD) or a mixture of oxygen and neon with some sodium and magnesium (an ONeWD). HeWDs are formed by very low mass stars or more probably in binary stars and are not of interest in this work. COWDs are the remnant cores of stars that lose their envelopes on the AGB. The Super-AGB stars develop an ONe core and are the source of ONeWD.

\subsection{Neutron Stars}
\begin{center}
``neutronium, but...'' \\
\textit{Scott, Star Trek, A Piece of the Action.}
\end{center}
Neutrons are unstable and free neutrons decay to a proton and electron with a half-life of around 8 minutes. They are however stable in a nucleus or neutron star which is effectively one giant nucleus held together and made stable by the gravitational binding energy. They are around 10km in radius and there are multiple models for their internal structure. It is thought that the innards might consist of a quark gluon mix rather than actual neutrons.

Whatever their structure they are supported by neutron degeneracy pressure that is similar to electron degeneracy pressure but since a neutron's mass is roughly 2000 times that of an electron a greater mass can be supported. The theoretical maximum mass for a neutron star is $3M_{\odot}$, above which the star collapses to a black hole. Neutron stars were initially believed to be hypothetical but the detection of pulsars and their interpretation as rapidly rotating neutron stars is the primary evidence of their existence. Their association with supernovae is evidenced by the presence of a pulsar in the Crab nebula. The nebula is the other type of supernovae remnant commonly referred to, that is the expanding ejecta from the explosion rather than the compact object. 

Neutron stars are formed by one of two mechanisms. In the lowest-mass stars to go supernova nuclear burning progresses as far as carbon burning to form an ONe core. If this core has a mass equal to $M_{\rm Ch}$ it collapses and once the central density reaches around $10^{9.8} {\rm g \, cm^{-3}}$ electrons are captured on to the nuclei present, most notably $^{24}$Mg. The removal of electrons removes the electron degeneracy pressure and the collapse is hastened until a neutron star is formed. The release of neutrinos in the formation of the neutron star is thought to lead to the transfer of energy to the envelope causing it to be ejected.

In more massive stars burning continues until a core of iron group elements is formed. Iron, being the most stable element, is the end of possible nuclear burning. Endothermic reactions occur including the photodisintegration of iron and others that lead to the core collapsing yet again to a neutron star and again it is thought the release of neutrinos leads to the ejection of the envelope and the supernovae. After the supernova and the envelope has been ejected the neutron star is left alone.

\subsection{Black Holes}

\begin{center}
\begin{tabular}{lc}
{  \textbf{Leela:}}&{  ``Oh my god we're heading straight for a black hole!''}\\
{  \textbf{Fry:}}&{  ``Talk about a mood killer.'' }\\ 
& \textit{  Futurama, A Flight to Remember.}\\
\end{tabular}
\end{center}

Determination of the remnant formed at the heart of a SN is a black art. The physics is extreme and poorly understood. However there are many prescriptions which rely on the conjecture that a more massive core leads to a more massive remnant. As the remnants become more massive neutron degeneracy pressure cannot support the collapsing core and a black hole is formed. The physics of black holes is deceptively simple, their escape velocity is the speed of light therefore not even light can escape from these objects, hence the name black holes. However they aren't really that black. Single black holes produce (a weak) Hawking radiation. Black holes in a binary however may accrete material from their companion. This leads to an accretion disc that gets very hot and therefore very bright at X-ray wavelengths. Many such binaries are observed and there are low-mass and high-mass X-ray binaries named after the mass of the donor star. These objects can only be formed in supernovae or by accretion on to neutron stars that then subsequently collapse.

The minimum mass for a black hole is classically $3M_{\odot}$ however it seems that they can form from remnants of only $2M_{\odot}$. The mass distribution of black holes is also interesting. We can only find black holes if they are accreting and hence luminous sources. So an important question is: what is the initial black hole mass function? Can we compare this to the observed population and work out whether most black holes are at the masses they were born with or have they accreted a substantial amount of material?

\section{Dissertation Outline}

\begin{center}
\begin{tabular}{lc}
{  \textbf{Data:}}&{   ``Do you think this is a wise course of action, sir?''}\\
{  \textbf{Picard:}}&{  ``We're about to find out, Data.''}\\ 
& \textit{ Star Trek: Nemesis.}\\
\end{tabular}
\end{center}
 
The rest of this dissertation is split into six chapters. In chapter 2 we discuss the stellar evolution code that is used to produce our models. We also detail the opacity tables that were constructed to enable the study that has been undertaken. In chapter 3 contains a detailed look at the structure of SNe progenitors, trying to estimate from their properties which SNe type they might give rise to. We use observations of SNe and their progenitors to help us reach our conclusions. Also in this chapter we examine the lowest mass of star to undergo supernova. This is a complicated area and the results greatly depend upon the physics used in our models. Chapter 4 contains a study into the various mass-loss prescriptions that are used and the effect they have on the progenitors of supernovae. This is split into three main sections: first we investigate the effect of including metallicity scaling for Wolf-Rayet stars; second we compare different mass-loss prescriptions at high metallicities ($0.001 \leq Z \leq 0.05$) and finally we compare different mass-loss prescriptions at low metallicities ($10^{-8} \leq Z \leq 0.001$). From this we choose a preferred scheme that best fits observations. 

Chapter 5 extends our study to include binary stars to search for low-luminosity progenitors and progenitors that have a different structure that we do not see from single stars. Chapter 6 combines our preferred results for all single star progenitors and discusses the implications of the observed progenitors to date for our models. We then combine the results from single stars with those from binaries so we can draw some conclusions about the relative fraction of single to binary stars and how closely we can match the observations that exist. We also discuss the distribution of progenitors on the HR diagrams indicating where single stars sit and discuss the relative position of binary stars. Finally we detail the remnant mass population and compare this to the limited observations that exist. In the conclusion, chapter 7, we summarise our results and list our final judgement on the progenitors of the SN types.

\chapter{Nuclear Reactions \& Opacities}
\begin{center}
``I  can't  change  the  laws of physics;  I've got to have  30 minutes.''\\
\textit{Scotty, Star Trek, The Naked Time.}
\end{center}

\section{The Cambridge Stars Code}
The Cambridge stellar evolution code was originally written by Peter Eggleton during his time as a PhD student; the first paper on this code was \citet{E71}. Since this time the code has been used, revised and adjusted by many people, in fact there are now at least four different stellar evolution codes that are based on the original. Because of this it is necessary to describe the features of the code used in this work. The full list of papers that detail the code are \citet{E71}, \citet{E72}, \citet{EFF73}, \citet{E73}, \citet{han94}, \citet{P95}, \citet{P97} and \citet{P98}.

The code is lightweight being less than 2000 lines of Fortran 77. It is simple to operate and to modify. This makes adding extra physics or routines to study wide ranging problems in stellar astrophysics easy and rapid. The code also runs remarkably fast on modern computers. Some time comparisons are detailed in appendix \ref{timex}. Stars can be evolved from main sequence to carbon burning in around 5 minutes. Therefore it is possible to produce large grids of stars with different masses, metallicities, mass-loss schemes, mixing schemes and varying other physics to study the effect on a star's evolution.

The essential features that make it unique among stellar evolution codes have survived its long life span. They are the use of a self-adaptive non-Lagrangian mesh, the treatment of both convective and semiconvective mixing as a diffusion process and the simultaneous and implicit solution of both the stellar structure equations and the diffusion equations for chemical composition.

The fact the Eggleton code solves for the structure, composition and the adaptive mesh simultaneously, an extremely elegant way to calculate the evolution of stars, is responsible for making the system so simple and adaptable. The calculations are fully consistent and remove possible errors when rapid composition changes might effect the structure. It also means little or no intervention is required to evolve a star.

In this chapter, after discussing the code, we shall detail two of the modifications we have made. First is the addition of more nuclear reactions and nuclei to the reaction network and second, to update and improve the opacity tables. Other additions, such as inclusion of various mass-loss rates and a new routine for following binary evolution, will be presented in later chapters.

\subsection{Assumptions}
As with any physical model there are always assumptions that have to be made to reduce the problem to a level where it might be realistically tackled. It is best to think that over time we refine our physical models and come closer to the truth of nature but that the truth is something we can never reach.

Nearly all stellar evolution packages share the approximation of stars as perfect spheres. This reduces the problem to one dimension with the structure varying in radius only. However real stars are not spherical and effects such as rotation and duplicity distort the structure. It is unlikely to be important however because, in the majority of cases, current models do fit observations. But there are some interesting cases that require the need to invoke aspherity \citep{b7b}. However including extra dimensions increases the complexity of the problem and its solution. For example, for a 1D star only 199 meshpoints are required for most simple stages of evolution. This increases with more dimensions and thus requires more computing power. The only known fully 3D code at the moment is Djehuty \citep{djehuty} although it is much slower than simpler 1D codes.

The other two primary assumptions are that the star is always in local thermodynamic equilibrium and in hydrostatic equilibrium. The first assumption does break down in the atmosphere of Wolf-Rayet stars. However this has little effect on the interior hydrostatic core. The second assumption is broken when dynamical effects occur, for example during rapid mass loss ($\dot{M}>0.01M_{\odot} {\rm yr}^{-1}$), burning in degenerate regions and rapid burning stages such as oxygen burning when nuclear reactions occur on a dynamical timescale. However these occur rarely in stellar evolution and we shall note when these features could effect the results.

\subsection{Equations}
The basic equations of stellar structure are taught in undergraduate courses and found in most astrophysics text books. Solving the coupled equations is not easy. Analytically assumptions must be made such as uniform composition, constant opacity or a simple equation of state. Calculation of realistic stellar structures is a demanding process. Let us start by listing the equations of stellar structure. They are derived in terms of radius which is not practical for calculating the evolution of stars due to contrasting length scales for the core and envelope. It is therefore common practice to recast the equations with mass as the variable over which the equations are solved. The equations are conservation of mass,
\begin{equation}
\frac{{\rm d} m}{{\rm d} r} = 4 \pi \rho r^{2}
\qquad {\rm or} \qquad
\frac{{\rm d} r}{{\rm d}m} = \frac{1}{4 \pi \rho r^{2}},
\end{equation}
hydrostatic equilibrium,
\begin{equation}
\frac{{\rm d}p}{{\rm d}r} = -\frac{G \rho m}{r^{2}}
\qquad {\rm or} \qquad
\frac{{\rm d} p}{{\rm d}m} = -\frac{Gm}{4 \pi r^{4}},
\end{equation}
energy production,
\begin{equation}
\frac{{\rm d} L_{r}}{{\rm d}r} = 4 \pi \rho r^{2} \epsilon
\qquad {\rm or} \qquad
\frac{{\rm d} L}{{\rm d}m} = \epsilon,
\end{equation}
radiative transport,
\begin{equation}
\label{radi}
\frac{{\rm d}T}{{\rm d}r} = - \frac{3}{16 \sigma} \frac{\kappa \rho}{T^{3}} \frac{L_{r}}{4 \pi r^{2}}
\qquad {\rm or} \qquad
\frac{{\rm d}T}{{\rm d}m} = - \frac{3}{16 \sigma} \frac{\kappa }{T^{3}} \frac{L_{r}}{16 \pi^{2} r^{4}},
\end{equation}
or convective transport,
\begin{equation}
\label{conv}
\frac{{\rm d}T}{{\rm d}r} = \frac{\gamma - 1}{\gamma} \frac{T}{p} \frac{{\rm d}p}{{\rm d}r}.
\qquad {\rm or} \qquad
\frac{{\rm d}T}{{\rm d}m} = - \frac{\gamma - 1}{\gamma} \frac{T}{p} \frac{Gm}{4 \pi r^{4}}.
\end{equation}
Also required is an equation of state. A simple example is
\begin{equation}
P=\frac{R \rho T}{\mu} + \sigma T^{4}.
\end{equation}
The variable $r$ is the radius of a sphere centred on the centre of the star, $m$ is the mass contained within that sphere, $\rho$ is the density at $r$, $p$ is the pressure, $L_{r}$ is the luminosity through the sphere, $\epsilon$ is the energy produced per unit mass of material, $T$ is the temperature, $\sigma$ is the Stefan-Boltzman constant and $\gamma$ is the adiabatic exponent.

The driving forces behind the evolution of stars are the nuclear reactions that alter the composition. Therefore we need equations that evolve the composition change due to nuclear reactions and mixing by convection or other processes.
\begin{equation}
\frac{\partial}{\partial m} \Big( \Sigma \frac{\partial X_{i}}{\partial m} \Big) = \frac{\partial X_{i}}{\partial t} + X_{i} R_{{\rm d},i} - R_{{\rm c},i}
\end{equation}
Where $X_{i}$ is the mass fraction of the element in question, $\Sigma$ is the diffusion constant or rate of mixing, $X_{i} R_{d,i}$ is the rate at which $X_{i}$ is destroyed by nuclear reactions and $R_{c,i}$ is the rate at which $X_{i}$ is created by nuclear reactions.

These coupled differential equations have no (known) analytic solution. To solve them we turn to numerical methods. Originally this meant sitting down with a slide rule and working as quickly as humanly possible becoming an original computer. Then, thanks to the invention of the electronic brains to crack codes in World War 2, the process became faster and reliable.

The process requires the differential equations to be recast as difference equations and the problem is solved over a finite mesh. The resolution of this mesh needs to be high enough that errors do not creep in when a section is under-resolved but low enough so as not to compromise calculation speed.

The difference equations are mass conservation
\begin{equation}
r^{3}_{k+1} - r^{3}_{k} = \Big( \frac{3}{4 \pi \rho} \frac{\partial m}{\partial k} \Big)_{k+\frac{1}{2}},
\end{equation}
hydrostatic equilibrium
\begin{equation}
\log P_{k+1} - \log P_{k} = - \Big( \frac{Gm}{4 \pi r^{4} P} \frac{\partial m}{\partial k} \Big)_{k+\frac{1}{2}},
\end{equation}
energy transport,
\begin{equation}
\log T_{k+1} - \log T_{k} = - \Big( \nabla \frac{Gm}{4 \pi r^{4} P} \frac{\partial m}{\partial k} \Big)_{k+\frac{1}{2}},
\end{equation}
where $\nabla=\frac{{\rm d} \log T}{{\rm d} \log P}$ and is calculated from radiative or convective theory in their respective regions. Then we have the equation for energy production,
\begin{equation}
L_{k+1} - L_{k} = \Bigg( \Big( \epsilon_{nuc}+\epsilon_{\nu}-T\frac{\partial S}{\partial t}+C_{P} T \frac{Gm}{4 \pi r^{4} P} (\nabla - \nabla_{a}) \frac{\partial m}{\partial t} \Big) \frac{\partial m}{\partial k} \Bigg)_{k+\frac{1}{2}},
\end{equation}
we are also required to solve for the mesh,
\begin{equation}
m_{k+1} - m_{k} = \Big( \frac{\partial m}{\partial k} \Big)_{k+\frac{1}{2}}
\end{equation}
and the composition,
\begin{eqnarray}
\Sigma_{k+\frac{1}{2}}(X_{i,k+1}-X_{i,k})-\Sigma_{k-\frac{1}{2}}(X_{i,k}-X_{i,k-1}) = \Big(\frac{\partial X_{i,k}}{\partial t} +R_{i,k} \Big) \Big(\frac{\partial m}{\partial k} \Big)_{k} \nonumber \\
- (X_{i,k+1} - X_{i,k}) \Big[ \frac{\partial m}{\partial t} \Big]_{k}+(X_{i,k}-X_{i,k-1}) \Big[ -\frac{\partial m}{\partial t} \Big]_{k+1}.
\end{eqnarray}
Where $\Sigma=K[\nabla_{r}-\nabla_{a}]^{2}M^{2}/t_{\rm nuc}$, $\nabla_{r} = \Big( \frac{{\rm d} \ln T}{{\rm d} \ln p} \Big)_{\rm rad \, eqm} = \frac{3}{16 \pi a c G} \frac{P\kappa}{T^{4}} \frac{L}{m}$and $\nabla_{a} = \Big( \frac{\partial \ln T}{\partial \ln p} \Big)_{S}$. Values in square brackets are set to zero if their value is negative, $t_{\rm nuc}$ is the current nuclear timescale, $M$ the star's current mass and $K$ is a suitable constant chosen as discussed in \citet{E72} to achieve mixing on a timescale similar to that from Mixing-length theory. Each of these equations is evaluated at points from the centre of the star to the surface. The location of these discrete points forms a mesh, the distribution of which depends on the mesh spacing function.

\subsection{The Mesh and its Spacing Function}

The mesh is dependent on the value of a variable, $Q$, which is a function of physical variables calculated for the star. Those used are the pressure, mass, temperature and radius. Mesh points are then placed with equal spacing in Q so that $\frac{{\rm d}Q}{{\rm d}k}$ is constant over the mesh at each timestep. The present version of the non-Lagrangian mesh-spacing is a function of local pressure $P(r)$, temperature $T(r)$, Lagrangian mass $m(r)$ and radius $r$. A stellar model is well represented by 199 meshpoints although in certain cases where the evolution is complex such as thermally pulsing AGB stars more are required. If an increase in resolution is required meshpoints cannot be added by hand. It requires either raising the total number of meshpoints in the mesh or more commonly refining the mesh-spacing function to move meshpoints into the troublesome region. The most common way to do this is to use the pressure at the point of interest because it is a monotonically decreasing variable from centre to the surface.

To make the process automatic the pressure at the burning shells is found for each model and input into the $Q$ equation,
\begin{eqnarray}
Q = \alpha_{1} \ln P + \alpha_{2} \ln(P+P_{\rm H}) + \alpha_{3} (\ln(P+P_{\rm He}) + \ln(P+P_{\rm C})))  \nonumber \\
+\ln \Big( \frac{(m/M)^{\frac{2}{3}}}{\alpha_{4}}+1 \Big) + \alpha_{5} \ln \Big( \frac{r^{2}}{c_{3}} +1 \Big) -\alpha_{6} \ln \Big( \frac{T}{T+\alpha_{7}} \Big)
\end{eqnarray}
Where $\alpha_{i}$ are constants that can be altered to increase or decrease the influence of each term on the mesh-spacing. The different subscripts to the pressures represent the pressures at that burning shell which are found by looking for a large composition gradient in the element being burnt. These pressure terms produce great changes in the value of Q when the two pressures are near equal thus many mesh points are put at these points. There are other ways of forming these equations to push mesh points between burning shells. However in this case with multiple burning shells we need to resolve these regions more accurately than the space between them. The counter example would be in thermally pulsing AGB stars where it is important to resolve the intershell convection zones accurately.

\subsection{Boundary conditions}

To solve any differential, or difference, equation boundary conditions are required. The conditions are applied at the surface and centre of the star. At the surface, where $k=1$, the boundary conditions are for the conservation of mass
\begin{equation}
\frac{{\rm d}m}{{\rm d}t} = \dot{M}_{\rm wind} + \dot{M}_{\rm binary},
\end{equation}
emission from a black body
\begin{equation}
L_{r}=\pi R^{2} \sigma_{\rm B} T_{\rm eff}^{4},
\end{equation}
the point at which photons can escape to infinity
\begin{equation}
P_{\rm gas}=\frac{2}{3} \frac{g}{\kappa} \Big(1-\frac{L}{L_{\rm Edd}} \Big).
\label{gasequation}
\end{equation}
We take the surface of the star to be the photosphere when the optical depth is $\frac{2}{3}$. The boundary conditions take account of radiation pressure to prevent super Eddington luminosities. In taking the outer mesh point to be the photosphere we assume in equation \ref{gasequation} that the gravity, $g$, is due to the mass of the star and therefore the photosphere has negligible mass. This is a valid assumption for most stars, any discrepancy will be greatest for red giants.

The simplest central boundary conditions would be $L=r=m=0$ treating the centre as an infinitesimal point. Currently the central point is modelled as a spherical region. Therefore the condition at the centre where $k=199$ are for mass conservation
\begin{equation}
m=-\frac{{\rm d}m}{{\rm d}k},
\end{equation}
the radius,
\begin{equation}
m=\frac{4}{3}\pi r^{3} \rho,
\end{equation}
and the central luminosity
\begin{equation}
L=(\epsilon - T \dot{S})m.
\end{equation}
With the equations and central boundary conditions it is possible with a suitable number of variables to solve the problem of stellar structure and evolution.

\subsection{Solution}
In the above equations we require 14 variables to model the stellar structure. The first eleven, already within the code, were $\ln T$, $\ln m$, $\frac{{\rm d}Q}{{\rm d}m}$, $\ln r$, $L$, $X({\rm ^{1}H})$, $X({\rm ^{4}He})$, $X({\rm ^{12}C})$, $X({\rm ^{16}O})$, $X({\rm^{20}Ne})$ and $\ln f$, a quantity related to electron degeneracy. To increase the composition network extra variables were added $X({\rm ^{14}N})$, $X({\rm ^{24}Mg})$ and $X({\rm ^{28}Si})$. Nitrogen was included to resolve problems with AGB and low-metallicity stars in the CNO cycle, while magnesium and silicon were included to follow later stages of nuclear burning.

The solution method is based on the Henyey technique \citep{henyey}. An initial model is input, usually from a previous calculation. If the model does not satisfy the equations the variables are relaxed to a solution, hence the alternate name for this technique, the relaxation method. The relaxation method iteratively calculates how to improve the values of the variables to achieve a better solution to the difference equations in a fully implicit manner. The iterations repeat until a certain level of accuracy in the solution is reached. Otherwise if a maximum number of iterations is reached the method is aborted since either it has insufficient resolution or the initial guess solution was too poor.

The most important features of making sure the code converges to a solution are timestep and mesh control; effectively the resolution in time and space. Unfortunately owing to the nonlinearity of the partial differential equations this is not yet an exact science and something of a black art. To determine the next timestep the code compares the changes in structure from the last timestep to a constant that determines the maximum change required between models. If a large time step is allowed then the code can evolve a star with few timesteps although accuracy is reduced. Using a smaller timestep can also produce problems but the main drawback is that more models are required for the same period of evolution.

By experimenting with time step control the number of breakdowns of the code can be reduced. This is necessary when computing large numbers of models. The main points of breakdown are ignition of nuclear burning in a highly degenerate region, burning shells close to the surface of a star, extreme mass-loss rates (greater than $10^{-3} M_{\odot} {\rm yr^{-1}}$) and short timescale nuclear burning reactions (i.e. Neon and Oxygen burning) requiring a low timestep (less than $10^{-4}{\rm yr}$).

\subsection{Convection and Mixing}
The treatment of convection was first described by \citet{E72}. The mixing is modelled as a diffusion process so we can correctly follow mixing due to convection and equilibrium semiconvection. The composition changes are calculated at the same time as the structure equations making this code unique. Convection also effects heat transport and must be considered carefully. To decide whether a meshpoint is stable to convection we must calculate two temperature gradients, the adiabatic gradient $\nabla_{{\rm a}}=\Big( \frac{{\rm d}\log T}{{\rm d}\log p} \Big)_{S}$ and the radiative gradient $\nabla_{{\rm r}}=\Big( \frac{{\rm d}\log T}{{\rm d}\log p} \Big)_{\rm radiative}$. These are related to equations \ref{radi} and \ref{conv} respectively.

We then use the classical Schwarzschild criterion for instability, normally expressed as
\begin{equation}
\nabla_{{\rm r}} > \nabla_{{\rm a}},
\end{equation}
for convective instability to occur. Otherwise the region is stable and heat transport is by radiation and no mixing occurs ($\Sigma=0$). If the system is convective then mixing is treated as a diffusion process with $\Sigma=(K(\nabla_{{\rm r}} - \nabla_{{\rm a}})^{2} M^{2}) / \tau_{\rm nuc}$ where $K$ is a large constant calculated from mixing length theory, $M$ the stellar mass and $\tau_{\rm nuc}$ the current nuclear timescale. Where $\nabla_{{\rm r}} - \nabla_{{\rm a}}$ is small but positive a semiconvective region occurs. Heat transport is as calculated from mixing length theory with the length over which mixing occurs a free parameter in terms of the pressure scale height $H_{p}$. The mixing length is set to $l= \alpha H_{p}$ as in the formalism of \citet{BV58}. We take $\alpha=2.0$ which gives a good fit to the Sun \citep{P97}.

However there is some evidence that extra mixing beyond that given by standard 1D convection is required in some cases as discussed by \citet{P97} and \citet{b7b}. Extra mixing can be included in the form of convective overshooting. We do not parameterise the overshoot by a fixed fraction of the pressure scale height. Instead we alter the classical Schwarzschild criterion above to
\begin{equation}
\nabla_{r} > \nabla_{a} - \delta,
\end{equation}
where
\begin{equation}
\delta = \frac{\delta_{{\rm ov}}}{2.5+20\zeta+16\zeta^{2}} 
\end{equation}
and $\zeta$ is the ratio of radiation to gas pressure and $\delta_{{\rm ov}}$ is the overshooting parameter that we vary to include convective overshooting or to remove it. Further details of the inclusion of this scheme in the Eggleton code can be found in \citet{SPE97}. By comparing models to observations of eclipsing binaries with an evolved component they found a best fit for $\delta_{{\rm ov}}$ to be $0.12$. This leads to overshooting lengths $l_{{\rm ov}}$ between $0.25$ and $0.32 H_{P}$ in stars of mass range $2.5$ to $6.5 M_{\odot}$. For more massive stars the overshooting length remains around $0.3 H_{P}$. Another way of quantifying the effect of overshooting is to measure the mass of the convective core of a ZAMS model. For the $6.5 M_{\odot}$ star the mass over which mixing occurs grows from $1.6$ to $2.1 M_{\odot}$. For a $10M_{\odot}$ star the growth is from $3.2$ to $4.1 M_{\odot}$ and for a $100M_{\odot}$ star from $83M_{\odot}$ to $87M_{\odot}$.

In our calculations we use $\delta_{{\rm ov}}=0.12$ to include convective overshooting and set $\delta_{{\rm ov}}=0$ to remove overshooting. The amount of overshooting affects the core mass during evolution. This affects the nature of the remnants and also the mass of stars that undergo second-dredge up. There is evidence from asteroseismology observations that overshooting does occur in massive stars \citep{aertsos}. However the extra mixing owing to convective overshooting in our models can be considered to mimic any mixing process such as rotation or gravity wave mixing, not just convective overshooting.


\subsection{Equation of State}
The equation of state (EoS) is described by \citet{EFF73} and \citet{P95}. The EoS is necessary to describe the relationship between pressure, density and temperature over the vast range of regimes that stars encounter. There are a number of ways of including the equation of state in a stellar evolution code. One is to interpolate in tabulated data. Within this code the EoS is approximated by ingenious formulae to an accuracy of $<0.1\%$. They provide the details of electron density, pressure, internal energy, pressure ionisation and Coulomb interactions.

\subsection{GridCheck the Babysitter}

\begin{center}
``I think there is a world market for maybe five computers.''\\
\textit{Thomas Watson (1874-1956), Chairman of IBM, 1943.} \\
\end{center}

\begin{figure}
\begin{center}
\includegraphics[height=160mm,angle=0]{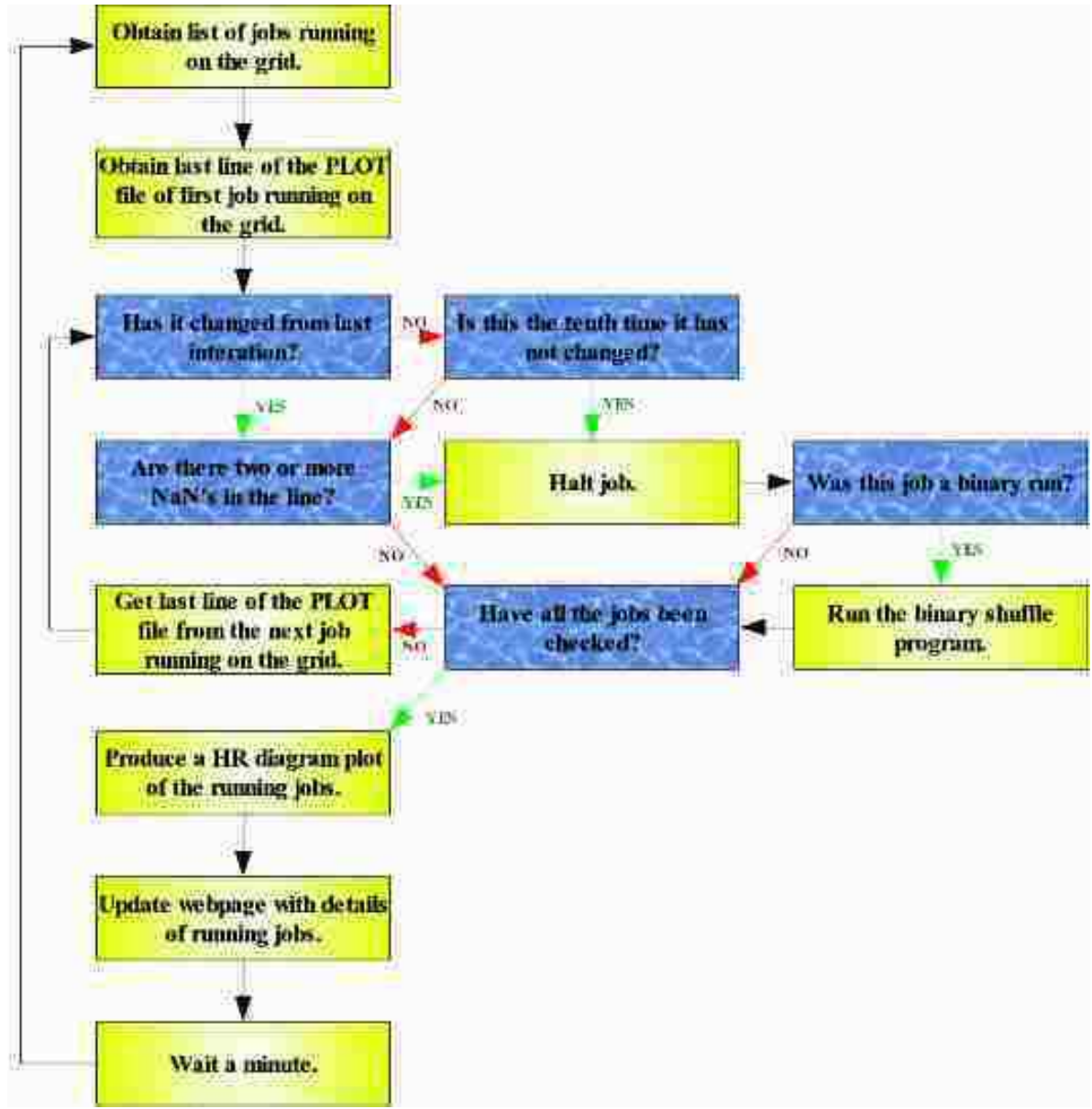}
\caption[Flow diagrams for the grid check program.]{Flow diagrams for the grid check program. The blue boxes are decision statements, red arrows are the paths when the outcome is false, the green true.}
\label{babysitter}
\end{center}
\end{figure}

The code is not perfect and occasionally numerical errors occur, usually when the mesh does not resolve a region that is evolving rapidly or an error occurs in the solution routine. This results in one of two problems, the code can get stuck in an infinite loop and outputs no more data or the variables go infinite and the code continues to run spewing out blank results indefinitely\footnote{This results in a disk full of data that's as useful as a pair of big fat pants.}. To get round this problem a babysitting program written in Perl was devised to check on the program while it runs, stopping it when it encounters a fault. Figure \ref{babysitter} shows the program structure.

This was necessary to investigate a large parameter space when a great number of stars must be evolved in parallel around the clock. It would be difficult for any human to continuously perform this task. To exploit the SunGrid at the IoA this program had to be developed along with the logical structure to evolve hundreds of stars automatically. It is now simple to change some of the parameters and see the effect on a group of models. While it is not close to the number of models that are possible in simplified population synthesis algorithms, e.g. \citet{HPT00}, it has the advantage of producing full stellar evolution models rather than approximations. This work would not have been possible without this program.

\section{The Composition Variables and Nuclear Reaction Network}
\begin{center}
``Change is the essential process of all existence.''\\
\textit{Spock, Star Trek, Let That Be Your Last Battlefield.}
\end{center}

\subsection{Producing the Network}
Stars derive their energy from nuclear fusion reactions. The first and longest phase, is that of hydrogen burning to helium. Next helium is transmuted into carbon and oxygen, the remaining phases are then carbon burning, neon burning, oxygen burning and finally silicon burning. Once an iron core is formed the star is doomed to supernova.

Not all these stages occur in all stars. Below certain mass limits the conditions required for reactions cannot be reached. Also mass loss can truncate the evolution of a star and expose the core so that it becomes a white dwarf before further reactions can occur.

To study these reactions in a stellar evolution code the abundances of elements must be variables. In the Eggleton code we are limited to the number of elements we can study without reducing the performance of the code because we solve for the composition simultaneously with the structure and each requires a second order equation. In the original code the elements taken to be variables were $^{1}$H, $^{4}$He, $^{12}$C, $^{16}$O and $^{20}$Ne. Nitrogen was dealt with as the missing mass fraction required for the sum of all elements to be one. The reaction scheme for these elements is listed in the first column of table \ref{scheme1}. Most of these reactions are correct, however there are approximations due to the small number of variables. Noticeably the nitrogen alpha addition and the later burning stages are contrived.

To make a more accurate and realistic network we added three extra composition variables. This does slow down the code but not excessively thanks to modern computing power. The full list of elements followed is now $^{1}$H, $^{4}$He, $^{12}$C, $^{14}$N, $^{16}$O, $^{20}$Ne, $^{24}$Mg and $^{28}$Si with the remaining fraction taken to be iron. The new reaction scheme is shown in the second column of table \ref{scheme1}. We retain the nitrogen alpha addition since while this reaction should lead to $^{22}$Ne only small amounts of this element are formed. Although it is an important neutron source in AGB stars it is likely to have a small effect on the structure. The later stages are now slightly improved and can be followed all the way to oxygen burning, although during these burning stages the code experiences numerical problems as the time step drops to less than $10^{-4}{\rm yr}$ and the assumption of hydrostatic equilibrium is likely to be broken with dynamical effects altering the progress of burning. In the new reaction scheme there are rates stored in the code that are not currently used. This is because they occur after oxygen burning which we currently can only see the ignition of.

\begin{table}
\begin{center}
\caption[Nuclear Reaction Network.]{Nuclear Reaction Network. Those schemes in blue had the rates included in the code but not used in the reaction network. The first five reactions, comprising the pp-chain, are assumed to be in transient equilibrium with each other so that the abundances of $^{4}$He and $^{7}$Be do not need to be followed explicitly.}
\label{scheme1}
\begin{tabular}{|l|l|}
\hline
Old network. & New Network.\\
\hline
$^{1}$H (p, $\beta^{+} \nu$) $^{2}$H (p,$\gamma$) $^{3}$He&$^{1}$H (p, $\beta^{+} \nu$) $^{2}$H (p,$\gamma$) $^{3}$He\\
$^{3}$He ($^{3}$He,2p) $^{4}$He&$^{3}$He ($^{3}$He,2p) $^{4}$He\\
$^{2}$He ($^{4}$He,$\gamma$) $^{7}$Be&$^{2}$He ($^{4}$He,$\gamma$) $^{7}$Be\\
$^{7}$Be (e$^{-}$,$\nu$) $^{7}$Li (p,$\alpha$) $^{4}$He&$^{7}$Be (e$^{-}$,$\nu$) $^{7}$Li (p,$\alpha$) $^{4}$He\\
$^{7}$Be (p,$\gamma$) $^{8}$B ($\beta^{+} \nu$) $^{8}$Be$^{*}$ ($\alpha$) $^{4}$He&$^{7}$Be (p,$\gamma$) $^{8}$B ($\beta^{+} \nu$) $^{8}$Be$^{*}$ ($\alpha$) $^{4}$He\\
$^{12}$C (p,$\beta^{+} \nu$) $^{13}$C (p,$\gamma$) $^{14}$N&$^{12}$C (p,$\beta^{+} \nu$) $^{13}$C (p,$\gamma$) $^{14}$N\\
$^{14}$N (p,$\beta^{+} \nu$) $^{15}$N (p,$\gamma$) $^{16}$O&$^{14}$N (p,$\beta^{+} \nu$) $^{15}$N (p,$\gamma$) $^{16}$O\\
$^{14}$N (p,$\beta^{+} \nu$) $^{15}$N (p,$\alpha$) $^{12}$C&$^{14}$N (p,$\beta^{+} \nu$) $^{15}$N (p,$\alpha$) $^{12}$C\\
$^{16}$O (p,$\beta^{+} \nu$) $^{17}$O (p,$\alpha$) $^{14}$N&$^{16}$O (p,$\beta^{+} \nu$) $^{17}$O (p,$\alpha$) $^{14}$N\\
$^{4}$He ($\alpha$) $^{8}$Be$^{*}$ ($\alpha$,$\gamma$) $^{12}$C&$^{4}$He ($\alpha$) $^{8}$Be$^{*}$ ($\alpha$,$\gamma$) $^{12}$C\\
$^{12}$C ($\alpha$,$\gamma$) $^{16}$O&$^{12}$C ($\alpha$,$\gamma$) $^{16}$O\\
$^{14}$N ($\alpha$,$\gamma$) $^{18}$F ($\frac{1}{2}\alpha$,$\gamma$) $^{20}$Ne&$^{14}$N ($\alpha$,$\gamma$) $^{18}$F ($\frac{1}{2}\alpha$,$\gamma$) $^{20}$Ne\\
$^{16}$O ($\alpha$,$\gamma$) $^{20}$Ne&$^{16}$O ($\alpha$,$\gamma$) $^{20}$Ne\\
{\color{blue}$^{20}$Ne ($\alpha$,$\gamma$) $^{24}$Mg}&$^{20}$Ne ($\alpha$,$\gamma$) $^{24}$Mg\\
&{\color{blue}$^{24}$Mg ($\alpha$,$\gamma$) $^{28}$Si}\\
$^{12}$C ($^{12}$C,$\alpha \gamma$) $^{20}$Ne&$^{12}$C ($^{12}$C,$\alpha \gamma$) $^{20}$Ne\\
{\color{blue}$^{12}$C ($^{12}$C,$\gamma$) $^{24}$Mg}&$^{12}$C ($^{12}$C,$\gamma$) $^{24}$Mg\\
{\color{blue}$^{12}$C ($^{16}$O,$\alpha \gamma$) $^{24}$Mg}&$^{12}$C ($^{16}$O,$\alpha \gamma$) $^{24}$Mg\\
{\color{blue}$^{16}$O ($^{16}$O,$\alpha \gamma$) $^{28}$Si ($\gamma$,$\alpha$) $^{24}$Mg}&$^{16}$O ($^{16}$O,$\alpha \gamma$) $^{28}$Si\\
$^{20}$Ne ($\gamma$,$\alpha$) $^{16}$O&$^{20}$Ne ($\gamma$,$\alpha$) $^{16}$O\\
{\color{blue} $^{24}$Mg ($\gamma$,$\alpha$) $^{20}$Ne}&$^{24}$Mg ($\gamma$,$\alpha$) $^{20}$Ne\\
&{\color{blue} $^{28}$Si ($\gamma$,$\alpha$) $^{24}$Mg}\\
\hline
\end{tabular}
\end{center}
\end{table}

\subsection{Testing the results, a $25M_{\odot}$ star}

\begin{figure}
\begin{center}
\includegraphics[height=140mm,angle=270]{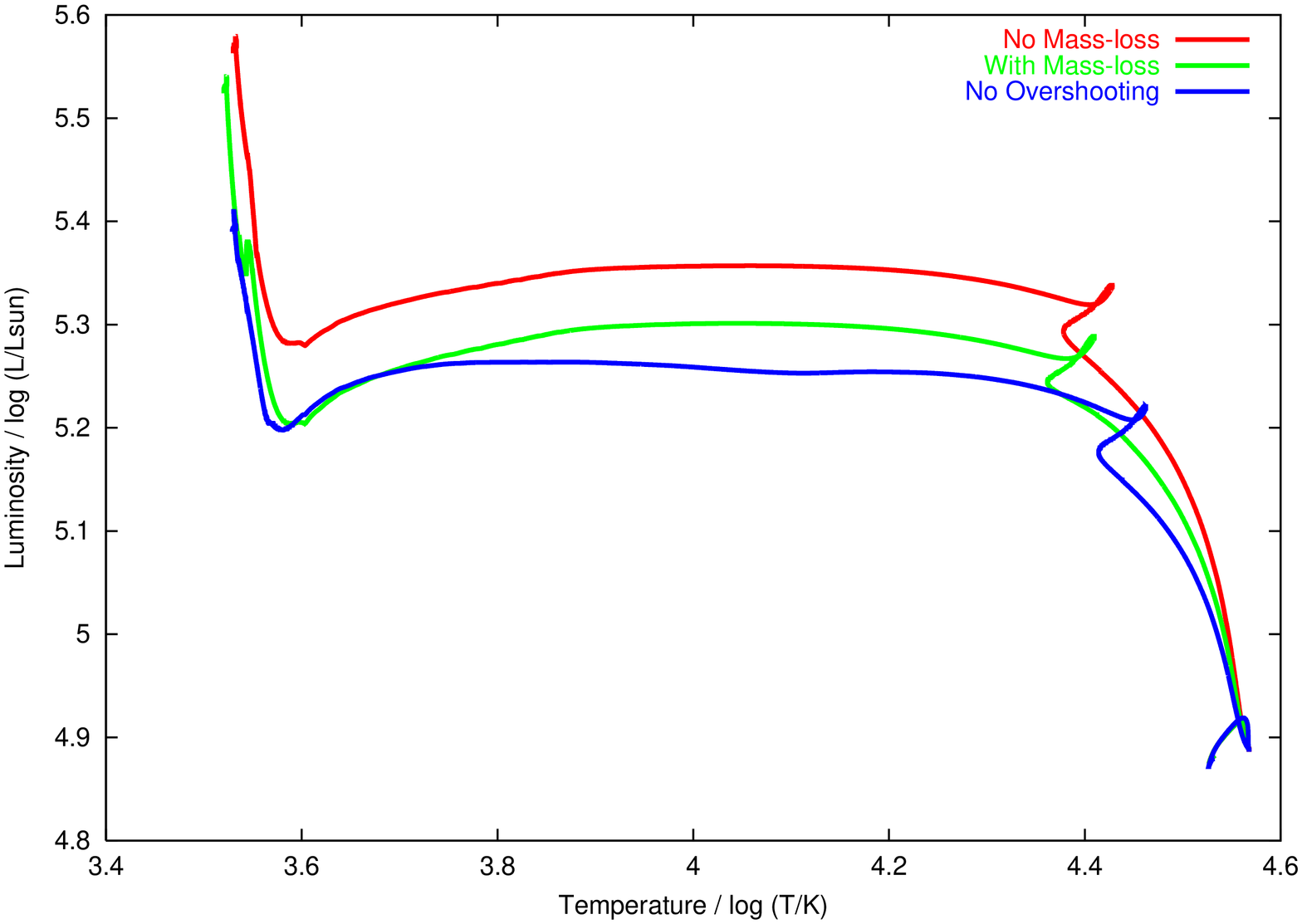}
\includegraphics[height=140mm,angle=270]{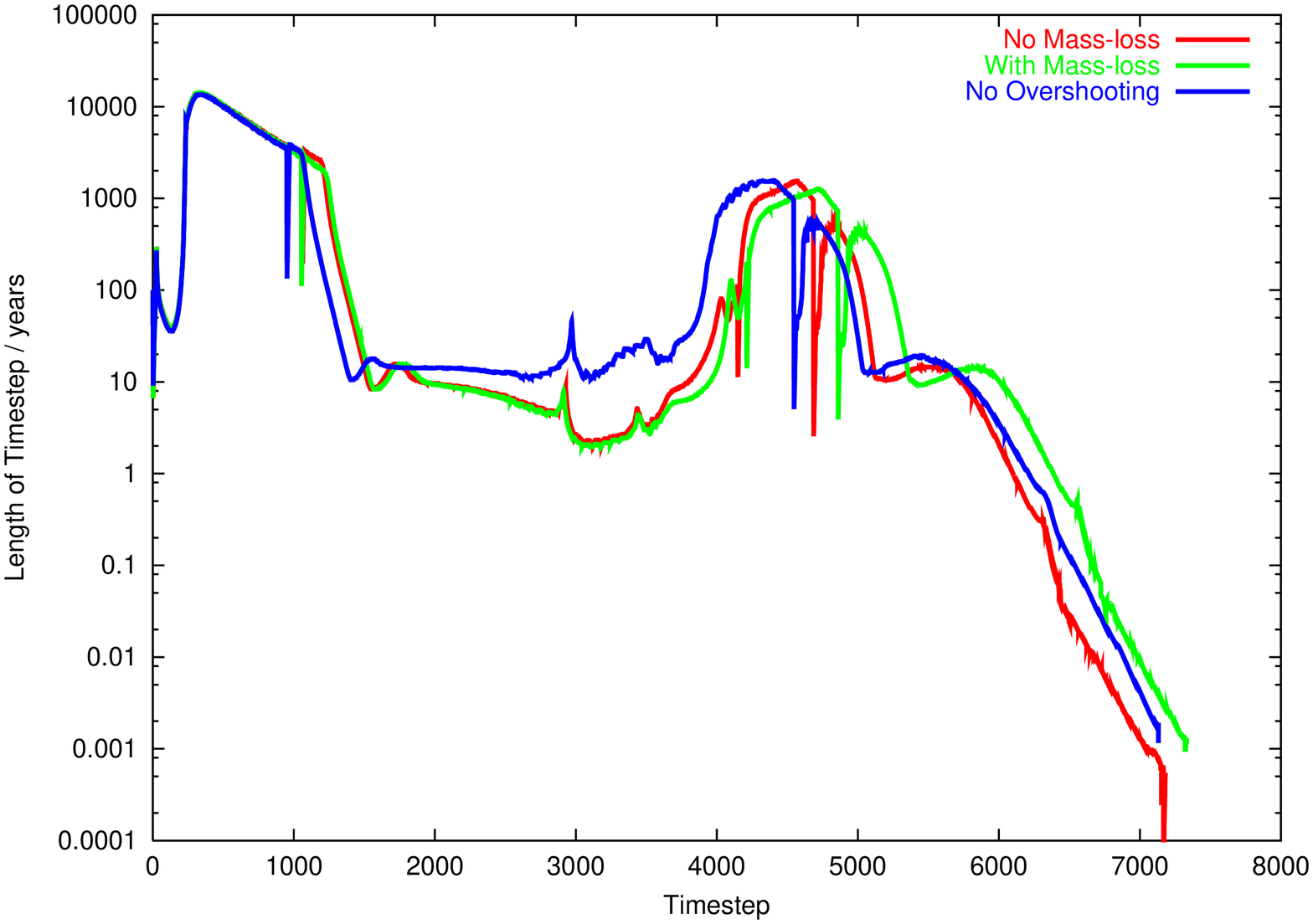}
\caption[The HR diagrams and timestep details for comparison of $25M_{\odot}$ star models.]{Comparison of a $25M_{\odot}$ stars.Top is the HR diagram and bottom is the evolution of the time step with model number.}
\label{f25-3}
\end{center}
\end{figure}

\begin{figure}
\begin{center}
\includegraphics[height=140mm,angle=270]{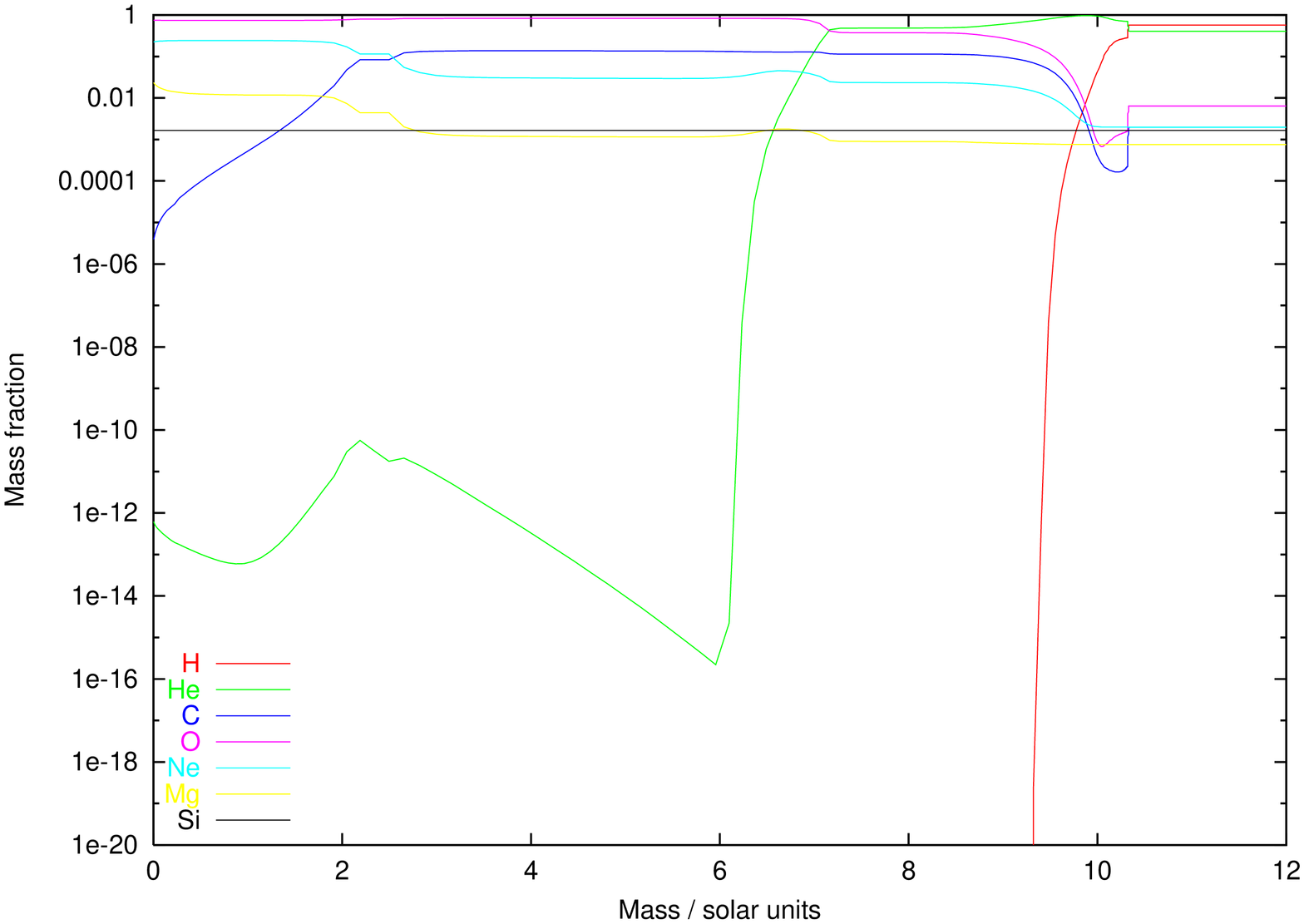}
\includegraphics[height=140mm,angle=270]{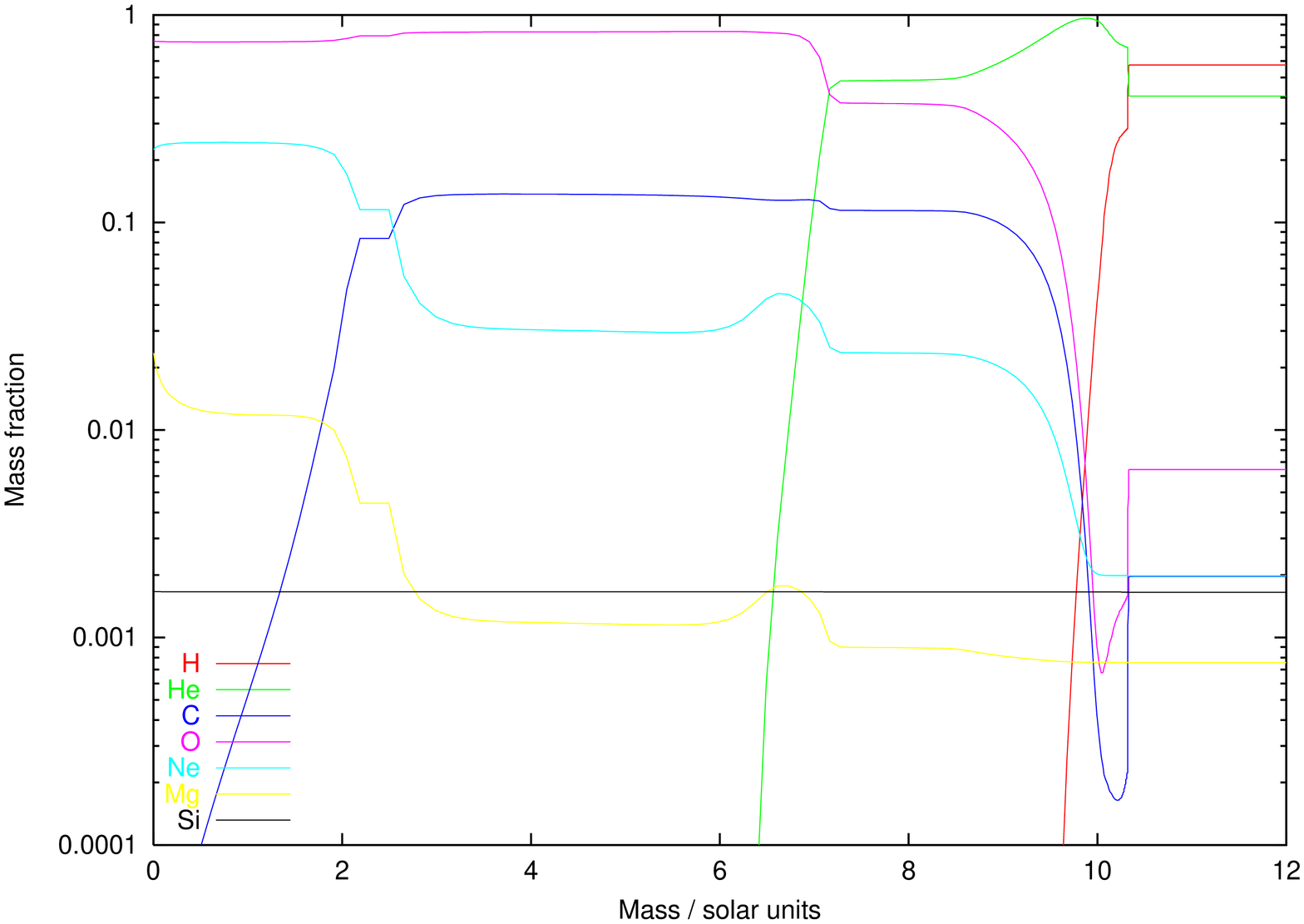}
\caption{The internal composition structure in the final calculated model of a $25M_{\odot}$ star with mass loss and convective overshooting, before the ignition of neon.}
\label{f25-1}
\end{center}
\end{figure}

\begin{figure}
\begin{center}
\includegraphics[height=140mm,angle=270]{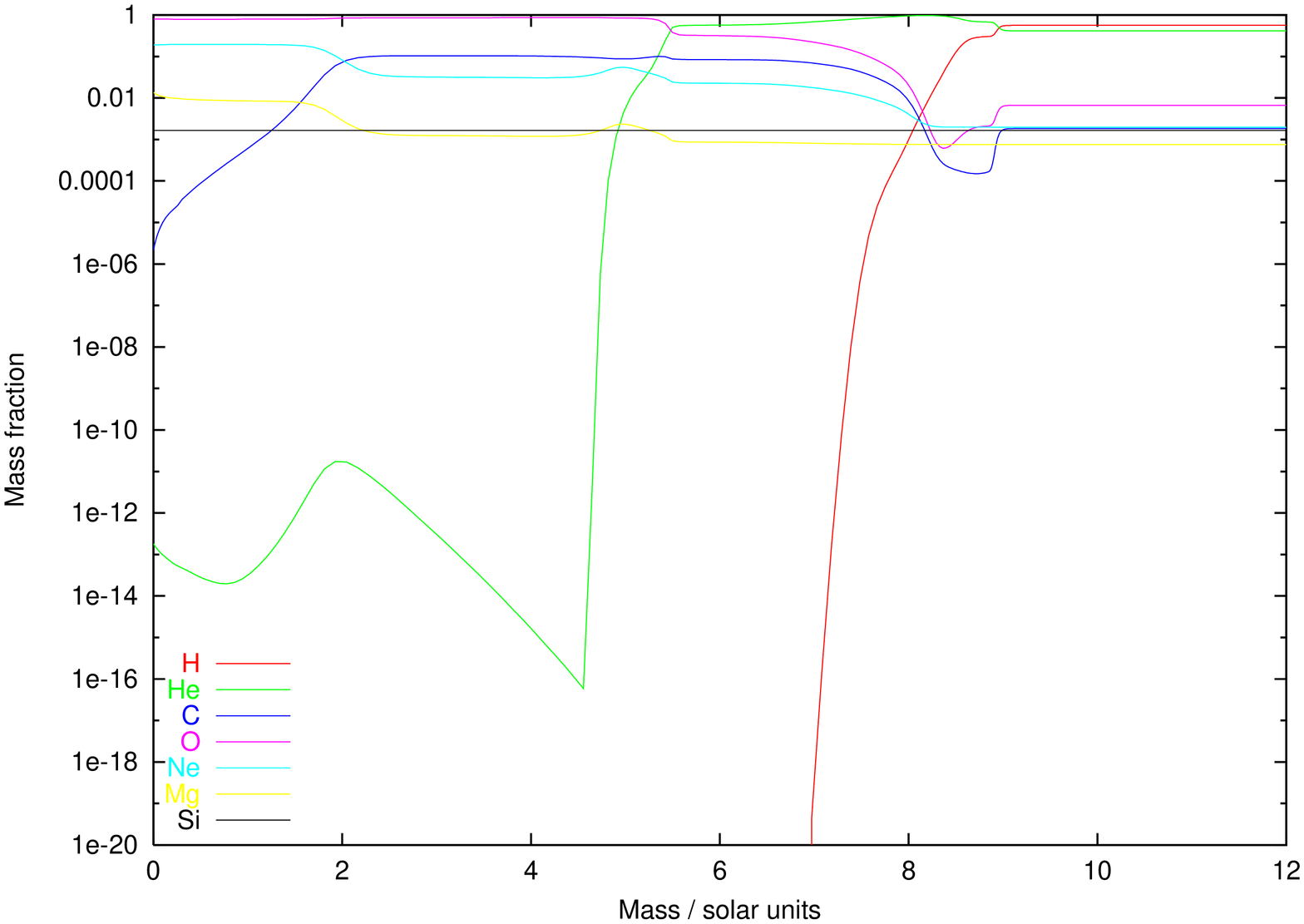}
\includegraphics[height=140mm,angle=270]{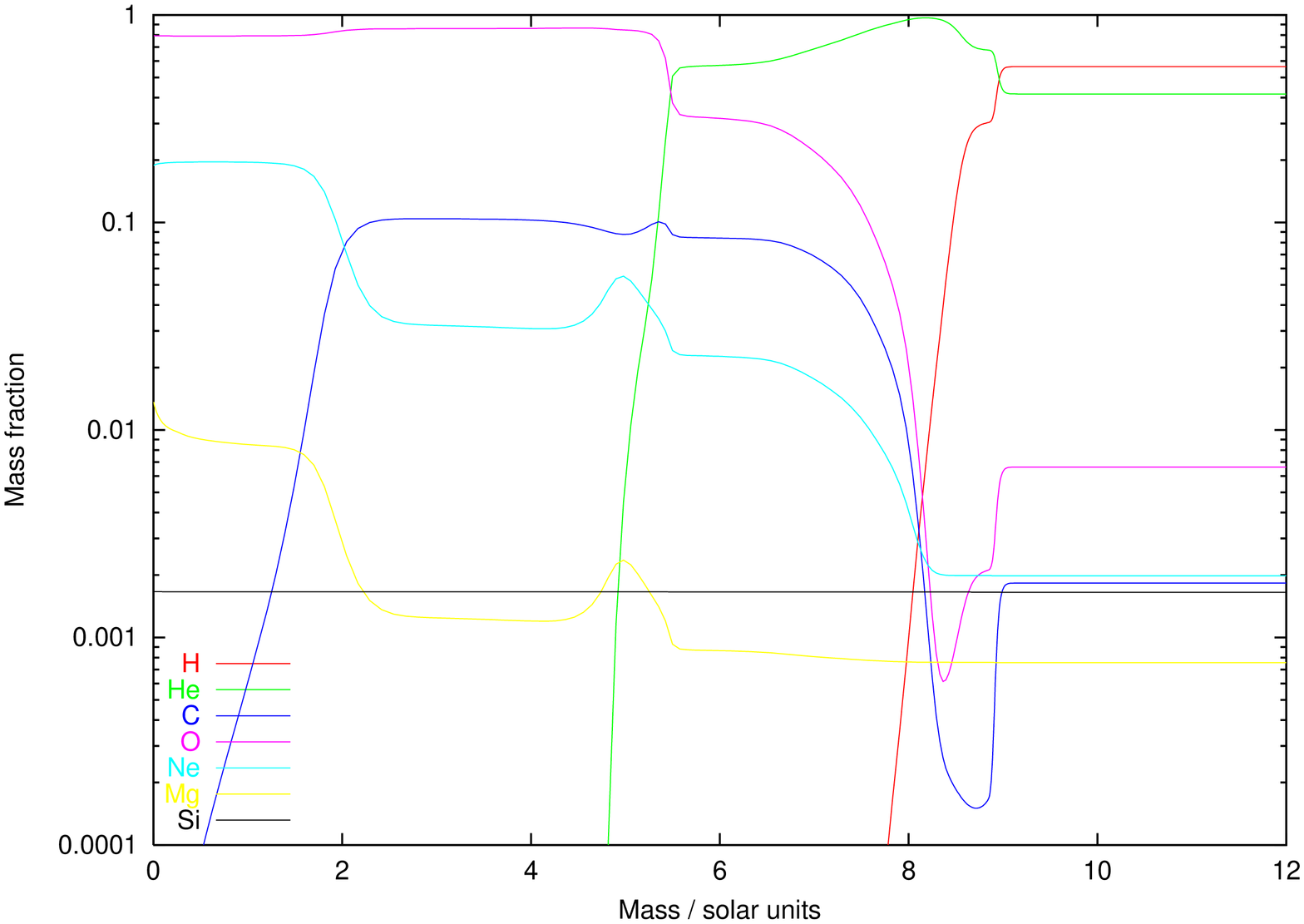}
\caption{The internal composition structure in the final calculated model of a $25M_{\odot}$ star with mass loss and without convective overshooting, before the ignition of neon.}
\label{f25-2}
\end{center}
\end{figure}

Figures \ref{f25-1}, \ref{f25-2} and \ref{f25-3} show the results of the evolution of a $25M_{\odot}$ star.  These models were evolved, the mass-loss rates taken from \citet{dJ}. The figures display the internal composition structure, evolution over the HR diagram and timestep versus model number. Both diagrams are similar and show the hydrogen, helium, carbon and neon burning shells. In none of the stars do we progress to oxygen burning. Although the star with no mass loss does ignite oxygen but the model fails soon afterwards. Figure \ref{f25-3} shows that for this model the timestep has dropped to extremely low values and our assumption of hydrostatic equilibrium is likely to break down. The code becomes unstable at low timesteps as terms in the energy equation are large with only a small difference between them.

In table \ref{25modelA} we compare our models to those of \citet{25Mmodel}, they use the mass-loss rates of \citet{NJ90} which are similar to those we have used but do lead to greater mass loss in models. Using these rates in our models produces similar core masses however the final mass drops to $12.2 \,M_{\odot}$. They do not use convective overshooting and full details of their treatment of convection are given in \citet{WHW02}. The closest agreement is with our third model without convective overshooting which leads to smaller values for the helium and CO cores. Our oxygen neon core mass is smaller because we are not able to evolve the model through to the last burning stages. The remnant masses are also of the same magnitude although our higher values are due to convective overshooting producing larger cores.

The differences that are there are due to the way we define the core masses, the nuclear rates adopted and the details convection. The rates of hydrogen and helium burning are well known in comparison to the later burning reactions so it is not surprising that the final core masses are different. The treatment of convection especially in the late stages of evolution is quite uncertain. During this time the convection will probably affect the energy production and vice versa. Our main conclusion is that while there are is some uncertainty in the physics of our models and we are not able to evolve through the final stages of evolution we do gain similar results to other codes for the earlier burning stages.

\begin{table}
\caption{Comparison of models for a 25$M_{\odot}$ star.}
\label{25modelA}
\begin{center}
\begin{tabular}{|l|cccc|}
\hline
		      & No Mass Loss & Mass Loss& Mass Loss	&Rauscher et al.\\
		      &	$\delta_{ov}=0.12$ & $\delta_{ov}=0.12$ & $\delta_{ov}=0.00$&(2002)\\
\hline
Initial mass  & 25.00		& 25.00		&25.00& 25.136\\
Final mass	  & 25.00		& 16.27		&18.40 & 13.079\\
He core mass  & 10.78		& 10.06		&8.47 & 8.317\\
CO core mass  & 7.32		& 6.96		&5.33 & 6.498\\
ONe core mass& 2.14		& 1.77		&1.66 & 2.443\\
Remnant mass  & 4.00		& 3.17		&1.49 & 1.961\\
\hline
\end{tabular}
\end{center}
\end{table}

\section{Opacity}
The main source of a star's energy is the nuclear fusion reactions occurring either in its core or in thin burning shells around the core. This energy is transported from the production site to the surface by radiative transfer or convection. The first transports energy in the form of photons, while the second is a cyclic macroscopic mass motion that carries the energy in bulk.

In regions stable to convection, radiative transfer leads to the equation of stellar structure,

\begin{equation}
\frac{dT}{dr} = - \frac{3 \overline{\kappa_{\rm R}} \rho}{16 \sigma T^{3}} \frac{L_{r}}{4 \pi r^{2}},
\end{equation}

where $L_{r}$ is the luminosity at radius $r$, $\rho$ the density, $\sigma$ the Stefan-Boltzmann constant, $T$ the temperature and $\overline{\kappa_{{\rm R}}}$ is the Rosseland mean opacity.

Opacity $\kappa$ is a measure of the degree to which matter absorbs photons. There are four main sources of opacity in stars

\begin{itemize}
\item Bound-bound transitions are the transitions of an electron in an atom, ion or molecule between energy levels which are accompanied by either the absorption or emission of a photon. Only a photon with the correct wavelength can cause a given transition, the process is wavelength dependent.
\item Bound-free transitions, or photoionisation, occur when an incoming photon has enough energy to ionise an atom or ion and free an electron. The reverse process is the capture of an electron by an atom or ion. It will not occur until photons above a threshold energy are available and falls off as $\nu^{-3}$ where $\nu$ is the frequency.
\item Free-free transitions are scattering processes which occur when an electron and photon interact near an atom or ion. The process is also known as bremsstrahlung. Again it is proportional to $\nu^{-3}$.
\item Electron scattering is wavelength independent at low temperatures where it is Thomson scattering. The electron, with its low cross-section, is a small target and so only dominates at high temperatures when most atoms are ionised. At very high temperatures relativistic effects are important and Compton scattering dominates.
\end{itemize}

From this list it is possible to see that calculating the opacity in a stellar model is a difficult process and depends on the composition, temperature and density of the material. Deep in the star local thermodynamic equilibrium (LTE) is achieved and an average for radiative transfer over all wavelengths requires the Rosseland mean opacity $\overline{\kappa_{{\rm R}}}$ \citep{RO24} expressed as

\begin{equation}
\frac{1}{\overline{\kappa_{{\rm R}}}} =\frac{ \displaystyle \int_{0}^{\infty} \frac{1}{\kappa_{\nu}} \frac{\partial B_{\nu}}{\partial T} d\nu }{\displaystyle \int_{0}^{\infty} \frac{\partial B_{\nu}}{\partial T} d\nu},
\end{equation}

where $\kappa_{\nu}$ is the opacity at frequency $\nu$, $T$ is the temperature and $B_{\nu}$ is the flux per unit area into unit solid angle per unit frequency in LTE.

The most comprehensive opacities available today are from the OPAL \citep{IR96} or OP \citep{OP} groups who have made detailed  models of the above processes. They provide tables of the Rosseland mean opacity variation with temperature, density and composition with the metal abundance usually scaled to solar compositions. \citet{IR93} took a step forward for the OPAL project team by providing tables that include mixtures enhanced in carbon and oxygen (C~and~O) relative to the base solar composition. We have incorporated their full range of composition tables in the \citet{E71} evolution code. In the last implementation of this code \citep{P95} only 10 of the 265 tables available were used. Now we have taken the opportunity to refine our models so as to accurately follow the changes of opacity which occur in the later stages of evolution. We can expect the new tables to affect many types of stars. While they may be small for the evolution of main-sequence stars and white dwarfs we expect larger differences to be found in AGB and Wolf-Rayet stars.

Asymptotic Giant Branch (AGB) stars undergo third dredge-up which mixes helium burning products to the surface and forms carbon stars. Using the enhanced mixture tables we shall be able to model the thermal pulses and envelope evolution more accurately. Wolf-Rayet (WR) stars are massive and have lost their hydrogen envelopes exposing the helium cores. As time progresses helium burning products are slowly exposed at the surface and in some cases the stars are eventually mostly composed of carbon and oxygen.

We shall present the method of opacity table construction and detail its implementation in the Eggleton stellar evolution code. We shall discuss the effects on main-sequence stars, red giants and white dwarfs. We shall present three tests of these tables. The first is the effect of including extra carbon and oxygen on the structure and evolution of a low-mass population-III star, the second is a $5{\rm M}_{\odot}$ thermally pulsing AGB star and the third a Wolf-Rayet star of $40{\rm M}_{\odot}$  with mass loss.

\section{Opacity Table Construction}

We start with the OPAL tables \citep{IR96} as the framework around which to construct the full tables in an approach similar to that employed by \citet{P95}. However we choose to use the variable ${\mathcal{R}}=\rho / T^{3}_{6}$ rather than the density. This has the advantage that our tables can be a third smaller in this dimension. This partly compensates for increased memory requirement of the full tables.

The OPAL tables extend in $\log_{10} (T/{\rm K})$ from $3.75$ to $8.70$ in steps of $0.05$ to $0.2$ and in $\log_{10} ({\mathcal{R}}/ {\rm g \, cm^{-3} MK^{-3}})$ from $-7$ to $+1$ in steps of $0.5$. We choose the range of the final tables to cover from $3$ to $10$ in $\log_{10} (T/{\rm K})$ with increments of $0.05$, and from $-8$ to $+7$ in $\log_{10} ({\mathcal{R}}/ {\rm g \, cm^{-3} MK^{-3}})$ with increments of $0.5$. This gives us tables of 141 by 31 elements.

To account for the effect of composition, the OPAL tables include typically 265 $T$~and~$\mathcal{R}$ grids for each metallicity. These tables are split into 5 groups with different hydrogen mass fractions X of $0.0$, $0.03$, $0.1$, $0.35$ and $0.7$. The tables in each group have different compositions in helium, carbon and oxygen.

The lowest temperature in the OPAL tables is $\log_{10} (T/{\rm K}) = 3.75$. Below this we use the tables of \citet{AF94}, which extend from $3$ to $4.1$ in $\log_{10} (T/{\rm K})$. Where the tables overlap they match well \citep{IR96}. It should be noted that the tables provided by \citet{AF94} do not include enhanced carbon and oxygen mixtures. Therefore at temperatures when $\log_{10} (T/{\rm K}) < 4.0$ our tables do not follow these mixtures. This is only important in the surface of AGB stars. We plan in future to deal with low temperature enhanced mixtures by including the work of \citet{M02} before application to the structure of AGB star envelopes.

To complete each table we fill the region from $8.70$ to $10$ in $\log_{10} (T/{\rm K})$, where electron scattering dominates, according to \citet{BY76}. We then combine with a full table of the effective opacity owing to electron conduction for which we use the fits of \citet{I75} to the tables of \citet{HL69} and \citet{Ca70}. We combine the radiative opacity with the effective conductive opacity by reciprocal addition,

\begin{equation}
\frac{1}{\kappa_{\rm eff}}=\frac{1}{\kappa_{\rm rad}}+\frac{1}{\kappa_{\rm cond}}.
\end{equation}

Before the complete tables are implemented in the stellar evolution code we perform one last process. We rescale the tables over the C and O plane to a more regular grid. In the OPAL grid the spacing used requires interpolation within irregular polygons, so to make the interpolation simpler and faster, we include extra tables in a similar but fully rectangular grid system. This means we have 61 tables for each hydrogen abundance or 305 tables for each metallicity.

To construct tables of different metallicity care must be taken since there are no corresponding OPAL tables for some of the low temperature tables and vice-versa. In these cases we construct two tables with the nearest two OPAL tables both including the same low temperature tables and interpolate. A similar method is applied for the OPAL tables. 

\section{Implementation into the code}

 The main drawback with implementation of the new tables is the increased memory requirement. Whereas before there were only 10 opacity tables there are now 305. The new tables are each a third of the size of the old tables so the overall increase is a factor of ten. This corresponds to a memory requirement increase from 17$\,$Mb to 170$\,$Mb but the typical specification of today's desktop computers comfortably accommodates this. One of the reasons why this memory requirement is so large is that we also store the spline coefficients for each of the tables. When we use the tables within the code we interpolate in the $\mathcal{R}$ and $T$ plane with bicubic splines. This ensures the smoothness of our tables. We then use linear interpolation in the 3 composition dimensions to derive the final opacity value at each point.

To test the effect of these tables we use three different interpolation schemes,

\begin{itemize}
\item Method A uses 5 tables with $X=0,\,0.03,\,0.1,\,0.35$ and $0.7$ and $Y=1-X-Z$. Where $X$ is the hydrogen mass fraction, $Y$ the helium mass fraction and $Z$ the initial, solar mixture, metallicity. We add a further two tables, both with $X=Y=0$, the first with $X_{\rm C}=1-Z$ and $X_{\rm O}=0$. The other with $X_{\rm C}=0$ and $X_{\rm O}=1-Z$. Where $X_{\rm C}$ and $X_{\rm O}$ are the mass fractions for the enhanced amount of carbon and oxygen above that included in the metallicity mass fraction so the total carbon mass fraction is $X_{\rm C} + Z({\rm C})$ and for oxygen $X_{\rm O} + Z({\rm O})$. This gives a total of 7 tables allowing simple interpolation in composition similar to the old method.
\item Method B is similar to method A but we include two extra tables for each hydrogen mass fraction. Again both have $Y=0$. One has $X_{\rm C}=1-X-Z$ and $X_{\rm O}=0$ and the other $X_{\rm C}=0$ and $X_{\rm O}=1-X-Z$. This leads to a total of 15 tables in all.
\item Method C incorporates all 305 tables in the opacity calculations. We have 61 tables for each hydrogen abundance. When interpolating in composition we use three variables $X$, $X_{\rm C}/(1-X-Z)$ and $X_{\rm O}/(1-X-Z)$. The hydrogen abundances are as for method A. We take values of $0,\,0.01,\,0.03,\,0.1,\,0.2,\,0.4,\,0.6$ and $1$ for the carbon and oxygen grid planes. On this grid we perform a three-dimensional linear interpolation.
\end{itemize}

\section{Testing \& Results}

\begin{table}[!b]
\caption[Variation of Evolutionary Time Scale with increasing carbon and oxygen mass fraction.]{Variation of Evolutionary Time Scale with increasing carbon and oxygen mass fraction. Here $t_{\rm A}, t_{\rm B}$ and $t_{\rm C}$ are the evolution times to the helium flash for methods A, B and C and $t_{\rm Z}$ is the evolution time for models with an equivalent mass of scaled solar metallicity.}
\label{ta}
\begin{center}
\begin{tabular}{|c|cccc|}
\hline
$(X_{\rm C}+X_{\rm 0})$ & $\frac{t_{\rm C}-t_{\rm A}}{t_{\rm A}}$ & $\frac{t_{\rm B}-t_{\rm A}}{t_{\rm A}}$ &  $\frac{t_{\rm B}-t_{\rm C}}{t_{\rm A}}$&  $\frac{t_{\rm Z}-t_{\rm C}}{t_{\rm C}}$\\
or $Z$ Mass  & /\% & /\% & /\%  & /\% \\
Fraction & & & & \\
\hline
$10^{-5}$ & 0.035 & 0.031 & 0.004 & 0.124\\
$10^{-4}$ & 0.355 & 0.311 & 0.045 & 1.134\\
$10^{-3}$ & 3.535 & 3.047 & 0.488 & 10.02\\
$10^{-2}$ & 32.94 & 31.49 & 1.449 & 57.54\\
\hline
\end{tabular}
\end{center}
\end{table}

The new tables give almost identical results to the old for main-sequence stars and red giants. The timescale of core helium burning is altered. From method A to method C it increases by a factor of 0.1\% for a $5{\rm M}_{\odot}$ star. This is reassuring. Current stellar evolution models describe well the evolution of the main sequence and the red giant branch. However we do find that the opacity  tables enhance numerical stability in the later stages of evolution because they resolve changes with composition in greater detail. This leads to a smoother variation in opacity.

A similar result is found for white dwarfs. Between methods~A~and~C for a carbon and oxygen white dwarf the radius increases by 0.2\% and cooling times are affected by about 1\% with the initial cooling faster with method C than for method A until later when the cooling times are longer with method C. While conditions in the atmosphere do pass through regions of the greatest difference between the two methods (the dotted line in figure \ref{fc}) there is not enough mass at these temperatures and densities to make a significant global difference.

\subsection{Polluting a $0.5{\rm M}_{\odot}$ Population-III Star}
\begin{center}
``Like Granny said, if you want a box hurled into the sun, you got to do it yourself.'''\\
\textit{Hermes, Futurama, The Farnsworth Paradox.}
\end{center}

For our first detailed test we take a zero metallicity $0.5 \, {\rm M}_{\odot}$ star and pollute it uniformly with carbon and oxygen. Such a star could have been formed just after the death of the first stars if the products of these stars were mainly carbon and oxygen. Or such stars could be made by the hypothetical industry of stellar engineering by an advanced civilizations. We choose such a low-mass star to limit the effect on the results of burning via the CNO cycle. Our initial model has the following properties, $X=0.7$, $Y=0.3$, $R=0.45 \, {\rm R}_{\odot}$, $T_{{\rm eff}}=4780\,{\rm K}$ and $L=0.96 \, {\rm L}_{\odot}$. The tables provided by OPAL do not include one with $X=0.75$, we could extrapolate to obtain the opacity but for this test we prefer to remain within the tables. From the initial model we perform three sub-tests. In the first we pollute the star at a constant C/O ratio of 1, then evolve each model and compare the time from zero-age main sequence to the helium flash for methods A, B and C. We also compare a sequence of models for stars of the same mass but with a scaled solar metallicity of the same mass fraction as the carbon and oxygen. These results are presented in table \ref{ta}. We see that with the pollution fraction increase, the timescale of hydrogen burning increases because the energy produced at the centre takes longer to reach the surface. However when comparing the difference between methods B \& C we see only a comparatively small increase in timescale. When we compare with the scaled solar models we see a greater difference. This is due to the much larger opacity of heavier elements such as iron.

\begin{table}
\caption{Variation in physical parameters relative to method A with increasing carbon mass fraction.}
\label{tb}
\begin{center}
\begin{tabular}{|cc|ccc|}
\hline
 & & $\Delta R / R$ & $\Delta T_{\rm eff} /T_{\rm eff} $ & $\Delta L / L$ \\
 & $X_{\rm C}$ & /\% & /\% & /\%\\
\hline
(B-A)/A & $10^{-6}$ & 0.000 & 0.000 & 0.005\\				
 & $10^{-4}$ & 0.046 & 0.071 & 0.384 \\	
 & $10^{-3}$ & 0.485 & 0.654 & 3.533 \\	
 &$10^{-2}$ & 4.124 & 3.439 & 21.094 \\
 \hline
(C-A)/A & $10^{-6}$ & 0.000 & 0.000 & 0.002 \\				
 & $10^{-4}$ & 0.041 & 0.060 & 0.326 \\	
 & $10^{-3}$ & 0.433 & 0.563 & 3.058 \\	
 & $10^{-2}$ & 4.168 & 3.457 & 21.219 \\
 \hline
(B-C)/A & $10^{-6}$ & 0.0000 & 0.0000 & 0.0023 \\				
 & $10^{-4}$ & 0.0046 & 0.0115 & 0.0574 \\	
 & $10^{-3}$ & 0.0619 & 0.0915 & 0.4743 \\	
 & $10^{-2}$ & -0.0441 & -0.0178 & -0.1250\\
\hline
\end{tabular}
\end{center}
\end{table}

\begin{table}
\caption{Variation in physical parameters relative to method A with increasing oxygen mass fraction.}
\label{tc}
\begin{center}
\begin{tabular}{|cc|ccc|}
\hline
 & & $\Delta R / R$ & $\Delta T_{\rm eff} /T_{\rm eff} $ & $\Delta L / L$ \\
 & $X_{\rm O}$ & /\% &  /\% & /\% \\
\hline
(B-A)/A & $10^{-6}$ & 0.000 & 0.000 & 0.000\\		
 & $10^{-4}$ & 0.009 & 0.023 & 0.113\\		
 & $10^{-3}$ & 0.108 & 0.209 & 1.054\\		
 & $10^{-2}$ & 1.140 & 1.692 & 8.721\\	
 \hline	
(C-A)/A & $10^{-6}$& 0.000 & 0.000 & 0.000\\		
 & $10^{-4}$ & 0.009 & 0.021 & 0.108\\		
 & $10^{-3}$ & 0.106 & 0.200 & 1.013\\		 
 & $10^{-2}$ & 1.179 & 1.710 & 8.859\\		
 \hline
(B-C)/A & $10^{-6}$ &0.0000&0.0000&0.0000\\		
 & $10^{-4}$ &0.0000&0.0023&0.0046\\		
 & $10^{-3}$ &0.0023&0.0092&0.0410\\		
 & $10^{-2}$ &-0.0387&-0.0181&-0.1386\\		
\hline
\end{tabular}
\end{center}
\end{table}

\begin{table}
\caption{Variation in physical parameters relative to method A with varying the C/O ratio.}
\label{td}
\begin{center}
\begin{tabular}{|cc|ccc|}
\hline
 & C/O & $\Delta R / R$ & $\Delta T_{\rm eff} /T_{\rm eff} $ & $\Delta L / L$ \\
 & Ratio & $/10^{-2}$ & $/10^{-2}$ & $/10^{-2}$\\
\hline
B-A & $\frac{1}{3}$ & 2.022 & 2.308 & 12.564 \\		
 & $\frac{1}{2}$ & 2.281 & 2.467 & 13.593 \\	
 & 1     & 2.806 & 2.770 & 15.375 \\			
 & 2     & 3.288 & 3.027 & 17.297 \\
 & 3     & 3.506 & 3.141 & 18.054 \\		
\hline
C-A & $\frac{1}{3}$ & 1.891 & 2.200 & 11.948 \\		
 & $\frac{1}{2}$ & 2.121 & 2.348 & 12.886 \\	
 & 1     & 2.685 & 2.687 & 13.080 \\			
 & 2     & 3.223 & 2.985 & 17.040 \\
 & 3     & 3.477 & 3.116 & 17.921 \\		
\hline
B-C & $\frac{1}{3}$ & 0.1309 & 0.1080 & 0.6162 \\		
 & $\frac{1}{2}$ & 0.1599 & 0.1191 & 0.7072 \\	
 & 1     & 0.1209 & 0.0829 & 0.4952 \\			
 & 2     & 0.0646 & 0.0424 & 0.2575 \\
 & 3     & 0.0289 & 0.0245 & 0.1322 \\		
\hline
\end{tabular}
\end{center}
\end{table}

With the second set of models we look at the radius, surface temperature and luminosity of the star on the main sequence as we pollute it with either only carbon or only oxygen (tables \ref{tb} and \ref{tc}). We use method A as the base to which we compare methods B and C. A similar trend is seen in the results in tables \ref{tb} and \ref{tc}. We therefore see it is important to include the affect of enhanced C and O mixtures before hydrogen exhaustion. Similar mixtures are encountered in the envelopes of carbon stars of very low metallicity.

Finally we consider the effect of varying the C/O ratio of the pollution (table \ref{td}) at a constant pollutant mass fraction of $0.001$. Table \ref{td} demonstrates the need to use method C to follow accurately the variation in opacity as the C/O ratio changes. This is important in AGB evolution during the thermal pulses when the C/O ratio changes gradually as carbon becomes dominant. The difference between methods B and C in this case comes about because of non-linear structure in the C and O opacities. It can only be resolved with the full set of tables.

\subsection{Thermal Pulses of an AGB Star}

Our thermally pulsing AGB model is of a $5{\rm M}_{\odot}$ star. To follow a number of pulses we include convective overshooting during the pre-AGB evolution. This leaves the star with a larger core which reduces the rate at which the thermal pulses grow in strength. This means we do not find dredge-up in our models but we can evolve through the pulses. We obtain the average pulse period as the arithmetic mean of the times between the first ten helium shell flashes. Results are recorded in table \ref{te}.

\begin{table}
\caption{Variation of timescales for a $5{\rm M}_{\odot}$ star.}
\label{te}
\begin{center}
\begin{tabular}{|c|cc|}
\hline
 & Age at 1st & Average Pulse \\
Method  & Pulse / $10^{8} {\rm yrs}$ & Period / ${\rm yrs}$ \\
\hline
A & 1.199679 & $2{\,}803$ \\
B & 1.193637 & $2{\,}802$ \\
C & 1.200724 & $2{\,}702$ \\
\hline
\end{tabular}
\end{center}
\end{table}

First we find that the different methods affect the age of the star when it starts to undergo thermal pulses. Second we find that methods A and B have similar interpulse periods while method C's is 3.6\% less. We attribute this to the change in opacity due to varying He/C/O ratios in the inter-shell region which alters the time evolution of the pulse.

\subsection{Evolution of a Wolf-Rayet Star}

We have used the mass-loss rates of \citet{dJ} and \citet{NL00} to manufacture a Wolf-Rayet star of solar metallicity and an initial mass of $40{\rm M}_{\odot}$. Some details are described by \citet{DT03}. Our tables are expected to make a difference because Wolf-Rayet stars become naked helium stars and helium burning products are mixed throughout and exposed at the surface of the star. We only use methods A and C and not method B because the carbon and oxygen mass fractions do not rise above those included in the base metallicity until after the hydrogen envelope has been removed and so our tables only begin to make a difference after this.

\begin{figure}
\begin{center}
\includegraphics[height=140mm,angle=270]{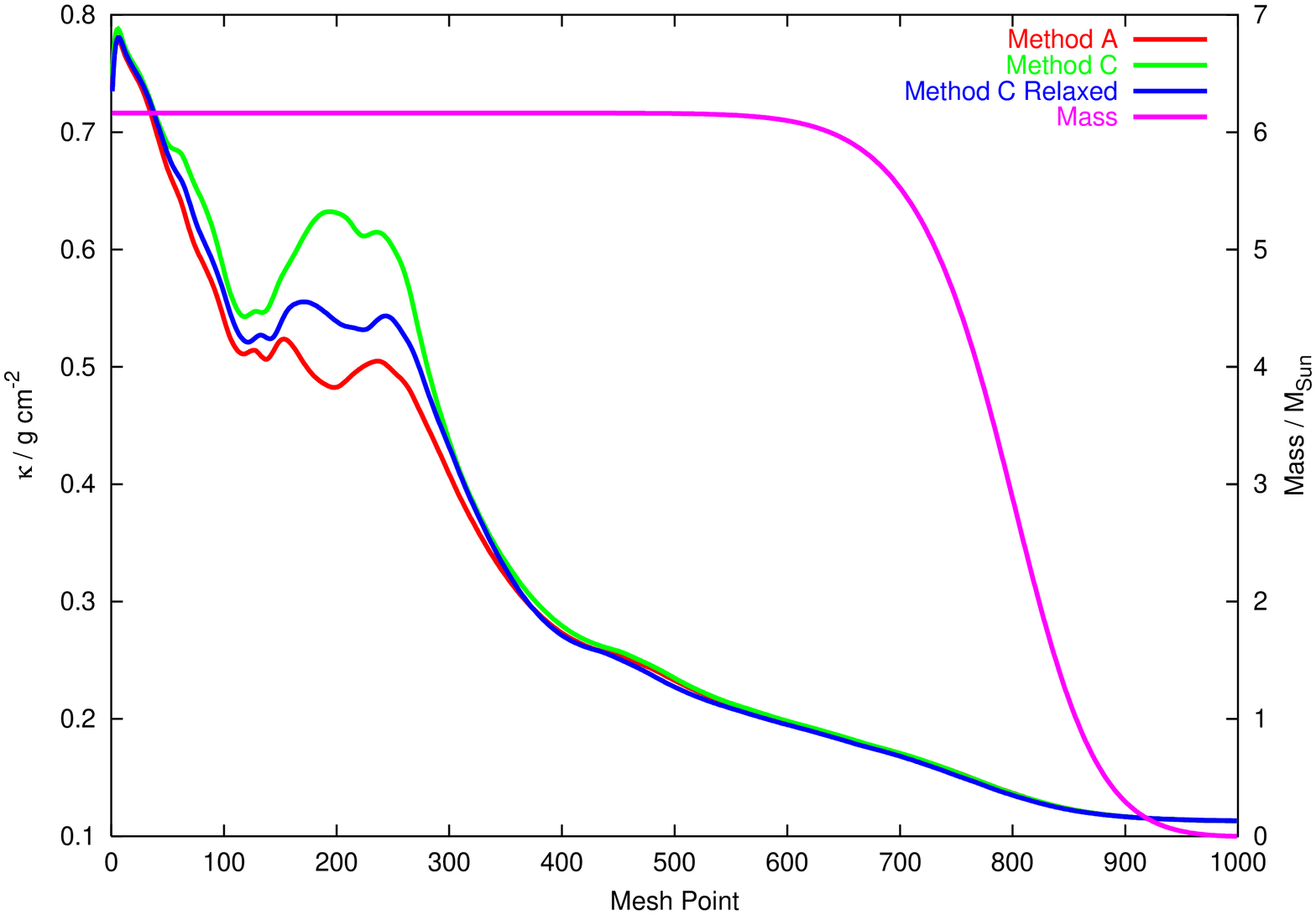}
\caption[The variation of opacity throughout the Wolf-Rayet star just after core carbon ignition.]{The variation of opacity throughout the Wolf-Rayet star just after core carbon ignition. The solid line is calculation from model A, the dashed line from model C and the dash-dotted lines from model C with hydrostatic equilibrium. The dotted line is the mass interior to the mesh point. Mesh point 0 is at the surface and 1000 is at the centre of the star.}
\label{fa}
\includegraphics[height=140mm,angle=270]{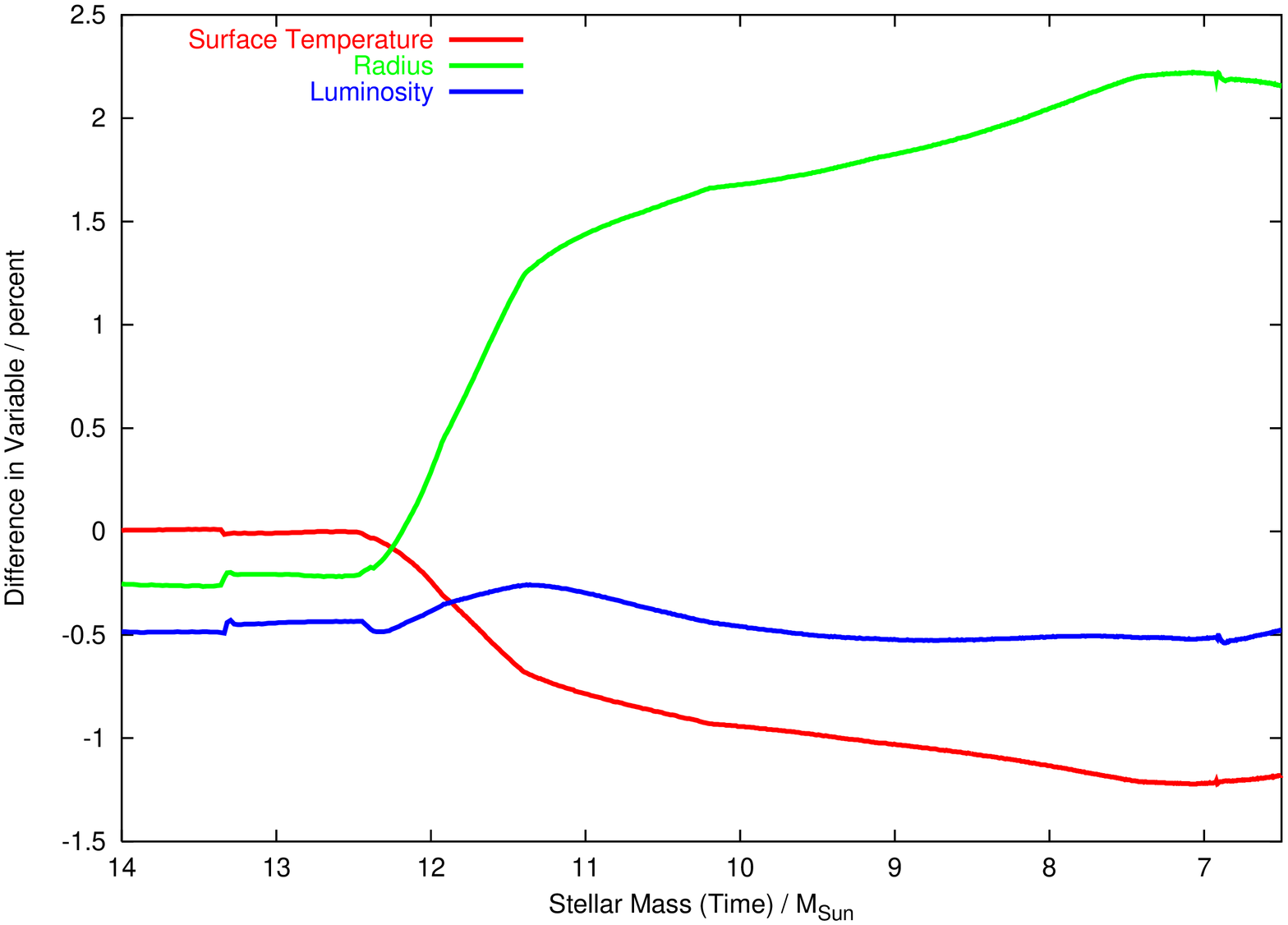}
\caption{The percentage difference in model parameters between methods A and C versus mass of the Wolf-Rayet star.}
\label{fb}
\end{center}
\end{figure}

In figure \ref{fa} we present the variation of opacity through the star, just after core carbon ignition, with mesh point. The solid line is with method A. We then take this structure and use method C to calculate the opacities to give the dashed line. This model is not in hydrostatic equilibrium and the dot-dashed line shows the relaxed new structure in hydrostatic equilibrium from method~C. Comparing the physical parameters of these calculations we find the main change to be an increase in radius of 3.0\%. The surface temperature has decreased by 1.5\%.

We then made two complete runs with similar initial conditions by methods~A~and~C. In figure \ref{fb} we display the difference between Wolf-Rayet evolution as the hydrogen envelope is removed. The greatest changes are in the radius and surface temperature. The star is in the WN phase between $14$ to $12 {\rm M}_{\odot}$ and then enters the WC phase. In the WN phase there are only small differences but upon entering the WC phase the differences build up so that the temperature is lower by around 2\% and the radius about 1\% larger. The luminosity also tends to be larger by around 0.5~~\%.

\begin{figure*}
\begin{center}
\includegraphics[height=160mm]{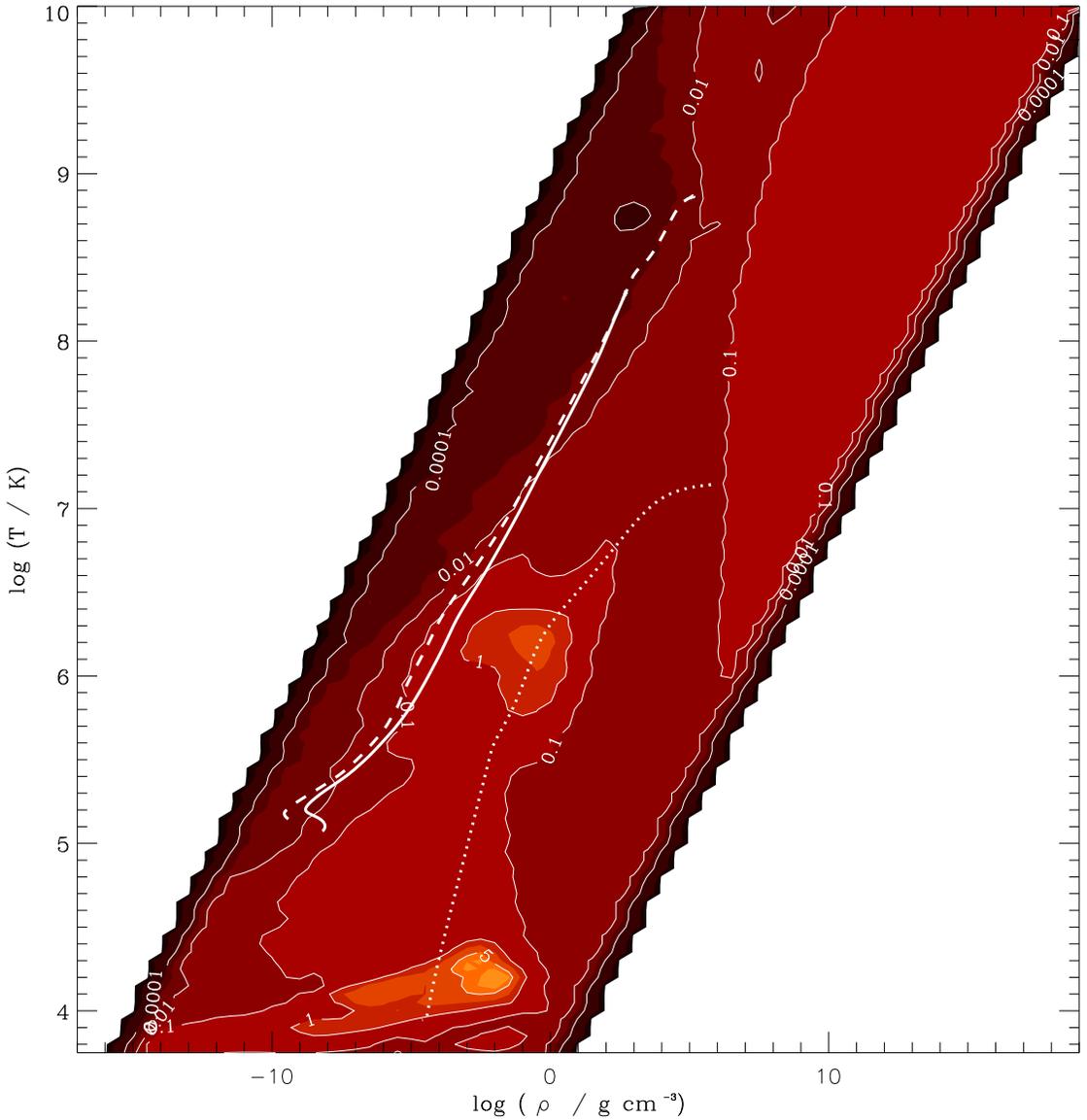}
\end{center}
\caption[Representation of were the new opacity tables have their greatest impact.]{The value plotted by the contours is $ \int_{S} (\kappa_{\rm C}-\kappa_{\rm A}) \, dX_{\rm C} \, dX_{\rm O} / \int_{S} \kappa_{0} \, dX_{\rm C} \, dX_{\rm O}$. Here $\kappa_{\rm A}$ and $\kappa_{\rm C}$ are the opacities calculated by methods A and C. The opacities are all calculated when $X=0$ and over the surface ($S$) defined by $(X_{\rm C}+X_{\rm O})=0$ to $1$. The vertical artifacts are from the wide spacing of the tables in $\rho$. We show two Wolf-Rayet models on this plot of the same star at different times, solid line $12{\rm M}_{\odot}$ and dotted $6.1{\rm M}_{\odot}$. This second model is the same as shown in figure \ref{fa}. We see that the largest differences in the models indeed occur as the star drifts into the region where the opacity difference is greatest. Also plotted is the dotted line for a $0.3{\rm M}_{\odot}$ pure C/O white dwarf model. While the line passes through regions of the diagram with greater values the white dwarf is only slightly affected because very little mass is in the regions with greatest difference.}
\label{fc}
\end{figure*}

Figure \ref{fc} is a plot of the difference in opacity between methods A and C over the $\rho$ and $T$ plane when there is no hydrogen present. The plot is of the mean difference in the opacity calculated by methods A and C at all possible carbon and oxygen abundances divided by the opacity when $Y=1-Z$. From the plot we see that the regions of greatest difference are small. The lines show where the Wolf-Rayet model calculated by method C lies at certain points of its evolution. The solid line is for the star at $12.3{\rm M}_{\odot}$ and the dashes is the same star at a later time when its mass has fallen to $6.1{\rm M}_{\odot}$. From figure \ref{fb} we see that the largest changes begin around the $12{\rm M}_{\odot}$ point. This is when the model has moved into the region where the differences between A and C are greatest.

We find the opacity tables alter the timescales for the late burning stages. Method C increases the helium burning timescale by 0.1\% while the carbon-burning time scale increases by 2.5\% relative to method A. These changes come about because the opacity varies with composition and effects the temperature structure and burning rates. While there is no large difference in a single variable the overall effect on structure and evolution is more significant.

\section{Conclusions}

Our major conclusion is that the changes induced by properly including opacities for varying C and O mixtures are small. However they are a good thing to include in stellar evolution codes because they help numerical stability by removing sharp changes in the variation of opacity with composition that are encountered when we use fewer tables and add another level of detail to models without much loss of computational speed.

The main computational cost is the extra memory to store the tables. Method C requires ten times the memory (about 200$\,$Mb) compared to methods A and B. However there is little computational speed cost: relative to method A, methods B and C require about 4\% more time for the evolution of a Wolf-Rayet star even though we must evaluate four times as many opacities for each $\mathcal{R}$ and $T$ value than with method A.

\newpage

\chapter{The Evolution of Single-Star Supernovae Progenitors}
\begin{center}
``Life is pleasant. Death is peaceful. It's the transition that's troublesome.''\\
\textit{Isaac Asimov.}
\end{center}


\section{Introduction}
The three main factors that determine whether a single star will go supernova, and the type of supernova it will become, are initial mass, composition and mass loss. These factors are linked, generally a more massive star experiences greater mass loss and higher metallicity also enhances mass loss. These factors affect the composition of the ejecta, the main detail to consider in deciding the resultant SN type. The outcome of evolution depends on a competition between mass loss and the nuclear reactions at the centre of the star. If the core collapse occurs before much mass is lost a type II SN occurs. However if mass loss is severe all hydrogen is removed and a type I SN occurs. In the most extreme cases mass loss can affect the nuclear evolution of the core and lead to smaller core masses.

Single stars lose mass in stellar winds. For red supergiant stars and Wolf-Rayet stars the mass loss mechanism is uncertain. There are a number of proposed mechanisms including magnetic fields, opacity effects and pulsations \citep{WRwinds1}. For Wolf-Rayet stars the most promising explanation currently seems to be optically thick winds \citep{WRwinds2}. 

Mass loss in stellar models is achieved by altering the mass boundary condition. The rate of mass loss at a particular point in a stars evolution is taken from analytical expressions derived from theory e.g. \citet{VKL2001} and \citet{KD2002} or empirically determined from observation e.g. \citet{dJ}, \citet{NJ90} and \citet{NL00}. The current problem is that there are a number of different prescriptions with different groups using their favoured rates. Also there is evidence that mass-loss rates may be lower than observed because clumping in the winds leads to overestimates of the mass ejected \citep{WRclump1}. The classic case of this is that the WR mass-loss rates have been lowered from initially determined values by about one third due to this effect. There is now growing evidence that clumping is also present in the mass loss of main sequence OB stars. This means that there is a tendency for mass-loss prescriptions to overestimate the mass loss from single stars. Binary stars are likely to be a source of extra or enhanced mass-loss in stars, rapid rotation could be another possibility.

In this chapter we describe the main types of single star progenitor predicted by theory, with close consideration of the minimum mass for a SN to occur. We then look in detail at progenitor models and discuss their evolution at three metallicities, solar, LMC and SMC. We discuss which details of these models we use to determine the type of SN a progenitor gives rise to. 

\section{Low mass progenitors}

\begin{center}
``Balderdash, I'll be the judge of who's cool using the coolometer! Good Lord!\\ I'm getting a reading of over 40 MegaFonzies!''\\
\textit{Professor Farnsworth, Bender should not be allowed on TV, Futurama.}
\end{center}

Stars with zero-age mass less than $20\,M_{\odot}$ give rise to red supergiant SN progenitors. These progenitor stars have cool surface temperatures and radii of many hundreds of solar radii. They should be the most common single-star progenitors observed. We can split them into two groups distinguished by the core-collapse mechanism, either iron core-collapse by photodisintegration or ONeMg core-collapse by electron capture.

\subsection{Normal SN Progenitors}
As the normal case we refer to the most common type of SN progenitor a red supergiant where the nuclear fusion reactions have progressed from hydrogen burning all the way to silicon burning ending in the formation of an iron group elements core. While the internal nuclear evolution has occurred the envelope has expanded to a star with a radius hundreds or thousands of times that of the sun. With this expansion the surface temperature of the envelope drops to relatively cool temperatures of less than $4000 {\rm K}$.

The pre-SN structure of the core has multiple shells with an iron core at the centre surrounded by the earlier nuclear fusion reactions burning in shells with the hydrogen burning shell outermost. After the formation of the iron core further nuclear reactions are endothermic so the core collapses by the photodisintergration of iron. The SN event is complex and includes much extreme physics \citep{snsim1,snsim2,snsim3,snsim4}. Collapse leads to the formation of a neutron star or black hole and releases a vast amount of energy in neutrinos that leads to the ejection of the mass surrounding the compact remnant. This gives rise to a type II SN. If little of the hydrogen envelope has been lost, a type IIP SN. These are SN that have a long plateau phase where the SN luminosity remains constant. This is because the photosphere which is located at the point of hydrogen reionisation remains at a constant radius. This is due to the ejecta expanding outward while the ionisation front moves inwards in mass.

Observations of the progenitor to SN2003gd have found it to be a red supergiant \citep{SJM03}. This and other data on the progenitors of type IIP SN from \citet{S03a} confirm that these normal progenitors do exist and maybe quite common.

\subsection{Super-AGB Progenitors}
There are two paths of evolution to become an AGB star. For lower masses the helium burning shell catches up with the hydrogen burning shell, while for more massive AGB stars (above $3M_{\odot}$) the convective envelope penetrates down to the CO core in second dredge up. This puts the hydrogen and helium burning shells in close proximity, an unstable arrangement that leads to thermal pulses rather than steady shell burning. Furthermore because the hydrogen burning shell is now at a higher temperature near the helium burning shell hydrogen burning occurs as a much greater rate and the luminosity of the star increases by at least a factor of 10. 

Most AGB stars have CO cores and do not ignite carbon. These go on to form CO White Dwarfs as the envelope is removed by a stellar wind during the thermally pulsing phase. If the envelope was not removed these stars would eventually form a degenerate CO core approaching $M_{\rm Ch}$ and carbon would ignite degenerately in the centre of the star leading to a thermonuclear SNe. The resultant SN would have observational characteristics of a type Ia SN with a type II mask. They are sometimes referred to as a type $1 \frac{1}{2}$ as suggested by \citet{type1.5}. 

However stars above about $7M_{\odot}$ experience carbon burning as detailed in \citet{IBEN1}, \citet{IBEN2} and \citet{IBENX}. The lowest mass stars in this range ignite carbon in a shell around the more degenerate central regions that have been cooled by neutrino losses. The burning shell then slowly burns inwards, raising the degeneracy of the inner shell before burning takes place. In more massive stars carbon ignites in the core and then burns outward in brief shell flashes rather than in a steady burning shell. This is due to the mild degeneracy of the CO core when the burning begins. Carbon burning leads to the formation of an ONe core. With the new core the star becomes a Super-AGB star. The carbon burning continues outwards until it reaches the base of the helium burning shell at which point it extinguishes. However the hydrogen and helium shells still continue burning outwards forming a CO mantle above the ONe core. It is likely that mass loss will lead to envelope loss and formation of an ONeWD.

In the cases with the most massive cores it is possible that the envelope is not lost before the core reaches $M_{\rm Ch}$. This can occur in the most massive stars when the core is close to $M_{\rm Ch}$ after second dredge-up. At this point the core collapses, however now there is no carbon to ignite and prevent collapse. Densities can be reached of around $\log (\rho / {\rm g \, cm^{-3}}) = 9.6-9.8$ where electron capture on to $^{24}$Mg can occur. This accelerates the core-collapse and formation of a neutron star and a type II SN. The only difference from other SN is that the core-collapse is via electron capture rather than iron photodisintigration.

We find because of our new nuclear reaction rates and opacity tables the mass ranges for different behaviour have shifted relative to previous studies. It is possible to summarise the behaviour as follows:

\begin{itemize}
\item $M_{\rm noOS} \le 7M_{\odot}$, $M_{\rm OS} \le 5M_{\odot}$, stars undergo second dredge-up and thermal pulses with a central CO core as a thermally pulsing AGB star. These lose their envelope and leave CO white dwarfs before their cores reach $M_{\rm Ch}$.
\item $M_{\rm noOS} \approx 8M_{\odot}$, $6M_{\odot} \le M_{\rm OS} \le 7M_{\odot}$, during or after second dredge-up carbon ignition occurs in a shell because the core is degenerate and has a temperature inversion caused by neutrino losses. The star then undergoes thermal pulses with an ONe core. Since the core mass after dredge-up is less than $M_{\rm Ch}$ these stars can lose their envelopes and form ONeWDs. These are Super-AGB stars.
\item $M_{\rm noOS} \approx 9M_{\odot}$, $M_{\rm OS} \approx 7.5M_{\odot}$, carbon ignites before second dredge-up (in a shell if the centre is degenerate). Thus at dredge-up there is a growing ONe core. If this can reach $M_{\rm Ch}$ before the envelope is lost then the star undergoes a SN. The outcome depends on the nature of the thermal pulses.
\item $M_{\rm noOS} \approx 10M_{\odot}$, $M_{\rm OS} \approx 8M_{\odot}$, the CO core is greater than $M_{\rm Ch}$ before dredge-up. However shell carbon burning (enhanced by a thin neon burning shell in the most massive stars of this type) drives a convection zone that reduces the size of the CO core to $M_{\rm Ch}$ so dredge-up can occur. This CO material is mixed with the envelope and increases the CO abundance at the surface during second dredge-up. After dredge-up the star has an ONe core of $M_{\rm Ch}$ which progresses to a SN by electron capture on to $^{24}{\rm Mg}$. These are Hyper-AGB stars.
\item $M_{\rm noOS} \ge 11M_{\odot}$, $M_{\rm OS} \ge 9M_{\odot}$, the helium core or CO core masses are too great for dredge-up to occur. The limiting mass for the helium core is $3M_{\odot}$ and for the CO core $1.5M_{\odot}$. Nuclear burning in these stars progresses until it cannot support the core and a SN occurs.
\end{itemize}

Where $M_{\rm OS}$ and $M_{\rm noOS}$ are the mass ranges with and without convective overshooting included. The luminosity of SN progenitors determines how many we can expect to find in current observational searches for progenitors. If Super-AGB stars do undergo SN then the luminosity of the lowest mass SN progenitors depends upon the occurrence of second dredge-up in the late stages of evolution \citep{Smartt2002}. In the mass range of interest second dredge-up can increase the final luminosity from $\log (L/L_{\odot}) < 4.6$ to $\log (L/L_{\odot}) > 5.2$. Therefore these stars, if they occur in nature, should provide a large population of luminous red supergiant progenitors. 

Most Super-AGB stars probably do not undergo SNe but lose their envelopes and form ONe white dwarfs before core-collapse can occur, as in the work of \citet{IBEN1} and \citet{IBEN2}.  Also the much higher luminosity increases mass loss. From our models it can take longer than $10^{5}$ years for the burning shells to advance and for conditions in the core to reach those for core-collapse. To lose their envelopes during this time a mass-loss rate of less than $10^{-4} M_{\odot} {\rm yr^{-1}}$ is required. This is not unreasonable for a luminous red supergiant. Also if the helium burning shell is unstable and thermal pulses do occur, the burning shell's advance is slowed and there is more time for mass loss to occur.

\begin{figure}
\begin{center}
\includegraphics[height=105mm,angle=0]{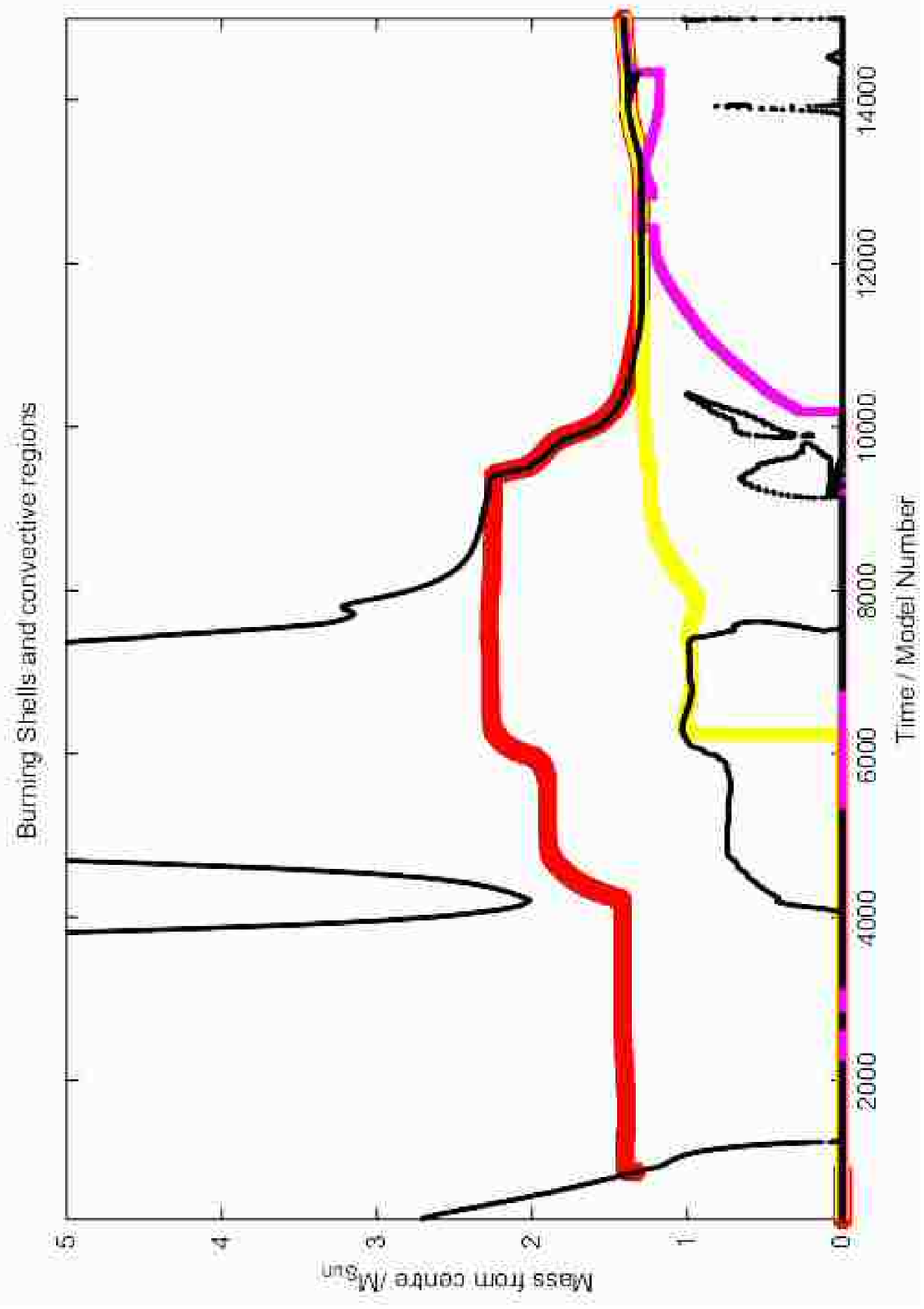}
\includegraphics[height=105mm,angle=0]{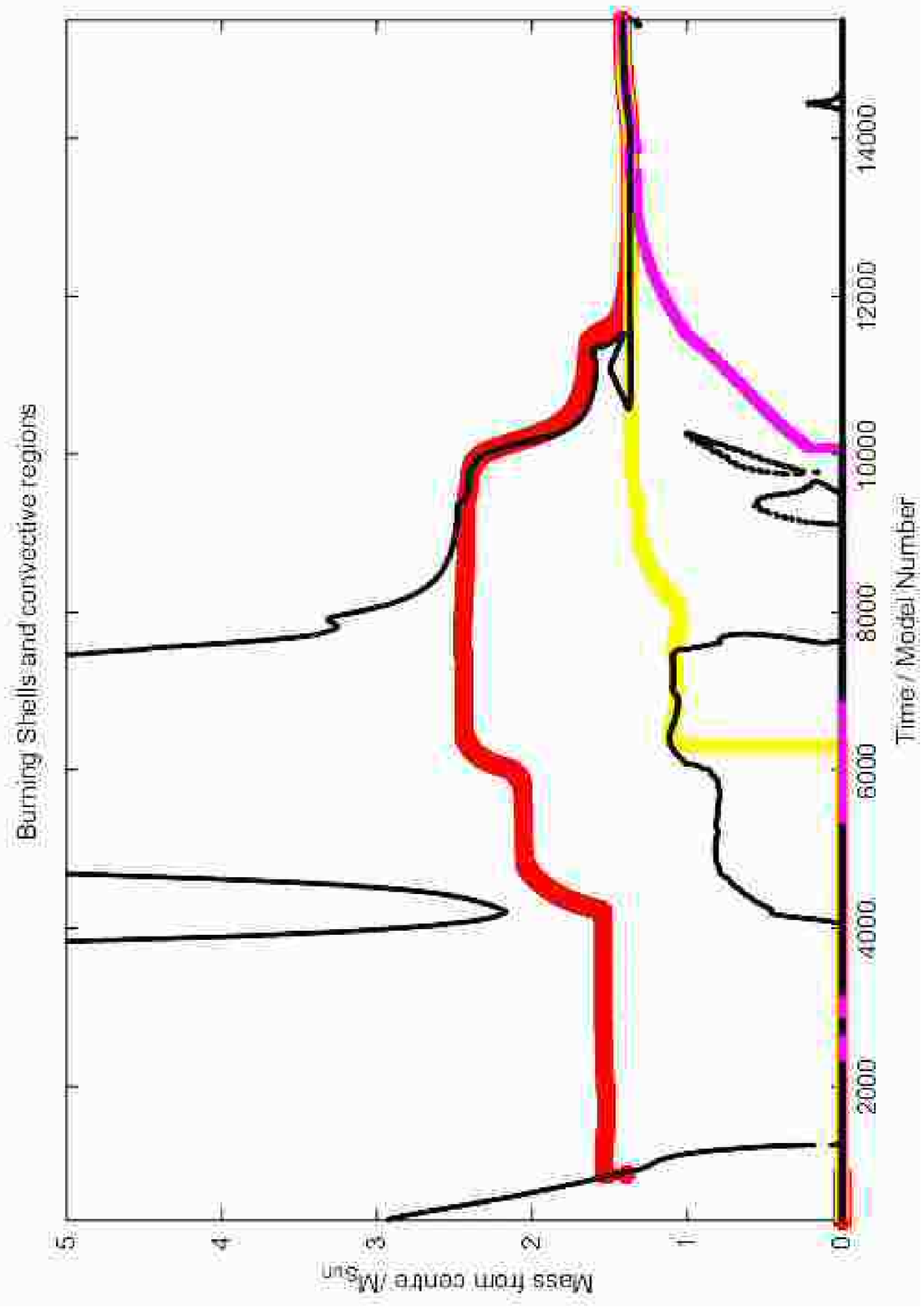}
\includegraphics[height=105mm,angle=0]{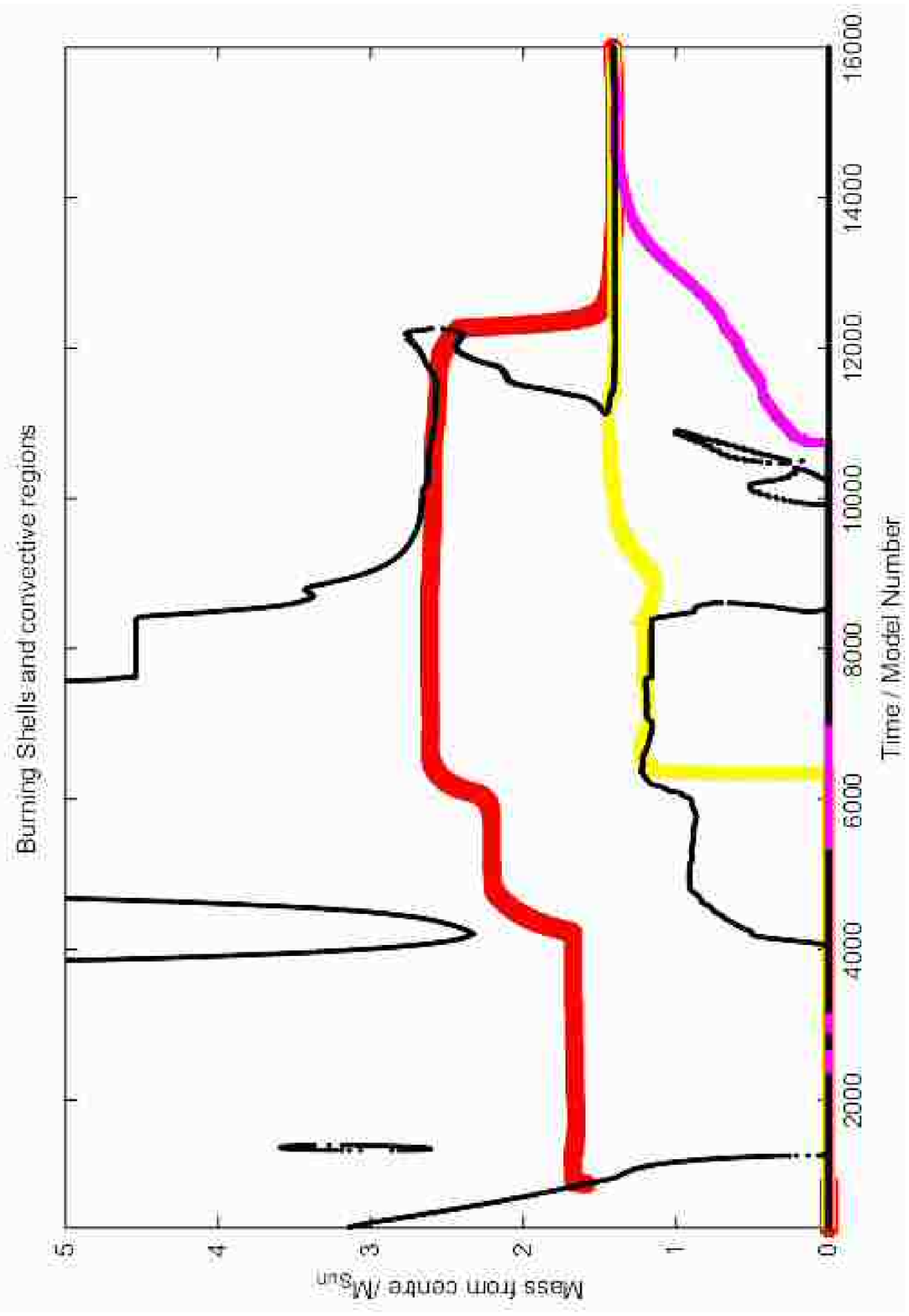}
\includegraphics[height=105mm,angle=0]{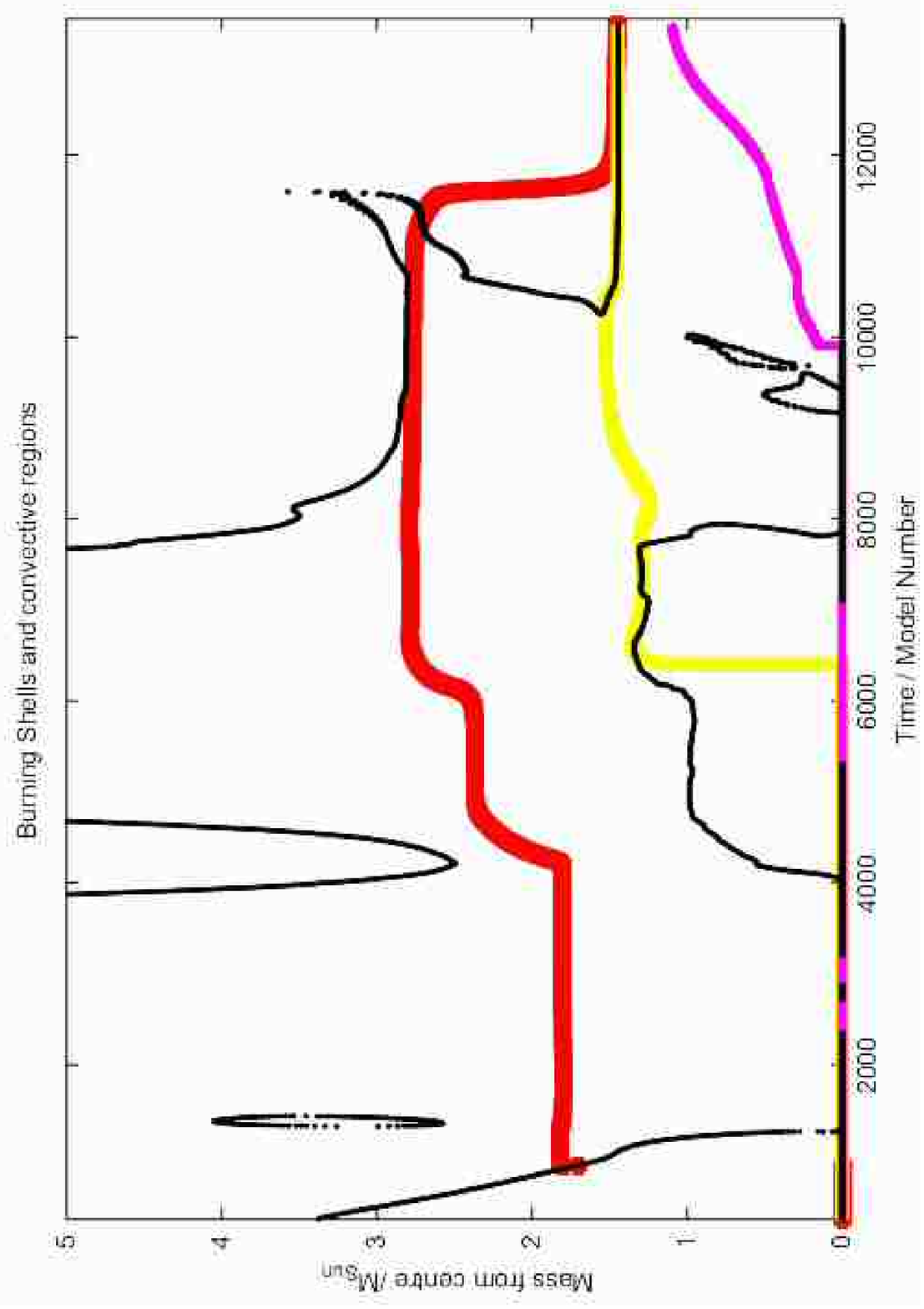}
\end{center}
\caption[The internal structure of 9, 9.5, 10 and 10.5$M_{\odot}$ stars with solar metallicity and no convective overshooting.]{The internal structure of 9, 9.5, 10 and 10.5$M_{\odot}$ stars with solar metallicity and no convective overshooting. The are shown above, top left, top right, bottom left, bottom right. Evolution continues up to the point the centres experience conditions for electron capture collapse. The thin black lines indicate convective regions, the red line indicates the hydrogen burning shell, the yellow line the helium burning shell and magenta line the carbon burning shell.}
\label{sagb1}
\end{figure}

However there is a class of stars, extreme Super-AGB or Hyper-AGB stars, that have CO cores greater than $M_{\rm Ch}$ when second dredge-up occurs and a convection zone between the hydrogen and helium burning shell forms to reduce the CO core mass to $M_{\rm Ch}$. Figure \ref{sagb1} shows examples of such stars of different masses at solar metallicity without convective overshooting. The $9M_{\odot}$ model undergoes second dredge-up and the core mass grows outwards and collapse occurs at just below $M_{\rm Ch}$. For the $9.5M_{\odot}$ model the CO core mass is just below $M_{\rm Ch}$ at second dredge-up and a small intershell convection zone forms as the convective envelope penetrates downwards. At $10$ and $10.5M_{\odot}$ the CO cores are now greater than $M_{\rm Ch}$ as second dredge-up occurs and the intershell convection zone that penetrates into the CO core reducing its mass is now much larger and eventually merges with the hydrogen envelope at which point helium burning products such as carbon and oxygen are mixed to the surface increasing the observed abundance.

The carbon burning in these stars can reach quite high temperatures, above $10^{9} {\rm K}$. This is high enough for some neon burning to occur so the cores acquire a larger fraction of magnesium. These stars do not have to wait for the hydrogen and helium shells to burn outward to reach $M_{\rm Ch}$. Neutrino losses cool the core and increase the central density to the point where electron capture can occur, we take this to occur at a central density of $\log (\rho / {\rm g \, cm^{-3}}) = 9.8$ \citep{ecapture1,ecapture2}. We define this as the point when the SN occurs and we end the evolution of the model.

\begin{figure}
\begin{center}
\includegraphics[height=130mm,angle=270]{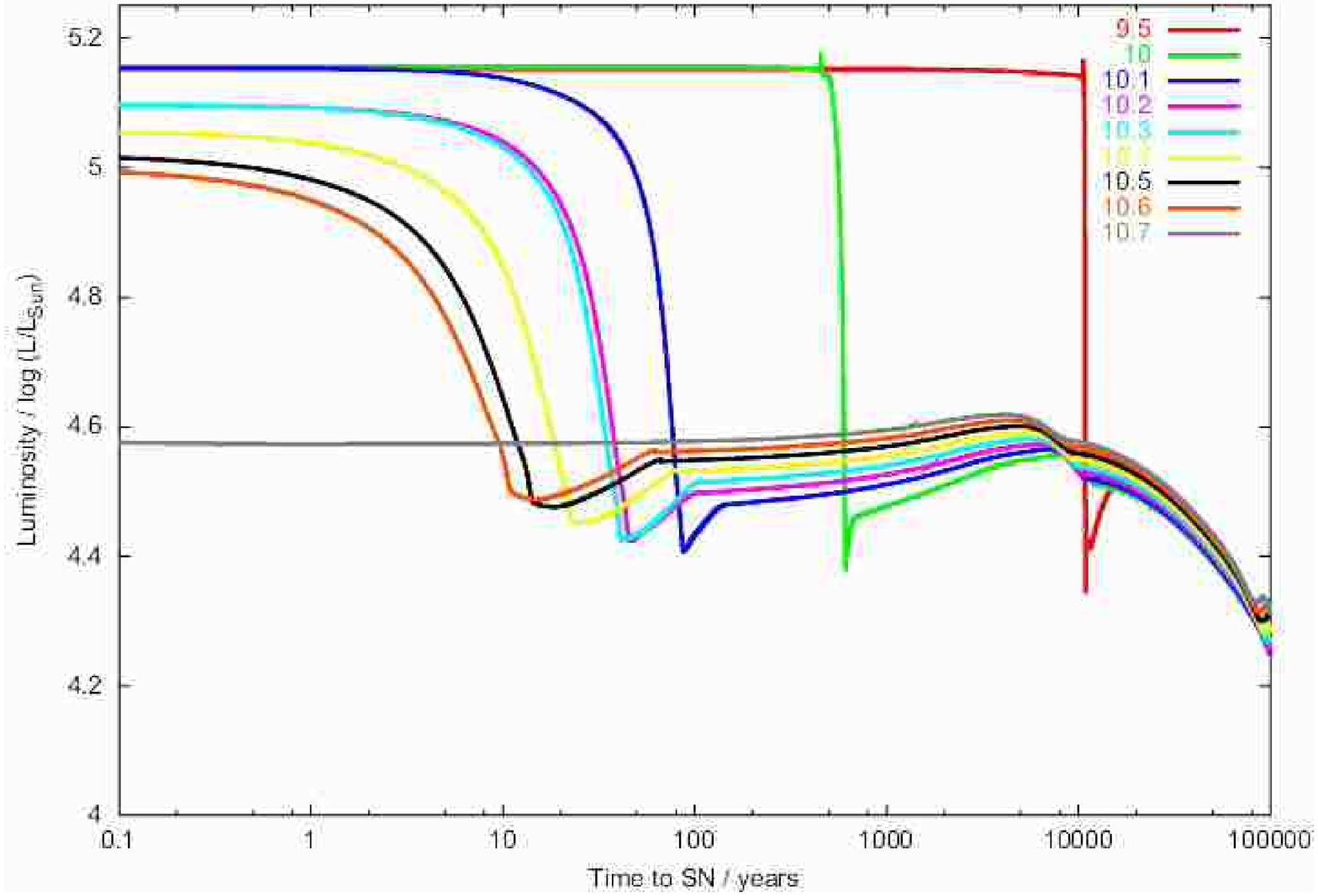}
\includegraphics[height=130mm,angle=270]{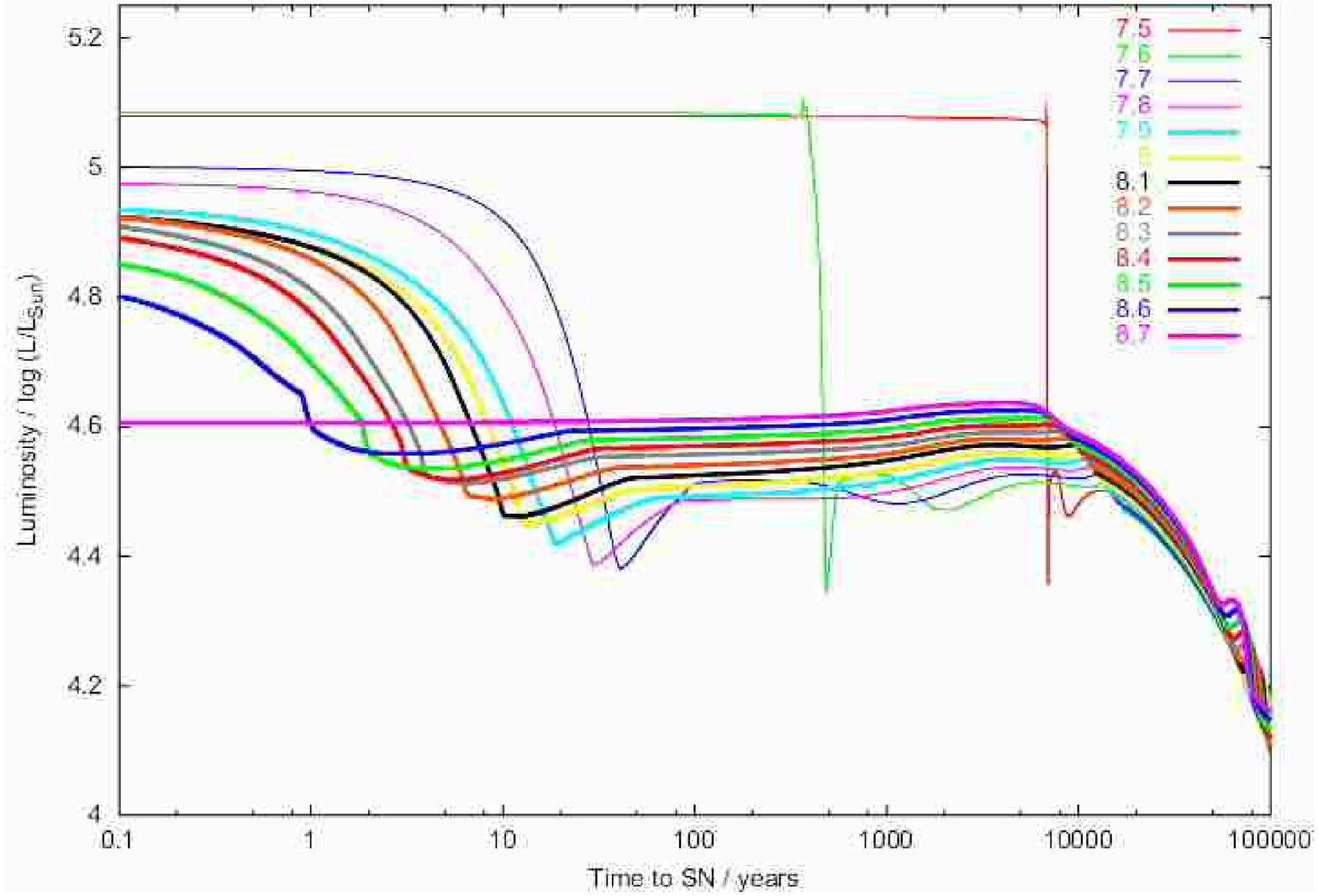}
\end{center}
\caption[The luminosity of solar metallicity SN progenitor models before the central density reaches the conditions required for core-collapse.]{The luminosity of solar metallicity SN progenitor models before the central density reaches the conditions required for core-collapse. The upper plot is without convective overshooting, the lower plot with convective overshooting. All units are in solar units.}
\label{sagb2}
\end{figure}

Extreme Super-AGB stars are difficult to model. The carbon burning shell flashes cause the timestep required to change unpredictably. In the later stages of the most massive extreme Super-AGB stars a neon burning shell ignites in the degenerate ONe core before collapse occurs and causes numerical instabilities that make further evolution troublesome. To experiment with this important range of masses we make some simplified models. We artificially limit the carbon and neon burning rates. This means they do not runaway in degenerate regions where there is no thermostatic control. They proceed to burn but more slowly than might be the case in reality. The effect can be see in figure \ref{sagb1} where the carbon burning shell moves out in a stable convective shell rather than the separate convective shells that occur when the burning proceeds via brief flashes of burning. These models provide an upper limit to the maximum mass of extreme Super-AGB stars. Neon burning in the ONe core is likely to have a small effect on preventing second dredge-up and at the highest masses the burning may progress all the way to the formation of a small iron core and collapse might not occur via electron capture.

In figure \ref{sagb2} we plot the luminosity of extreme Super-AGB models with and without convective overshooting against the time before the central density is reached when electron capture reactions would initiate. In these models we have not included mass loss. It has little effect on the final outcome of these models. The lowest mass models are on the limit of Super-AGB to extreme Super-AGB stars and may eject their envelope to form an ONeWD. The more massive models increase in luminosity 100 years to 1 year before SN. This is because before second dredge-up the stars have more massive cores which leads to a denser core after second dredge-up. Therefore less time is required to achieve the conditions for core collapse.

The luminosity rise from second dredge-up is substantial in all cases. Interestingly in figure \ref{sagb2} just before the increase there is a slight fall in the lightcurve. This is due to the hydrogen burning moving as the convective envelope penetrates into the helium core. The most massive star included in figure \ref{sagb2} does not experience second dredge-up and is the lowest luminosity progenitor that is possible. The only red supergiant progenitor observed, SN2003gd, had a pre-SN luminosity of $\log (L/L_{\odot}) = 4.3 \pm 0.3$. The lowest luminosity progenitors are within $1\sigma$ of this value. We can be sure that the progenitor did not go through second dredge-up because the observations of the progenitor were taken less than one year before the SN. Progenitors with a lower luminosity than this model might exist as an outcome of binary evolution or we might consider how to prevent second dredge-up. One way to do this would be by making convection less efficient so the process of second dredge-up is slower.

After observing a number of SN progenitors, we shall be able to determine more about the low mass end of stars that go SNe. If we make observations of fields containing prospective SN progenitors we shall observe that between 9-17\% of all SN progenitors will undergo second dredge-up before SN, with overshooting the percentage is 10-20\% with the IMF of \citet{KTG93}. The exact values depend on the range of masses for S-AGB stars to occurs. The range of Super-AGB stars are sensitive to the strength of convection used in our models. If we inhibit convection we find second dredge-up does not occur before core collapse.

This is a promising observational test of whether electron capture SNe, and S-AGB progenitors exist. The only other possibility is small differences in the nucleosynthesis products in these SN. In normal SN progenitors the collapsing core is surrounded by shells of silicon, oxygen, neon, carbon and helium while in electron capture SN there is only the hydrogen envelope so, important differences may exist, although it appears that a similar amount of $^{56}$Ni is produced in the SN of Super-AGB stars \citep{crazyidea}.

\section{High mass progenitors}

In stars of $M>20M_{\odot}$ black holes can be formed in SNe rather than neutron stars and mass loss becomes severe enough to effect the evolution of the star and remove a large fraction of hydrogen or in the most extreme cases all hydrogen is lost from the star. This loss leaves the core of the star exposed as a naked helium star. These massive stars which are loosing mass in a strong wind are called Wolf-Rayet stars. These are the last type of SN progenitor. Wolf-Rayet stars are classified by their spectra and there are three main classes WN, WC and WO in which different elements are observed. They represent the products of different stages of nuclear burning. WN stars still have some hydrogen and nitrogen is the dominant feature of the spectrum. WC stars start exposing the products of low-temperature helium burning and carbon begins to dominate. Finally WO stars expose the higher temperature helium burning where oxygen becomes the dominant element.

For single stars WR stars are probably formed at $M>30M_{\odot}$ at solar metallicity. However there is a complication in that binary interactions can lead to mass-loss events such as Roche Lobe overflow and common envelope evolution so that stars lower than this limit can lose their hydrogen and become Wolf-Rayet stars. We must be careful when looking at mass-loss rates not to pick a prescription that reproduces the lowest mass WR stars in the sky. These are likely to be in binaries so our single star limit must be higher.

The stars that retain hydrogen in the progenitor will give rise to type II SN. However they are more likely to be type IIL, IIn or IIb because they do not have extended hydrogen envelopes or a dense circumstellar environment with which the ejecta can interact. If there is enough hydrogen for a type IIP SN it may be very different if a black hole is formed at the centre of the collapsing core rather than a neutron star as suggested by \citet{lowLIIP}. Although an alternative scenario proposed by \citet{chuggi} is that these low luminosity IIP SN could also arise from lower mass progenitors with initial mass between 6 and $10 \, M_{\odot}$.

Stars that have lost all their hydrogen give rise to type I SNe, either type Ib or Ic since Ia is a thermonuclear event. SN Ib have helium in their spectra while Ic SN do not. There is some uncertainty in what difference in progenitors give rise to the difference between Ib or Ic SNe and since most progenitor models still retain some helium there is probably a continuum between the two objects but we shall return to this dilemma later.

\section{Mass-loss rates}

Mass loss is the biggest factor in determining the mass at which type I SN occur rather than type II. In chapter~4 we investigate different mass-loss prescriptions and their effect on supernovae progenitors over a wide range of metallicities and initial stellar masses. For the models below we use our preferred mass-loss rates for local metallicities ($10^{-3} < Z < 0.05$) and the inclusion of extra mixing in the form of convective overshooting. We shall examine 3 metallicites, solar, LMC and SMC ($Z=0.02, 0.008$ and $0.004$). 

When deciding which rates to use it is sensible to opt for empirical mass-loss rates if we aim to model nature. The mechanism for mass loss in OB stars is radiatively driven winds and theoretical predictions agree with observed values \citep{VKL2001}. WR theoretical rates do exist but the WR star mass-loss mechanism is not known for certain. It could be driven by radiation driving optically thick winds, pulsations or opacity \citep{WRwinds1,WRwinds2}. Because of this uncertainty it is currently best to use empirical rates for WR stars.

We split our mass-loss rates into pre-WR and WR because the nature of mass loss changes once the hydrogen envelope has been removed and observations show quite different rates. WR stars have greater mass loss than an OB star of the same luminosity. Our preferred rates are essentially the NL rates from \citet{DT03} but with the inclusion of the rates of \citet{VKL2001} for the OB stars.

\textbf{Pre-WR Evolution:} We use primarily the mass-loss rates of \citet{dJ}, here after JNH. The JNH rates are old and complex but considered to be the most accurate \citep{Crow2001}. The rates are dependent on surface luminosity and temperature and are derived from a large pool of observations. We apply the commonly adopted scaling with metallicity of $\dot{M}(Z) = \dot{M}(Z_{\odot}) \times (Z/Z_{\odot})^{0.5}$ as in \citet{H03} and \citet{DrayThesis}. This scaling arises from the assumption that stellar winds are line driven and with lower surface opacity at lower metallicity there are weaker winds. However while there is agreement that mass loss scales in this form there is a range of suggested values for the exponent. We investigate the effect of this scaling in chapter 4. We supplement these rates by with the rates of \citet{VKL2001} for OB stars. These include a theoretically predicted metallicity scaling.

\textbf{WR Evolution.} WR evolution begins just before the hydrogen envelope is completely removed from the star. The definition of when a model enters each stage is arbitrary. We use those defined by \citet{DrayThesis}.

\begin{itemize}
\item WNL: When $X < 0.4$ and $T_{\rm eff} > 10^{4}$.
\item WNE: When $X <0.001$ and $(X_{\rm C}+X_{\rm O})/Y < 0.03$.
\item WC: When $X <0.001$ and $0.03 < (X_{\rm C}+X_{\rm O})/Y < 1.0$.
\item WO: When $X <0.001$ and $(X_{\rm C}+X_{\rm O})/Y > 1.0$.
\end{itemize}

When $X$ is the surface hydrogen abundance by mass, $Y$, $X_{\rm C}$ and $X_{\rm O}$ are the surface abundance by number of helium, carbon and oxygen. During WN and WC evolution we use the rates of \citet{NL00} that depend on luminosity and surface abundance. They find that these depend strongly on composition as well as luminosity. During WO evolution we use a constant rate as in \citet{DT03}. One straightforward test of this mass-loss prescription is to compare the number ratios of WR stars to O stars, and the WR subtype ratios. \citet{DT03} performed this test in detail and found good agreement with observed ratios. The one exception was the WN/WR ratio for which this theoretical value is too low. There are a number of plausible reasons for this such as the effect of binaries on mass-loss or misclassification of some WN stars because spectral features change with metallicity. There is evidence that WR mass-loss rates should be scaled with the initial metallicity \citep{WRZscale}, we return to this in chapter 4 but here we include the scaling with metallicity as for pre-WR evolution of $\dot{M}(Z) = \dot{M}(Z_{\odot}) \times (Z/Z_{\odot})^{0.5}$.

\section{The Evolution of SN Progenitors}

\begin{figure}
\begin{center}
\includegraphics[height=220mm,angle=0]{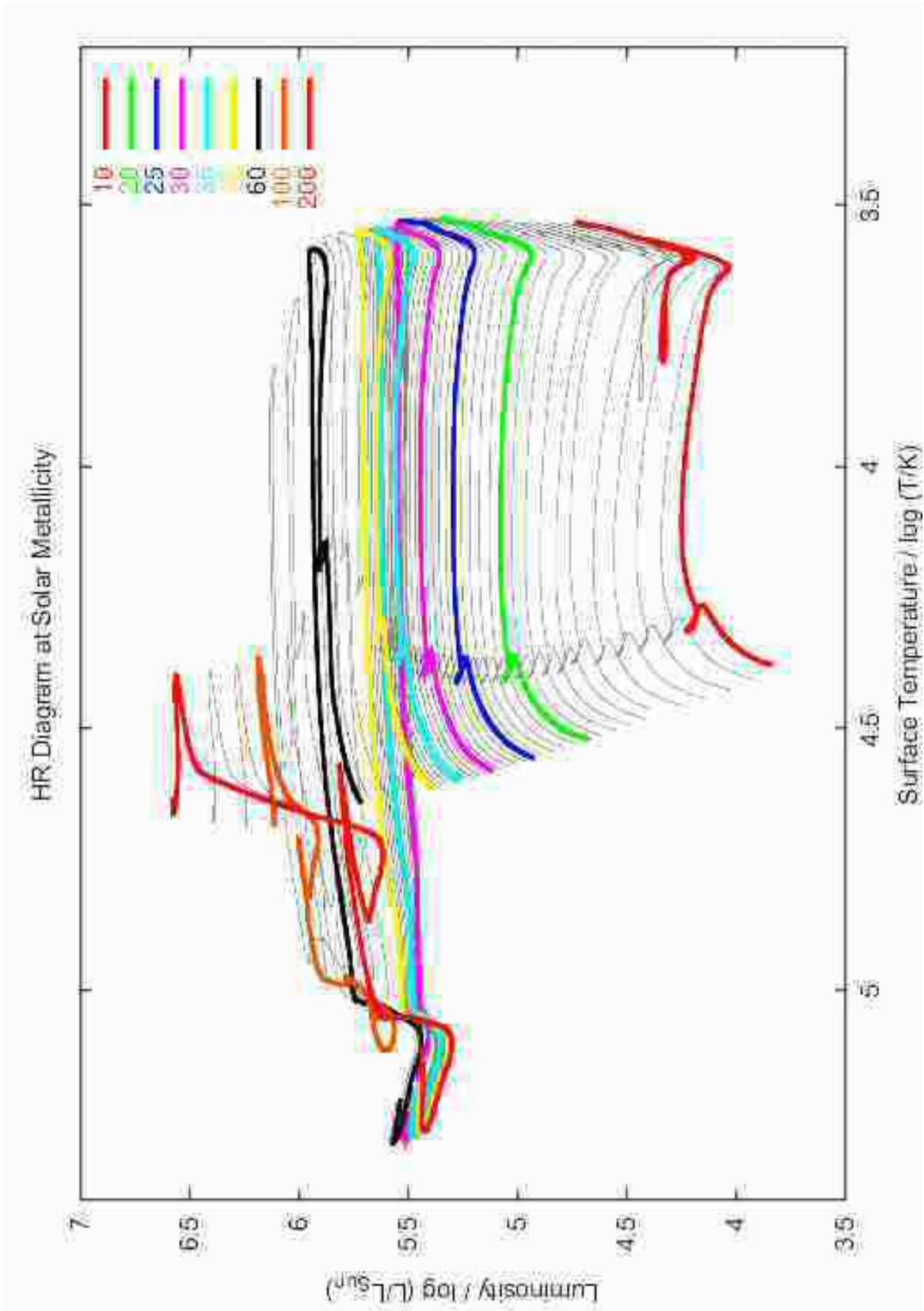}
\caption{Solar metallicity HR diagram. On these diagrams we plot all our models. Selected models have their mass indicated the lighter grey lines are the other models.}
\label{solarHR}
\end{center}
\end{figure}

\begin{figure}
\begin{center}
\includegraphics[height=220mm,angle=0]{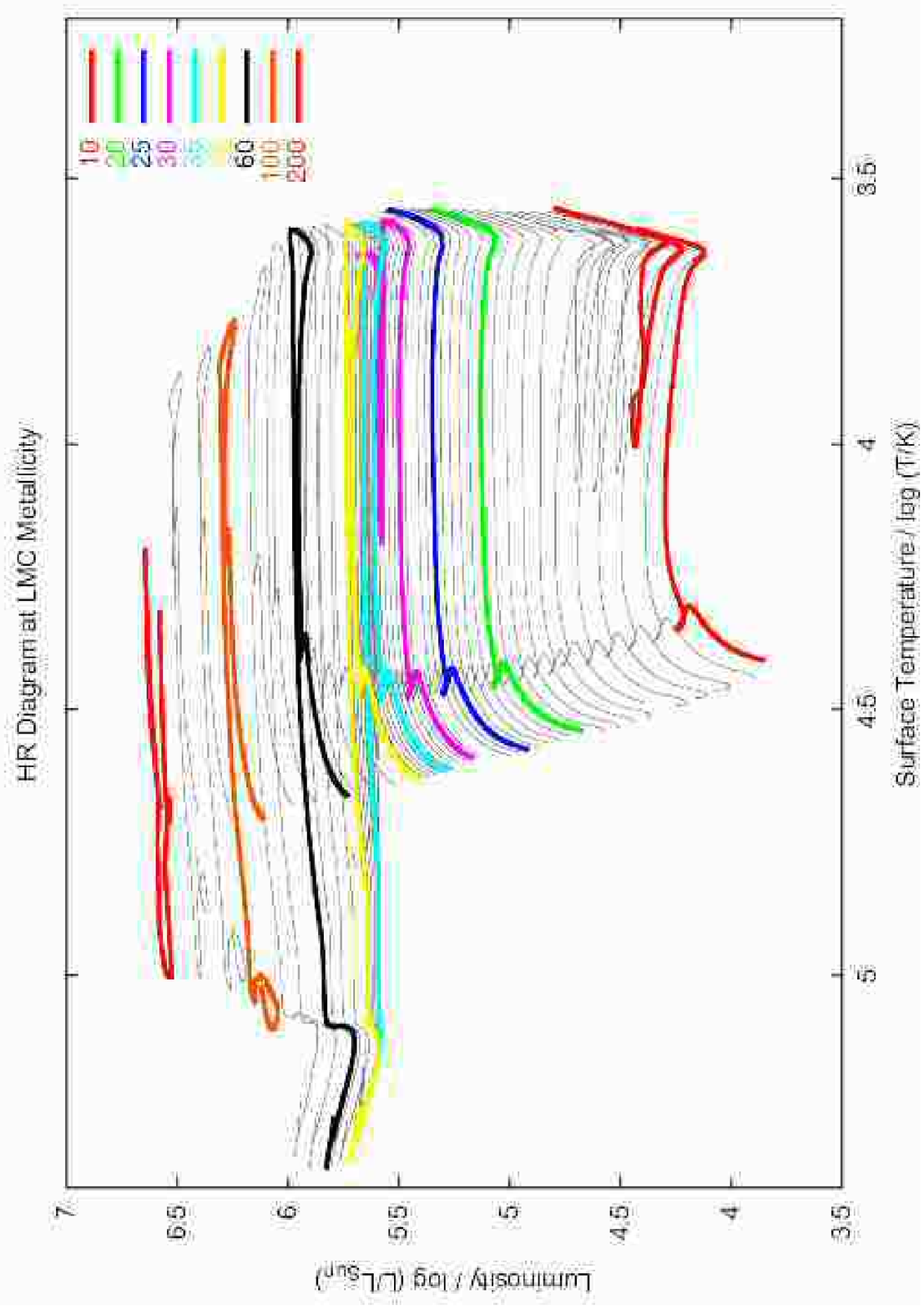}
\caption{LMC metallicity HR diagram. On these diagrams we plot all our models. Selected models have their mass indicated the lighter grey lines are the other models.}
\label{lmcHR}
\end{center}
\end{figure}

\begin{figure}
\begin{center}
\includegraphics[height=220mm,angle=0]{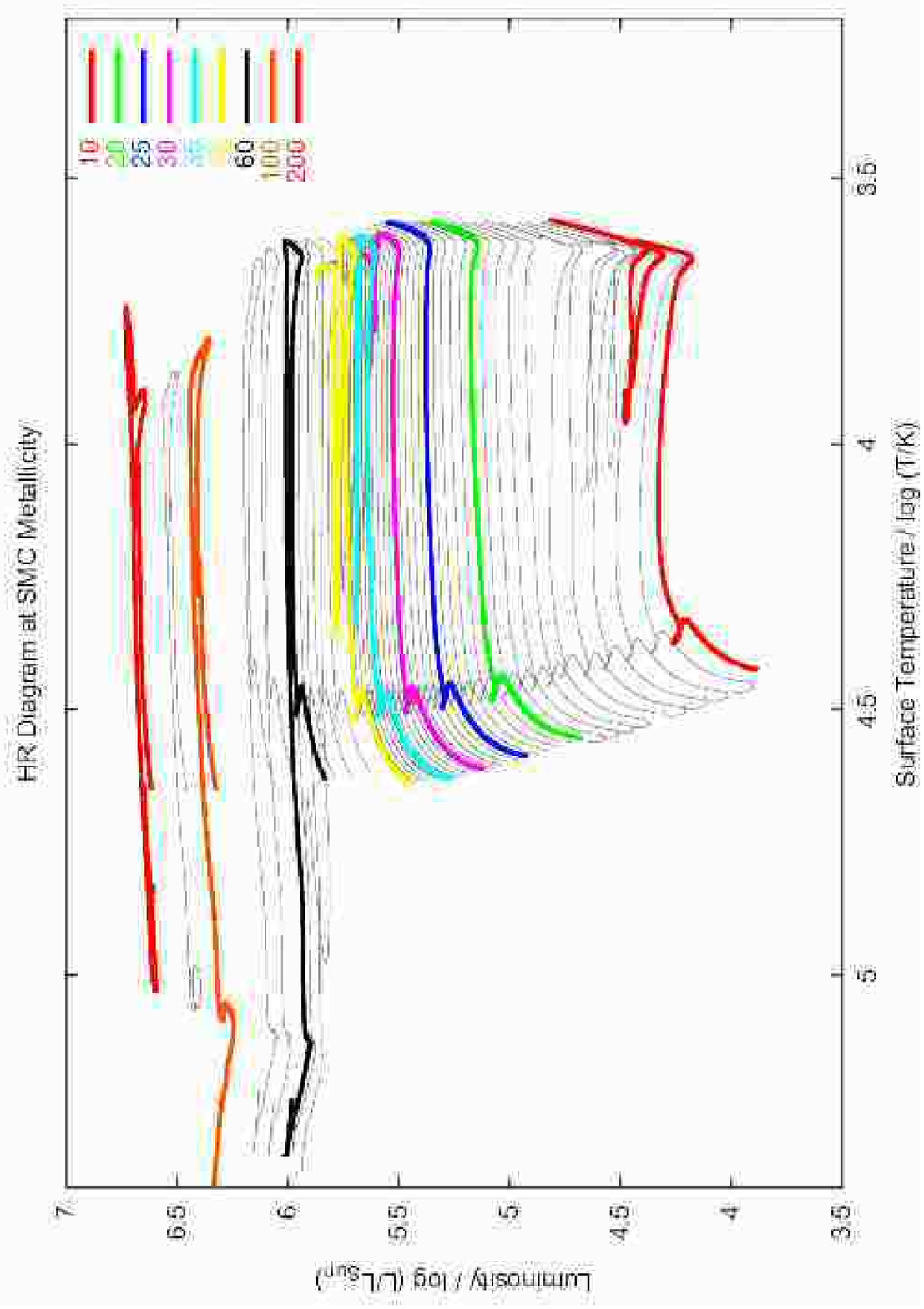}
\caption{SMC metallicity HR diagram. On these diagrams we plot all our models. Selected models have their mass indicated the lighter grey lines are the other models.}
\label{smcHR}
\end{center}
\end{figure}

We can now closely examine the variation between SN progenitors over a range of masses. We look in this case at solar, LMC and SMC metallicities and masses from $10 - 200 M_{\odot}$. We do not go to lower masses to avoid the electron capture SN region. We have included convective overshooting in these models. The compiled HR diagrams for the stars are shown in figures \ref{solarHR}, \ref{lmcHR} and \ref{smcHR}. After evolution on the main sequence stars move over the Hertzprung gap before becoming a red supergiant. Higher mass stars then try to go up the supergiant branch but mass loss drags them back across the HR diagram to the blue side and they become Wolf-Rayet stars evolving along the WR sequence. At the upper end of the mass scale the stars become WR stars straight from the main sequence. For the LMC and SMC samples these transitions happen at higher initial masses than in the solar case. 


\begin{figure}
\begin{center}
\includegraphics[height=100mm,angle=270]{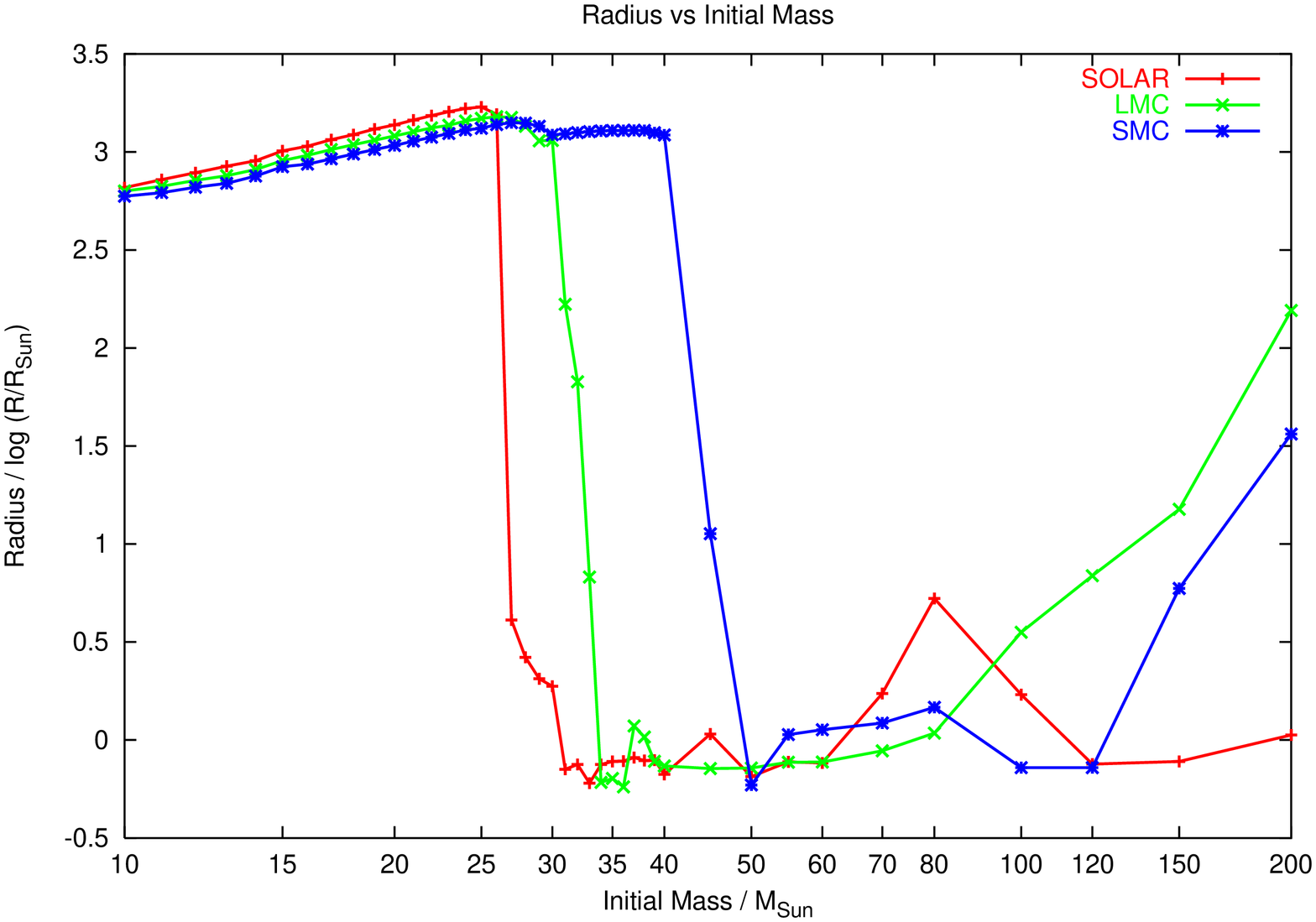}
\includegraphics[height=100mm,angle=270]{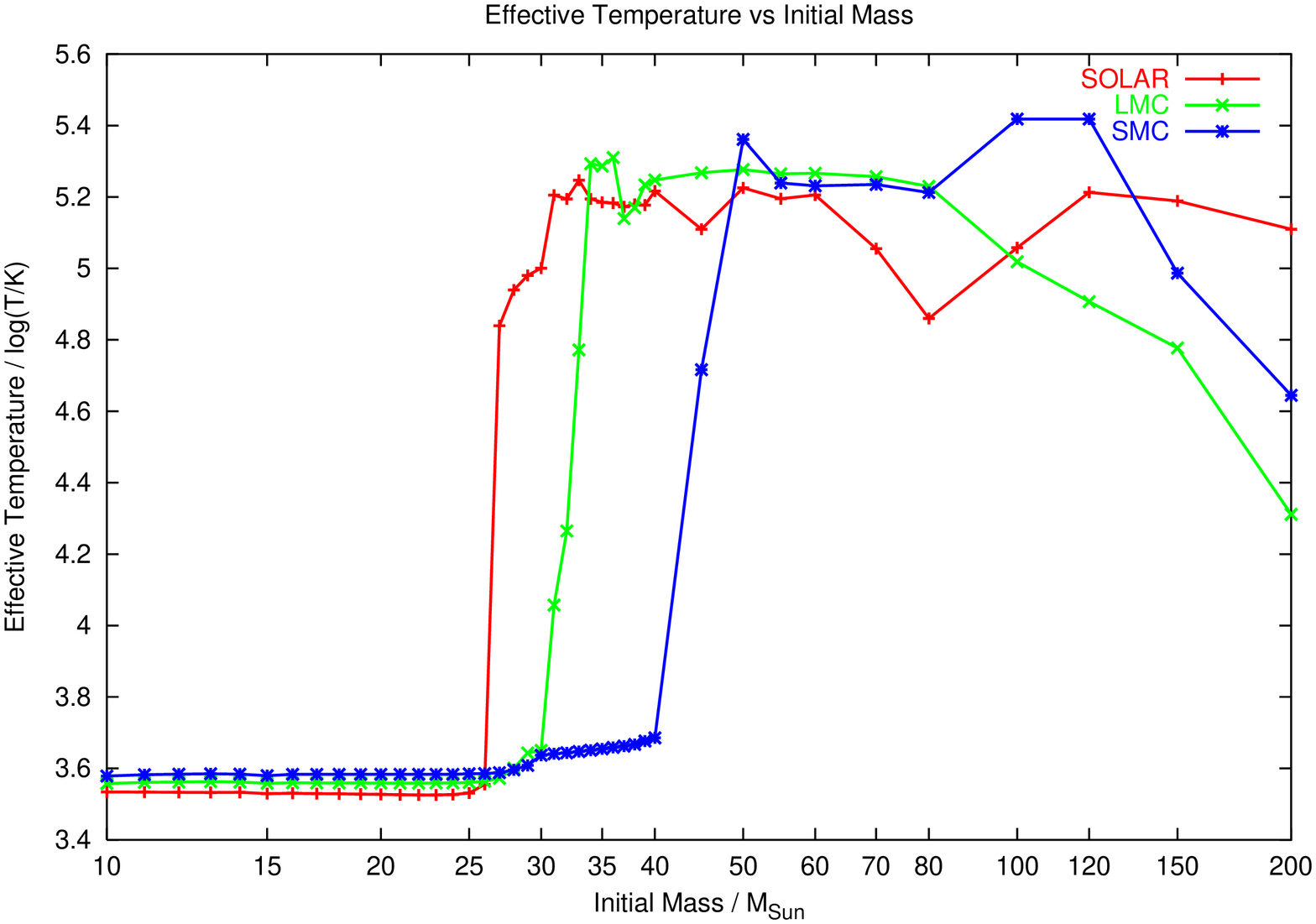}
\includegraphics[height=100mm,angle=270]{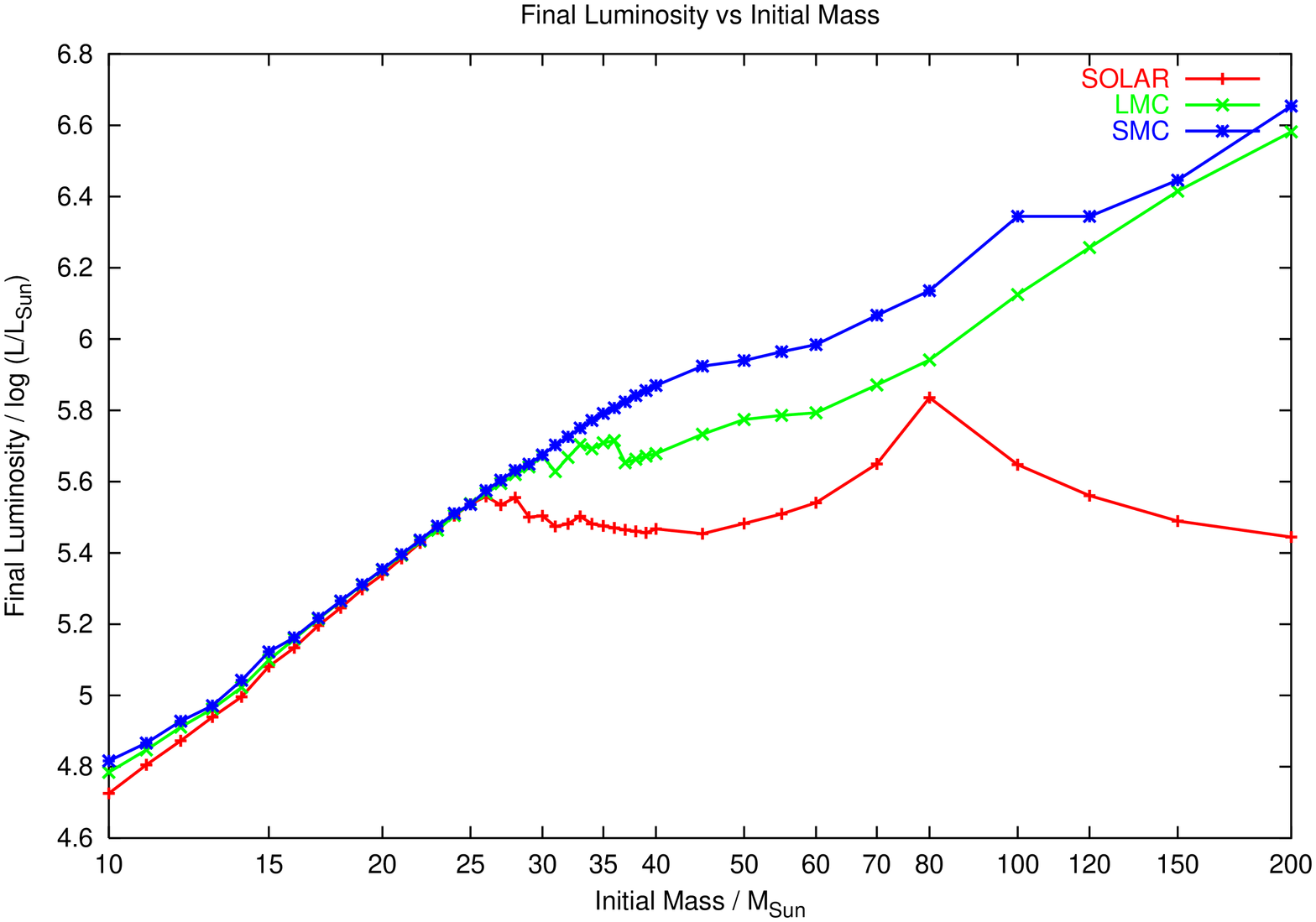}
\end{center}
\caption{The comparison between the different metallicities of the final radius, surface temperature and luminosity of our models versus initial mass.}
\label{3mods1}
\end{figure}

\begin{figure}
\begin{center}
\includegraphics[height=130mm,angle=270]{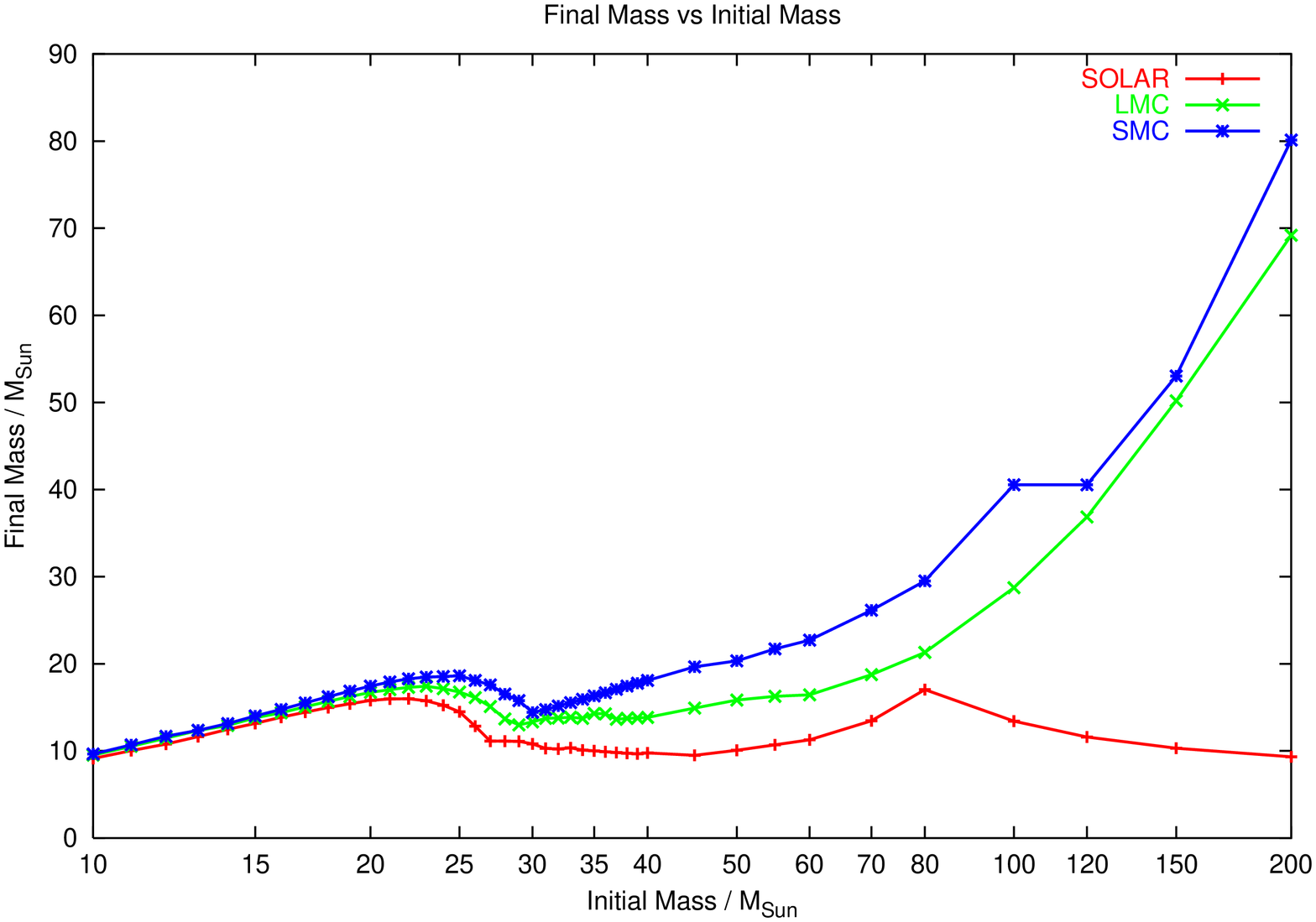}
\includegraphics[height=130mm,angle=270]{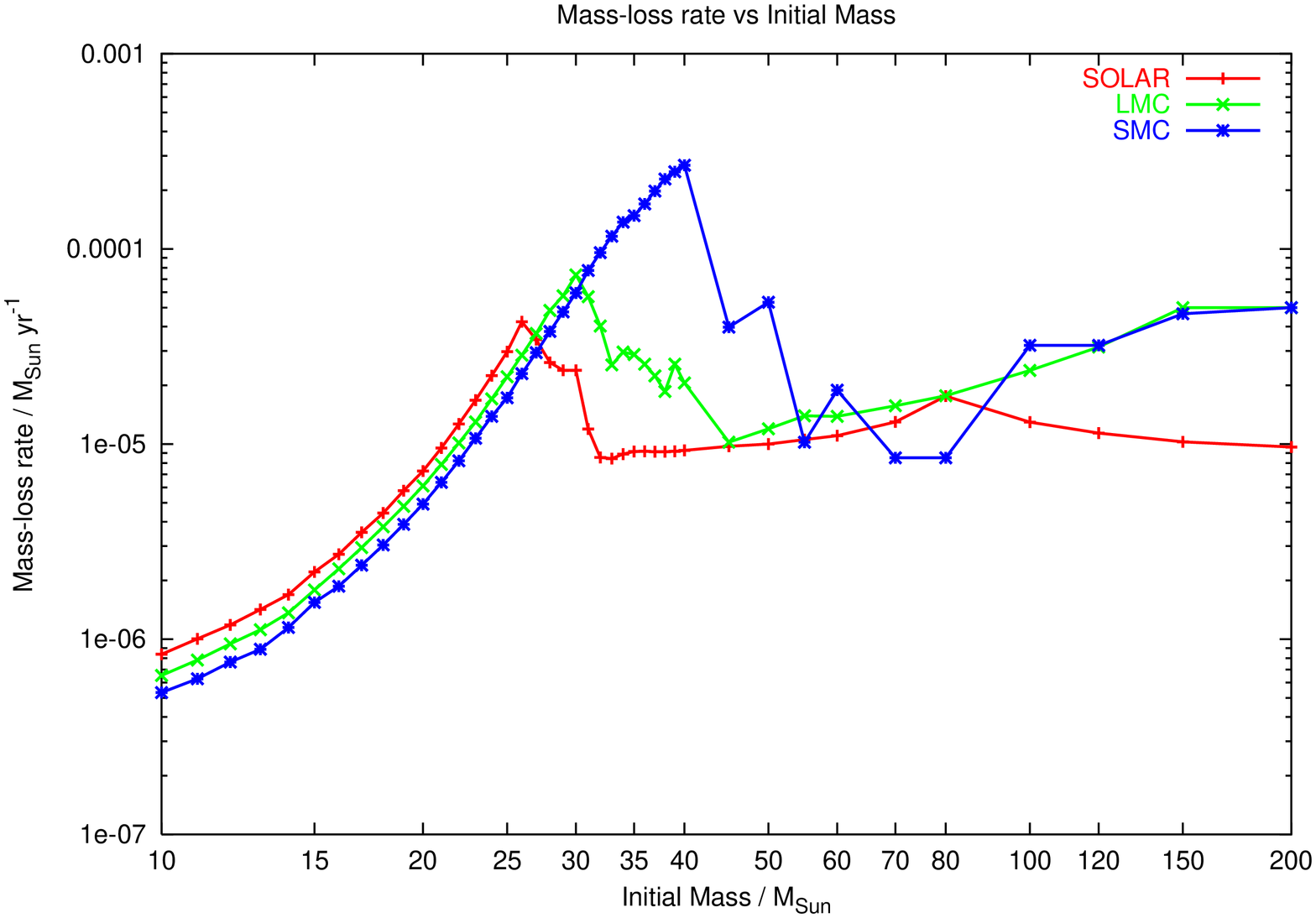}
\end{center}
\caption{The comparison between the different metallicities of the final mass and mass-loss rate versus initial mass}
\label{3mods2}
\end{figure}

\begin{figure}
\begin{center}
\includegraphics[height=70mm,angle=0]{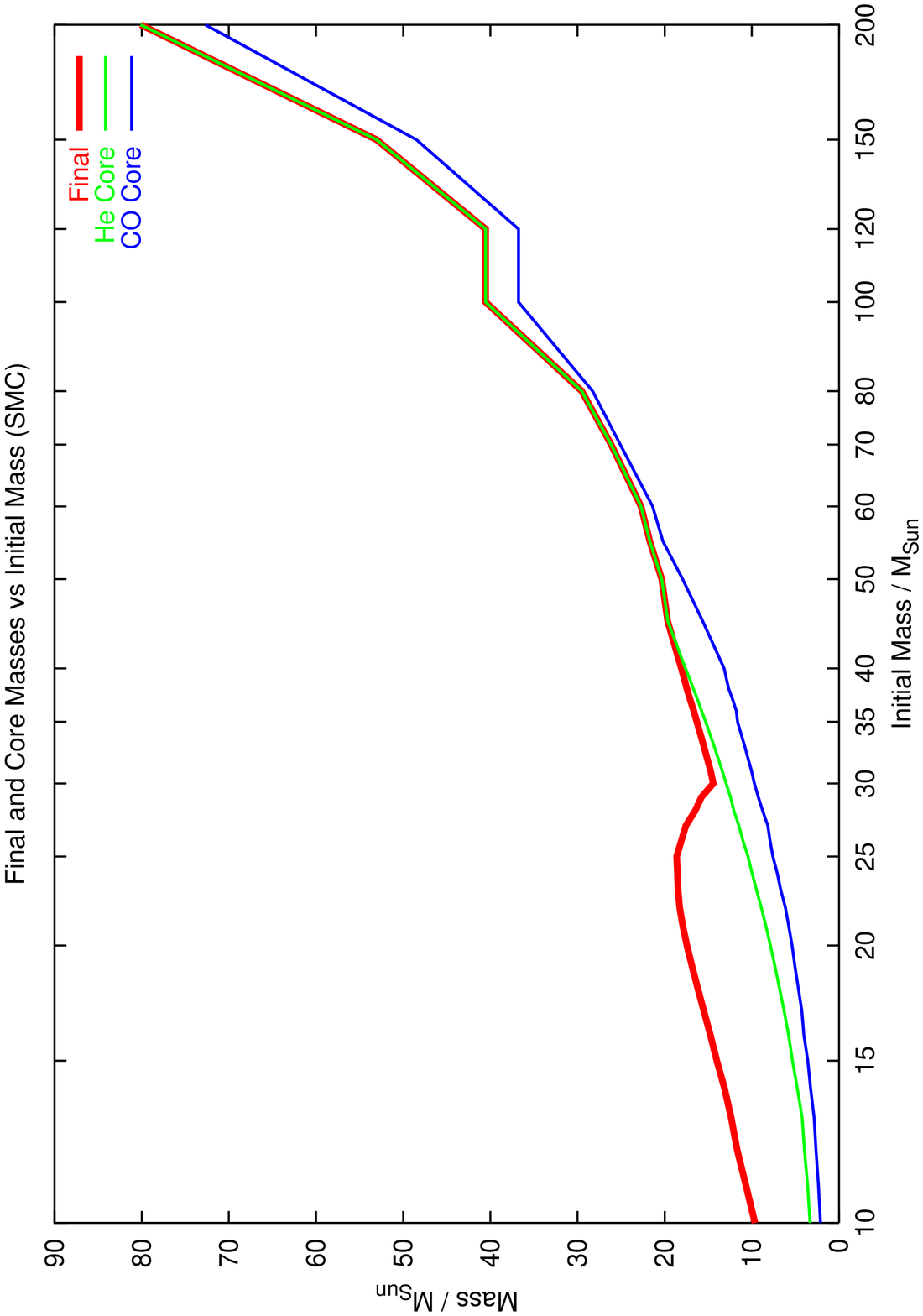}
\includegraphics[height=70mm,angle=0]{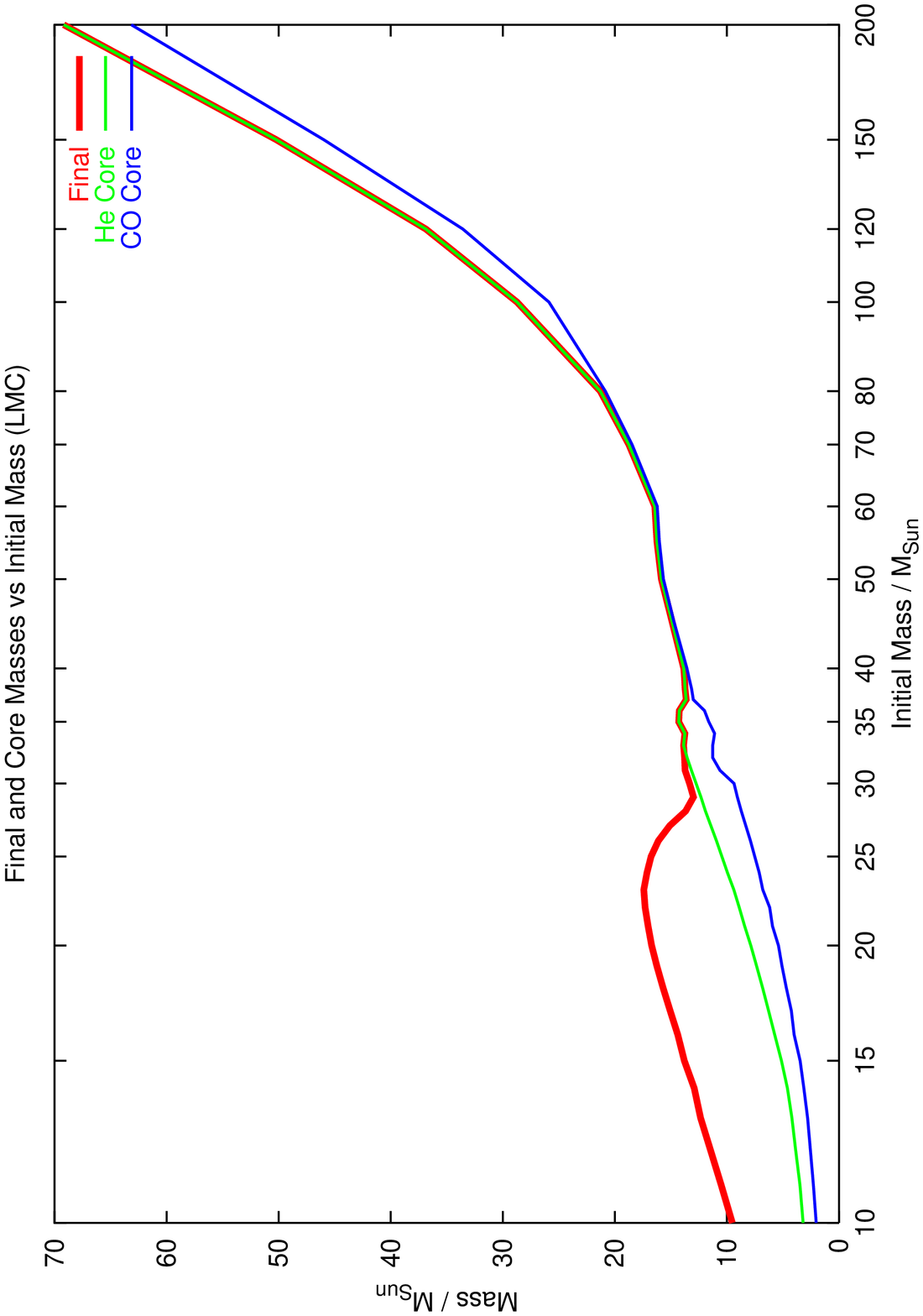}
\includegraphics[height=70mm,angle=0]{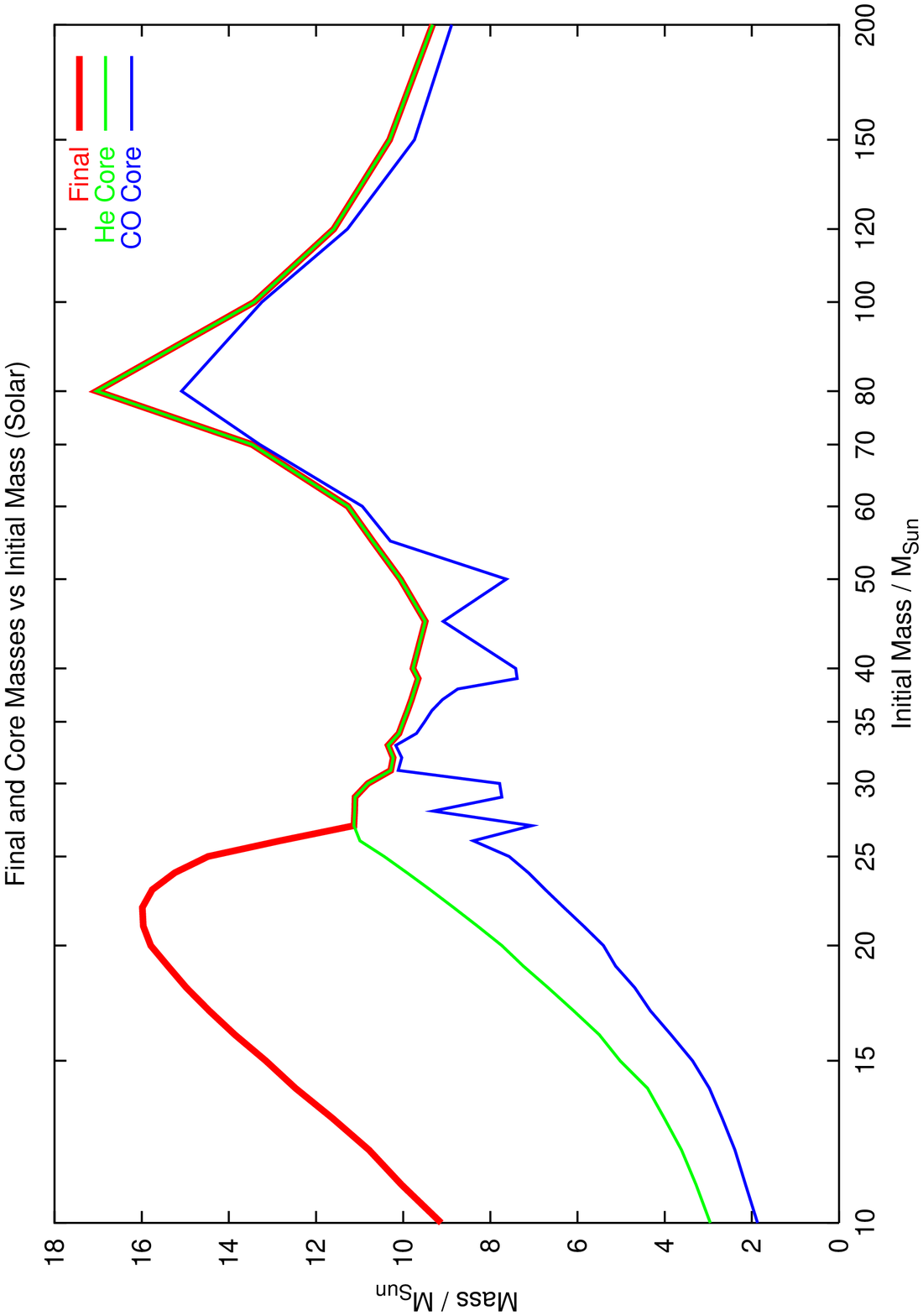}
\includegraphics[height=70mm,angle=0]{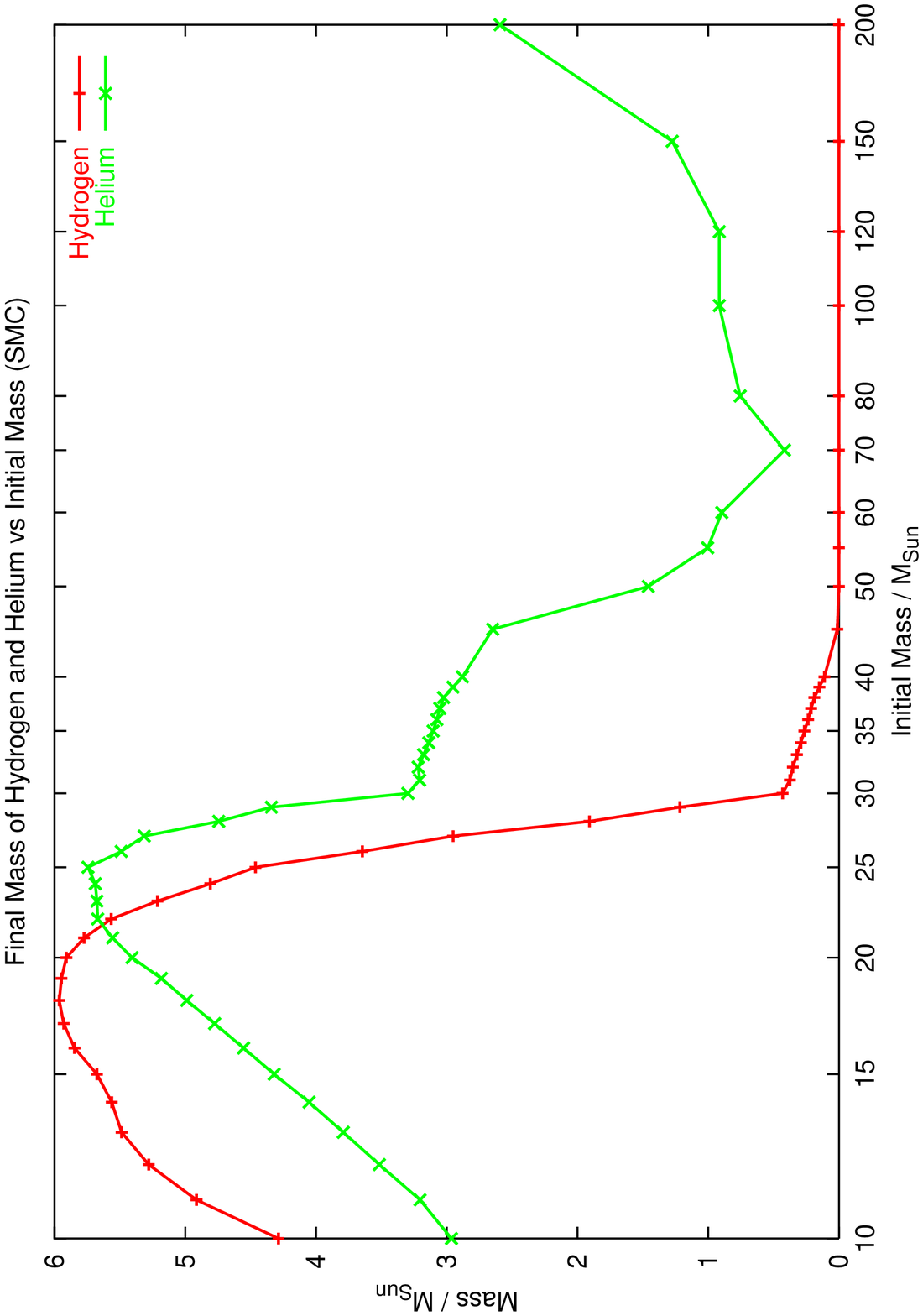}
\includegraphics[height=70mm,angle=0]{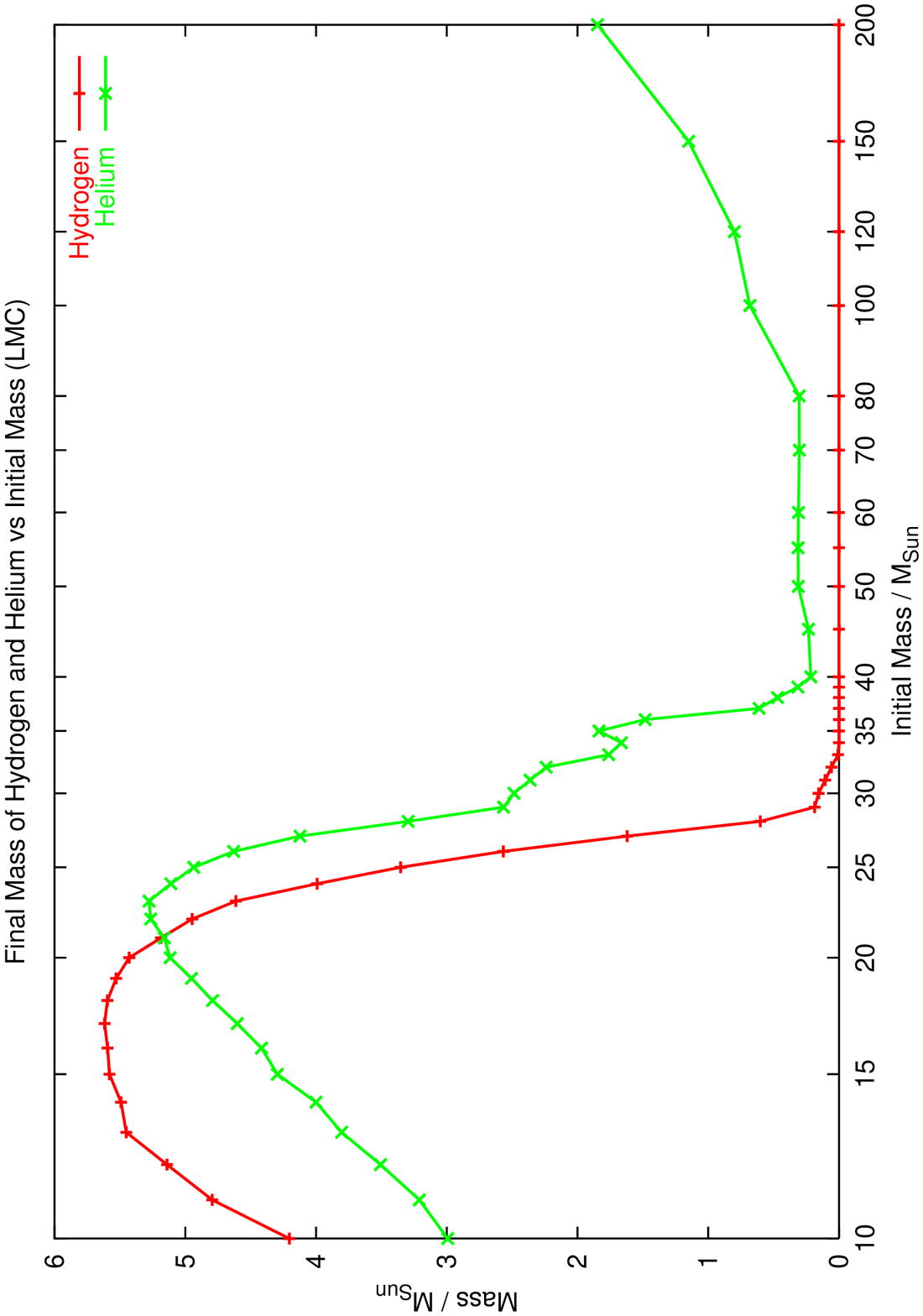}
\includegraphics[height=70mm,angle=0]{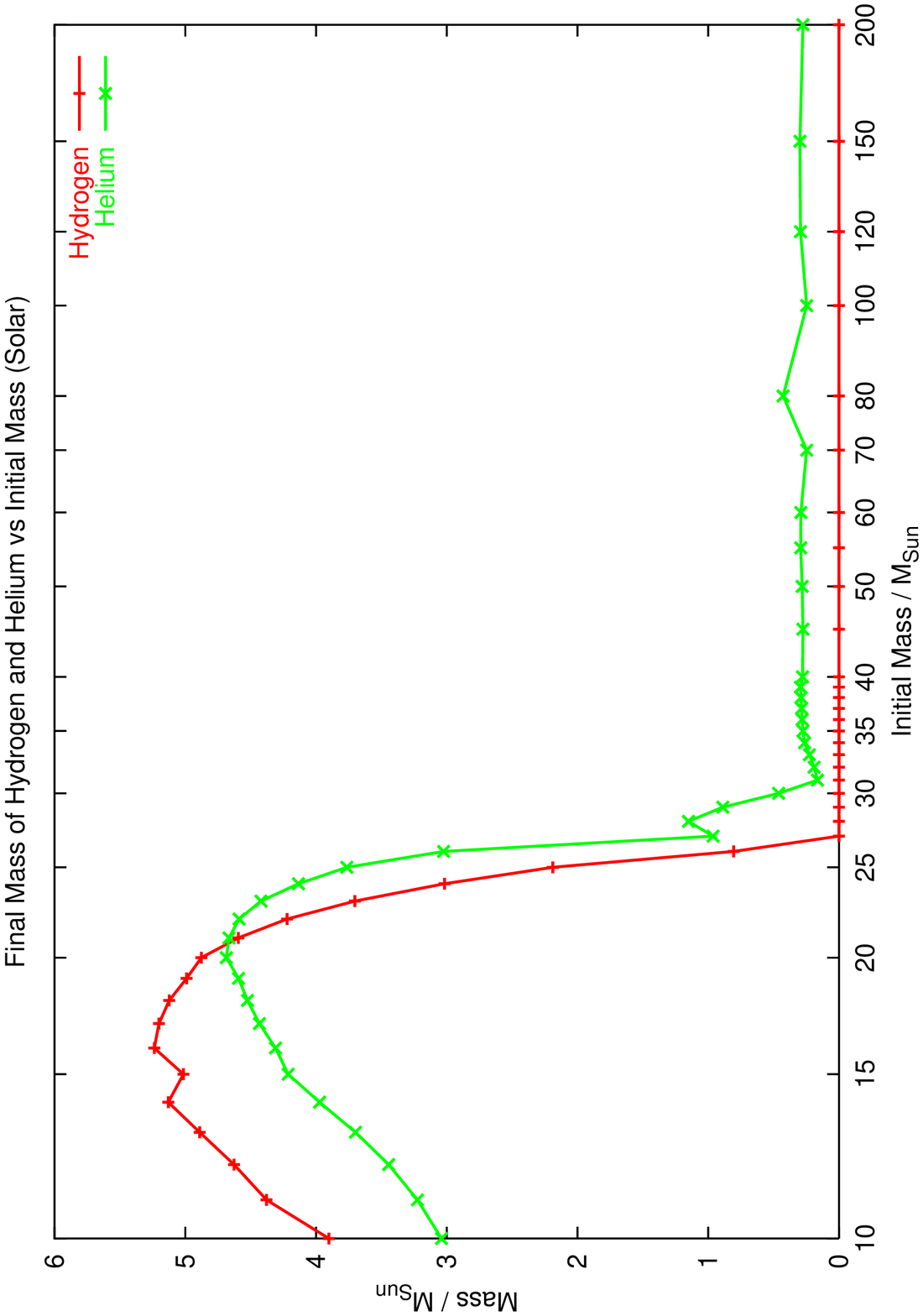}
\includegraphics[height=70mm,angle=0]{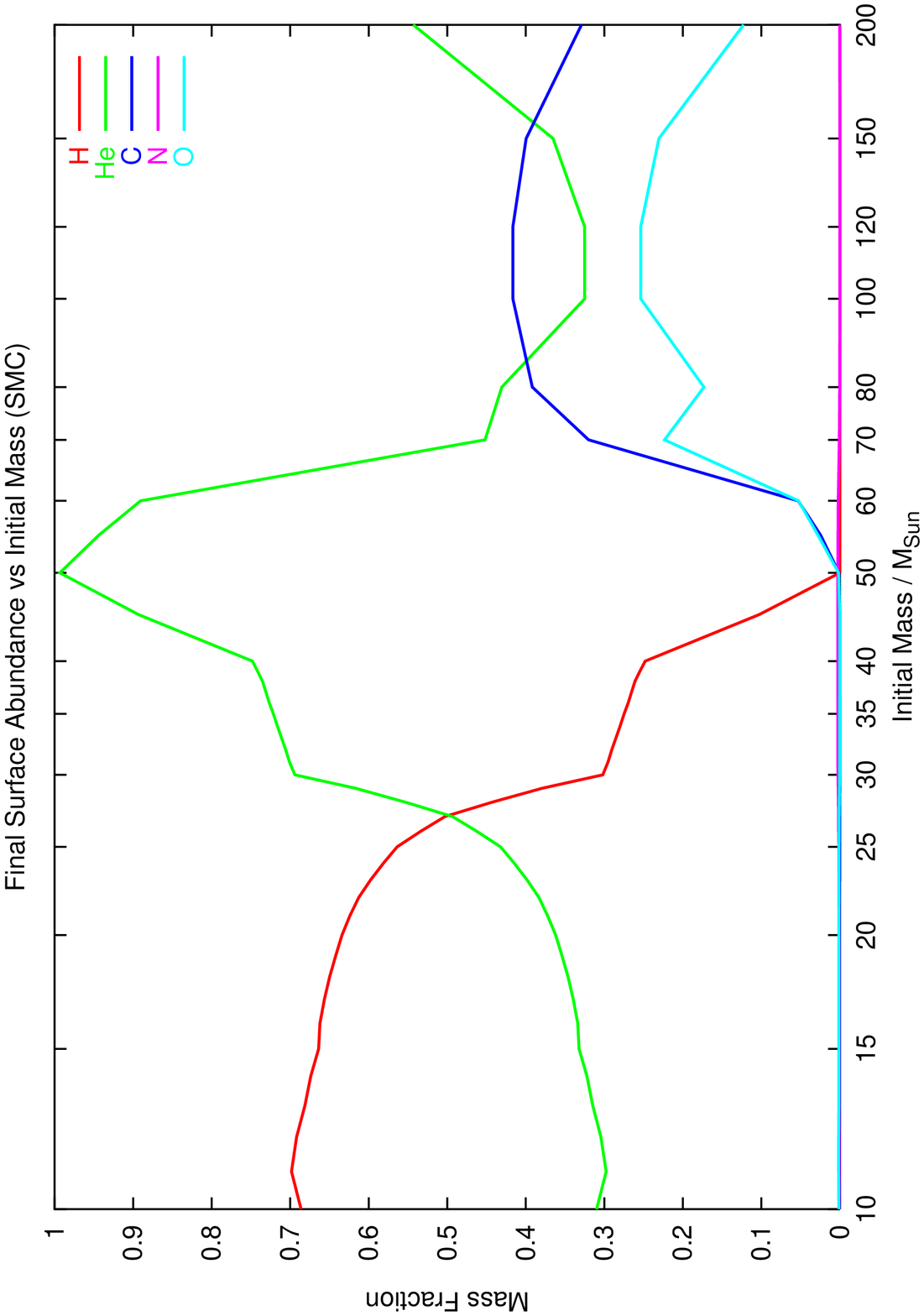}
\includegraphics[height=70mm,angle=0]{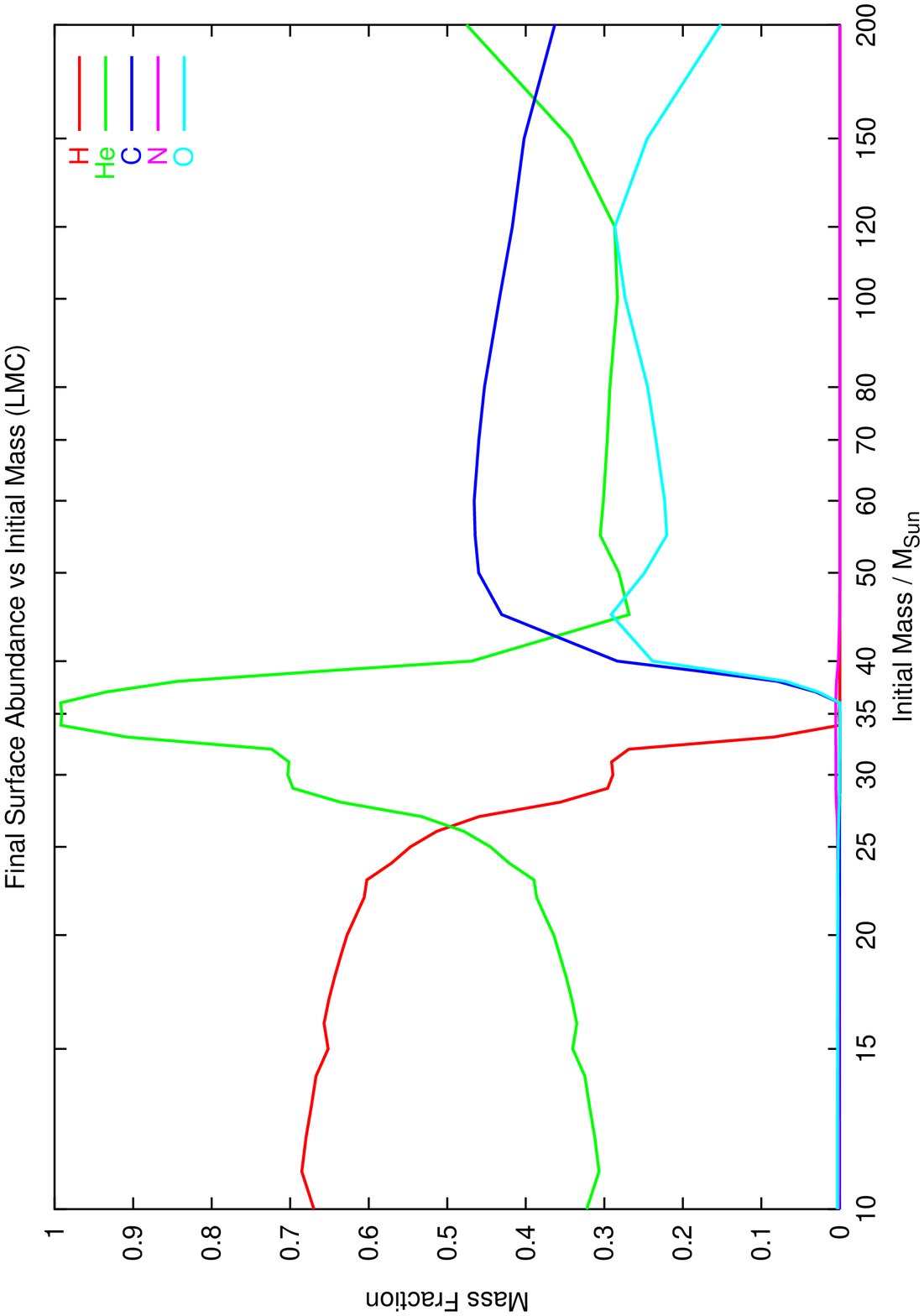}
\includegraphics[height=70mm,angle=0]{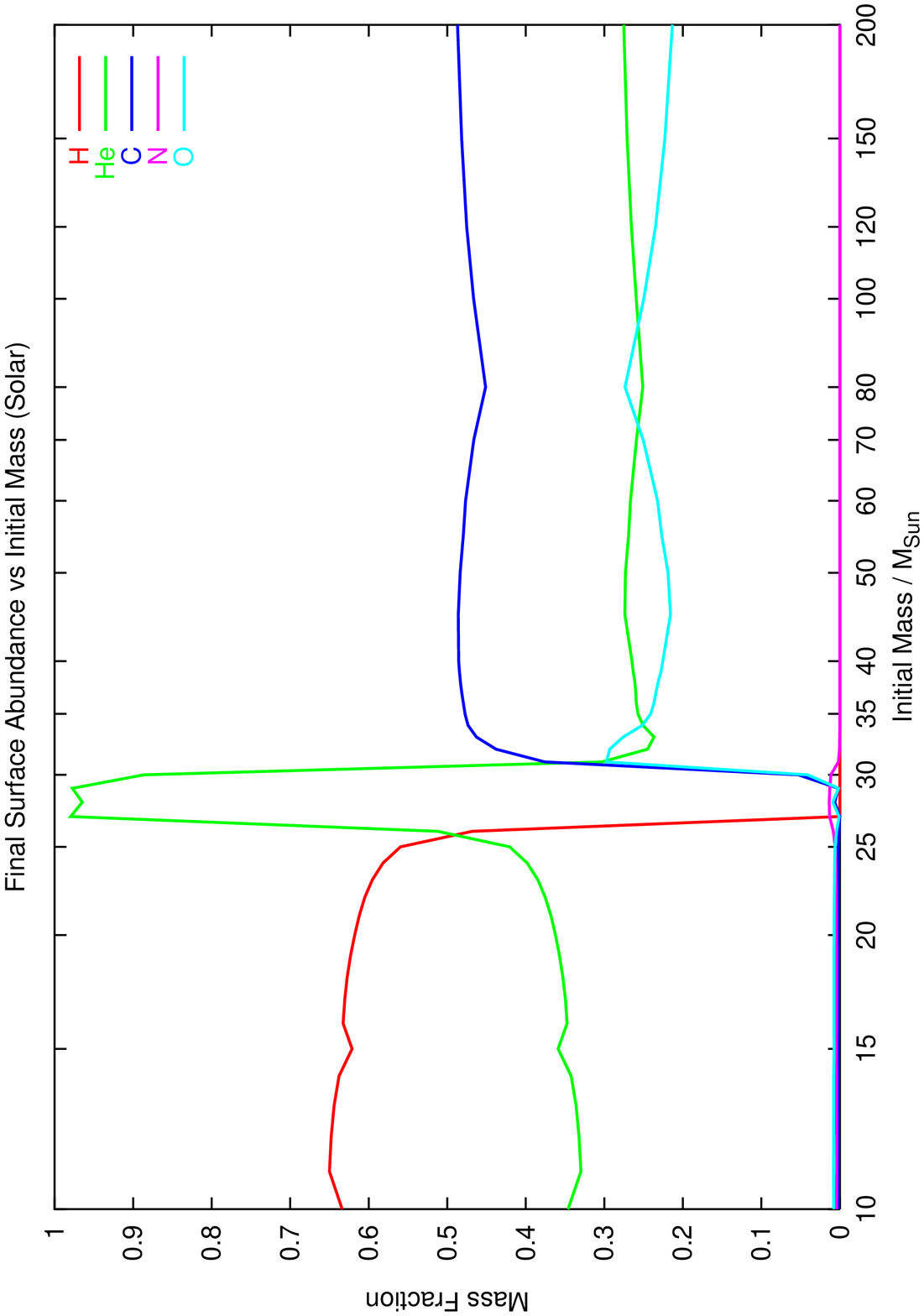}
\end{center}
\caption{The details of the final surface composition and core structure of our models.}
\label{3mods6}
\end{figure}

Figures \ref{3mods1}, \ref{3mods2} and \ref{3mods6} display various details of these stars after core carbon burning or neon burning in some cases. There are many trends with mass and metallicity. Mass ranges of certain features tend to become spaced out over wider regions at lower metallicity. For example in figure \ref{3mods1} at solar metallicity there is a sharp changeover in radius, there are supergiants and then with the increment of $1M_{\odot}$ the stellar radius drops to around that of the Sun, while for the LMC and SMC there is a smoother transition. Figure \ref{3mods2} displays the final mass and the final mass-loss rate of the stars. These show that in the most massive stars with hydrogen envelopes there are extreme winds. Higher final masses for stars only become possible at lower metallicities owing to the reduced mass-loss rates. The maximum final mass-loss rates increases with decreasing metallicity. This is because with less mass lost on the main sequence the stars spend longer ascending the supergiant branch and become more luminous which increases the mass loss from these stars.	

Figure \ref{3mods6} looks at the internal composition of the different stars. What we must consider carefully here is when the changeover between type Ib and Ic might occur. The helium rich progenitors become more prevalent at low metallicity and there seems to be a very clear definition between the two. Although at solar metallicity even with models at every integer mass it is difficult to resolve the helium rich progenitors that might give rise to the Ib SN, from those deficient in helium that give rise to the Ic. However to complete the type I progenitor story we must look at binaries in chapter 5.

\begin{figure}
\begin{center}
\includegraphics[height=130mm,angle=270]{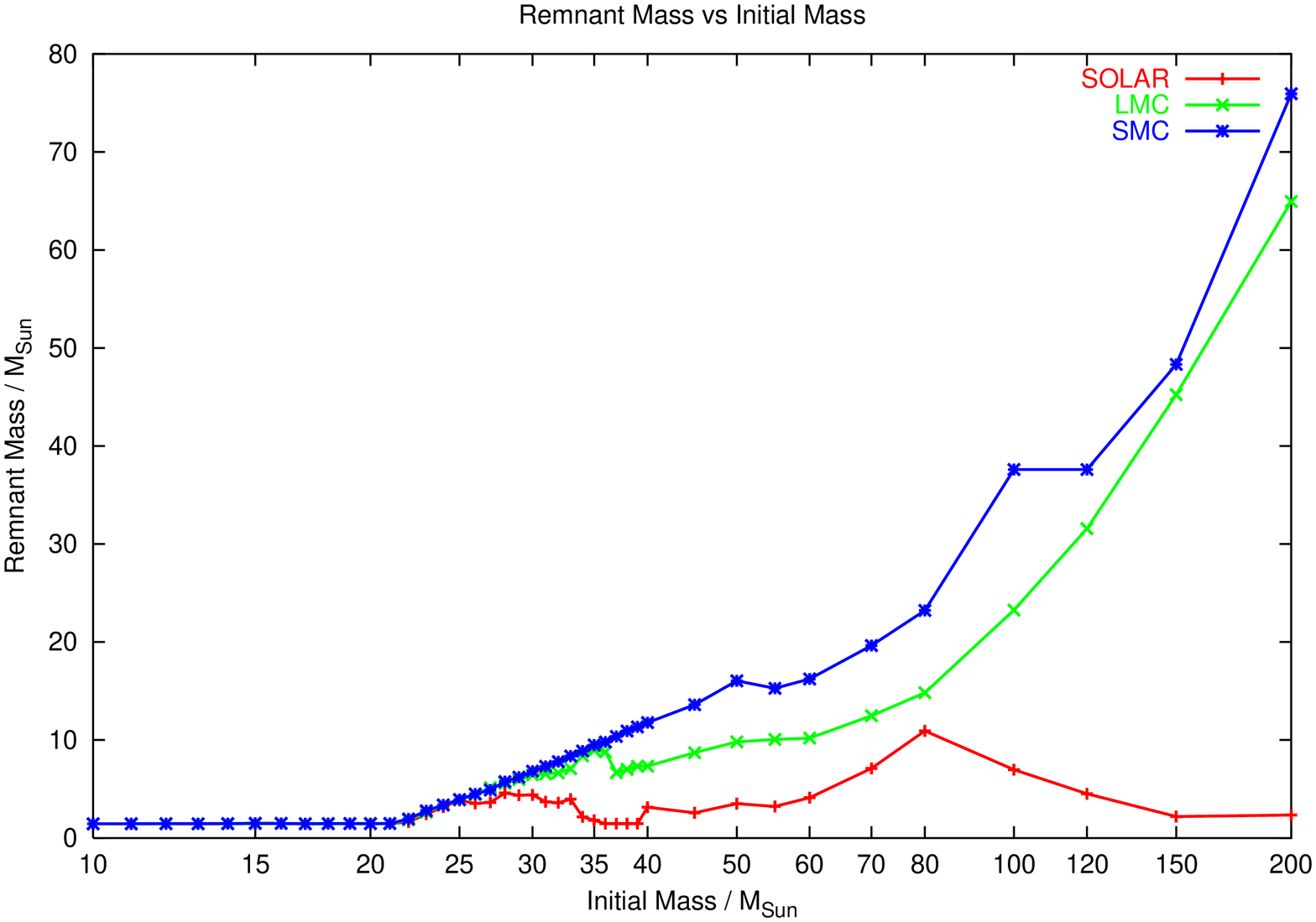}
\includegraphics[height=130mm,angle=270]{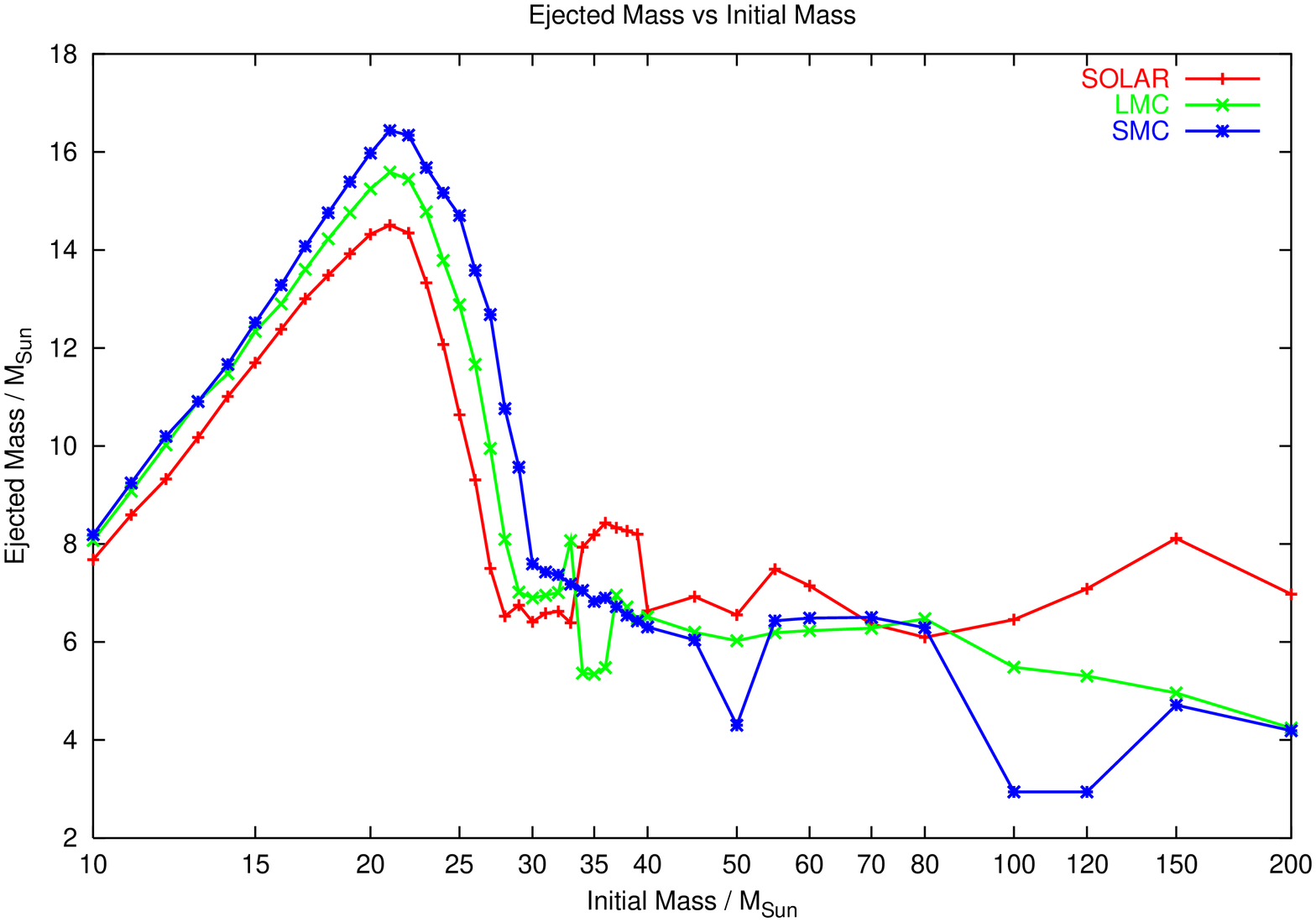}
\end{center}
\caption{Estimates of the remnant and ejected masses from our models.}
\label{3mods3}
\end{figure}

Figure \ref{3mods3} shows estimates of the remnant and ejected masses if the stars were to go SN. These are obtained with the binding energy of the star. We assume a neutron star is formed at the centre of the star after core collapse of mass $M_{\rm Ch}=1.44M_{\odot}$. This produces about $10^{46} {\rm J}$ of energy from the release of gravitational binding energy in neutron star formation. We then assume a hundredth of this energy is transferred into the envelope by some unknown mechanism. The current suggestion is the transfer occurs via neutrinos released when the proto-neutron star forms that are thermalised within the envelope and the mantle of the core.

We integrate the binding energy of the star from the surface towards the centre until we reach $10^{44} J$. The envelope outside this region is ejected with the remaining amount forming the remnant.
\begin{equation}
\int^{M_{*}}_{M_{\rm rem}} \Big( U-\frac{GM}{r(M)} \Big) dM \, = \, 10^{44} {\rm J} 
\label{remnantequation}
\end{equation}

If we have a remnant mass $M_{\rm rem} > 2M_{\odot}$ the remnant is a black hole. This is a very rapid method to estimate the remnant and ejected masses quickly. It must be remembered however that the remnant masses are at best a lower bound as we use a very simple model for a complex event. From the figures it is possible to see that there are only very large remnants possible at low metallicities. Otherwise commonly there are black holes that are a few solar masses and neutron stars. Interestingly there is a maximum possible ejected mass in SN. This is important to compare to the calculated ejected masses from SN observations such as those of \citet{iiplength}.

With these models we can choose which features we need to consider in deciding the SN type a certain progenitor produces. But first we must investigate how SNe are separated into their various types.

\section{Charting the change between SN types: Observation}

Type I SNe do not have hydrogen in their spectra and are split into types Ia, Ib and Ic. Type Ia are thermonuclear events from low-mass stars ($M_{\rm initial}<8M_{\odot}$). They are identified by the presence of silicon lines but they are not core-collapse events so are not of interest to us here. Types Ib and Ic are SN with WR-star progenitors \citep{EW88}. Comparing their spectra we find Ib SN to have helium while Ic do not and this is how they are distinguished. An example of a type Ib SN is SN1984I \citep{sn1984i} while an example type Ic is SN1987M \citep{sn1987m}. From the presence or absence of helium in the spectrum we could jump to the assumption that Ib progenitors must have helium while Ic progenitors do not. However it's not that simple. In Ib SNe the helium lines do not become prevalent until after a few weeks from maximum light and simulations show that helium can be difficult to see if it is only present in trace abundances \citep{Ibccalcs}. There is also some evidence that Ic SNe do have helium present in the progenitor \citep{moreIc}. We also know that the light curve and spectra of a SN depends on the size of the star. Helium giants will give rise to quite different SN to those from helium dwarfs. In all the models presented we find a small proportion of helium. So unless all Ic SNe are the results of a helium giants after a common envelope phase where they lose their helium envelope, we must determine which stars become Ib or Ic SNe. Binary stars models are required for a full understanding.

Type I SN types are poorly defined. They are defined by what they lack rather than what they have. For both they do not have hydrogen or silicon lines while Ic also has no helium. Indeed observations include a large number of SNe that supposedly fit into this category, including the newly identified hypernovae with energies ten times that of normal SNe, SNe that have been associated with Gamma-ray bursts and also fainter SNe events. The main reason for this is that there are not enough high-quality observations of type Ib and Ic events to recognise the different types of SNe. However there are groups around the world that will be collecting vast samples of data on core-collapse events to improve our observational understanding \citep{cccp,panstars}. In this work we can tackle this problem from the theoretical end, describing what is possible from our current understanding of stellar evolution. This should help classify and recognise new SN categories in future.

Type II SNe are better defined by what they do have. Most importantly they do have hydrogen in their spectra. The main types are IIP, IIL, IIb, IIn and IIpec. Type IIP have a long plateau phase where after the peak luminosity the SN luminosity remains constant owing to the photosphere remaining at a constant radius. A well observed SN of this type was SN1999em \citet{sn1999em}. The plateau occurs because while the ejecta expand outward the ionisation front moves inwards in mass. The photosphere is located at the point of hydrogen reionisation. A massive hydrogen envelope is required for the progenitor. Red supergiants fit the requirement well.

IIL SNe have a linear decay in their light curves after maximum light mainly powered by radioactive decay of $^{56}$Co. An example is SN1980K \citep{sn1980k}. Single-star progenitors of this type of SN are probably initially more massive than the IIP progenitors, close to the boundary of the type I SNe and have undergone more extreme mass-loss. We should note that IIL SNe are also possible from binary evolution. Type IIb are a step closer to type I SNe. These objects were first discovered looking like a type II SN. However, after a few weeks, they change to type Ib objects. Therefore the objects have only very small hydrogen envelopes. The classic example of this type is SN1993J \citep{sn1993j}, one of the three SN with an identified progenitor.

Type IIn are distinguished by narrow lines in their spectra, SN1998S was a SN of this type \citep{sn1998s}. The narrow lines indicate slow expansion velocities. Either this is due to a low explosion energy or a dense circumstellar environment where the ejecta is tarried by its interactions with circumstellar material. Either this is material ejected from a common envelope phase or a strong wind from the progenitor star. From radio observations of SNe it is possible to put limits on the strength of the wind required to be between $10^{-5}$ and $10^{-3} M_{\odot} {\rm yr^{-1}}$. There are indications that IIn SNe are related to other SN types with features that are common to the other type II SN such as a plateau or linear lightcurve. It is also possible to split IIn SN into their two possible progenitor systems. For example a IIP SN with a low explosion energy would have a plateau phase with narrow lines which would have a P-Cygni profile which is evidence of an expanding hydrogen envelope. However if the ejecta interacts with a dense circumstellar environment the lines will not have the P-Cygni profile.

We should note there are also impostor type IIn events. One class of these are luminous outbursts from luminous-blue variable stars or WR stars, examples of such events are 1961V and 1997bs. These are not SN. The other class is type Ia supernovae in dense circumstellar environments. An example is SN2002ic where a type Ia SN occurred in a hydrogen rich environment.

Finally we have IIpec, where we put the type II SN flotsam and jetsam. SN1987A is the most common example and can give some idea of what makes a SN peculiar. The progenitor star, Sk -69 202, was a blue supergiant, there is evidence for various mass loss events over the time of evolution and there is now a large amount of evidence that points towards a binary merger of two stars \citep{podsi92}. The surrounding environment is highly complex and the star was relatively compact and massive. Also the core structure may be quite unique due to the merger.

\section{Charting the change between SN types: Theory}

From stellar models we can draw some conclusions about SNe. However the only way to know for sure what type of SN a model produces would be to model the SN itself, the expansion of the envelope, hydrodynamics of the core, detailed nucleosynthesis and interactions with circumstellar material. We would have to simulate the light curve and spectra. Such models exist but there are a number of parameters, such as the amount of mixing that can be varied to produce different results. There are some analytical models but care must be taken in their use. For example \citet{iiplength} and \citet{nadzy} found very large ejected masses that are at odds with our models and the direct limits deduced by Smartt et al. (2003,2004).

To predict the SN a model gives rise to we use analytic methods detailed below and specific details about the models. We also use the ejected mass and remnants mass estimated from equation \label{remnantequation}. Then we compare to observations. The best observations are of SN which have identified progenitors. There are only three observed progenitors to date and there are another six where observations exist to put limits on the properties of the progenitor stars \citep{S03a}.

We do not produce a theoretical scheme to name SN. Such a scheme would have limited use and it is preferable to estimate what we will find in observations so we use the observational scheme as a starting point. To estimate what we might see we need a number of details from our models. They are mostly those that have already been plotted in figures \ref{3mods1}, \ref{3mods2}, \ref{3mods6} and \ref{3mods3}. But there are some details we need that are not directly available from our models so we must make some analytical estimate of their value.

One of these is the remnant and ejected masses as discussed earlier. However to determine whether a type II SN has a plateau phase or not we use the following expressions from Popov~(1993),
\begin{equation}
\Lambda=11.74 \frac{\kappa_{0.34}^{2} M_{10}^{3/2} T_{\rm ion,5054}^{4}}{E_{51}^{1/2} R_{0,500}},
\end{equation} 
\begin{equation}
t_{\rm plateau} \approx 99 \frac{\kappa_{0.34}^{1/6} M_{10}^{1/2} R_{0,500}^{1/6}}{E_{44}^{1/6} T_{\rm ion, 5054}^{2/3}} {\rm days}.
\end{equation}

In these equations $\kappa_{0.34}$ is the opacity in multiples of $0.34 {\rm g \, cm^{-2}}$, $R_{0,500}$ the radius in multiples of $500 R_{\odot}$, $T_{\rm ion,5054}$ the hydrogen ionisation temperature in multiples of 5054K and $E_{44}$ the explosion energy in multiples of $10^{44} {\rm J}$. The mass, $M_{10}$, is the mass ejected in the SN in multiples of $10M_{\odot}$. Our calculated values for plateau lengths agree with values from observations. In all our calculations we set $\kappa_{0.34}=E_{44}=T_{\rm ion, 5054}=1$. It is thought that the explosion energy is not a constant between supernovae so there will be some scatter around our values. However $\Lambda$ has the greatest dependence on the energy and not the plateau length. For a plateau phase to occur $\Lambda \gg 1$. Comparing models to these limits we set the limiting value $\Lambda \ge 2$. The plateau length must be sufficient that it can be observed. We take this limit to be $30$ days. $\Lambda$ is a variable that defines the strength of the plateau phase and it gives the limit that there must be enough mass in the envelope compared to the radius of the envelope.

We list the information we use for SN type determination in table \ref{snclmain}. If we think a model might give rise to a IIP SN we denote it as IIP$_{80,3.0}$. This means that it will be a type IIP SN with a plateau length of 80 days and the second number is the value of $\Lambda$ as defined above. Or for type Ibc we denote it as Ib$_{3}$c$_{0.95}$ or Ib$_{0.1}$c$_{0.2}$ with the subscript on the b equal to the mass of helium present and the c subscript the helium mass fraction. 

\begin{table}
\label{snclmain}
\caption{Theoretical SN Classification Scheme.}
\begin{center}
\begin{tabular}{|l|l|}
\hline
Label	& Indication \\
\hline
I		& No hydrogen present.\\
II		& Hydrogen present.\\
\hline
a		& Thermonuclear explosion mechanism.\\
e		& Electron capture core collapse.\\
\hline
P$_{x,y}$	& Hydrogen present and light curve plateau length of $x$ days. \\
			& $y$ is the value of $\Lambda$ for the progenitor.\\
L$_{x,y}$		& If $M_{\rm env} \le M_{\rm core}({\rm He})$, $x =M_{\rm env}/M_{\odot}$ and $y=M_{\rm ej}/M_{\odot}$.\\
b$_{x}$		& If type II $x=M_{\rm H}/M_{\odot}$, if type I $x=M_{\rm He}/M_{\odot}$.\\
c$_{x}$		& If type I, $x$ is the helium mass fraction.\\
n$_{x,y}$	& If the progenitor model has a strong pre-SN wind, $x$ is the mass-loss rate in $10^{-5} M_{\odot} {\rm yr^{-1}}$\\
				& and $y$ is the total mass lost during evolution in $M_{\odot}$.\\
\hline
(i)		& Pre-SN radius,  $R< R{\odot}$.\\
(ii)	&  $R_{\odot} < R < 10R{_\odot}$.\\
(iii)	& $10R_{\odot} < R < 100R{_\odot}$.\\
(iv)	& $100R_{\odot} < R < 1000R{_\odot}$.\\
(v)	&$R > 1000R{_\odot}$.\\
\hline
\end{tabular}
\end{center}
\end{table}

From our example models we have calculated these factors and listed them in tables \ref{solartable}, \ref{lmctable} and \ref{smctable}. For each model we have listed the values relevant for all subtypes. By doing this it is easier to decide over which mass ranges different SNe might occur. We start by calculating whether the star is type IIP. If it is not then we describe it as IIL this seems to occur when the hydrogen envelope mass is less than about $6M_{\odot}$ when we apply the limit from the value of $\Lambda$. This is a little larger than the $2M_{\odot}$ used by \citet{H03} but is similar to the upper limit of $5M_{\odot}$ determined by the lightcurve models of \citet{crazyidea}. However the plateau length is still quite large therefore these SNe could be the low luminosity IIP SNe that have been observed by \citet{lowLIIP} rather than IIL SNe. These SNe also form black holes in their centres so the SN might be different. For type IIb SN we will say these occur when the hydrogen envelope mass is less than $1M_{\odot}$. Simulations of IIb SNe indicate that while there is some uncertainty the value is of this order \citep{IIba,IIbb,IIbc}.

Progenitors of type II SNe eject most of their mass as the star ascends the red supergiant branch. Because of this, during the SN, there could be substantial mass in the circumstellar environment in the most massive type II progenitors.

\begin{table}
\caption{Theoretical SN details for solar metallicity models with overshooting.}
\label{solartable}
\begin{center}
\begin{tabular}{|l|l|}
\hline
 $M_{\rm i}$	&	Solar	\\
$/ M_{\odot}$	&\\
\hline
10 & IIP$_{ 91, 4.3 }$L$_{ 6.2, 7.7 }$(iv)\\
11 & IIP$_{ 98, 4.5 }$L$_{ 6.8, 8.6 }$(iv)\\
12 & IIP$_{ 103, 4.5 }$L$_{ 7.2, 9.3 }$(iv)\\
13 & IIP$_{ 109, 4.6 }$L$_{ 7.6, 10.2 }$(iv)\\
14 & IIP$_{ 115, 4.7 }$L$_{ 8.1, 11.0 }$(iv)\\
15 & IIP$_{ 120, 4.2 }$L$_{ 8.1, 11.7 }$(v)\\
16 & IIP$_{ 125, 4.2 }$L$_{ 8.4, 12.4 }$(v)\\
17 & IIP$_{ 130, 3.9 }$L$_{ 8.4, 13.0 }$(v)\\
18 & IIP$_{ 134, 3.6 }$L$_{ 8.3, 13.5 }$(v)\\
19 & IIP$_{ 137, 3.3 }$L$_{ 8.2, 13.9 }$(v)\\
20 & IIP$_{ 140, 3.1 }$L$_{ 8.1, 14.3 }$(v)\\
21 & IIP$_{ 142, 2.7 }$L$_{ 7.7, 14.5 }$(v)\\
22 & IIP$_{ 143, 2.3 }$L$_{ 7.1, 14.3 }$n$_{ 1.3, 6.0 }$(v)\\
23 & IIP$_{ 139, 1.9 }$L$_{ 6.4, 13.3 }$n$_{ 1.7, 7.2 }$(v)\\
24 & IIP$_{ 133, 1.4 }$L$_{ 5.3, 12.1 }$n$_{ 2.2, 8.8 }$(v)\\
25 & IIP$_{ 125, 0.9 }$L$_{ 4.0, 10.6 }$n$_{ 3.0, 10.5 }$(v)\\
26 & IIP$_{ 115, 0.3 }$L$_{ 1.8, 9.3 }$b$_{ 0.81 }$n$_{ 4.2, 13.2 }$(v)\\
27 & Ib$_{ 0.96 }$c$_{ 0.98 }$n$_{ 3.4, 15.9 }$(ii)\\
28 & Ib$_{ 1.15 }$c$_{ 0.96 }$n$_{ 2.6, 16.9 }$(ii)\\
29 & Ib$_{ 0.89 }$c$_{ 0.98 }$n$_{ 2.4, 17.9 }$(ii)\\
30 & Ib$_{ 0.46 }$c$_{ 0.89 }$n$_{ 2.4, 19.2 }$(ii)\\
31 & Ib$_{ 0.16 }$c$_{ 0.30 }$n$_{ 1.2, 20.7 }$(i)\\
32 & Ib$_{ 0.19 }$c$_{ 0.24 }$(i)\\
33 & Ib$_{ 0.23 }$c$_{ 0.24 }$(i)\\
34 & Ib$_{ 0.26 }$c$_{ 0.25 }$(i)\\
35 & Ib$_{ 0.28 }$c$_{ 0.26 }$(i)\\
36 & Ib$_{ 0.28 }$c$_{ 0.26 }$(i)\\
37 & Ib$_{ 0.29 }$c$_{ 0.26 }$(i)\\
38 & Ib$_{ 0.29 }$c$_{ 0.26 }$(i)\\
39 & Ib$_{ 0.29 }$c$_{ 0.26 }$(i)\\
40 & Ib$_{ 0.28 }$c$_{ 0.26 }$(i)\\
45 & Ib$_{ 0.28 }$c$_{ 0.27 }$(ii)\\
50 & Ib$_{ 0.28 }$c$_{ 0.27 }$(i)\\
55 & Ib$_{ 0.29 }$c$_{ 0.27 }$n$_{ 1.1, 44.3 }$(i)\\
60 & Ib$_{ 0.29 }$c$_{ 0.27 }$n$_{ 1.1, 48.7 }$(i)\\
70 & Ib$_{ 0.25 }$c$_{ 0.26 }$n$_{ 1.3, 56.5 }$(ii)\\
80 & Ib$_{ 0.43 }$c$_{ 0.25 }$n$_{ 1.8, 63.0 }$(ii)\\
100 & Ib$_{ 0.25 }$c$_{ 0.26 }$n$_{ 1.3, 86.6 }$(ii)\\
120 & Ib$_{ 0.29 }$c$_{ 0.27 }$n$_{ 1.1, 108.4 }$(i)\\
150 & Ib$_{ 0.30 }$c$_{ 0.27 }$n$_{ 1.0, 139.7 }$(i)\\
200 & Ib$_{ 0.28 }$c$_{ 0.28 }$(ii)\\
\hline
\end{tabular}
\end{center}
\end{table}

\begin{table}
\caption{Theoretical SN details for LMC metallicity models with overshooting.}
\label{lmctable}
\begin{center}
\begin{tabular}{|l|l|}
\hline
 $M_{\rm i}$	&	LMC		\\
$/ M_{\odot}$	&\\
\hline
10 & IIP$_{ 92, 4.7 }$L$_{ 6.3, 8.1 }$(iv)\\
11 & IIP$_{ 99, 5.2 }$L$_{ 7.0, 9.1 }$(iv)\\
12 & IIP$_{ 105, 5.4 }$L$_{ 7.6, 10.0 }$(iv)\\
13 & IIP$_{ 111, 5.7 }$L$_{ 8.2, 10.9 }$(iv)\\
14 & IIP$_{ 115, 5.5 }$L$_{ 8.3, 11.5 }$(iv)\\
15 & IIP$_{ 121, 5.2 }$L$_{ 8.7, 12.3 }$(iv)\\
16 & IIP$_{ 125, 4.9 }$L$_{ 8.6, 12.9 }$(iv)\\
17 & IIP$_{ 130, 4.7 }$L$_{ 8.8, 13.6 }$(v)\\
18 & IIP$_{ 134, 4.5 }$L$_{ 8.9, 14.2 }$(v)\\
19 & IIP$_{ 138, 4.3 }$L$_{ 8.9, 14.8 }$(v)\\
20 & IIP$_{ 142, 4.1 }$L$_{ 8.9, 15.2 }$(v)\\
21 & IIP$_{ 144, 3.7 }$L$_{ 8.6, 15.6 }$(v)\\
22 & IIP$_{ 145, 3.4 }$L$_{ 8.4, 15.4 }$n$_{ 1.0, 4.7 }$(v)\\
23 & IIP$_{ 142, 3.1 }$L$_{ 8.0, 14.8 }$n$_{ 1.3, 5.6 }$(v)\\
24 & IIP$_{ 139, 2.5 }$L$_{ 7.2, 13.8 }$n$_{ 1.7, 6.9 }$(v)\\
25 & IIP$_{ 135, 2.0 }$L$_{ 6.3, 12.9 }$n$_{ 2.2, 8.2 }$(v)\\
26 & IIP$_{ 129, 1.4 }$L$_{ 5.2, 11.7 }$n$_{ 2.9, 9.9 }$(v)\\
27 & IIP$_{ 119, 0.9 }$L$_{ 3.7, 10.0 }$n$_{ 3.7, 11.9 }$(v)\\
28 & IIP$_{ 105, 0.3 }$L$_{ 1.8, 8.1 }$b$_{ 0.60 }$n$_{ 4.8, 14.3 }$(v)\\
29 & IIP$_{ 95, 0.1 }$L$_{ 0.7, 7.0 }$b$_{ 0.19 }$n$_{ 5.8, 16.0 }$(v)\\
30 & IIP$_{ 94, 0.1 }$L$_{ 0.6, 6.9 }$b$_{ 0.16 }$n$_{ 7.3, 16.7 }$(v)\\
31 & IIP$_{ 69, 0.5 }$L$_{ 0.6, 7.0 }$b$_{ 0.11 }$n$_{ 5.7, 17.3 }$(iv)\\
32 & IIP$_{ 59, 0.3 }$L$_{ 0.2, 7.0 }$b$_{ 0.06 }$n$_{ 4.0, 18.2 }$(iii)\\
33 & IIb$_{ 0.01 }$n$_{ 2.5, 19.1 }$(ii)\\
34 & Ib$_{ 1.66 }$c$_{ 0.99 }$n$_{ 3.0, 20.3 }$(i)\\
35 & Ib$_{ 1.83 }$c$_{ 0.99 }$n$_{ 2.9, 20.7 }$(i)\\
36 & Ib$_{ 1.48 }$c$_{ 0.99 }$n$_{ 2.6, 21.7 }$(i)\\
37 & Ib$_{ 0.61 }$c$_{ 0.93 }$n$_{ 2.2, 23.4 }$(ii)\\
38 & Ib$_{ 0.47 }$c$_{ 0.84 }$n$_{ 1.9, 24.3 }$(ii)\\
39 & Ib$_{ 0.31 }$c$_{ 0.66 }$n$_{ 2.6, 25.2 }$(i)\\
40 & Ib$_{ 0.21 }$c$_{ 0.47 }$n$_{ 2.1, 26.2 }$(i)\\
45 & Ib$_{ 0.23 }$c$_{ 0.27 }$n$_{ 1.0, 30.1 }$(i)\\
50 & Ib$_{ 0.31 }$c$_{ 0.28 }$n$_{ 1.2, 34.2 }$(i)\\
55 & Ib$_{ 0.31 }$c$_{ 0.31 }$n$_{ 1.4, 38.7 }$(i)\\
60 & Ib$_{ 0.31 }$c$_{ 0.30 }$n$_{ 1.4, 43.6 }$(i)\\
70 & Ib$_{ 0.30 }$c$_{ 0.30 }$n$_{ 1.6, 51.2 }$(i)\\
80 & Ib$_{ 0.30 }$c$_{ 0.29 }$n$_{ 1.8, 58.7 }$(ii)\\
100 & Ib$_{ 0.68 }$c$_{ 0.28 }$n$_{ 2.4, 71.3 }$(ii)\\
120 & Ib$_{ 0.80 }$c$_{ 0.29 }$n$_{ 3.1, 83.1 }$(ii)\\
150 & Ib$_{ 1.15 }$c$_{ 0.34 }$n$_{ 5.0, 99.8 }$(iii)\\
200 & Ib$_{ 1.85 }$c$_{ 0.48 }$n$_{ 5.0, 130.8 }$(iv)\\
\hline
\end{tabular}
\end{center}
\end{table}
\begin{table}
\caption{Theoretical SN details for SMC metallicity models with overshooting.}
\label{smctable}
\begin{center}
\begin{tabular}{|l|l|}
\hline
 $M_{\rm i}$	&	SMC	\\
$/ M_{\odot}$	&\\
\hline
10 & IIP$_{ 92, 4.9 }$L$_{ 6.3, 8.2 }$(iv)\\
11 & IIP$_{ 99, 5.7 }$L$_{ 7.1, 9.2 }$(iv)\\
12 & IIP$_{ 105, 6.0 }$L$_{ 7.7, 10.2 }$(iv)\\
13 & IIP$_{ 109, 6.2 }$L$_{ 8.1, 10.9 }$(iv)\\
14 & IIP$_{ 114, 6.0 }$L$_{ 8.4, 11.7 }$(iv)\\
15 & IIP$_{ 121, 5.7 }$L$_{ 8.7, 12.5 }$(iv)\\
16 & IIP$_{ 125, 5.8 }$L$_{ 9.0, 13.3 }$(iv)\\
17 & IIP$_{ 130, 5.6 }$L$_{ 9.2, 14.1 }$(iv)\\
18 & IIP$_{ 134, 5.5 }$L$_{ 9.4, 14.8 }$(iv)\\
19 & IIP$_{ 139, 5.3 }$L$_{ 9.5, 15.4 }$(v)\\
20 & IIP$_{ 142, 5.1 }$L$_{ 9.6, 16.0 }$(v)\\
21 & IIP$_{ 146, 4.8 }$L$_{ 9.5, 16.4 }$(v)\\
22 & IIP$_{ 146, 4.5 }$L$_{ 9.4, 16.3 }$(v)\\
23 & IIP$_{ 144, 4.0 }$L$_{ 9.0, 15.7 }$n$_{ 1.1, 4.5 }$(v)\\
24 & IIP$_{ 143, 3.6 }$L$_{ 8.5, 15.2 }$n$_{ 1.4, 5.5 }$(v)\\
25 & IIP$_{ 141, 3.3 }$L$_{ 8.2, 14.7 }$n$_{ 1.7, 6.4 }$(v)\\
26 & IIP$_{ 137, 2.5 }$L$_{ 7.1, 13.6 }$n$_{ 2.3, 7.9 }$(v)\\
27 & IIP$_{ 133, 2.0 }$L$_{ 6.1, 12.7 }$n$_{ 2.9, 9.4 }$(v)\\
28 & IIP$_{ 122, 1.3 }$L$_{ 4.5, 10.8 }$n$_{ 3.8, 11.5 }$(v)\\
29 & IIP$_{ 114, 0.8 }$L$_{ 3.3, 9.6 }$n$_{ 4.8, 13.2 }$(v)\\
30 & IIP$_{ 100, 0.3 }$L$_{ 1.5, 7.6 }$b$_{ 0.43 }$n$_{ 5.9, 15.6 }$(v)\\
31 & IIP$_{ 99, 0.2 }$L$_{ 1.3, 7.4 }$b$_{ 0.38 }$n$_{ 7.8, 16.3 }$(v)\\
32 & IIP$_{ 99, 0.2 }$L$_{ 1.3, 7.4 }$b$_{ 0.35 }$n$_{ 9.6, 16.8 }$(v)\\
33 & IIP$_{ 98, 0.2 }$L$_{ 1.2, 7.2 }$b$_{ 0.32 }$n$_{ 11.6, 17.5 }$(v)\\
34 & IIP$_{ 97, 0.2 }$L$_{ 1.1, 7.1 }$b$_{ 0.29 }$n$_{ 13.7, 18.1 }$(v)\\
35 & IIP$_{ 96, 0.1 }$L$_{ 1.0, 6.8 }$b$_{ 0.26 }$n$_{ 14.8, 18.7 }$(v)\\
36 & IIP$_{ 96, 0.1 }$L$_{ 0.9, 6.9 }$b$_{ 0.24 }$n$_{ 17.0, 19.3 }$(v)\\
37 & IIP$_{ 95, 0.1 }$L$_{ 0.8, 6.7 }$b$_{ 0.21 }$n$_{ 19.8, 19.9 }$(v)\\
38 & IIP$_{ 94, 0.1 }$L$_{ 0.8, 6.5 }$b$_{ 0.19 }$n$_{ 22.8, 20.5 }$(v)\\
39 & IIP$_{ 93, 0.1 }$L$_{ 0.6, 6.4 }$b$_{ 0.15 }$n$_{ 24.8, 21.2 }$(v)\\
40 & IIP$_{ 91, 0.0 }$L$_{ 0.5, 6.3 }$b$_{ 0.11 }$n$_{ 26.8, 21.9 }$(v)\\
45 & IIP$_{ 41, 0.0 }$L$_{ 0.0, 6.0 }$b$_{ 0.01 }$n$_{ 4.0, 25.4 }$(iii)\\
50 & Ib$_{ 1.46 }$c$_{ 0.99 }$n$_{ 5.3, 29.7 }$(i)\\
55 & Ib$_{ 1.00 }$c$_{ 0.94 }$n$_{ 1.0, 33.3 }$(ii)\\
60 & Ib$_{ 0.90 }$c$_{ 0.89 }$n$_{ 1.9, 37.3 }$(ii)\\
70 & Ib$_{ 0.42 }$c$_{ 0.45 }$(ii)\\
80 & Ib$_{ 0.76 }$c$_{ 0.43 }$(ii)\\
100 & Ib$_{ 0.92 }$c$_{ 0.33 }$n$_{ 3.2, 59.5 }$(i)\\
120 & Ib$_{ 0.92 }$c$_{ 0.33 }$n$_{ 3.2, 79.5 }$(i)\\
150 & Ib$_{ 1.28 }$c$_{ 0.37 }$n$_{ 4.6, 97.0 }$(ii)\\
200 & Ib$_{ 2.59 }$c$_{ 0.54 }$n$_{ 5.0, 119.9 }$(iii)\\
\hline
\end{tabular}
\end{center}
\end{table}

\begin{table}
\begin{center}
\caption[The Plateau lengths from S-AGB and Extreme S-AGB Progenitors.]{The Plateau lengths from S-AGB and Extreme S-AGB Progenitors. The first set are models without convective overshooting, the second set are models with convective overshooting.}
\label{listIIe}
\begin{tabular}{|lc|lc|}
\hline
Initial    &   &Initial &\\
Mass        &Without	&Mass   &With\\
$/ M_{\odot}$ &Overshooting&$/ M_{\odot}$  & Overshooting \\
\hline
9.5	&IIP$_{105, 3.0}$L$_{8.1}$    &7.5&IIP$_{90,2.1}$L$_{6.1}$    \\
10   	&IIP$_{104,3.0}$L$_{8.6}$    &7.7&IIP$_{92,2.2}$L$_{6.3}$    \\
10.1	&  IIP$_{105,3.0}$L$_{8.7}$   &7.9&IIP$_{93,2.4}$L$_{6.5}$    \\
10.2	&  IIP$_{106,3.2}$L$_{8.8}$    &8.1&IIP$_{90,3.2}$L$_{6.7}$    \\
10.3	&  IIP$_{107,3.3}$L$_{8.9}$   &8.3&IIP$_{91,3.4}$L$_{6.9}$    \\
10.4	& IIP$_{107,3.4}$L$_{9.0}$  &8.4&IIP$_{91,3.6}$L$_{7.0}$    \\
10.5	&  IIP$_{105,3.8}$L$_{9.1}$   & 8.5&  IIP$_{90,4.1}$L$_{7.1}$    \\
10.6	&  IIP$_{103,4.5}$L$_{9.2}$   &8.6&IIP$_{89,4.5}$L$_{7.2}$    \\
10.7	& IIP$_{99,10.7}$L$_{7.3}$   &8.7&IIP$_{86,6.5}$L$_{6.1}$    \\
\hline
\end{tabular}
\end{center}
\end{table} 
While we do not specifically identify any model as a IIn SN we list the final mass-loss rate and total mass lost in the winds during the evolution if the pre-SN wind is stronger than $10^{-5} M_{\odot} {\rm yr^{-1}}$. This enables evaluation as to whether these stars will be observable by radio observations or the strength of any IIn signature in the SNe. SNe IIn probably have an underlying SN of another type. For now we separate type Ib and Ic by choosing so that if $Y_{\rm surface} > 0.5$ we have a Ib SN, otherwise we have a Ic. 

\section{Naming the SN from the progenitors}
We list the details of our models in terms of the theoretical SN scheme in tables \ref{solartable}, \ref{lmctable} and \ref{smctable}. From these we list the ranges of SN types with metallicity in table \ref{mylist}. Comparison with observed ratios is not possible because we have not included binaries although the ratio of type II to type Ibc is in the correct region for spiral galaxies. However we should first notice that there are not enough IIL SNe to fit the observed value of 10-50\% of II SNe being IIL \citep{IILstats,cap1} observed today. To resolve this without binaries we would have to lower the mass range for IIL SNe or include some IIe SNe in the IIL group. There are also candidates for the type IIn SN. In the range of stars we have labelled as IIL there are stars that have extremely high mass-loss rates before the SN. These stars provide a very dense circumstellar medium that absorbs some of the energy in the expanding envelope. Although, as with all the determinations, there is a continuum of objects. 

Looking at the type I SN we see that for the models all hydrogen deficient SN retain some helium. However the surface abundance does vary. Therefore it is possible that this is the source of the difference between the two SN types. All the type I progenitors are very small and therefore tightly bound so only a little mass will be ejected to provide a little display so some of the more massive type Ib stars may not be observed at all.

In table \ref{newobse} we use the values from tables \ref{solartable} and \ref{lmctable} to constrain the mass ranges or the progenitors. Note that from this list we have removed SN1987A and SN1993J due to these systems being confirm as in binaries. SN1980K we discuss below due to the disparity between out models and the observational limits. With the remaining SNe we are able to fit them reasonably well with out models. Assuming our LMC models are at roughly the same as half solar metallicity.

Returning to the lower end of the mass range we take the minimum for the IIP region as the upper mass from the IIe SNe from extreme Super-AGB stars. We separate out the IIe SN for the specific reason that these stars may have strange SN properties. Table \ref{listIIe} shows that they have similar properties to our IIP models. However with overshooting the case for these having a plateau weakens as the models with the longest period to SN after second dredge-up will lose a large amount of mass, reducing the envelope mass. These models could in fact undergo a type IIL SN, as suggested by \citet{crazyidea}. 

\begin{table}
\begin{center}
\caption[Comparing values from models to observations of progenitors.]{Comparing values from models to observations of progenitors. Adapted from the table in \citet{S03a} and details from \citet{SJM03}.}
\label{newobse}
\begin{tabular}{|l|ccccc|cc|}
\hline
     & Observed &         &Observed&     & Observed & Theoretical   & Theory\\
     &   SNe & Observed&Stellar& Initial  & Ejected  & Mass Range   & Ejected\\
SN   &  Type   & $Z/Z_{\odot}$&Type &Mass $/M_{\odot}$& $Mass /M_{\odot}$ &$M_{\odot}$&Mass $/M_{\odot}$\\
\hline
2002ap & Ic     & 0.5     & WR?     &	$<40$ &2.5-5 &  38-40  &  $<6$ \\
1980K  & IIL    & 0.5     & ?       &	$<20$ &  -   &  25-28  &  6-12\\
2003gd & IIP    & 0.5     & M       & $8$     & -    &  8      &  6.5\\
2001du & IIP    & ~1      & G-M     &	$<15$ &-     & 8.7-15  &  6-12\\
1999em & IIP    & 1-2     & K-M     &	$<15$ &5-18  & 8.7-12  &  6-10\\
1999gi & IIP    & ~2      & G-M     &	$<12$ &10-30 & 8.7-12  &  6-10\\
\hline
\end{tabular}
\end{center}
\end{table} 

\begin{table}
\begin{center}
\caption[Relative populations for SN types from theoretical models with convective overshooting.]{Relative populations for SN types from theoretical models with convective overshooting. The figures are obtained from tables \ref{solartable}, \ref{lmctable} and \ref{smctable} applying our determination on SN types.}
\label{mylist}
\begin{tabular}{|l|cc|cc|cc|}
\hline
SNe Type  & Solar	&	&	LMC	&	&	SMC	&	\\
	&Mass Range&Population&Mass Range&Population&Mass Range&Population\\
	&$/M_{\odot}$& &$/M_{\odot}$& &$/M_{\odot}$& \\
\hline
IIe		&	7.5-8.6	&21.6\%	&	6.5-7.5&	22.5\%	&	6-7&	24.0\%	\\
IIP		&	8.7-22&	63.3\%	&	7.6-25&67.9\%&	7.1-27& 68.7\%	\\
IIL		&	23-26&	3.4\%	&	26-28 &1.7\%	&	28-35 &2.7\%	\\
IIb		&		-&	0\%	&	29-32& 1.6\%	&	36-40 &1.3\%  \\
Ib		&	27-30&	2.5\%	&	33-39 &1.8\%	&	45-200  &2.8\%\\
Ic		&	31-200&	8.9\%	&	40-200& 4.4\%	&	- & 0\%\\
\hline
\end{tabular}
\end{center}
\end{table} 

If we include overshooting then S-AGB stars have smaller hydrogen envelopes. Making it is easier for mass-loss to remove the envelope to the point where a type IIL SNe occurs. Table \ref{listIIe} also provides further evidence with the lower values of $\Lambda$ for the overshooting models. Therefore to search for the progenitors of electron-capture SNe we should search for IIL SNe as well as IIP.

There is one SN that provides a tentative hint that electron captures do give rise to IIL SN, 1980K. This was a type IIL SN and even though the progenitor was not detected the limit placed on its mass was $M<20M_{\odot}$. There are three possible interpretations. First that the stellar winds must be high to remove the envelope of a star between $15$ and $20M_{\odot}$, enough for a IIL to occur, second that a binary interaction occurred removing the mass or third that the initial mass was in fact quite low and we have an electron capture event. 

There is a further piece of information to consider. \citet{crazy2} found evidence for a change in the mass loss 10,000 years before SN over a period of about $4000$ years. This information fits most comfortably with the binary or electron capture models. From figure \ref{sagb2} we can see that a $9.5M_{\odot}$ star ($7.5M_{\odot}$ with overshooting) fits the required change in mass loss history. A strong wind over 10,000 years would have reduced the envelope mass to a great degree and a type IIL SN would be the most likely outcome. The derived mass-loss rate is about $2 \times 10^{-5} M_{\odot} {\rm yr}^{-1}$. This would be enough for both types of star to remove sufficient envelope for the SN to be IIL in the required time.

The binary model is the only other viable alternative. The problems with the binary model is that the mass loss must be relatively constant and must occur at a very specific time close to the time of SN. Thus we would require a stable non-conservative mass transfer phase, something that may be difficult but not impossible to achieve.

\section{Conclusion}

In this chapter we have shown that single stars provide a restricted number of possibilities for types of SNe. It is also apparent that the definition of which progenitors give rise to which SNe is rather arbitrary. This problem arises because there are not clear cut changes but a continuum. At high metallicities this continuum is compressed in the range of masses over which it occurs and is only resolved at lower metallicities. 

We have also described electron capture SN progenitors that evolve rapidly just before the SN owing to the event of second dredge-up. There may also be some current observational evidence that these electron capture events do occur. Only observational searches for progenitors will find these stars.

Stellar wind mass-loss rates are probably not as large as commonly assumed. It is likely that the requirements for a IIL SN are not as restrictive as previously thought or that the transition objects are currently misidentified as IIpec or IIn before the IIL behaviour dominates.

Wolf-Rayet stars are quite similar at solar metallicity. Most end up as WC stars. With lower metallicity the reduced mass loss slows down WR evolution through the WR types. This means the fate of the most massive stars changes with metallicity. 

With this detailed study we must next move on to study how changing the mass-loss scheme changes the ratios we derive and how these change over a wider metallicity parameter space.

\chapter{Mass-Loss Prescriptions \& Progenitor Populations}

\begin{center}
``The thing's hollow -- it goes on for ever -- and -- oh my God -- it's full of stars!''\\
	\textit{Dave Bowman, 2001: A Space Odyssey, by Arthur C. Clarke.}
\end{center}

\section{Introduction}

There are many examples of similar studies to that in the previous chapter, \citet{WHW02} or \citet{vanb03} for example. However each one uses a different mass-loss prescription and no one has investigated the difference in models between different prescriptions using the same stellar evolution code. Because of this and the basic uncertainty in the mass-loss rates from massive stars a detailed comparison of how the evolution of massive stars depends on the mass-loss prescription is required. Therefore in this chapter we will compare a number of different prescriptions to determine the effect on the SN progenitor population. We will use observations to constrain which mass loss prescription we should use.

The mass-loss rates we study are broken down into pre-WR and WR rates. We also study the effect of scaling the mass-loss rates with metallicity by a factor of $(Z/Z_{\odot})^{n}$ for various $n$ rather than the usually assumed $0.5$ taken from the work of \citet{K87} and \citet{K91}. This scaling arises from the assumption that stellar winds are line driven and with lower surface opacity at lower metallicity there are weaker winds. However while there is agreement that mass-loss scales in this form there is a range of suggested values for the exponent. We have tested the sensitivity to changes in this scaling by producing a grid with $n=0.7$, the exponent suggested by Vink et al. As expected a larger exponent increases variation with metallicity: the type II/Ibc SN boundary varies with metallicity to a greater degree. Previous authors have commonly assumed that this scaling applies at all stages of evolution even though there is some evidence that it varies with spectral type. This seems likely because the exponent assumes the winds are line driven while mass loss from giants is driven by some quite different, unestablished mechanism. If we do not scale the mass loss for giant stars ($\log(T_{\rm eff}/{\rm K}) <3.7$) the result is that the minimum initial mass for WR star formation decreases with decreasing metallicity. This is contrary to observations of WR stars in the LMC and SMC \citep{masseyetal2000}. This provides evidence that mass-loss from giants must scale with metallicity.

We also investigate whether the scaling with initial metallicity should be included for WR stars and the effect this has on our results. We should note that in this thesis we study a broad range of metallicity, high ($1 \times 10^{-3} \le Z \le 0.05$) and  low ($10^{-8} \le Z \le 10^{-3}$). We include an overlap to see how different schemes in the two regimes match.

First we present the various mass-loss rates we use. We next discuss the effect of including the metallicity scaling for WR mass-loss rates and then compare a large number of mass loss rates at high and low metallicity. Finally we present the results from our preferred mass loss rates and compare the differences caused by removing convective overshooting.

\section{The Mass-loss Rates}
In order to produce the various mass-loss prescriptions we combine various rates. Some of these are designed for low metallicity only and this requires us to be careful in which regions we apply the rates.

The pre-WR mass-loss rates we use are
\begin{itemize}
\item JNH: \citet{dJ}. Despite being old these are accurate and detailed \citep{Crow2001}. They give a mass-loss rate for a specific place in the HR diagram. They are quite complex with over 20 terms in the equation but the extra detail increases the accuracy.
\item NJ: \citet{NJ90}. These rates are based on the same data as for JNH but they are simpler with only 3 terms in the mass-loss equation. This does provide some advantages. They are much easier to extrapolate from than the JNH rates and thus do not have such a limited coverage of the HR diagram. This rate generally leads to greater mass loss than the JNH rates.
\item Vink: \citet{VKL2000} and (2001). These are theoretical rates that match well to observed values listed in \citet{dJ}. The model is only valid over a metallicity range of $\frac{1}{30}<(Z/Z_{\odot})<3$.
\item Krki: \citet{KD2002}. These are theoretical rates for the low metallicity regime. They seem to indicate a break in the mass-loss rates at around $Z=10^{-5}$. We have only used this rate in our low metallicity regime.
\end{itemize}
The Vink and Krki rates are only valid for OB stars so they are combined with the NJ and JNH rates.

When $X_{\rm surface} < 0.4$ and $T_{\rm eff} > 10^{4}{\rm K}$ a star becomes a WR star. The WR mass-loss rates we use are
\begin{itemize}
\item NL: \citet{NL00}. These rates are the most recently derived WR mass-loss rates from observations. There are some theoretical rates for WR stars but they do not agree well. These also include the rate for WO stars used by \citet{DrayThesis}.
\item Vbev: \citet{vanbevfull} and \citet{vanbmassloss}. This rate is extremely simple depending only on luminosity and does not consider the change of mass loss with WR stellar type.
\item WL: \citet{WL1999}. These are effectively the rates from \citet{Langer} but lowered by a factor of $3$ to account for clumping of the winds.
\end{itemize}

The analytical expressions used for these mass-loss rates are detailed in appendix A. We combine them in various ways to produce results similar to those in chapter 3 but with coarser resolution in mass because we here have a wider range of metallicity. The combinations we use at high metallicity are
\begin{itemize}
\item HG: NJ+WL, from \citet{H03}.
\item VB: NJ+Vbev, based on the work of \citet{vanbmassloss}.
\item NJ: NJ+NL, for comparison with the JNH rates.
\item NJV: NJ+Vink+NL, to observe the effect of the Vink rates.
\item JNH: JNH+NL, the rates of \citet{DrayThesis}.
\item JNHV: JNH+Vink+NL, to observe the difference of the Vink rates.
\end{itemize}
At low metallicity we also use
\begin{itemize}
\item NJK: NJ+Krki+NL, to observe the effect of the Krki rates.
\item JNHK: JNH+Krki+NL, to observe the effect of the Krki rates.
\end{itemize}

\begin{figure}
\begin{center}
\includegraphics[height=79mm,angle=0]{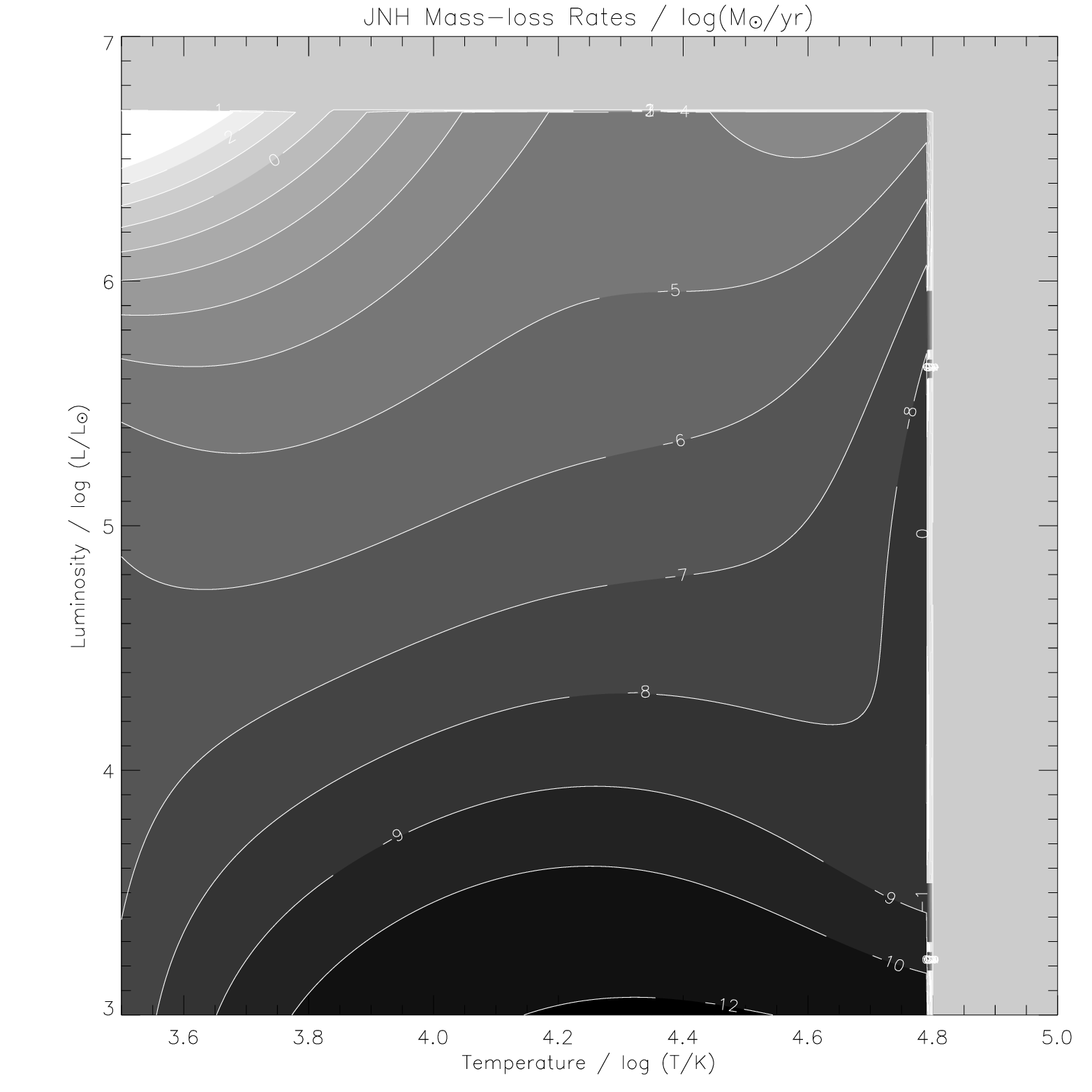}
\includegraphics[height=79mm,angle=0]{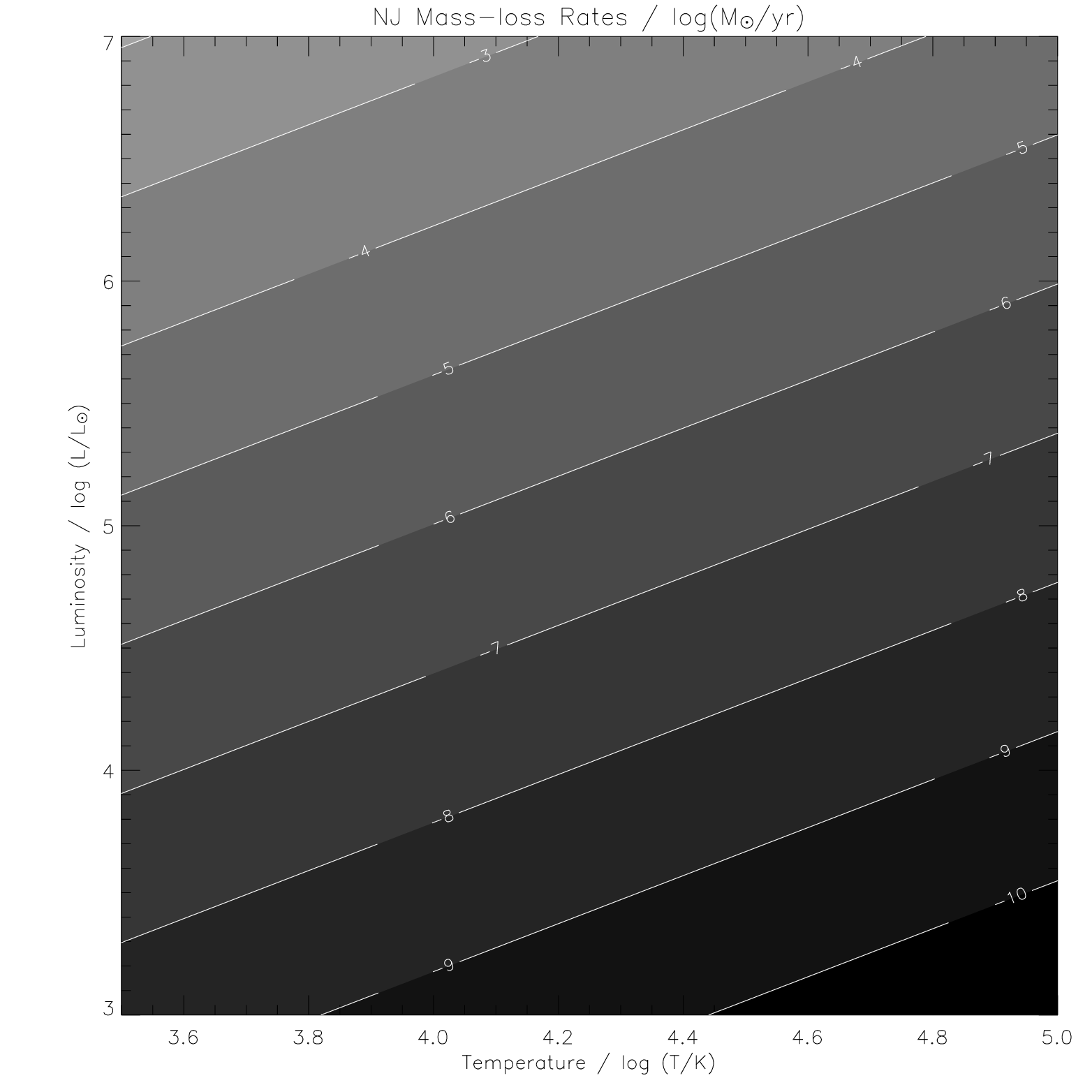}
\end{center}
\caption[The JNH (left) and NJ (right) mass-loss rates on the HR diagram.]{The JNH (left) and NJ (right) mass-loss rates on the HR diagram. The NJ rates are calculated for a $10M_{\odot}$ star. Changing the mass of the stars for which the rates are calculated has only a small effect on the rates.}
\label{rates}
\end{figure}

\section{Models details}
In this work we use 15 zero-age main-sequence models that have masses from $10$ to $200M_{\odot}$ with a uniform composition $X=0.75-2.5Z$ and $Y=0.25+1.5Z$, where $X$ is the mass fraction of hydrogen, $Y$ that of helium and $Z$ is the initial metallicity. We have also created models with alternate compositions: $X=0.7$, $Y=0.3-Z$ and $X=0.76-3Z$, $Y=0.24+2Z$. With constant hydrogen abundance the variation with metallicity is accentuated. For example 2nd dredge-up occurs at higher masses with increasing metallicity. Changing the primordial hydrogen abundance from 0.75 to 0.76 as in the second case only produces small changes in our results.

Unless stated otherwise all models are evolved with convective overshooting included as described in chapter 2 and are evolved with 499 mesh points. Some problems were encountered in modelling the highest mass stars at high metallicity. There is a small range of mass where large helium stars with high mass-loss rates are formed and these cannot be evolved. The physical parameters move outside of the region covered by the opacity tables, to high temperature and low density. Linear extrapolation of the tables does not work and the density continues to decrease. The star is expanding and becoming a high mass helium giant. For most of the sets of models we have interpolated over mass to complete the grid of models. If the models are at the extremities of our range we have not extrapolated into that region. The method to produce models for this regime is to limit the mass-loss rate during the WR phase to a constant value that gives models of similar mass and only slightly more massive than neighbouring models and similar to the expectations from interpolation. This indicates the interpolation is an acceptable compromise. The range of initial masses is around $80-100M_{\odot}$ at solar metallicity. This range decreases with increasing metallicity and increases with lower metallicity. When $Z \le 0.001$ the problem region moves above the range of mass studied. These stars indicate that there may be a maximum WR mass-loss rate because the mass loss for these stars is so severe it causes them to almost evaporate. Alternatively it could mean that the mass loss should be greater and that the low-density regions actually should be ejected. To properly investigate this theoretical rates for WR stars are required to better understand this feature and discover if it is real or not. 

\section{Supernovae Type and Remnant Determination}

We use similar methods to those in chapter 3 but for comparison we use simpler schemes. For a star with a model at the end of core carbon burning we say it will become a type II SNe if there is any hydrogen in its envelope. Furthermore we know observationally that type IIP SNe have retained a good proportion of their original hydrogen. This hydrogen causes the light curve to follow a plateau powered by a moving hydrogen ionisation front. Type IIL SNe only retain a small fraction of their hydrogen and the light curve decays linearly. We adopt the scheme of \citet{H03} with type IIP SNe until the mass of hydrogen in the star drops below $2.0 M_{\odot}$ when type IIL SNe occur

When there is no hydrogen the SN will be type I. As already detailed it is difficult to discriminate between type Ib and Ic. Therefore we adopt a scheme to use the surface helium abundance for the moment. If $Y_{\rm surface} >0.7$ a type Ib SN occurs; if $0.3 < Y_{\rm surface} < 0.7$ we denote it as a type Ibc due to uncertainty and finally if $Y_{\rm surface} <0.3$ type Ic. These are arbitrary boundaries but are simple enough that we can gain some understanding of the relative populations.

We adopt from \citet{H03} a method of determining the strength of a type Ibc SN to quantify the number of SNe that have no observable display and are thus unseen. This allows us to estimate ratios of type II to type Ibc SNe. The largest progenitors are thought to give no display because the core is so massive that, even with a large explosion energy, nothing escapes the forming black hole. The exception would be if a jet driven SNe occurs that makes a black hole and produces an observable display \citep{MWH01}. We take the ranges from \citet{EW88} for the strength of the display:

\begin{itemize}
\item $M_{\rm Core}({\rm He}) > 15M_{\odot}$, no display.
\item $15M_{\odot} > M_{\rm Core}({\rm He}) > 8M_{\odot}$, faint SN.
\item $8M_{\odot} > M_{\rm Core}({\rm He}) > 5M_{\odot}$, possibly faint SN.
\item $M_{\rm Core}({\rm He}) < 5M_{\odot}$, bright SN.
\end{itemize}

Determination of the remnant formed at the heart of a SN is a black art. The physics is extreme and poorly understood. However there are many prescriptions which rely on the conjecture that a more massive core leads to a more massive remnant. We consider here three types of stellar remnant, white dwarfs, neutron stars and black holes.

White dwarfs come in three flavours. They can be made of helium (a HeWD), a mixture of carbon and oxygen (a COWD) or a mixture of oxygen and neon with some sodium and magnesium (an ONeWD). HeWDs are formed by very low mass stars or more probably in binary stars and are not of interest to us here. COWDs are also not formed in SNe but are the remnant cores of stars that lose their envelopes on the asymptotic giant branch (AGB). At solar metallicity stars in the mass range $0.8$ to $7 M_{\odot}$ (no convective overshooting) produce a COWD while larger stars, up to $10.5M_{\odot}$ develop an ONe core during the early AGB. Some of these lose their envelope and form an ONeWD. The most massive undergo a SN driven by electron capture on to magnesium. The dividing line between these two is of great interest but is somewhat uncertain and depends on the mass-loss rate and the details of the thermal pulses in AGB stars as discussed in the previous chapter.

In our selected mass range ($10-200M_{\odot}$) all stars undergo a SN and form a neutron star or black hole. To distinguish between neutron stars and black holes there are a number of different prescriptions. We compare that of \citet{H03} with our own.

\citet{H03} use a simple system based on the helium core mass at the end stage of evolution. This is useful because the helium burning life-time is slower than for the later stages of burning so the core mass increases by only a small amount in the late burning stages. However the method relies on setting the values for different remnants by using simulations of SNe. The values used are
\begin{itemize}
\item Neutron Star: $M_{\rm Core}({\rm He}) < 8M_{\odot}$.
\item Black Hole (by fall back): $8M_{\odot} < M_{\rm Core}({\rm He}) < 15M_{\odot}$.
\item Black Hole (directly): $M_{\rm Core}({\rm He}) > 15M_{\odot}$.
\end{itemize}
These values are based on the SN models of \citet{Fry99}. By using this scheme we can compare directly with \citet{H03}. We should also note that SNe that form a black hole directly also have no display, while those SN forming black holes by fall back will be faint unless a jet driven SNe occurs producing a black hole and an observable display.

\section{WR Mass-loss rates: scaling with initial metallicity}
The first matter to investigate is that of the scaling of WR mass-loss rates with metallicity. WR stars have higher mass-loss rates than OB stars of the same mass. The main difference between OB and WR stars is that the WR stars have very little hydrogen or no hydrogen at all with the surface dominated by helium or in later WR types carbon and oxygen. This affects the number of lines for radiatively driven winds. This provides evidence that the composition affects mass-loss. 

It is known that mass loss is affected by metallicity in OB stars. Our question is, `if OB winds depend on initial metallicity do WR winds too?' This would mean that the metals in the initial metallicity have a strong effect on the mass-loss mechanism of WR stars, perhaps providing evidence that WR winds are still linked to radiative driving. There is observational evidence for this too, \citet{WRZscale} see a scaling of $(Z/Z_{\odot})^{0.5}$ from observations of WR stars in the LMC. From these tentative observations it is necessary to determine how this affects the evolution and final end states of massive stars. For this study we use the rates of JNH, Vink and NL. We then vary the metallicity scaling of the NL rates, setting the exponent to 0, 0.5 and 0.7. 

\begin{figure}
\begin{center}
\includegraphics[height=75mm,angle=270]{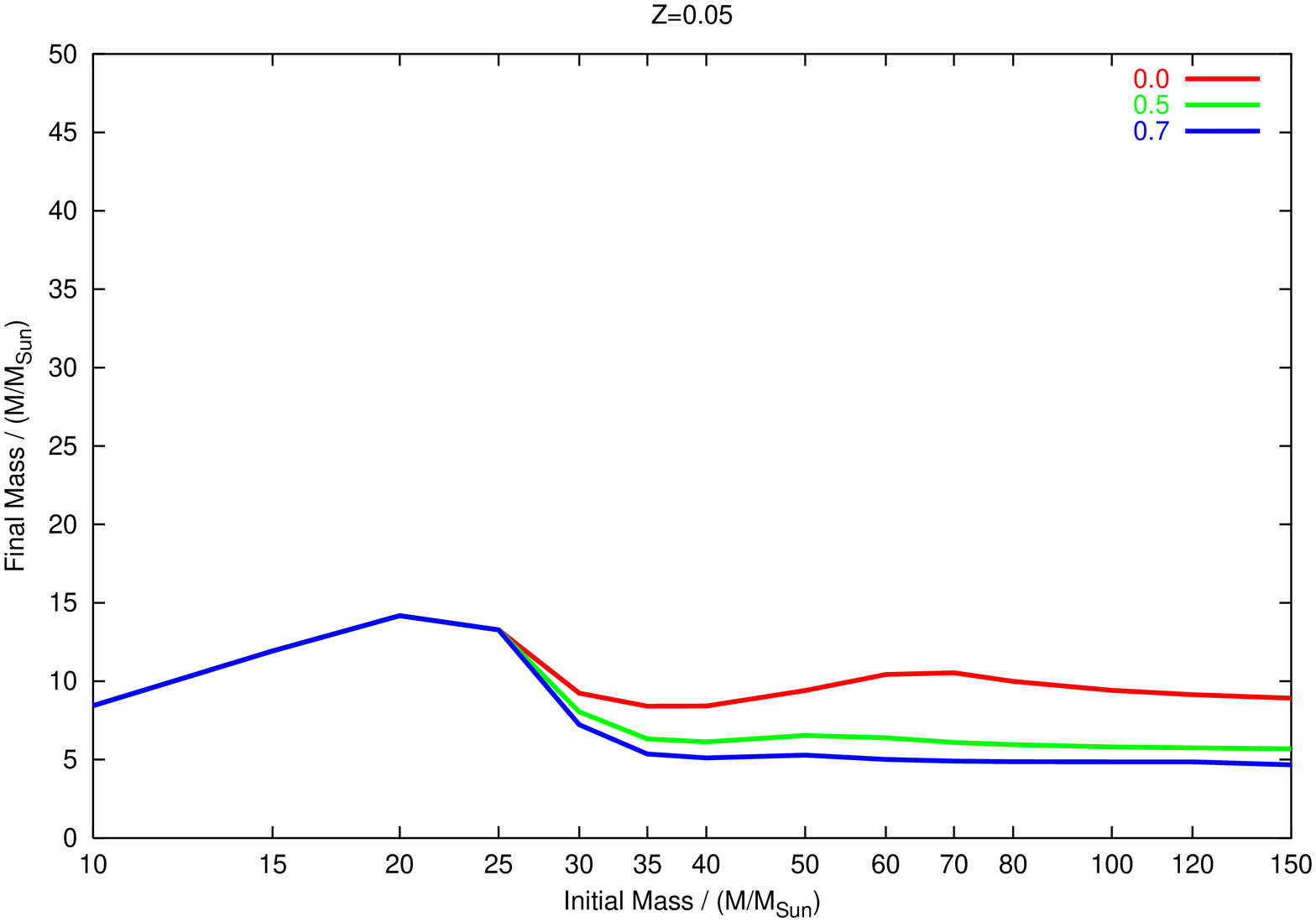}
\includegraphics[height=75mm,angle=270]{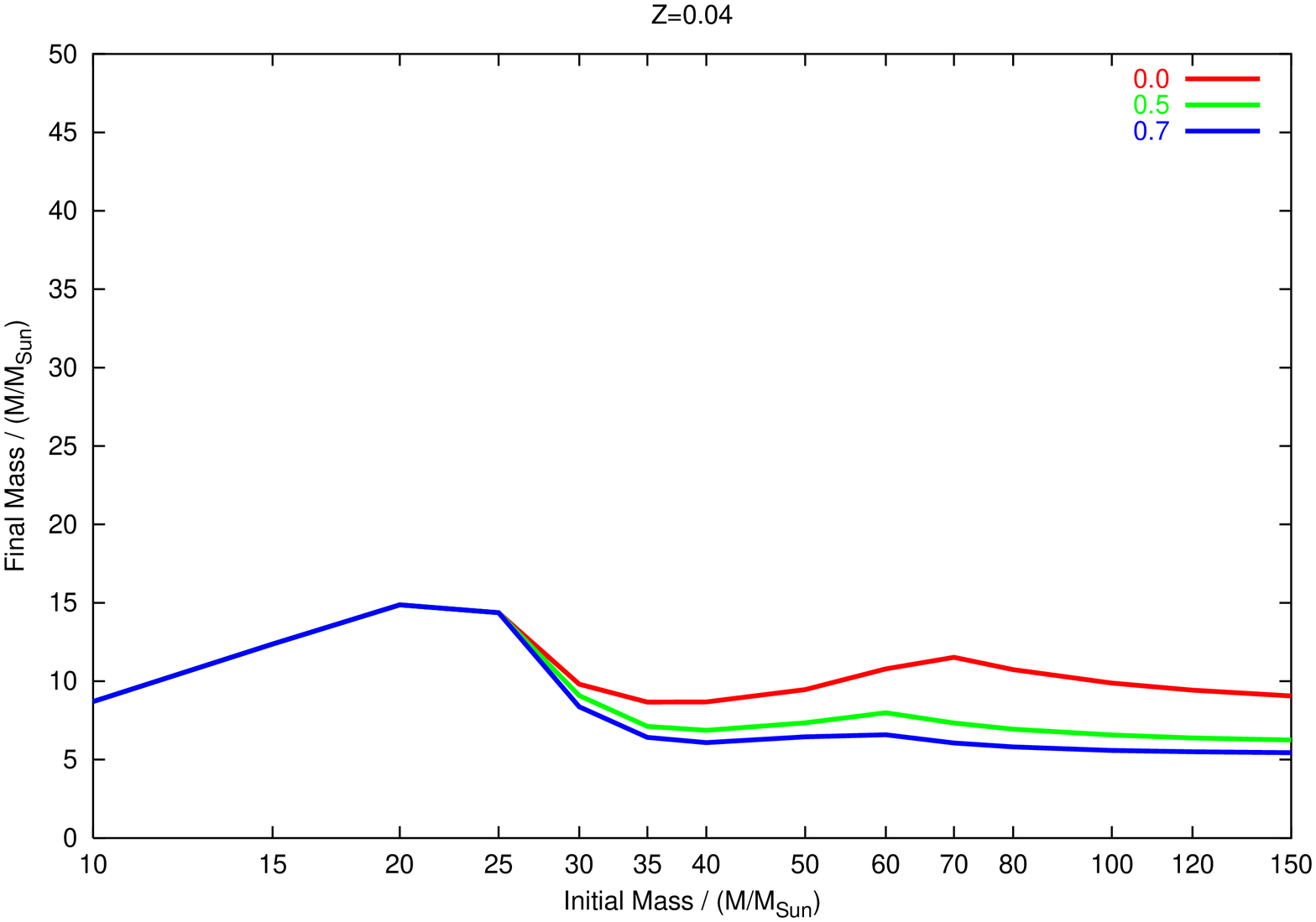}
\includegraphics[height=75mm,angle=270]{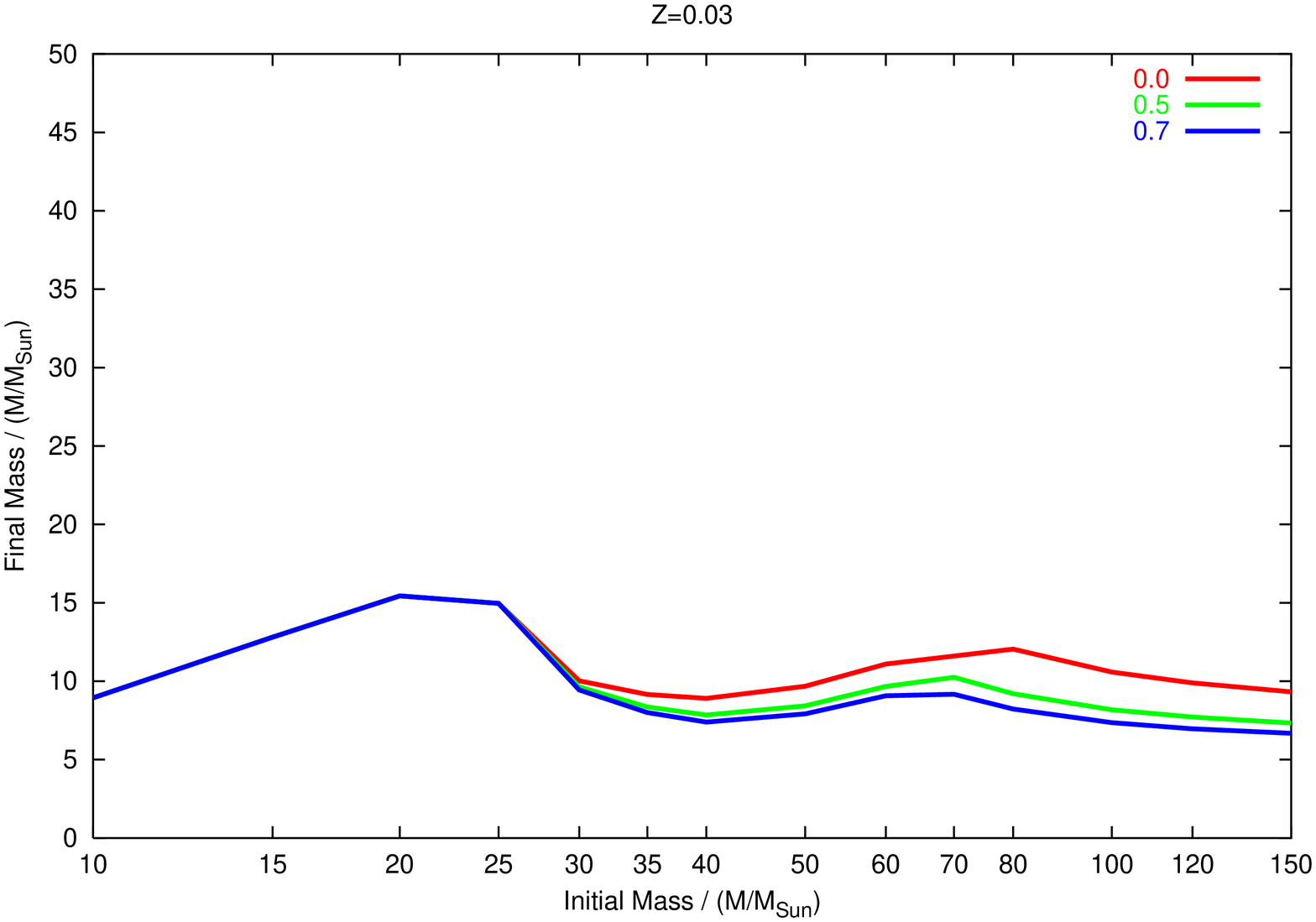}
\includegraphics[height=75mm,angle=270]{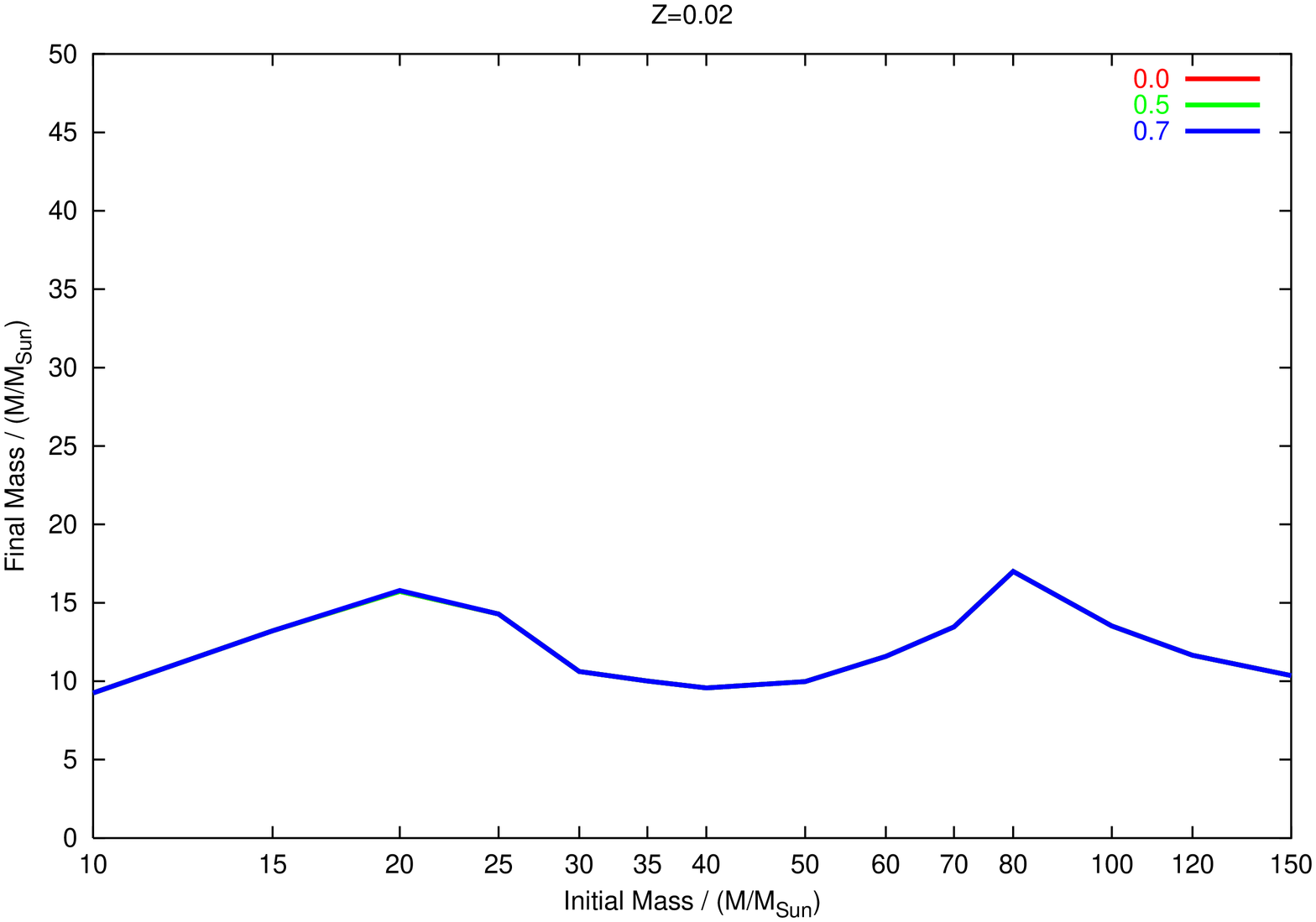}
\includegraphics[height=75mm,angle=270]{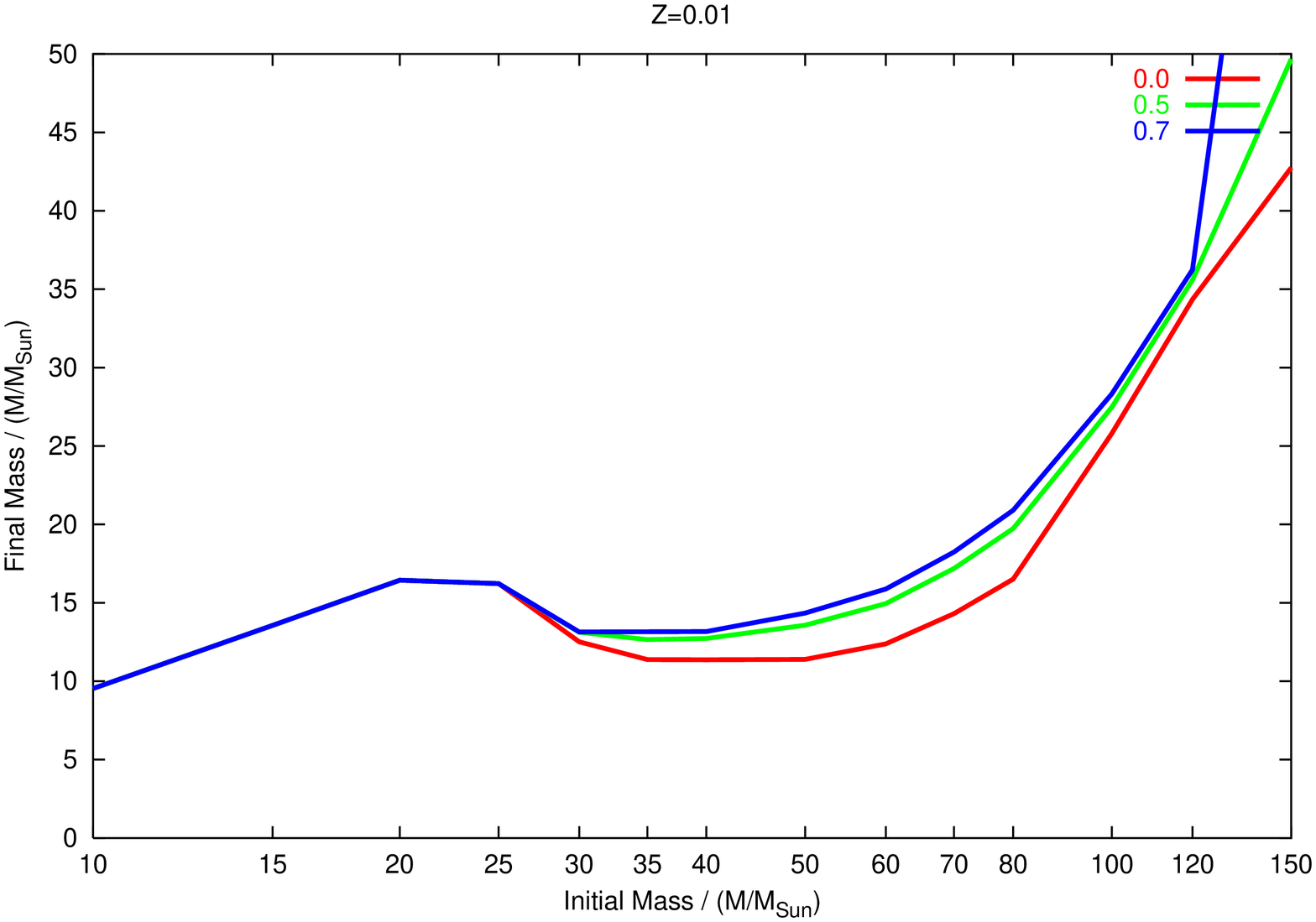}
\includegraphics[height=75mm,angle=270]{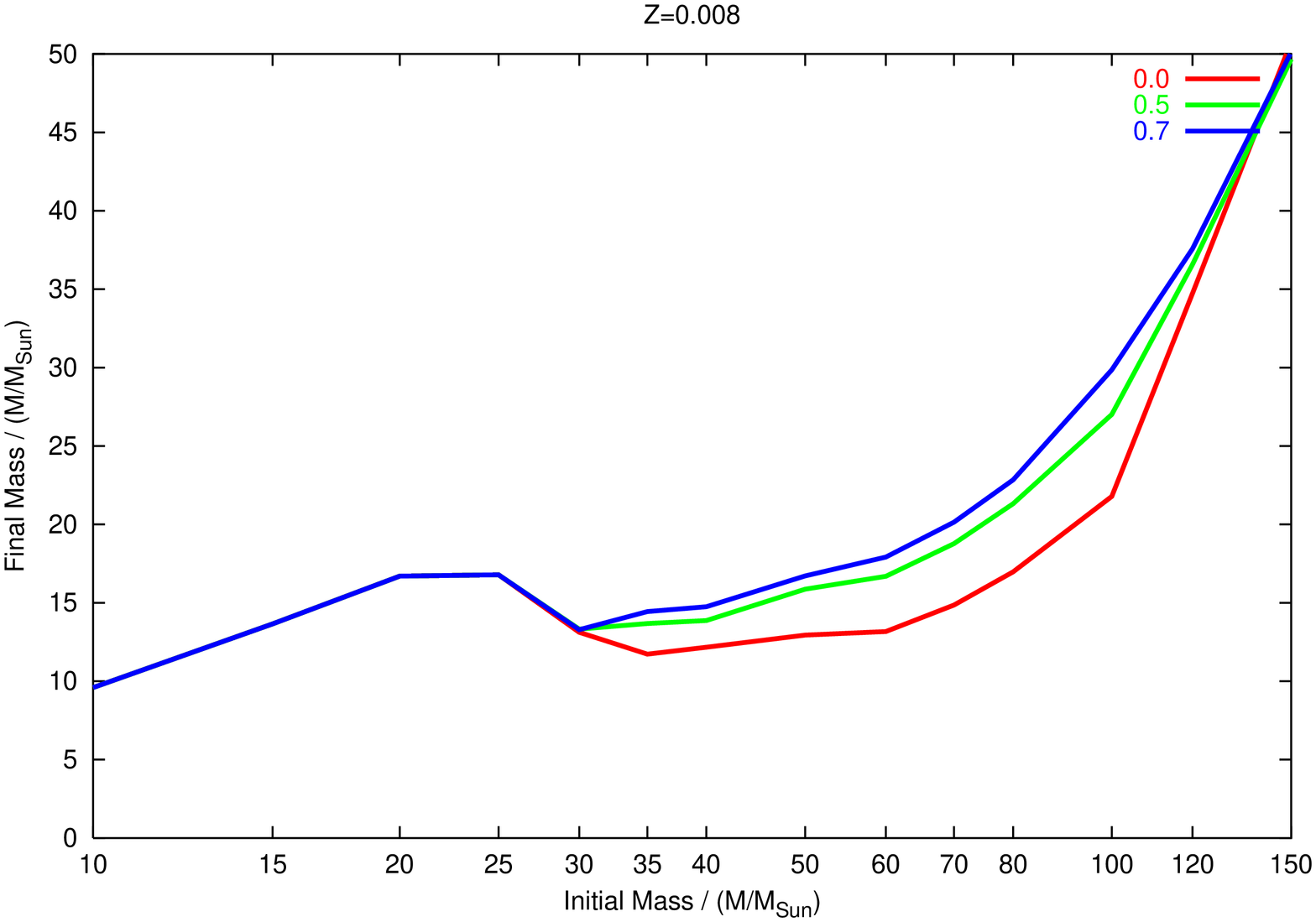}
\includegraphics[height=75mm,angle=270]{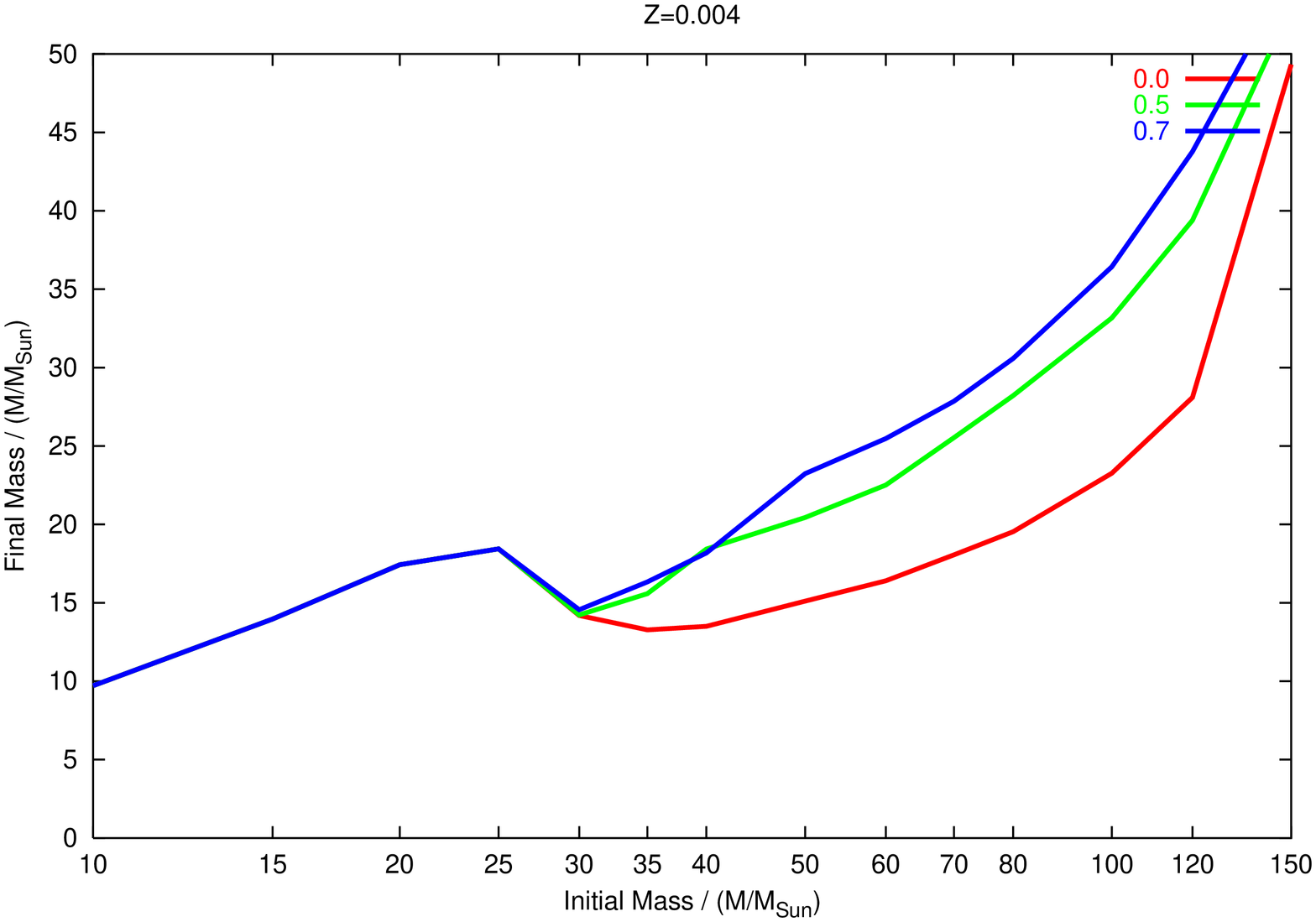}
\includegraphics[height=75mm,angle=270]{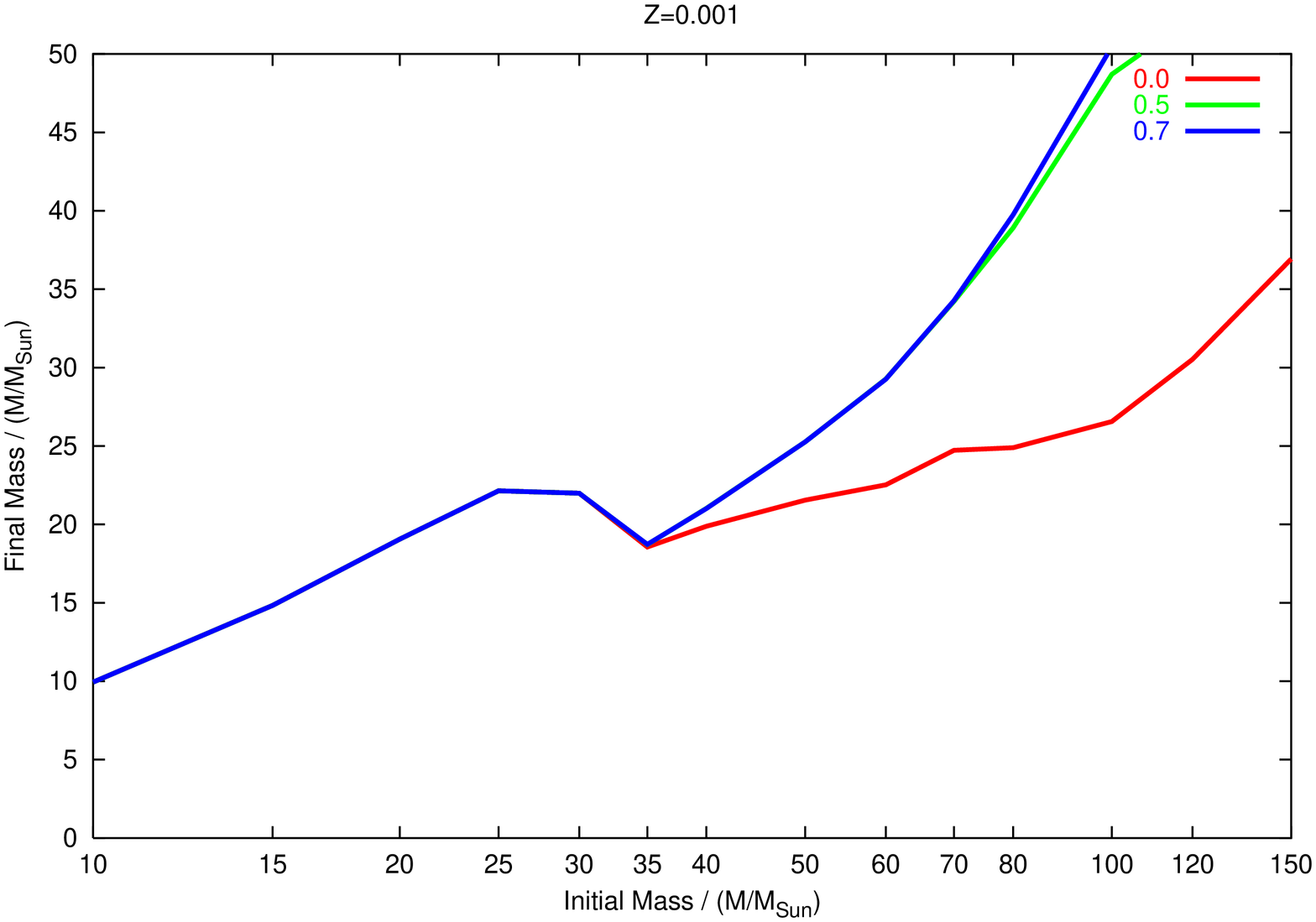}
\end{center}
\caption[The final mass versus initial mass for different WR mass loss scaling laws at different metallicities.]{The final mass versus initial mass for different WR mass loss scaling laws at different metallicities. The numbers in the legend are the value of the exponent, $n$, in the initial metallicity scaling $\dot{M}(Z)=\dot{M}(Z_{\odot})(Z/Z_{\odot})^{n}$. }
\label{wrfinal}
\end{figure}

\citet{masseyetal2000} determined the limits for WR stars to form at different metallicities by studying the turn-off ages of clusters with WR stars. The limits are $M_{{\rm min,} \odot} > 25M_{\odot}$, $M_{{\rm min, LMC} } > 30M_{\odot}$ and $M_{{\rm min,SMC} } > 70M_{\odot}$. They also derive lower limits for the formation of WC stars but these are all limited by small number statistics. The WC minimum masses are higher than the WR minimum masses and increase at lower metallicity. Our models must therefore agree with this result. But we must consider the fact that the observations are for a very small sample of stars and thus have large errors associated with them. Some of the observed stars are also likely to be binary stars.

\begin{figure}
\begin{center}
\includegraphics[height=75mm,angle=270]{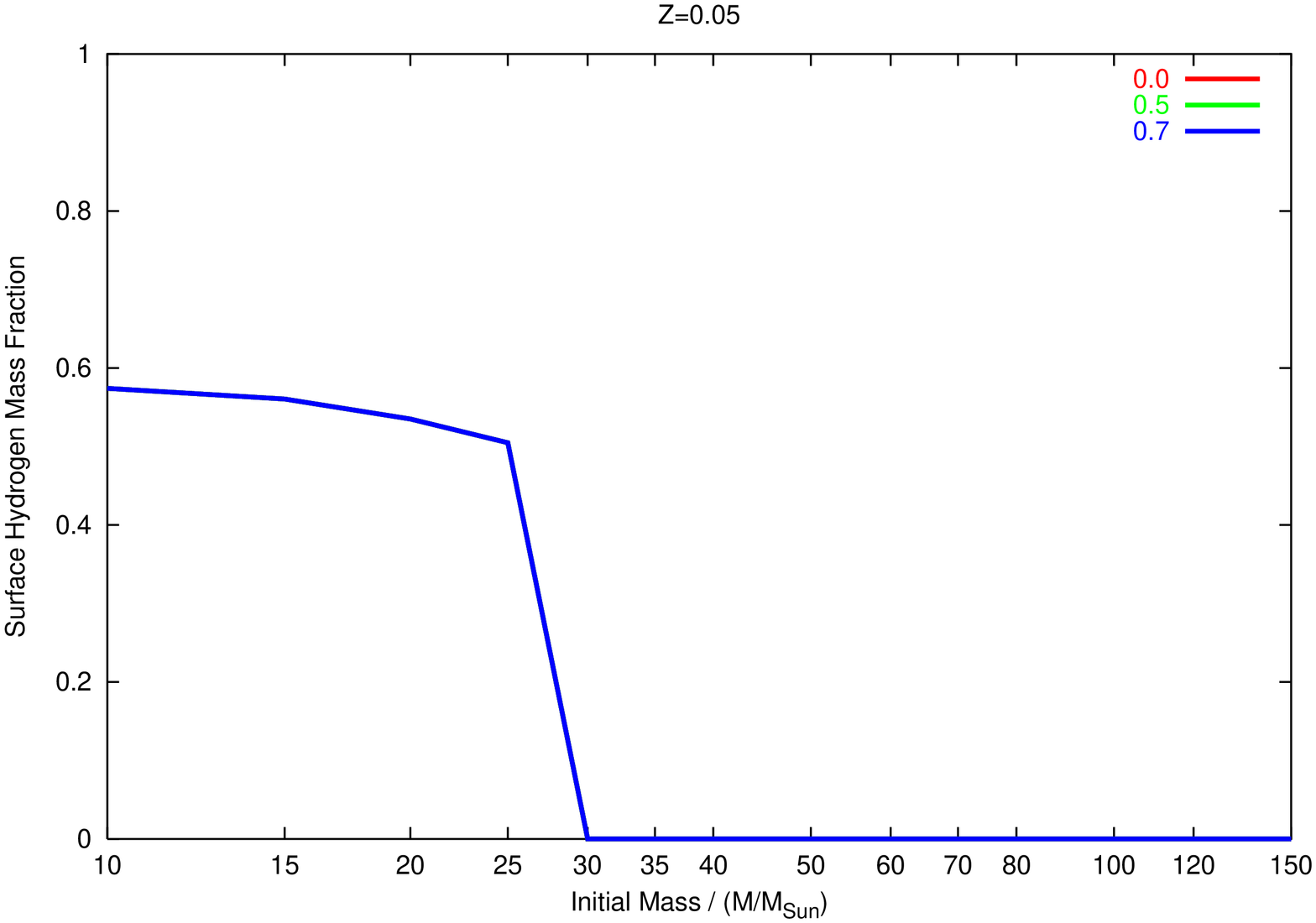}
\includegraphics[height=75mm,angle=270]{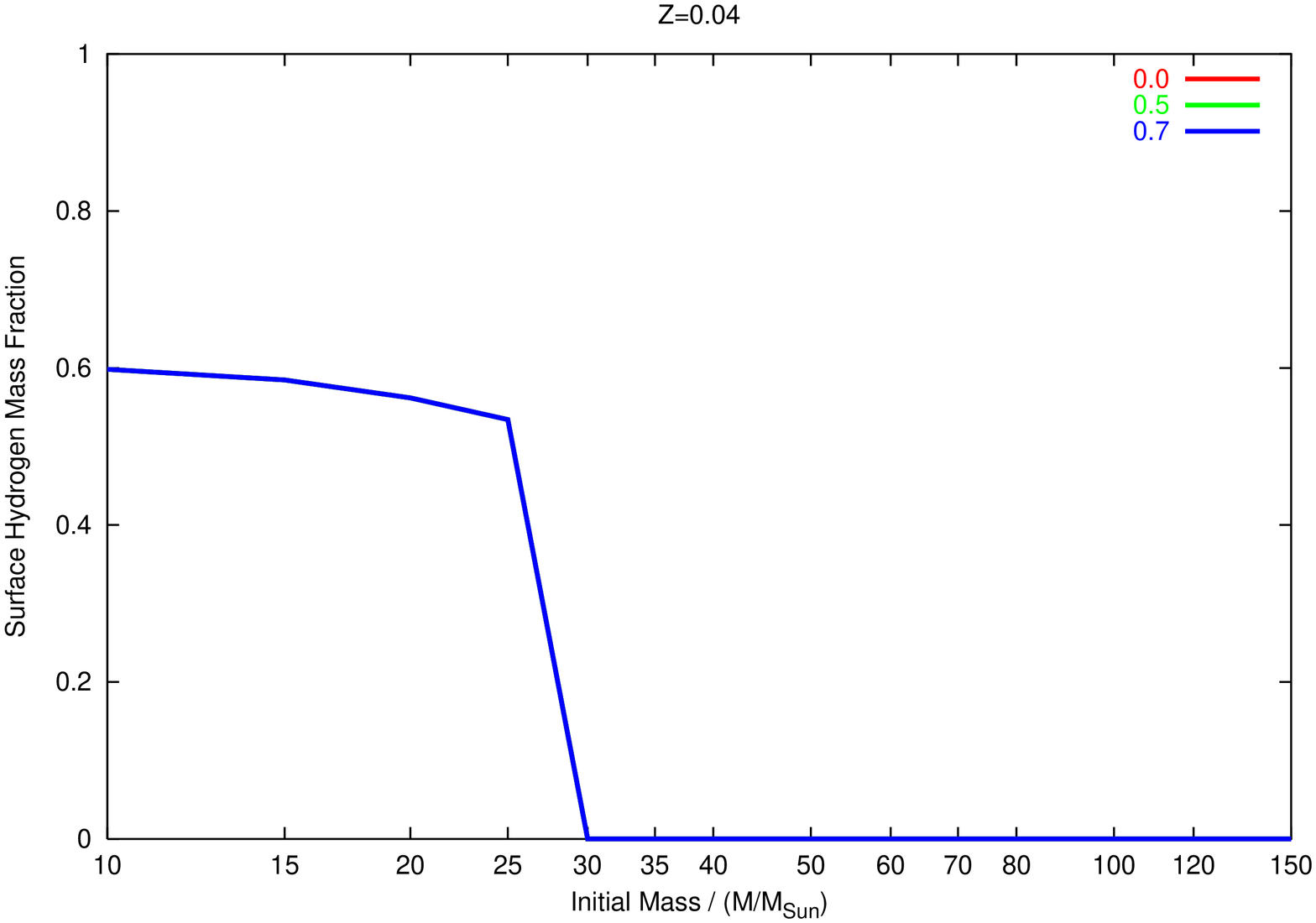}
\includegraphics[height=75mm,angle=270]{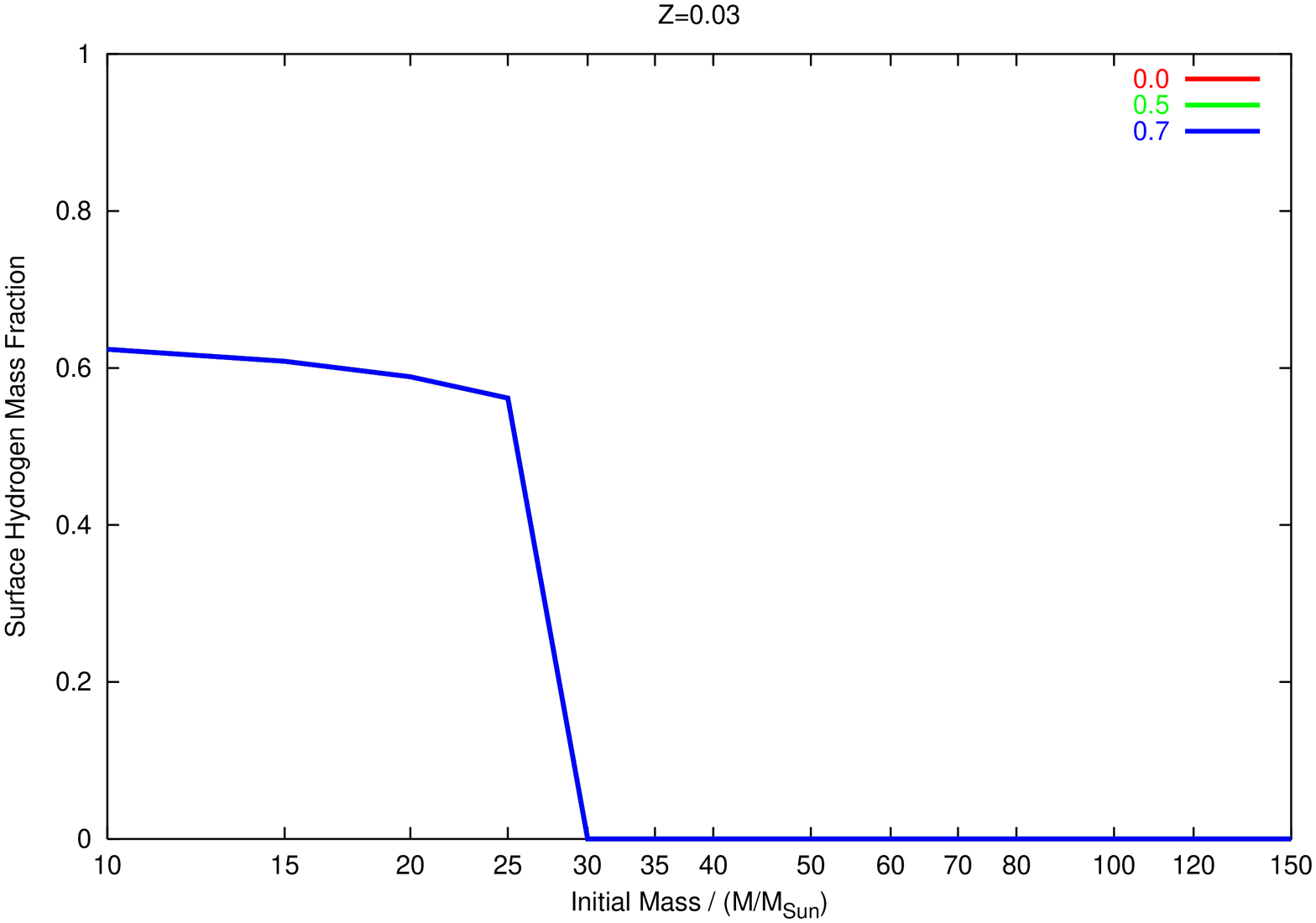}
\includegraphics[height=75mm,angle=270]{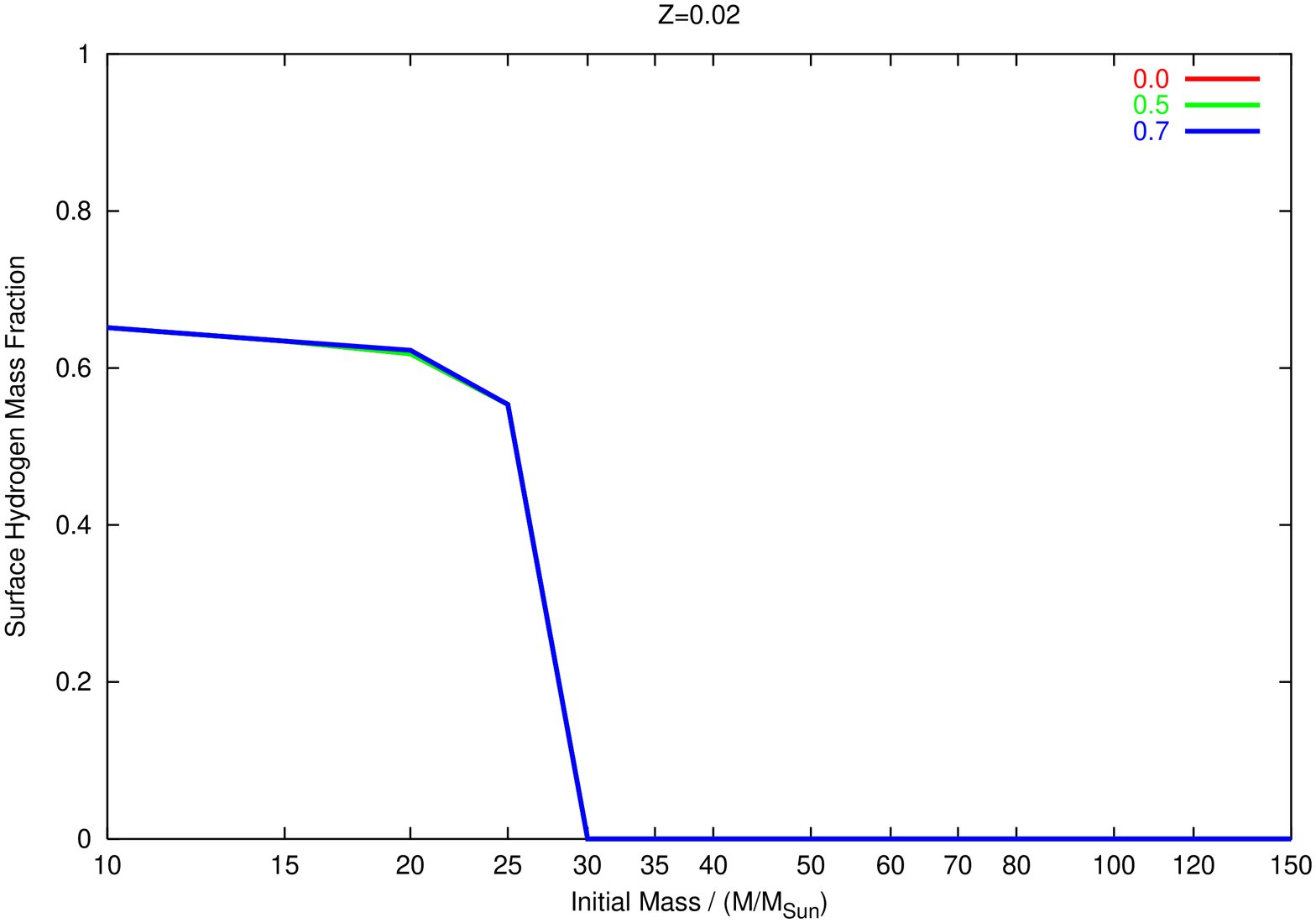}
\includegraphics[height=75mm,angle=270]{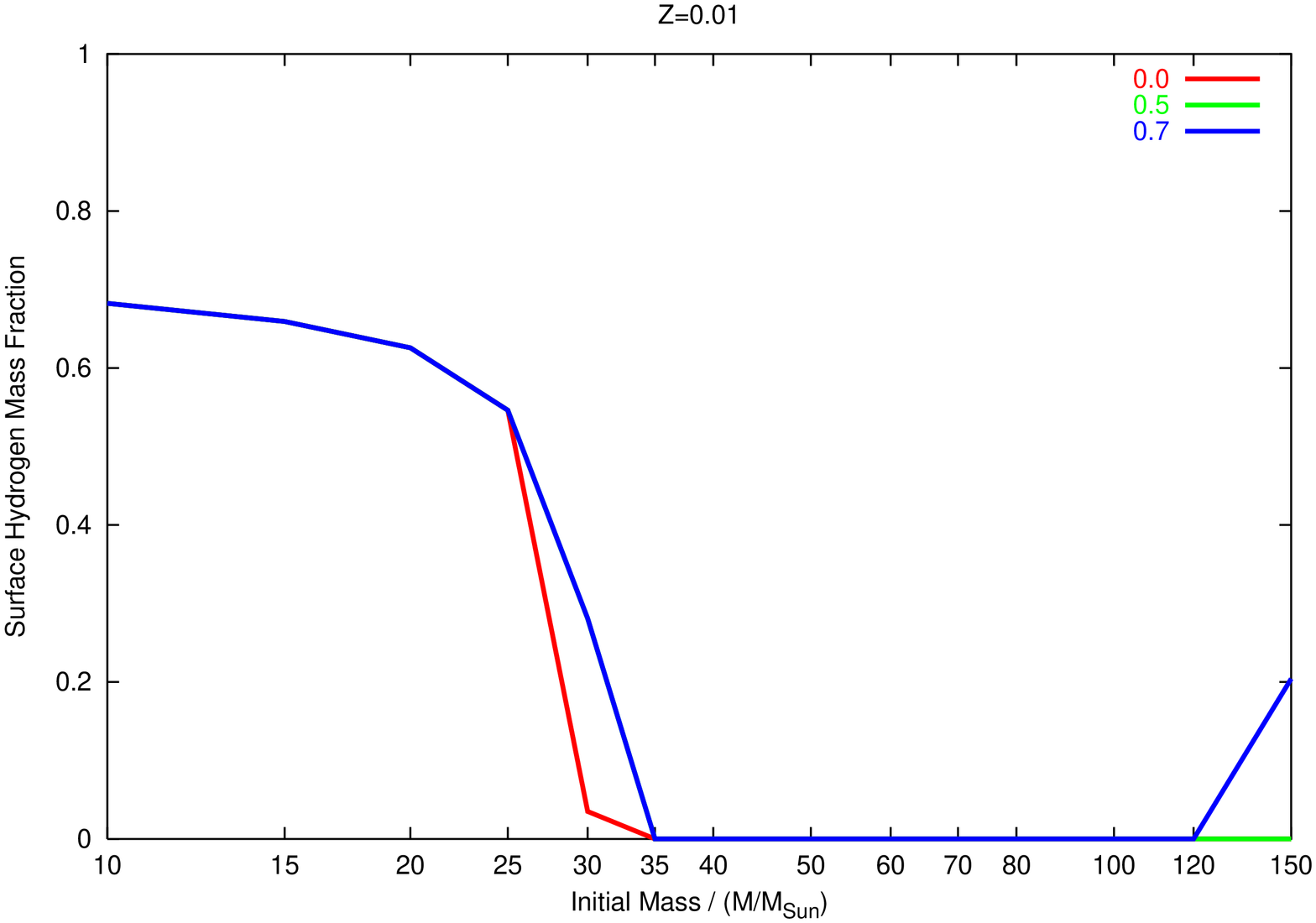}
\includegraphics[height=75mm,angle=270]{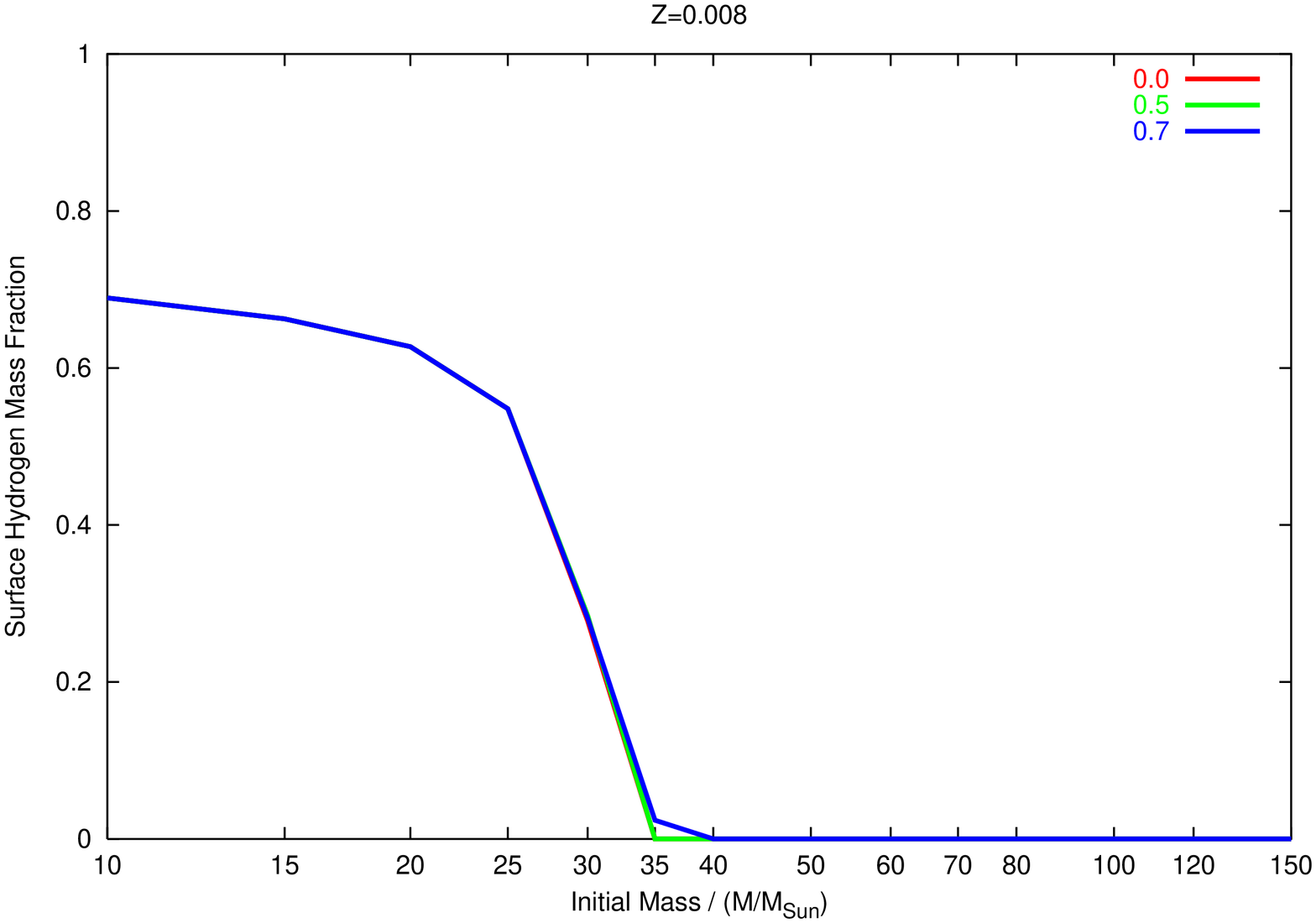}
\includegraphics[height=75mm,angle=270]{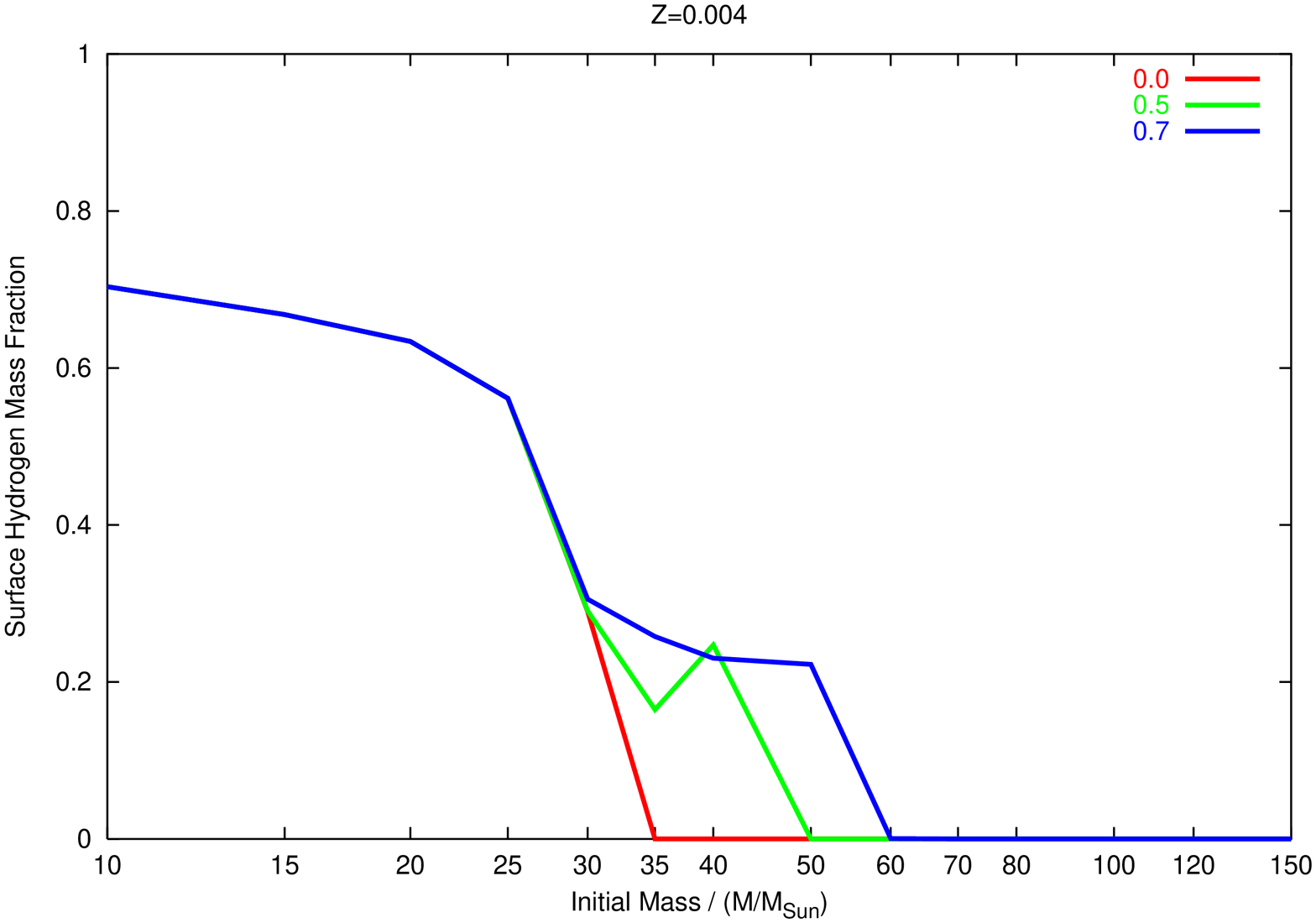}
\includegraphics[height=75mm,angle=270]{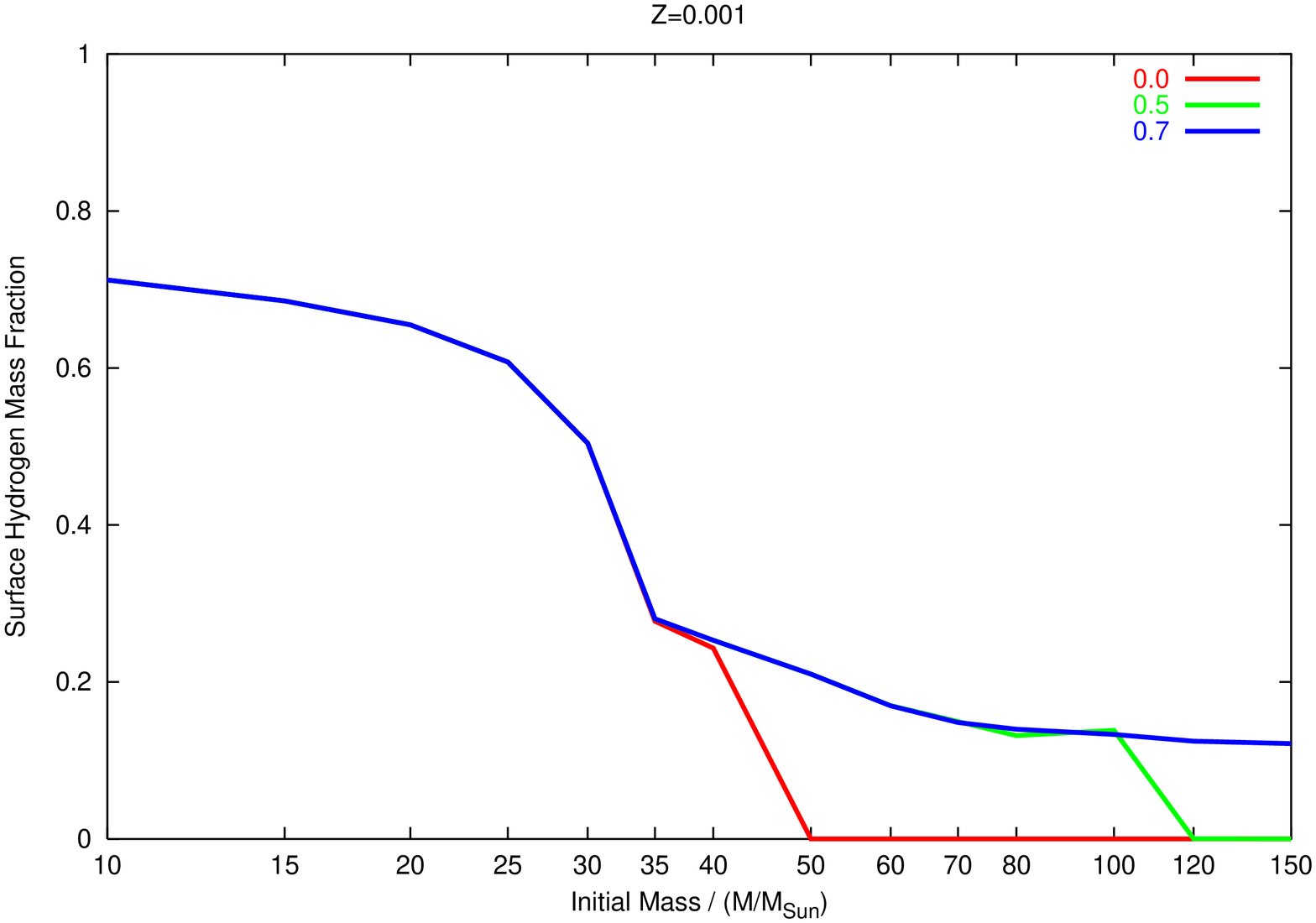}
\end{center}
\caption[The final hydrogen mass fraction on the surface versus initial mass for different WR mass loss scaling laws at different metallicities.]{The final hydrogen mass fraction on the surface versus initial mass for different WR mass loss scaling laws at different metallicities. The numbers in the legend are the value of the exponent, $n$, in the initial metallicity scaling $\dot{M}(Z)=\dot{M}(Z_{\odot})(Z/Z_{\odot})^{n}$.}
\label{wrX}
\end{figure}

\begin{figure}
\begin{center}
\includegraphics[height=75mm,angle=270]{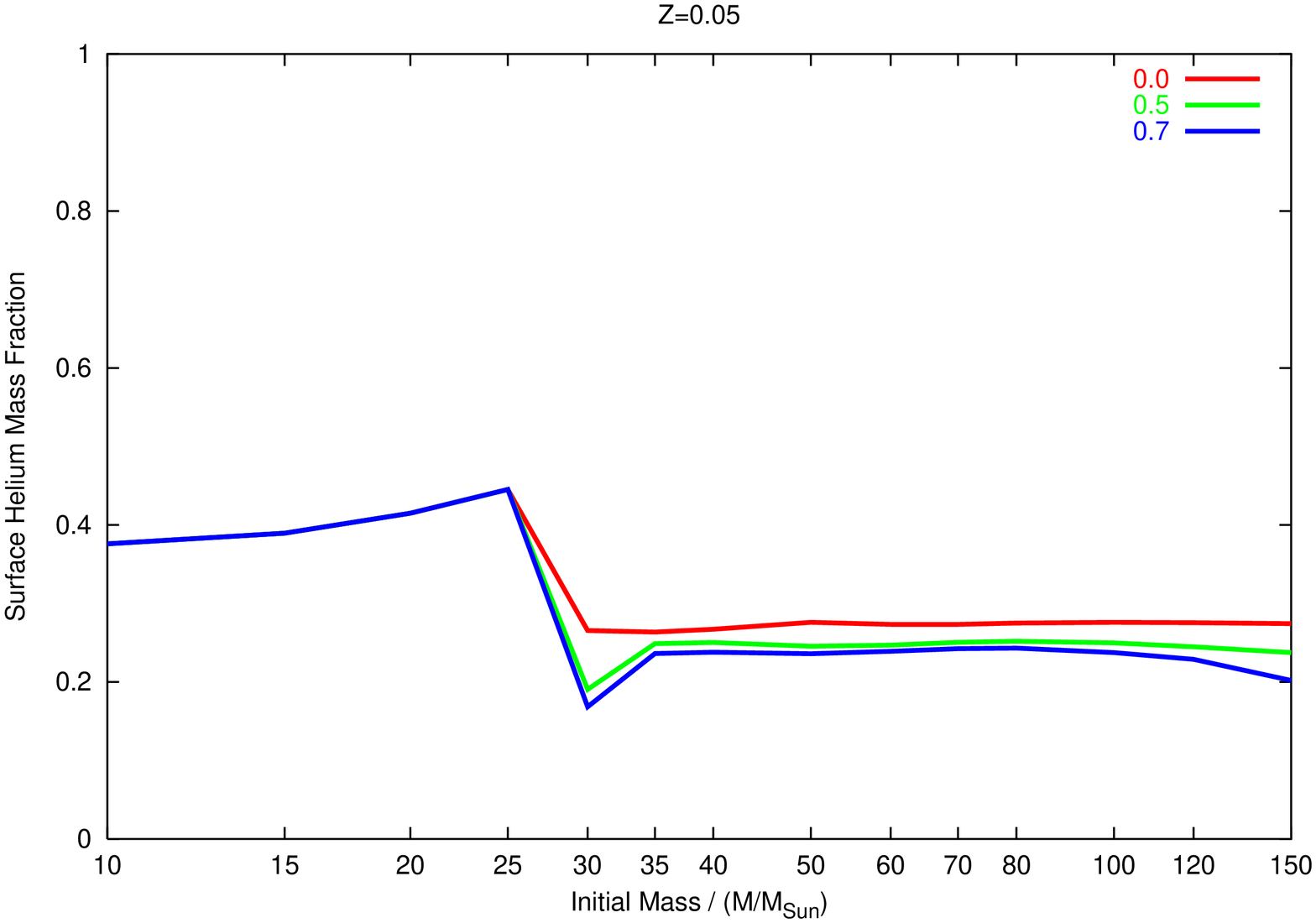}
\includegraphics[height=75mm,angle=270]{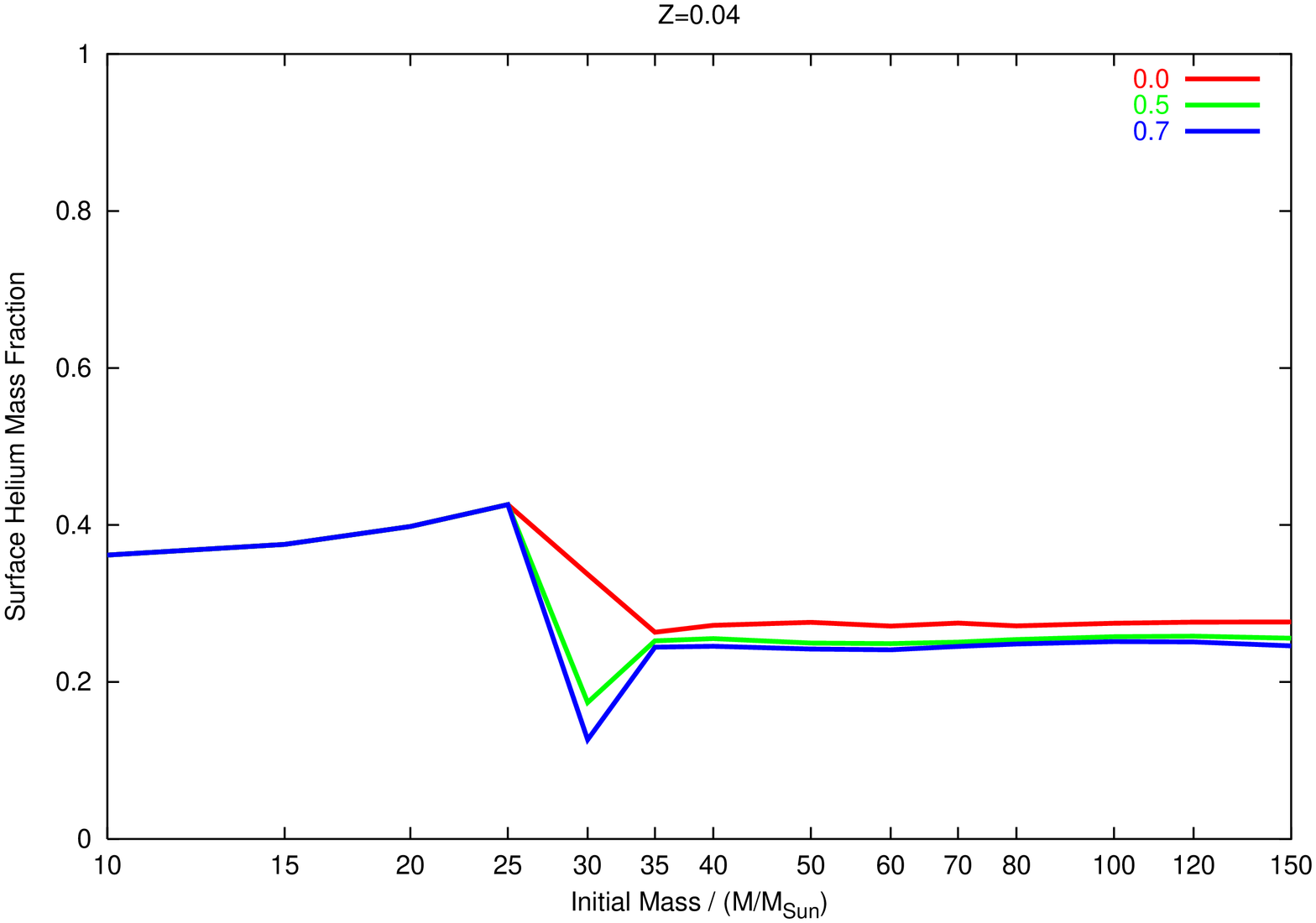}
\includegraphics[height=75mm,angle=270]{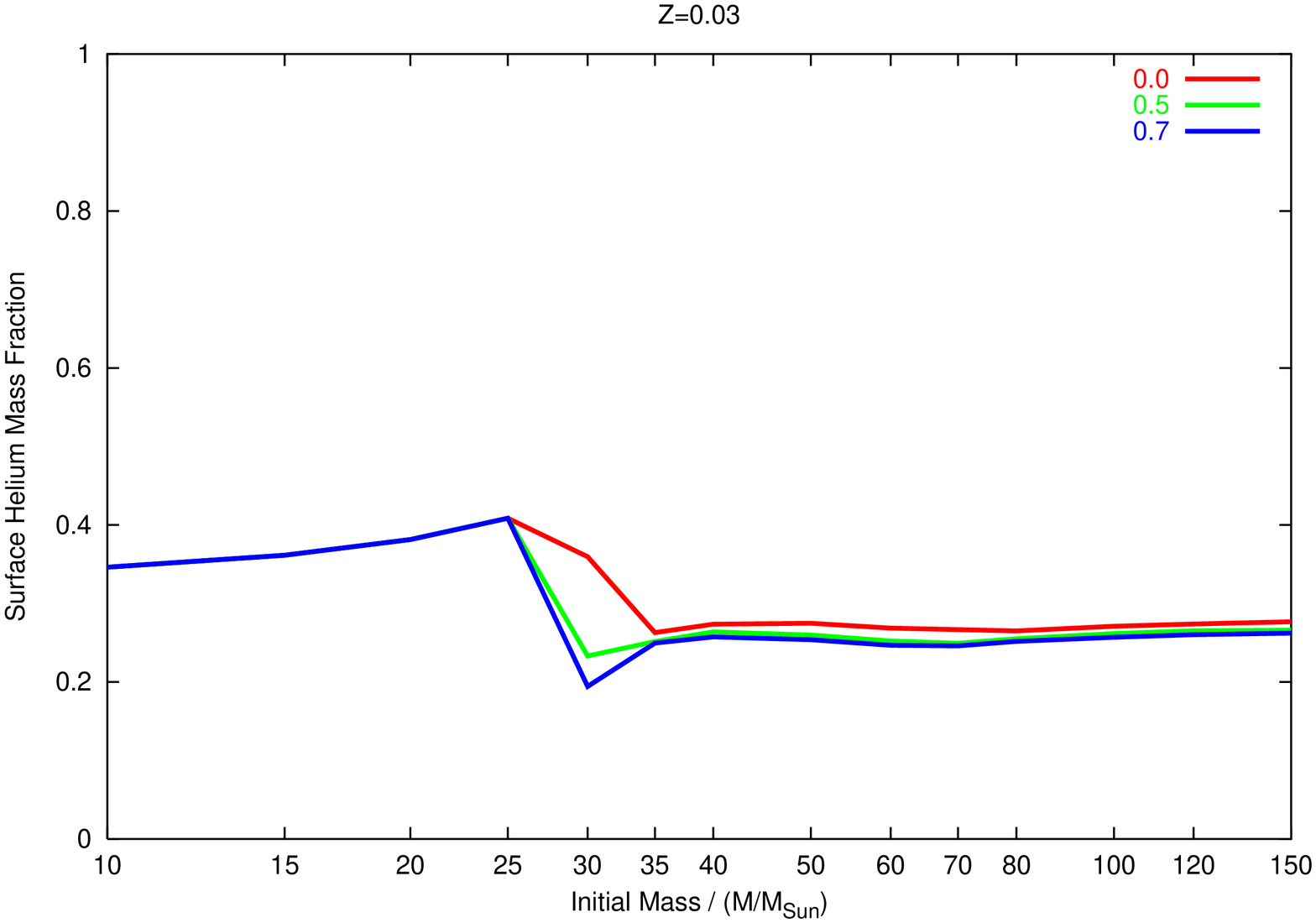}
\includegraphics[height=75mm,angle=270]{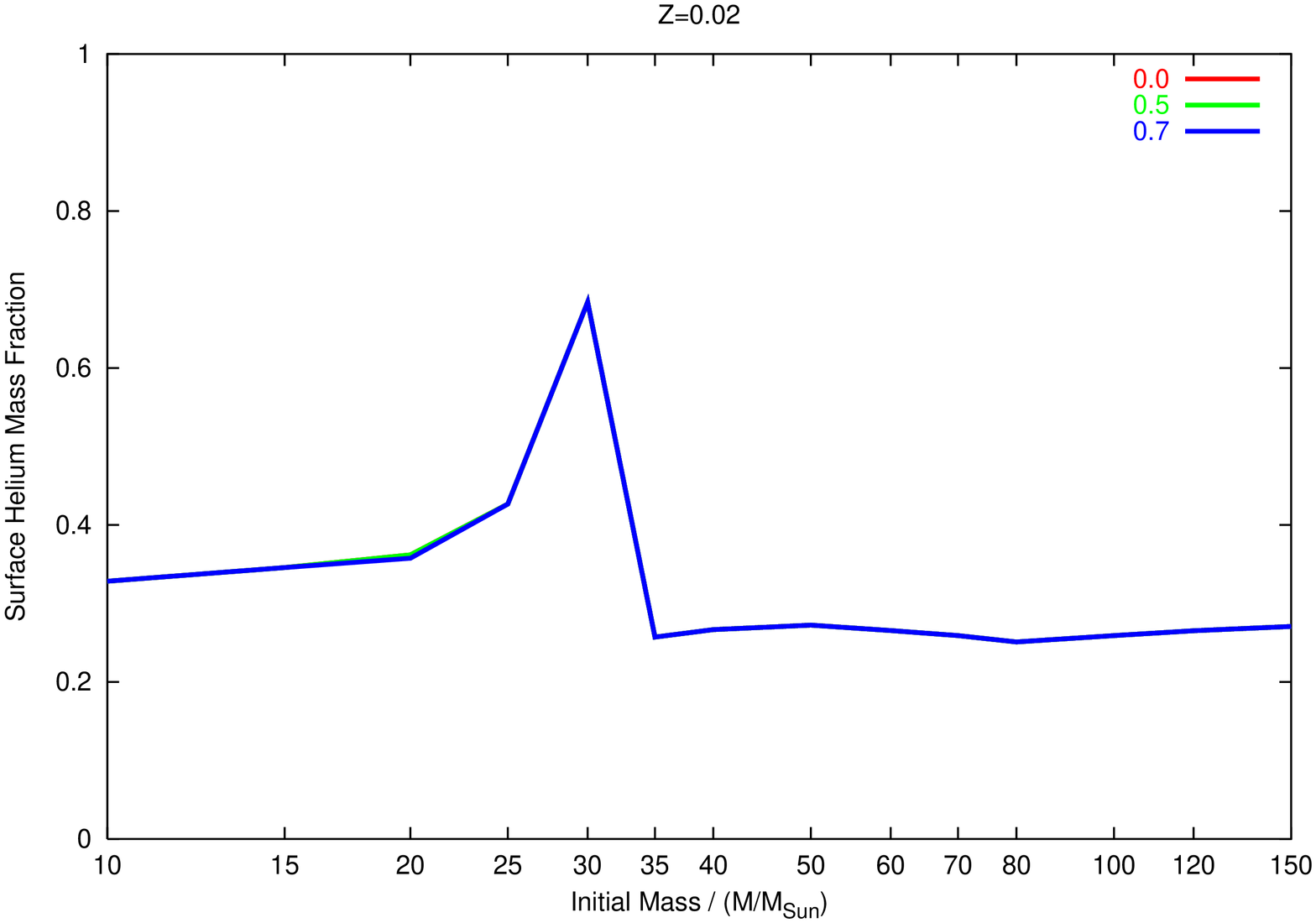}
\includegraphics[height=75mm,angle=270]{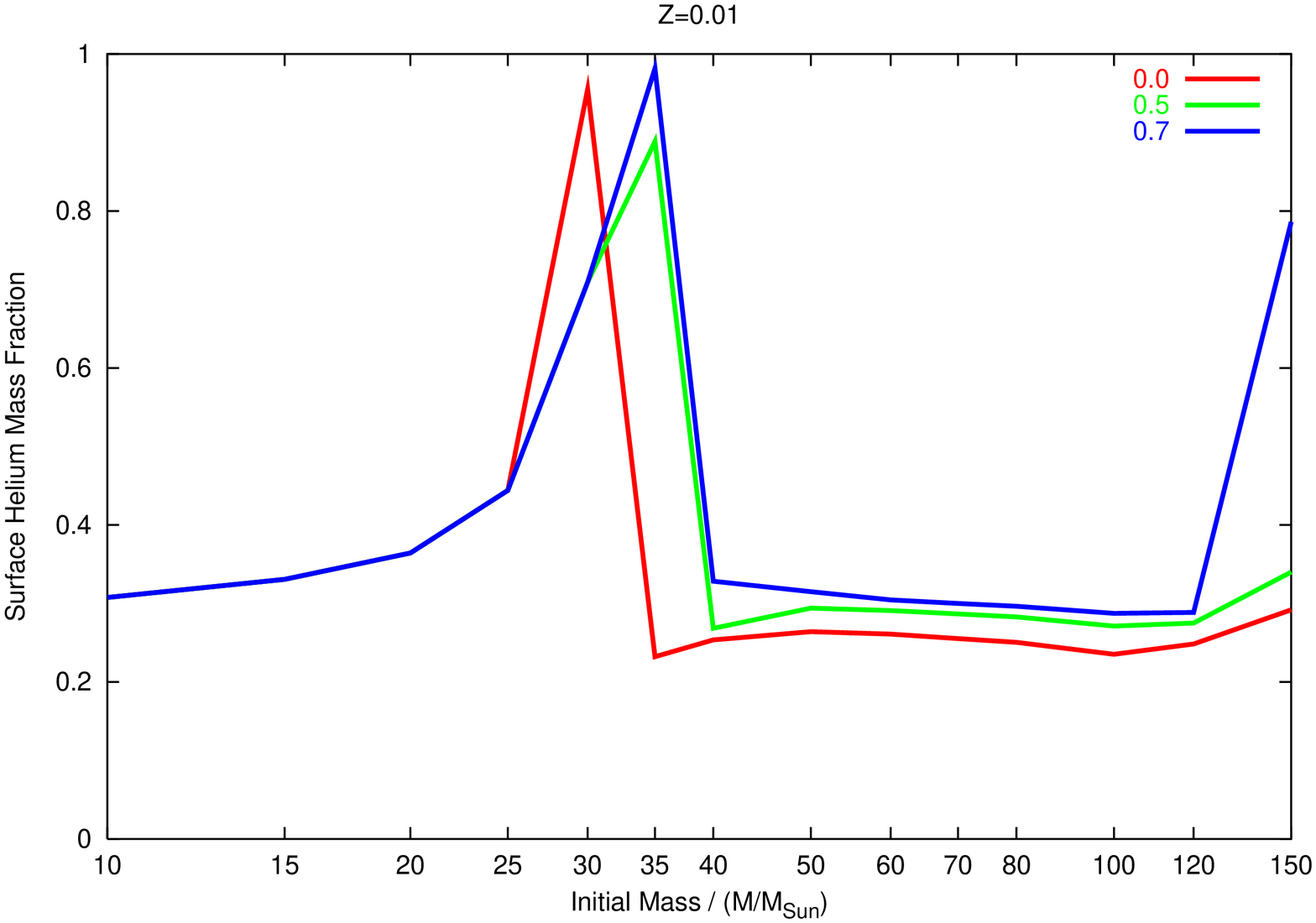}
\includegraphics[height=75mm,angle=270]{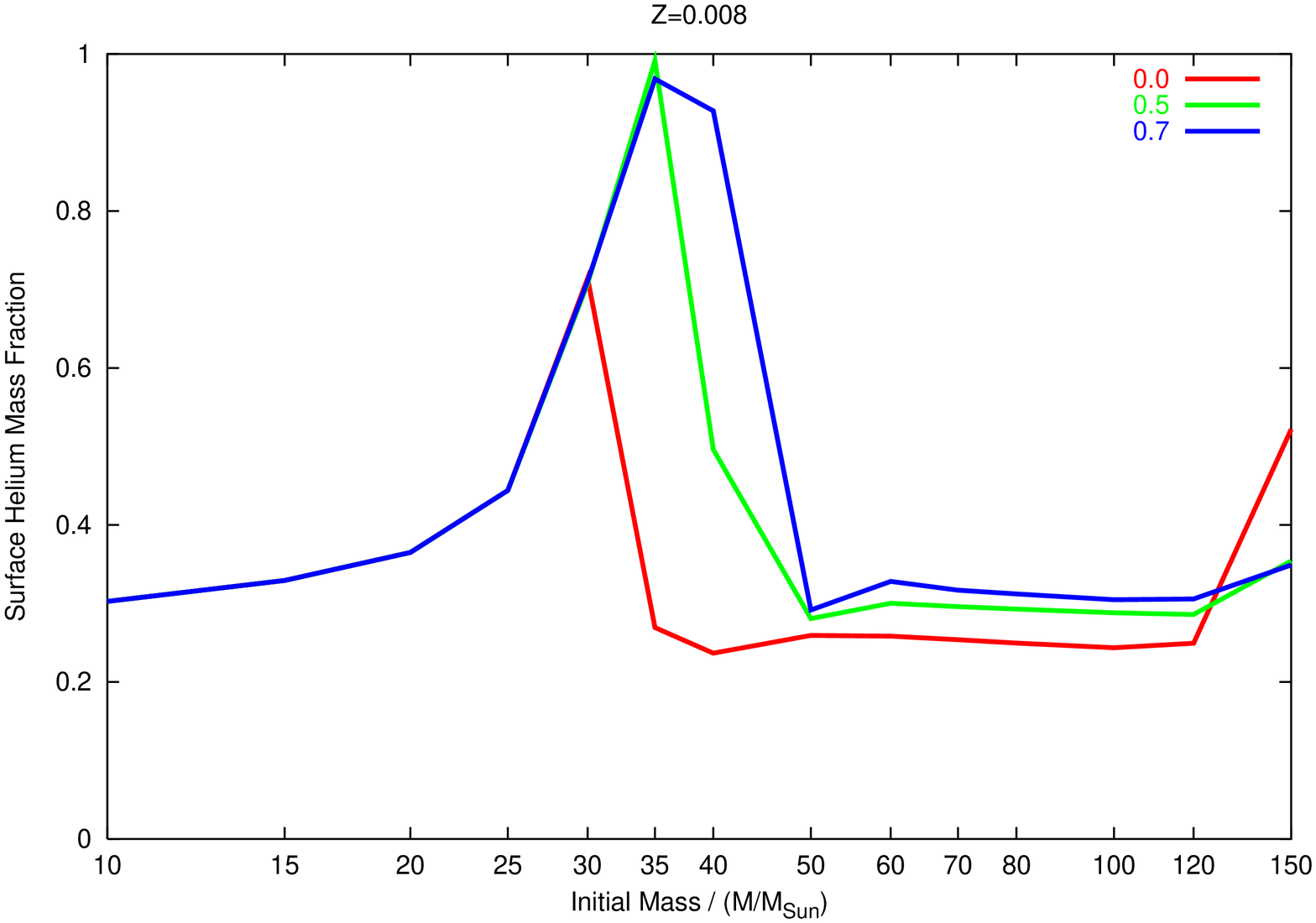}
\includegraphics[height=75mm,angle=270]{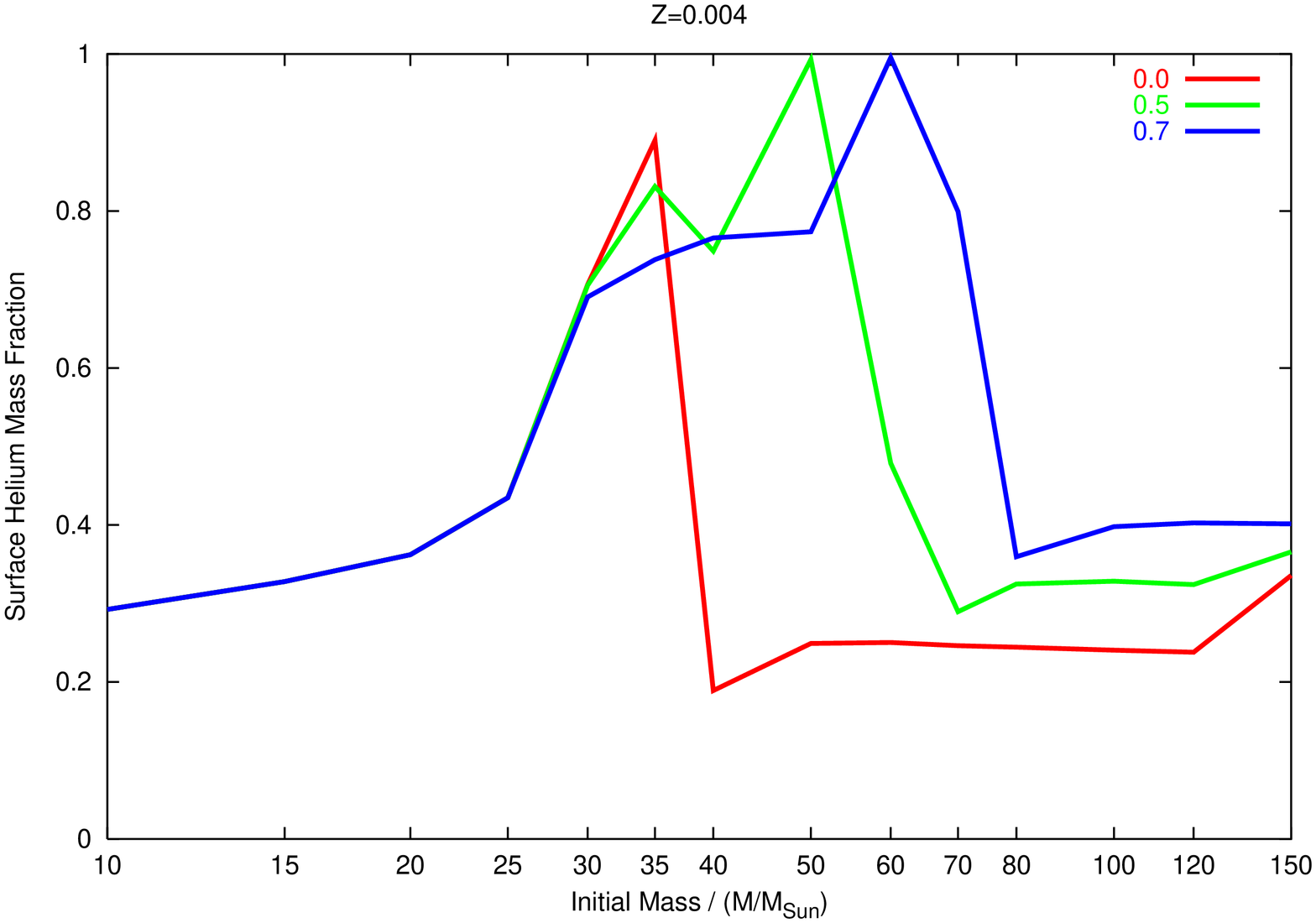}
\includegraphics[height=75mm,angle=270]{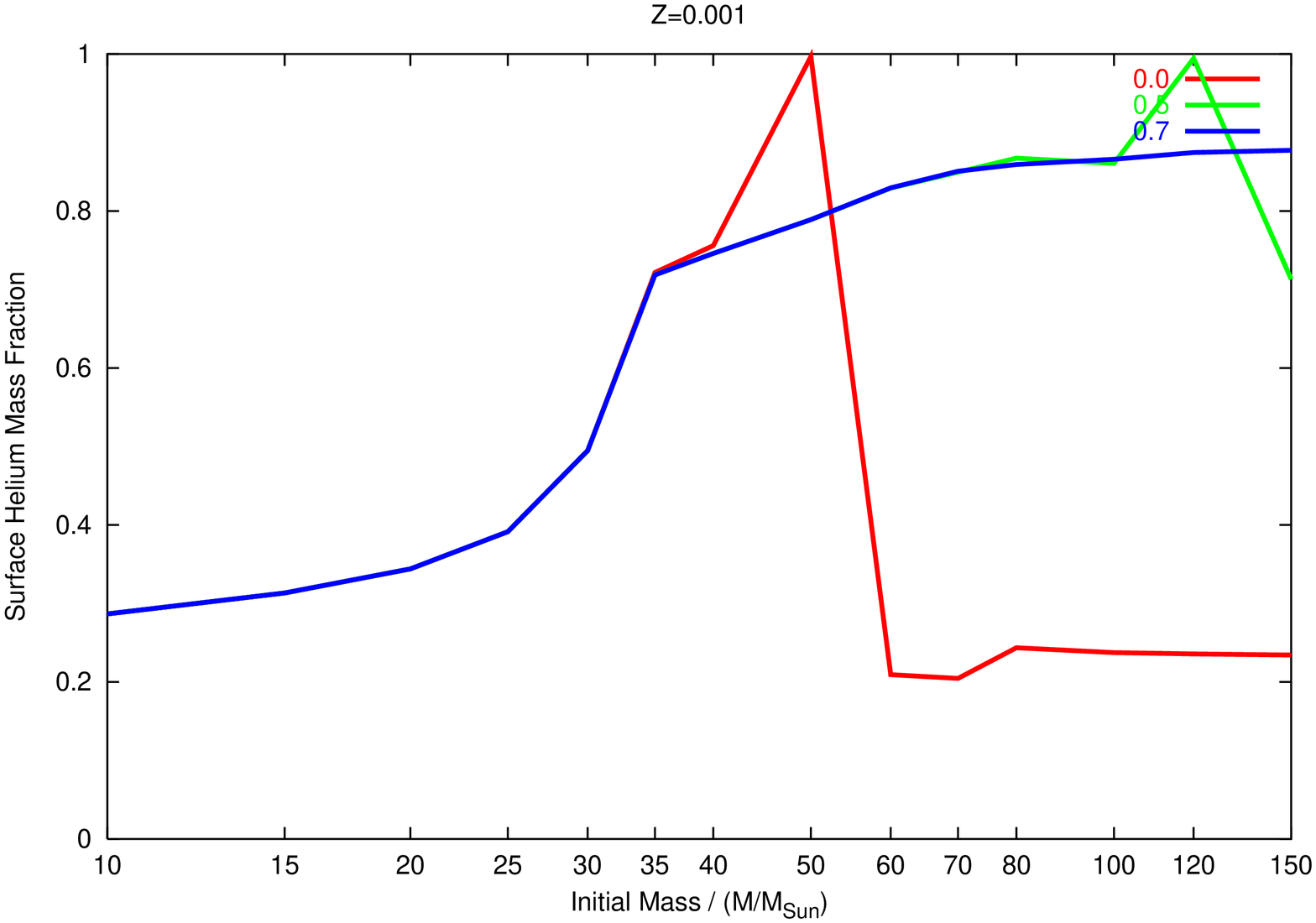}
\end{center}
\caption[The final helium mass fraction on the surface versus initial mass for different WR mass loss scaling laws at different metallicities.]{The final helium mass fraction on the surface versus initial mass for different WR mass loss scaling laws at different metallicities. The numbers in the legend are the value of the exponent, $n$, in the initial metallicity scaling $\dot{M}(Z)=\dot{M}(Z_{\odot})(Z/Z_{\odot})^{n}$.}
\label{wrY}
\end{figure}

\begin{figure}
\begin{center}
\includegraphics[height=75mm,angle=270]{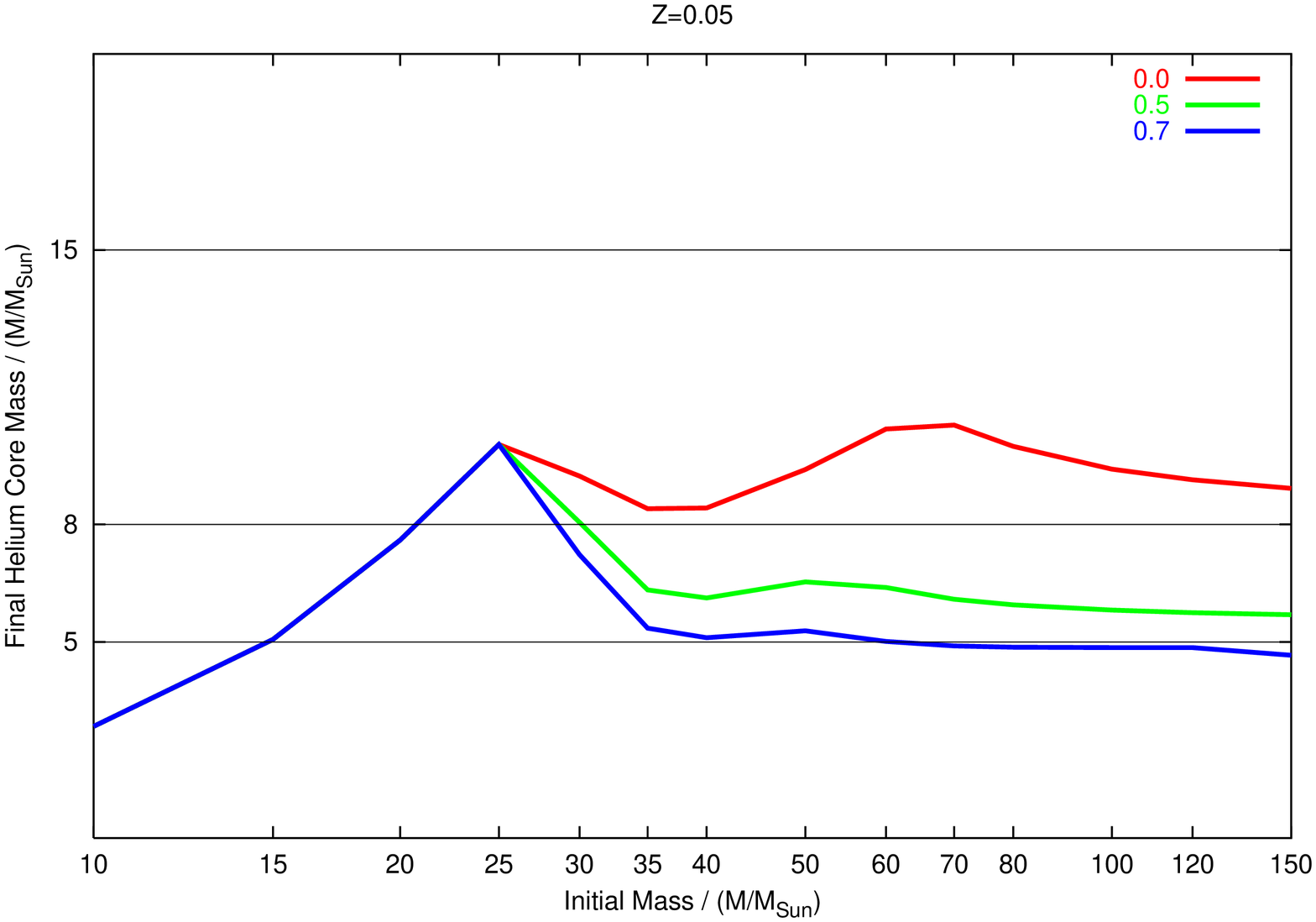}
\includegraphics[height=75mm,angle=270]{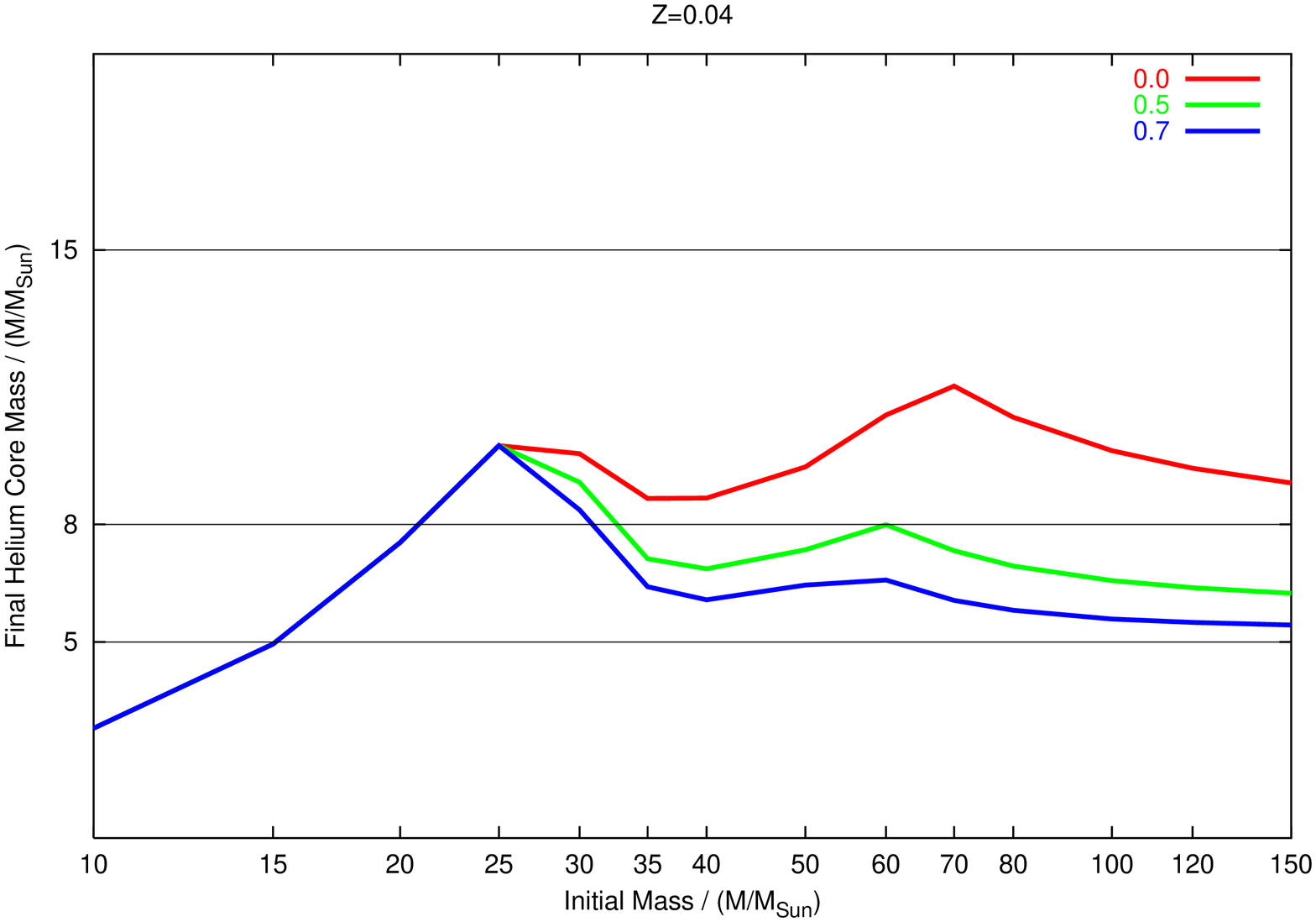}
\includegraphics[height=75mm,angle=270]{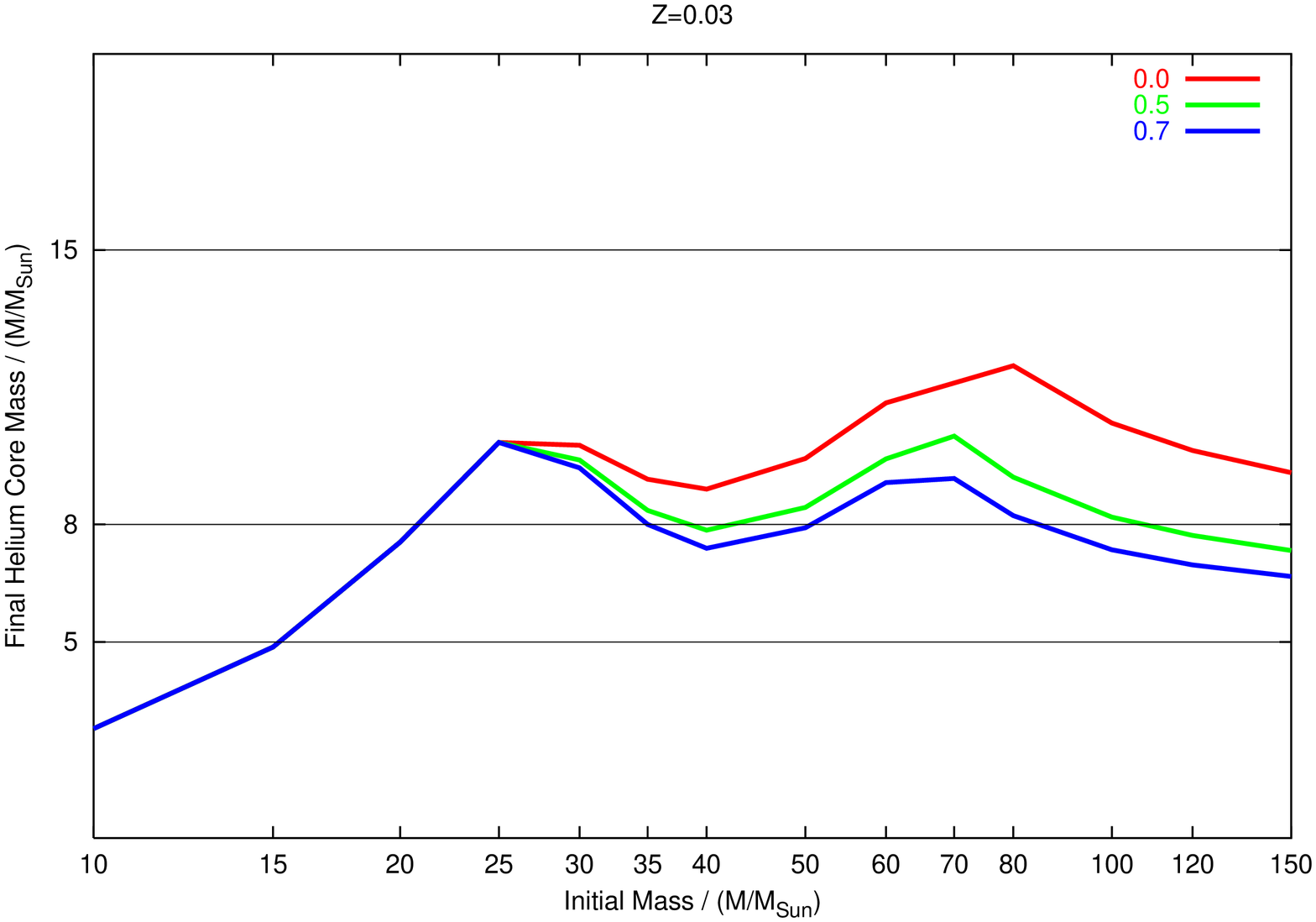}
\includegraphics[height=75mm,angle=270]{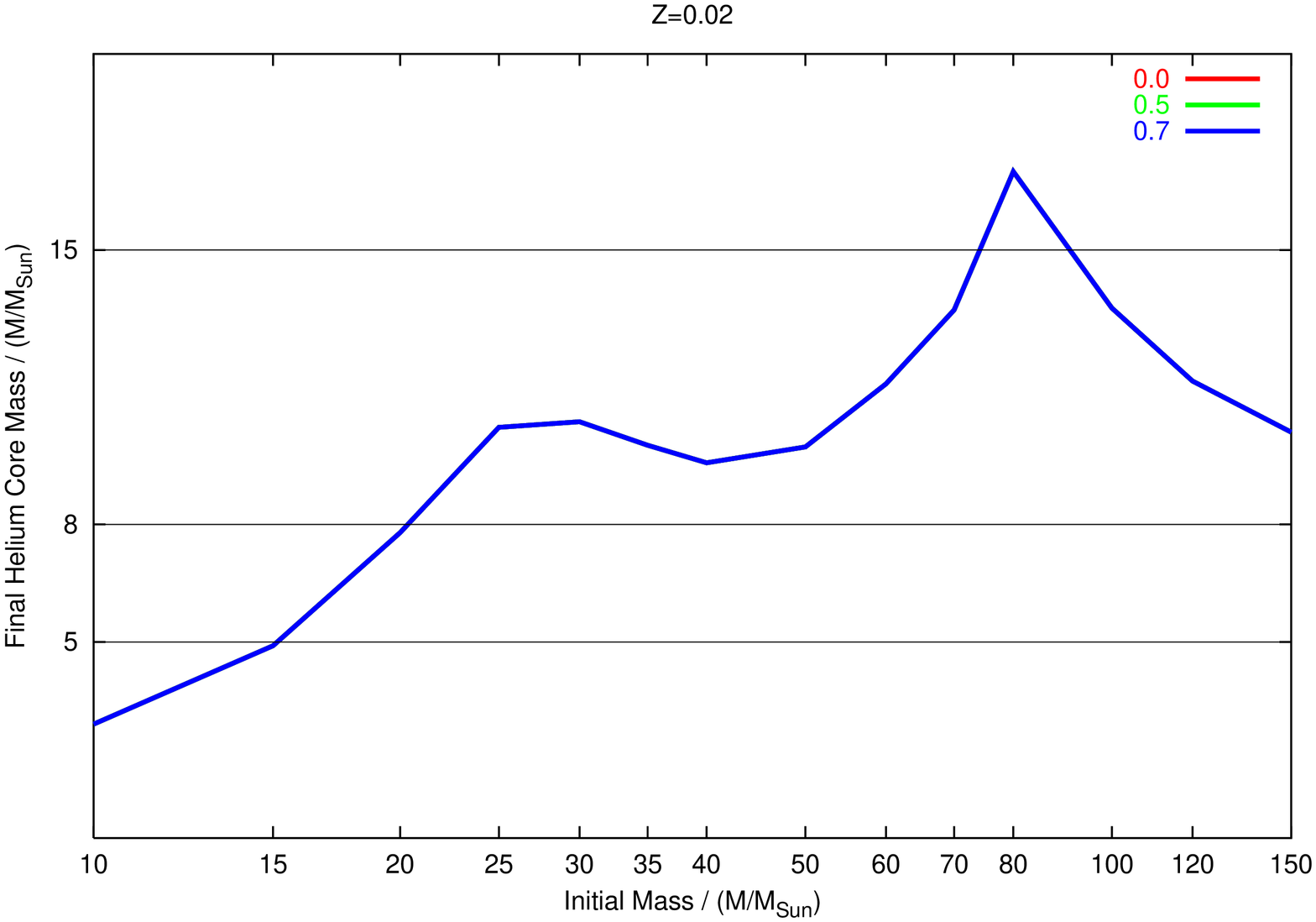}
\includegraphics[height=75mm,angle=270]{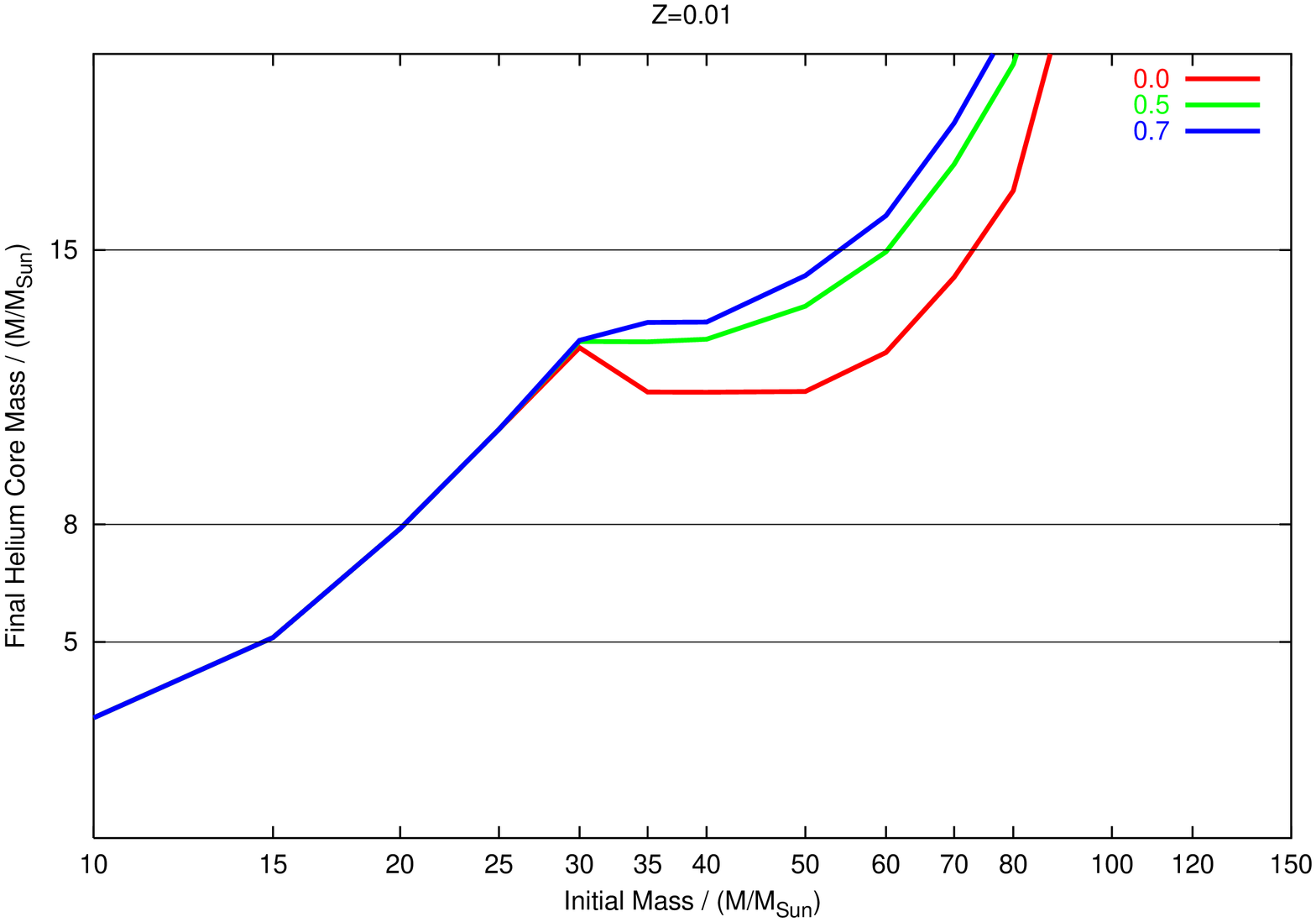}
\includegraphics[height=75mm,angle=270]{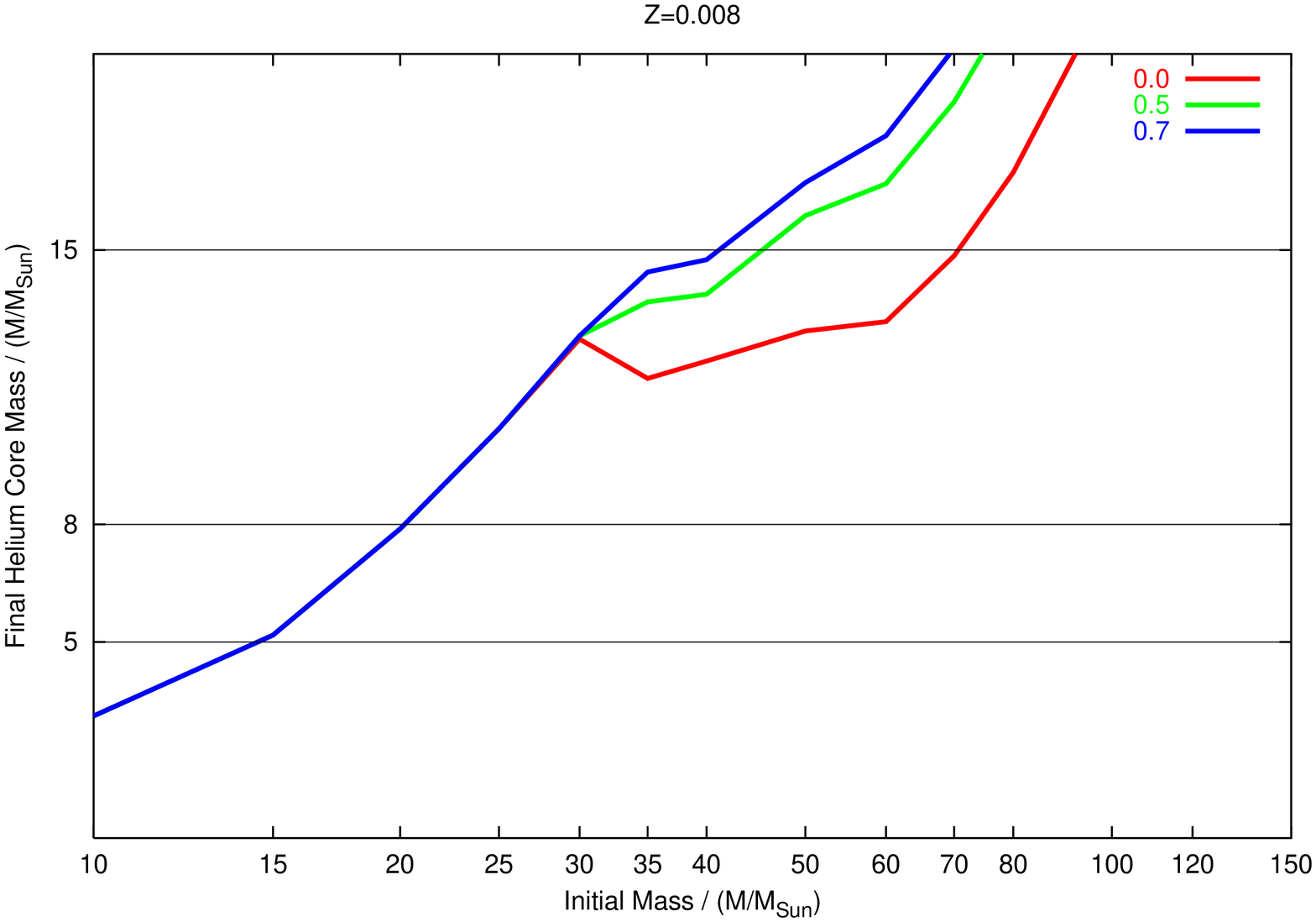}
\includegraphics[height=75mm,angle=270]{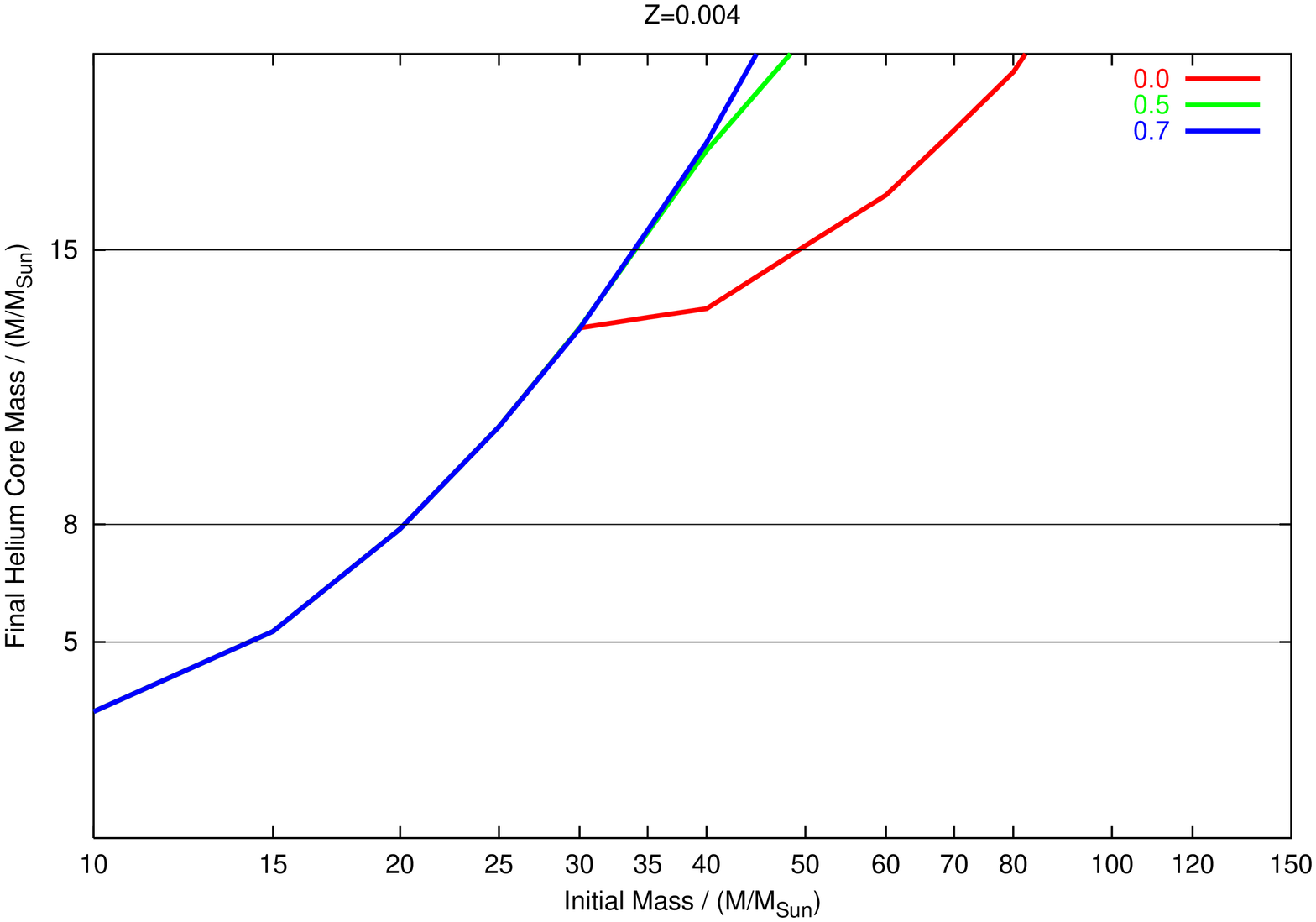}
\includegraphics[height=75mm,angle=270]{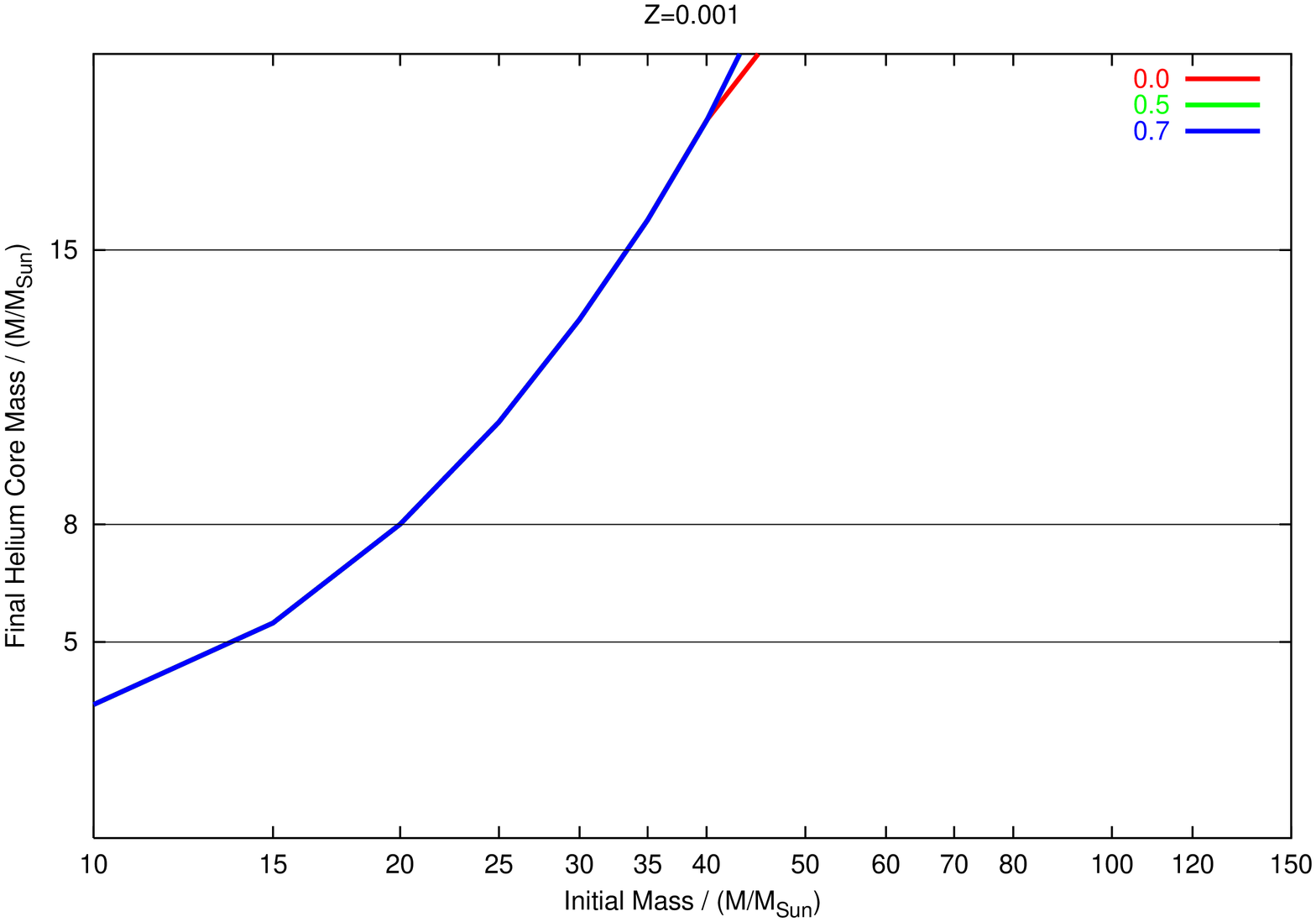}
\end{center}
\caption[The final helium core mass versus initial mass for different WR mass loss scaling laws at different metallicities.]{The final helium core mass versus initial mass for different WR mass loss scaling laws at different metallicities. The numbers in the legend are the value of the exponent, $n$, in the initial metallicity scaling $\dot{M}(Z)=\dot{M}(Z_{\odot})(Z/Z_{\odot})^{n}$.}
\label{wrcore}
\end{figure}

In figure \ref{wrfinal} we present the mass just before the SN for the stars. Figures \ref{wrX} and \ref{wrY} show the hydrogen and helium mass fractions and figure \ref{wrcore} shows the final helium core mass that we can use to determine the remnant type. One thing that is evident from these figures is that the effect of scaling the mass loss follows some general trends but can also be non-linear in some places. 

The final masses of the pre-SN provide our main limits on the mass loss at metallicities above solar. In general without scaling the remnant masses are higher and all WR stars that go SN are more massive than $5M_{\odot}$. Therefore, according to the above prescription, this means that we see no bright Ibc from single stars. When the scaling exponent is 0.7 rather than 0.5 we find that the resultant masses are slightly lower than that with 0.5.

When we move our attention to metallicities lower than solar the picture becomes quite different. The remnant masses without the scaling are consistently lower than when it is included. There is little difference between the results achieved when using different values for the scaling exponent. At the lowest metallicities the effects are weak due to the later on set of the WR phase because pre-WR mass loss is also lower at lower metallicity and gives rise to initially more massive WR stars so there is more mass for the WR mass loss to remove.

A similar tale is told by the mass fractions of helium and hydrogen. A larger range of WR stars at lower metallicity retain more hydrogen than their higher metallicity cousins with the same mass. This indicates quite strong evolution of WR type ratios with metallicity especially if the mass-loss rates do scale with initial metallicity. Indeed below SMC metallicity there will be very few WC stars. At solar and above nearly all WR stars end their evolution as WC or WO stars with low surface helium while at lower metallicity a growing fraction retain more helium at a high mass fraction, $Y>0.9$. The graphs also show strong evidence for two populations of WR SN progenitors. This could be the type Ib and Ic division.

The helium core mass in figure \ref{wrcore} is important for determining remnants. At the highest metallicities, first we can see with the scaling there are no black holes although without the scaling there are black holes formed by fall back. Also we can see black holes cannot form directly until below solar metallicity. Then black holes formed become more massive at lower metallicity if the scaling exists. Therefore limits on this scaling could be possible from measuring the black hole initial mass function and how it depends on metallicity. However this is likely to be a very difficult to measure.

We have compared the ratios of WR types from these models against observations. The models with no scaling give a constant ratio of WC/WR and WC/WN which is at odds with observations. The models with scaling do roughly match the trend as can be seen in figure \ref{wrratios}. Therefore observation leads us to include the extra scaling for WR stars. Models with the extra scaling of mass loss with initial metallicity do agree with the observations of \citet{masseyetal2000} that the limit for WR and WC stars increases with lower metallicity. However the direct measurement of WR mass loss by \citet{WRZscale} provides the most convincing argument. A measure will also be possible from the ratio of type I/II SNe and the ratio of type Ib/Ic against composition. This requires more observations. In all the mass-loss schemes we now include the scaling of WR mass loss with metallicity using an exponent of $0.5$. 

\section{Comparing Prescriptions at High Metallicity}
\begin{figure}
\begin{center}
\includegraphics[height=75mm,angle=270]{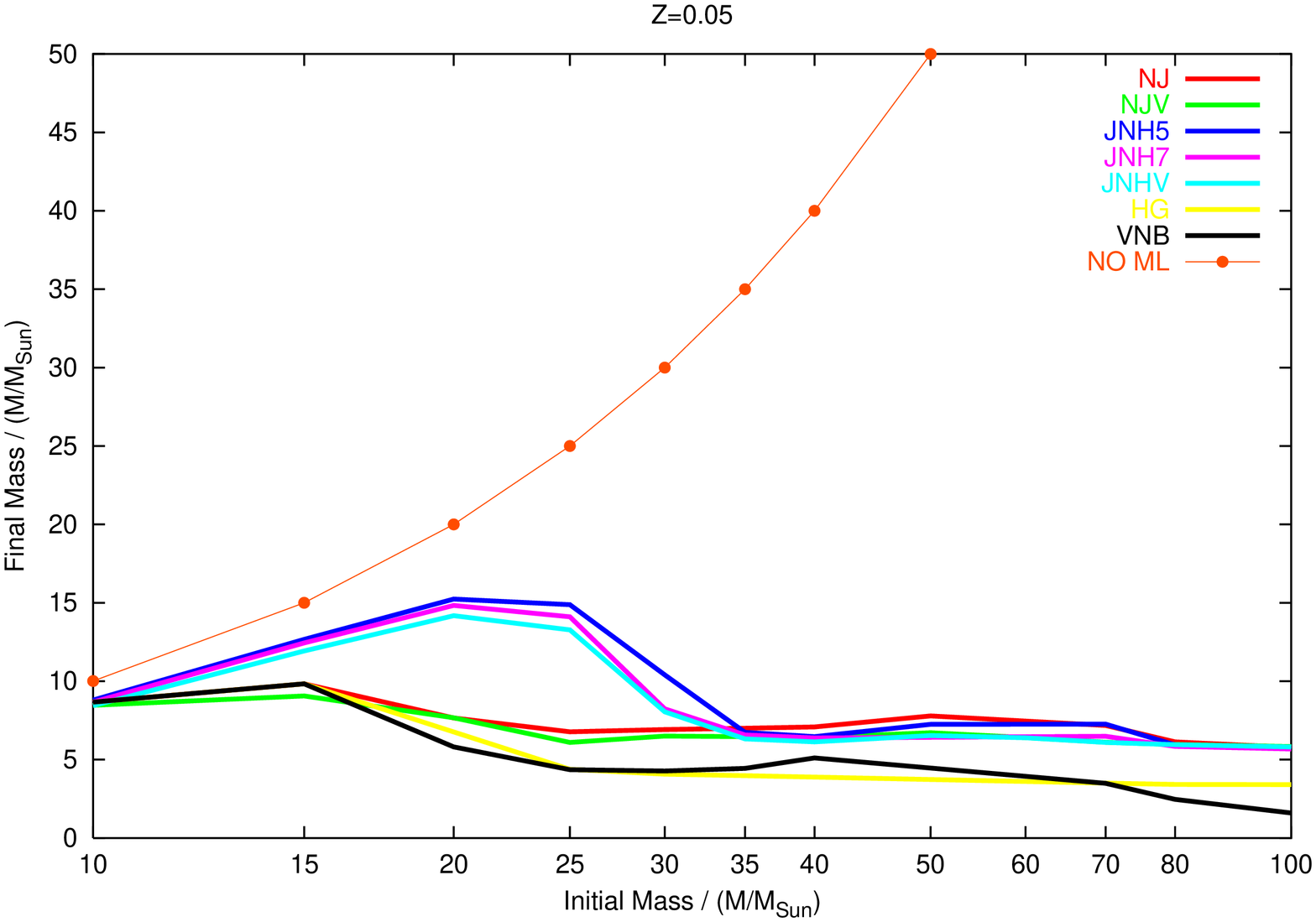}
\includegraphics[height=75mm,angle=270]{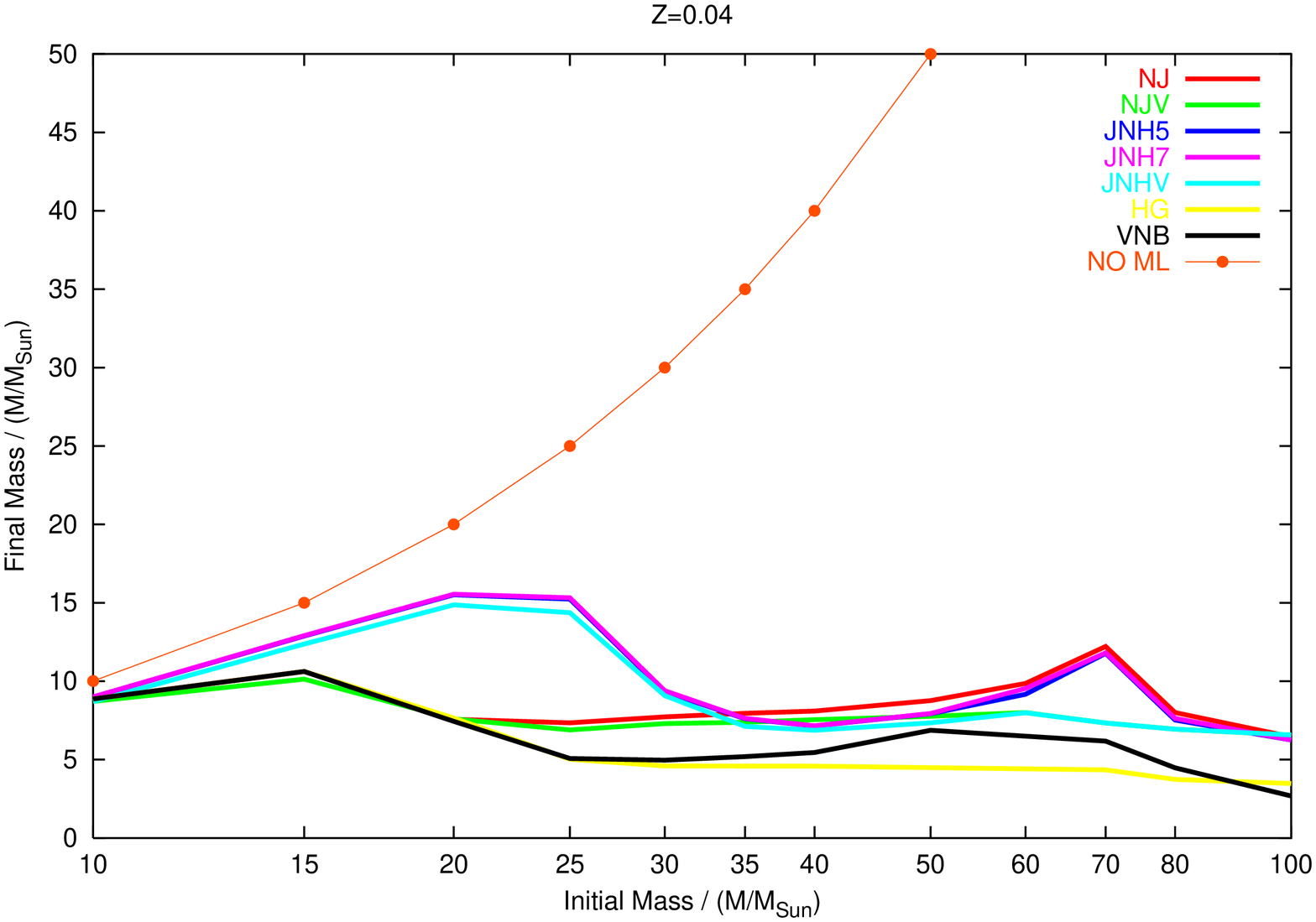}
\includegraphics[height=75mm,angle=270]{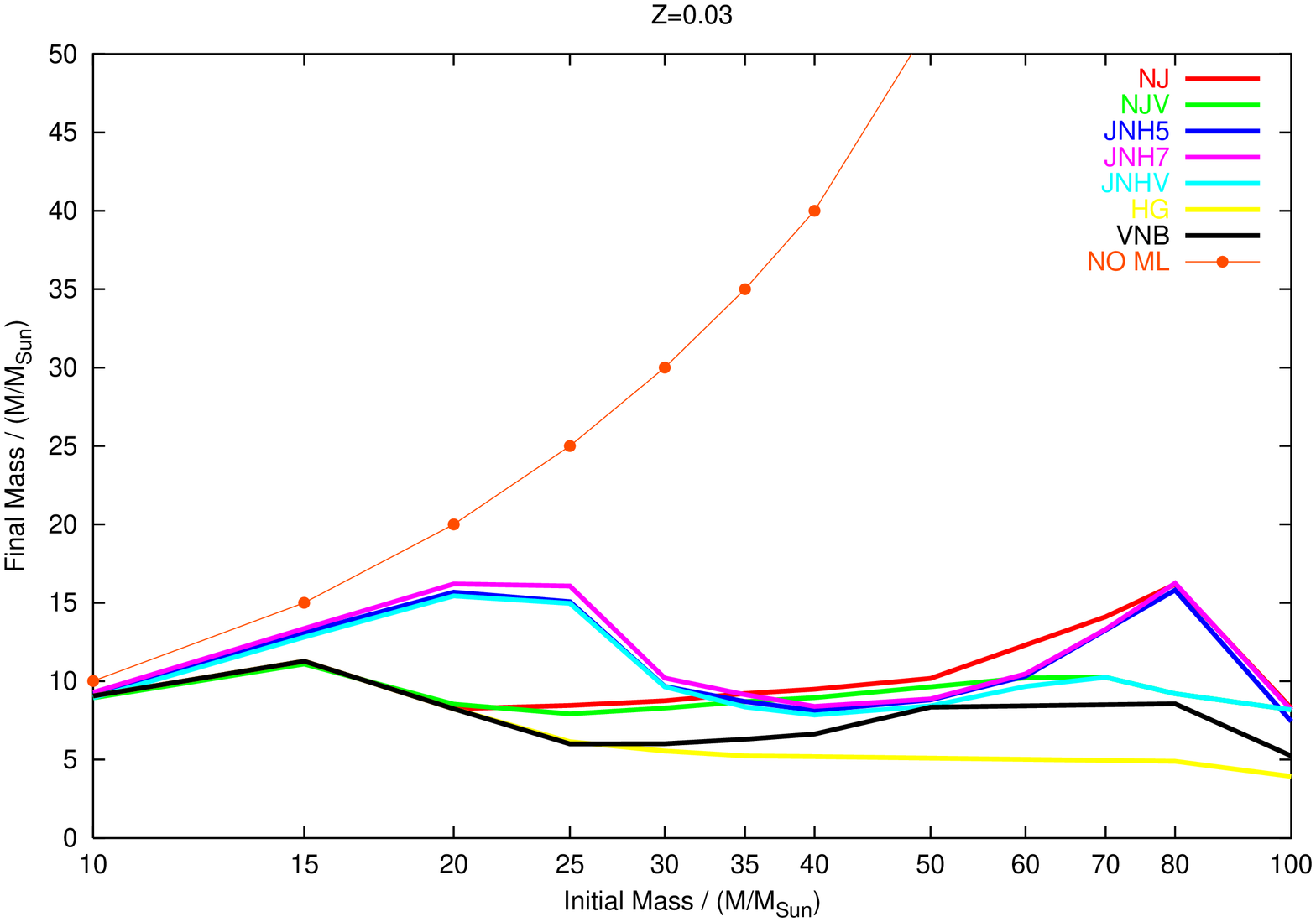}
\includegraphics[height=75mm,angle=270]{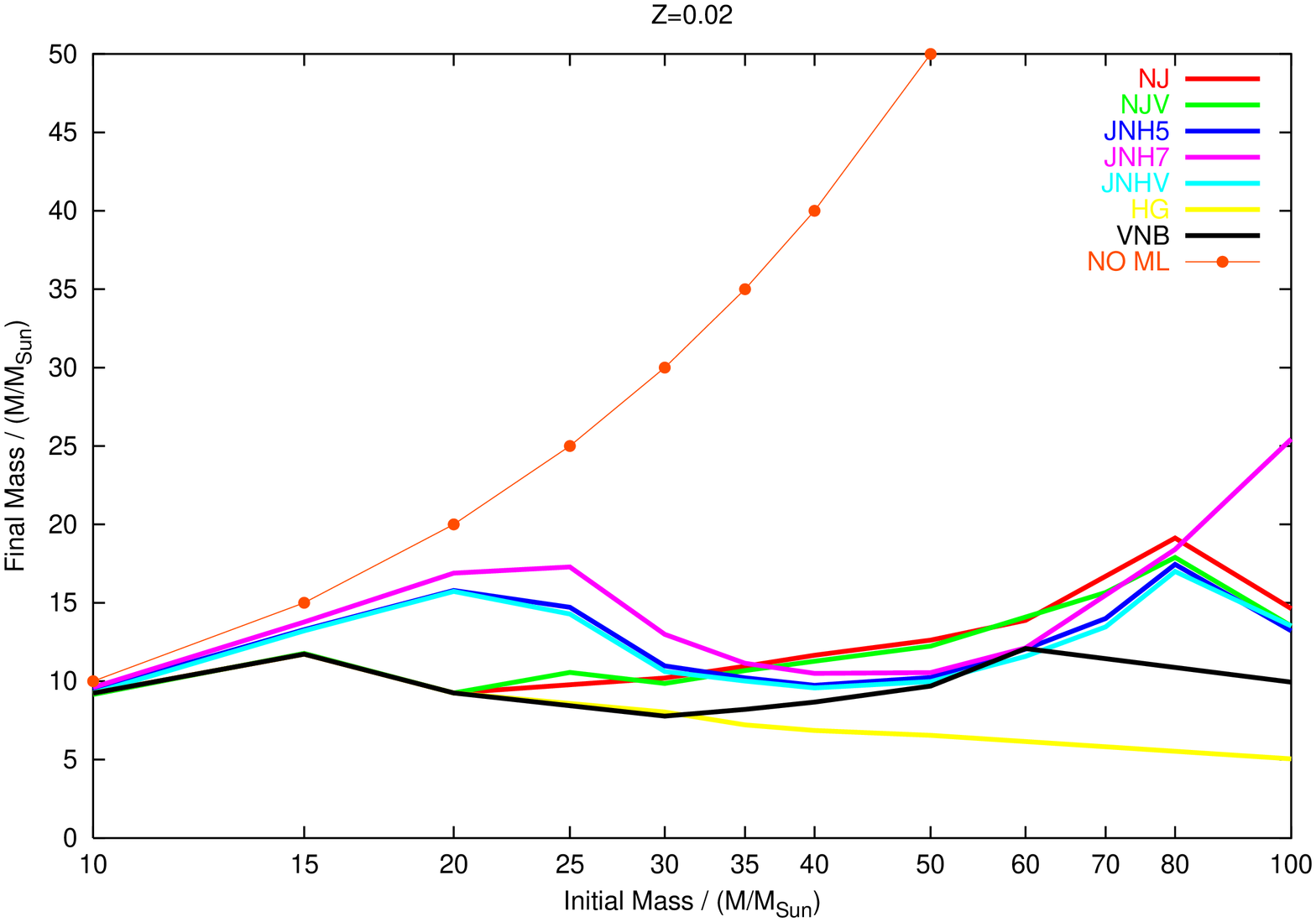}
\includegraphics[height=75mm,angle=270]{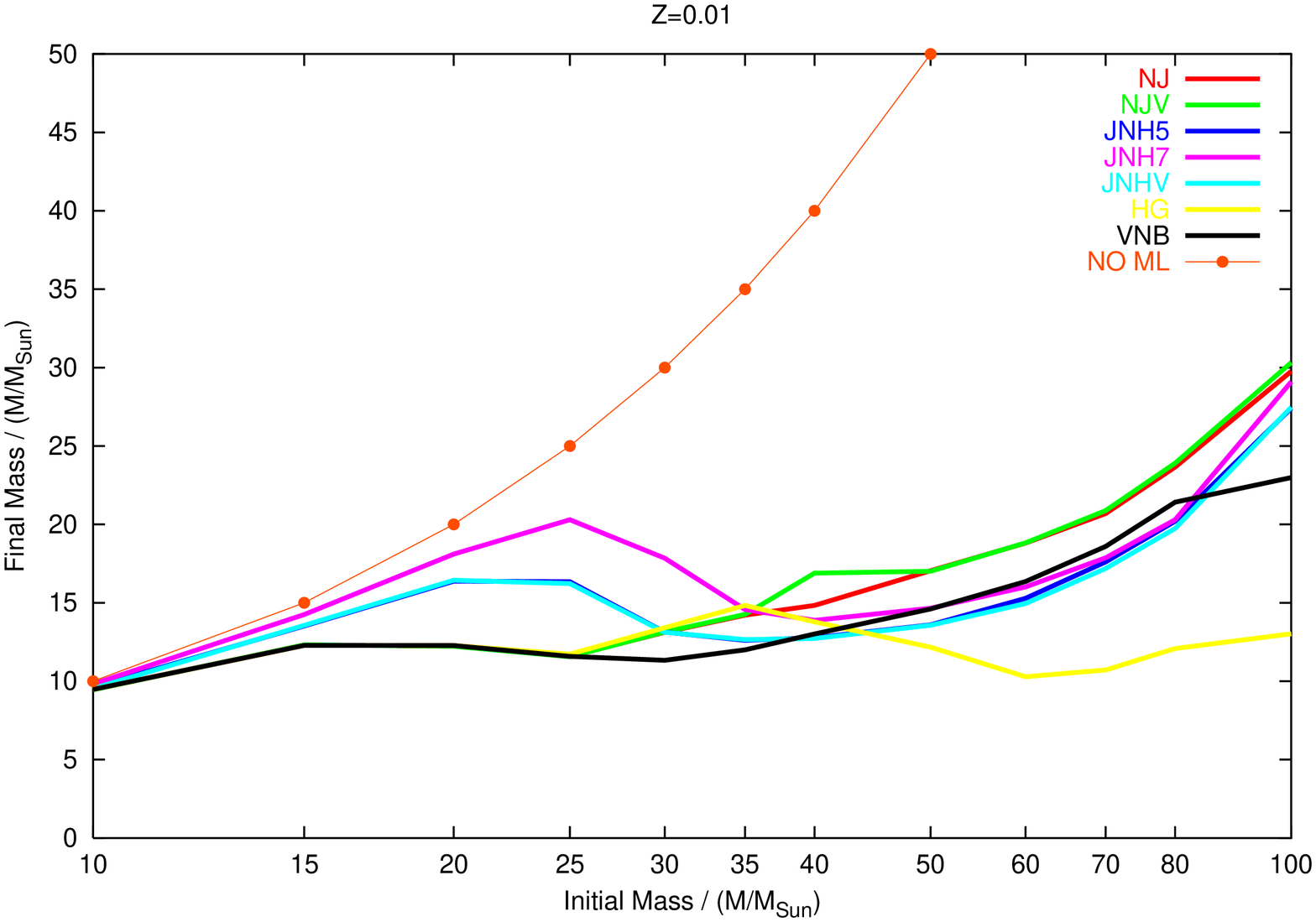}
\includegraphics[height=75mm,angle=270]{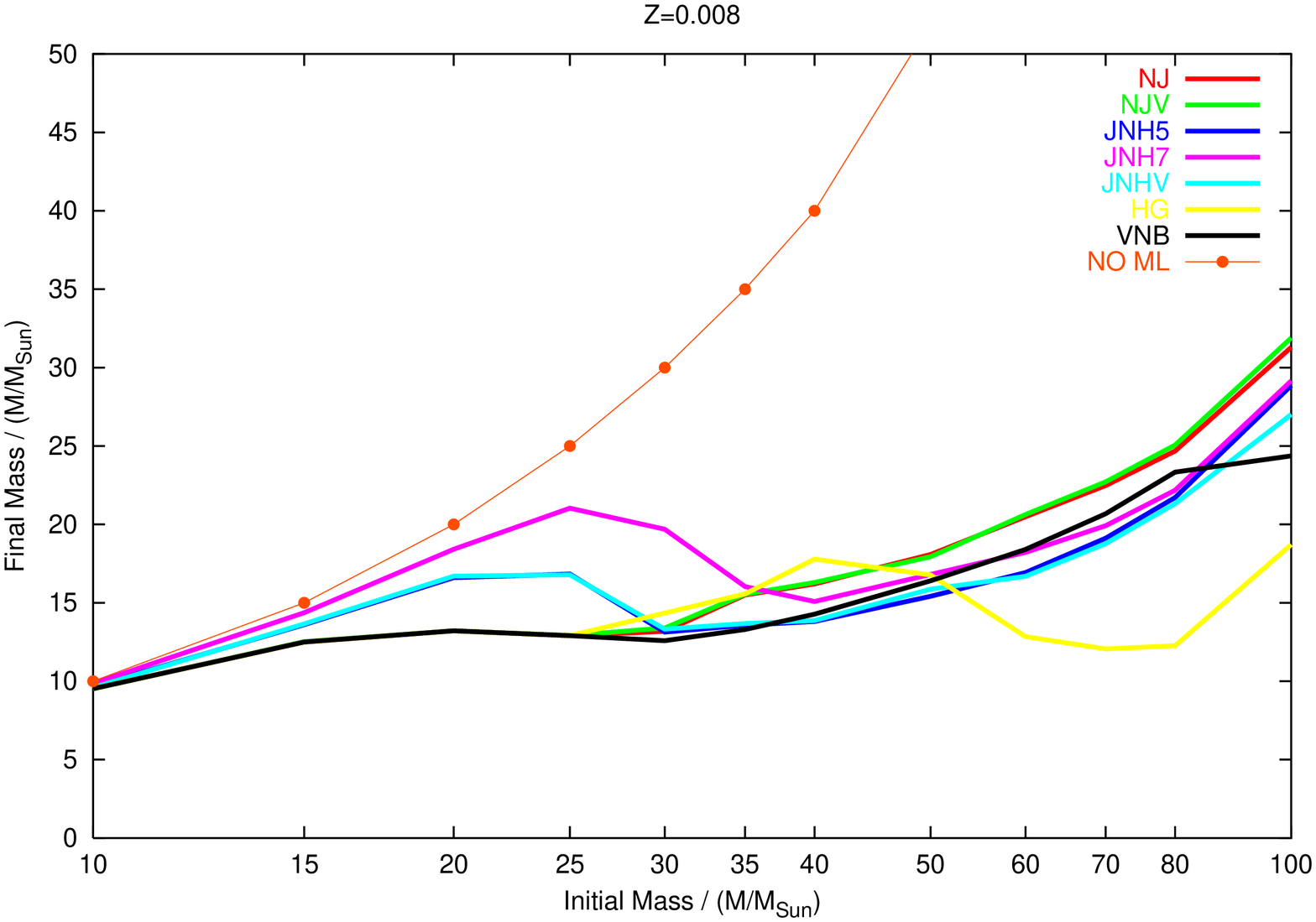}
\includegraphics[height=75mm,angle=270]{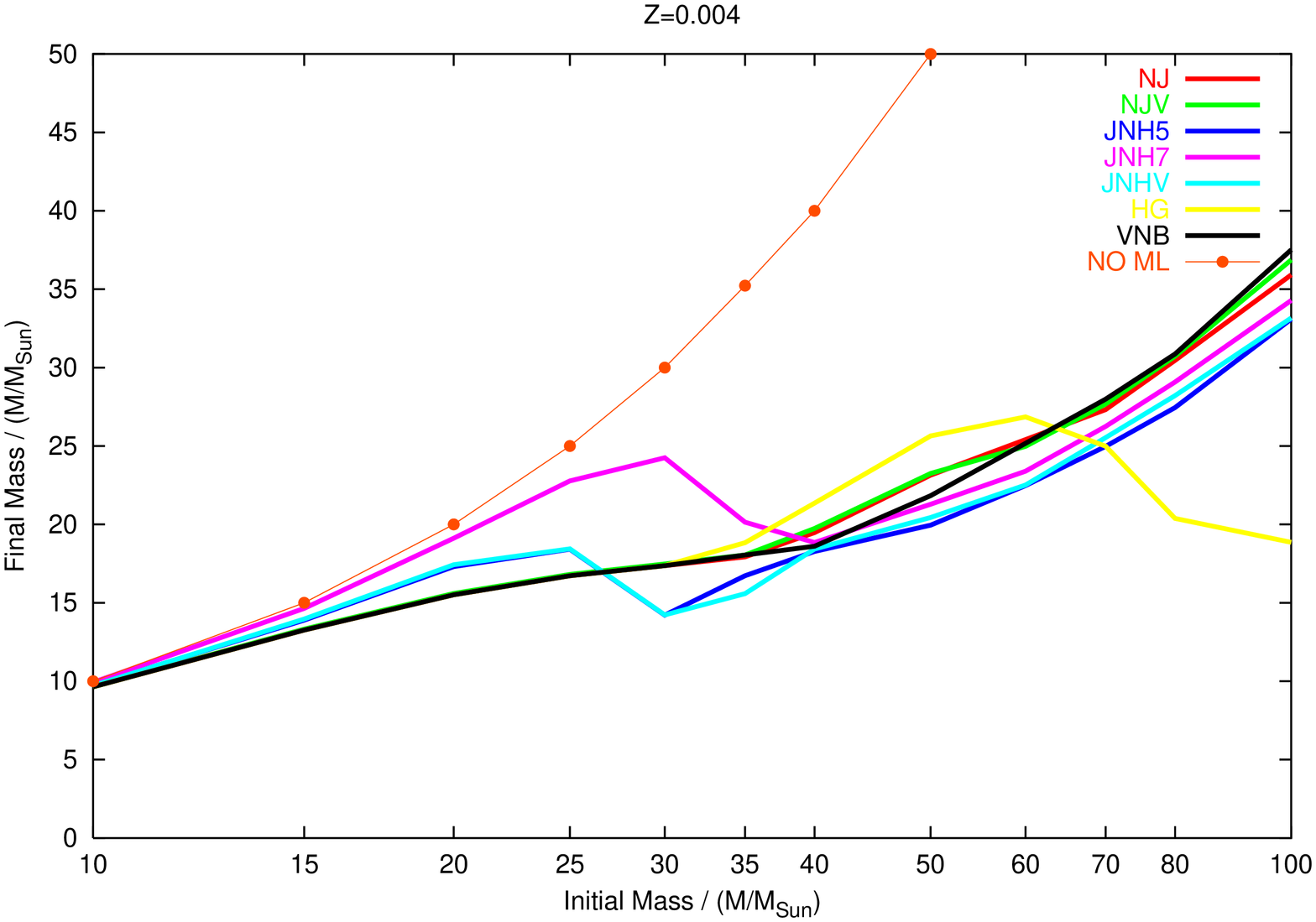}
\includegraphics[height=75mm,angle=270]{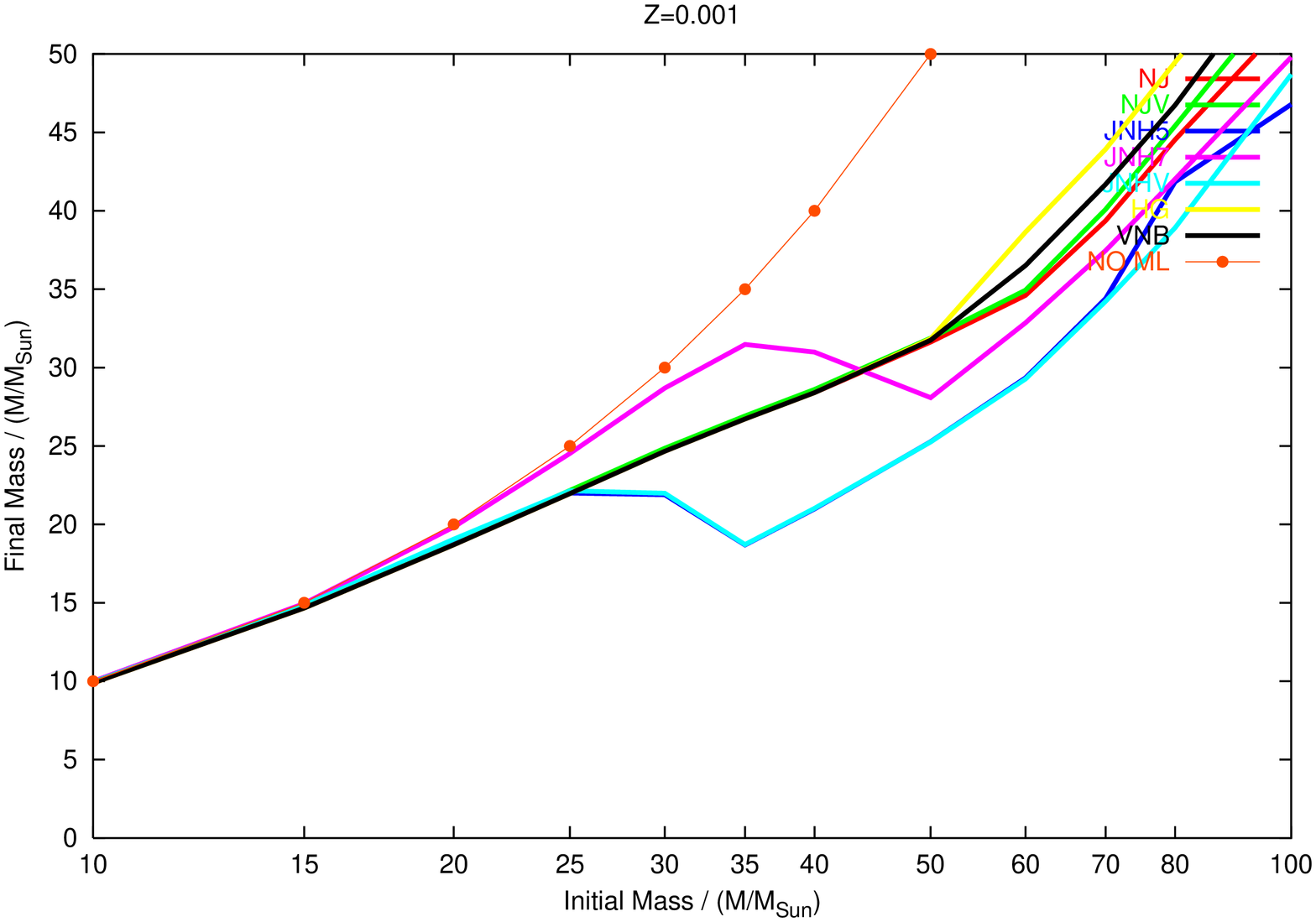}
\end{center}
\caption{The final mass versus initial mass for different mass-loss schemes at different metallicities.}
\label{allfinal}
\end{figure}

We now compare a number of commonly used mass-loss schemes. We also include a set of models that have zero mass loss.  We also show results with JNH rates with different scaling of the mass-loss with metallicity. JNH5 has a scaling of $(Z/Z_{\odot})^{0.5}$ while JNH7 has a scaling of $(Z/Z_{\odot})^{0.7}$.

Two of the schemes we study are similar to those used by \citet{H03} and \citet{vanbevfull}. For the HG mass-loss scheme we use the NJ rates for pre-WR and the rates of \citet{WL1999} for the WR phase, while for the VNB rates we use the NJ rates which are similar to the VNB rates for OB stars. We have experimented with their full mass-loss scheme but find it non-trivial to implement due to the large disparity in their mass-loss rates. The NJ rates do provide a similar magnitude of mass loss and prevent the occurrence of giant stars at solar metallicity with $M_{\rm bol} > -9.5$. This is argued to be necessary because red giants of this luminosity are not observed in the solar neighbourhood. We do use their WR mass loss rate despite its extreme simplicity because it does not consider the effect of varying surface composition or WR type indicated by \citet{NL00}. For all WR mass loss we scale the rate with the initial metallicity such that $\dot{M}_{\rm WR}(Z)=\dot{M}(Z_{\odot})(Z/Z_{\odot})^{0.5}$.

\begin{figure}
\begin{center}
\includegraphics[height=75mm,angle=270]{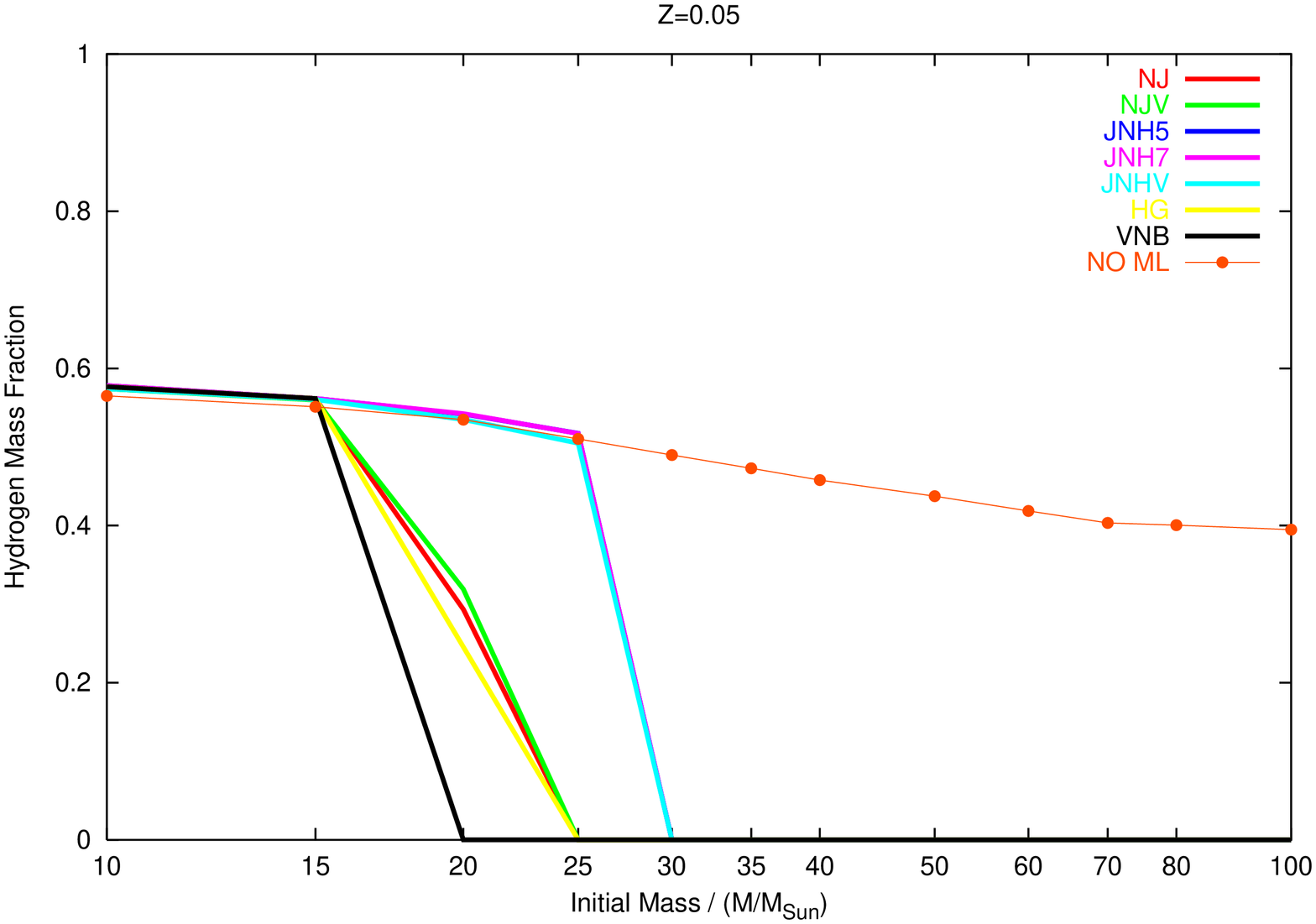}
\includegraphics[height=75mm,angle=270]{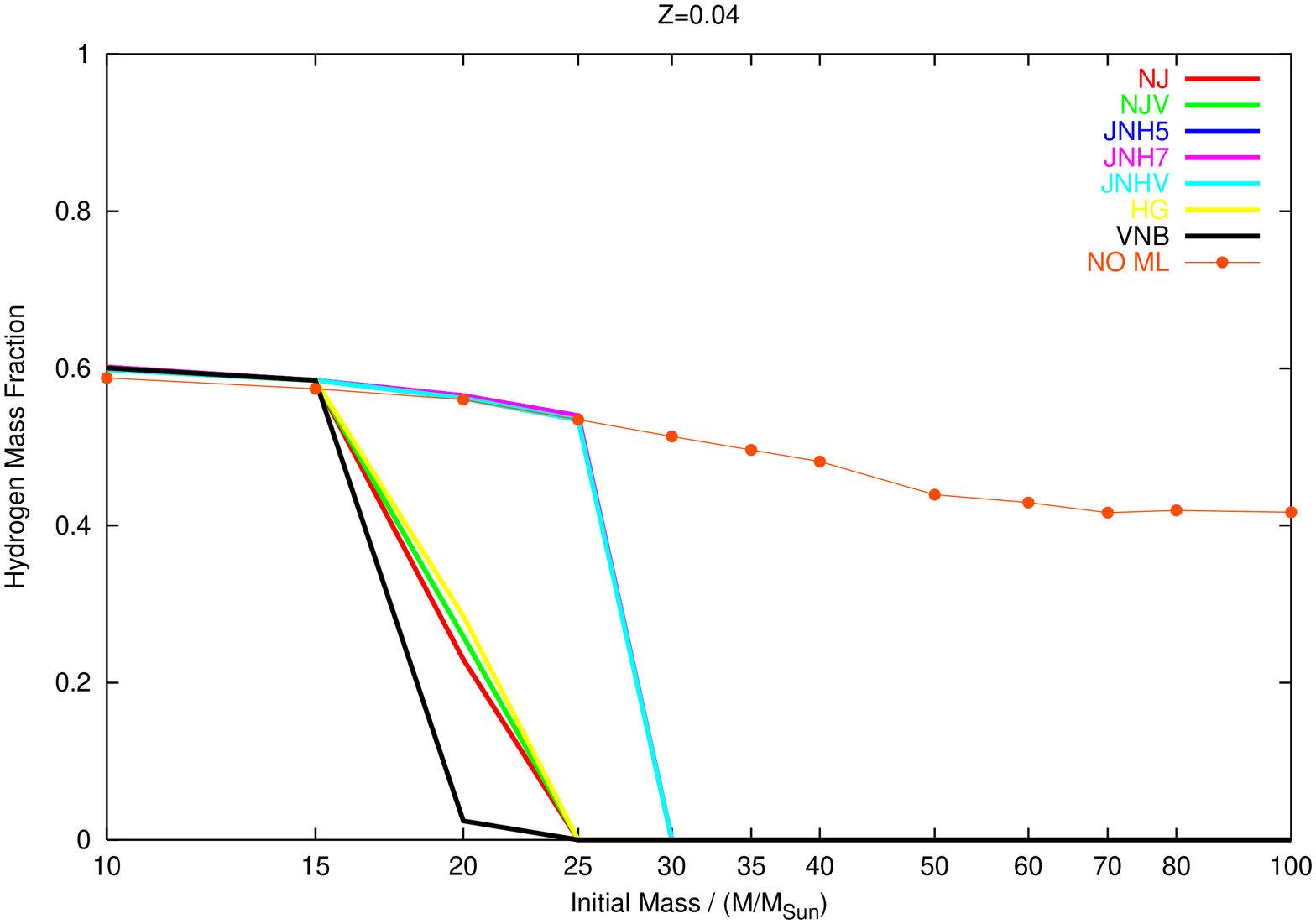}
\includegraphics[height=75mm,angle=270]{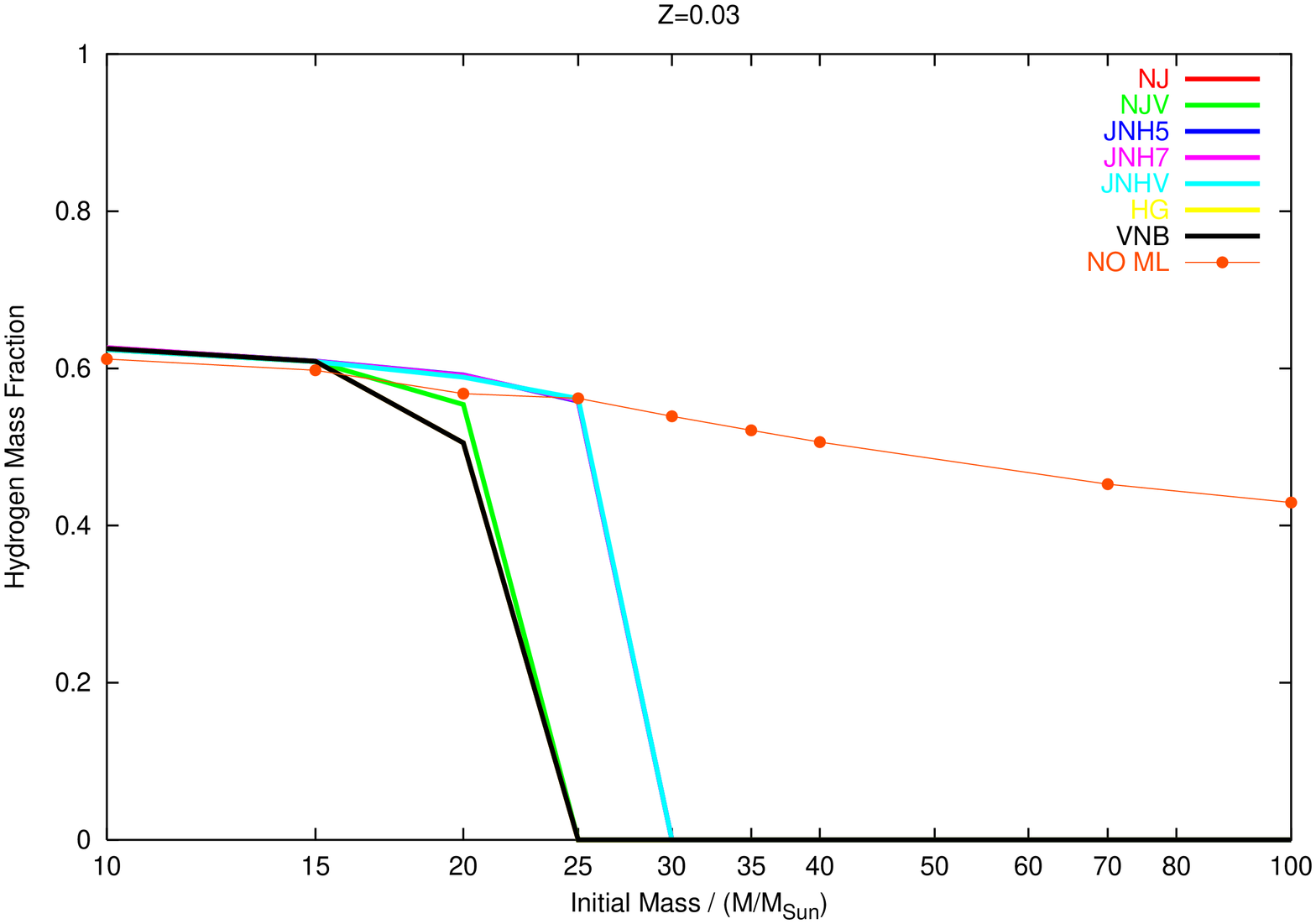}
\includegraphics[height=75mm,angle=270]{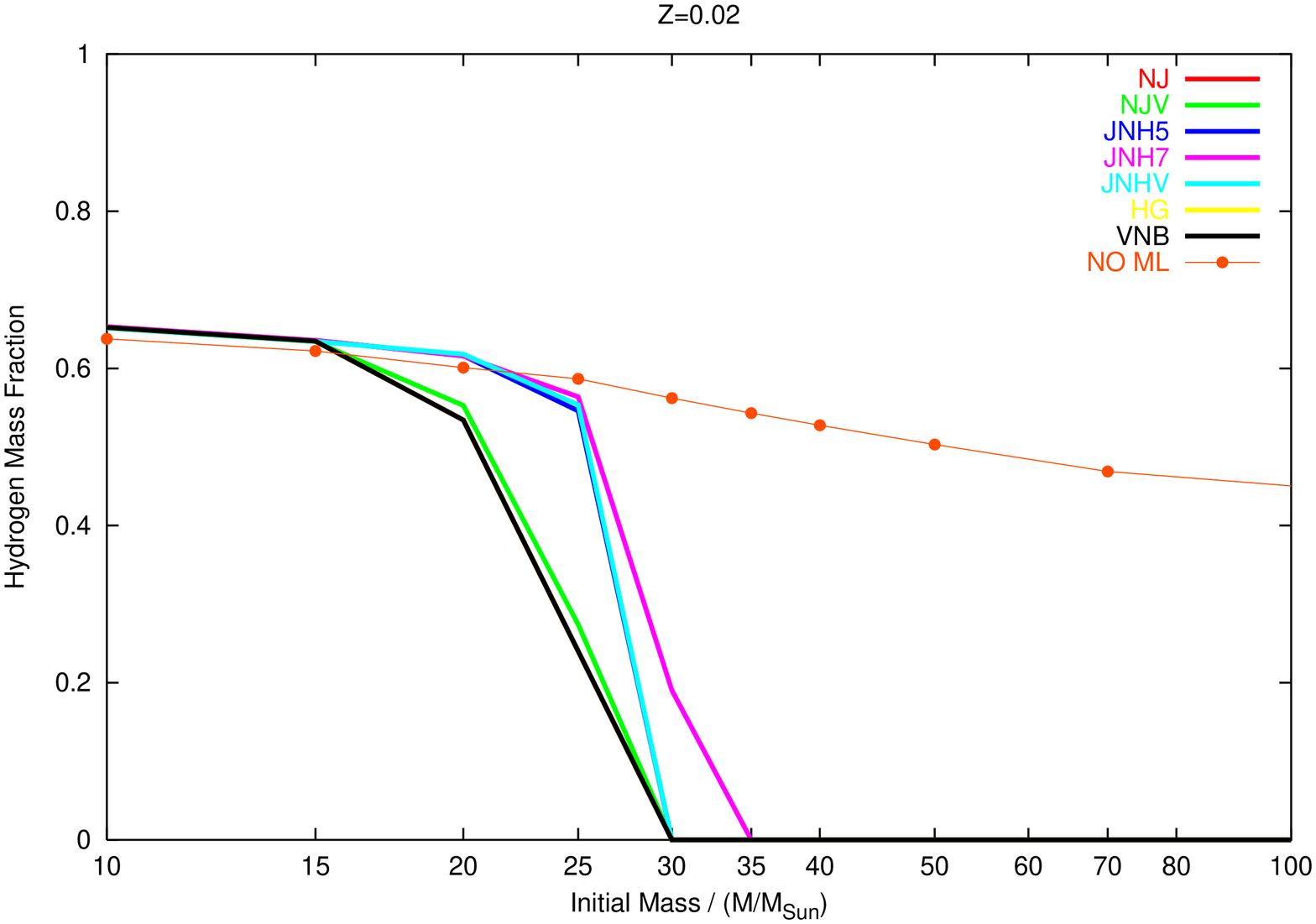}
\includegraphics[height=75mm,angle=270]{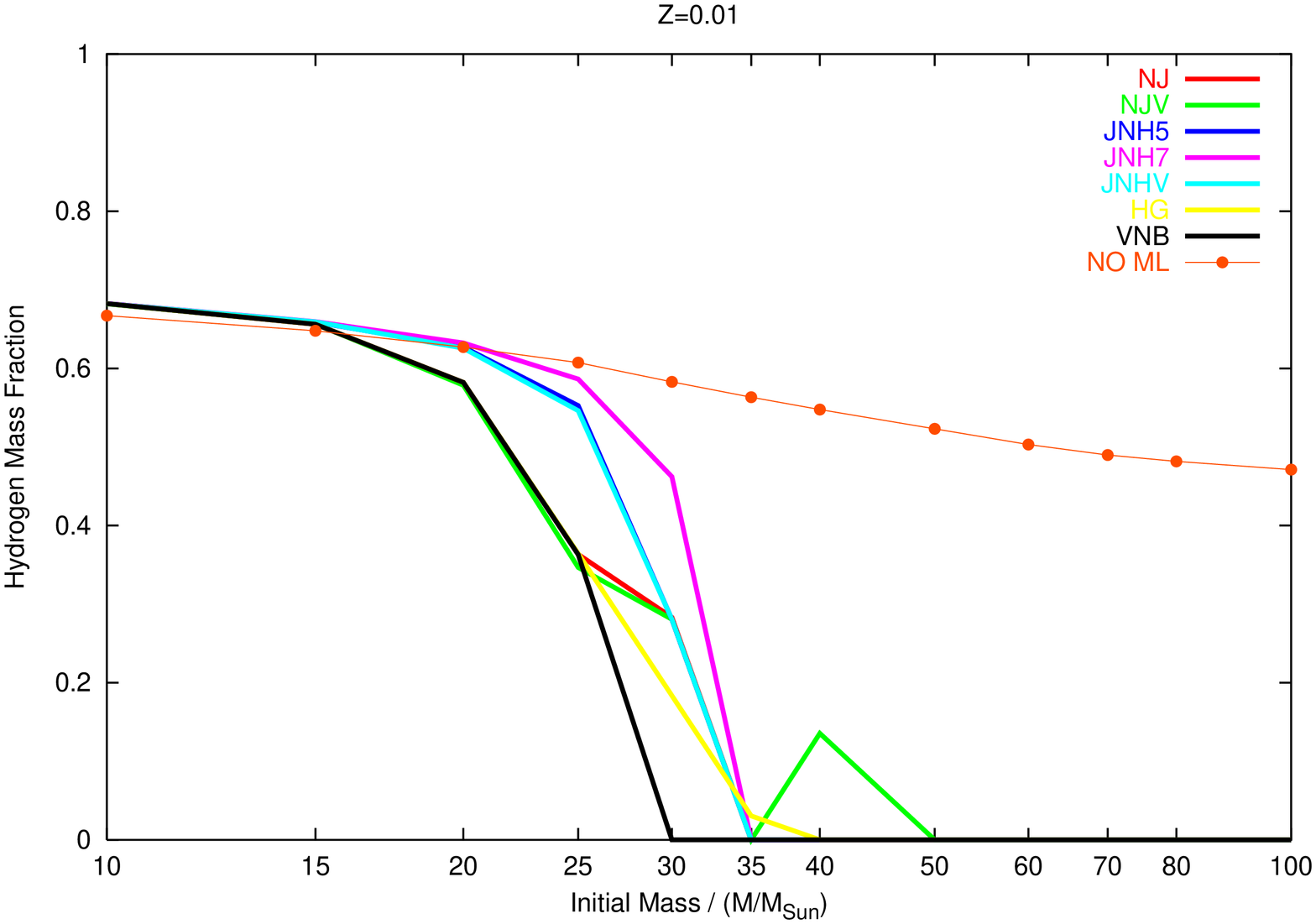}
\includegraphics[height=75mm,angle=270]{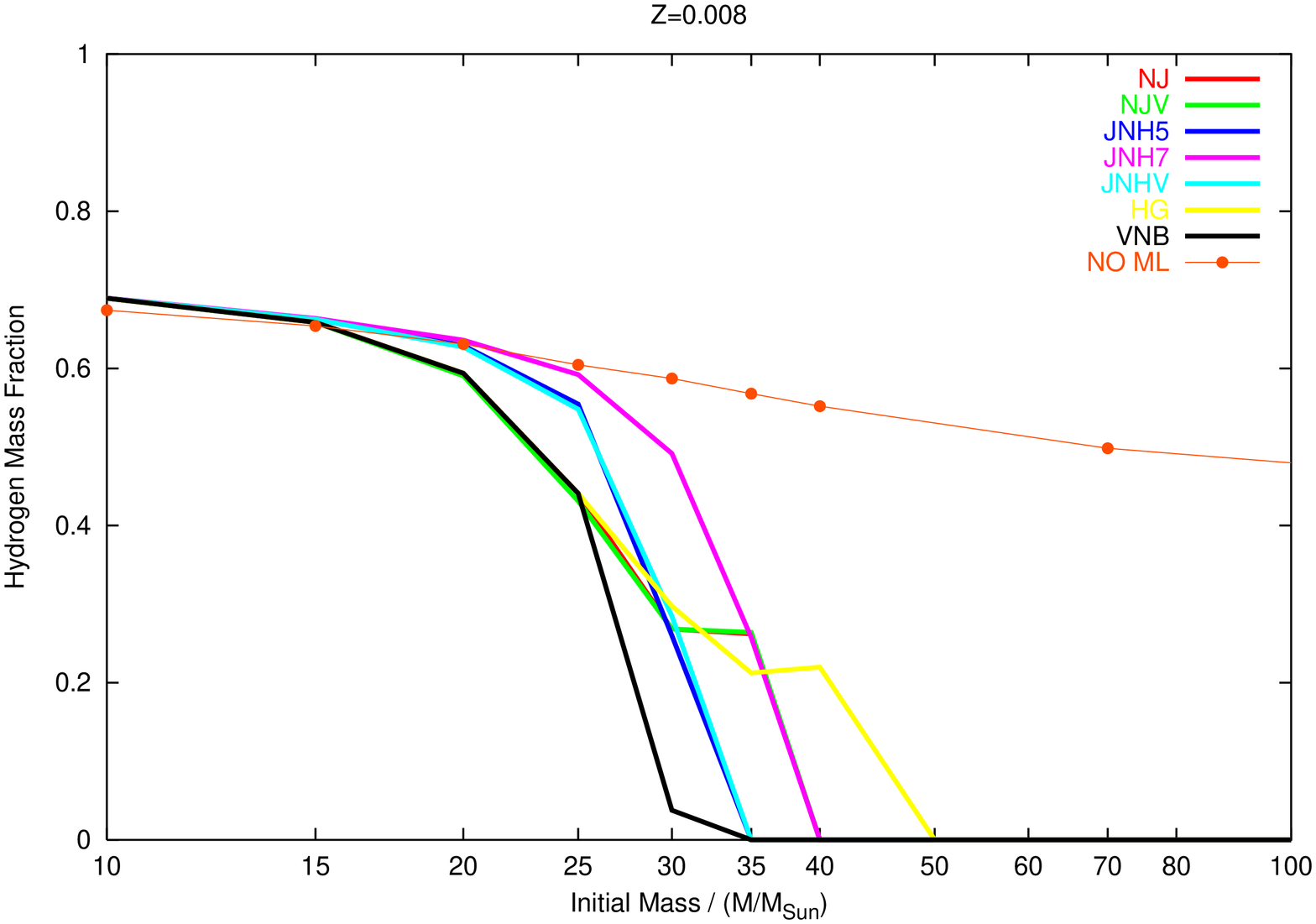}
\includegraphics[height=75mm,angle=270]{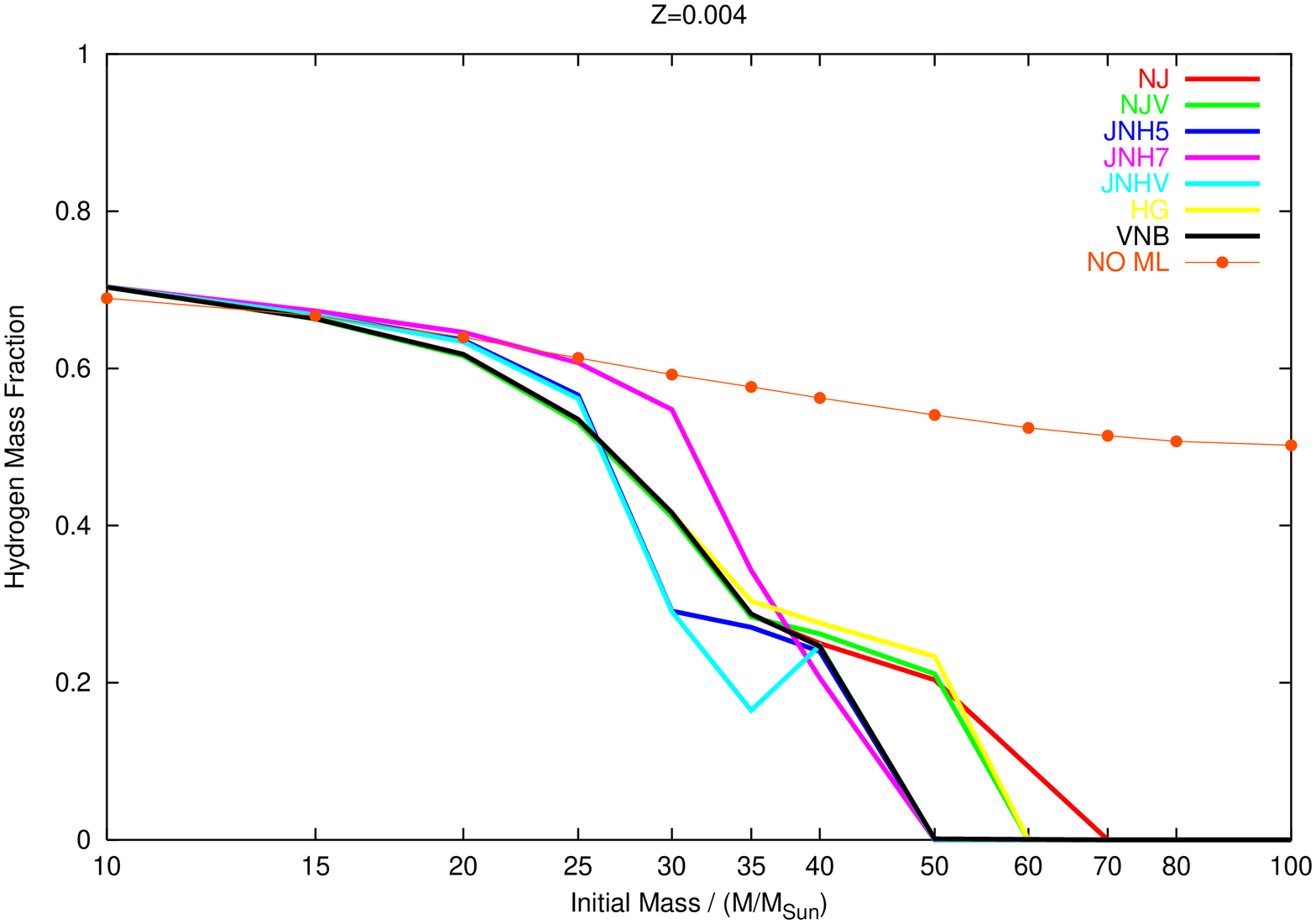}
\includegraphics[height=75mm,angle=270]{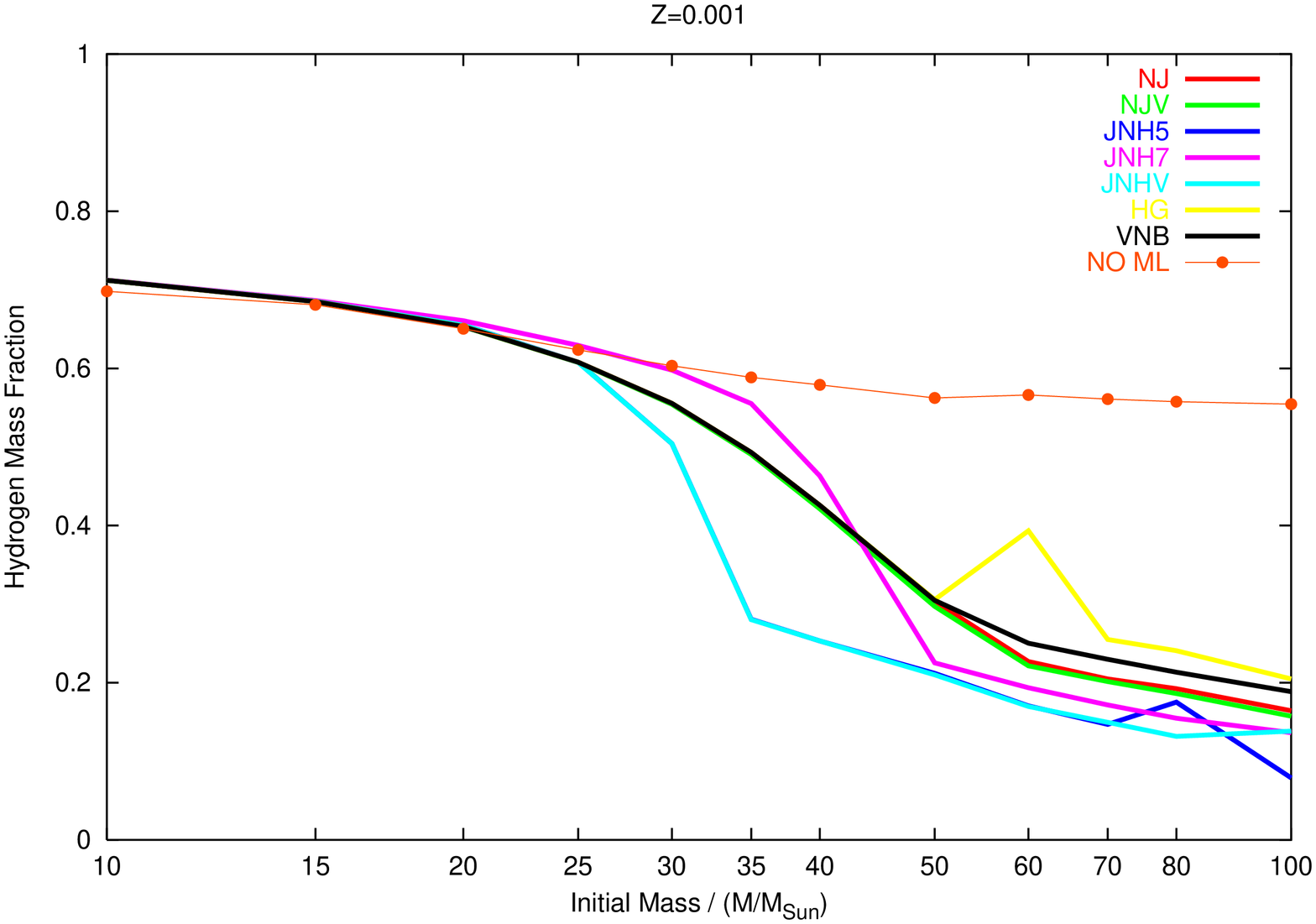}
\end{center}
\caption{The final surface hydrogen mass fraction versus initial mass for different mass-loss schemes at different metallicities.}
\label{allX}
\end{figure}

\begin{figure}
\begin{center}
\includegraphics[height=75mm,angle=270]{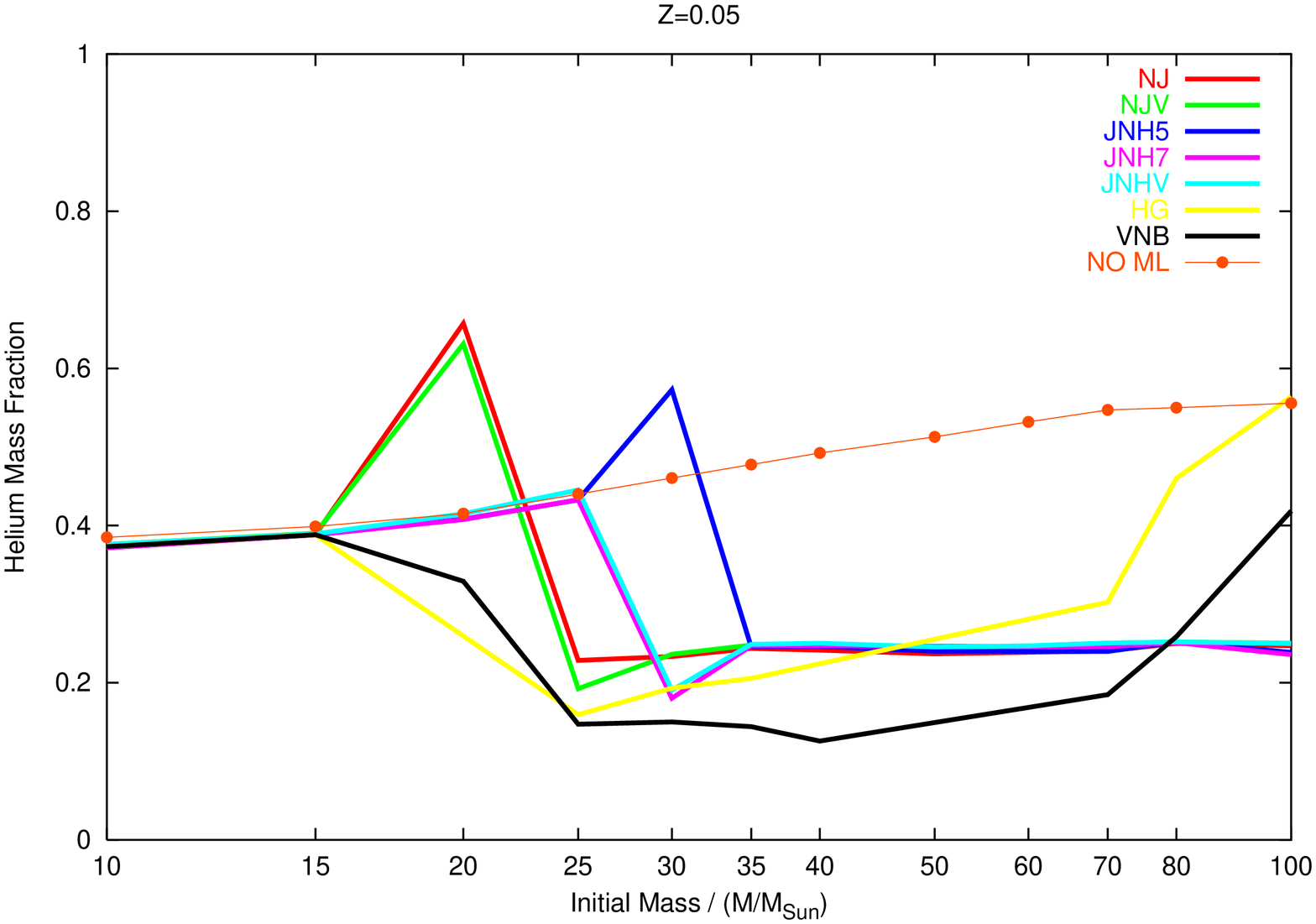}
\includegraphics[height=75mm,angle=270]{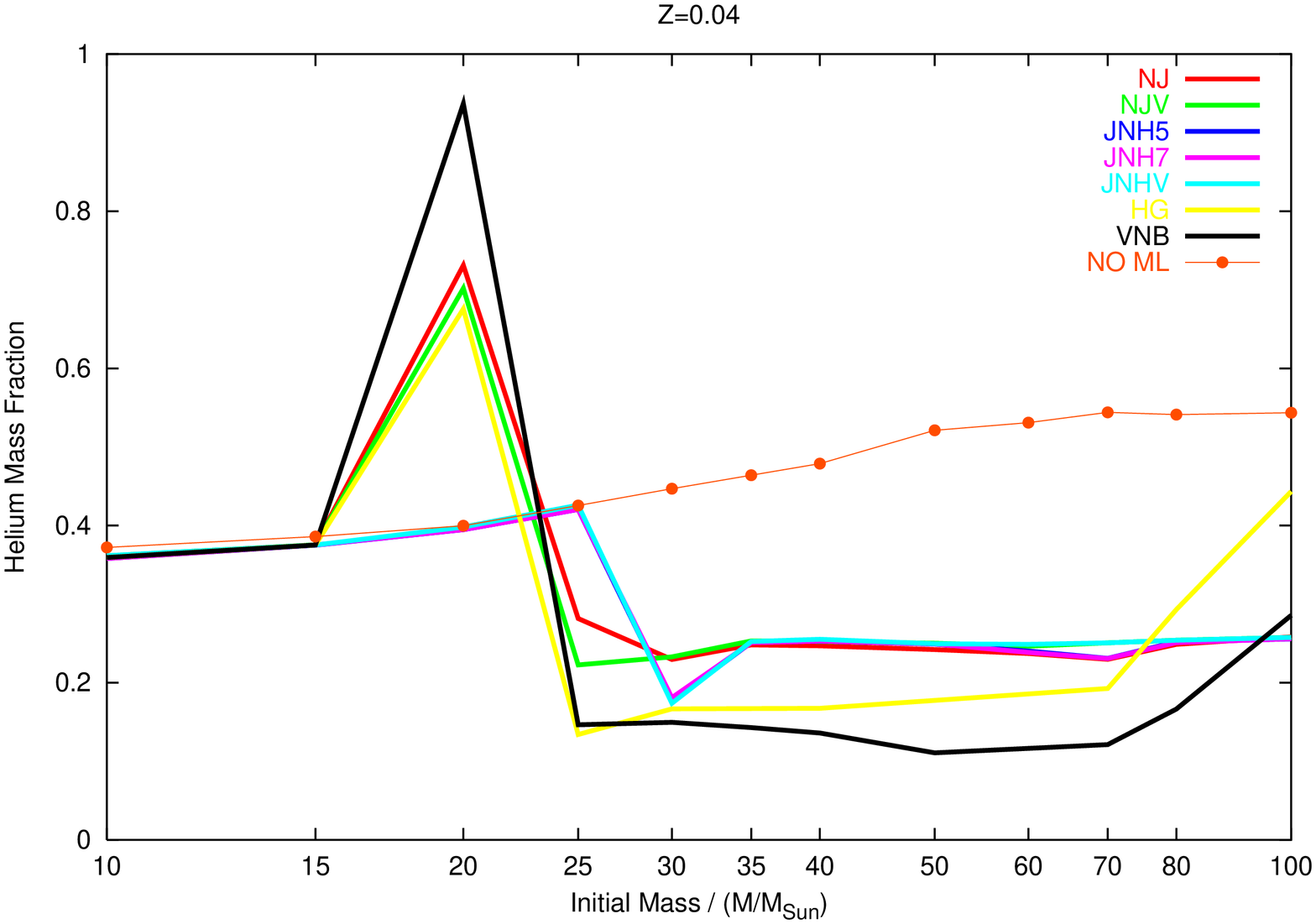}
\includegraphics[height=75mm,angle=270]{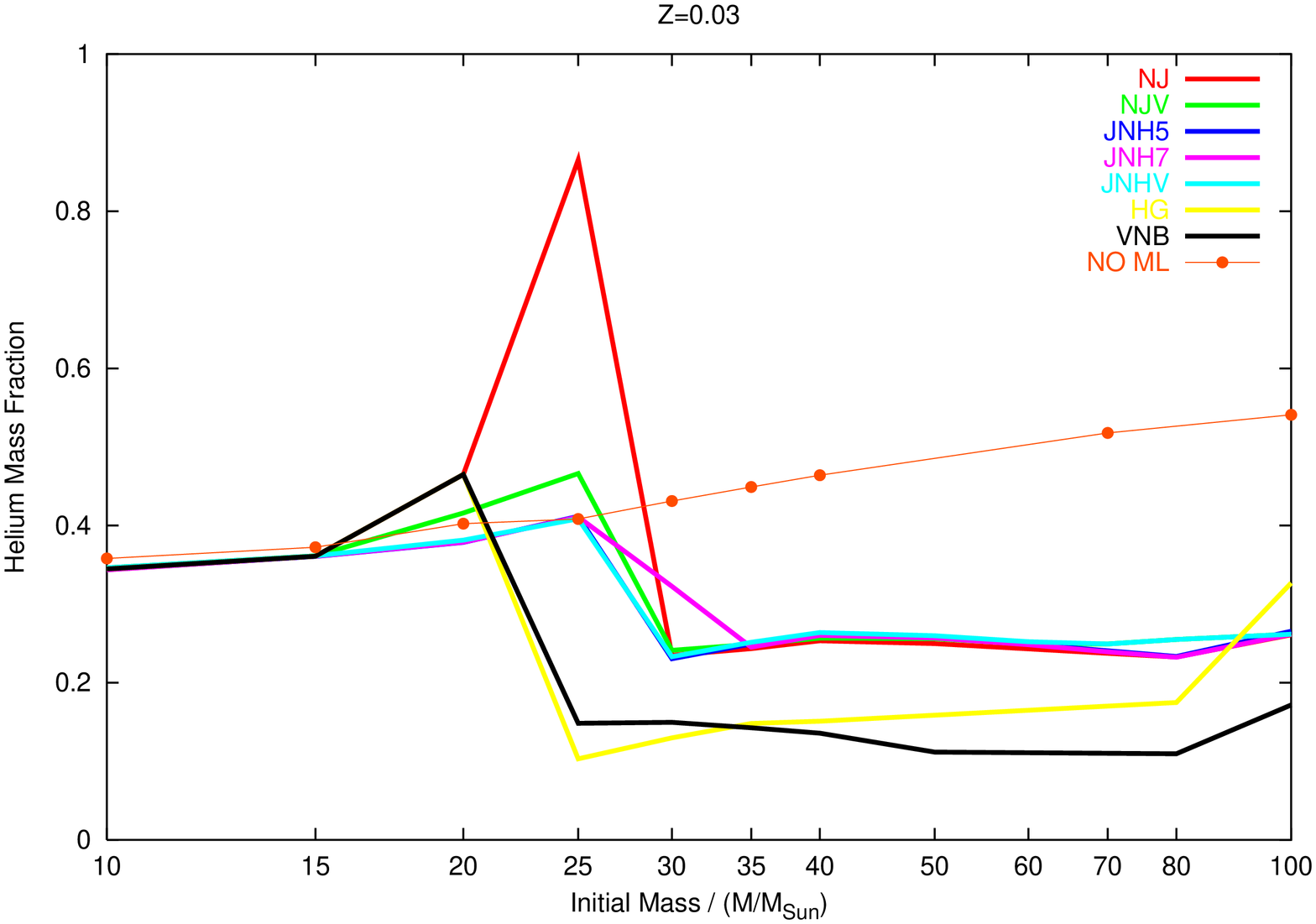}
\includegraphics[height=75mm,angle=270]{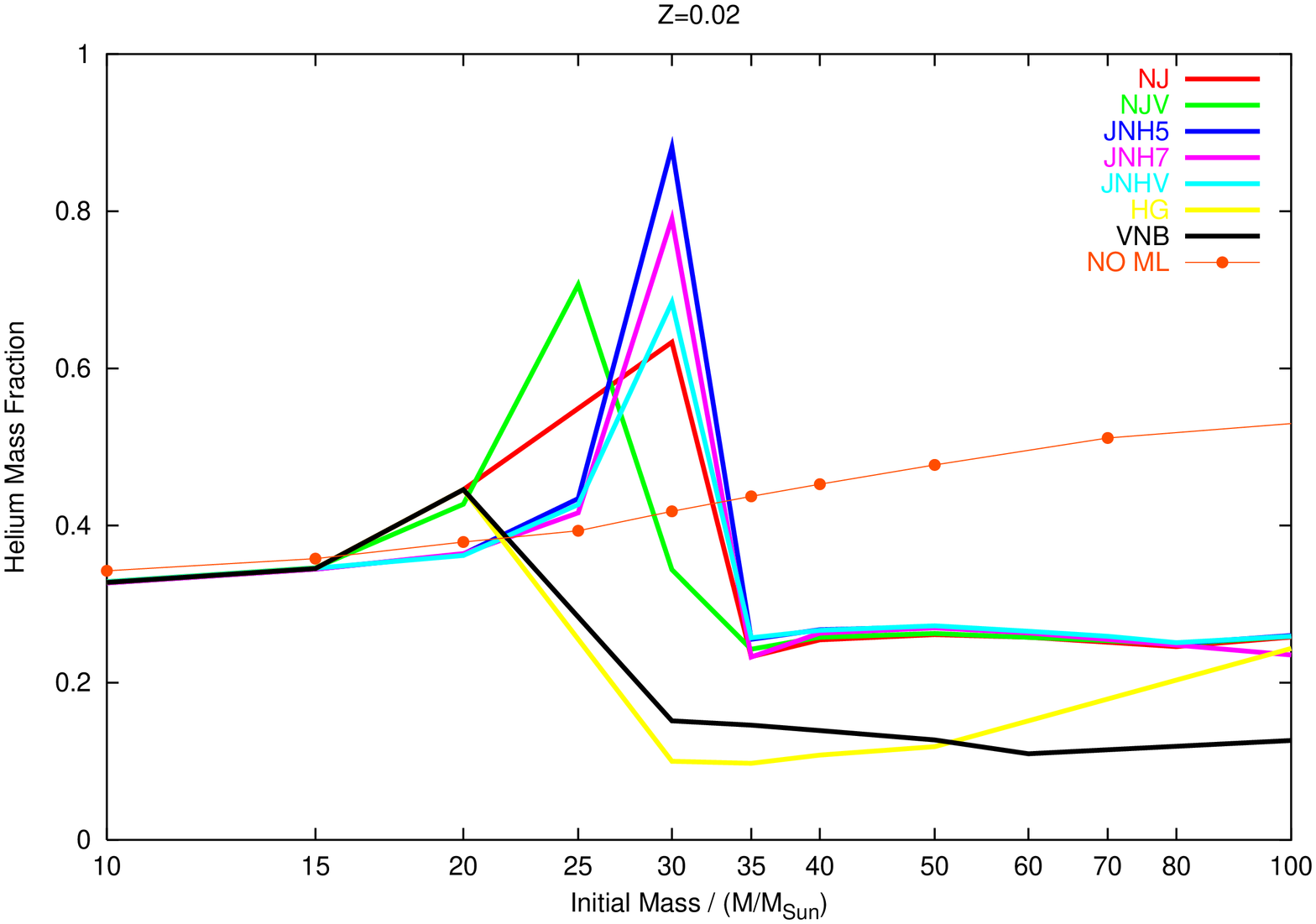}
\includegraphics[height=75mm,angle=270]{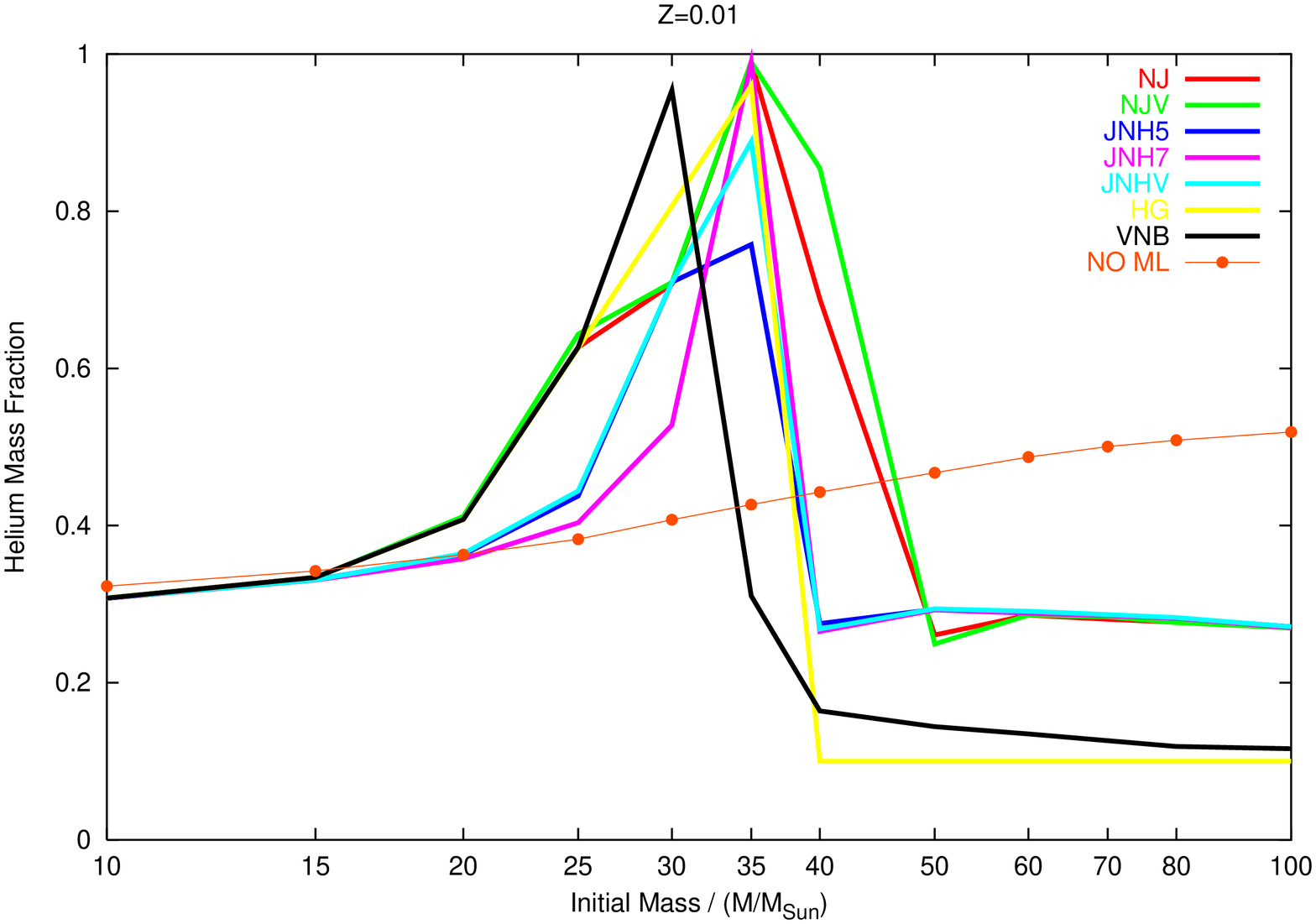}
\includegraphics[height=75mm,angle=270]{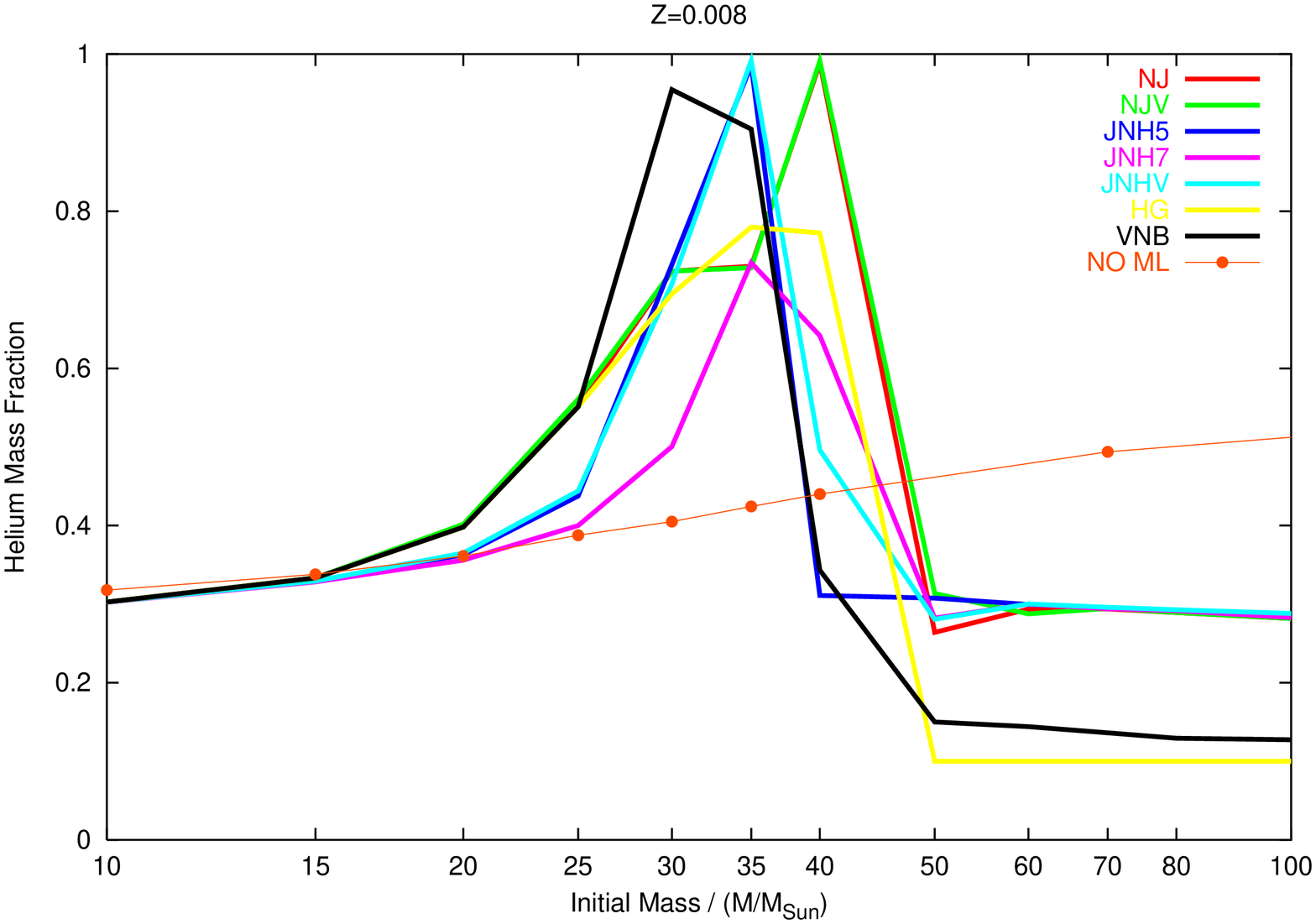}
\includegraphics[height=75mm,angle=270]{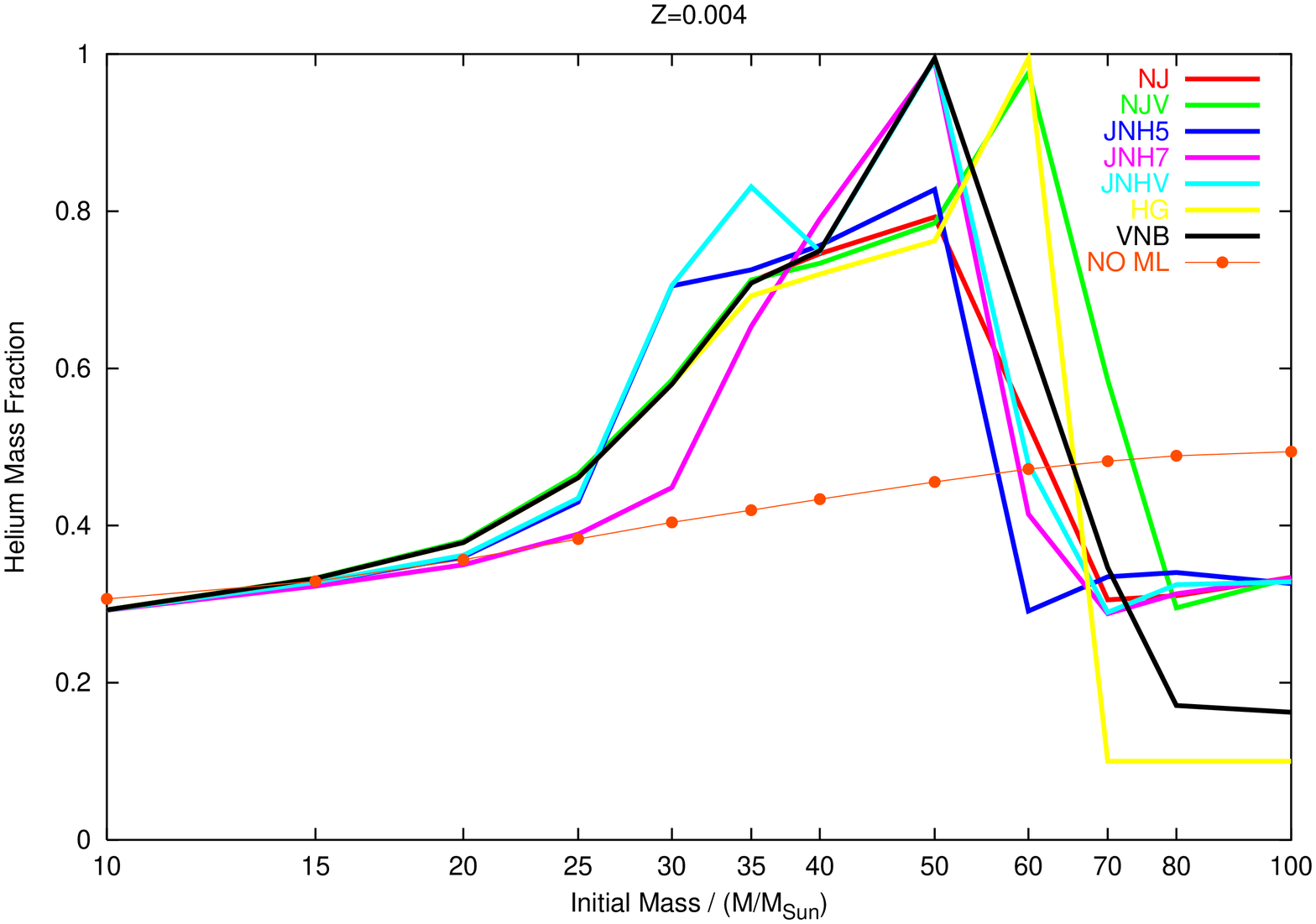}
\includegraphics[height=75mm,angle=270]{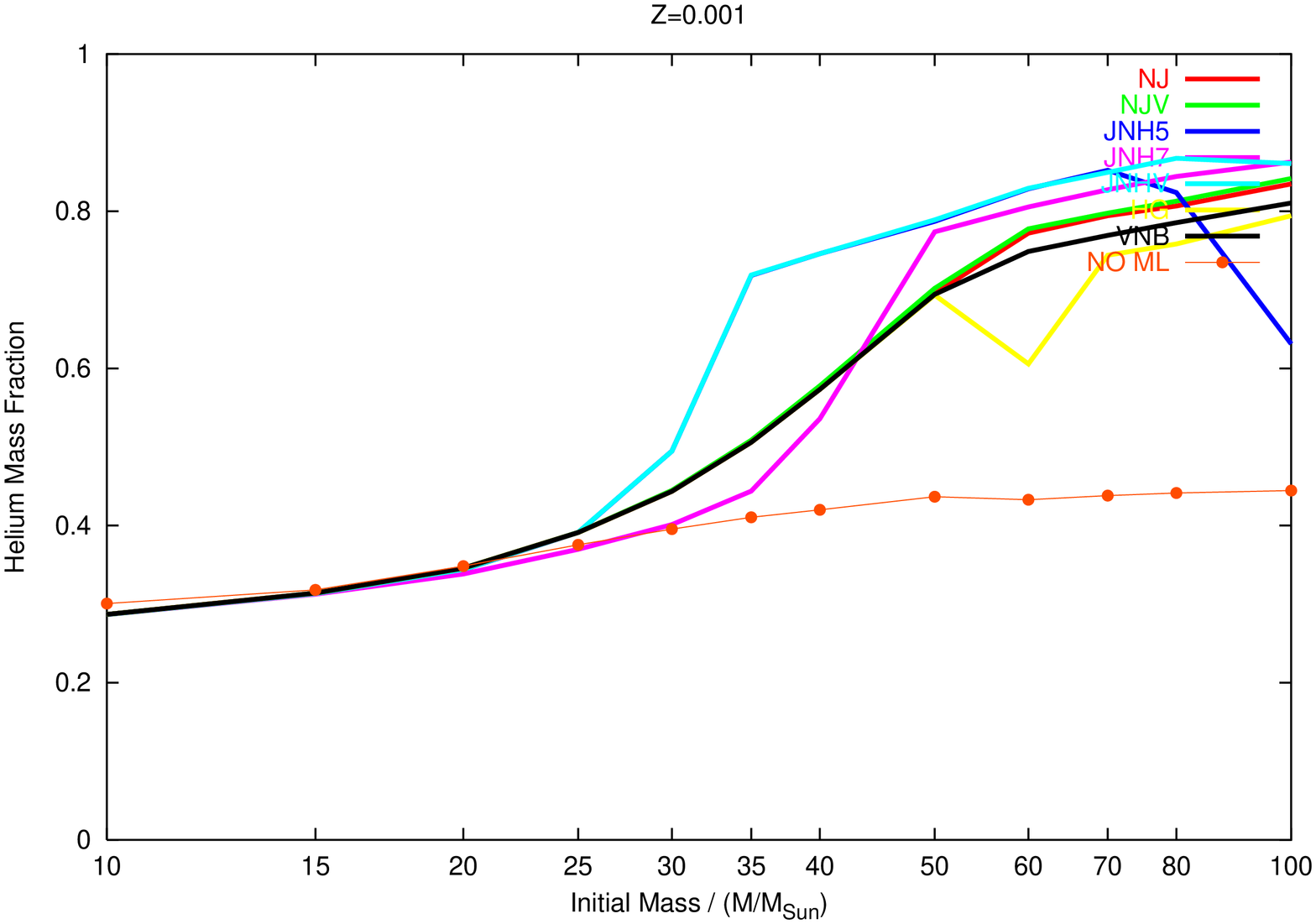}
\end{center}
\caption{The final surface helium mass fraction versus initial mass for different mass-loss schemes at different metallicities.}
\label{allY}
\end{figure}

\begin{figure}
\begin{center}
\includegraphics[height=75mm,angle=270]{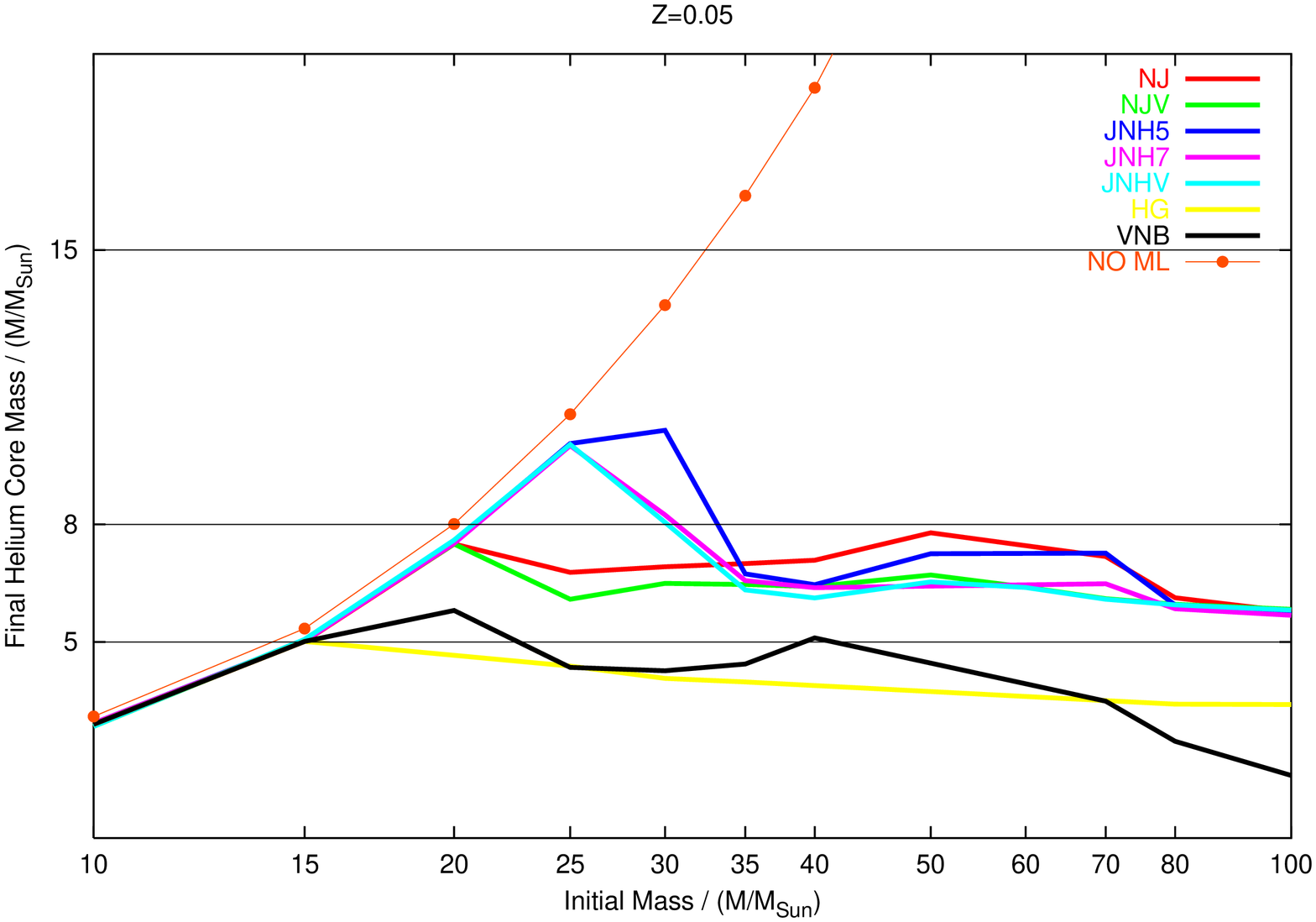}
\includegraphics[height=75mm,angle=270]{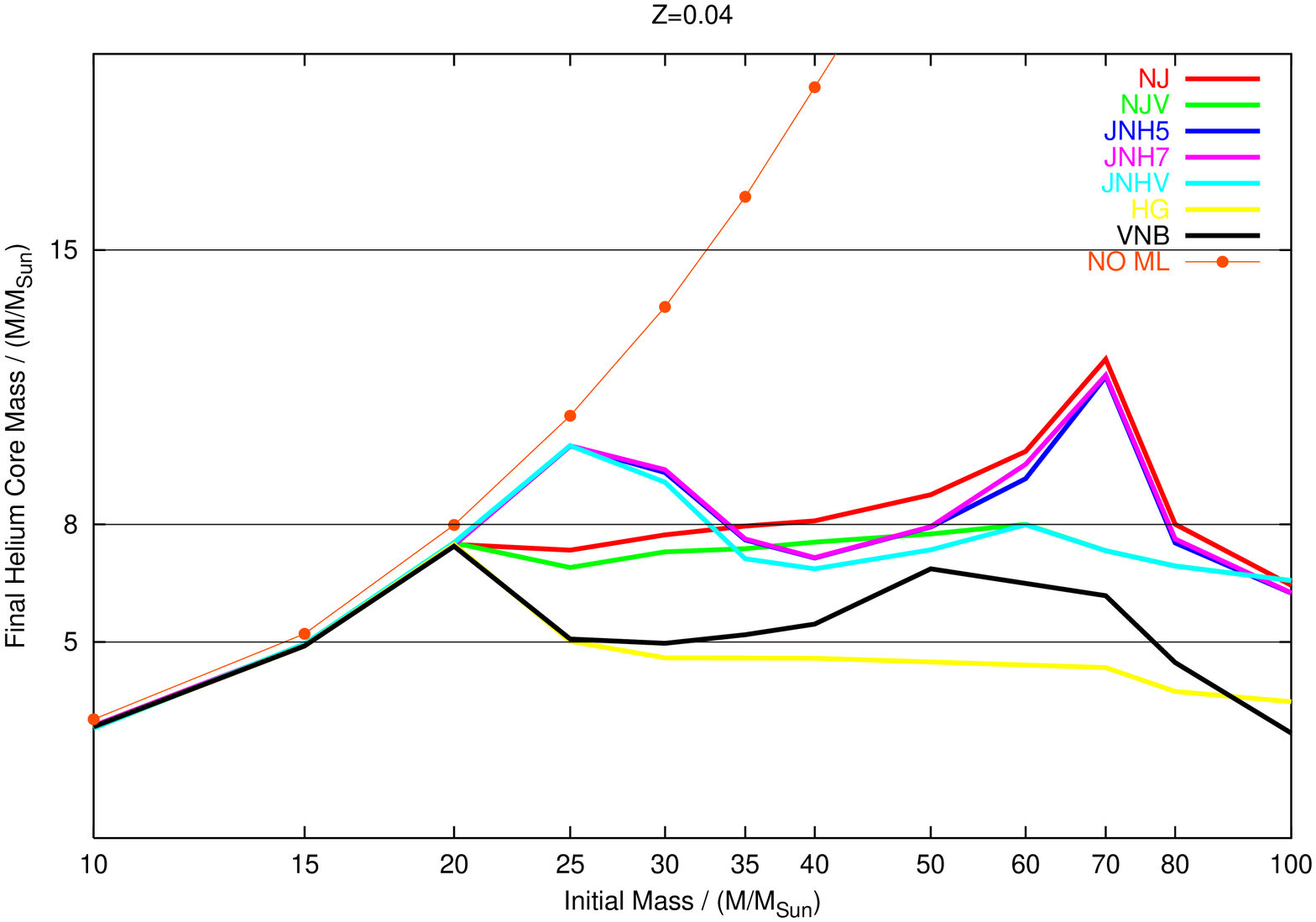}
\includegraphics[height=75mm,angle=270]{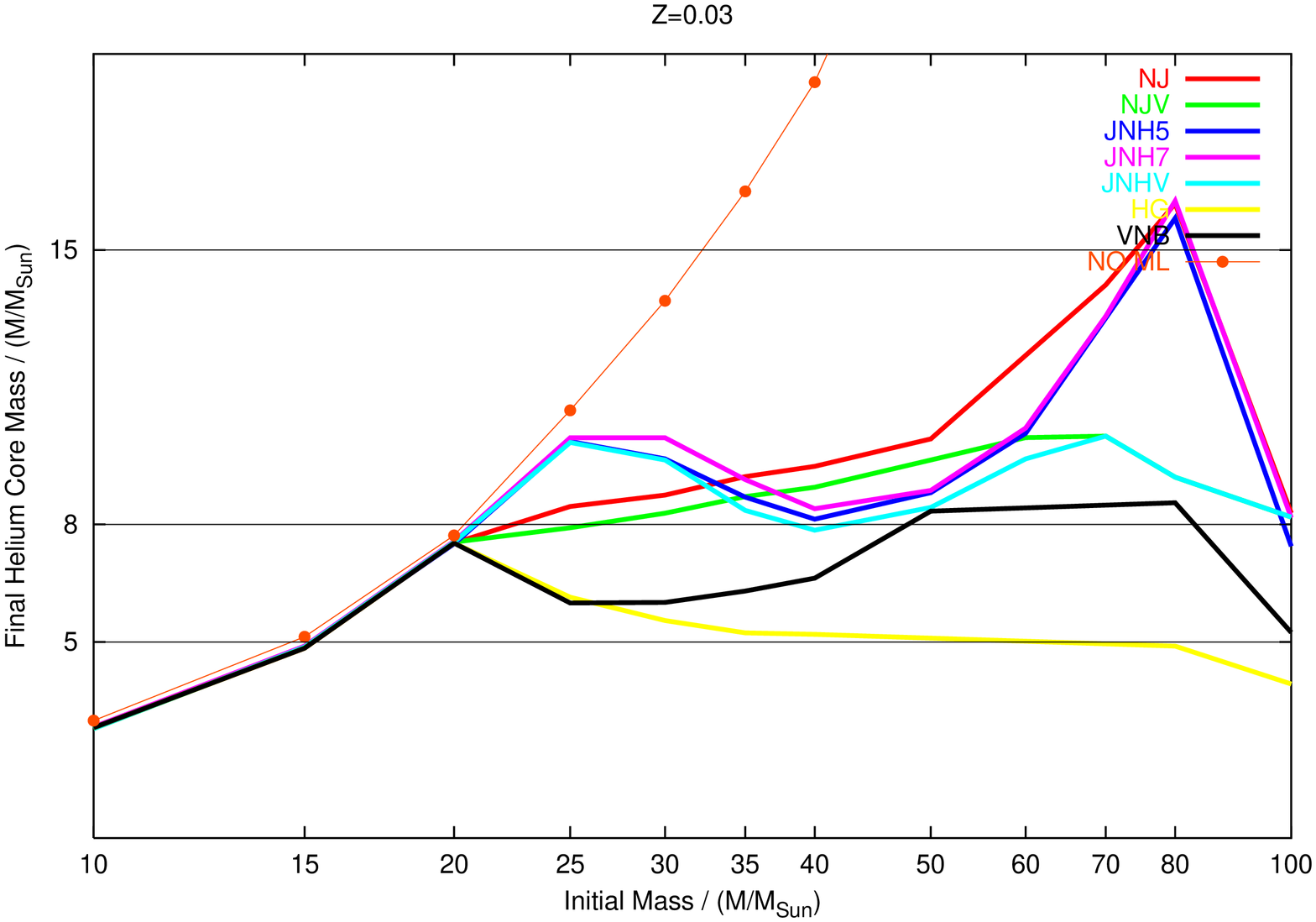}
\includegraphics[height=75mm,angle=270]{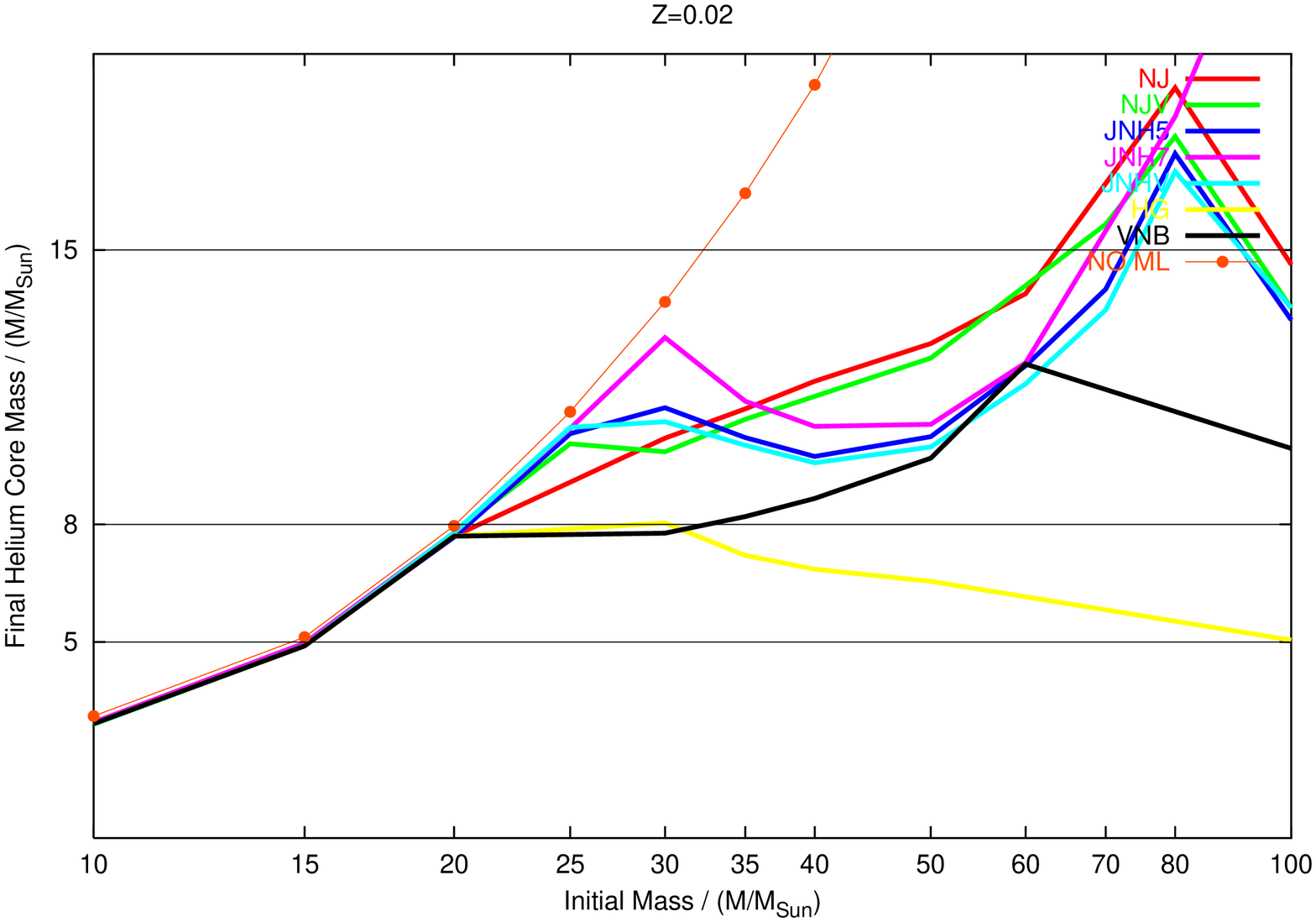}
\includegraphics[height=75mm,angle=270]{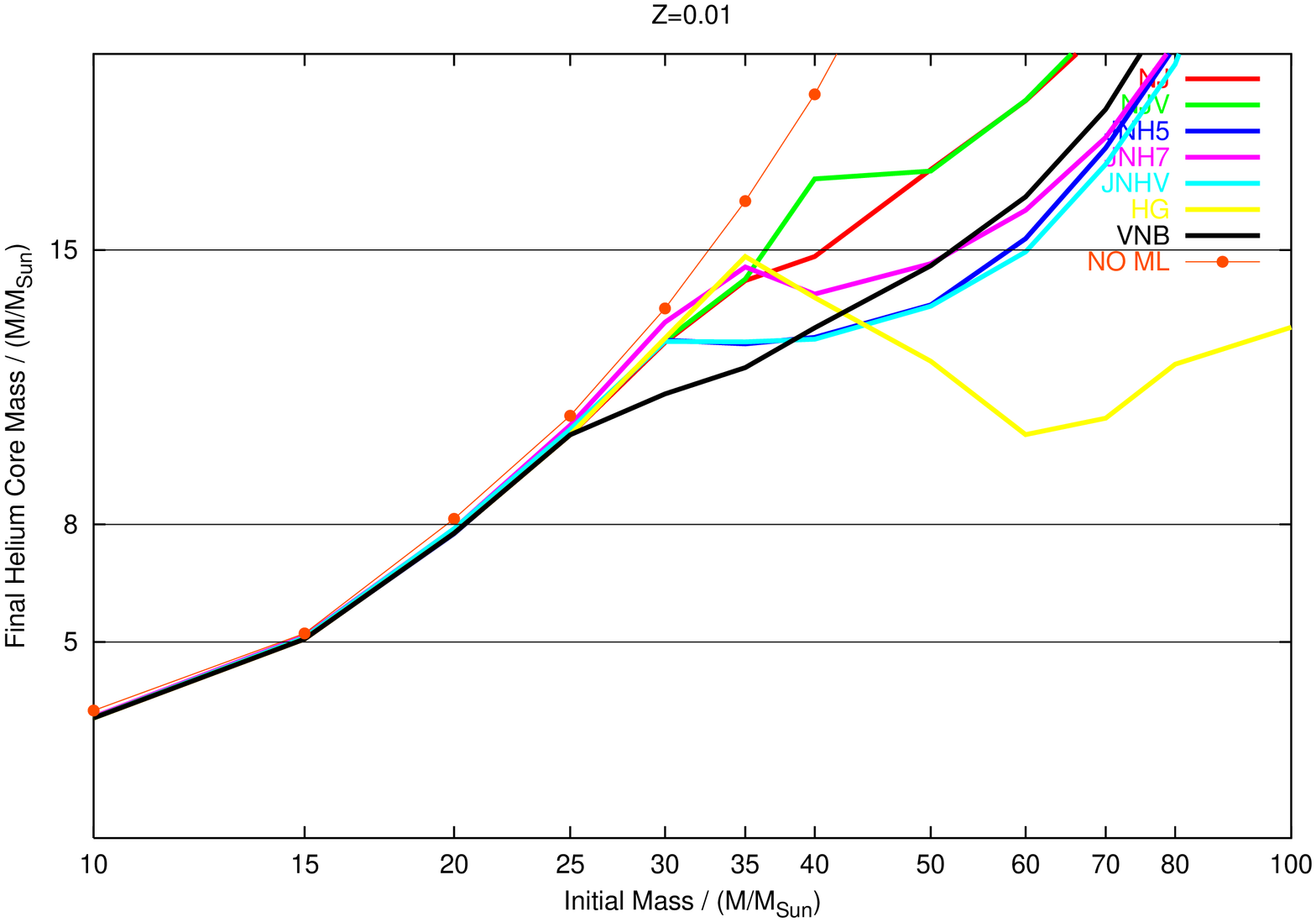}
\includegraphics[height=75mm,angle=270]{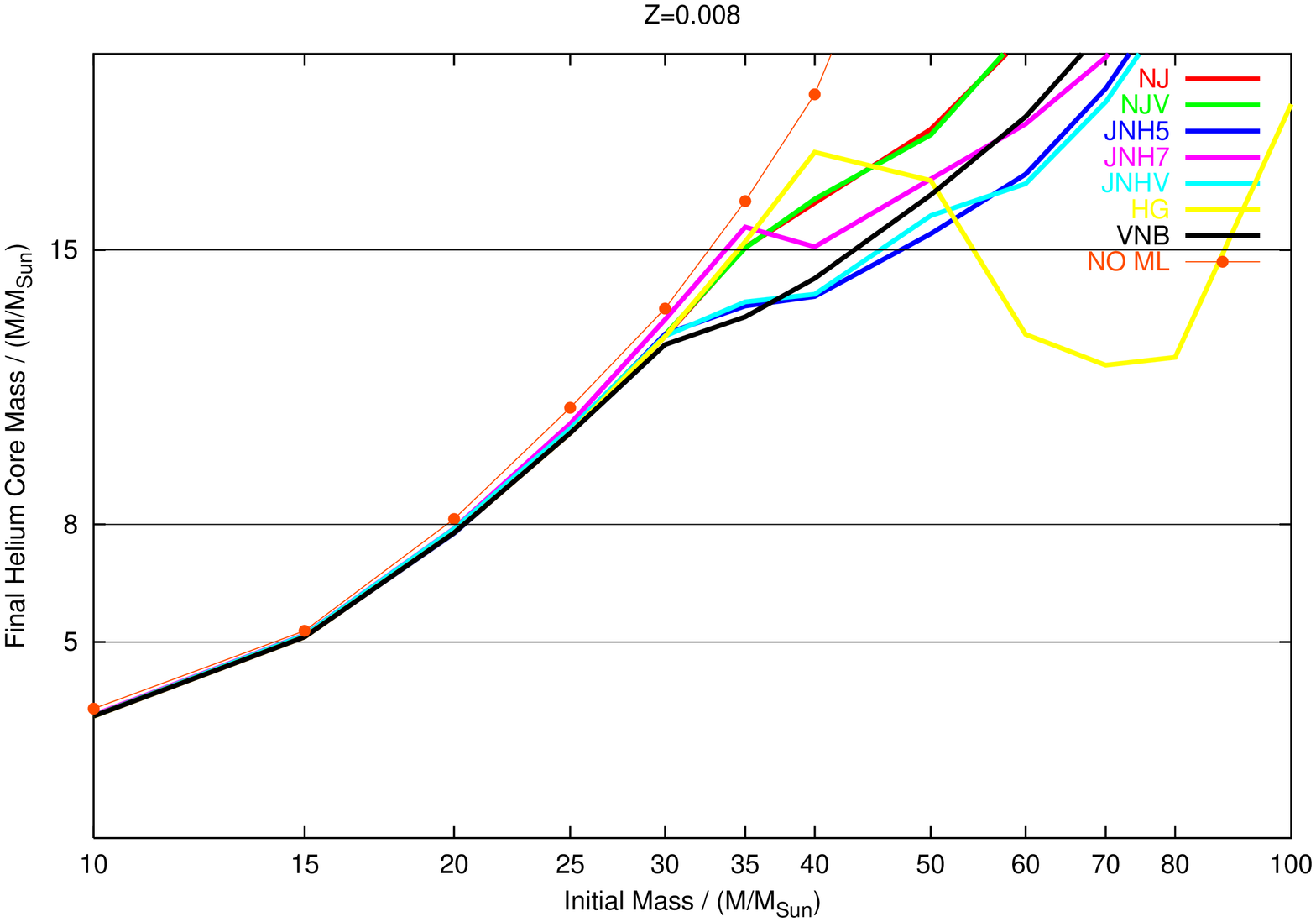}
\includegraphics[height=75mm,angle=270]{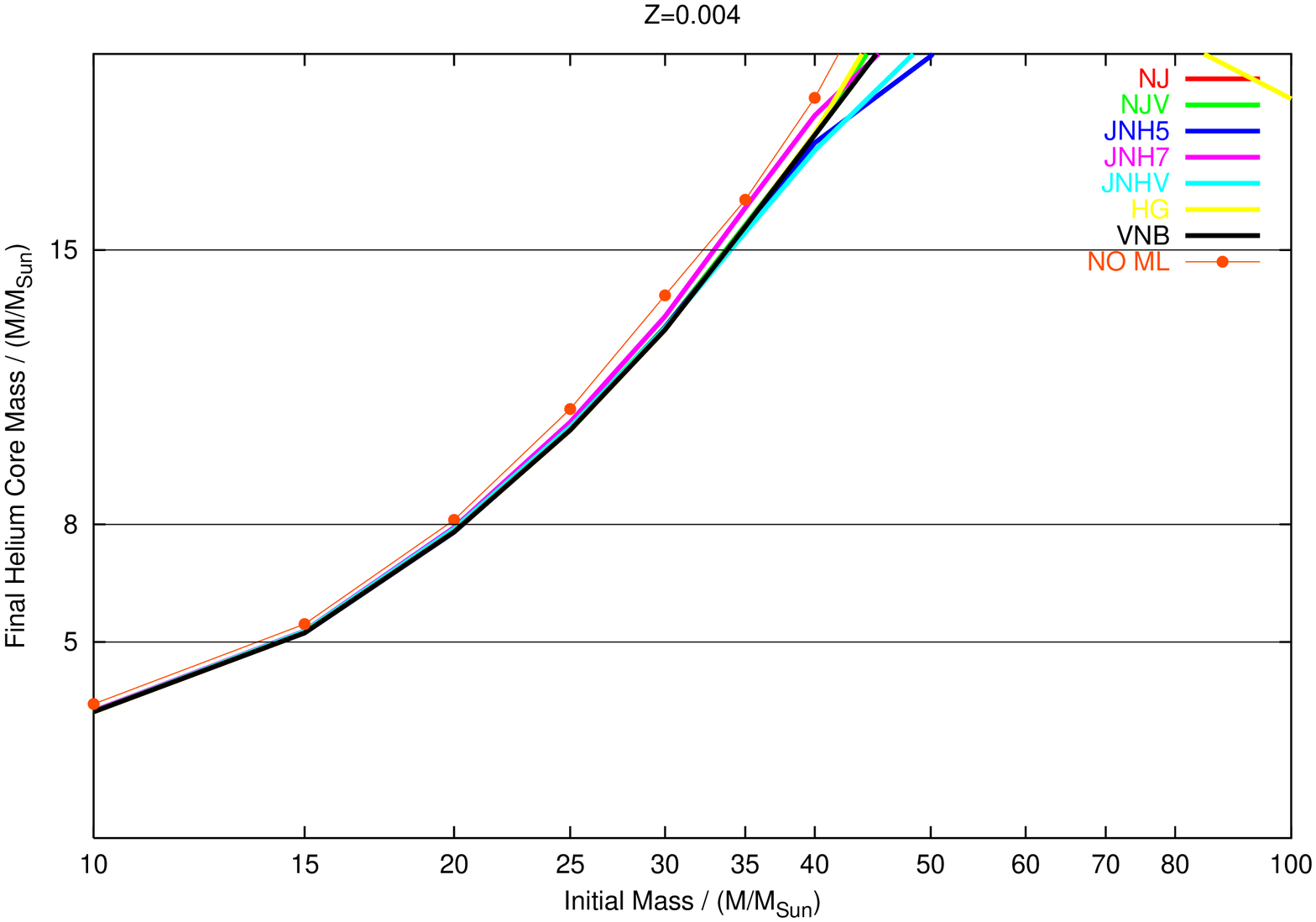}
\includegraphics[height=75mm,angle=270]{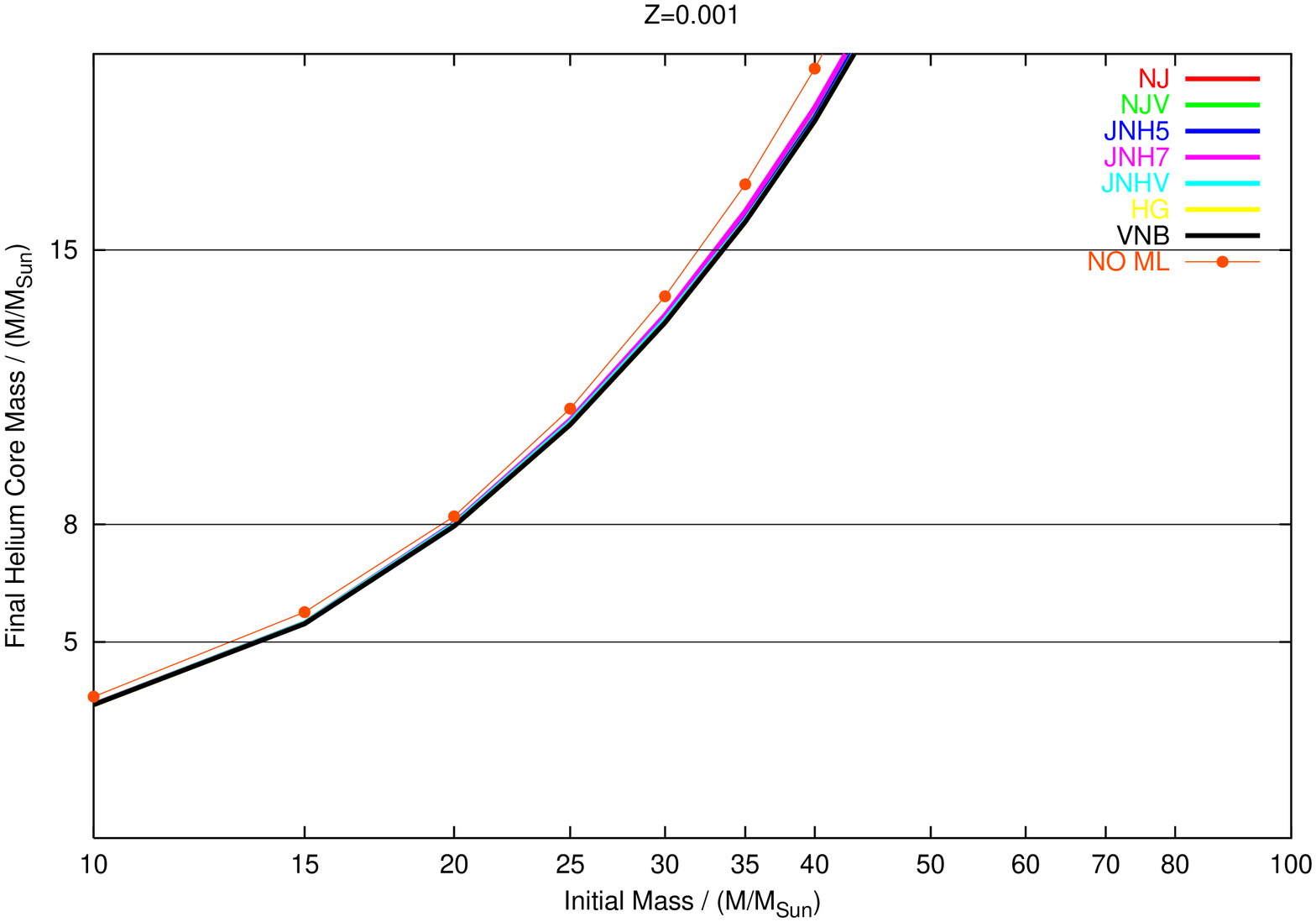}
\end{center}
\caption{The final helium core mass versus initial mass for different mass-loss schemes at different metallicities.}
\label{allcore}
\end{figure}

\begin{figure}
\begin{center}
\includegraphics[height=75mm,angle=270]{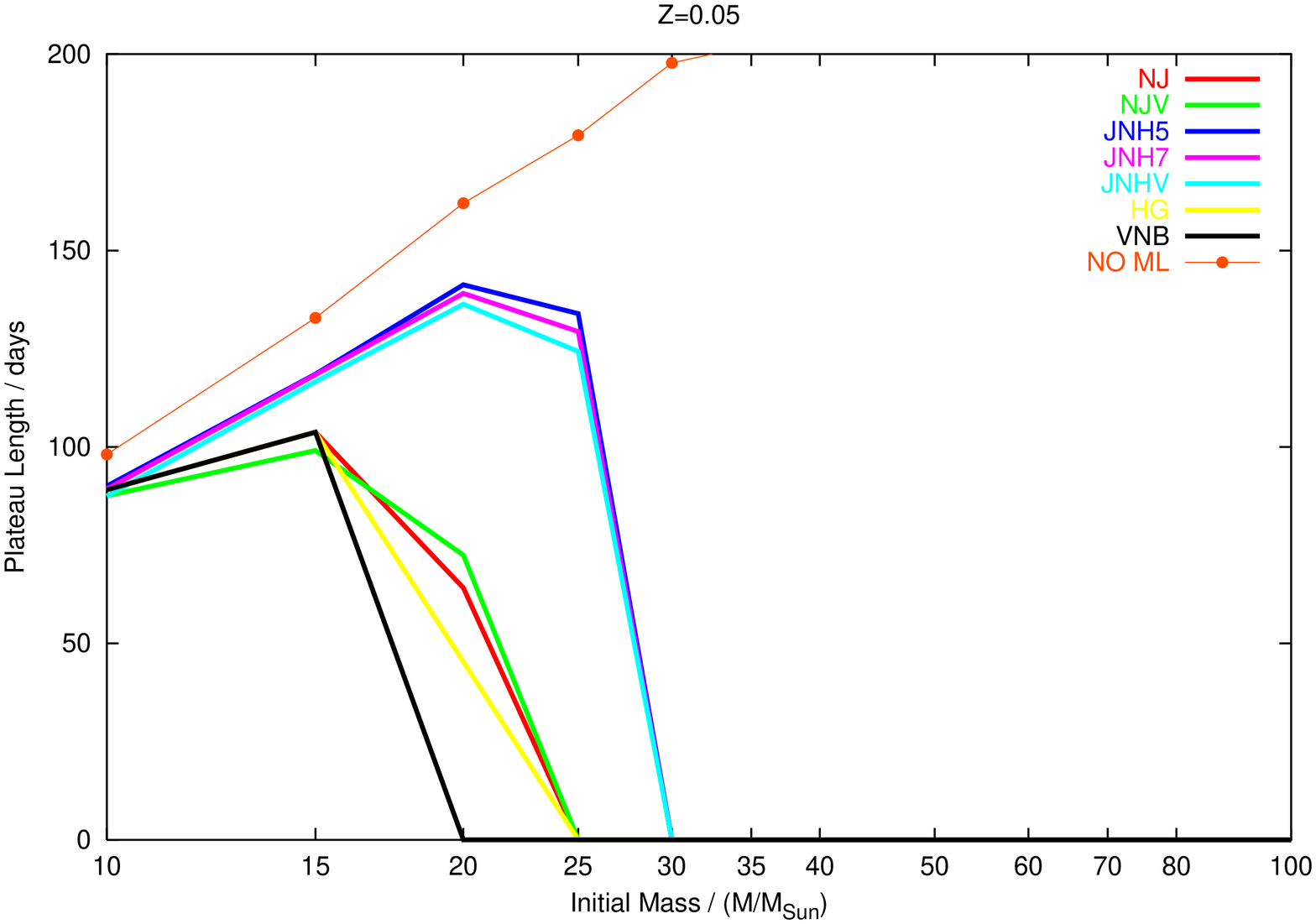}
\includegraphics[height=75mm,angle=270]{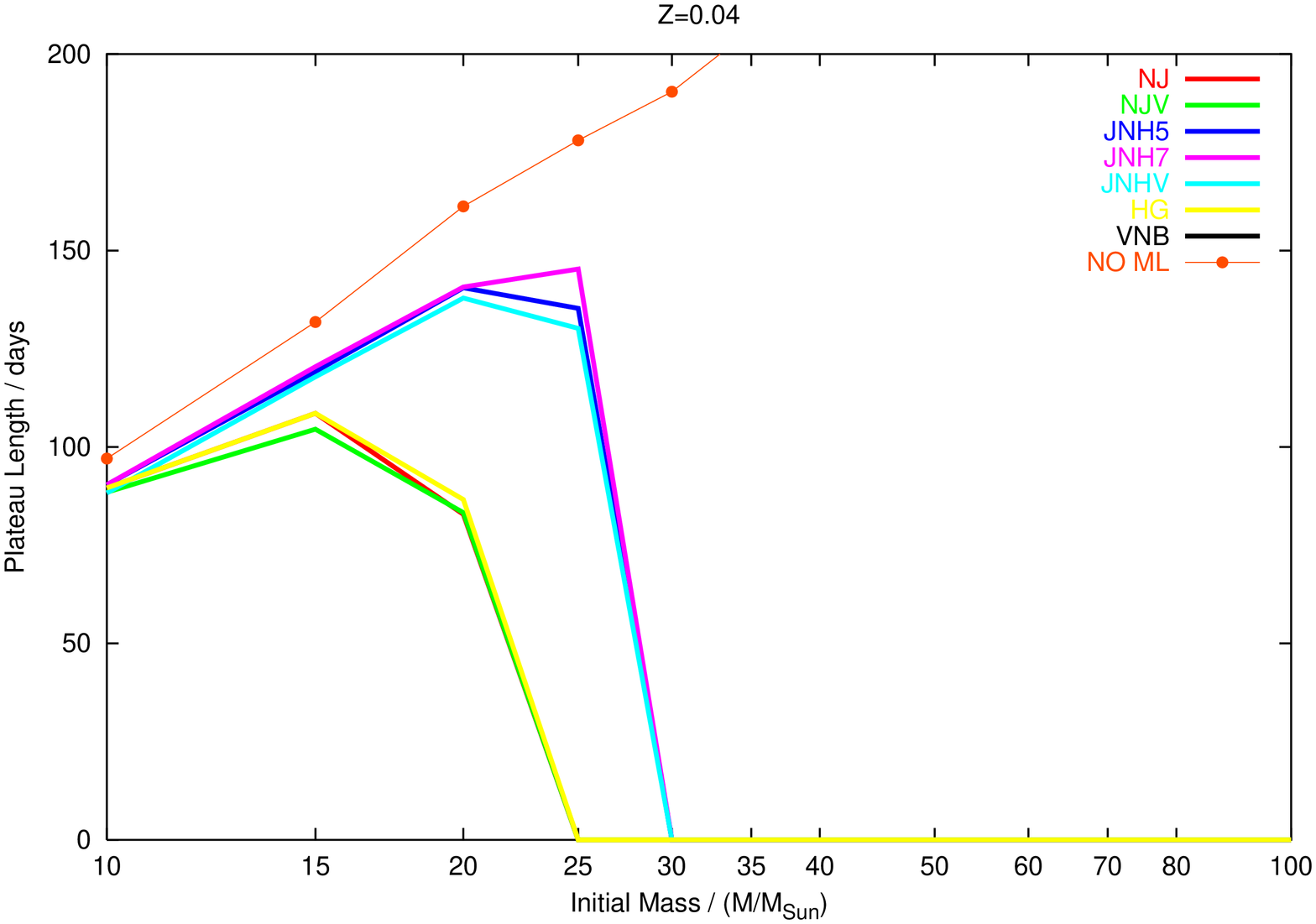}
\includegraphics[height=75mm,angle=270]{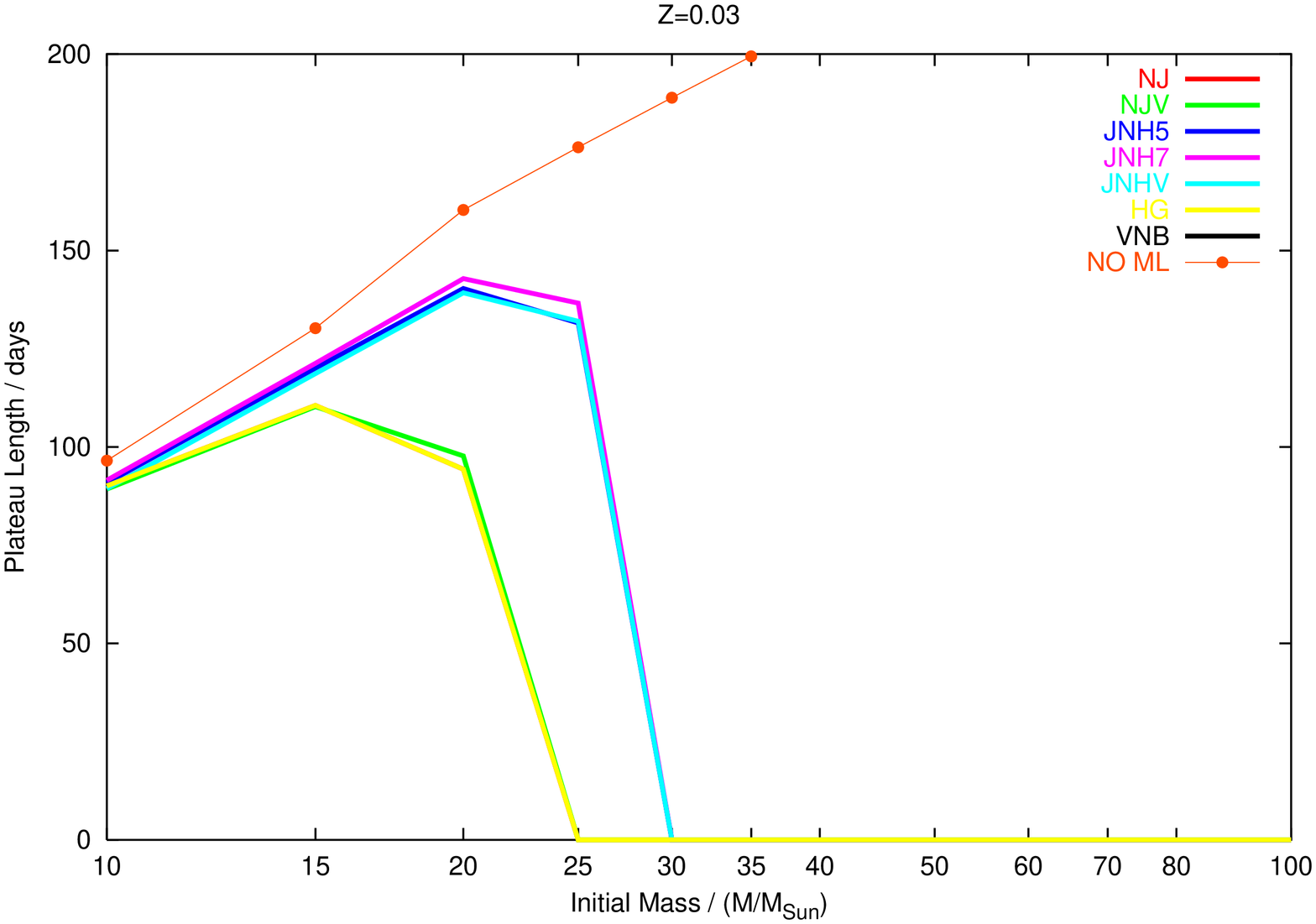}
\includegraphics[height=75mm,angle=270]{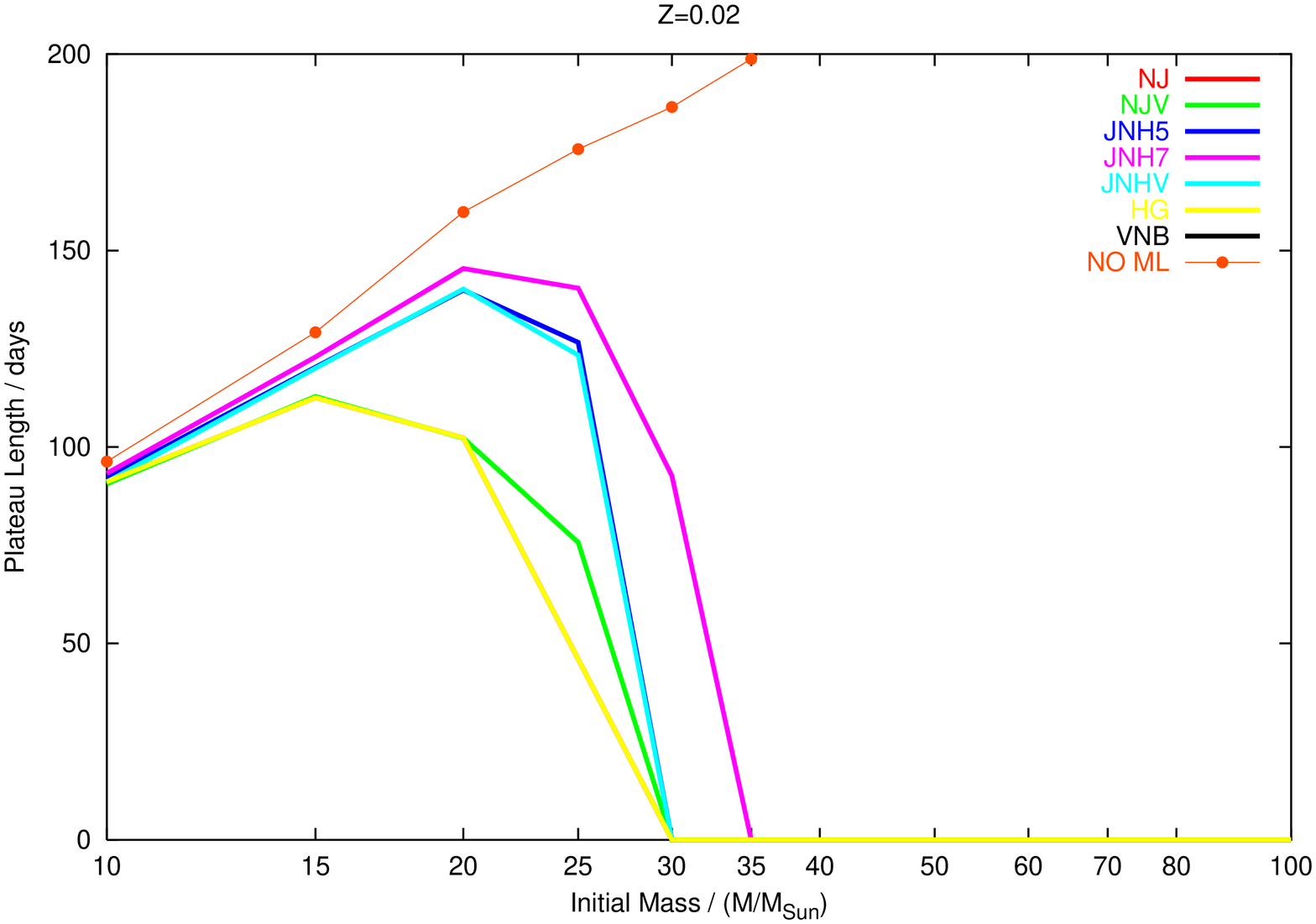}
\includegraphics[height=75mm,angle=270]{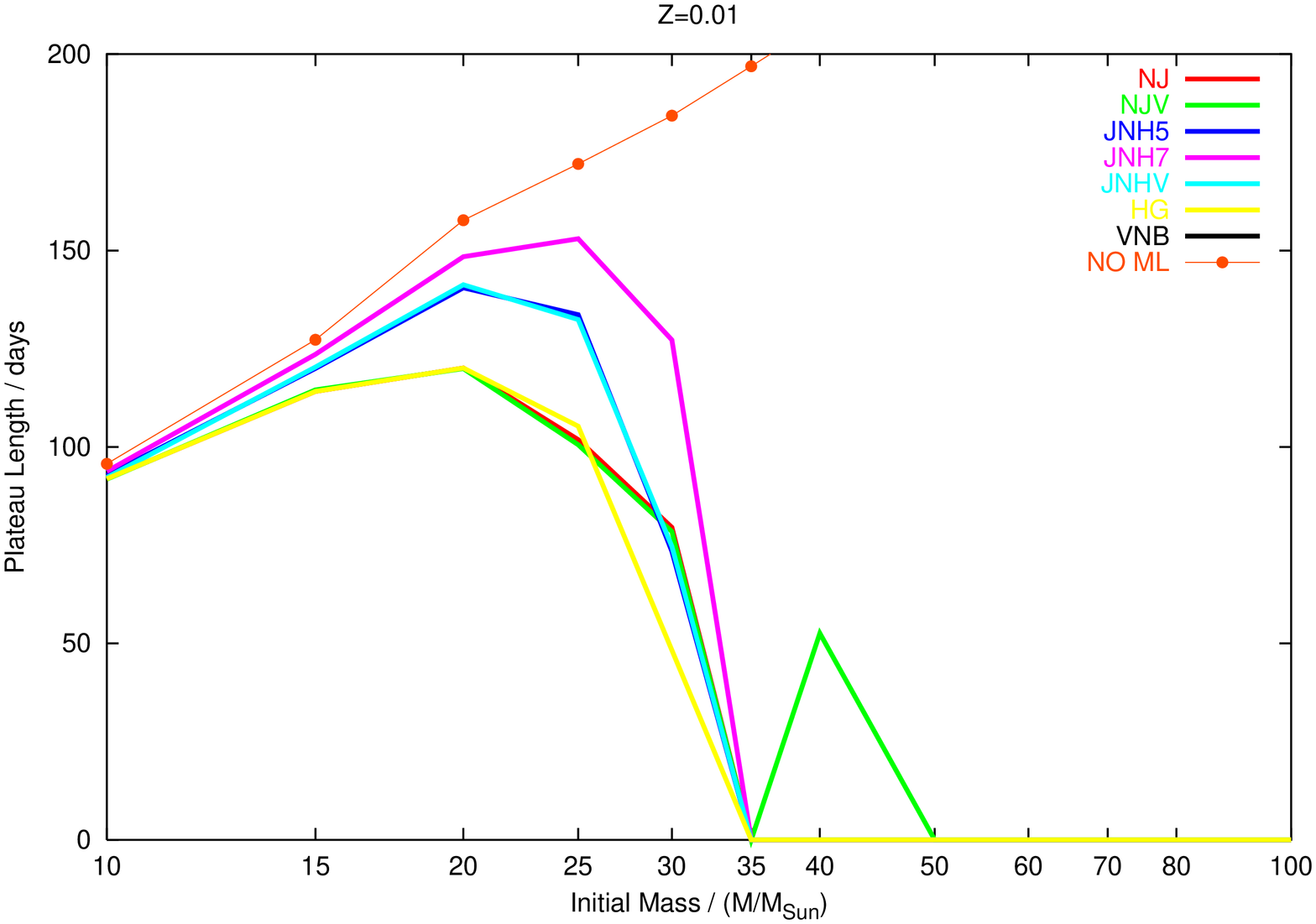}
\includegraphics[height=75mm,angle=270]{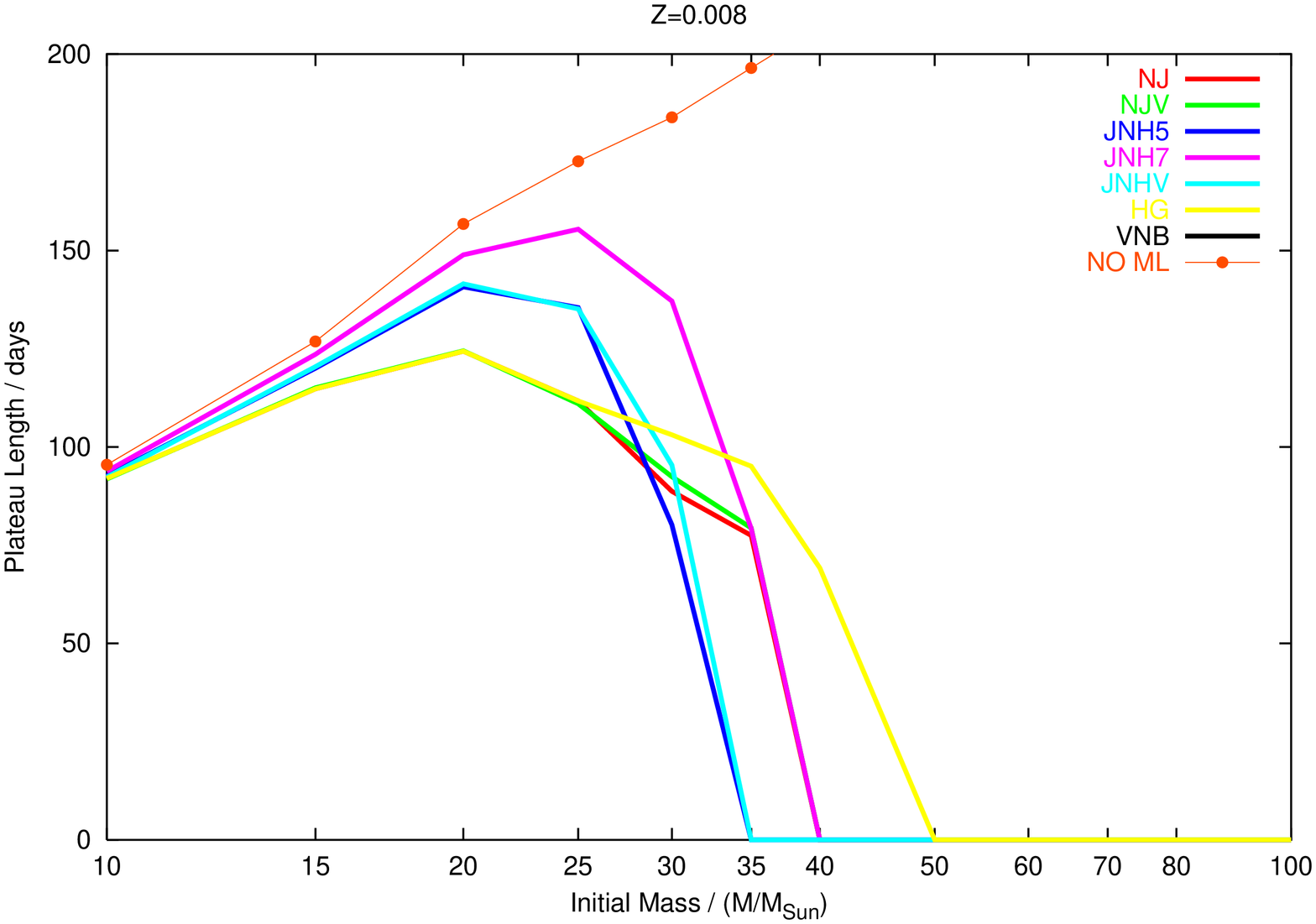}
\includegraphics[height=75mm,angle=270]{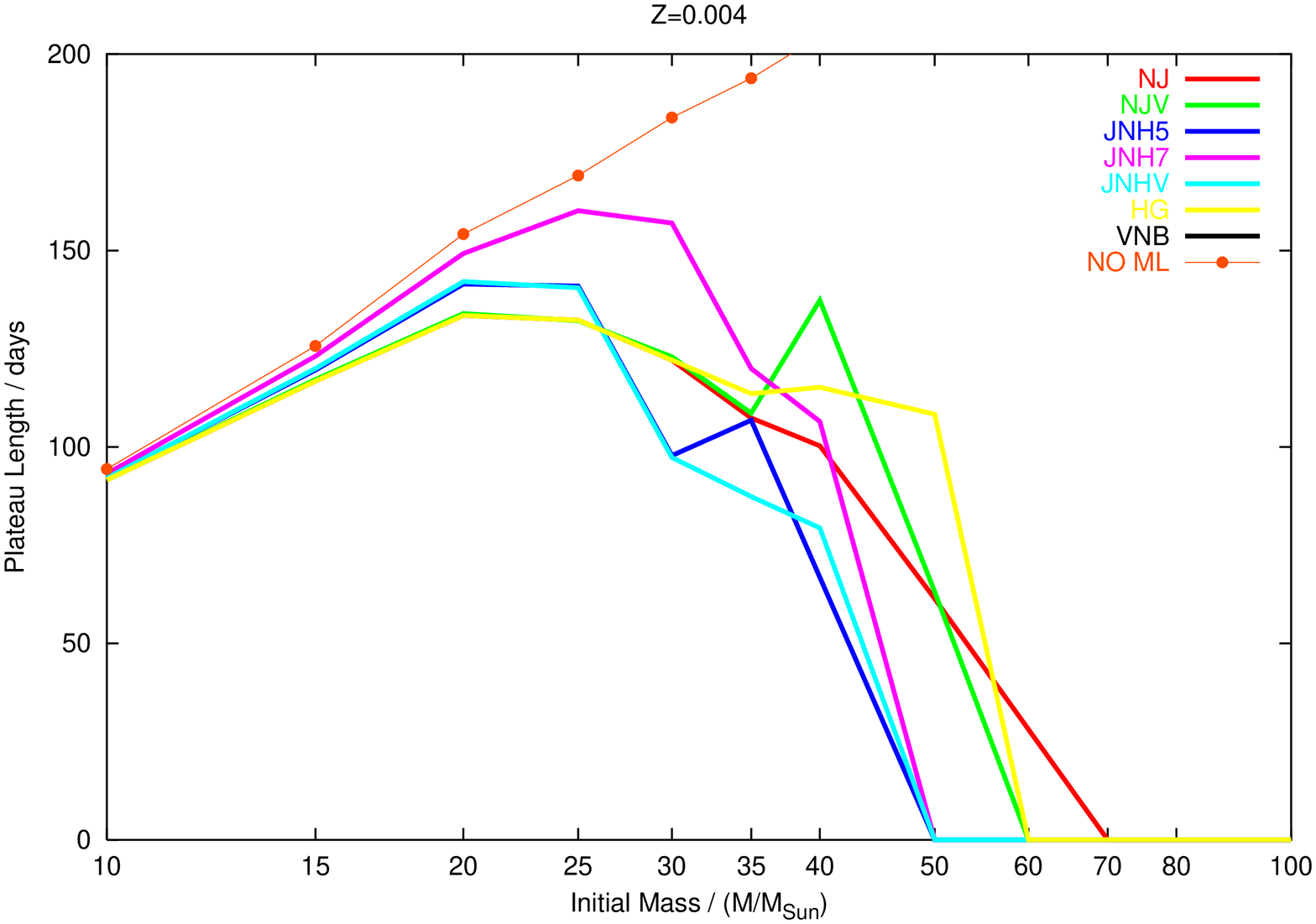}
\includegraphics[height=75mm,angle=270]{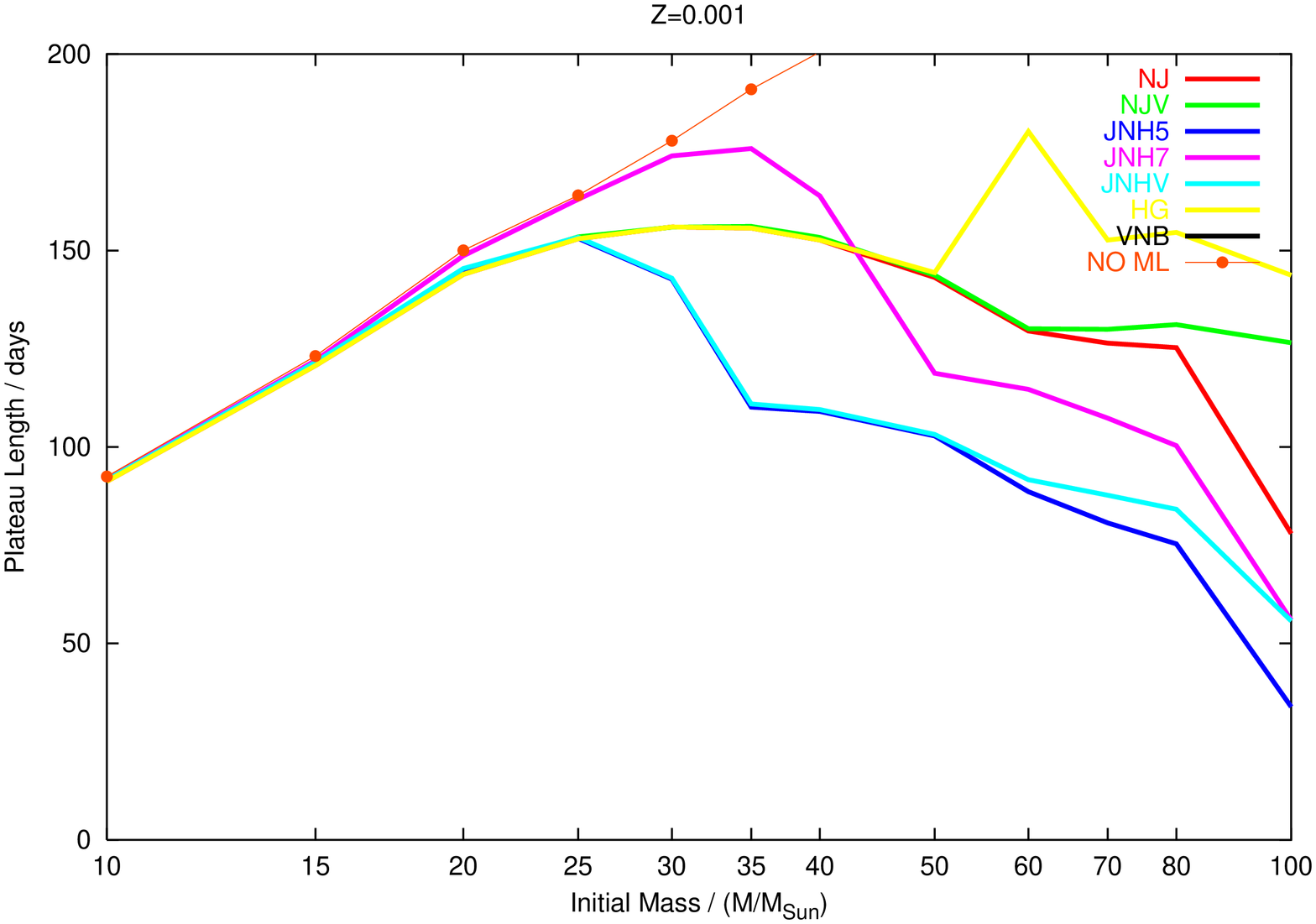}
\end{center}
\caption{The plateau length duration for SN with IIP light curves versus initial mass for different mass-loss schemes at different metallicities.}
\label{allplateau}
\end{figure}

We shall analyse the following details from our models, the final mass in figure \ref{allfinal}, the surface hydrogen and helium mass fractions in figures \ref{allX} and \ref{allY}, the final helium core mass in \ref{allcore} and the length of the type II plateau phase in figure \ref{allplateau}. From these we can draw conclusions on the differences between the mass loss rates.

The final mass data shows interesting agreements between the different stars. In all the diagrams there is a clear split between those with NJ and those with JNH rates. Around solar metallicity and above this behaviour is clearly seen when $M_{\rm initial} < 30M_{\odot}$. The JNH rates give rise to higher mass progenitors than the NJ rates for type~II SNe. As the metallicity drops the difference alters so that the NJ rates give rise to more massive progenitors. The reason for these differences is the extra structure in the JNH rates that effectively leads to lower mass loss for stars than the NJ rates. However, at lower metallicity, the stars are more luminous and this can lead to more mass loss for the most massive stars because they end up in the region with greatest mass loss as shown in figure~\ref{rates}. The mass loss becomes severe despite the metallicity scaling. This demonstrates the only problem of the JNH rates which were measured at solar metallicity. At different metallicities the position of the evolution tracks of the star shuffles due to different opacities and different initial hydrogen and helium composition. Near to solar this effect is small but it escalates as the metallicity becomes more extreme. The NJ rates, because of their simplicity, suffer less from this problem although it is their simplicity that makes them less accurate at near solar metallicities where mass loss is too great. The only way to resolve these problems is further observations of the mass loss from stars at different metallicities and theoretical mass-loss rates for all stars.

The second region of interest in these diagrams is the region when $M_{\rm initial} > 30M_{\odot}$ at around solar metallicity and above. They are all WR despite having different pre-WR mass-loss rates. At the highest metallicity there are two populations. Those using the NL rates have higher final WR masses than those using alternative WR mass loss rates. As the metallicity moves toward solar we find that the VNB rates move into agreement with results from the others while the HG rates continue to have lower pre-SN masses. At the lowest metallicity the difference becomes small due to the metallicity scaling. The most discrepant result seems to be that of HG because of the estimate of lower WR masses. The mass-loss rate they may be using has not been reduced enough from the rates of \citet{Langer} however the result is not discrepant to more than a factor of 2 at solar metallicity. The final point to note is that by comparing the JNH results, we can also see that the larger exponent also has its greatest effect at the lowest metallicities. The addition of the Vink rates to the NJ and JNH rates makes only small changes.

In \ref{allX} and \ref{allY} it is easier to see over what ranges we see the WR stars. The mass of hydrogen and helium has a tight correlation to the mass fraction which means we can deduce the amount of hydrogen and helium in the models from these plots. From the hydrogen plot at solar metallicity and above there is a clear separation between the NJ and JNH rates with different minimum masses at which all the hydrogen is removed from a star. Immediately below solar metallicity the results start to agree more closely until the relation is inverted at the lowest metallicities where JNH produces more hydrogen deficient stars than NJ. From studying the helium graphs at solar and above once the hydrogen has been removed all the models agree to a large degree but with the HG and VNB results showing the lowest helium fraction due to greater mass loss. Below solar metallicity the difference in the helium fraction grows until below SMC metallicity when the JNH rates all have more helium than all the NJ based rates. This is because the JNH rates remove more hydrogen from the star than the NJ rates so in the latter much of the surface mass is still hydrogen leading to lower helium abundance.

The helium core mass details in figure \ref{allcore} confirm the above conclusions. It is apparent that there are different maximum metallicities at which black holes will form directly, with the HG rates needing a metallicity of below $Z = 0.01$ for this to occur. They also require a lower metallicity, lower than solar, to form black holes by fall back. The other schemes have direct black holes or fall back black holes at metallicities higher than these rates. This again indicates that the HG rates are higher on average than others. For all rates very massive helium cores are not possible until SMC metallicity and below so in these regions we find the most massive black holes. In view of the lowest helium core masses only the HG and VNB rates have WR stars below $5M_{\odot}$ that will give rise to bright Ib events.

Our final figure \ref{allplateau} of plateau duration for IIP SN splits the JNH and NJ rates again with a clear distinction between the two. The JNH rates give rise to a maximum plateau length of 140 days while the NJ rates give rise to a maximum of 110 days at solar and above. These increase below this range due to the greater amount of mass left in the stars as the stellar winds weaken. The longest plateaus are achieved by the JNH7 rates. The remaining rates all tend to agree at the lowest metallicities. In table \ref{averages} we compare these values with those from our two main contending schemes.

\begin{table}
\begin{center}
\caption{The average values for the plateau duration of the IIP SN light curve.}
\label{averages}
\begin{tabular}{|l|ccc|}
\hline
Source		&	Average Value	& Shortest	& Longest \\
			& 		/days		&/days		&/days\\
\hline
Nadyozhin	&	$105 \pm 15$	&	80		&	140	  \\
Hamuy		&	$131 \pm 21$	&	105		&	171		\\
\hline
JNHV Solar	&		104			&   90		& 140	\\
JNHV LMC	&		105			&	90		& 140	\\
JNHV SMC	&		104			&	90		& 140	\\
\hline
NJV Solar	&		97			&	90		& 112\\
NJV LMC		&		98			&	90		& 120\\
NJV SMC		&		104			&	90		& 132\\
\hline
\end{tabular}
\end{center}
\end{table} 

Both of these agree with the results but do favour JNH with the longer possible plateau lengths. We should bear in mind that more massive envelopes lead to more luminous plateaus so there is also a selection effect on these statistics and they are for a small sample of SN from \citet{iiplength}. Also in calculating the plateau lengths we have used a constant explosion energy. Despite this the results for the JNH models agree best with these observations, apart from JNH7 at low metallicity. While NJ rates only agree at the SMC metallicity and below, they lack plateau lengths of around $140$ days. We really need further observations to provide details of how the plateau length changes with metallicity to choose one of these schemes over the other based on these details, especially since they will also have strong winds that might give a type IIn appearance if they are in the early WR phase.

In summary there are some important discrepancies between mass-loss schemes and there are some important similarities. Which prescription is best is testable by SN and WR observations. The rates which seem to provide the best results are the JNH5 and JNHV rates. The latter are our preferred rates at the higher metallicities covered here and we shall investigate them further below.

\section{Comparing Prescriptions at Low Metallicity}

We now move our analysis to the lowest metallicities. These are closer to population III stars where $Z=0$ however they still have evolution similar to the stars at higher metallicities. The biggest divergence of evolution of low metallicity stars occurs around $Z=10^{-12}$. At this level the CNO abundance is too small for the CNO bi-cycle to support the star from collapse so the star continues to contract until helium ignites and forms enough carbon that is then processed to nitrogen and oxygen by hydrogen burning which can then support the star. The mass-loss rates for these stars however remain largely unknown. For the hot OB stars we can use the Vink rates although we are far below the regime they were derived for. The only rates that do exist are those of \citet{KD2002}. These extend far down into the extremely low metallicity regime. At these low metallicities there are few WR stars so their rates are not as important here. In this section therefore we must compare the NJ and JNH rates combined with the theoretical rates for OB stars.
\begin{figure}
\begin{center}
\includegraphics[height=70mm,angle=0]{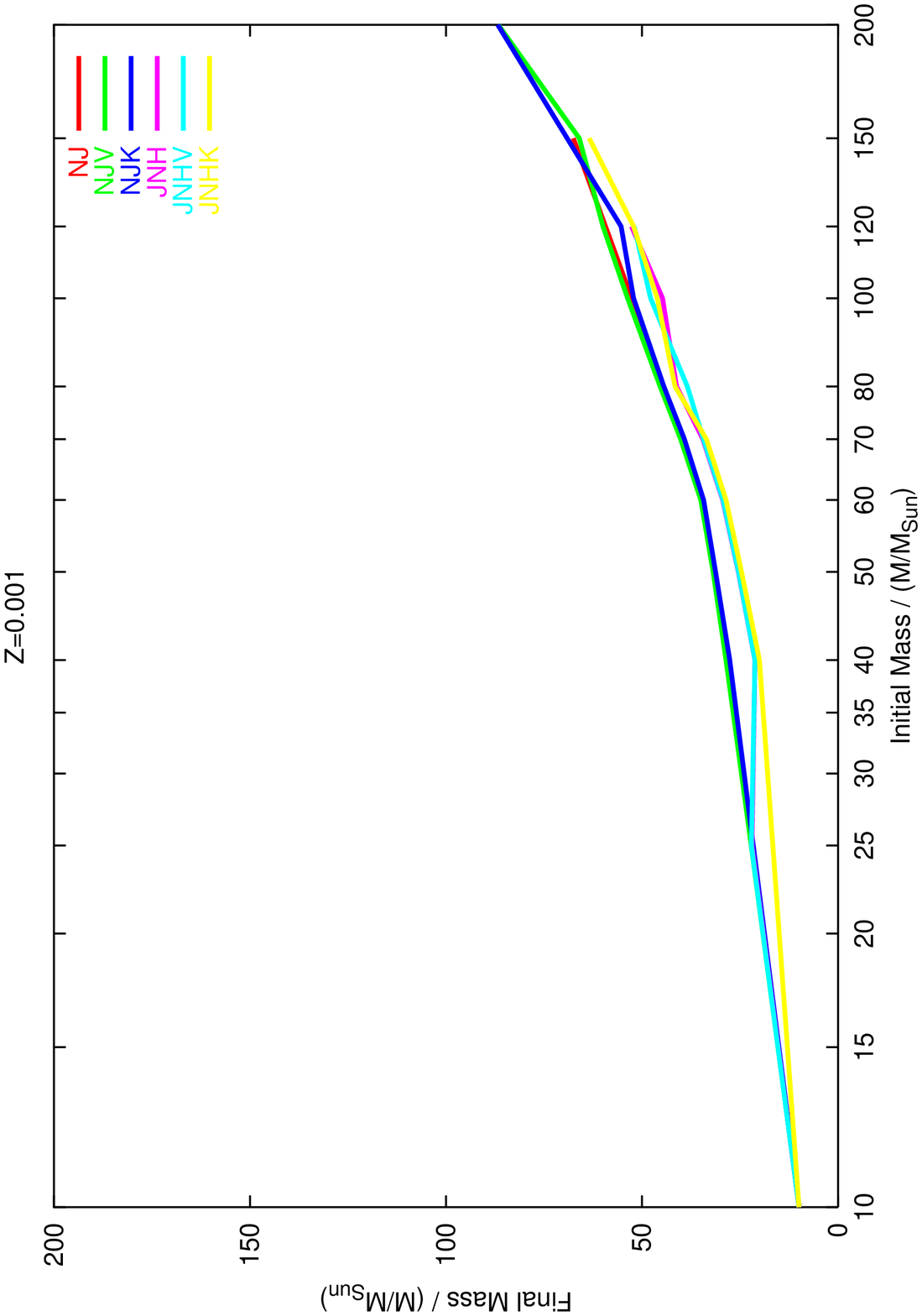}
\includegraphics[height=70mm,angle=0]{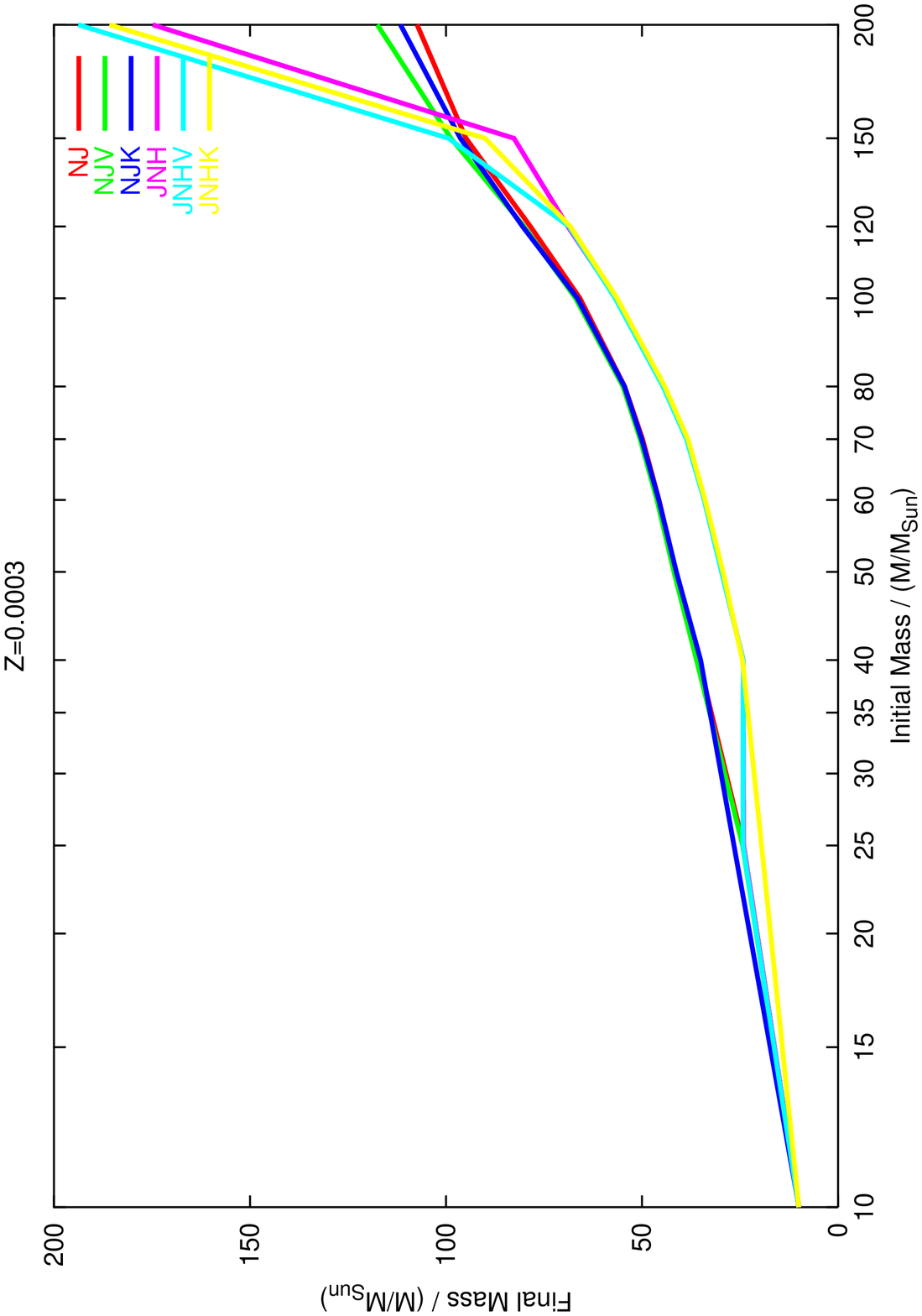}
\includegraphics[height=70mm,angle=0]{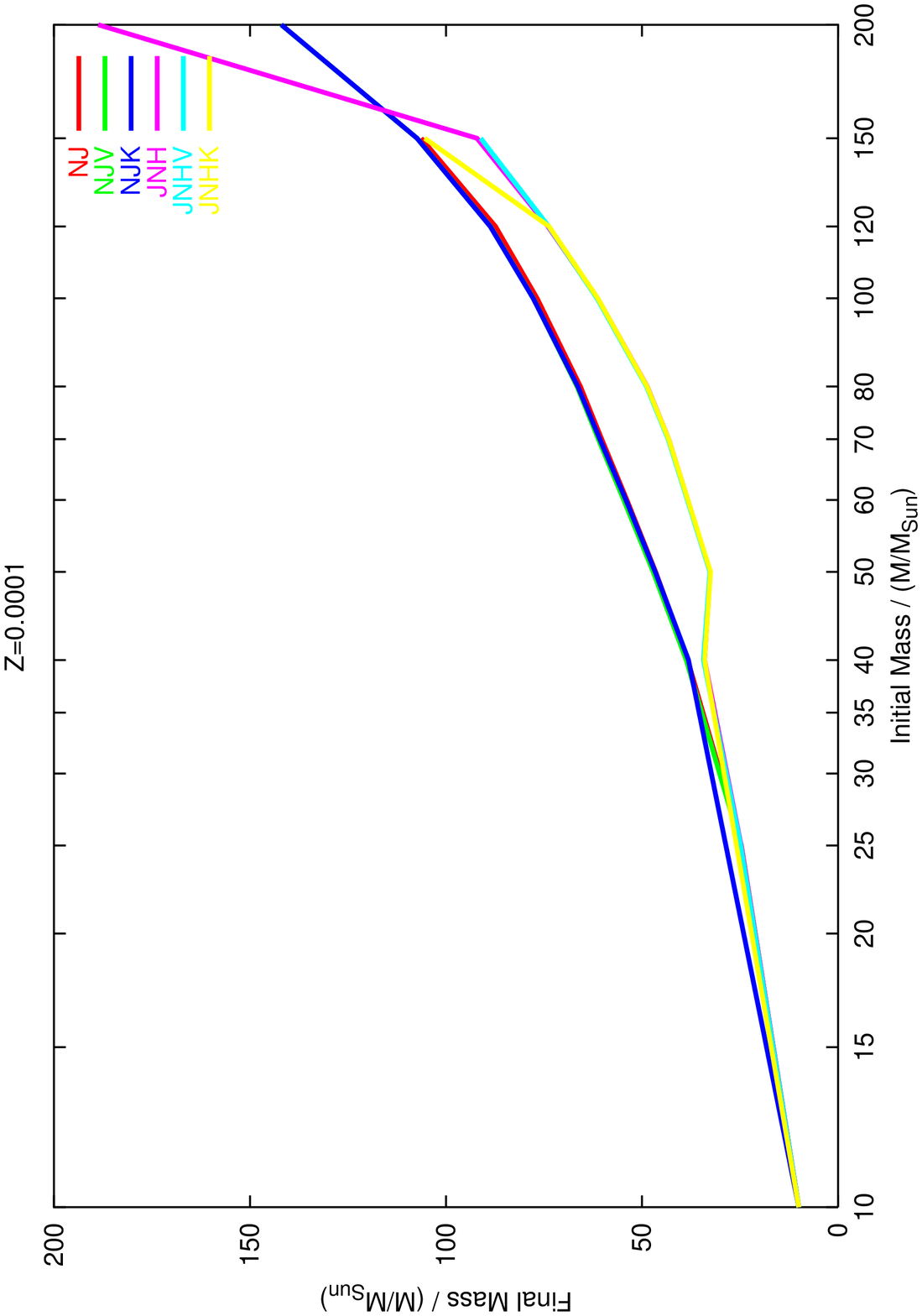}
\includegraphics[height=70mm,angle=0]{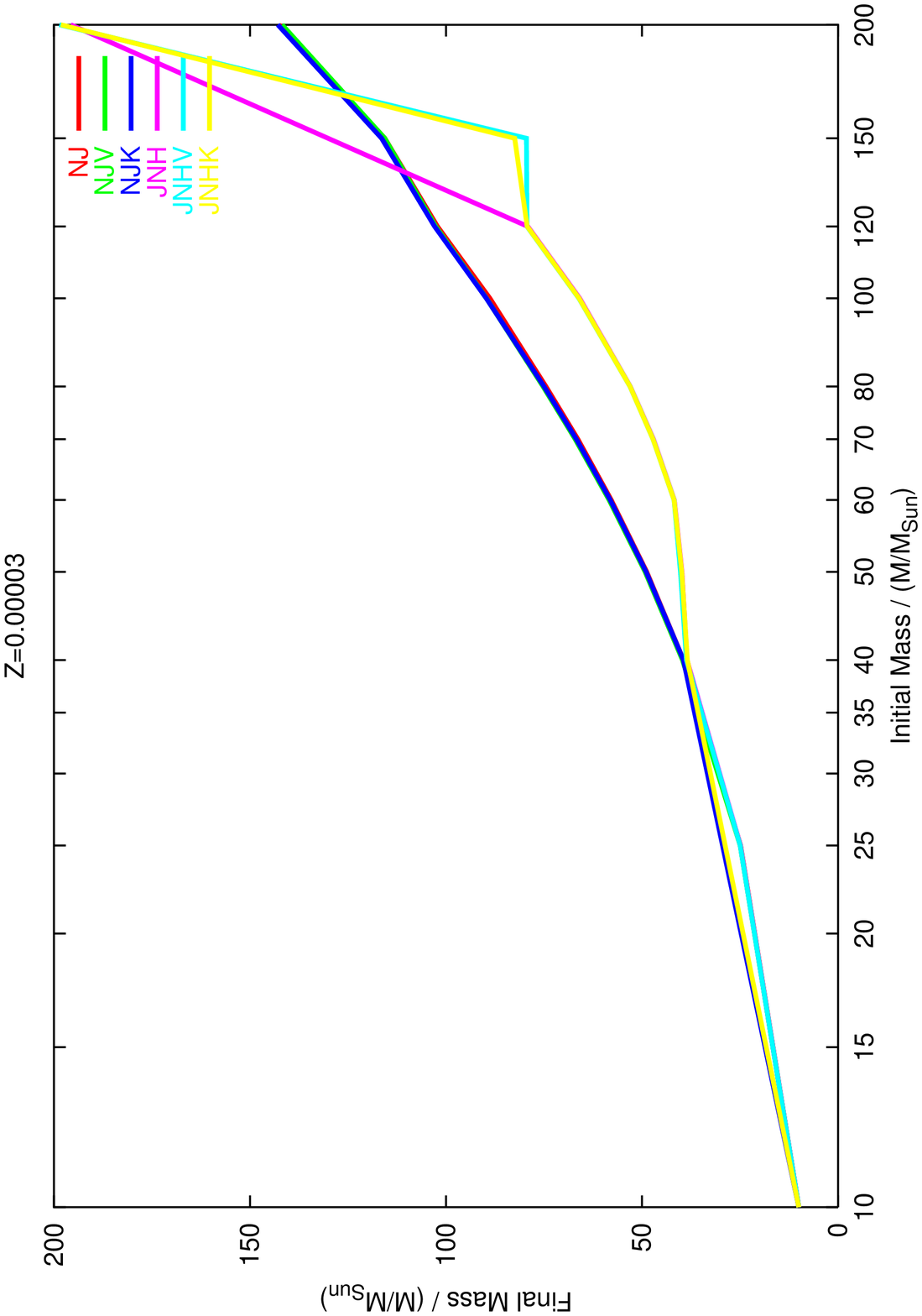}
\includegraphics[height=70mm,angle=0]{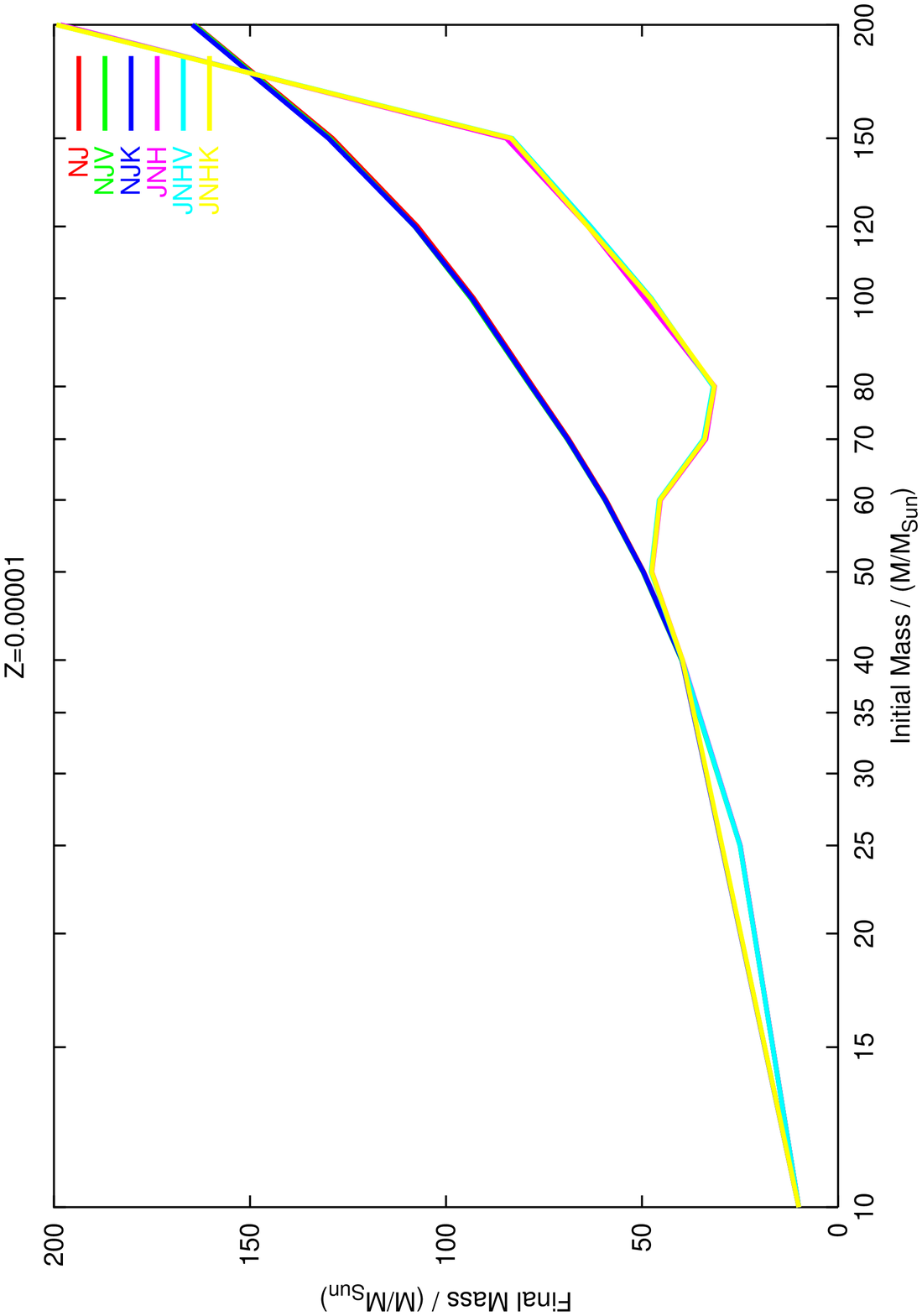}
\includegraphics[height=70mm,angle=0]{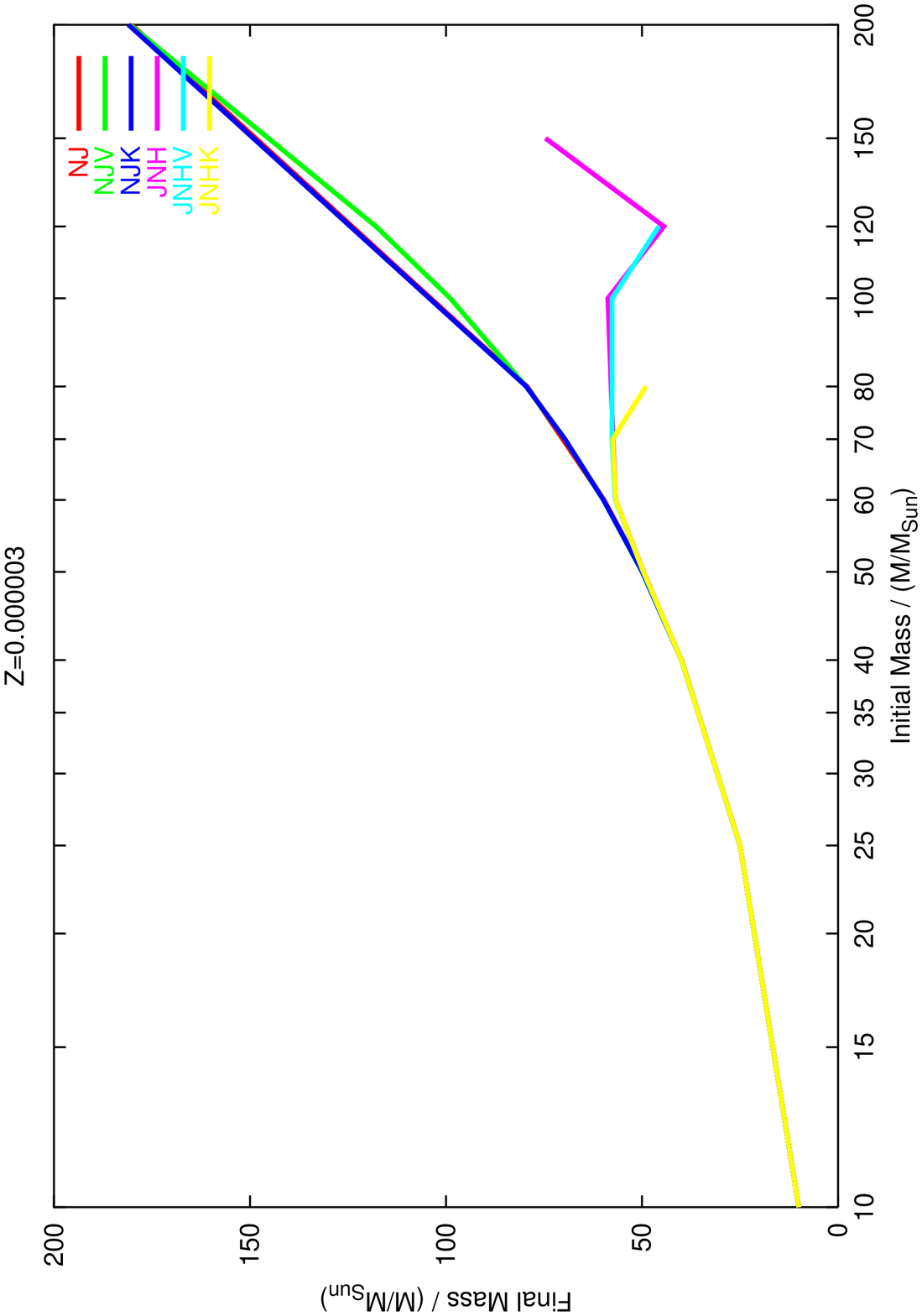}
\includegraphics[height=70mm,angle=0]{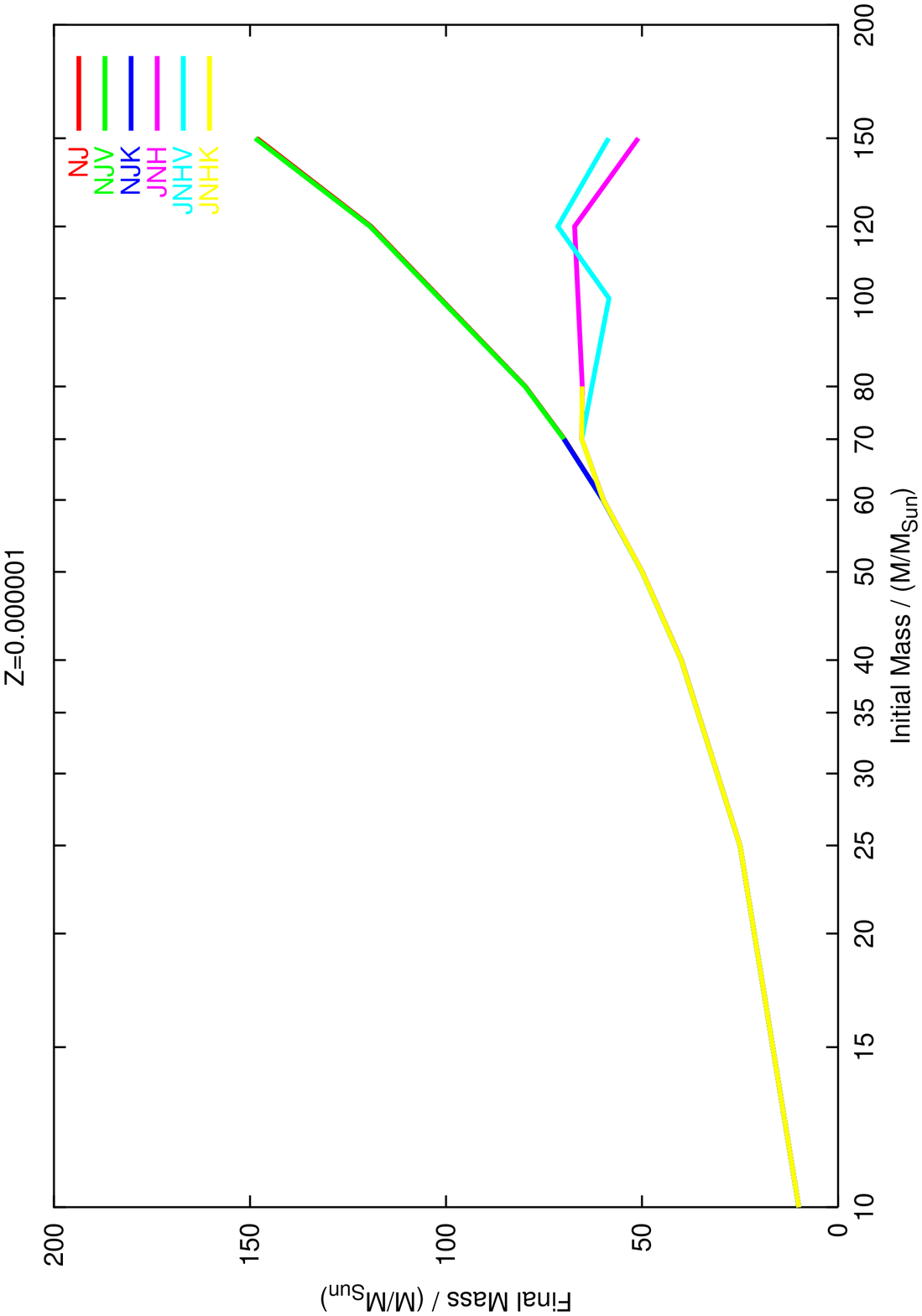}
\includegraphics[height=70mm,angle=0]{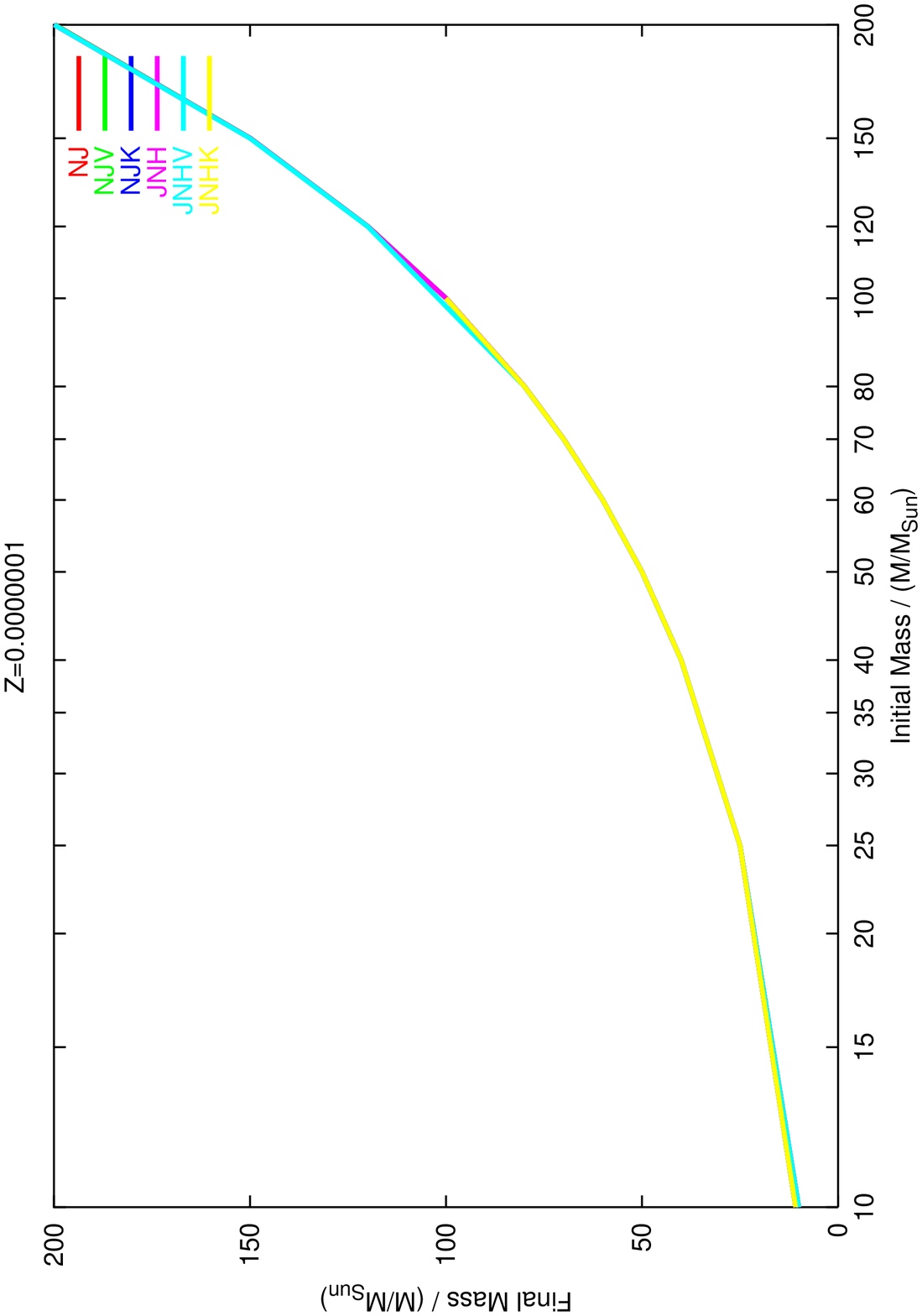}
\includegraphics[height=70mm,angle=0]{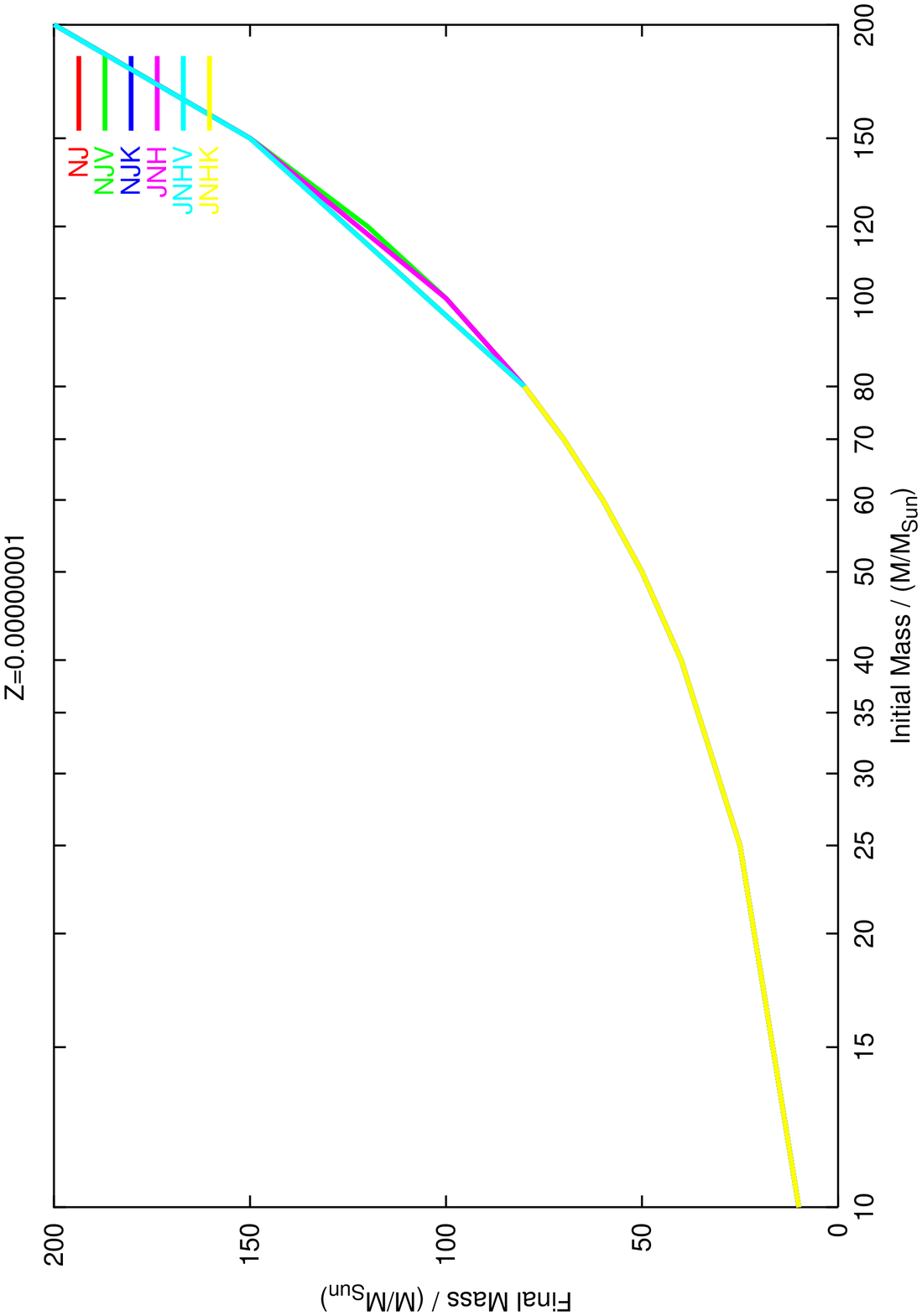}
\end{center}
\caption{The final mass versus initial mass for different mass-loss schemes at different low metallicities.}
\label{lowfinal}
\end{figure}
\begin{figure}
\begin{center}
\includegraphics[height=70mm,angle=0]{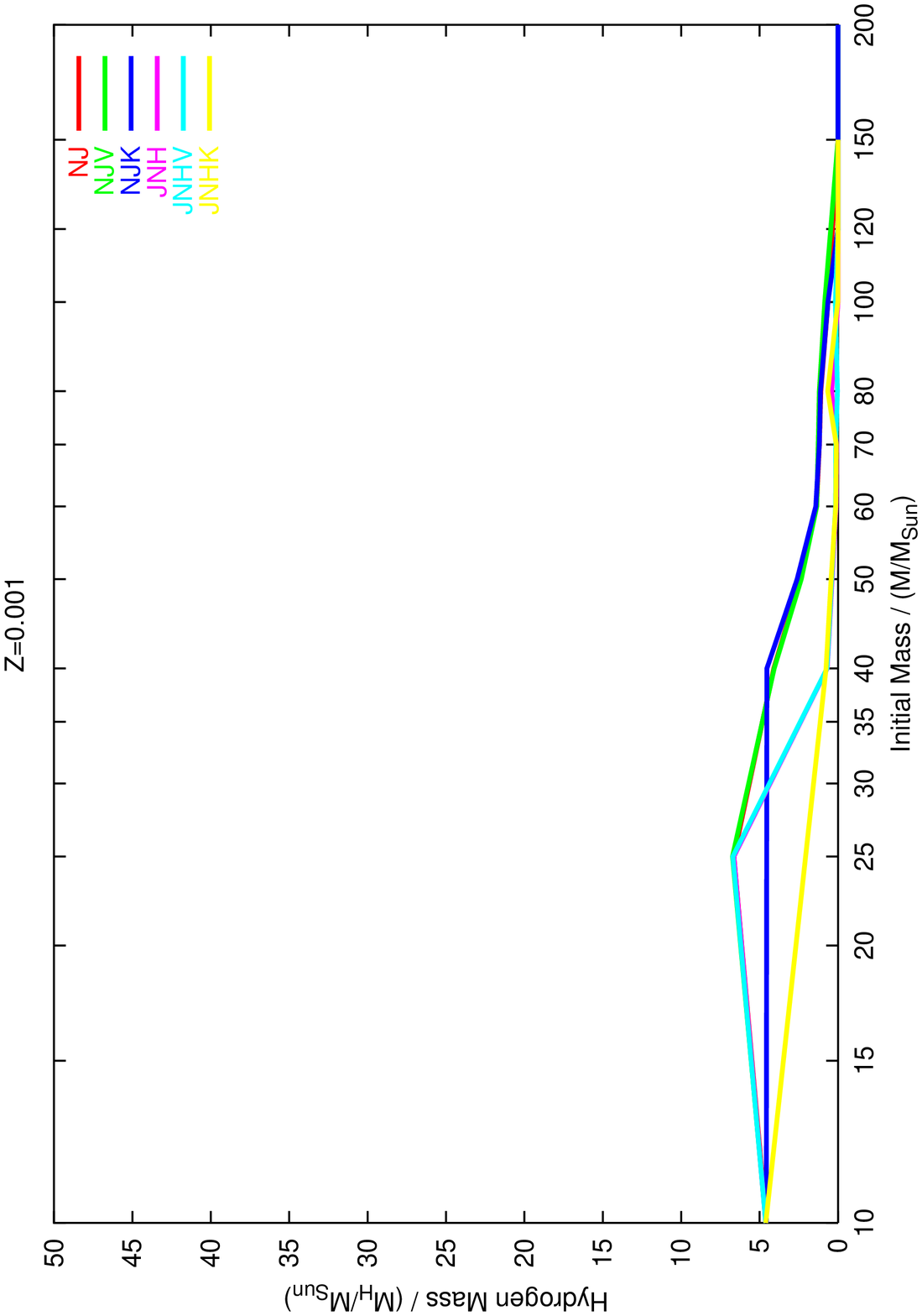}
\includegraphics[height=70mm,angle=0]{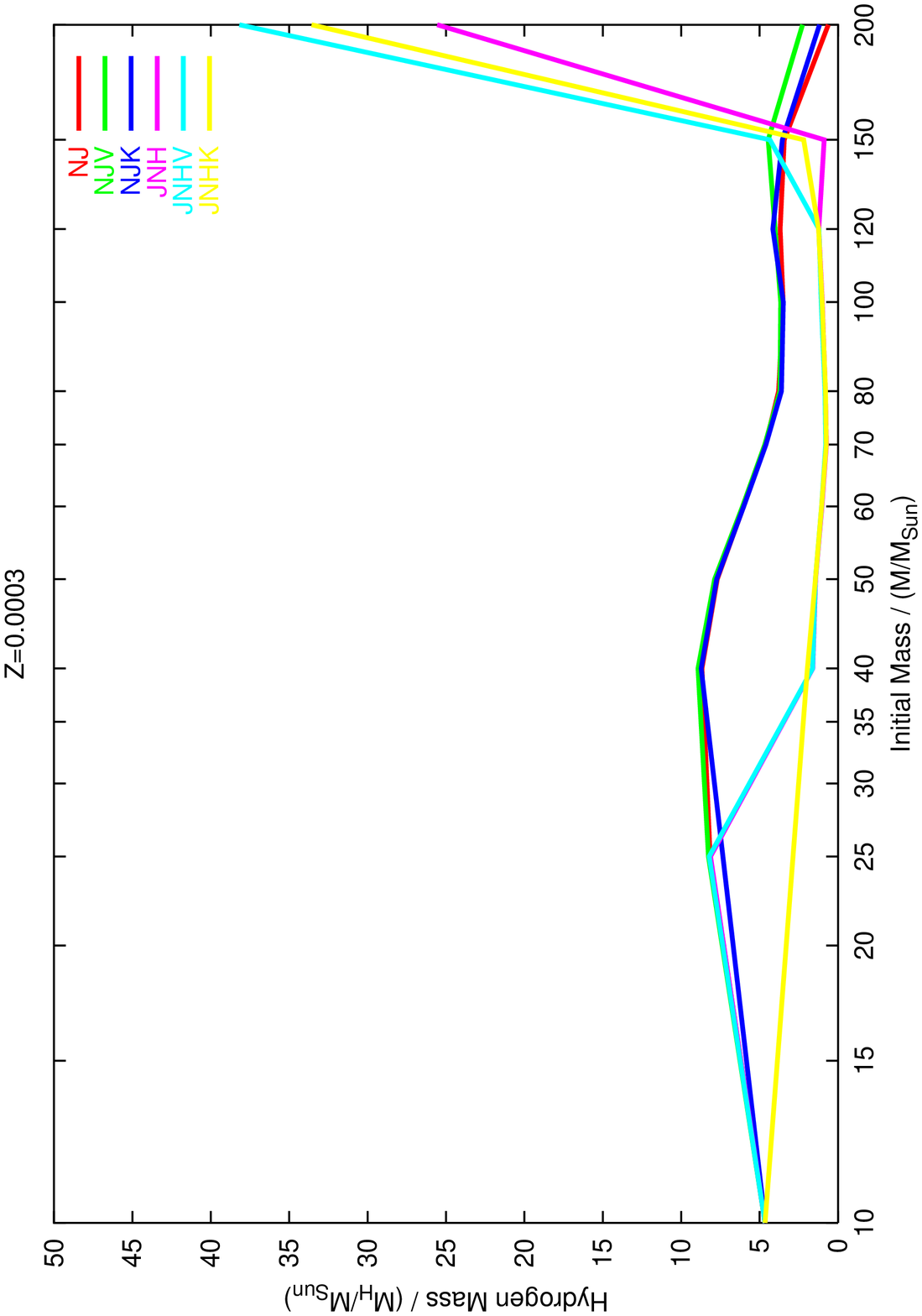}
\includegraphics[height=70mm,angle=0]{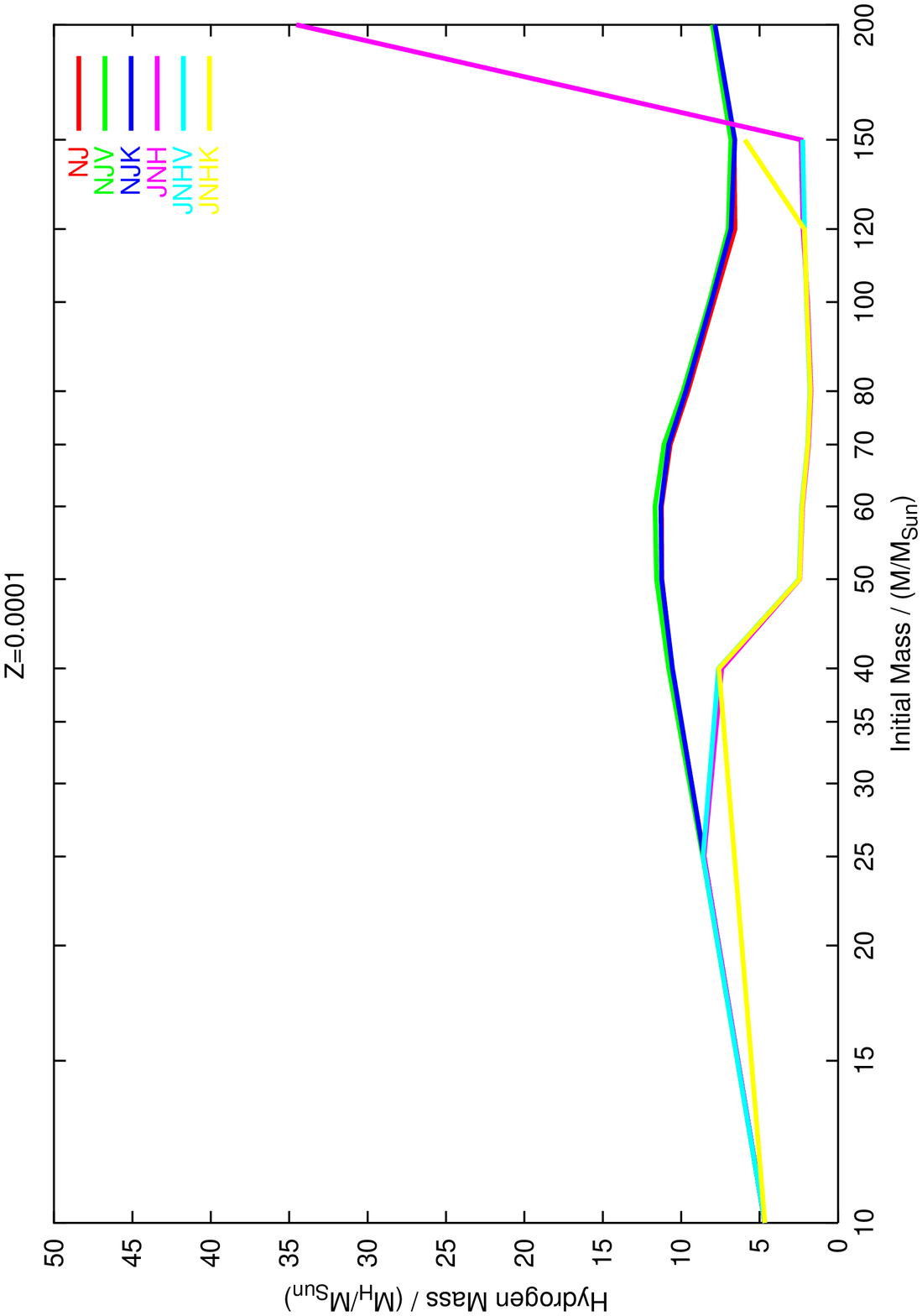}
\includegraphics[height=70mm,angle=0]{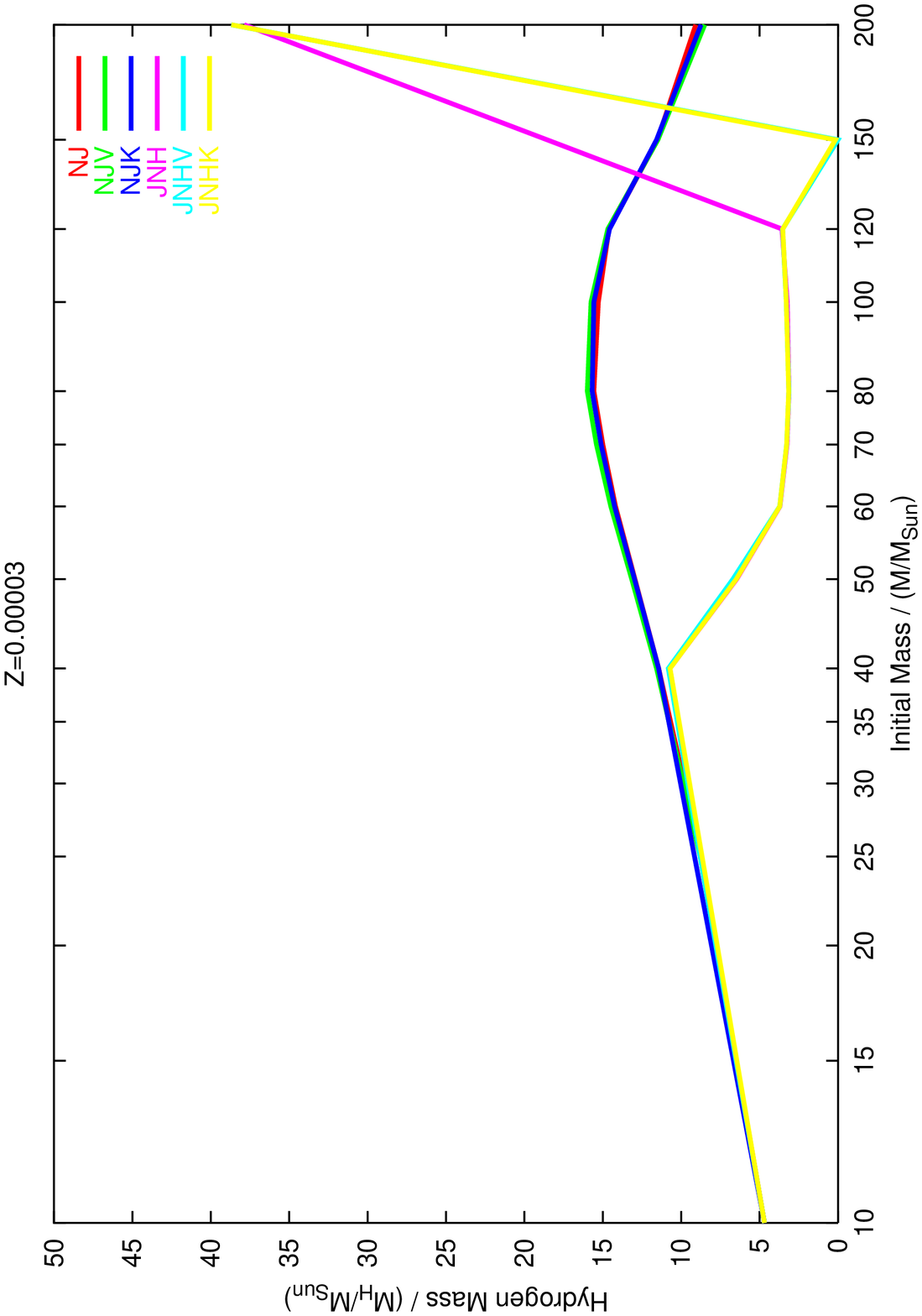}
\includegraphics[height=70mm,angle=0]{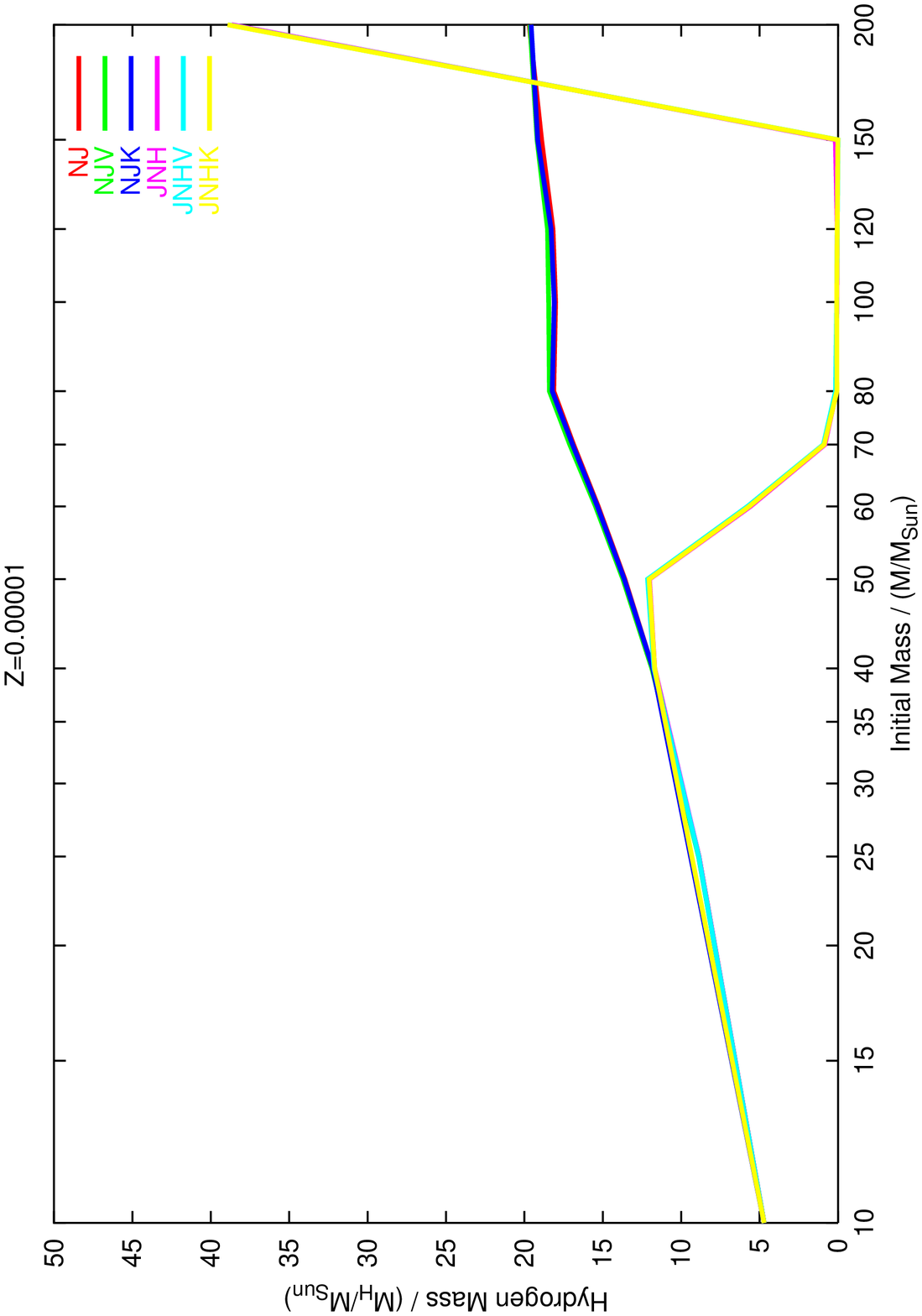}
\includegraphics[height=70mm,angle=0]{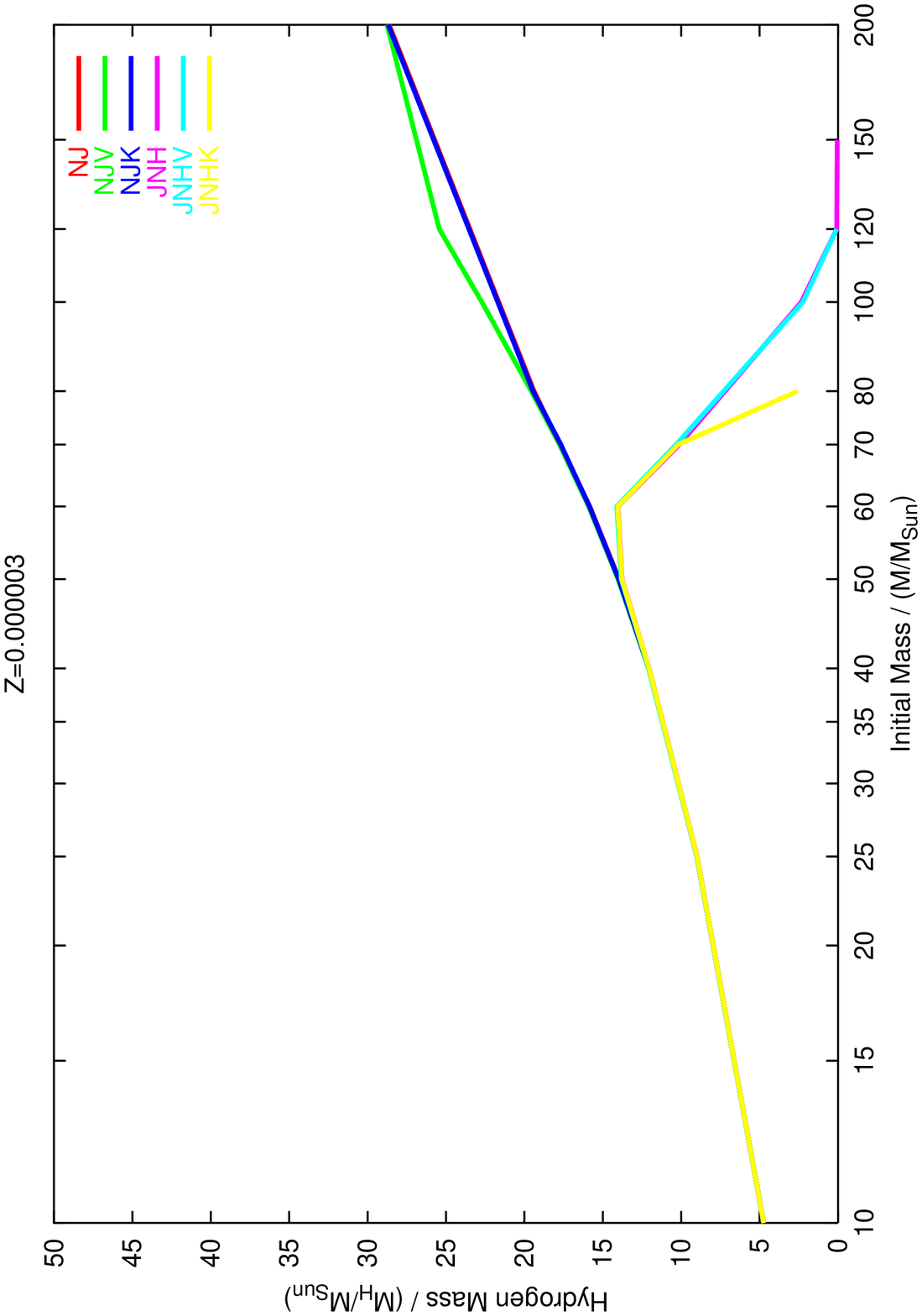}
\includegraphics[height=70mm,angle=0]{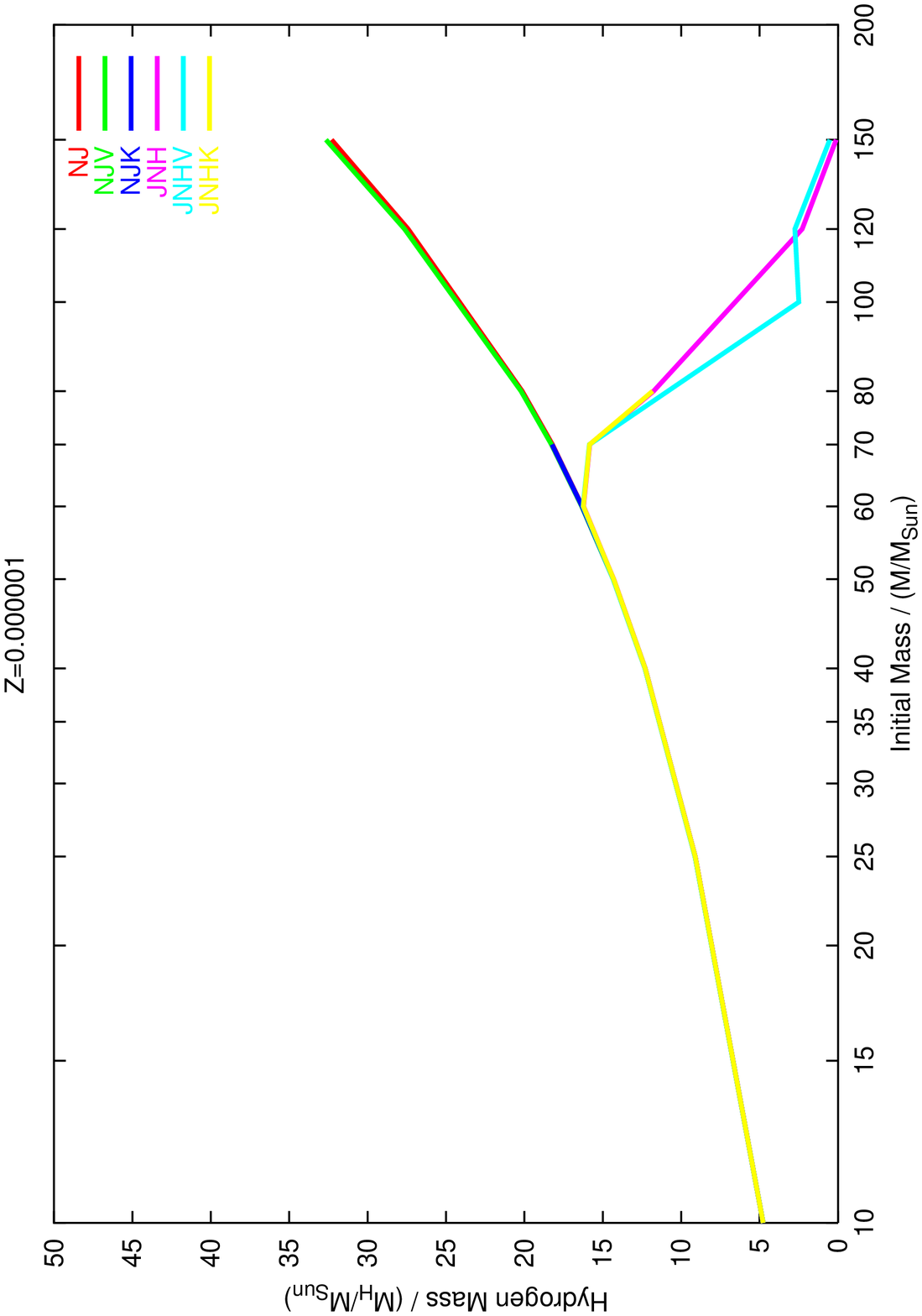}
\includegraphics[height=70mm,angle=0]{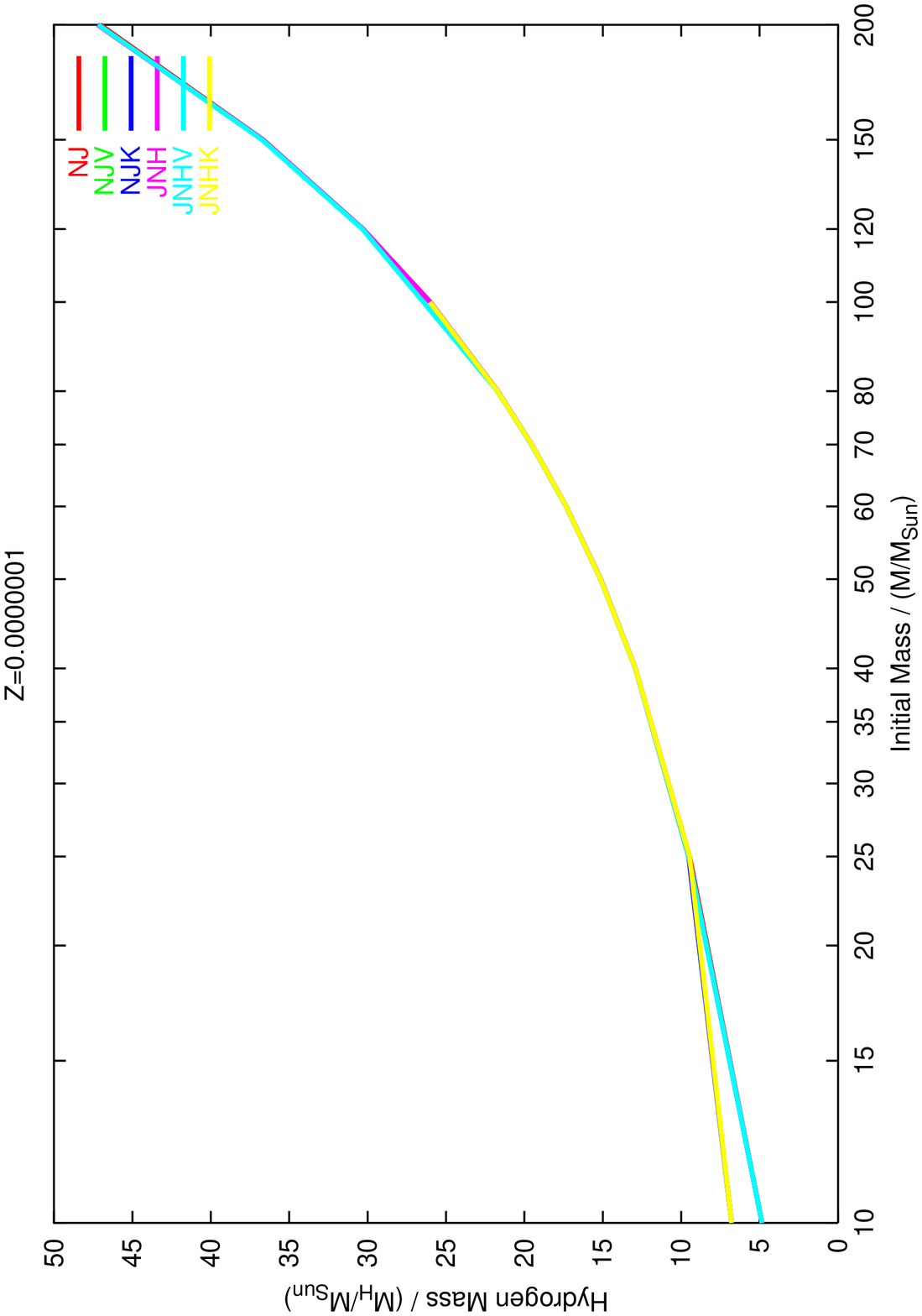}
\includegraphics[height=70mm,angle=0]{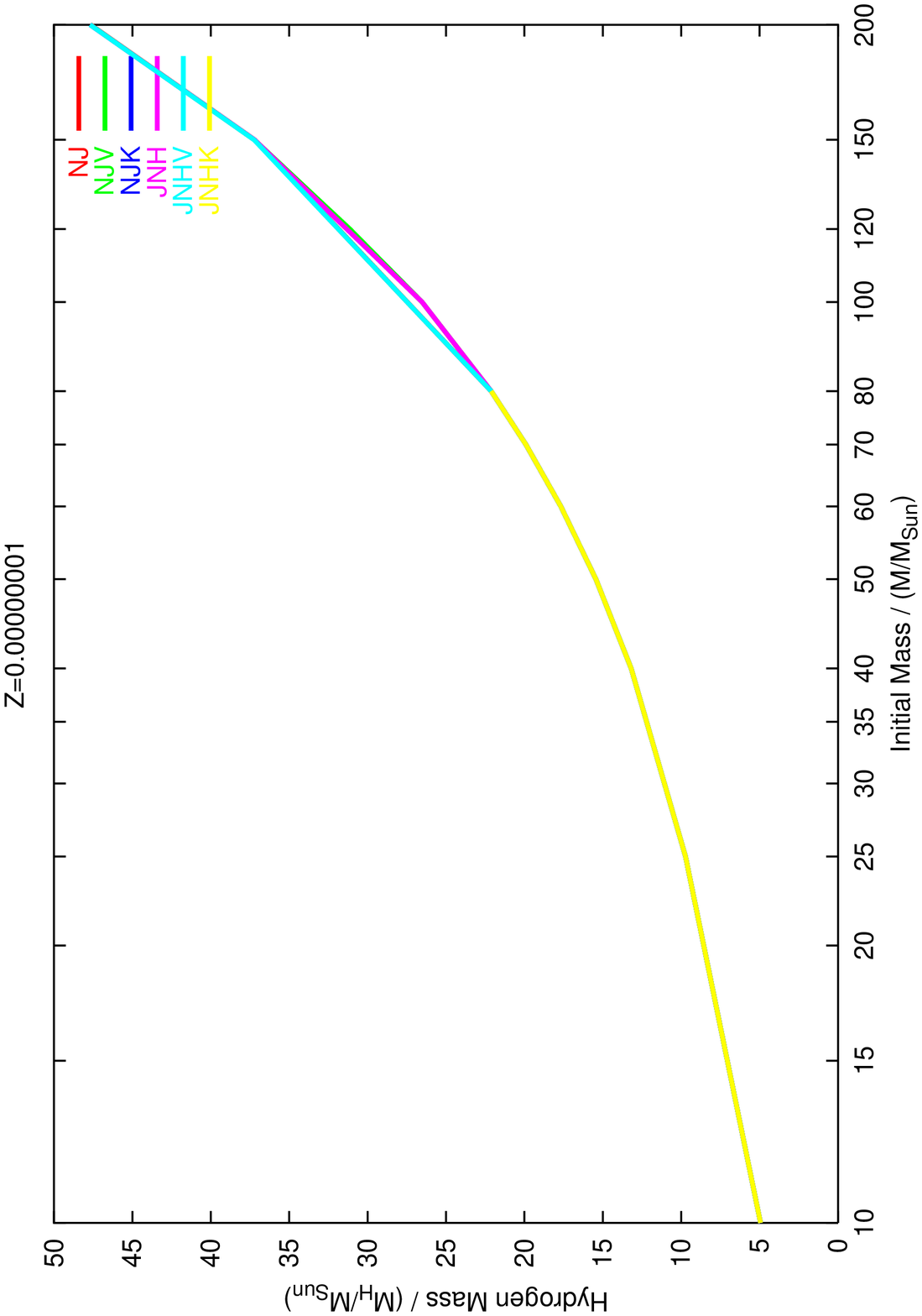}
\end{center}
\caption{The final mass of hydrogen versus initial mass for different mass-loss schemes at different low metallicities.}
\label{lowH}
\end{figure}

To compare we use the final masses and final amount of hydrogen present as in figures \ref{lowfinal} and \ref{lowH}. In the final masses we see that when $Z=0.001$ the results all agree well but then upon lowering metallicity we find they exhibit peculiar behaviour. The $200M_{\odot}$ JNH models retain most of their mass as do stars below about $70M_{\odot}$. The stars in between however lose a large proportion of their initial mass and hydrogen down to the lowest metallicities. This is because the star's position on the HR diagram is moving into either the region where the mass-loss rates are greatest or outside the valid region of the JNH rates. This means that the results are inaccurate because we are not applying the rates in the correct way. Either we could move the rates so a star starts at the same position on the rates diagram or just use the NJ rates that provide similar results for lower masses. Interestingly there are only minimal differences between the Krki and Vink rates.

Over this range of metallicity we use the NJ rates supplemented with the Krki rates. The NJ rates avoid the problems of the JNH rates and the Krki rates are derived specifically for this range of metallicity.  Unfortunately it is difficult to obtain observational constraints so we are reduced to using our preferred rates. Note however that we have shown that the differences are minor. Therefore we are now able to combine our preferred mass-loss rates from the two grids for coverage over the entire range from $Z=10^{-8}$~to~$0.05$.

\section{Our preferred mass loss prescriptions}
Combining the above rates we can produce grids that represent the evolution of single SN progenitors over a large range of metallicity. While we could plot any details of the progenitors we concentrate on a small group and briefly examine their shape on the grids. We display the final mass, luminosity, surface temperature, radius, plateau length, helium core mass, SN type, estimated ejected and remnant masses, hydrogen envelope mass, mass of hydrogen and helium and discuss each result in detail and compare the overshooting and no overshooting cases. After presenting the results of this test we describe some tests of our models from SN type ratios, SN progenitors and WR ratios. In appendix \ref{moreofthebest} we show more details of these models and in appendix \ref{moreoftheother} we show the same graphs for the NJV scheme at high metallicity, the second scheme on our list.

Our maps tend to broadly agree with those of \citet{H03}. Owing to our specific coverage in metallicity, we obtain more structure within the map. All of the structure is due to the mass-loss prescriptions we have used in producing the models. It is important to note however that while the broad nature of these maps are correct some of the fine structure would in practice be blurred out by different helium content, rotation, magnetic fields and other mixing processes. We assess our maps for each variable individually.

\begin{figure}
\begin{center}
\includegraphics[height=79mm,angle=0]{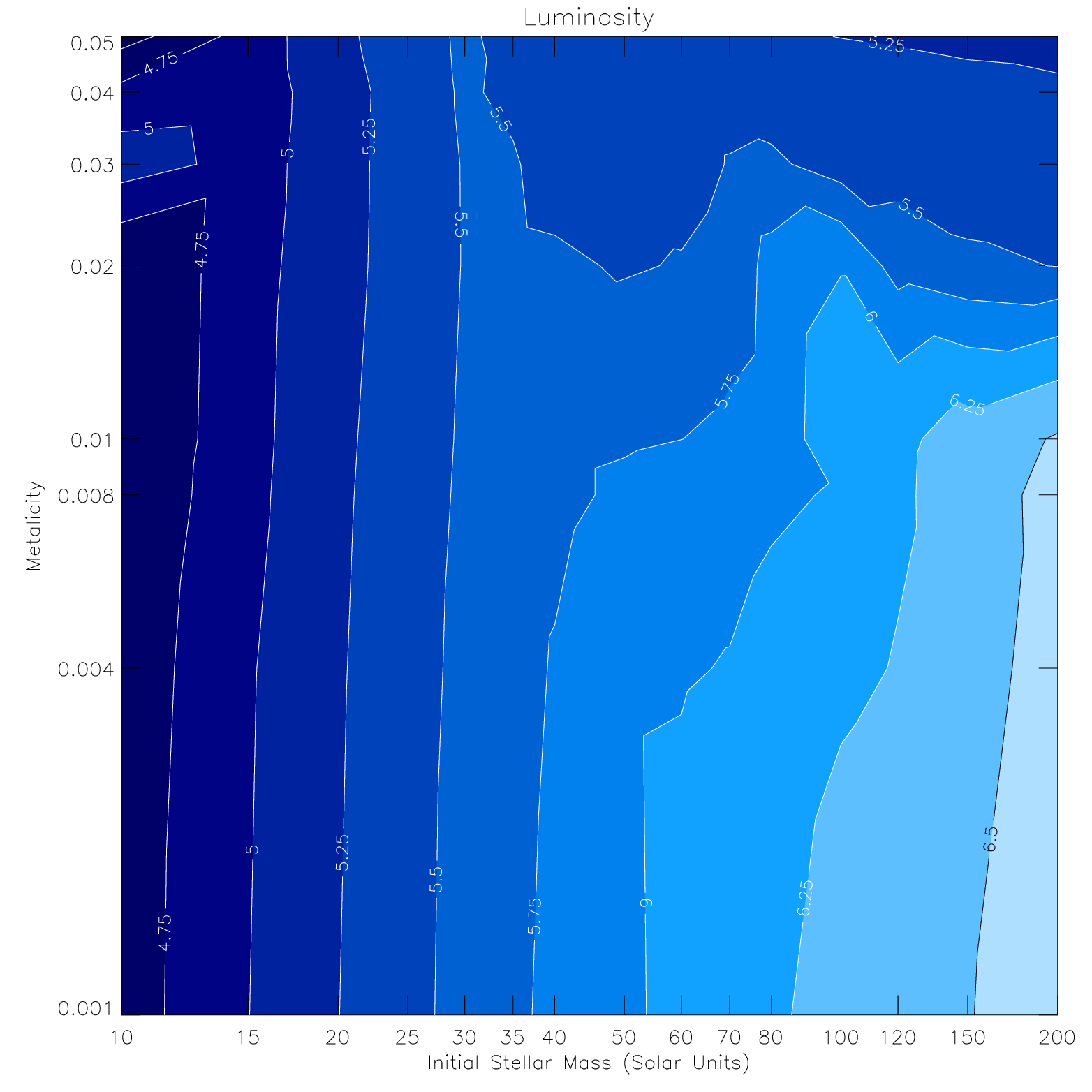}
\includegraphics[height=79mm,angle=0]{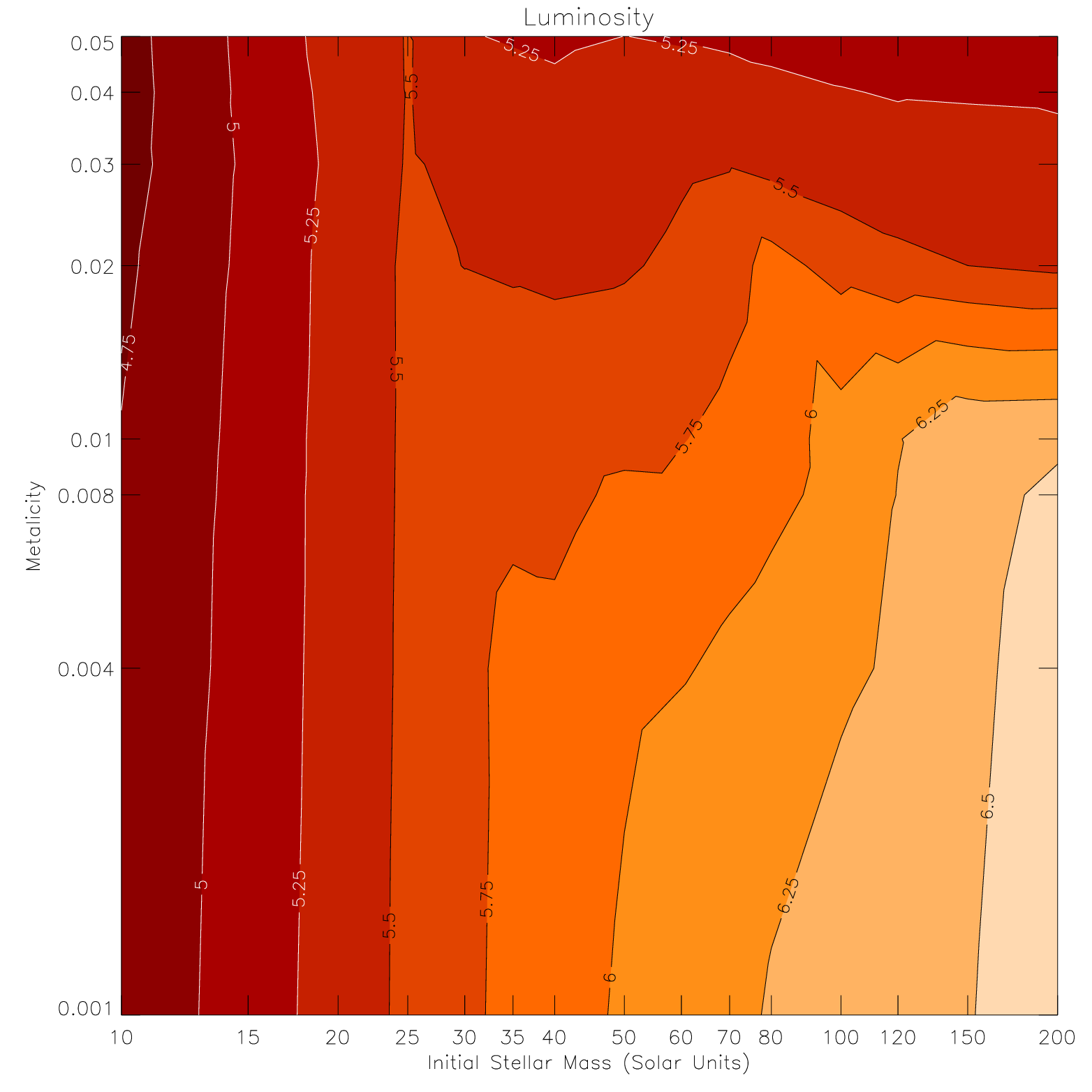}
\includegraphics[height=79mm,angle=0]{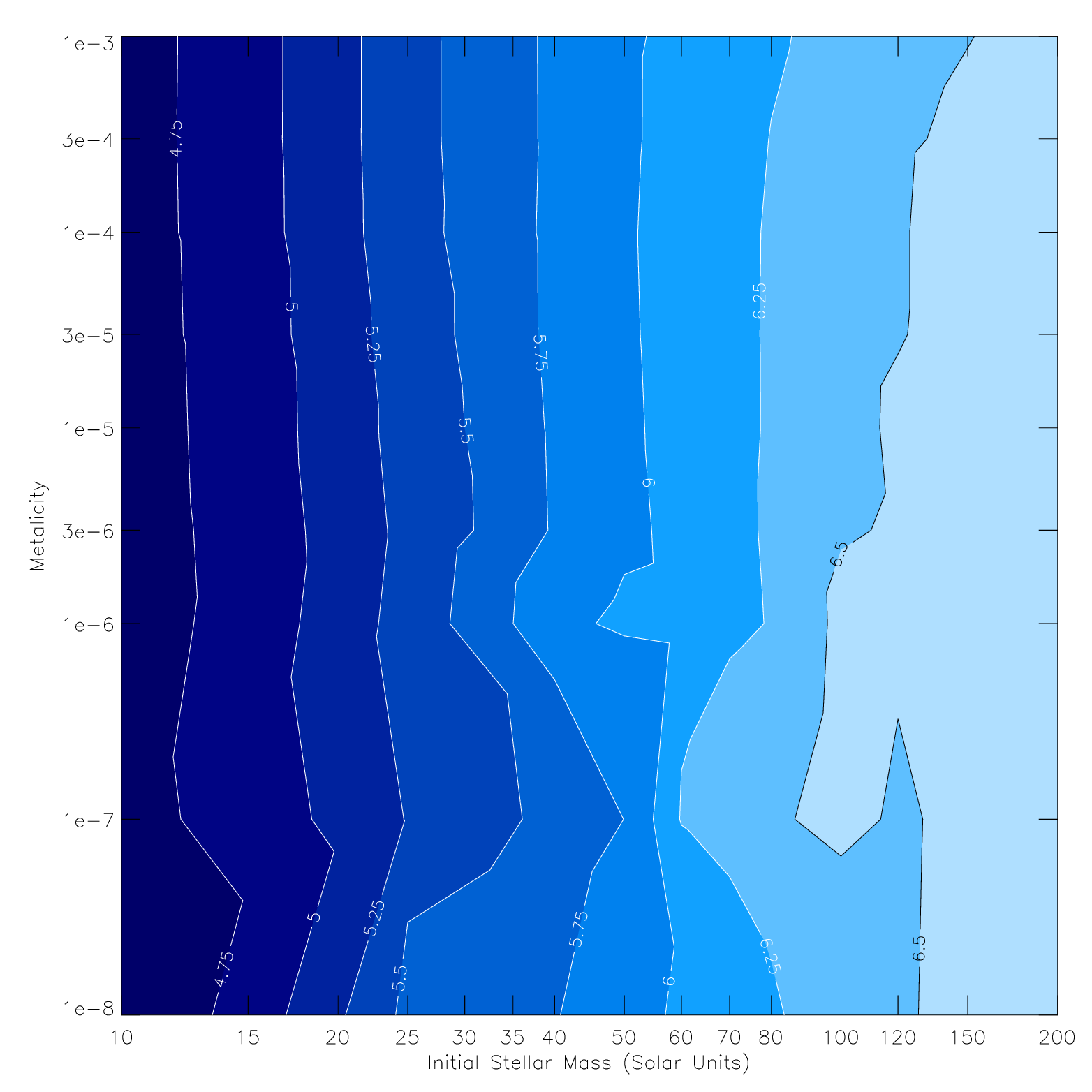}
\includegraphics[height=79mm,angle=0]{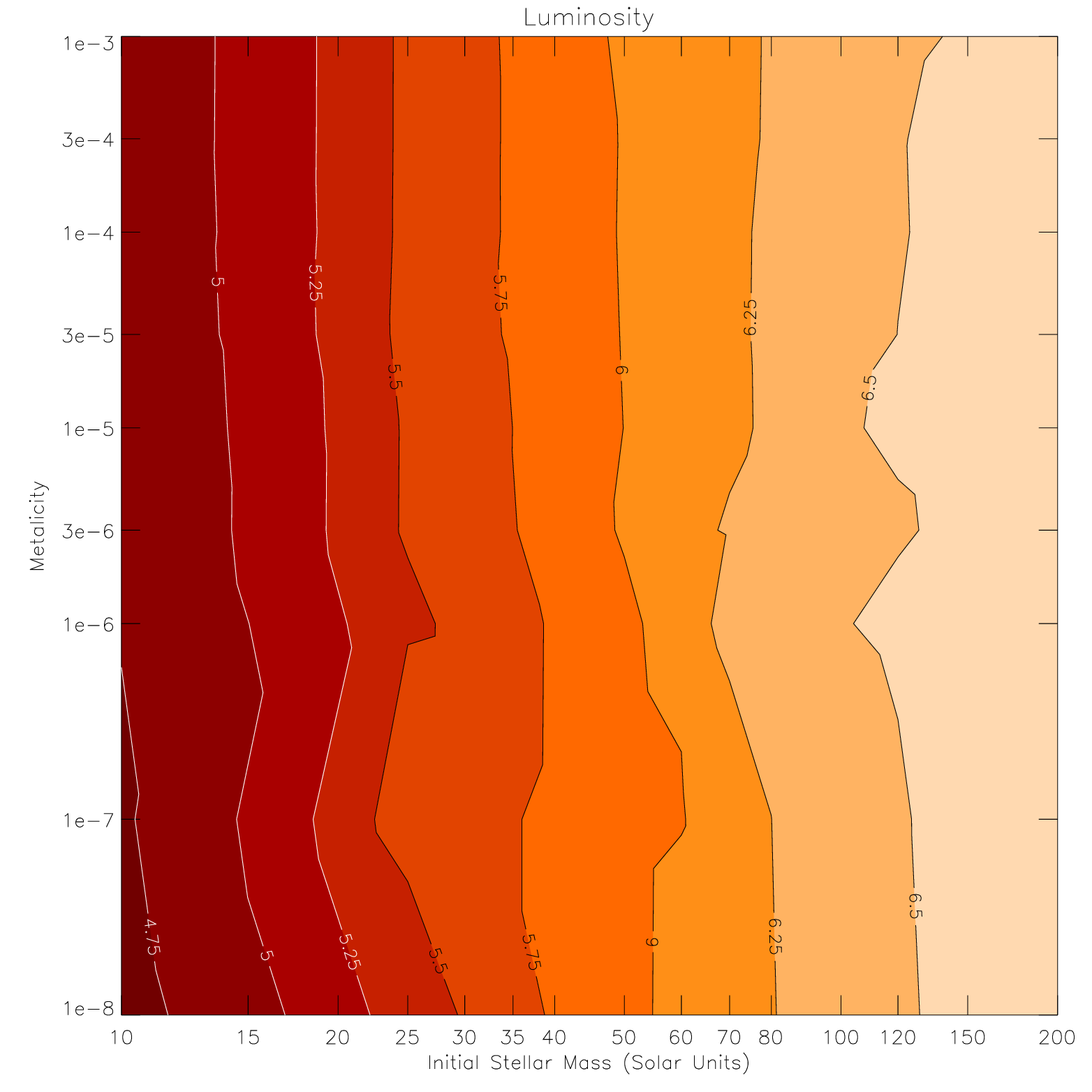}
\caption[Contours of pre-SN luminosity.]{Contours of pre-SN luminosity. Contours are in $\log (L/L_{\odot})$. Left no overshooting, right overshooting.}
\label{mapA}
\end{center}
\end{figure}

\subsection{Luminosity, Figure \ref{mapA}}
In the no overshooting diagram we can see in the upper left corner stars that undergo 2nd dredge-up. For models at lower metallicity and for models with overshooting we do not go to low enough masses to resolve this region. We will cover this in more detail later. The main feature in the diagram is the structure in the upper right corner at high metallicity. As mass increases we see the last few red giant progenitors are quite bright but are then replaced by WR stars and the pre-SN luminosity drops abruptly. The luminosity increases again as higher mass WR stars form. At the highest initial masses these WR stars once more become less massive and therefore less luminous. This structure becomes less pronounced at lower metallicities owing to the lower mass-loss rates. At low metallicity there is limited mass-loss and the main structure at the lowest metallicities is due to changes in the nature of burning reactions at the lowest temperatures.

\begin{figure}
\begin{center}
\includegraphics[height=79mm,angle=0]{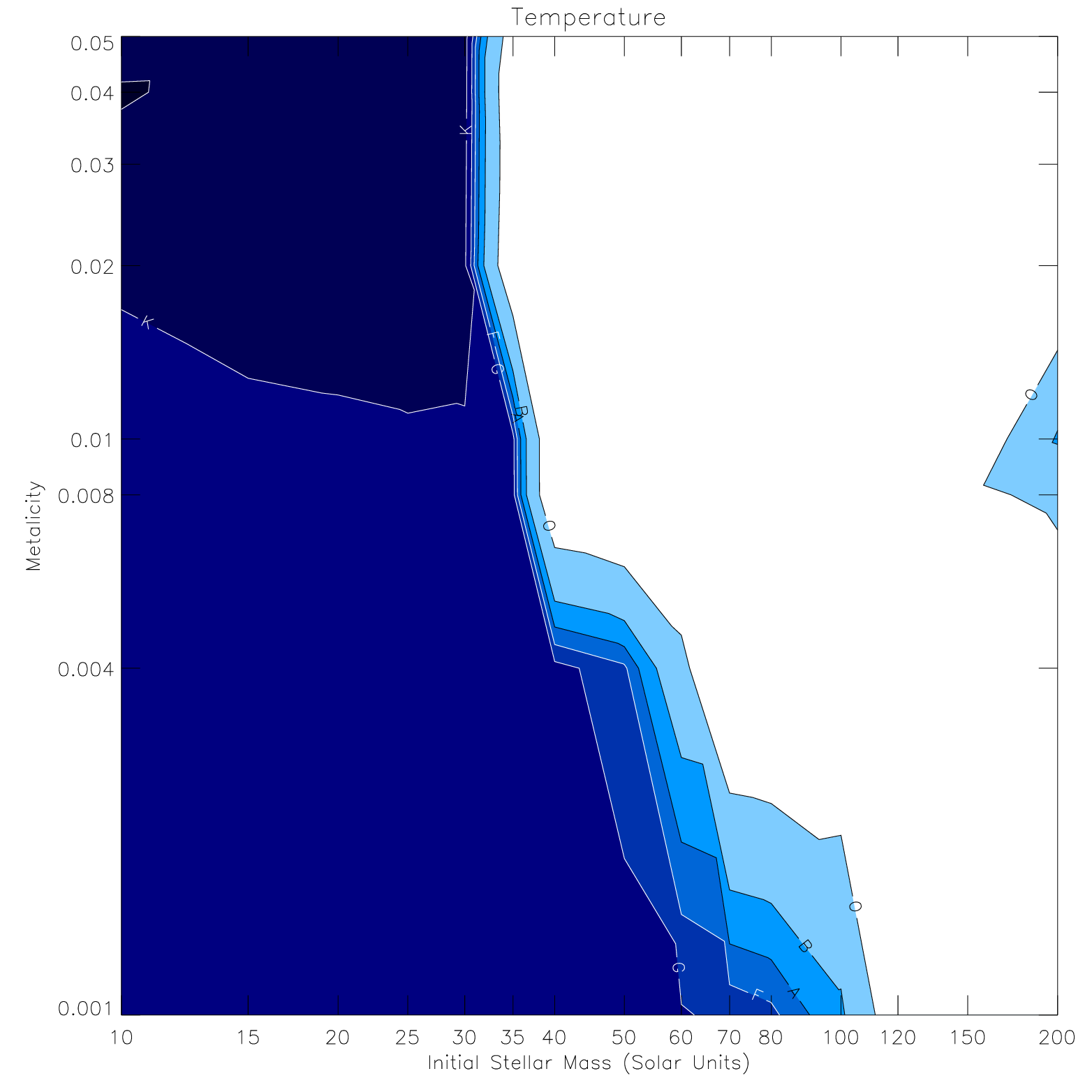}
\includegraphics[height=79mm,angle=0]{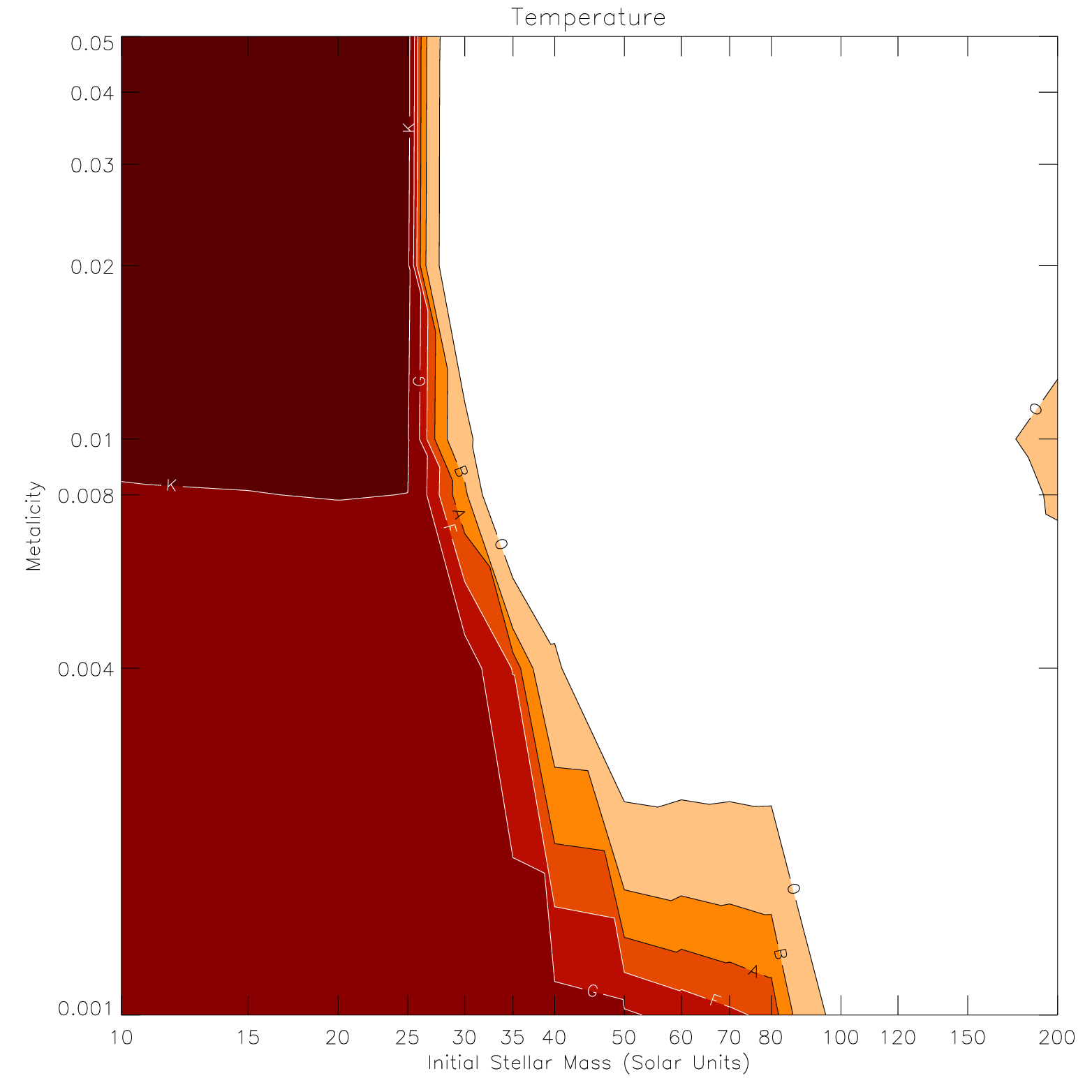}
\includegraphics[height=79mm,angle=0]{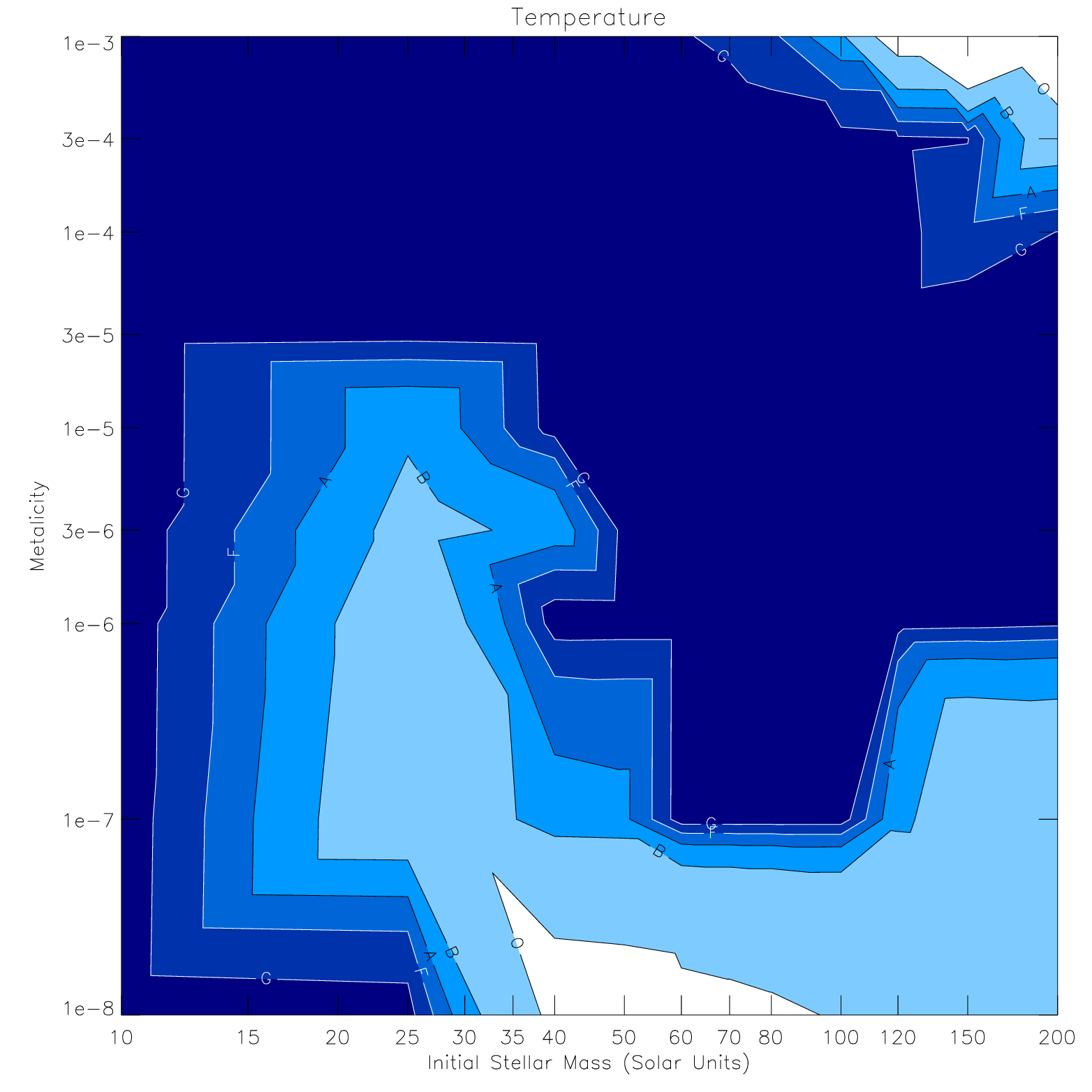}
\includegraphics[height=79mm,angle=0]{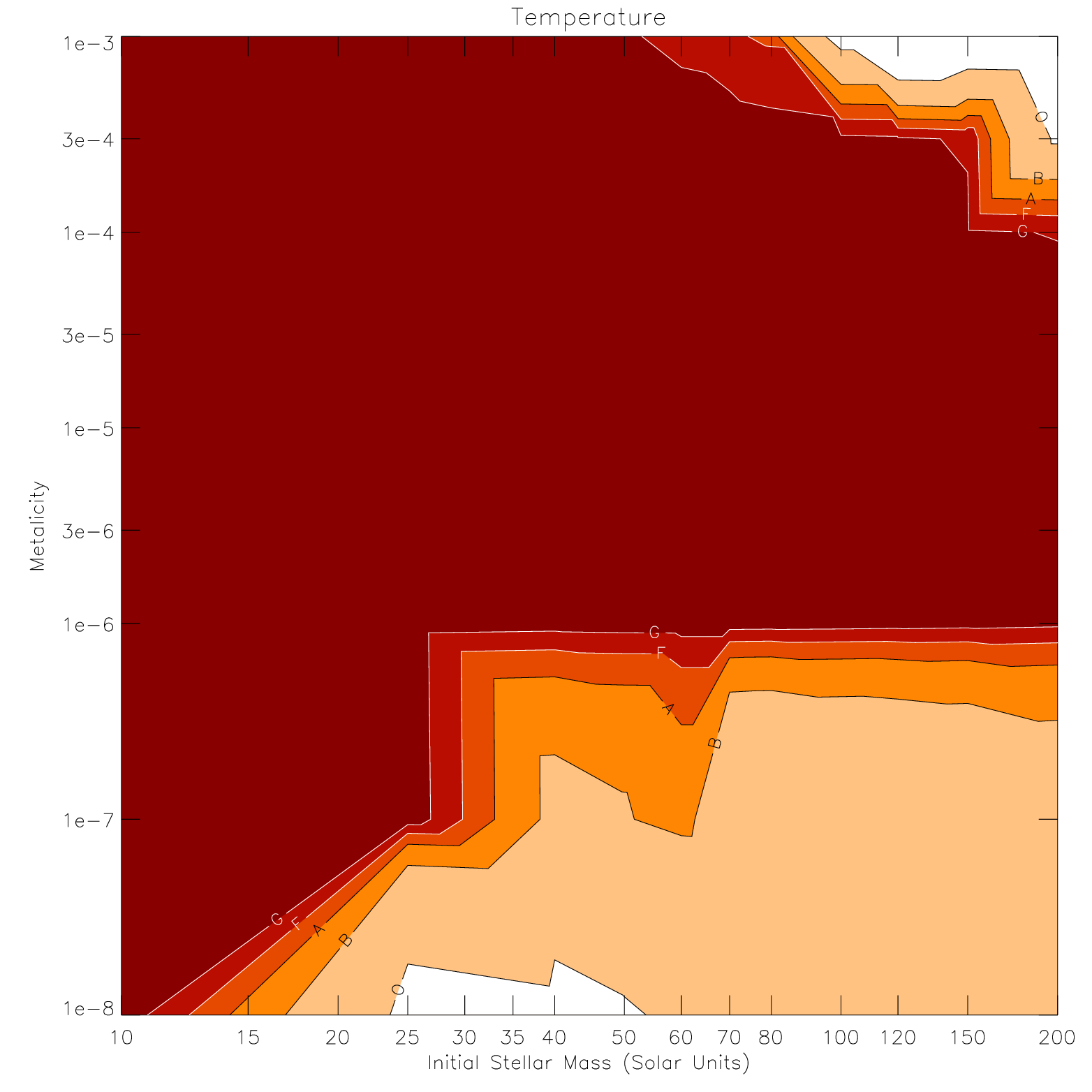}
\caption[Contours of pre-SN spectral type.]{Contours of pre-SN spectral type. Left no overshooting, right overshooting. The effective temperatures taken for each spectral type are as follows, $T_{\rm eff}({\rm O})> 27000{\rm K}$, $T_{\rm eff}({\rm B})> 11000{\rm K}$, $T_{\rm eff}({\rm A})> 7600{\rm K}$, $T_{\rm eff}({\rm F})> 6000{\rm K}$, $T_{\rm eff}({\rm G})> 4900{\rm K}$, $T_{\rm eff}({\rm K})> 3600{\rm K}$ and $T_{\rm eff}({\rm M})< 3600{\rm K}$.}
\label{mapB}
\end{center}
\end{figure}

\subsection{Spectral type, Figure \ref{mapB}}
High metallicity results show a strong difference between giant stars and the hotter WR progenitors. As metallicity decreases this change over becomes more gradual. Towards the middle right section there are slightly cool WR stars. These are those WR stars for which to evolve the model, we have had to adopt a limited constant mass-loss rate.

Low metallicity has the remnants of the previous behaviour in the upper right corner however there is an interesting change at the lowest metallicity with the slightly hotter progenitors. This is because the opacity is low enough to prevent the formation of red giants.

\begin{figure}
\begin{center}
\includegraphics[height=79mm,angle=0]{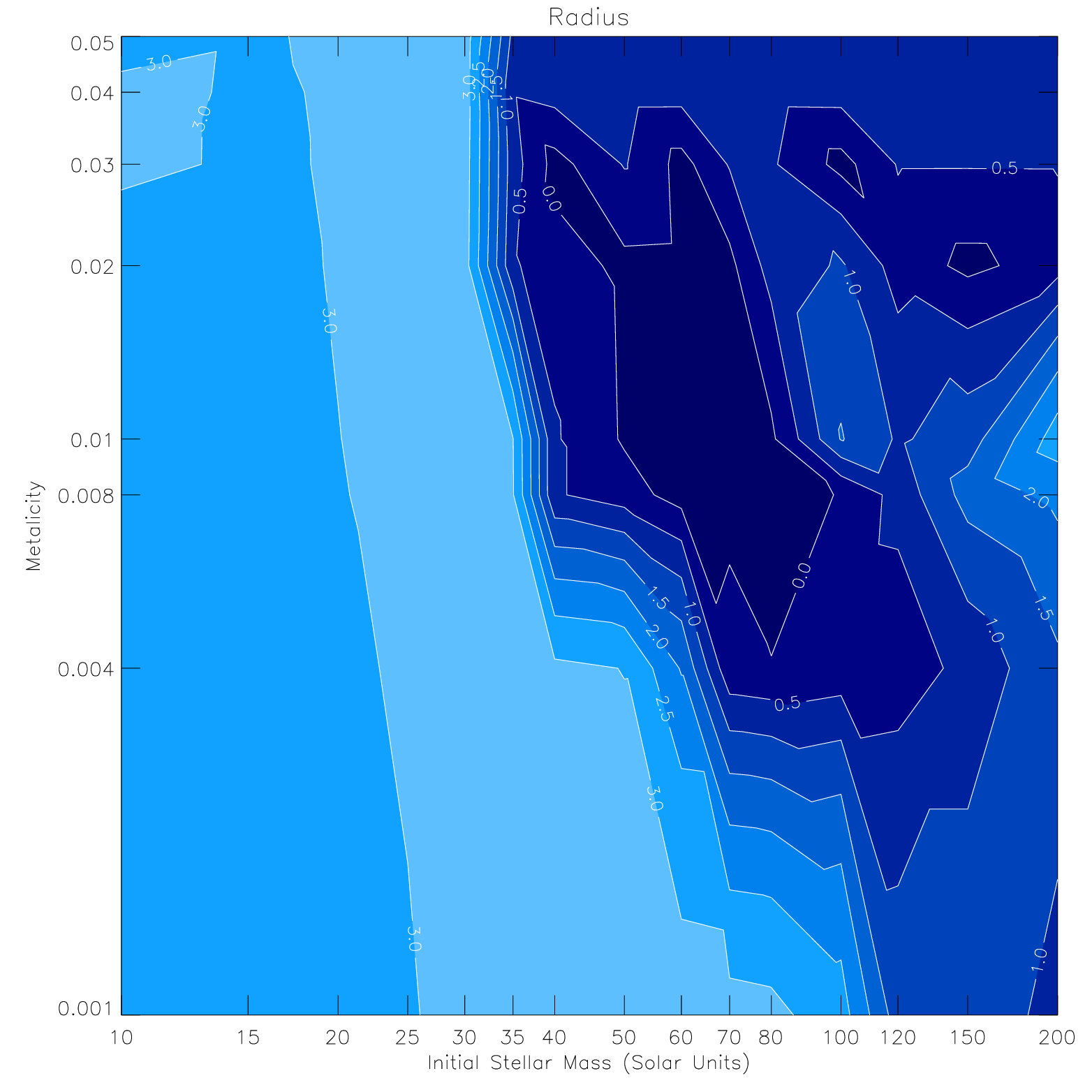}
\includegraphics[height=79mm,angle=0]{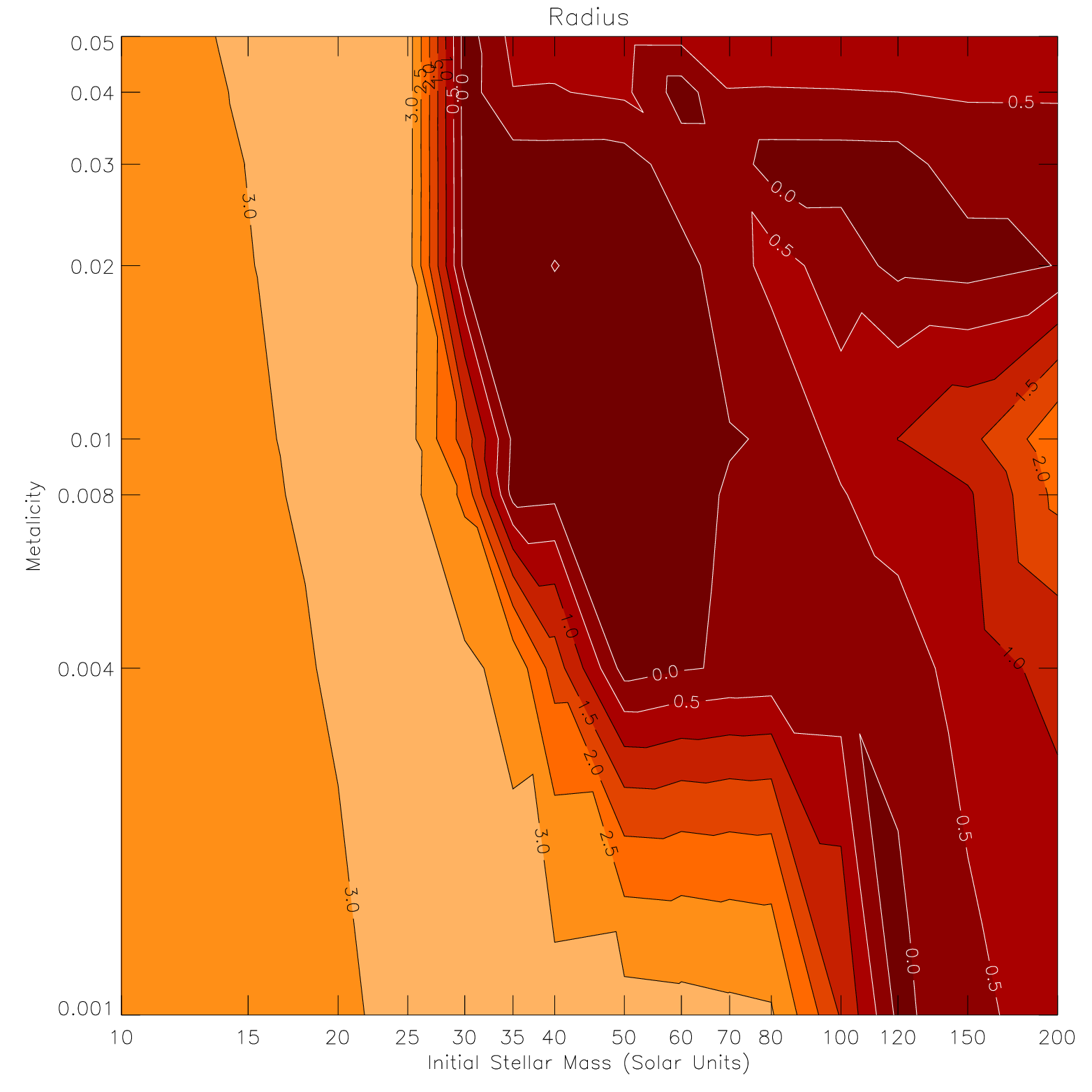}
\includegraphics[height=79mm,angle=0]{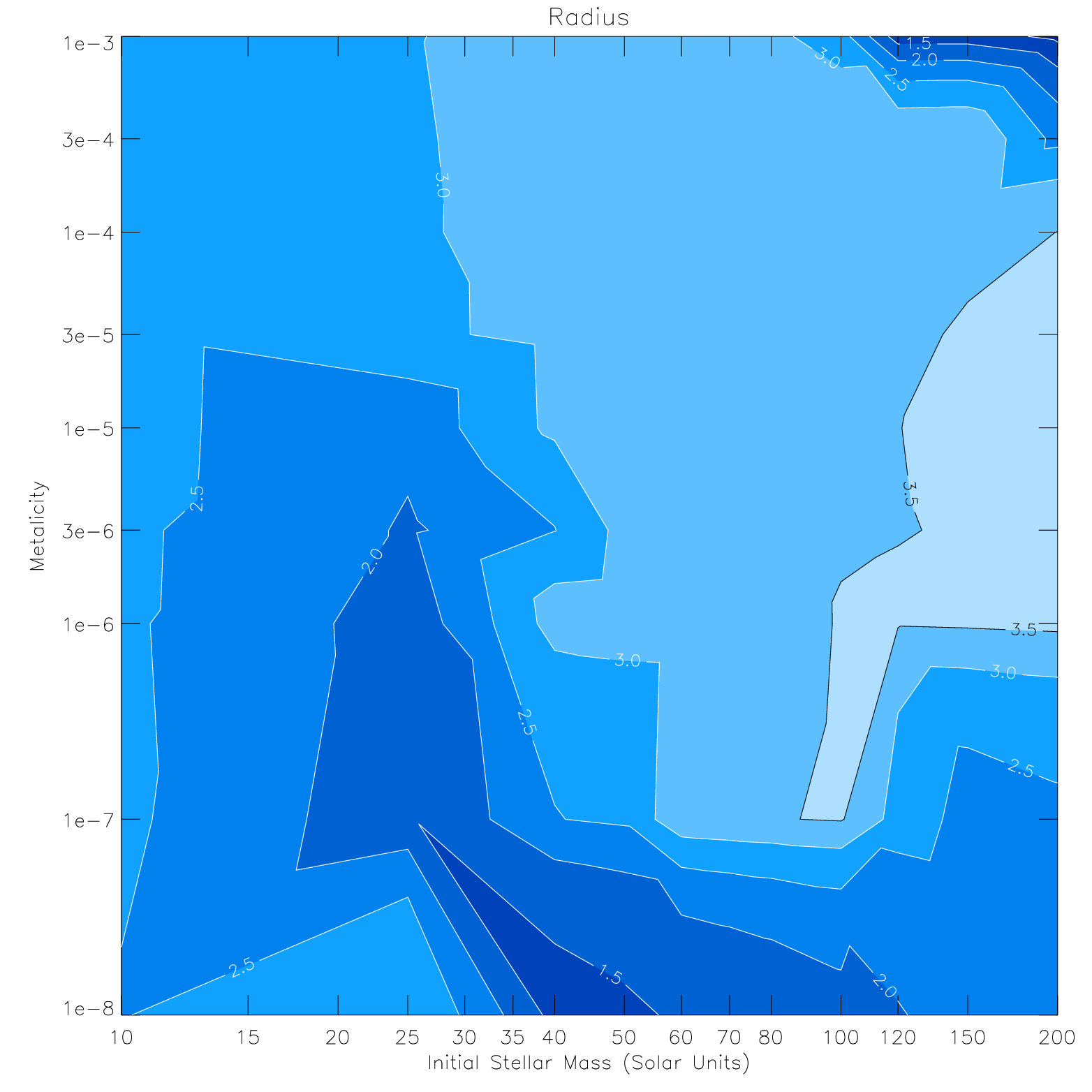}
\includegraphics[height=79mm,angle=0]{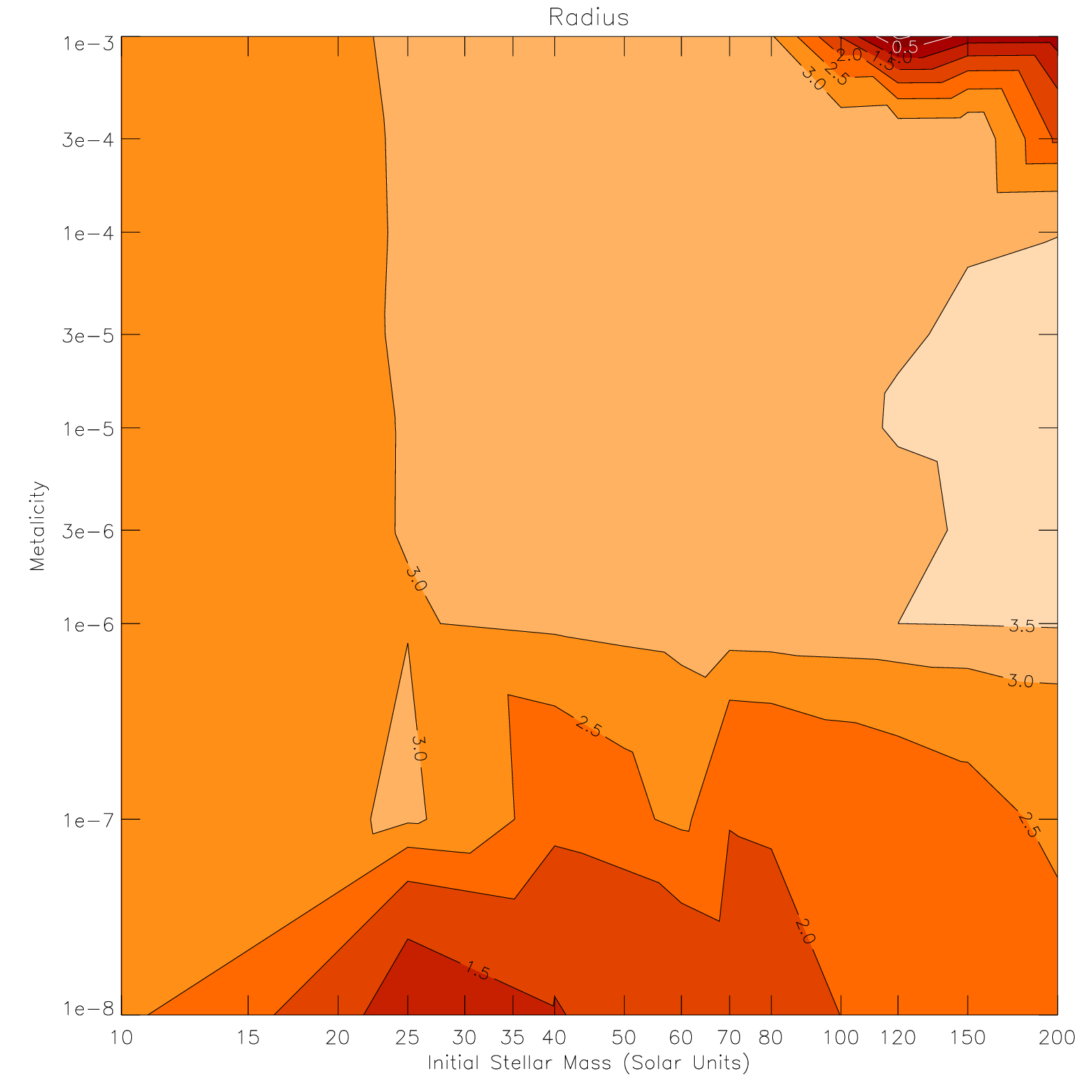}
\caption[Contours of pre-SN radis.]{Contours of pre-SN radis. Contours are in $\log (R/R_{\odot})$. Left no overshooting, right overshooting.}
\label{mapC}
\end{center}
\end{figure}

\subsection{Radius, Figure \ref{mapC}}
The radius grids reflect the results of surface temperature. It should be noted that with convective overshooting the WR models are generally more compact and in both cases WR stars become less compact towards lower metallicities. The low metallicity grid also shows the interesting increased compactness at the lowest metallicity for some stars.

\begin{figure}
\begin{center}
\includegraphics[height=79mm,angle=0]{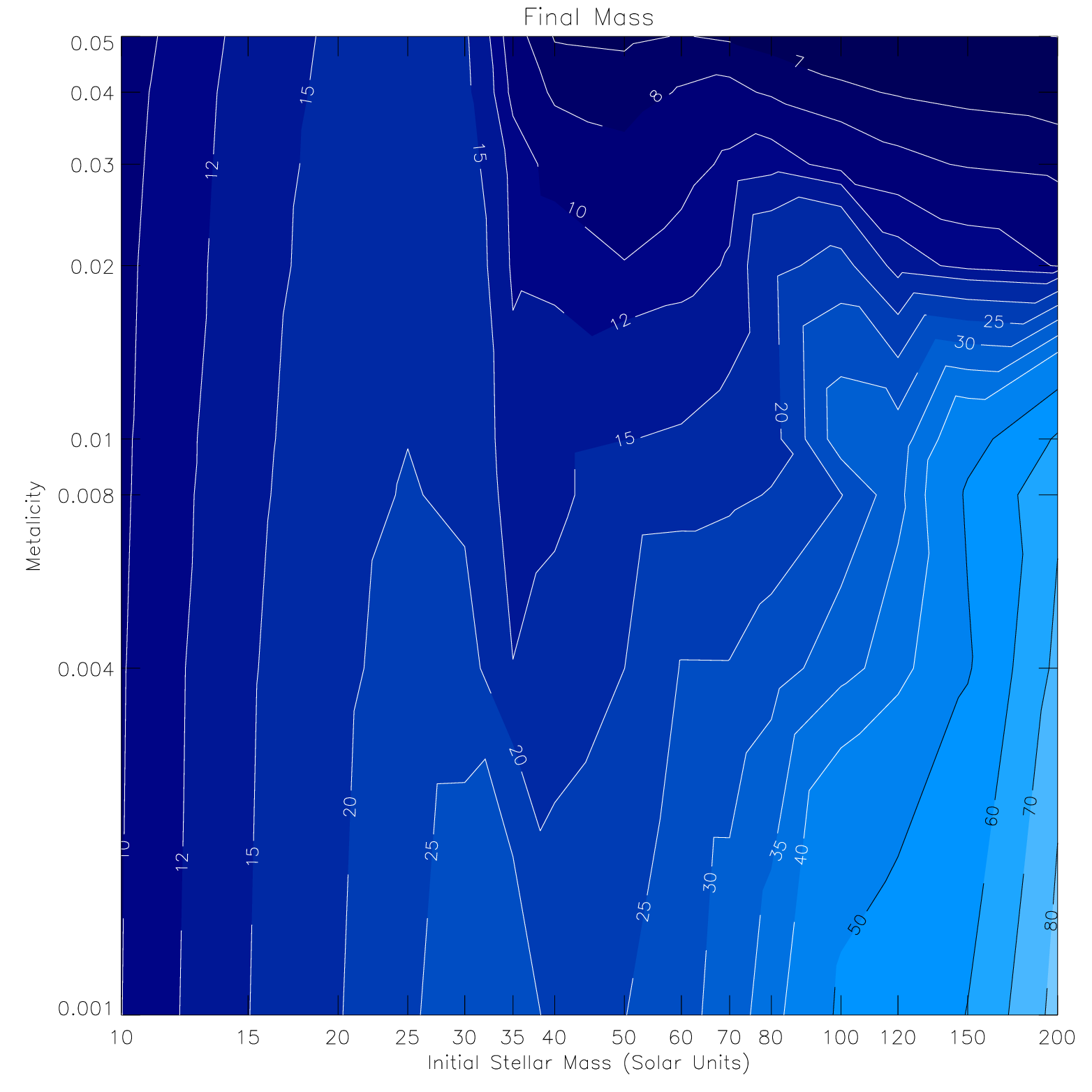}
\includegraphics[height=79mm,angle=0]{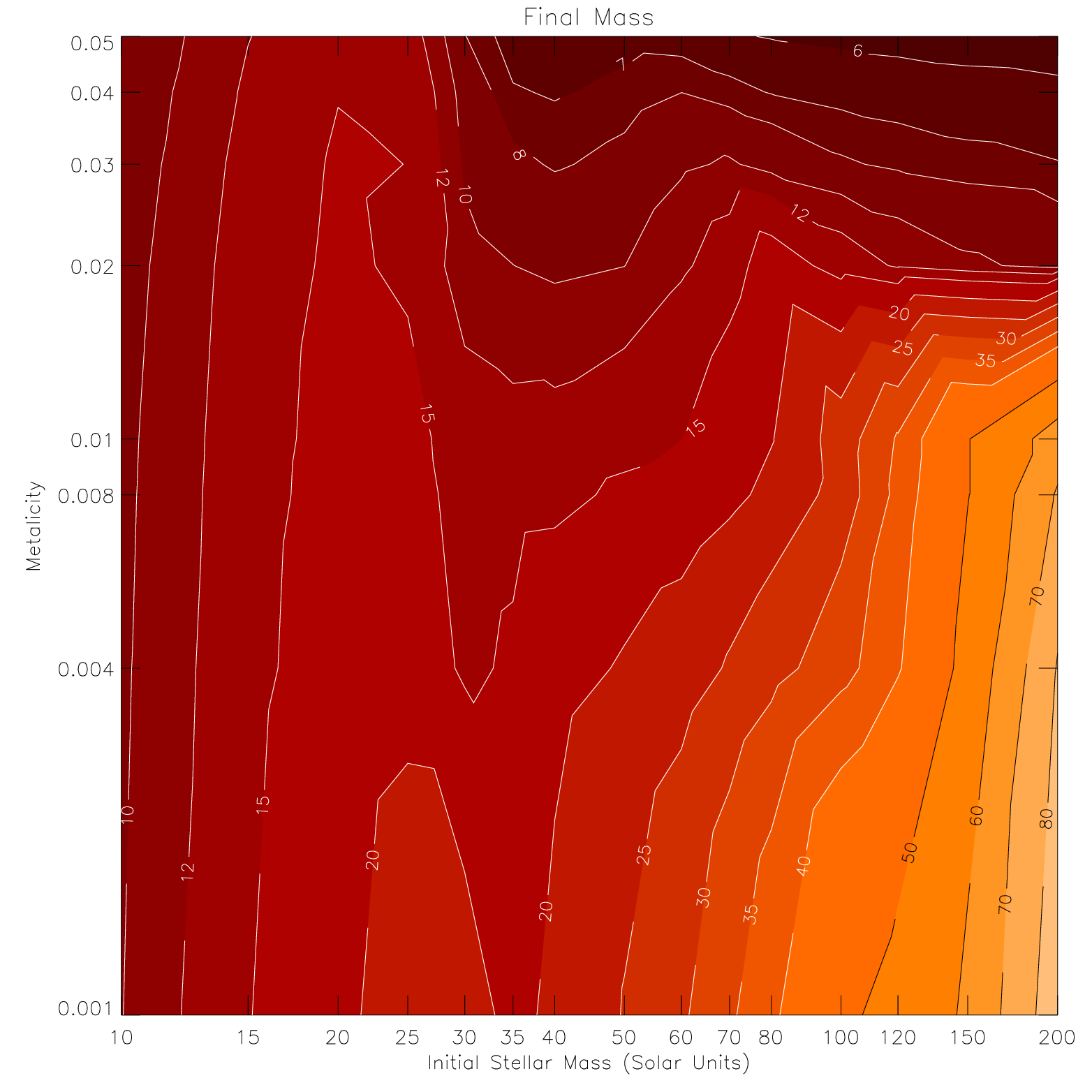}
\includegraphics[height=79mm,angle=0]{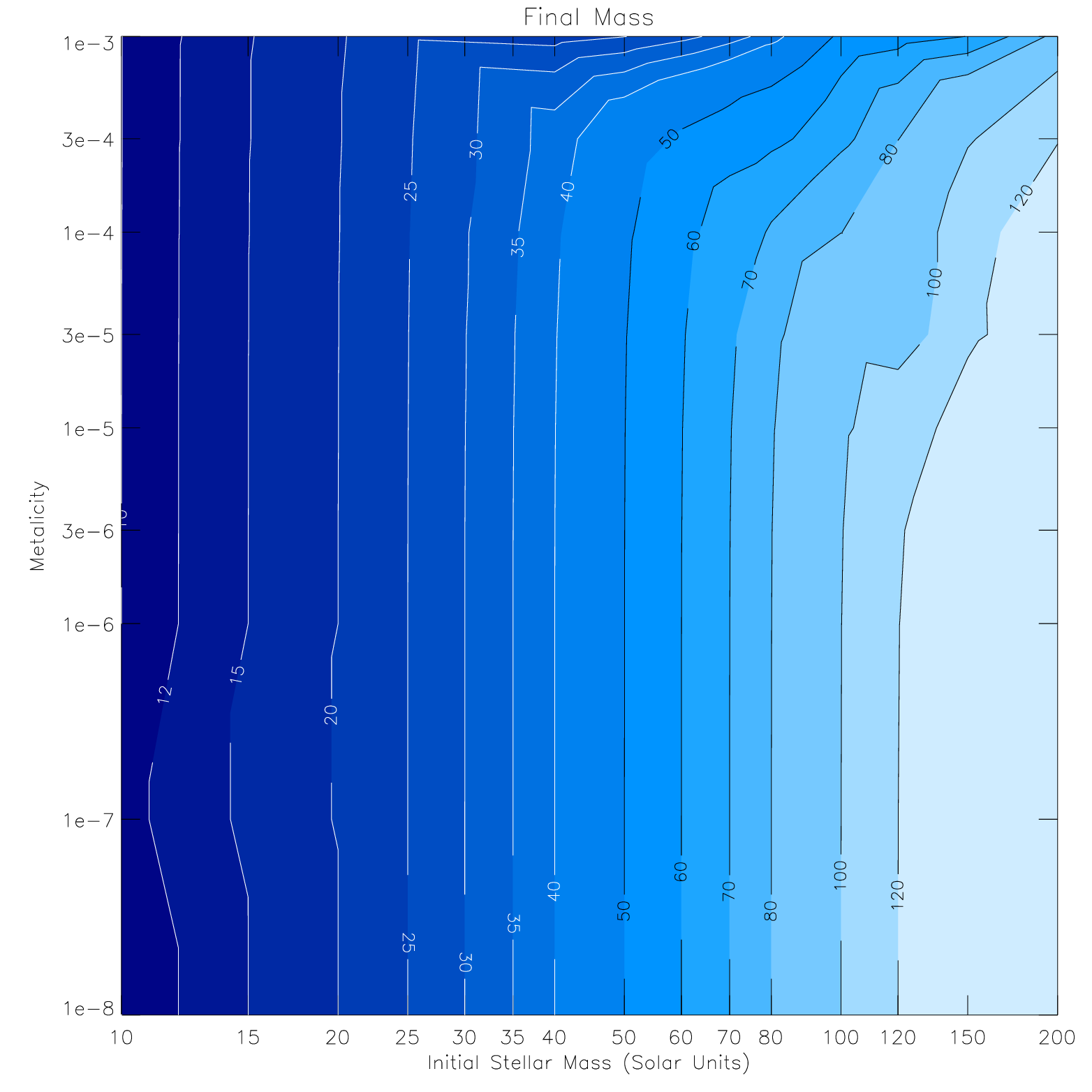}
\includegraphics[height=79mm,angle=0]{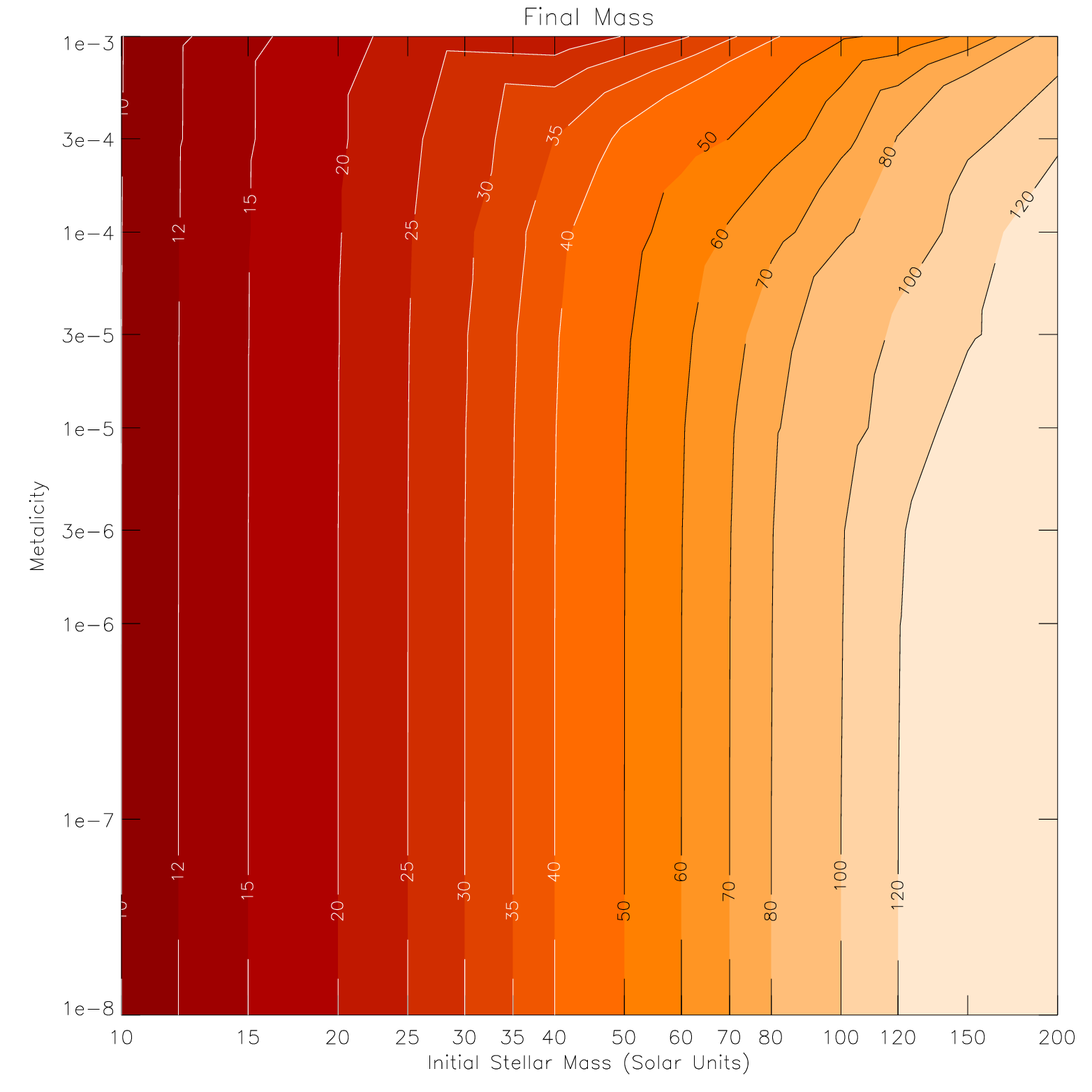}
\caption[Contours of pre-SN mass.]{Contours of pre-SN mass. Contours are in $(M/M_{\odot})$. Left no overshooting, right overshooting.}
\label{mapD}
\end{center}
\end{figure}

\subsection{Final Mass, Figure \ref{mapD}}
The final mass diagram has very similar structure to the luminosity diagram and again reflects how mass loss is more severe in the high-metallicity region and drops off as metallicity decreases. The low metallicity graph shows less structure and is essentially similar to that with no mass loss.

\begin{figure}
\begin{center}
\includegraphics[height=79mm,angle=0]{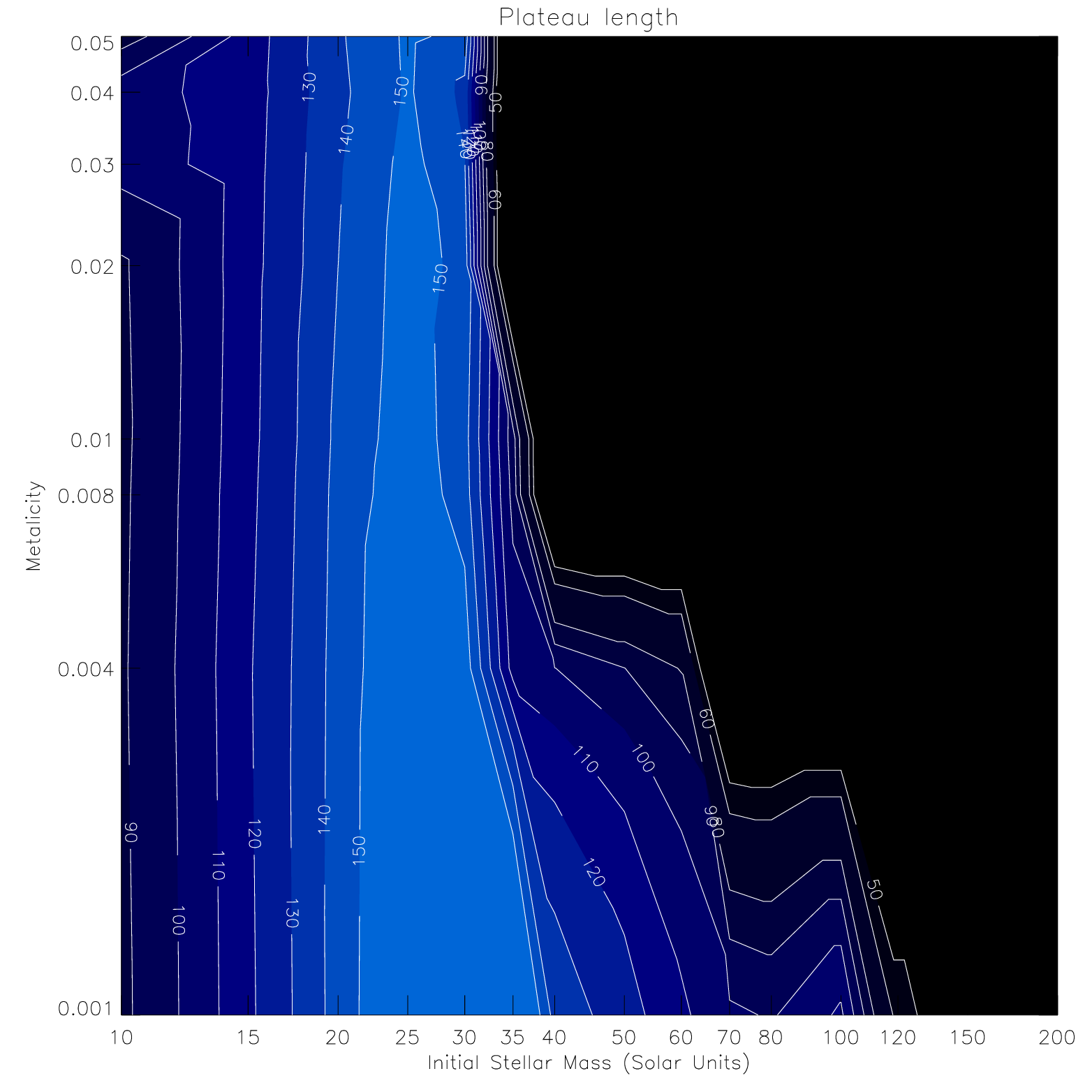}
\includegraphics[height=79mm,angle=0]{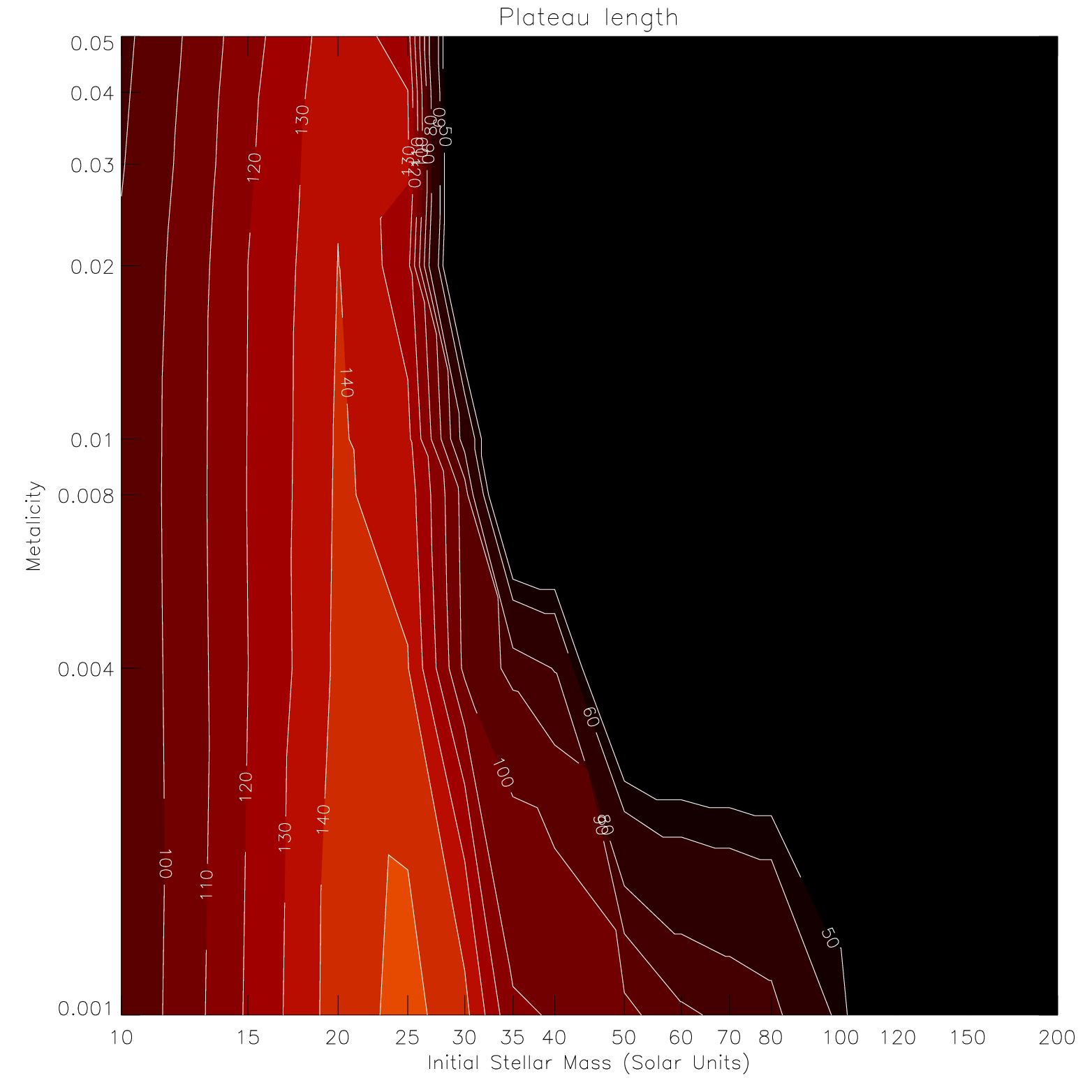}
\includegraphics[height=79mm,angle=0]{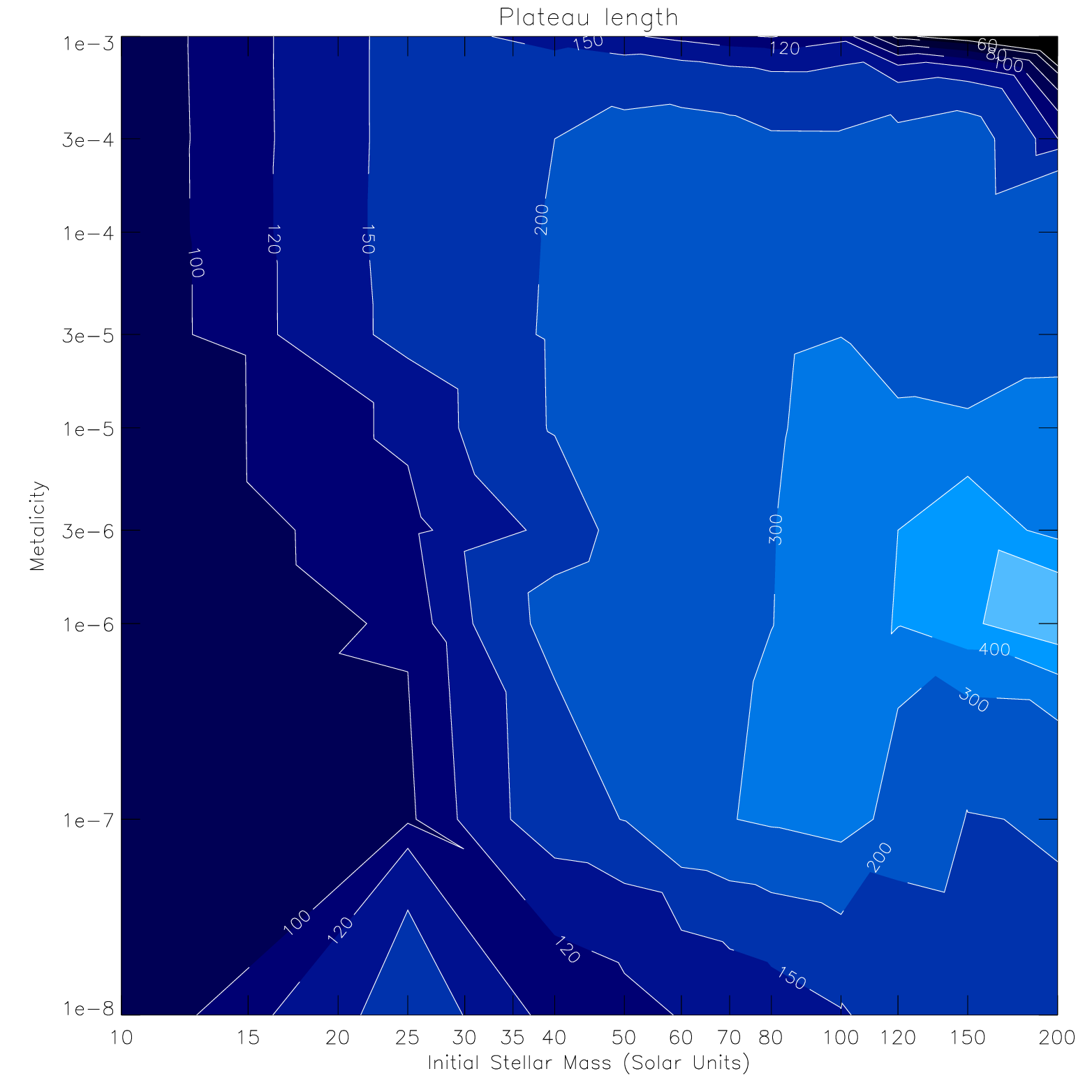}
\includegraphics[height=79mm,angle=0]{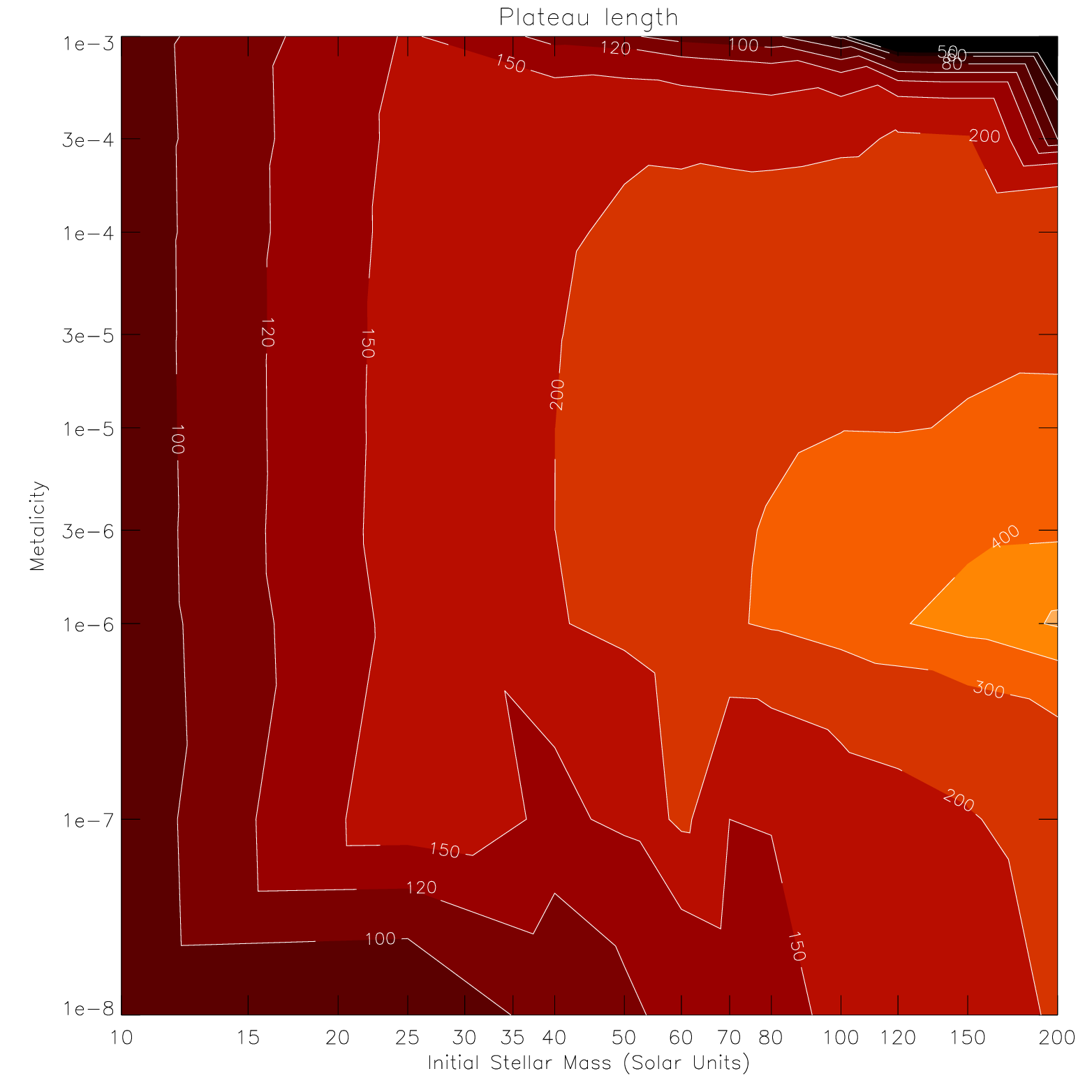}
\caption[Contours of plateau length.]{Contours of plateau length. Contours are in days. Left no overshooting, right overshooting.}
\label{mapE}
\end{center}
\end{figure}

\subsection{Plateau Length, Figure \ref{mapE}}
The plateau length at high metallicity shows the limited region for IIP SNe. The plateau length is on average higher for the no overshooting case than when it is removed. This provides a statistical test which probably indicates that most stars do have some form of convective overshooting in their cores.

Low metallicity grids tell a very different story with extremely long plateaus possible. However these are at too high a red-shift to observe today. However since their duration is longer there is more of a chance of spotting one. At the lowest metallicity here the plateau lengths drop off quickly due to the increasing compactness of the stars.

\begin{figure}
\begin{center}
\includegraphics[height=79mm,angle=0]{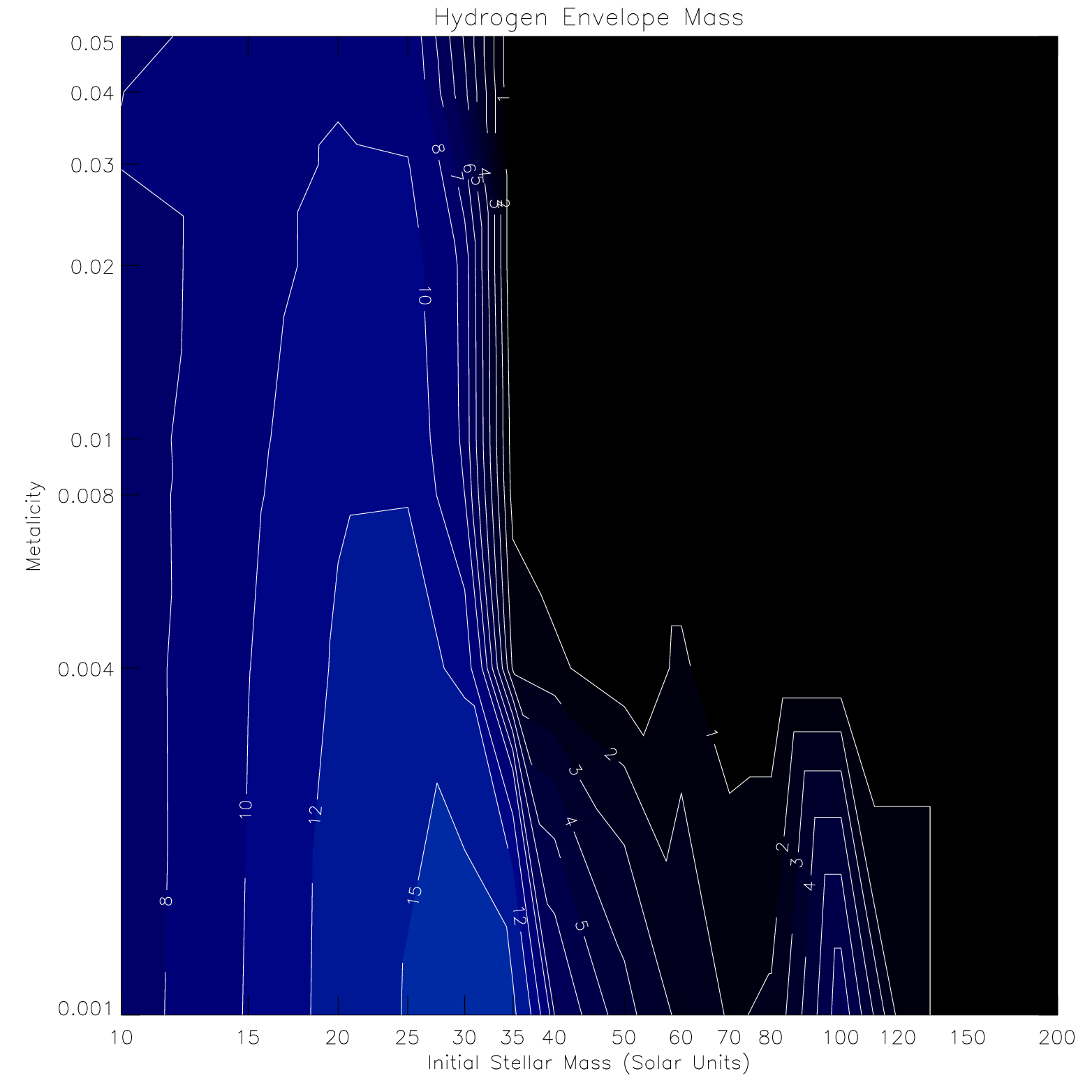}
\includegraphics[height=79mm,angle=0]{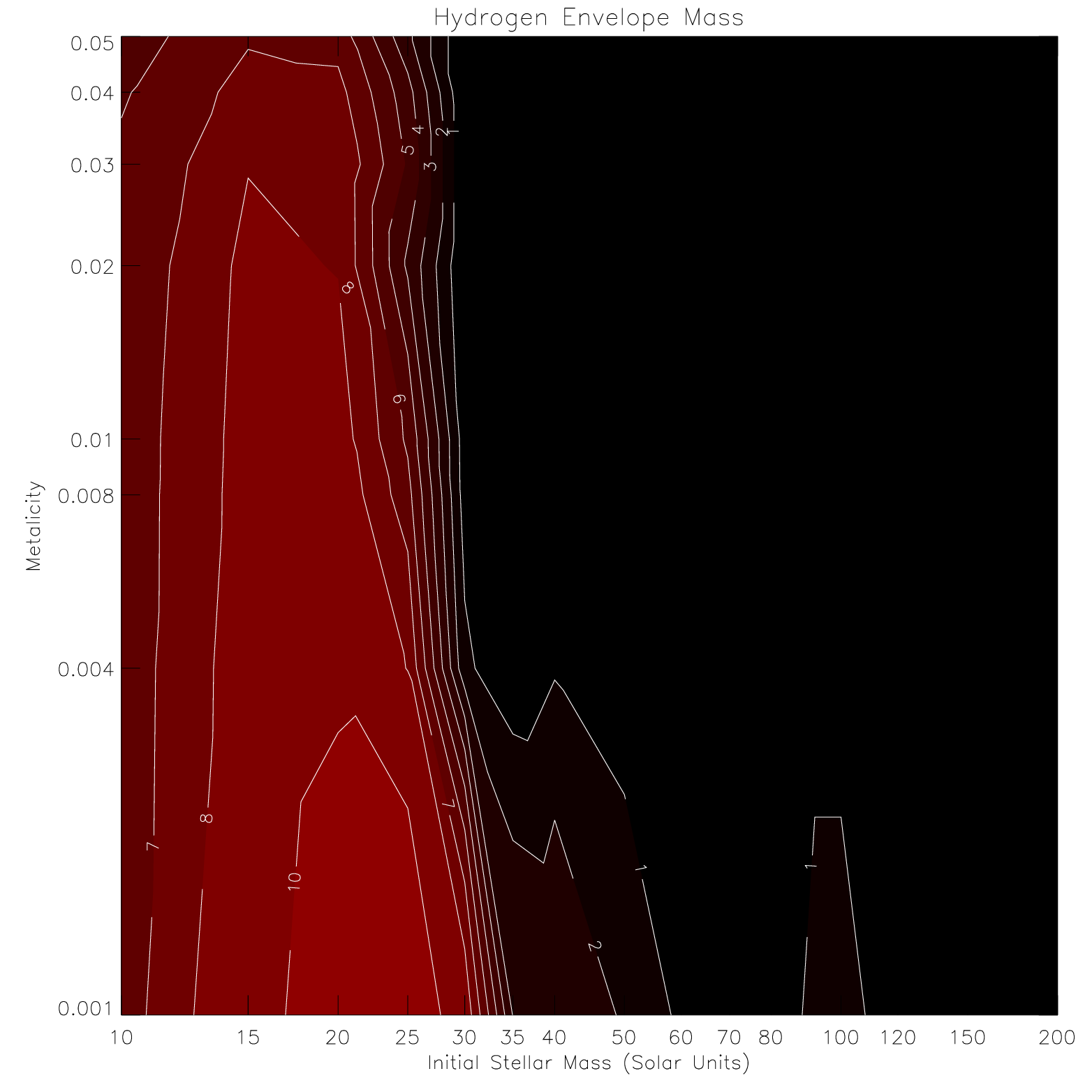}
\includegraphics[height=79mm,angle=0]{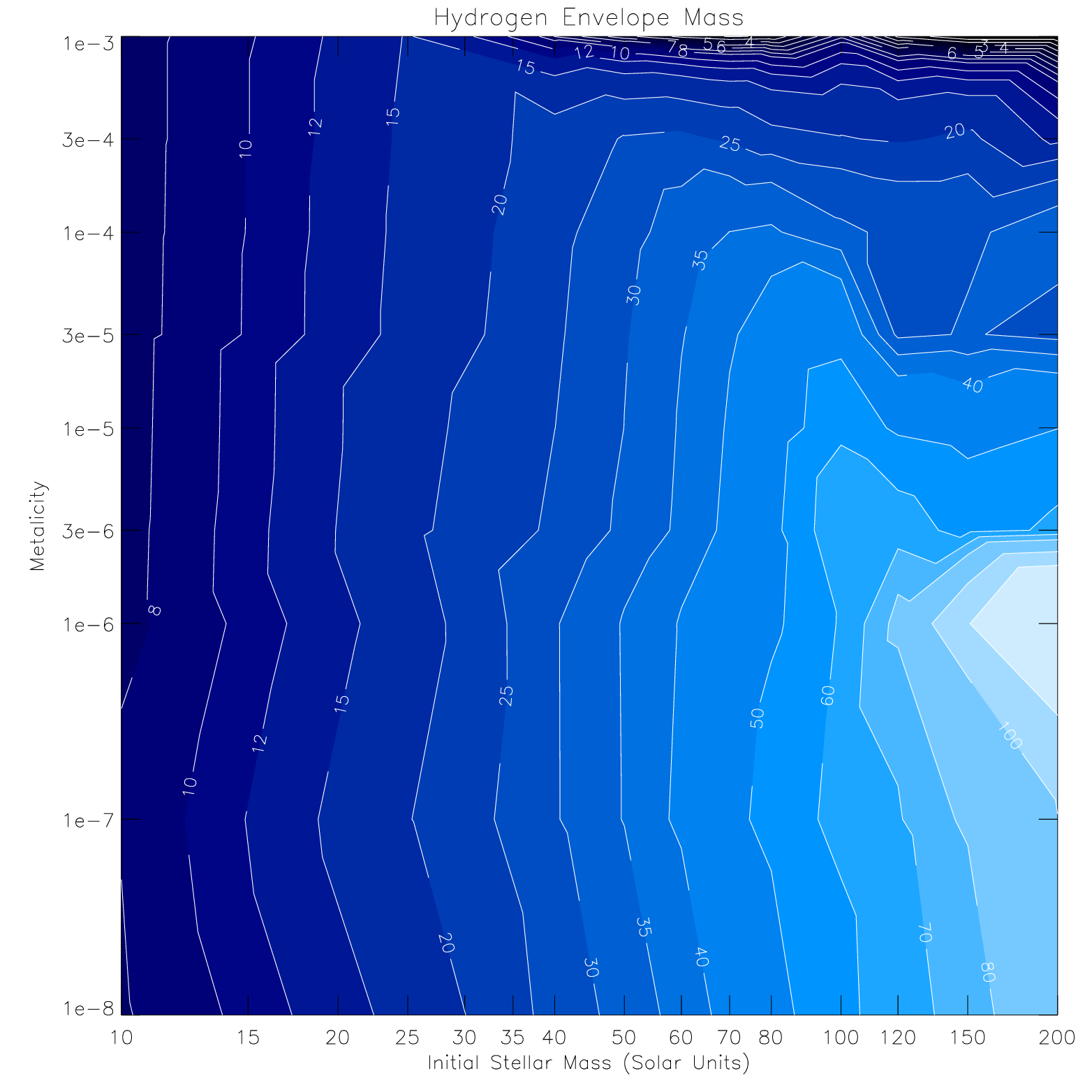}
\includegraphics[height=79mm,angle=0]{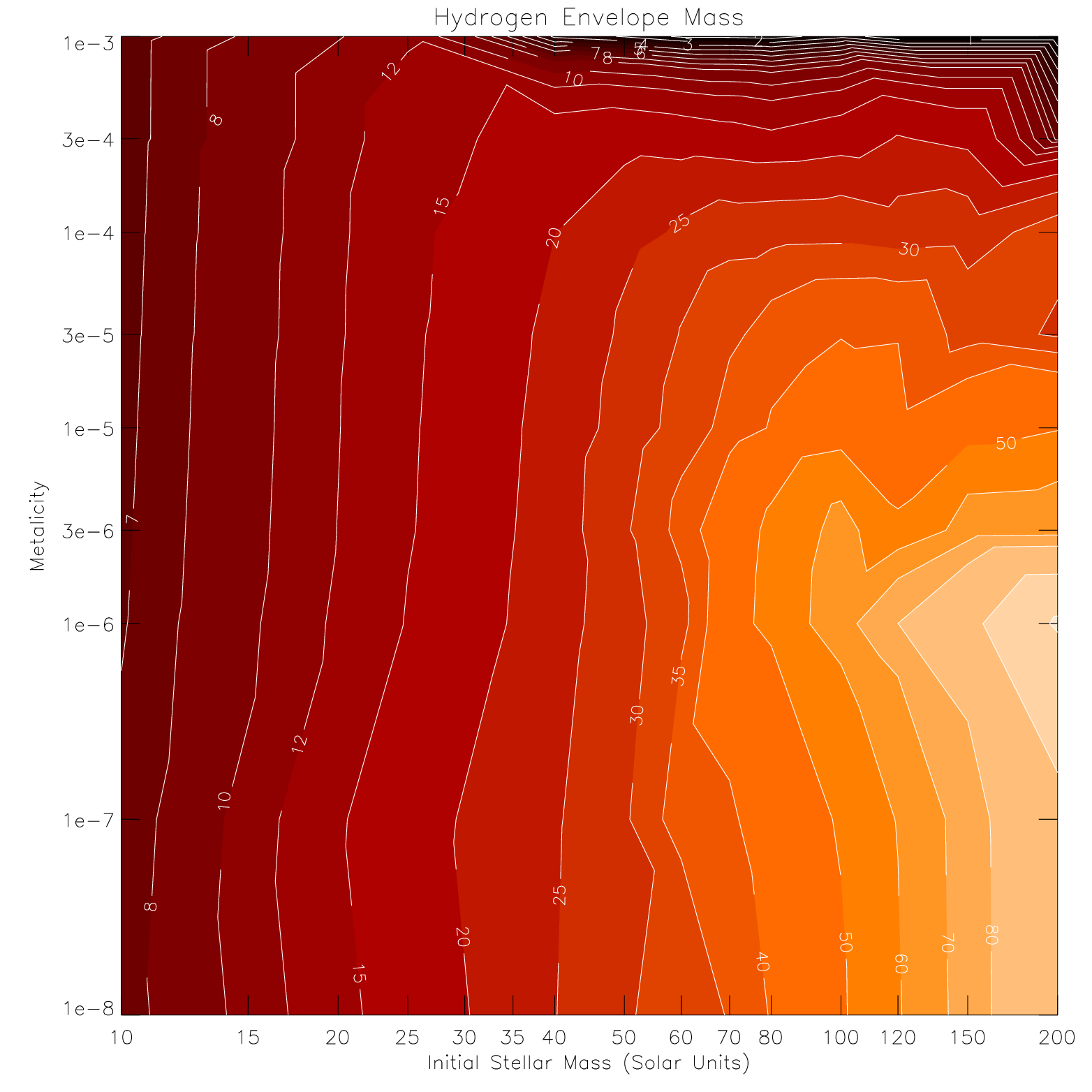}
\caption[Contours of hydrogen envelope mass.]{Contours of hydrogen envelope mass. Contours are in $(M/M_{\odot})$. Left no overshooting, right overshooting.}
\label{mapF}
\end{center}
\end{figure}
\begin{figure}
\begin{center}
\includegraphics[height=79mm,angle=0]{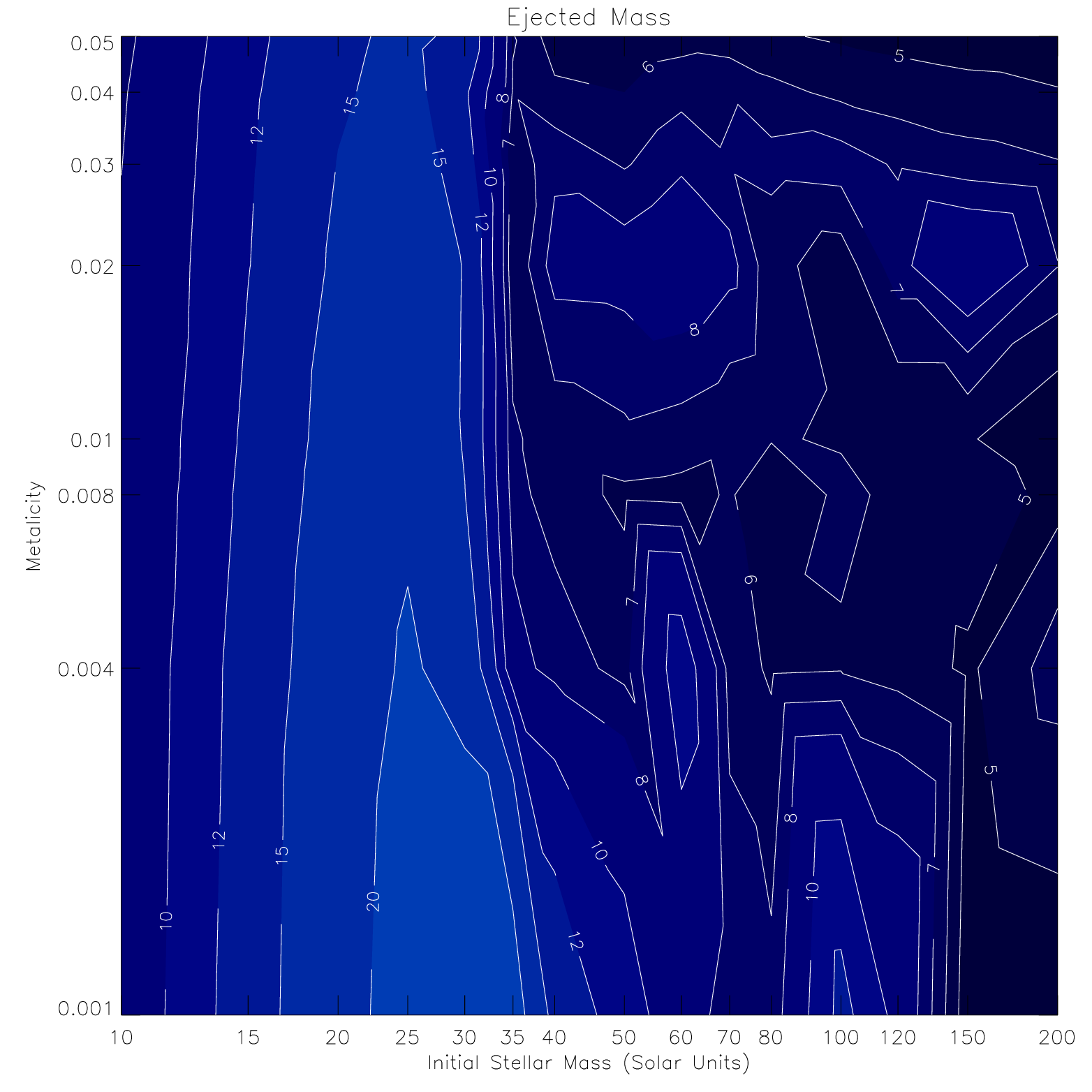}
\includegraphics[height=79mm,angle=0]{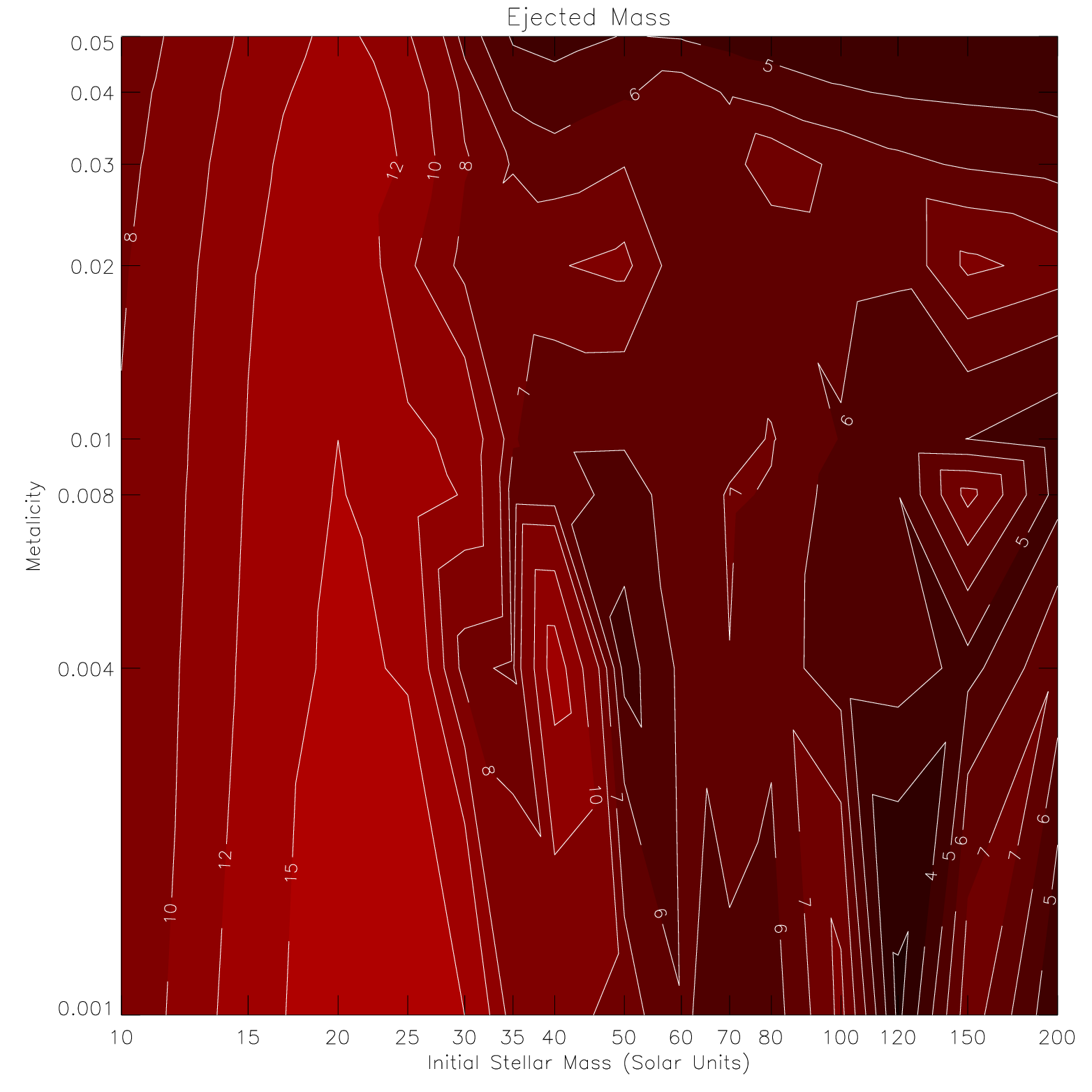}
\includegraphics[height=79mm,angle=0]{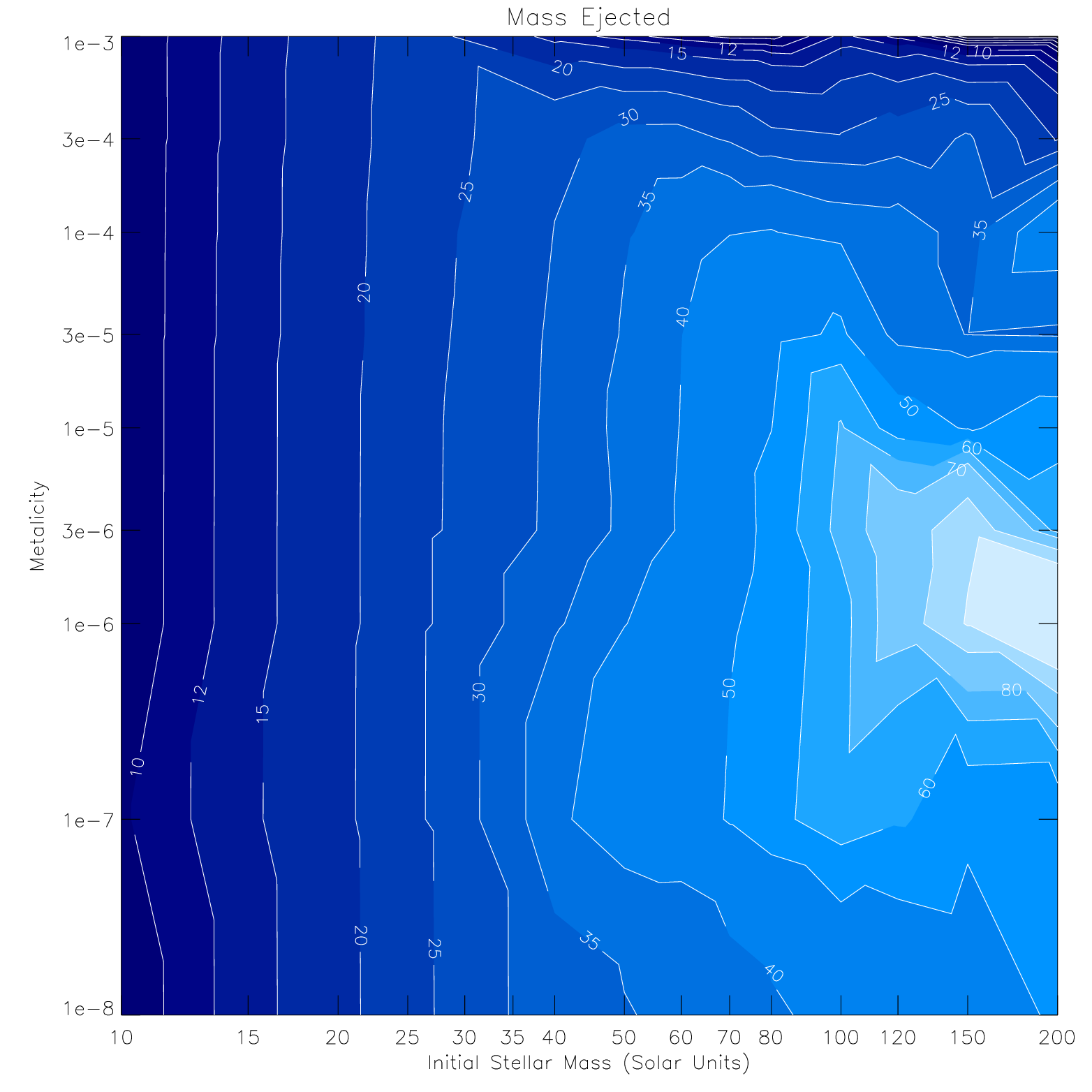}
\includegraphics[height=79mm,angle=0]{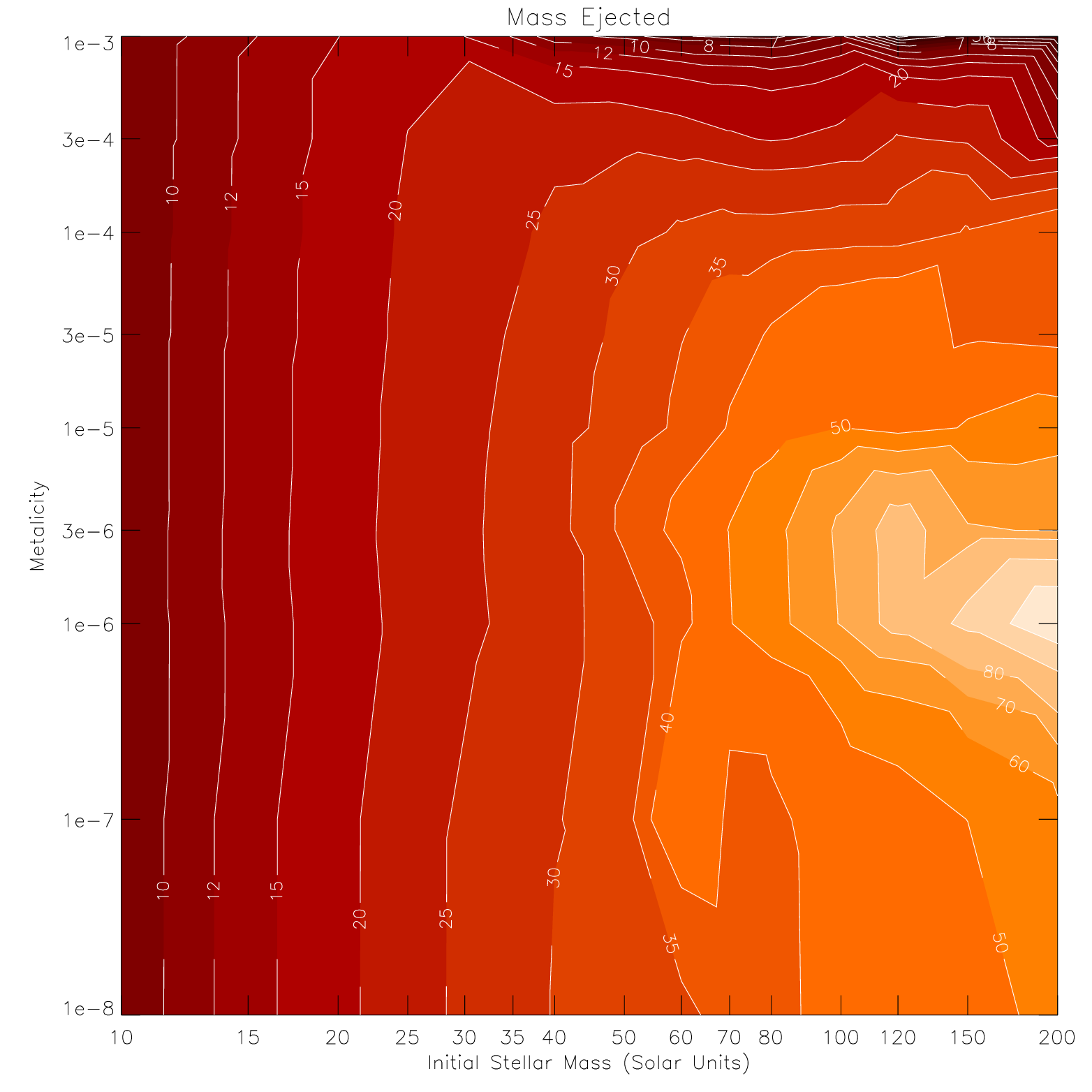}
\caption[Contours of mass ejected in SN.]{Contours of mass ejected in SN. Contours are in $(M/M_{\odot})$. Left no overshooting, right overshooting.}
\label{mapG}
\end{center}
\end{figure}
\begin{figure}
\begin{center}
\includegraphics[height=79mm,angle=0]{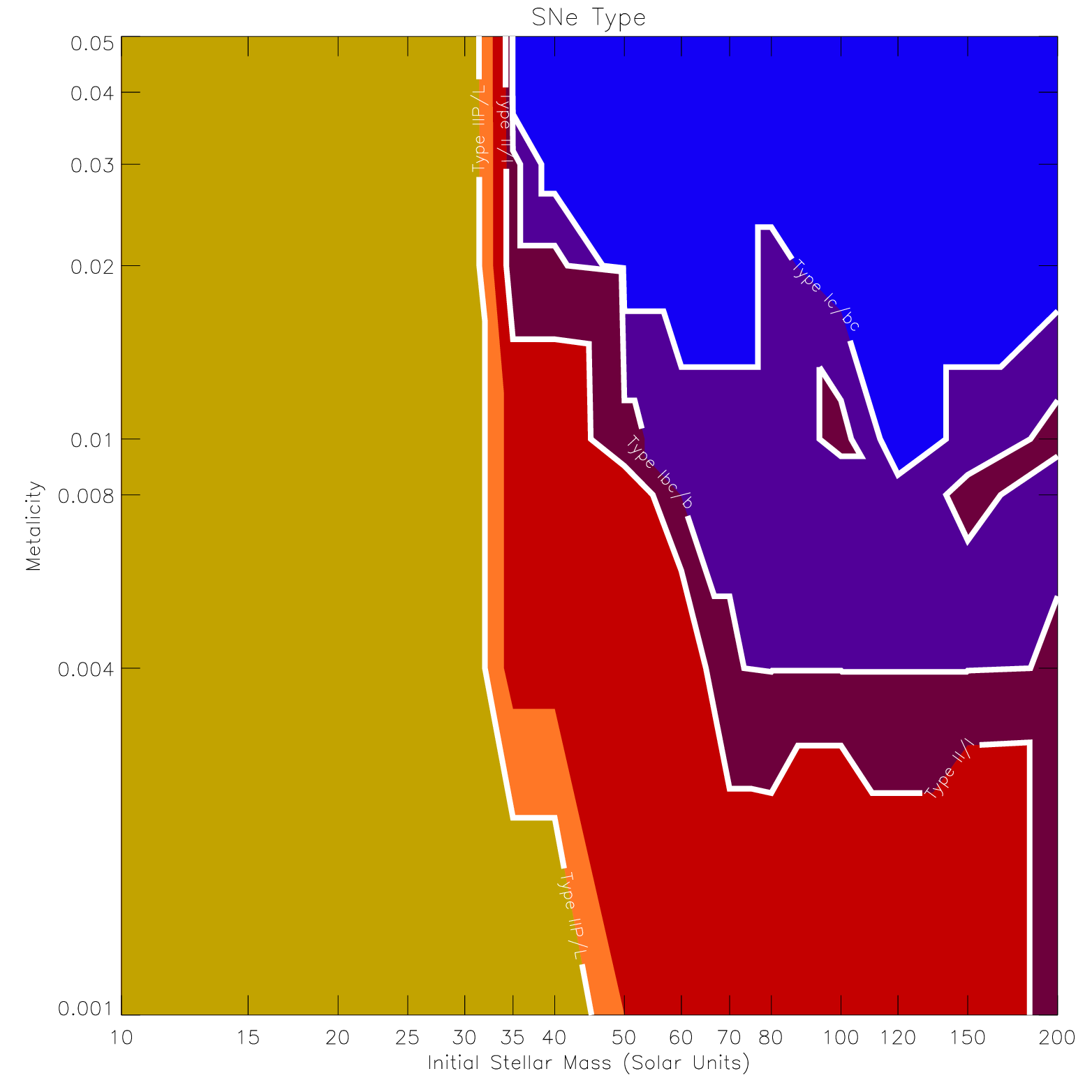}
\includegraphics[height=79mm,angle=0]{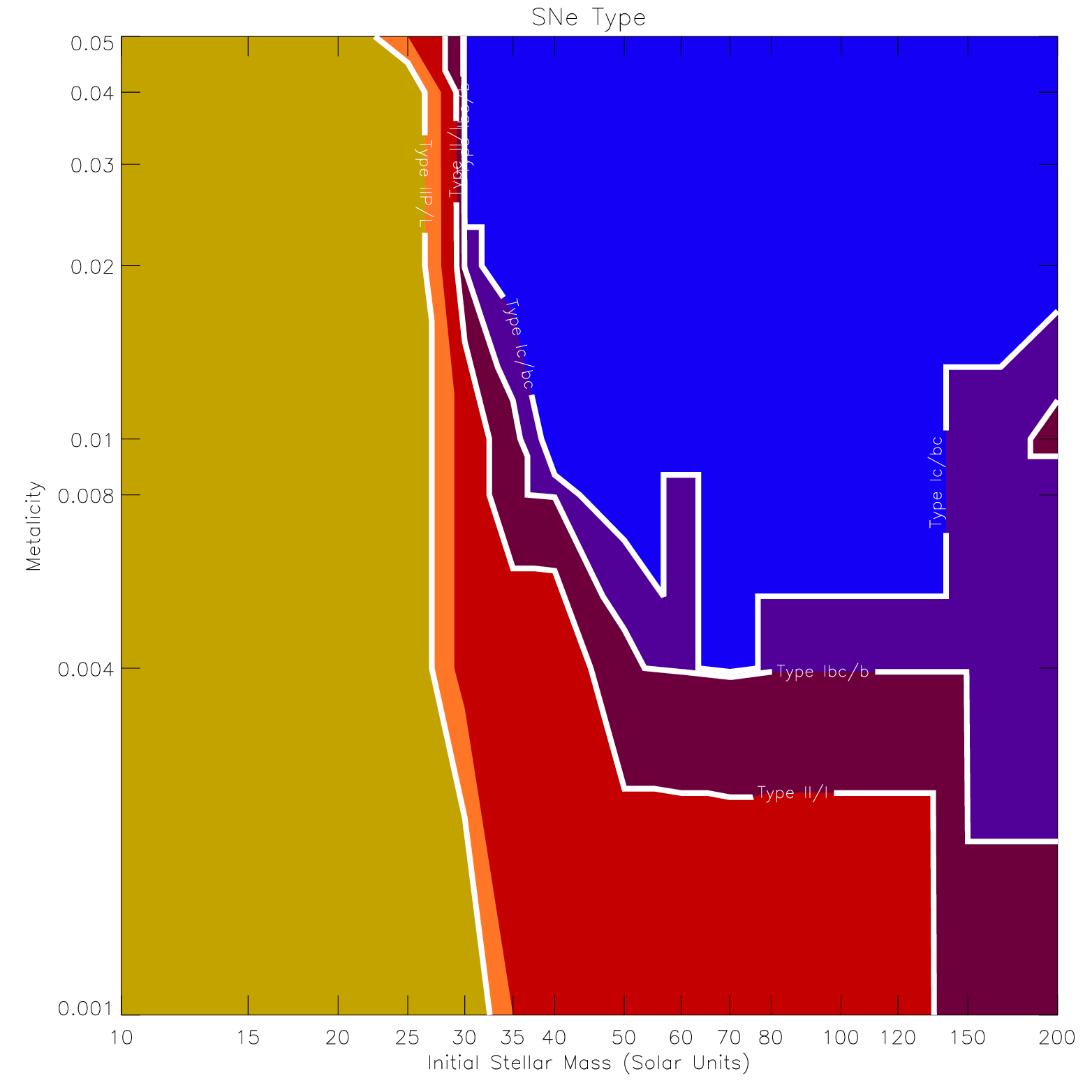}
\includegraphics[height=79mm,angle=0]{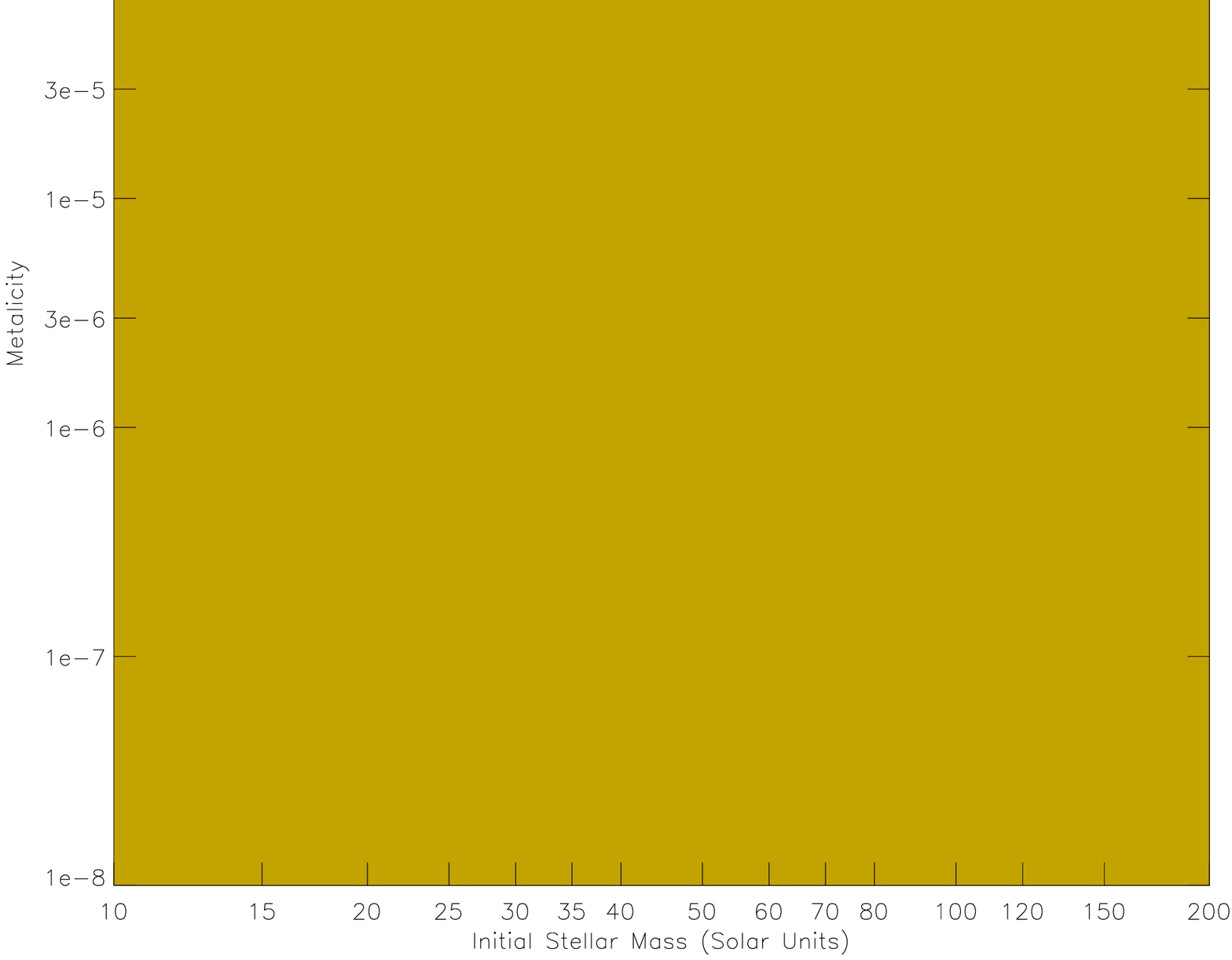}
\includegraphics[height=79mm,angle=0]{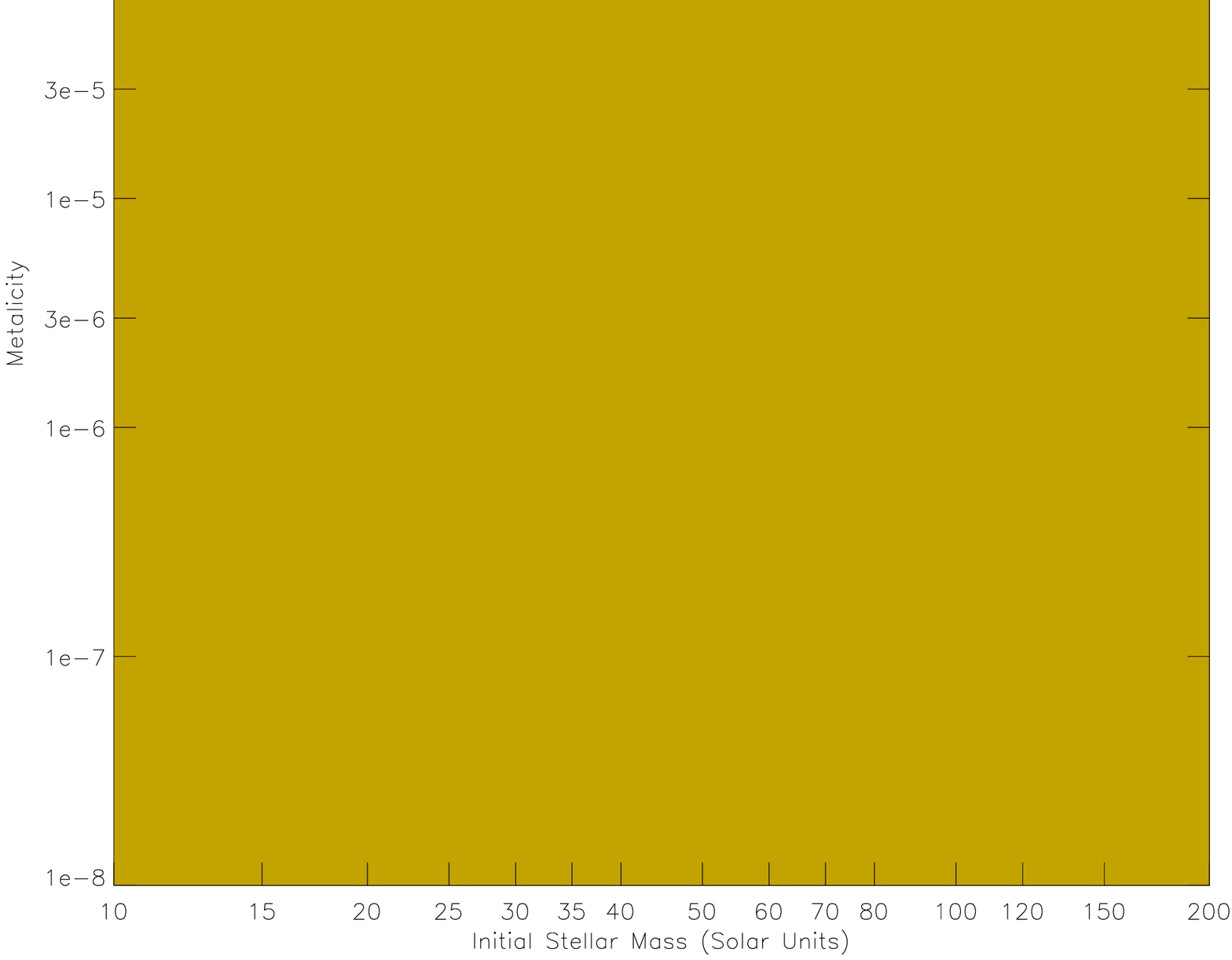}
\caption[Contours of SNe type.]{Contours of SNe type. Green - IIP, orange - IIL, red - IIb, blue - Ic, purple - Ibc and brown - Ib. Left no overshooting, right overshooting.}
\label{mapH}
\end{center}
\end{figure}
\begin{figure}
\begin{center}
\includegraphics[height=79mm,angle=0]{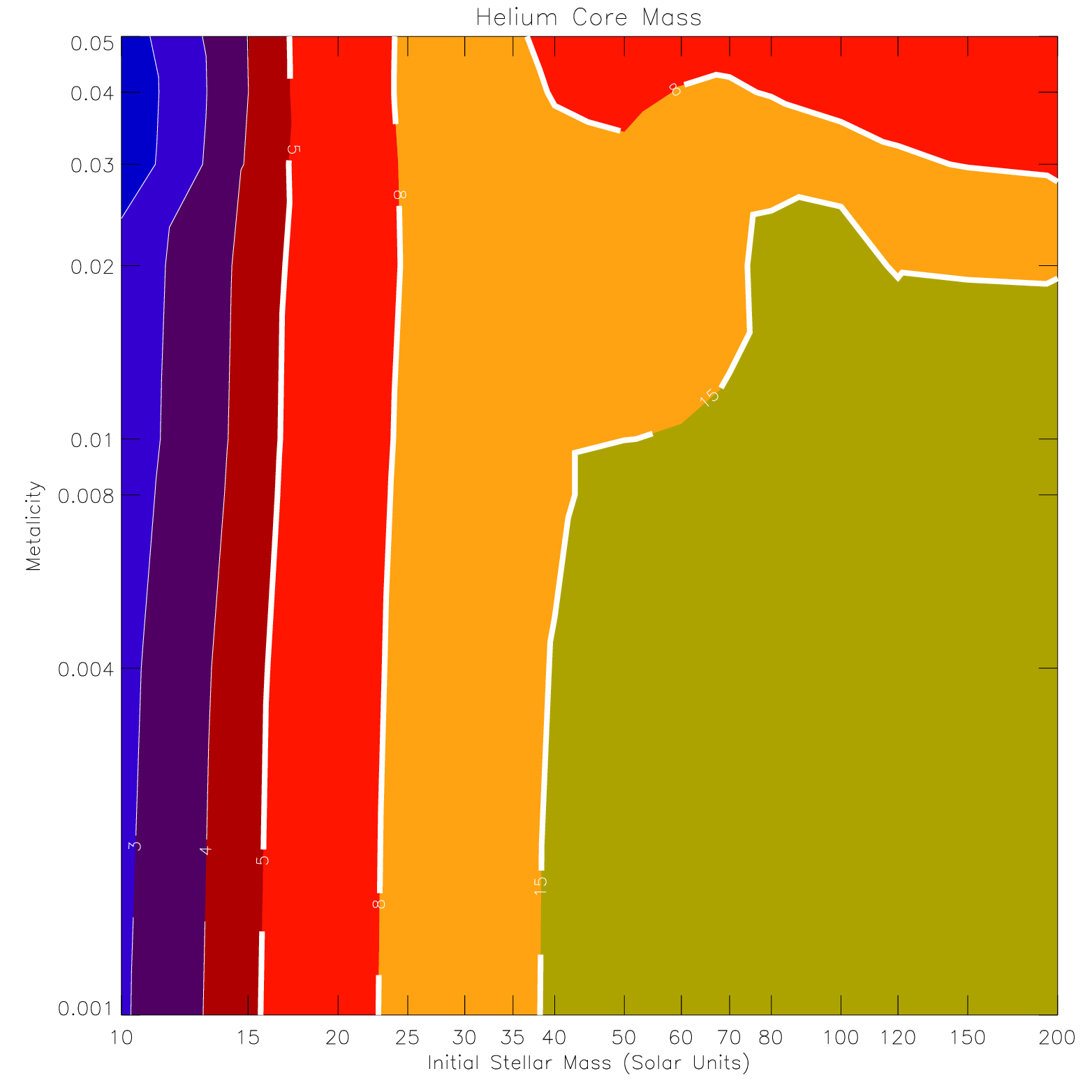}
\includegraphics[height=79mm,angle=0]{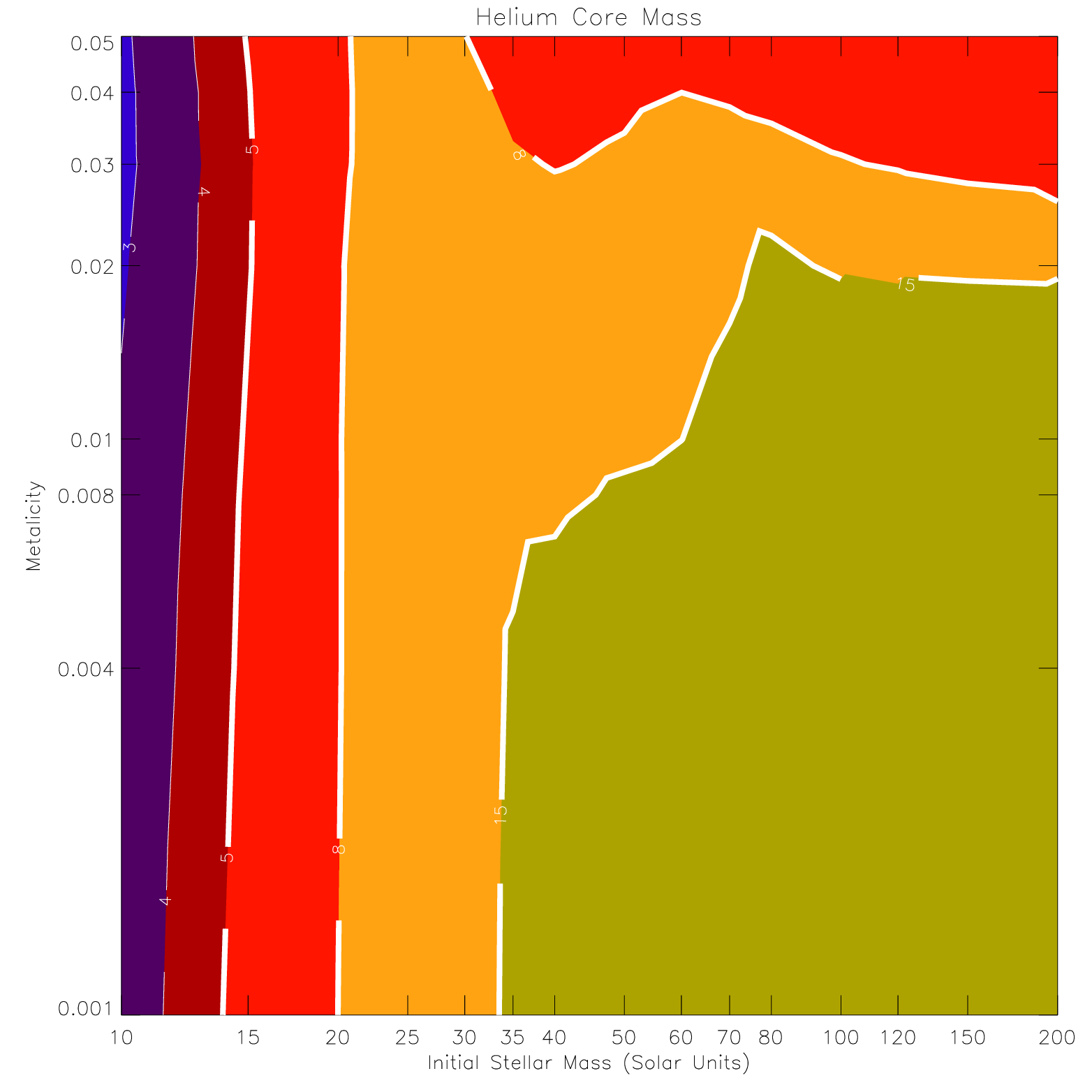}
\includegraphics[height=79mm,angle=0]{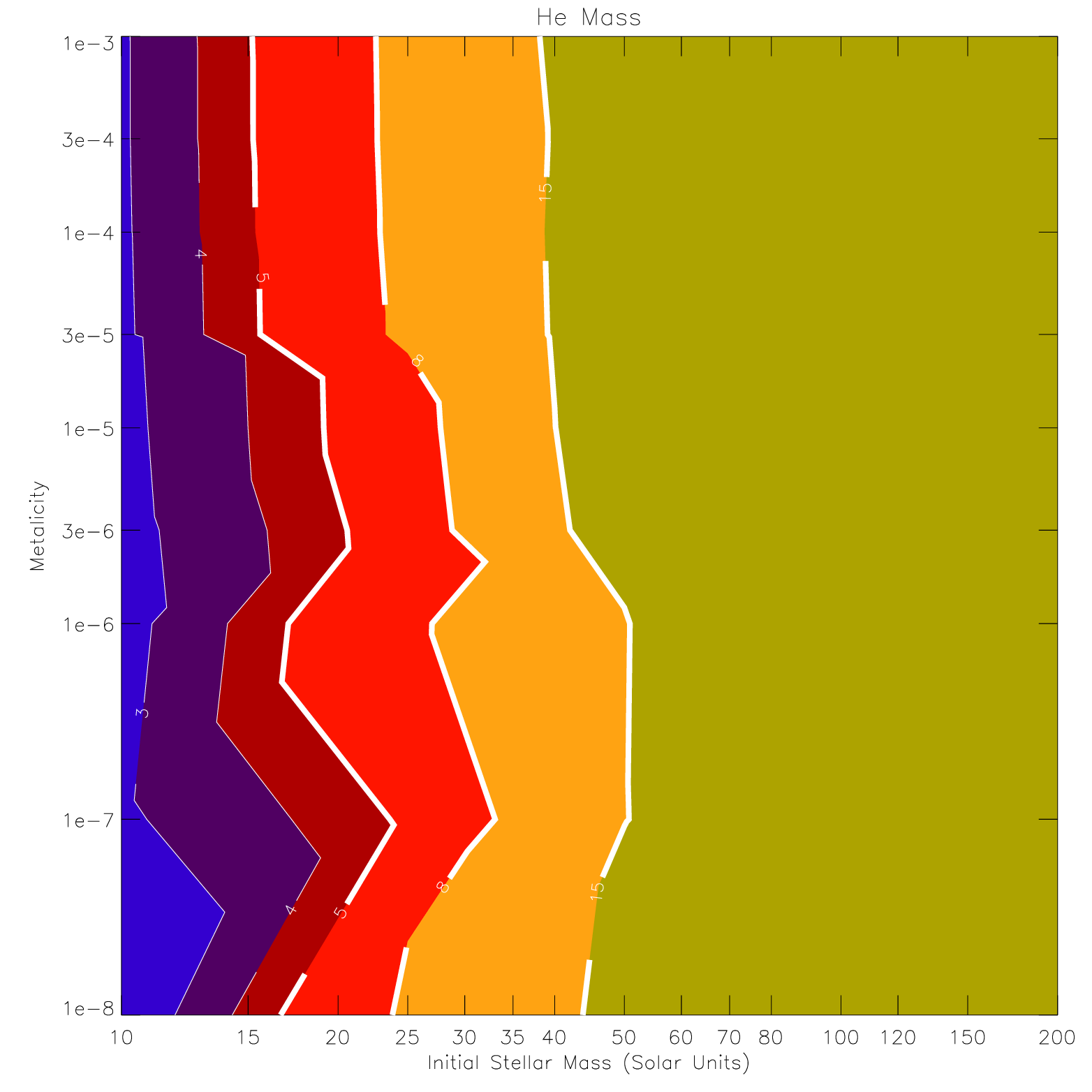}
\includegraphics[height=79mm,angle=0]{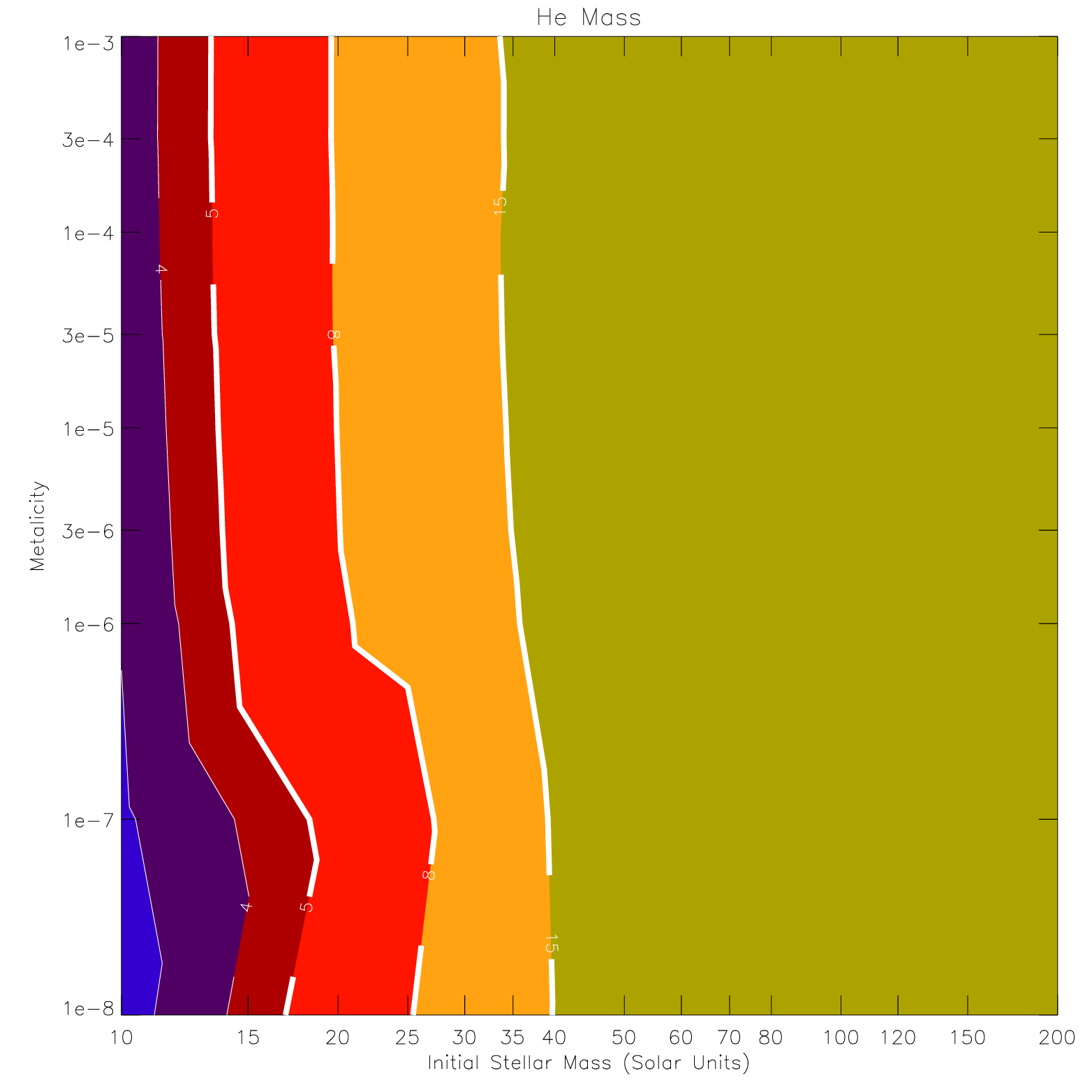}
\caption[Contours of remnant of helium core mass]{Contours of remnant of helium core mass. Numerical contour values are helium core mass in $M_{\odot}$. Green - direct black holes, yellow - fall back black hole, red and other colours - neutron stars. Left no overshooting, right overshooting.}
\label{mapI}
\end{center}
\end{figure}
\begin{figure}
\begin{center}
\includegraphics[height=79mm,angle=0]{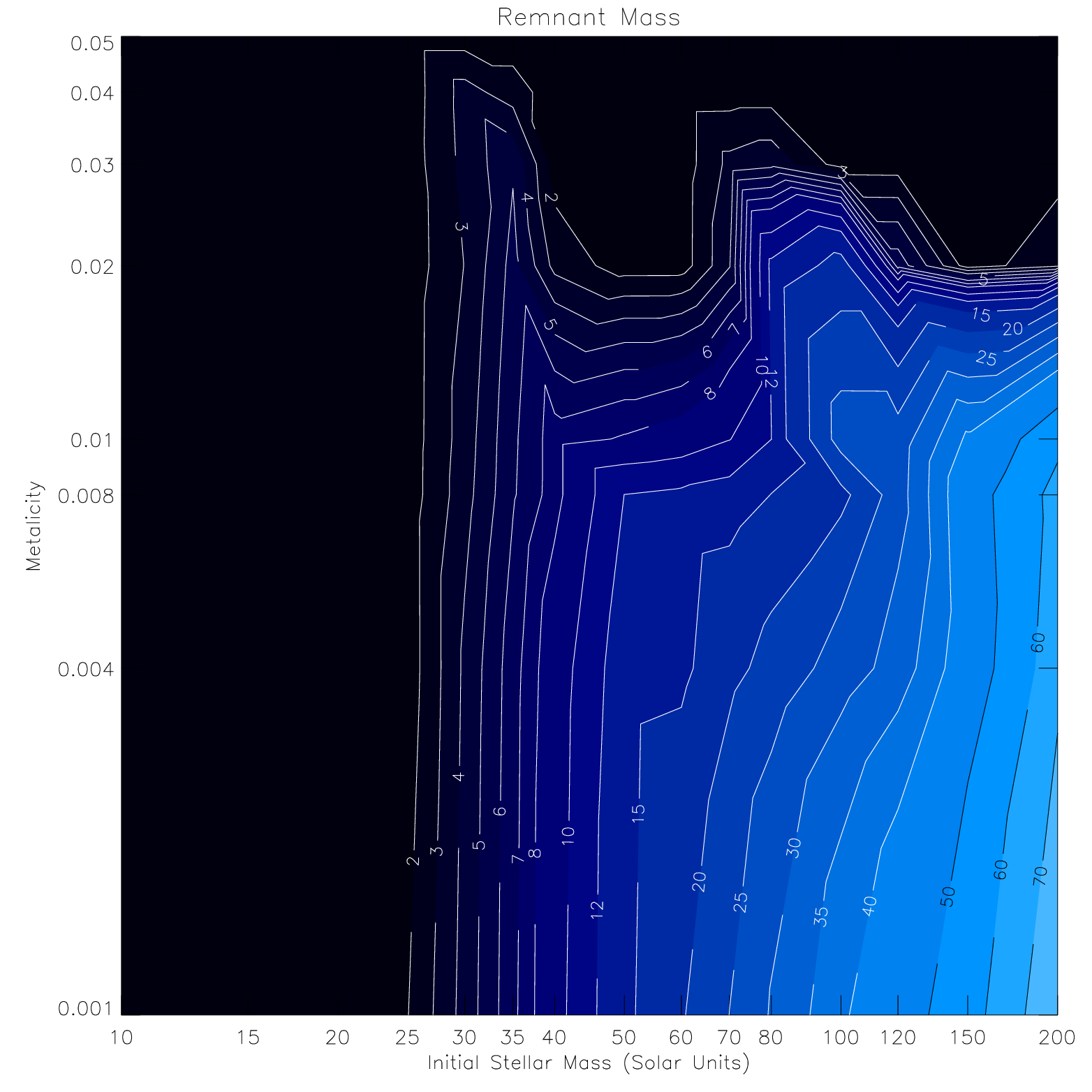}
\includegraphics[height=79mm,angle=0]{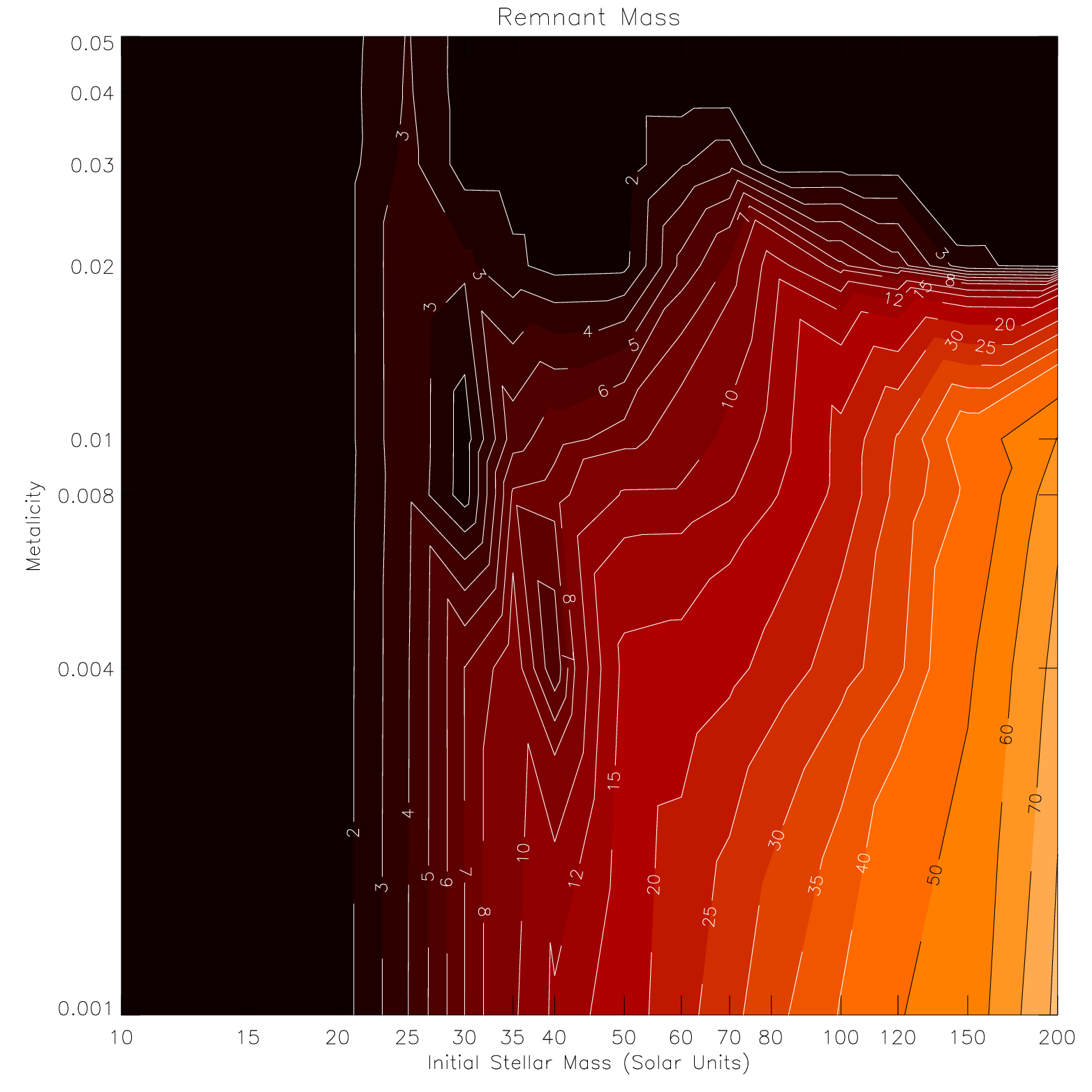}
\includegraphics[height=79mm,angle=0]{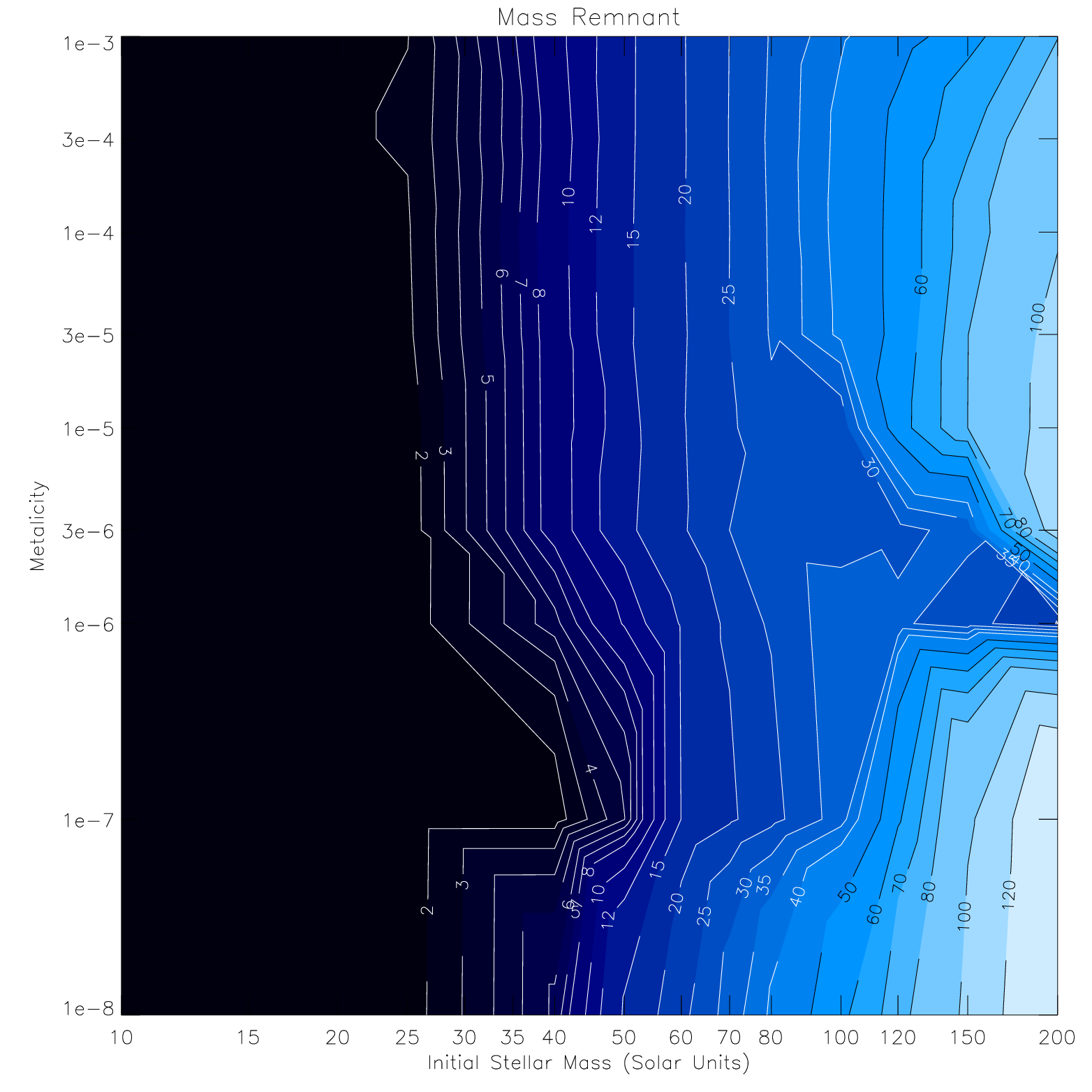}
\includegraphics[height=79mm,angle=0]{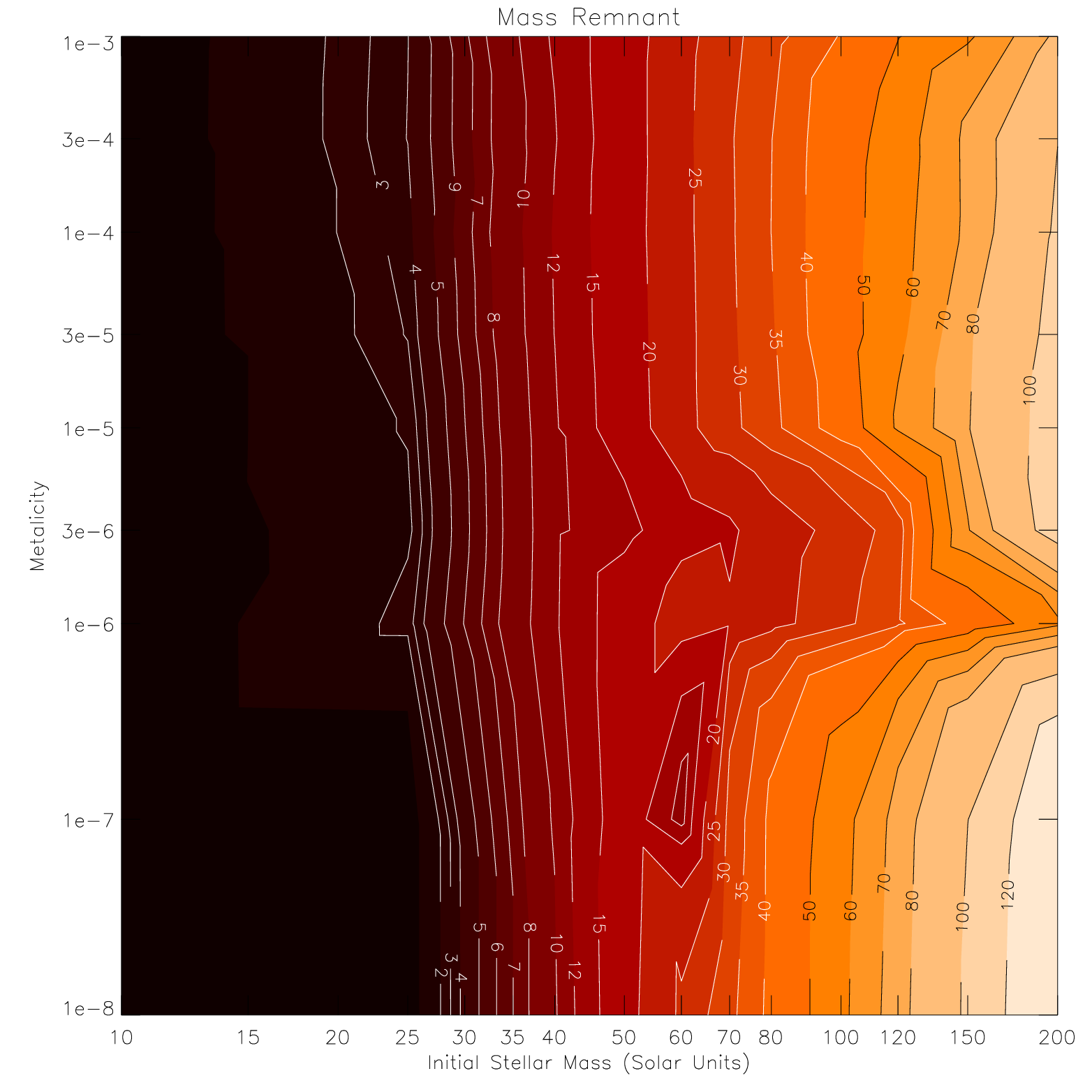}
\caption[Contours of SN remnant mass.]{Contours of SN remnant mass. Contours are in $(M/M_{\odot})$. Left no overshooting, right overshooting.}
\label{mapJ}
\end{center}
\end{figure}

\subsection{Hydrogen Envelope Mass, Figure \ref{mapF}}
The high metallicity grid mimics the plateau length diagram closely. This also shows there are maxima at different metallicities of how massive the envelopes are and therefore how much envelope can be ejected. Although on top of this we must also consider the outer mantle of the core that is also ejected. The low metallicity plot shows that at lower metallicities there is more hydrogen and so smaller cores.

\subsection{Ejected Mass, Figure \ref{mapG}}
The mass ejected by the SNe is also interesting because it demonstrates that a WR star ejects very little mass because it is denser and so more tightly bound than a red giant progenitor. Also it is possible to see, along with figure \ref{mapF}, that for red giants a large fraction of the core mantle is ejected. In the low-metallicity diagrams the compact nature of the stars gives a more tightly bound envelope so less material is ejected.

\subsection{SNe Type, Figure \ref{mapH}}
The SNe type diagram demonstrates the different nature of the grids with and without overshooting well with a clear shift in the minimum mass for a type I SN. The line for the changeover from IIP to IIL SNe is quite constant. However the line separating IIL and type I SNe moves more and there is a cut-off in type I SN at just below $Z=0.001$. The separation between Ib and Ic is quite difficult to observe. At the lowest initial masses for type I SNe there is likely a band of Ib SNe whose width increases with decreasing metallicity. However at LMC metallicity and below there are no more Ic SNe. The low metallicity grids are structurally less captivating.

Using Figure \ref{mapI} and the details of section 4.4, we can identify the mass regime where SN Ibc become unseen. Those SNe that form black holes directly have no display, unless a jet driven SNe forms a black hole and produces an observable display. Unseen SNe affect the ratio of type II/Ibc SNe observed. At below solar metallicities all massive progenitor type Ibc SNe are unseen. We find very few bright type Ibc SNe in our sample because only at the highest metallicity do WR stars have helium cores of less than $5M_{\odot}$.

\subsection{SNe Remnant, figures \ref{mapI} \& \ref{mapJ}}
In shape figure \ref{mapI} is similar to those of \citet{H03} although we do not go to zero metallicity. The change between the no-overshooting and overshooting cases is minor with only the transitions moving down in mass while the shape remains the same. Noticeably at high metallicity and mass with overshooting it is possible to see that neutron stars are formed rather than black holes. This is similar to the findings of \citet{H03}. The transitions from neutron star to black hole are at similar positions in \citet{H03} although our value with overshooting is lower by $5M_{\odot}$.

Comparing to the energy method for determining the remnant in figure \ref{mapJ} we find the same transition point between neutron stars and black holes just below $25M_{\odot}$ without overshooting and $21M_{\odot}$ with overshooting. The structure is also quite similar although it is difficult to decide when direct black hole formation may occur rather than fall back on to a black hole. This would require a denser remnant and our $10M_{\odot}$ remnant line seems to follow the He core mass remnant region for direct black holes fairly well. While looking at these diagrams it is sensible to ask where different types of gamma-ray bursts (GRBs) may occur as discussed by \citet{H03}. It is thought that GRBs require a progenitor with a small radius (a few $R_{\odot}$) so the jets that form from the material accreted on to the central black hole can punch their way through the stellar envelope. If the stellar envelope is too large the jets dissipate and lead to a normal SN. Therefore any of our models that loose their hydrogen envelopes to become WR stars and also form a black hole may also be the progenitors of GRBs. It will be interesting to determine how many stars that fit this requirement give rise to an unseen SN compared to those that give a bright display. This could lead to an observational check if it were known how many GRBs have an associated SN. The remnant masses here should be considered lower limits to the possible remnant mass. The final mass increases during the course of carbon burning so if the model is truncated before then a misestimation might occur, for instance the feature at high masses at $Z=10^{-6}$.

\section{Discussion}

We have now presented the results from our many models and presented our preferred results. But how do these preferred results and all the other tables compare with observations? We have already had one test, the comparison of theoretical plateau lengths against observations. However we now discuss more observational tests and implications for the remnant mass population.

\subsection{Convolution with an IMF - SNe Ratios}
Observations tell us that in S0a--Sb galaxies, the ratio of Type II supernovae over Type Ibc supernovae is $3.82 \pm 2.71$. In Sbc--Sd galaxies, it is between $6.14 \pm 3.96$ \citep{cap1}. At solar metallicity we find the ratio without overshooting to be 8.2, and with overshooting to be 9.0 for a \citet{KTG93} IMF. We get slightly lower values with the NJV rates of 6.9 and 7.8. These are higher than the value of 5 calculated by \citet{H03} but we have a lower minimum mass for a star to go SN. Generally the theory agrees with observation but this basic picture must be altered at lower metallicities because most type Ibc SNe become faint or have no display thereby raising this ratio. However binary systems are likely to dominate Ibc supernova production at low metallicity and this complicates prediction of the SNe type ratio.

\subsection{Observations of SN Progenitors}
\citet{S03a} compiled details on the sample of well studied SN progenitors (shown in Table \ref{ptable}). We can calculate various details to check against these observations. Most match well with our overshooting models, with the JNHV models giving the best agreement. The one problem is that the various types of type II supernovae do not match our IIP predictions. This may be due these stars being binaries, for example recent observations \citep{Maund04} have confirmed the progenitor to 1993J was part of a binary. There are also strong arguments that 1987A was part of a binary \citep{podsi92}. Except for 1993J the ejected masses also match up well to those calculated from observations. As this list grows in length and detail more tests against our maps will become possible.

\begin{figure}
\begin{center}
\includegraphics[height=100mm,angle=270]{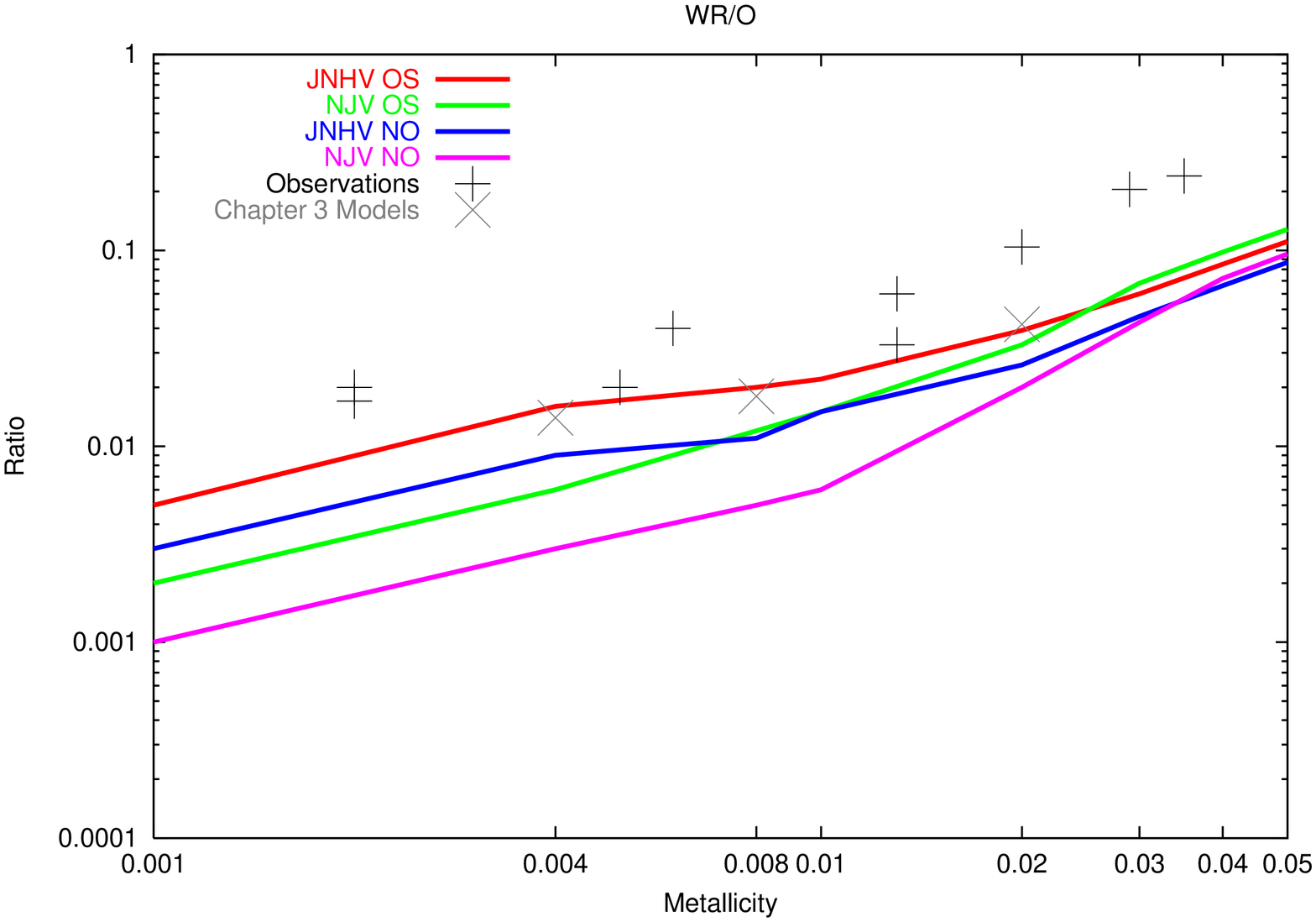}
\includegraphics[height=100mm,angle=270]{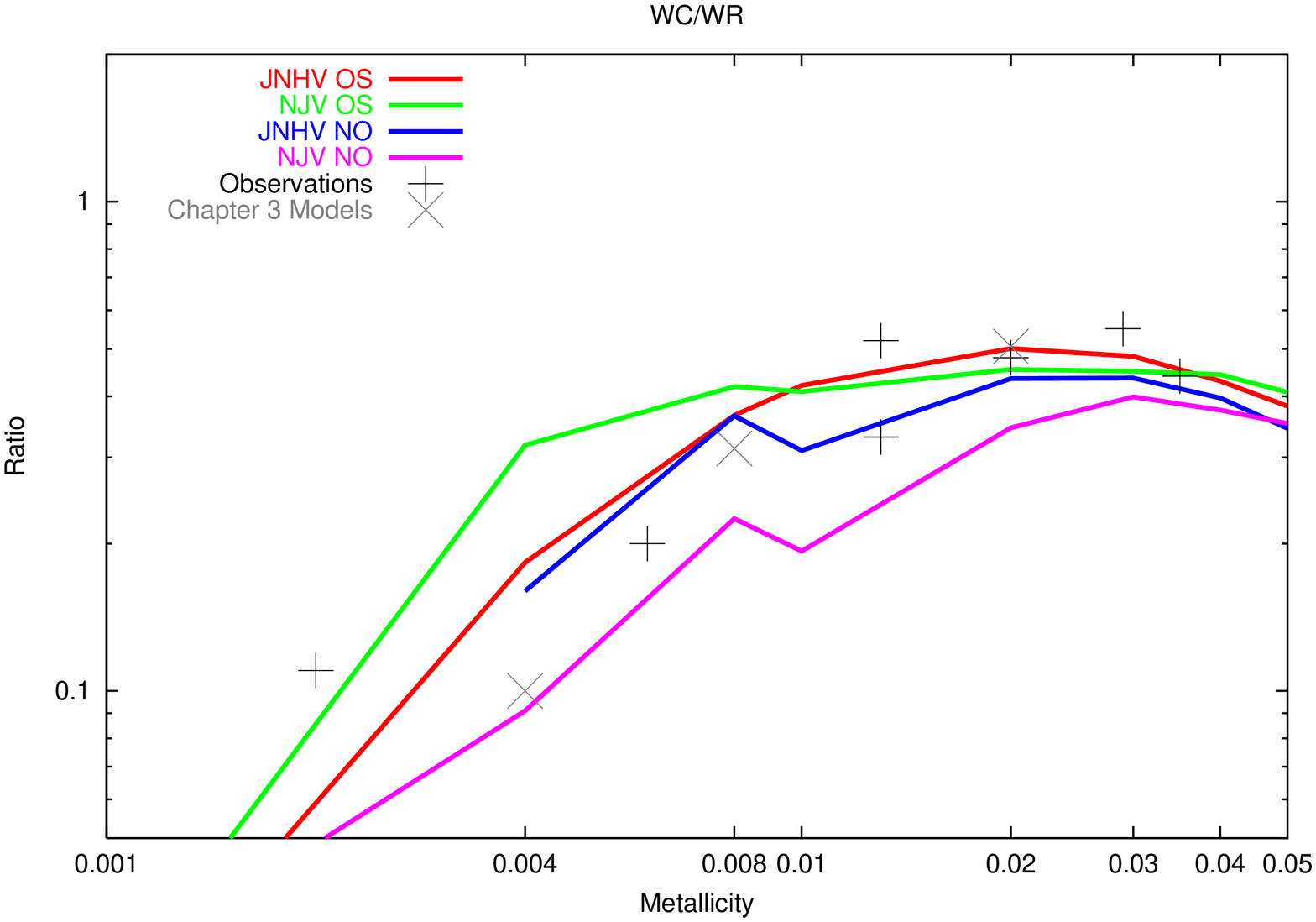}
\includegraphics[height=100mm,angle=270]{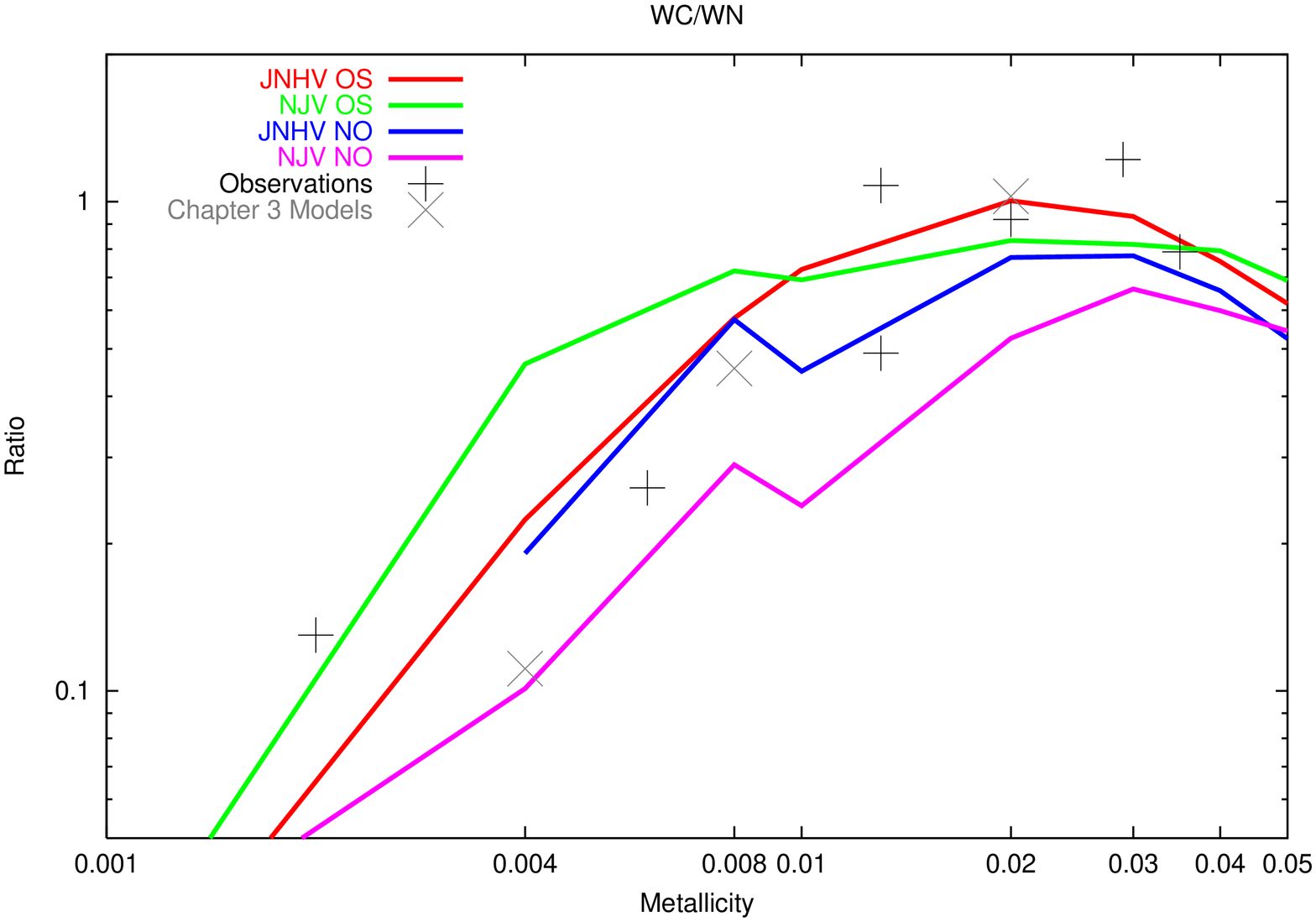}
\caption[Comparison of predicted WR type ratios with observations.]{Comparison of predicted WR type ratios with observations. Lines are calculated from the grid above. `$\times$' are from calculations on the models in chapter 3 and `$+$' are from observations in \citet{MM}.}
\label{wrratios}
\end{center}
\end{figure}

\subsection{WR Ratios}
In figure \ref{wrratios} we show the predicted WR type ratios from models with observations. Assuming constant star formation and using the IMF of \citet{KTG93} we determine how long our models spend in each of the WR types and then determine the relative populations. The theoretical lines are approximate owing to the low resolution in initial mass of our grids. However the JNHV OS results do agree with values from chapter 3 the minor differences are due to the lower resolution in mass of our grids in this chapter. These models do agree well with the observations. The general trend is for the WC/WR and WC/WN ratios to agree better than the WR/O ratio. This is expected because binaries produce WR stars from a lower initial mass due to increasing the opportunities for mass-loss therefore increasing the number of WR stars relative to O stars. But the ratios between WR types would be constant since this depends mass-loss once a star becomes a WR star and being in a binary will only have a slight affect on this. All the mass loss prescriptions provide reasonable agreement however the JNHV with overshooting is closest overall if we also consider the models from chapter 3.

\subsection{Remnant Mass Population}
Figure \ref{mapJ} shows that, at the highest metallicities, the remnants are all of very low mass, due to a maximum remnant mass cut off, with neutron stars dominant. Then as metallicity decreases more massive remnants become possible. Very massive $(M \ge 100M_{\odot})$ black holes can form in SN however this can only happen at low metallicity ($Z \le 10^{-3}$). The population of such black holes is very small because they are only formed in the most massive stars. At higher metallicities massive black holes are not formed because mass loss limits the size of the SN progenitor and thus the size of the remnant. Any black holes observed to be very massive in high metallicity environments must therefore have grown via accretion from a binary companion or must have formed from an earlier population of stars.

\section{Conclusion}

In this chapter we have presented results from numerous stellar evolution calculations with various mass-loss rates. We have highlighted how changing the mass loss affects the final outcome of these stars. We have also discussed in detail our favoured mass-loss prescriptions using the rates of \citet{dJ} with the rates of \citet{VKL2001} for pre-WR evolution. These we compare with models that have essentially the rates of \citet{NL00} with the constant WO rate from \citet{DrayThesis}. Except for the Vink rates which have their own scaling, we scale all the mass-loss rates by $(Z/Z_{\odot})^{0.5}$. We have shown that the rates agree with three sets of observations. Although to constrain our work more we require more data in these observations to reduce the limited number statistics affecting the results.

\begin{table}
\begin{center}
\caption[Comparing values from models to observations of progenitors.]{Comparing values from models to observations of progenitors. Adapted from the table in \citet{S03a} and details from \citet{SJM03}.}
\label{ptable}
\begin{tabular}{|l|ccccc|cc|}
\hline
     & Observed &         &Observed&     & Observed &Maps SN     &Ejected \\
     &   SNe & Observed&Stellar& Initial  & Ejected  &Type       &Mass / $M_{\odot}$\\
SN   &  Type   & $Z/Z_{\odot}$&Type &Mass $/M_{\odot}$& Mass $/M_{\odot}$&JNHV, NJV& JNH, NJV\\
\hline
2002ap & Ic     & 0.5     & WR?     &	$<40$ &2.5-5 &Ic, Ic    &6-7, 7\\
1987A  & II-pec & 0.5     & B3Ia    &  $20$   &15    &IIP, IIL  &15,11\\
1993J  & IIb    & 2       & K       &  $17$   &3     &IIP, IIP  &12, 8\\
1980K  & IIL    & 0.5     & ?       &	$<20$ &  -   &IIP, IIPL&$<12$, $<10$\\
2003gd & IIP    & 0.5     & M       & $8$     & -    &IIP, IIP  &6, 6\\
2001du & IIP    & ~1      & G-M     &	$<15$ &-     &IIP, IIP  &$<12$, $<10$\\
1999em & IIP    & 1-2     & K-M     &	$<15$ &5-18  &IIP, IIP  &$<12$, $<10$\\
1999gi & IIP    & ~2      & G-M     &	$<12$ &10-30 &IIP, IIP  &$<10$, $<9$   \\
\hline
\end{tabular}
\end{center}
\end{table}

\chapter{Binary Progenitors}
	
\begin{center}
``First things first, but not necessarily in that order.''\\
\textit{The 4th Doctor Who, Meglos.}
\end{center}

\section{Introduction}
Even though astronomers would like to believe that stellar evolution is done and dusted nothing could be further from the truth. The previous chapter has illustrated one problem, mass loss, the other great unknowns are rotation, magnetic fields and duplicity.

Rotation and magnetic fields are complex issues and affect the evolution of single stars although \citet{mm03} find they might cancel each other out. A binary companion probably has the greatest effect on a star's evolution. Its presence provides the possibility of mass loss, mass gain and other binary specific interactions (e.g. irradiation, colliding winds, surface contact, gravitational distortion) and hence a greater scope for interesting evolution scenarios. SN2003gd only just fits with our theoretical models as none of our single-star models have a low enough luminosity for an exact match. Therefore we undertook this study to try and find lower luminosity type IIP SN progenitors from binary evolution. 

The basic physics of binaries is deceptively simple until we see the strange and bizarre situations that require complex physics to model. The classic example of a problem binary is Algol. In this system it appears the less massive star is more evolved than the more massive star. The explanation is that mass transfer occurred to swap the mass ratio. However when attempts are made to model this system not all the mass transferred is accepted by the secondary star and a fraction is lost from the system \citep{algolpaper}.

In this chapter we first describe our treatment of binaries in the stellar evolution code. This also includes a discussion of what we haven't included in the code that produced our main results and the differences we can expect if we were to. Then we break up our binary models into similar mass ranges and discuss them in detail, highlighting changes in the evolution outcomes over these ranges. This study is similar to others performed by \citet{podsibin1}, \citet{vanb03} and \citet{izzy}. The first of these is most similar to our own although we use a more up-to-date stellar evolution code. The others use a binary population synthesis which does not use full stellar evolution calculations but instead uses analytic expressions that are fit to the results of detailed models. This has the advantage that they are far more rapid but not all the details of the stars' evolution are known.

To gain an overall picture of the evolution of SNe progenitors we need to have some idea of how duplicity affects them. We discuss the SN progenitors only possible from binaries and not single stars. Because the models also provide details of the evolution of massive binaries we discuss the possibility of X-ray binaries, from mass-transfer events with a compact remnant.

\section{Modelling Binaries}
To truly model a binary a code would have to follow both stars simultaneously. There is a problem however for systems with stars that do not have mass ratios close to 1 in that the process becomes costly in CPU time. The more massive star evolves with shorter time steps and this wastes CPU time on the evolution of the secondary with a longer evolution timescale. In our code we evolve one star at a time. The entire evolution of the primary, say 10,000 models, takes up less than 1000 models when we evolve the secondary. 

The best system would be to evolve one star until something interesting happens then switch back and forth evolving the stars to best save computer time. The system used for our models is a unique method for dealing with the secondary star in the system.

\subsection{Basic Details}
The first piece of the binary physics jigsaw is Kepler's laws. These detail the orbits of the planets around the Sun but they also apply to stars in orbit around each other. The two stars revolve around each other held together by gravity but cannot merge because of angular momentum. Modelling the stars as point masses in a circular orbit we set the centripetal force equal to the gravitational force between them,

\begin{equation}
F_{\rm centri} = \mu a \omega^{2} = \frac{G M_{1} M_{2}}{a^{2}} = F_{\rm grav},
\end{equation}

where $\mu=\frac{M_{1}M_{2}}{M_{1}+M_{2}}$ the reduced mass, $M_{1}$ is the mass of the primary and more massive star, $M_{2}$ the mass of the secondary star, $a$ the orbital separation and $\omega$ is the orbital period. From rearranging the equations and using the fact that $\omega = \frac{2 \pi}{P}$ we arrive at Kepler's law,

\begin{equation}
\Big( \frac{P}{2 \pi} \Big)^{2} = \frac{ a^{3}}{G(M_{1} + M_{2})},
\end{equation}

where $P$ is the period. During the evolution of the binary as a first assumption we can take the angular momentum to be conserved. If this is true then

\begin{equation}
\frac{\rm d}{{\rm d}t}(J)=\frac{\rm d}{\rm dt}(\mu a^{2} \omega) =0.
\end{equation}

With some simple analysis (see appendix \ref{binaryequationstuff}) we can find that mass transfer affects the orbit. For example if the more massive star is transferring mass to the less massive star then the orbital separation shrinks, however if the less massive star transfers mass the separation expands. This only occurs if mass transfer is conservative; if mass is lost from the system such as in a wind then the escaping material also carries away angular momentum. This leads to the widening of the orbit. For the most massive stars that have severe winds the orbit can widen significantly and prevent interactions.

\subsection{Dealing with the Secondary}

We could treat the secondary as a point source with no variables other than mass. This is acceptable for compact objects but is not satisfactory for a star of similar mass to the primary. Our solution was to use the single stellar evolution equations of \citet{HPT00}. These are mostly used in binary population synthesis calculations such as \citet{HPT02}. They are much faster than stellar evolution models, taking a fraction of a second to calculate the evolution of one star rather than tens of minutes.

Using equations enables us to estimate the details of the secondary to check whether it fills its own Roche Lobe and calculate the winds from the star and the amount of angular momentum lost. Once the primary's evolution is complete we recalculate the secondary's evolution with the full stellar evolution code, using the mass loss and gain from the analytic calculations during the primary's evolution. When we reach the point in the secondary evolution when the primary undergoes a SN we reset the binary calculations using a compact remmant with the remnant mass from the primary's calculation.

\subsection{Roche lobe overflow}

In a binary system the equipotential surfaces are not spherical but are distorted by the gravity of the companion star and by the rotation of the system. Solving for the equipotential surfaces in a corotating frame the shape of the equipotential surfaces can be found. If the stars are small compared to the size of the orbital separation these surfaces are spherical but as the star grows in comparison to the separation the shape is distorted to a greater degree becoming more ellipsoidal. The main mass transfer event in binaries occur when the radius of the star is of the order of the orbital separation. A star's surface will eventually reach the L1 Lagrangian point where the gravity of both stars cancels exactly. If a star expands beyond this point then material begins to flow towards the other star. The method to include this in the code is to define a Roche lobe radius, $R_{\rm L1}$, such that the sphere of this radius has the same volume as the material held within the Roche Lobe defined by the equipotential surface passing through the L1 point. Therefore when the star has an outer radius greater than the Roche lobe radius it will transfer material onto the other star. We use the Roche lobe radius given by \citet{E83},
\begin{equation}
\frac{R_{\rm L1}}{a}=\frac{0.49 q_{1}^{2/3}}{0.6 q_{1}^{2/3}+\ln (1+q_{1}^{1/3})},
\end{equation}
where $q_{1}=\frac{M_{1}}{M_{2}}$. It is accurate to within 2\% for the range $0<q_{1}<\infty$. When $R_{1} > R_{\rm L1}$ Roche lobe overflow (RLOF) occurs and we have mass transfer in our system. We use a second simple equation to calculate the rate at which mass is lost from the primary,
\begin{equation}
\dot{M}_{\rm 1R}=F(M_{1})[\ln (R_{1}/R_{\rm L1})]^{3} M_{\odot} {\rm yr^{-1}},
\end{equation}
where
\begin{equation}
F(M_{1}) =3 \times 10^{-6} [min(M_{1},5.0)]^{2}
\end{equation}
and is chosen by experiment to ensure mass transfer is steady \citep{HPT02}. We use this for all RLOF calculations unless common envelope evolution occurs which requires a different scheme.

Once we have taken the mass away from the primary we transfer it to the secondary but not all the mass is necessarily accreted. In the code the accreting matter has the same temperature and entropy as the star's surface. However this is a simplification since the accreted matter will cause the star to expand due to the increased mass of the star increasing energy production. The star will relax to a smaller radius but only on a thermal timescale, therefore we assume that the star's maximum accretion rate is determined by its current mass and its thermal timescale. We define a maximum accretion rate for a star such that $\dot{M}_{\rm 2} \le M_{2} / \tau_{\rm KH}$ where $\tau_{\rm KH}$ is the Kelvin-Helmholtz timescale. If the accretion rate is greater than this then the additional mass is lost from the system so the star only increases in mass at a rate that is determined by its thermal timescale. If the secondary is of lower mass it has a longer thermal timescale than its more massive primary. Efficient transfer is only seen if the two stars are of nearly equal masses so the thermal timescales are similar and the mass ratio changes soon after mass transfer begins so the orbit widens. If the masses are dissimilar we find that most of the mass is lost from the system and the separation of the orbit increases but if the increase is not rapid enough a common envelope phase quickly occurs.

\subsection{Common Envelope Evolution}

If RLOF occurs, but doesn't arrest the expansion of the mass losing star, growth continues until the radius is greater than the binary separation and the secondary is engulfed in the envelope of the primary. This is common envelope evolution (CEE). When this occurs it is thought that the envelope is lost by some unknown dynamical process but the energy for the envelope ejection is transferred from the orbital energy thus shrinking the orbital separation in the process. There is a chance that the two stars may coalesce before the envelope is ejected. Evidence suggests that the progenitor of SN1987A underwent such a merger \citep{podsi92}. The more common outcome is the two stars ending up in a tighter orbit, commonly one helium star, and one main sequence star. If the secondary then evolves a second CEE phase can occur with the outcomes of a merger or a very compact binary.

We require a simple prescription for this complex event. Commonly the energy required to eject the envelope is calculated and this is subtracted from the orbital energy to provide the new separation of the orbit. If there is not enough energy in the orbit to eject the envelope the stars merge. There is some uncertainty in the exact ratio of these two energies because there are other sources of energy that can be put into ejecting the envelope such as the energy from nuclear reactions and reionisation energy of the hydrogen in the envelope. Therefore an arbitrary constant is included to account for the other sources of energy. There are some other prescriptions based on angular momentum, for example \citet{Nelemenstout}.

In our models when CEE occurs we initiate a CEE wind to remove matter as quickly as possible, mass-loss rates of $10^{-3} \, M_{\odot} {\rm yr^{-1}}$ are possible but such rapid mass loss can cause the evolution to be unstable. Therefore we use a rate of $10^{-4} \, M_{\odot} {\rm yr^{-1}}$. The higher mass-loss rate restricts any thermal response from the star  and the envelope is lost in a short period of time, of the order of $1000\, \rm years$, the lower rate only changes the result slightly and allows more stable evolution of the star.

To remove the mass we assume the energy comes from the orbit. For our model we assume that the secondary accretes no mass unless it is a compact object which is an acceptable approximation since our calculations show most common envelopes occur in binaries with mismatched masses where the thermal timescales are quite different.

To determine how the mass loss effects the separation we take the small amount of energy, $\delta E_{\rm binding}$ to remove the mass, $\delta M_{1}$ from the surface as
\begin{equation}
\label{e1}
\delta E_{\rm binding} = -\frac{GM_{1} \delta M_{1}}{R_{1}}.
\end{equation}
We simplify the orbital energy and assume most of the orbital energy is stored in the core of the primary and the secondary. Therefore, by the Virial theorem, the orbital energy is
\begin{equation}
E_{\rm orbit} = -\frac{GM_{\rm 1C}M_{2}}{2a}.
\end{equation}
Differentiating this we find,
\begin{equation}
\label{e2}
{\rm d}E_{\rm orbit} = \frac{GM_{\rm 1C}M_{2}}{a^{2}}{\rm d}a.
\end{equation}
Equating equation \ref{e1} and \ref{e2} and rearranging we find,
\begin{equation}
{\rm d}a = \frac{a^{2}}{R_{1}} \frac{M_{1}}{M_{2}} \frac{{\rm d}M_{1}}{M_{\rm 1C}}
\end{equation}

In this formalism ${\rm d}M_{1}$ is negative so the orbit shrinks. We stop CEE once we have reached the helium core mass before the CEE phase began, at this point we have removed the hydrogen envelope. This is slightly different to the standard approach to just remove the envelope and calculate the final separation using the assumption that $E_{\rm binding}=\alpha_{\rm CE} (E_{{\rm orb,}f}-E_{{\rm orb,}i})$, obtaining the final separation from the final orbital energy. $\alpha_{\rm CE}$ is a free parameter to fix the efficiency of the energy transfer. We could not use this process for a full stellar evolution code as we cannot just remove the envelope! In comparison our method is equivalent to assuming $\alpha_{\rm CE}=1$. Although as we do not remove the envelope in one go the effective result is that the effective $\alpha_{\rm CE}$ slightly less than 1 and our final orbit is larger than from the standard approach. Binary population synthesis codes also use another parameter $\lambda_{\rm CE}$ to quantify the effect of structure of the envelope on the binding energy. We do not have to consider this since with a detailed model we can calculate the binding energy accurately. But in our model we calculate the binding energy of the material begin removed which is perhaps closer to reality.

\subsection{Compact Objects}
These are white dwarfs, neutron stars and black holes. There are changes for when we evolve with these objects. First the maximum accretion rate is derived from the Eddington accretion limit. For black holes we allow accretion ten times this limit to reflect the absence of a surface.

The other main difference is that we allow compact objects to accrete matter during CEE. The phase is so short and the Eddington rate so constraining we find only little change in the mass of the compact objects.

\subsection{Other things: tides, wind accretion, SN, etc...}
We have also included wind accretion in the formalism of \citet{HPT02}. However we find it irrelevant in SN progenitor systems because a slow wind for a long time is required to see an effect and only AGB stars fit these requirements. WR stars may have the strongest winds but they are too fast for a secondary to accrete much material. Winds also carry away angular momentum from the system. We assume any mass lost takes away angular momentum from the orbit.

We deal with SN very simply. We evaluate the ejected mass and if more than half the total mass of the binary is ejected the system is disrupted and the secondary evolves as a single star. Otherwise the secondary evolves with the remnant of the primary and we leave the separation at the separation of the last model for the primary. This gives us the worst case scenario for subsequent evolution. Based on observations of free neutron stars with high velocities it is thought that in SN neutron stars receive a kick . We have not included this in our simple model but it will increase the eccentricity of the post-SN system and disrupt more of these binaries.

We have experimented with tides however we find they are numerically unstable because the code is evolving on a thermal timescale rather than a dynamical timescale near RLOF. They do increase the chance of common envelope evolution but there is a competing effect if the mass transfer during RLOF is inefficient and large amounts of mass are lost from the system. Their inclusion is likely to increase the number of Ib SN from stars that have lost their hydrogen envelopes.

We have not included thermohaline mixing where heavier material is accreted on top of light material and then mixes into the envelope even if there is no convection at the surface. Most of our mass-transfer is inefficient and occurs while the primary ascends the red giant branch and so still has a large fraction of hydrogen in the envelope. However, as we shall discuss, some stars below $15M_{\odot}$ undergo a second RLOF event when the star becomes a helium giant. This is likely to affect our results slightly because thermohaline mixing is thought to be quite efficient occurring on a thermal timescale \citep{thaline2}.

Finally all these models are evolved at solar metallicity with convective overshooting and 199 meshpoints. The winds for the stars use the JNH rates for pre-WR evolution and the NL rates for WR evolution.

\subsection{Mass Ratio and Separation}
We are limited in our resolution in this study by computing time. On the SunGrid machines that we are able to use a single star takes one and a half hours to evolve and if we wish to have an answer in a week then if we use 20 computers simultaneously we have a maximum number of required models of about $2000$\footnote{This means 1000 binaries since there are two stars per binary.}. With a binary population synthesis code this would take about 4 minutes. In mass ratio we choose our grid to be $q=\frac{M_{2}}{M_{1}}=0.1$, $0.3$, $0.5$, $0.7$ and $0.9$. In separation we study $10^{1.5} \le a/R_{\odot} \le 10^{4}$. The lower limit is arbitrary and we assume any stars below this merge during the main sequence and evolve as single stars. The upper limit comes about from experiment to find the separation when there is no interaction. The initial separations we use are $\log (a/R_{\odot}) =$ 1.5, 1.75, 2, 2.2, 2.4, 2.6, 2.8, 3, 3.25 and 4. This gives us fairly good resolution over the entire range of binaries. For a $10M_{\odot}$ primary and with $q=0.5$ this is a range of periods from 5 days to 82 years. Our primary masses are $M_{1}=$5, 6, 7, 8, 9, 10, 11, 12, 13, 15, 20, 25, 30, 40, 60, 80, 100, 120, 150 and 200. We concentrate the resolution at the low end of our range to pick up the minimum mass SNe. The number of models in this grid is $2 \times 5 \times 10 \times 20 = 2000$ hitting our above target. In future more models could be run but for the moment we do not want to take too much time since the code is still a prototype.

\section{Evolution of Massive Binaries}
We will now describe the evolution of the binaries in our study. We will pay particular interest to the types of interaction and how these might affect the SN progenitors. The evolution of binaries is extremely non-linear, similar behaviour may occur over large ranges of masses and separation while a tiny change will cause a quite different outcome.

The mass transfer interactions are our main interest in for the evolution of binaries, the primary losing mass and the secondary gaining mass. This will radically affect the outcome of evolution. It may change the SN type, prevent the SN or produce a SN where none would have occurred.

\subsection{$5 \le M_{1}/M_{\odot} < 9$}
\label{thelowmassones}
In this range we only see a primary SN if $M_{1} \ge 8M_{\odot}$. Any star below this mass evolves towards the AGB phase and if there is no interaction will lose its envelope during thermal pulses. The interactions that occur are roughly equal occurrences of CEE and RLOF for the primary while secondaries experience more RLOF than CEE. This is because during primary evolution the masses are more mismatched so mass transfer is unstable while for the secondaries they have more similar masses to the remnants so mass transfer is stable in general. 

It is common for there to be two interactions during the evolution of the primaries. Once they lose their hydrogen envelopes by CEE or RLOF they become helium dwarfs then later after core helium burning they become helium giants expanding to the point where they can again interact. The second interaction is stable RLOF as the masses become roughly equal after the first interaction. After losing mass these stars form a COWD. The secondary may also experience this behaviour with the remnant from the primary. This behaviour only occurs if the mass is above about $3M_{\odot}$ otherwise the star never becomes a helium giant after the first mass transfer event. 

SNe that do occur depend on the mass of our stars. At $7M_{\odot}$ some SNe do occur but they are mostly Ib and IIL when interactions have stripped the envelope and prevented 2nd dredge-up and the AGB phase. We therefore have helium stars more massive than $M_{\rm Ch}$ that may go SN. We have set the minimum Ib mass at $1.6 \, M_{\odot}$. From $8M_{\odot}$ and above we see many more SNe because the stars would normally go SN if they were single stars. Most SNe are type Ib if an interaction occurs and IIP if there is no interaction. There is a small band of IIL SNe separating the two regions. If we only see Ib and IIP and no IIL or IIb then we assume that the stars in between the separation where the change over occurs give rise to IIL SNe. The closest binaries prevent SNe occurring by removing the hydrogen envelope at first dredge-up leaving helium stars too low in mass to undergo core-collapse. 

The secondaries are far more intriguing. Many of the secondaries in this range are not massive enough to undergo SNe if they were single stars. Despite this some of them do. When the mass ratio is close to one we find that mass transfer is very efficient because the thermal timescales of the two stars are similar. This can increase the mass of the secondary to a point where it becomes massive enough to go SN. We find these stars when $q \ge 0.7$ and at separations of a few $100R_{\odot}$. The more massive the primary the greater the separation must be. We find less than 10\% of systems exhibit this behaviour. For this to occur the total mass of the binary must be a few solar masses greater than the minimum mass for a single star SN.

Such binaries are similar to those discussed by \citet{DRK02} but these stars do not undergo a second interaction when the secondary evolves. The reason for this comes from angular momentum conservation. Because the mass transfer is so efficient the angular momentum of the system is almost fully conserved. In these systems, soon after the mass transfer occurs, enough mass has been accreted on to the secondary for it to be the more massive star. As the less massive primary now continues to transfer mass the orbit widens. Now the orbit is wider it prevents interaction between the secondary and the remnant of the primary. The secondary is now around $8M_{\odot}$ and evolves towards a type IIP SN. The main interest in these stars are that they appear to have a lower luminosity than single stars however upon closer inspection our models have failed just as second dredge-up is occurring. They are likely to continue and form extreme S-AGB stars. The helium core masses are lower than for single stars of the same mass because they were initially low-mass stars. However the CO and ONe cores are above $M_{\rm Ch}$ as are those of the extreme S-AGB stars mentioned in chapter 3. The details are listed in table \ref{lowL} and all fit the observed values of SN2003gd at this time. However if they do experience 2nd dredge-up this will move them far outside of this region.

\subsection{$9 \le M_{1}/M_{\odot} < 15$}
In this range we still find the two binary interactions during the evolution of the primary although by $13M_{\odot}$ they become rare. Primaries still interact by RLOF and CEE in roughly equal probability while for the secondaries there are still more RLOF events than CEE events. We also now find none of the secondary SNe discussed above because the secondaries of similar mass are massive enough that they would go SN without aid from their primary. From $10M_{\odot}$ upwards the primaries always undergo some form of SN. If there has been interaction these are Ib SN if not IIP. The band of SN progenitors where only a small fraction of mass has been removed for a IIL SN is again small. However this is likely to be the most common source of IIL SNe because the IMF favours low-mass stars.

The Ib SNe in this region are very different to the Ib and Ic SNe seen from single stars. They are normally helium giants rather than dwarfs due to their lower mass. Their surface mass fractions of helium are over 0.9 so this would easily be observed in the SN. It is possible therefore that there are two types of Ib SN, giants and dwarfs that are less and more massive respectively. If we use the prescription of \citet{H03} discussed in chapter~4 to determine how bright these SNe should be many of these Ib SN are highly luminous, were with our preferred rates we do not see any bright Ibc SNe from single stars at solar metallicity. This can be understood from the fact that the smaller stars have more tightly bound envelopes than the larger giants. Therefore dwarfs eject less mass than a similar sized giant with the same explosion energy. More ejected matter will provide a more visible display.

\subsection{$15 \le M_{1}/M_{\odot} < 30$}
We now start to see more CEE events than RLOF events in this range. We also see the end of the primary undergoing two interactions with the secondary because the helium stars formed from the first mass-loss event are too massive to become giants. In contrast the secondary can still exhibit this behaviour.

The total fraction of interactions now increases with even the very wide binaries, $R>10^{3}R_{\odot}$, now interacting because the massive giants become extremely large. Most interactions still lead to a Ib or Ic SN. Very few binaries are only slightly stripped of their envelopes to leave hydrogen remaining and give rise to a IIL SN.

Stars stripped of their hydrogen envelope in this region also give rise to WR stars and therefore increase the WR/O ratio, although only the most massive are likely to progress far along the WR evolution track to WC stars. For that the WR star must be massive enough to have a strong wind of its own; this will therefore mainly increase the fraction of WN stars.

\subsection{$30 \le M_{1}/M_{\odot} < 100$}
Stars in this range would normally become WR stars and mostly give rise to Ic SN if they were single stars. Most interactions take the form of CEE during the RGB stage of evolution which is when these stars would normally lose their envelopes in very strong winds. This makes evolution in this region highly uncertain. The combination of strong wind mass loss as well as CEE mass loss rapidly accelerates the formation of the WR star but does not change the final outcome. 

If the interactions occur just after the main sequence, evolution is altered to a greater degree because the helium core has only just formed. Removal of the envelope at this point gives rise to a slightly less massive WR star although the evolution still leads to the same end point.

Evolution starts to change at $80M_{\odot}$ when the mass loss becomes extreme and the time on the RGB is extremely brief. This means the binaries must be closer to interact because the separation is increased by the loss of angular momentum in the stellar wind and the star does not grow to a large radius on the RGB.

\subsection{$100 \le M_{1}/M_{\odot} \le 200$}
The evolution of these stars is less spectacular. Mass loss is so great that, unless the stars have a separation below $100R_{\odot}$, there is no interaction at all, even during the evolution of the secondaries! At the time of the primary SN the separation is normally greater than $1000R_{\odot}$. Therefore we know at solar metallicity that all the interesting binary SN progenitors must occur at masses less than $100M_{\odot}$. That is still quite a large range to study.

\section{Oddities}

We now discuss some of the oddities and outcomes of the above evolution that are important for SN progenitors.

\subsection{Low Luminosity Progenitors}
As discussed in section \ref{thelowmassones} we did find some progenitors that did have final low luminosities but it is possible that they may actually undergo second dredge-up before core collapse. However this is not certain and will require more study of these systems. All these stars fit well with the observations of SN2003gd, the low luminosity red giant progenitor. One feature that is likely to affect these results of these models is the composition of the material we accrete onto the secondary. Currently we accrete material with the same composition as the star itself but in reality this would be slightly different and may even prevent second dredge-up.

The only progenitors that were of significantly lower luminosity were those of Ib SN where having removed the hydrogen envelopes. The removal of hydrogen burning as an energy source decreases the luminosity. None of these models could explain SN2003gd since it was a type IIP SN but they do have implications for observations of low mass Ib progenitors.

\begin{table}
\caption{The final details of the IIP low luminosity and possible S-AGB progenitors.}
\label{lowL}
\begin{center}
\begin{tabular}{|ccc|cccc|}
\hline
$M_{1}/M_{\odot}$  &  $M_{2}/M_{\odot}$  &  $a_{\rm i}/R_{\odot}$ & $\log (L/L_{\odot})$& $\log (T_{\rm eff}/{\rm K})$& $\log (R/R_{\odot})$& $M_{\rm final}/M_{\odot}$\\ 
\hline         
          5    &    4.5  &      100   &    4.52 & 3.54 & 2.71  &   8.1		\\
          5    &    4.5  &   158.49   &    4.46 & 3.54 & 2.67  &   7.7\\
\hline
          6    &    4.2  &   158.49   &     4.60& 3.53 & 2.75  &    8.2\\
          6    &    4.2  &   251.19   &    4.52 & 3.54 & 2.71  &    7.9\\
          6    &    5.4  &   251.19   &     4.60& 3.54 & 2.75  &    8.7\\
\hline
          7    &    4.9  &   398.11   &    4.46 & 3.54 & 2.68  &    7.7\\
          7    &    6.3  &   398.11   &    4.57 & 3.54 & 2.73  &    8.4\\
\hline
          8   &     7.2  &   251.19   &    4.51 & 3.54 & 2.69  &    7.9\\
\hline
\end{tabular}
\end{center}
\end{table}

\subsection{Type I progenitors}
Of main interest to type I progenitors is the fact that we find another class of distinct progenitors that have no hydrogen. The three types are as follows,
\begin{itemize}
\item $Y_{\rm surface}<0.3$, $R/R_{\odot}<10$ and $M>6M_{\odot}$, from single stars $M_{\rm initial} \ge 31M_{\odot}$.
\item $Y_{\rm surface}>0.3$, $R/R_{\odot}<10$ and $M>6M_{\odot}$, from single stars $28 \le M_{\rm initial} < 31M_{\odot}$.
\item $Y_{\rm surface}>0.9$, $R/R_{\odot}>10$ and $M<6M_{\odot}$, from single stars $M_{\rm initial} < 25M_{\odot}$.
\end{itemize}

There is probably a continuum of objects between the first two types but the third type are fairly homogeneous although the more massive they are the smaller their radius. Our dilemma is now how to assign these to the observed types of Ib and Ic? This is difficult and anything we say is a guess. Many more detailed observations of type Ibc SNe are required to have a large enough sample to begin to distinguish the SNe from different progenitor types. Further the three types above do not include the more massive ($M>20M_{\odot}$) helium-rich progenitors at lower metallicity. However they may be too massive to have a bright display, ejecting little mass, and will be difficult to observe.

The guess we have made is that the main determinant is the fraction of helium in the surface since it is a good indication of the amount of helium in the progenitor. Therefore we stick with the Ib, Ic and Ibc definitions we made in the previous chapters. SN1993J was a type IIb SN and provides some evidence for this. The progenitor was a subgiant with a large fraction of helium in the envelope. In late times it had the appearance of a Ib SN so take away the hydrogen and we move towards a typical Ib SN where we need a large star and high surface helium fraction. However there is probably a continuum of objects and more detailed observations may indicate the observable differences in the SNe of these objects.

\subsection{Type II progenitors with stripped envelopes}
We find very few stars that retain hydrogen after interaction with a companion. Most retain only a tiny fraction of hydrogen. This could be due to having very strong limits on common envelope interaction. However this is necessary because the amount of hydrogen a star needs to become a giant is very small. \citet{podsibin1} found that IIL SNe were most common when the mass ratio was close to 1. The resolution of our study is quite poor and this has probably limited out ability to find these objects. Our total percentage of IIL/IIb SN however does reasonably match the results of \citet{podsibin1} despite its approximate nature. In our calculations below we take IIL SN to be all those that lose the majority of their envelopes but retain a small fraction, such that $M_{\rm H}<2M_{\odot}$, and become type IIL SNe. We make no differentiation between IIL and IIb with our data.

\subsection{ULXs}
ULXs are ultra luminous X-ray objects observed in other galaxies. They appear to be compact objects accreting at super Eddington rates onto a stellar mass black hole. The initial suggestion was that they are intermediate mass black holes (IMBH) with masses $100-1000M_{\odot}$. The problem with this is that there are not enough IMBHs in star forming regions where ULXs seem to be concentrated. \citet{king04} suggests that ULXs are normal X-ray binaries but are accreting at super-Eddington rates owing to thermal timescale mass transfer. By observing the Cartwheel galaxy they place limits on the system requirements for an ULX. These are that $M_{2}>15M_{\odot}$ and that, because the emission is thought to be observed along a jet axis, there must be many more of them, around 3000 in the Cartwheel galaxy. The perfect source would be massive stars in a CEE phase involving a black hole.

CEE is known to occur with white dwarfs because there are many double helium dwarf systems known \citet{Nelemenstout}. It is hypothesised that neutron stars can also undergo CEE with main-sequence or giant stars and if the two coalesce a giant is formed with a central neutron star, Thorne-\.Zytkow objects \citep{TZ}. However little is said about the interaction of a massive star and a black hole if a CEE occurs. When RLOF occurs on to a black hole an accretion disc which is hot enough to produce X-rays is formed . The more material the black hole accretes the greater the X-ray luminosity. These X-rays affect the structure of the donor star. \citet{PP02} calculate the effect the accreting luminosity has on the shape of the donor. They show the structure is shrunk and the star is pushed away from the black hole. Material moves further out towards the second Lagranian point. Therefore it is feasible that black holes can avoid CEE because rapid mass transfer on to a black hole produces X-rays that prevent or limit the rate of mass transfer. This is probably the true limit to accretion on to black holes rather than the Eddington limit.

Such systems are limited to those where the primaries form black holes initially. The stars must then be close enough after the first SN to interact. Such stars are found in the range $20<M_{1}/M_{\odot}<80$. Less massive stars do not form black holes, more massive stars form systems that are too wide. At the low-mass end of this range few stars have a large enough secondary to fit the requirements of \citet{king04} while at the higher-mass end most of the stars interact but some become WR stars and thus the situation is further complicated. To fully study this region of possible systems a more accurate treatment of CEE and accretion on to black holes is required.

\subsection{GRBs}

GRBs are thought to be closely associated with SNe since SN1998bw and GRB980425 were observed to be so associated. There are a few others where it is thought the two events are also related. Binaries are the most likely progenitor for two reasons. First in the collapsar model, where a black hole is formed and an accretion disc forms to fuel the black hole in its production of the jets that give rise to the observed display, the star must be rotating rapidly enough to form the accretion disc otherwise there is not enough angular momentum to form a disk and material accretes onto the central black hole spherically. Binary systems that have undergone a CEE phase fit the bill since the orbit is shifted to smaller radius and more rapid rotation if we assume the stars become tidally locked \citep{izzy}.

The second reason is that if the progenitors are WR systems these are extremely dusty and dirty environments with a lot of circumstellar material from the WR wind. A standard WR wind is not observed in the afterglows. If we assume the GRB jets are produced along the rotation axis we can clear this material away but if not the gravity of the secondary star and the rotation may be necessary to clear a region through which the jet can shine.

Importantly if we make the assumption that a black hole must form directly for a GRB to occur then we would not see any such systems at solar metallicity. At LMC metallicity we would see an initial mass of $35M_{\odot}$ or greater, while at SMC metallicity the mass limit drops further. Therefore it is likely that we need a low-metallicity environment for GRBs. This agrees with observations with the lowest redshift GRBs being GRB980425 with a redshift of $z= 0.0085$ and GRB031203 with a redshift of $z=0.1055$. Then also the CEE evolution would be more common at low metallicity because the winds from the stars are lower so there would be more massive giants that can engulf their companions and undergo CEE to shrink their orbit and spin them up.

\section{Binary SN Population Results}

Rather than trying to fit individual systems with our binary models we look at the SN ratios that we predict. For each primary mass we work out how many of the systems give rise to SNe. The remainder give rise to white dwarfs or are lost in a merger event. To calculate the relative rates we use a few different mass ratio and separation distributions. We use three main mass distributions; the first is flat in $q$, whilst the others are based on the distribution of \citet{Hog} where the probability is flat when $q<0.3$ and above this range the probability is proportional to $q^{-2}$. This therefore favours low-$q$ systems and is used by \citet{vanb03} and we denote it as Hog. It does however go against the observational evidence that most massive star binaries have components near equal mass. Although it will be difficult to detect low-mass secondaries to massive stars because by homology relations $L \propto M^{3}$ so a secondary with a tenth the mass it is three orders of magnitude fainter. To see the effect of biasing similar masses in binaries we use an Antihog distribution which is flat when $q>0.7$ but proportional to $(1-q)^{-2}$ above this range.

For separation we use a flat distribution in $\ln a$ and compare this to that derived from \citet{Hog} where the probability of a separation is proportional to $a^{-1.7}$. Again we use a Antihog distribution to favour large separation systems. Combining all these systems we are able to draw some conclusions about how close we can get to the observed SN ratios.

For comparison we use the observational details from \citet{cap1} and \citet{cap2} that the current ratio of type II SNe to type Ibc SN is $5.00 \pm 3.45$ and that the rate of IIP to IIL SNe is the range roughly equal. However \citet{IILstats} suggests that IIL SNe could be less common as low as only a tenth of all type II SNe. We can also place some limits by counting the number of different SN that are have been observed. If we do this we find that there are roughly twice as many IIP SN than IIL. There are no firm limits on the ratio of type Ib to Ic, mainly because the distinction between the two types is poorly defined.

The results are given in table \ref{ch51}. We list the percentage of all binaries that undergo type II SN, type Ibc SN, merge or produce white dwarfs. We also list the ratios of some of the SN types. We combine all the possible separation and mass distributions to compare the affect of varying these. Tables \ref{ch52} and \ref{ch53} present the same information but for the primary and secondary star separately. It is possible to see the different trends from the two halves of the binaries.

When we compare between the different distributions we see that the main effect is the separation distribution. This is because with the separation we have more resolution and the range is much wider than for mass ratio. The mass ratio distribution however can provide fine tuning. A bias to high or low mass binaries only affects the ratios to a small degree. The largest impact is on the overall percentage of SN from binaries, as would be expected because if we bias to low-mass secondaries most of these will have too low a mass for a SN to occur. 

When comparing separation distributions we see that the flat and Hog distributions are similar in the ratios. Also the flat distribution has relatively more SN and the II/Ibc ratio is close to unity. The Hog distribution removes a number of SN and also removes a large fraction of type II SN by biasing close binaries that interact. It also features more mergers which eliminates the second SNe from secondaries.

The II/Ibc ratio is perhaps the most interesting. Only the Antihog separation distribution results are within the error bars of the observed systems. This gives only one conclusion, single stars are very important. \citet{vanb03} states that most massive stars are likely to be in binaries and thus single-star studies are only of academic interest. However if our models are to be believed this means that in fact we must include a relatively large fraction of single stars to produce what is observed. Thus massive single star evolution is of much interest.

The Ib/Ic ratio in all systems is greater than one. This indicates that there are more Ibs than Ics. This can be understood in that Ibs are produced by the more populous low-mass stars while the Ics come from higher mass stars. Observations of this ratio are poorly constrained. Also massive type I progenitors might not have any observable display. The details of type I SNe are also confused by many different progenitors that fit the requirements for these SN.

The IIP/IIL ratio is the most worrying. For none of our systems can we get lower than a ratio of 2.12. There are two solutions to this problem. Either there are fewer IIP SNe in our models than we think; some are IIn SN and some are IIpec, so it is possible to reduce this ratio if we say around a quarter to a half of what we think are type IIPs are something else so our ratio is the ratio of all other type II SN to IIL SN. The value for this ratio \citet{podsibin1} provide is $ 4.75$. All are within the $\frac{1}{10}$ to $\frac{1}{2}$. One problem with observations are how complete are they? If there are low luminosity IIP SNe that form black holes they could be missed and thus decrease the fraction of IIL SNe by increasing the total type II population. The other less favourable scenario is to produce more type IILs from low mass S-AGB stars that go SN. 

To decide which schemes are the best is difficult. Those which match the observations best are those with a flat separation distribution in $\ln a$ with a flat or Antihog mass ratio distribution.

\begin{table}
\caption{Results for evolution of primary and secondary.}
\label{ch51}
\begin{center}
\begin{tabular}{|l|l|cccc|ccc|}
\hline
 & &SNII &SNIbc&Merge&White& & &\\
$q=\frac{M_{2}}{M_{1}}$&$a/R_{\odot}$&/  \%& /  \%&/\%&Dwarfs/\%&Ib/Ic&II/Ibc&IIP/IIL\\
\hline
Flat& Flat               &     14.01  & 14.14   &     2.72 &  69.13  &    3.62 &  0.99  &      3.43\\
Hog &Flat                &     11.76  & 12.17   &      3.8 &  72.27  &    3.41 &  0.97  &      4.16\\
Antihog& Flat            &     17.56  & 15.93   &     2.38 &  64.13  &    3.55 &   1.1  &      3.09\\
\hline
Flat& Hog                &      6.37  & 18.52   &     7.67 &  67.43  &    4.82 &  0.34  &      2.69\\
Hog& Hog                 &      5.98  & 15.98   &    10.74 &  67.31  &    4.47 &  0.37  &      4.95\\
Antihog& Hog             &      8.73  & 20.75   &     6.77 &  63.74  &     4.7 &  0.42  &      2.12\\
\hline
Flat& Antihog            &     20.79  &  8.31   &     0.45 &  70.46  &    1.78 &   2.5  &      5.89\\
Hog& Antihog             &     17.84  &  7.31   &     0.62 &  74.23  &    1.71 &  2.44  &      6.45\\
Antihog& Antihog         &     24.45  &  9.22   &     0.38 &  65.95  &    1.71 &  2.65  &      5.57\\
\hline
\end{tabular}
\end{center}

\caption{Results for evolution of primary.}
\label{ch52}
\begin{center}
\begin{tabular}{|l|l|cccc|ccc|}
\hline
 & &SNII &SNIbc&Merge&White& & &\\
$q=\frac{M_{2}}{M_{1}}$&$a/R_{\odot}$&/  \%& /  \%&/\%&Dwarfs/\%&Ib/Ic&II/Ibc&IIP/IIL\\
\hline
Flat& Flat               &      8.75  &  10.9   &        0 &  30.35  &    3.37 &   0.8  &      3.87\\
Hog& Flat                &      9.22  &  10.3   &        0 &  30.48  &    3.21 &  0.89  &      4.45\\
Antihog& Flat            &      8.75  & 11.24   &        0 &  30.01  &    3.34 &  0.78  &      3.59\\
\hline
Flat& Hog                 &      3.7  &  14.6   &        0 &   31.7  &    4.13 &  0.25  &      4.38\\
Hog &Hog                  &     4.96  & 13.81   &        0 &  31.23  &    4.05 &  0.36  &      6.41\\
Antihog &Hog              &     3.59  &  14.9   &        0 &  31.51  &    3.81 &  0.24  &      3.87\\
\hline
Flat &Antihog             &     14.2  &  6.39   &        0 &  29.41  &    1.73 &  2.22  &      6.05\\
Hog& Antihog              &    14.28  &  6.15   &        0 &  29.57  &    1.65 &  2.32   &     6.51\\
Antihog&Antihog          &    14.25  &  6.53   &        0 &  29.23  &    1.75 &  2.18   &     5.79\\
\hline
\end{tabular}
\end{center}

\caption{Results for evolution of secondary.}
\label{ch53}
\begin{center}
\begin{tabular}{|l|l|cccc|ccc|}
\hline
 & &SNII &SNIbc&Merge&White& & &\\
$q=\frac{M_{2}}{M_{1}}$&$a/R_{\odot}$&/  \%& /  \%&/\%&Dwarfs/\%&Ib/Ic&II/Ibc&IIP/IIL\\
\hline
Flat& Flat                &     5.27 &   3.24   &     2.72 &  38.77  &    4.71 &  1.62    &    2.87\\
Hog& Flat                 &     2.54 &   1.87   &      3.8 &  41.79  &    4.91 &  1.36   &     3.33\\
Antihog& Flat             &     8.81 &   4.69   &     2.38 &  34.12  &    4.14 &  1.88    &     2.7\\
\hline
Flat& Hog                 &     2.67 &   3.92   &     7.67 &  35.73  &   10.75 &  0.68    &    1.58\\
Hog &Hog                  &     1.01 &   2.17   &    10.74 &  36.08  &   10.56 &  0.47   &     2.03\\
Antihog& Hog              &     5.14 &   5.85   &     6.77 &  32.23  &    9.79 &  0.88   &     1.51\\
\hline
Flat &Antihog             &     6.59 &   1.93   &     0.45 &  41.04  &    1.93 &  3.42   &     5.58\\
Hog &Antihog              &     3.56 &   1.16   &     0.62 &  44.67  &     2.1 &  3.08   &     6.23\\
Antihog &Antihog          &    10.21 &    2.7   &     0.38 &  36.72  &    1.63 &  3.79   &     5.28\\

\hline
\end{tabular}
\end{center}
\end{table}

\section{Conclusion}

We have produced some results from simple binary evolution calculations of massive stars. From the results we have made predictions of the percentage outcomes and SN ratios. These fail to match observation indicating that we do need to include a non-negligible fraction of single stars to match observations. Also the ratio of IIP/IIL ratios is of particular interest to us. We also find only a few low luminosity binary SN progenitors that still have their hydrogen envelopes. Although there is uncertainty in the final evolution of these objects since they are possibly examples of extreme Super-AGB stars that may undergo second dredge-up before core collapse and therefore be very luminous prior to SN.

\chapter{Mixing the Progenitors}

\begin{center}
``Typical human, you can always count on them to mess things up.''\\
\textit{The 7th Doctor Who, Remembrance of the Daleks.}
\end{center}
\section{Introduction}
In this penultimate chapter we bring together the results from previous chapters to produce an overall picture of our work. First we add details on the lowest-mass progenitors, using results for S-AGB progenitors from chapter 3, to the results from chapter 4. On top of this data we plot details of observed progenitors to date and discuss the implications of our models for observations for the progenitors of SNe. Next we use all our single star data to discuss the implications for observation of progenitors and the mass distribution of the remnant black holes.

Finally we combine the results from chapter 5 with the detailed single star models from chapter 3 to gain some idea on the relative populations of progenitors from single and binary stars, trying to fix the relative population to observed SN ratios.

\section{The full range of single star progenitors}

\begin{figure}
\begin{center}
\includegraphics[height=79mm,angle=0]{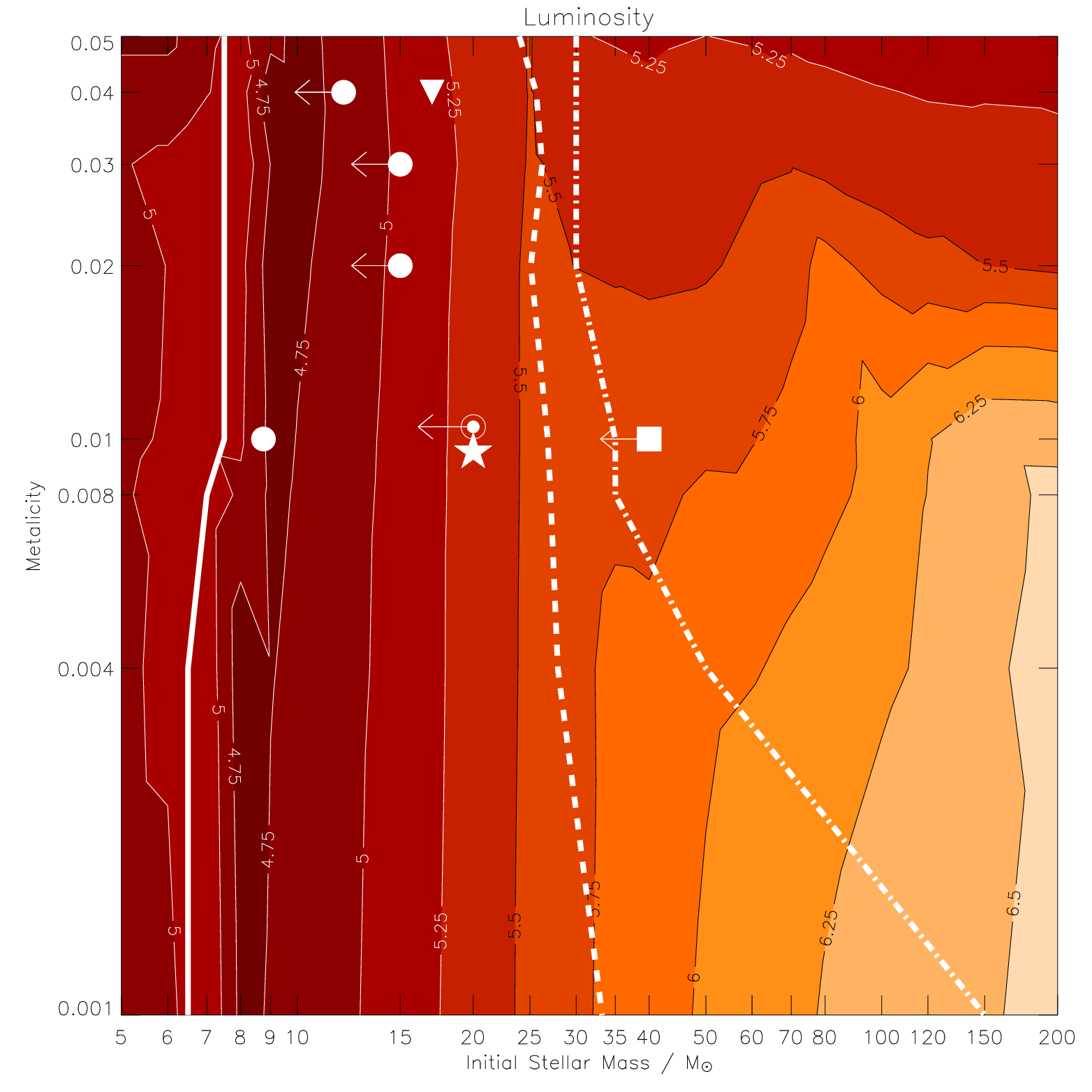}
\includegraphics[height=79mm,angle=0]{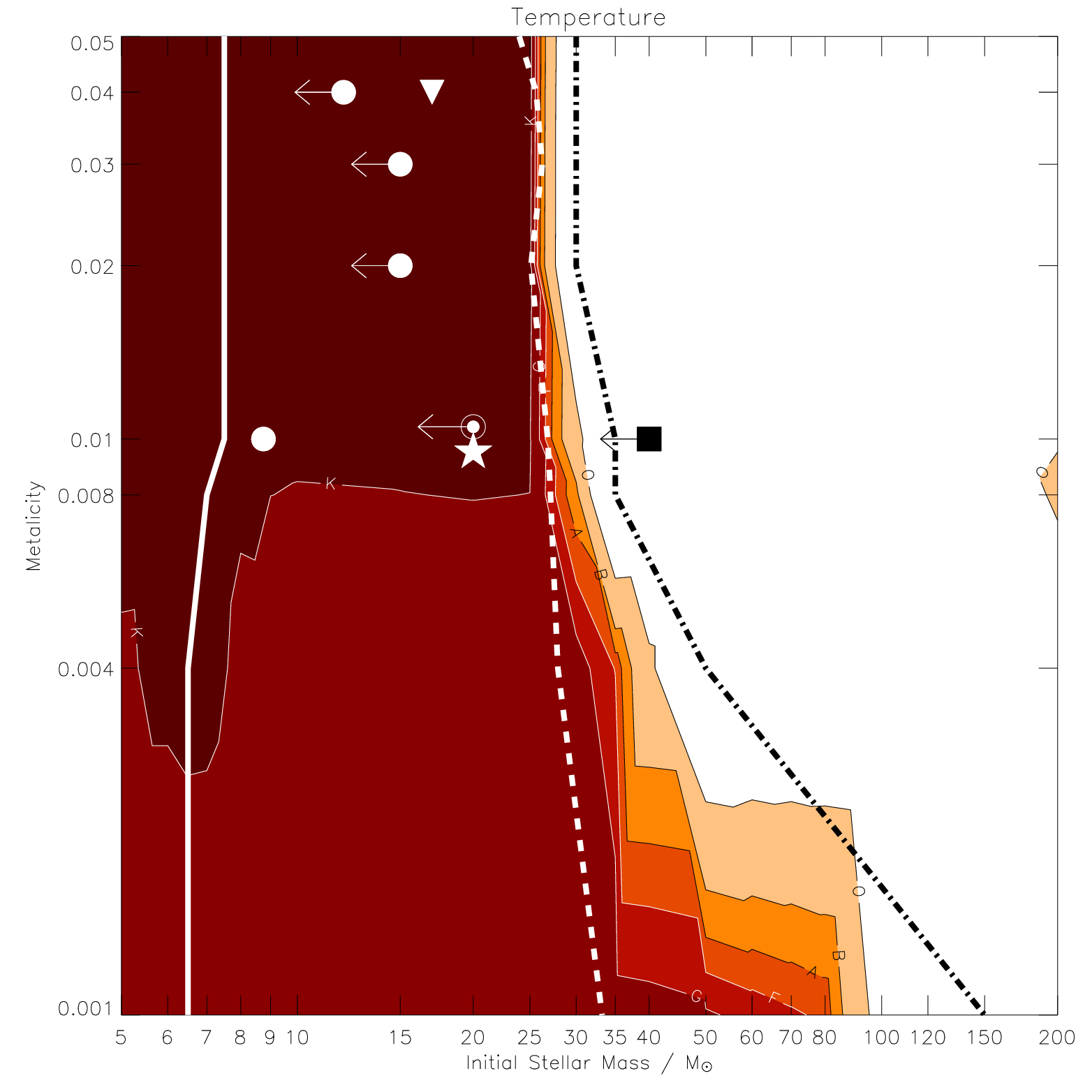}
\includegraphics[height=79mm,angle=0]{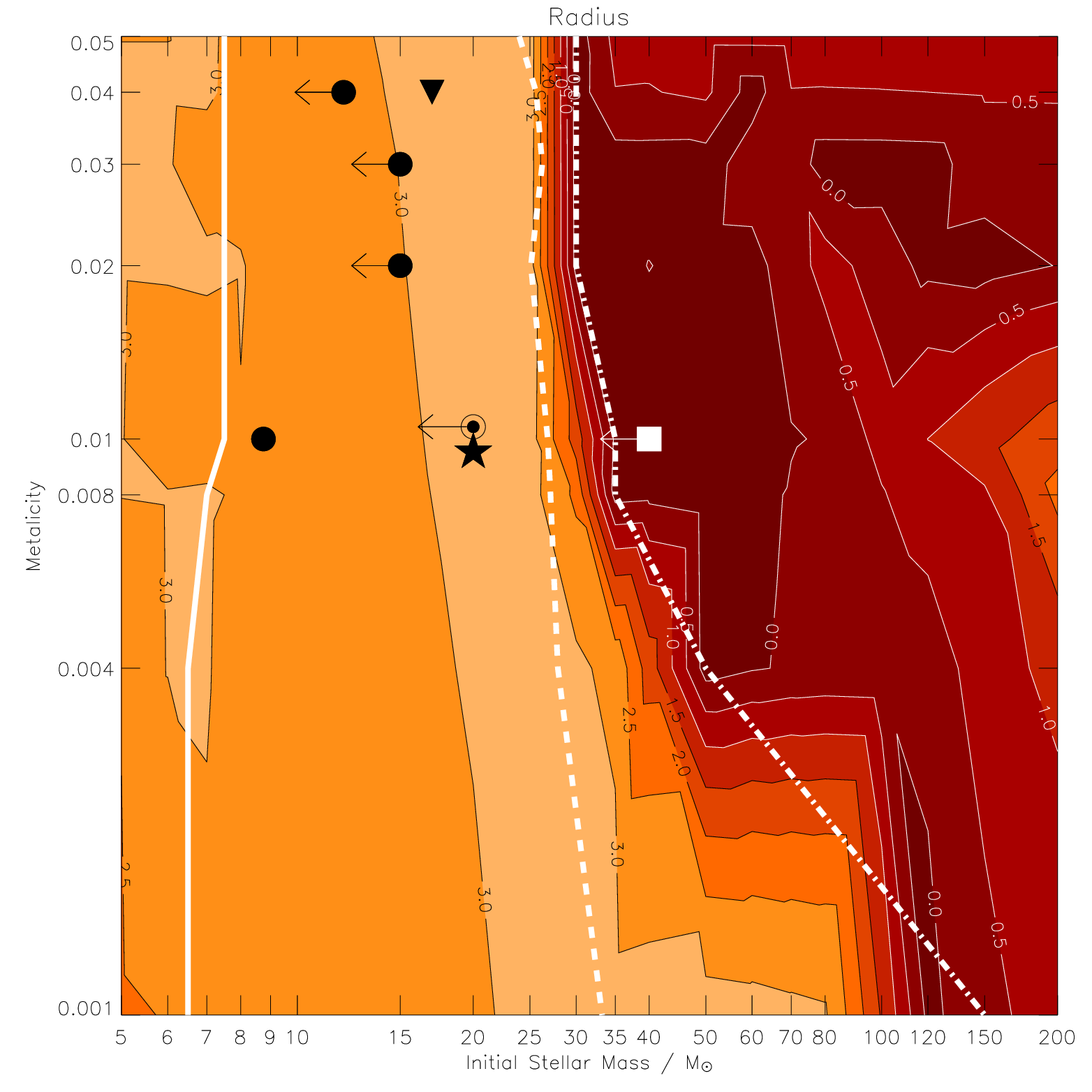}
\includegraphics[height=79mm,angle=0]{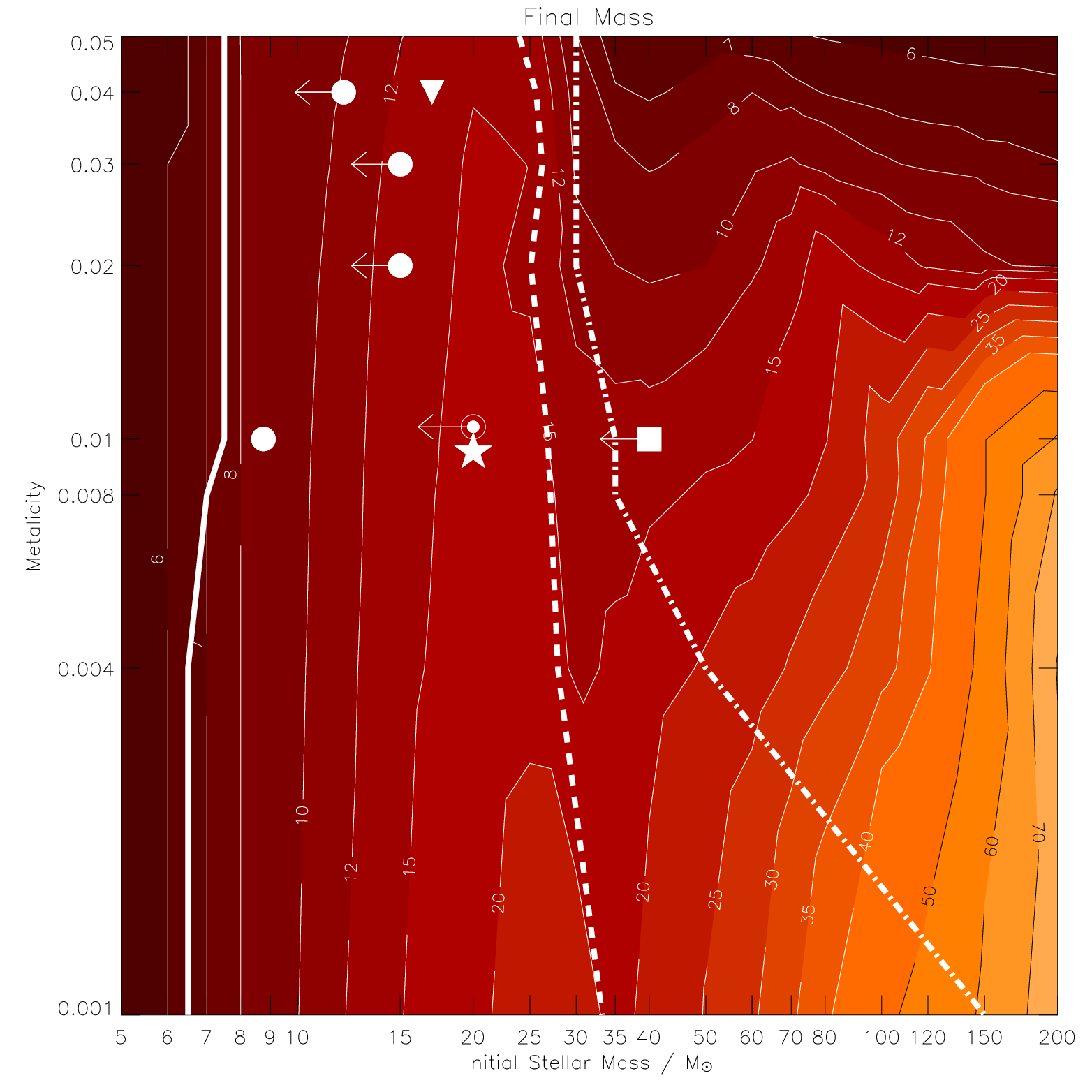}
\caption[Grids for comparison with observations of SN progenitors, luminosity, spectral type, radius and pre-SN mass.]{Grids for comparison with observations of SN progenitors. For all grids the thick solid white line is the minimum mass for SNe. The dashed line separates type II SNe with $M_{\rm H}>2M_{\odot}$ from those with $M_{\rm H}<2M_{\odot}$. The dashed-dotted line separates type II and type I SNe. Top-left the contours are for pre-SN luminosity in $\log(L/L_{\odot})$. Top-right contours are for stellar type determined by surface temperature. Bottom-left the contours are for pre-SN radius in $\log(R/R_{\odot})$. Bottom-right contours are for pre-SN mass of the progenitor in $M_{\odot}$. The filled circles represent IIP SN, the circle with filled centre a IIL, the upright triangle IIn, the inverted triangle IIb, the star symbol IIpec and the square a Ic. The IIP SNe are from top to bottom 1999gi, 1999em, 2001du and 2003gd. The IIL SN is 1980K. The IIb SN is 1993J. The IIpec is 1987A. The Ic SN is 2002ap.}
\label{map6-1}
\end{center}
\end{figure}

\begin{figure}
\begin{center}
\includegraphics[height=79mm,angle=0]{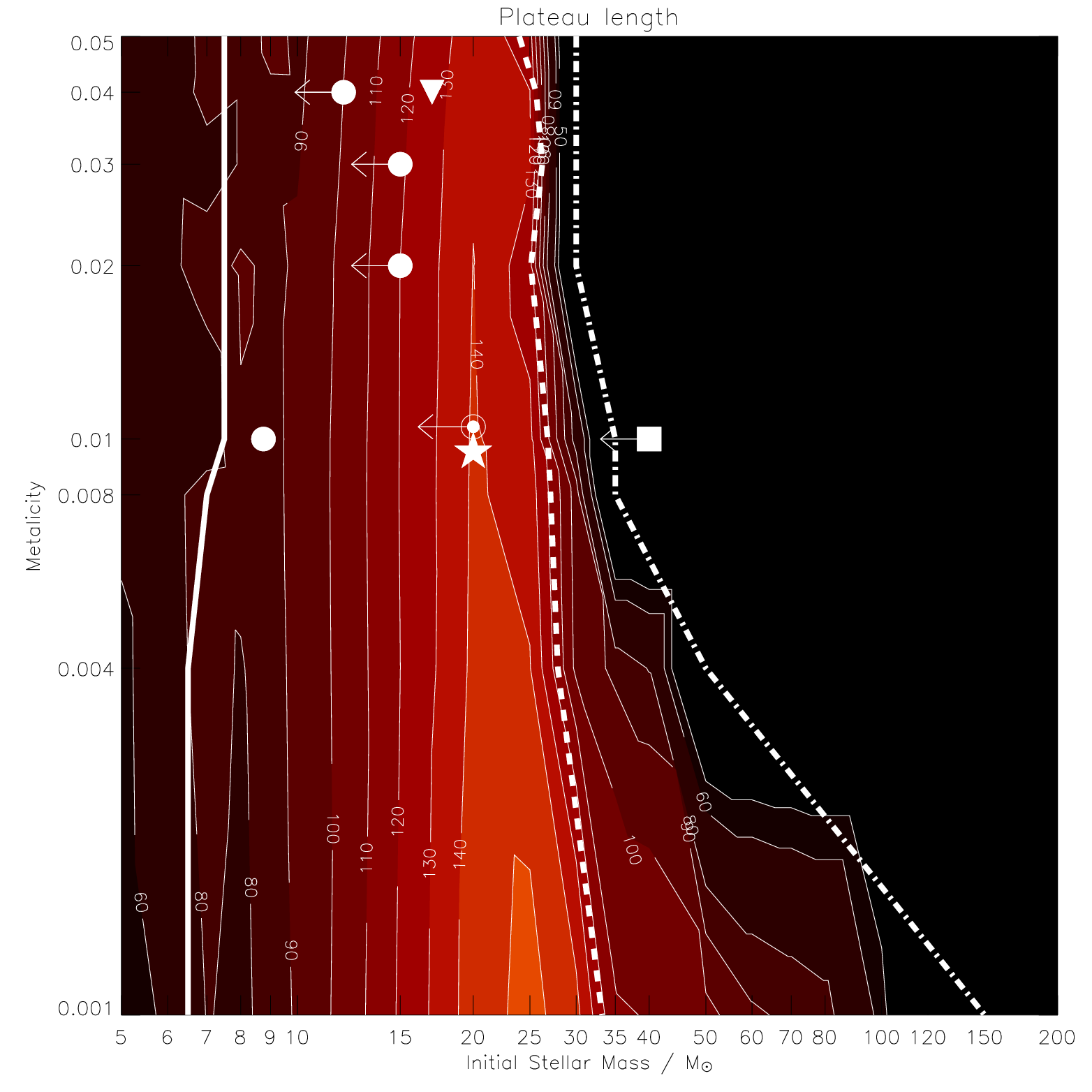}
\includegraphics[height=79mm,angle=0]{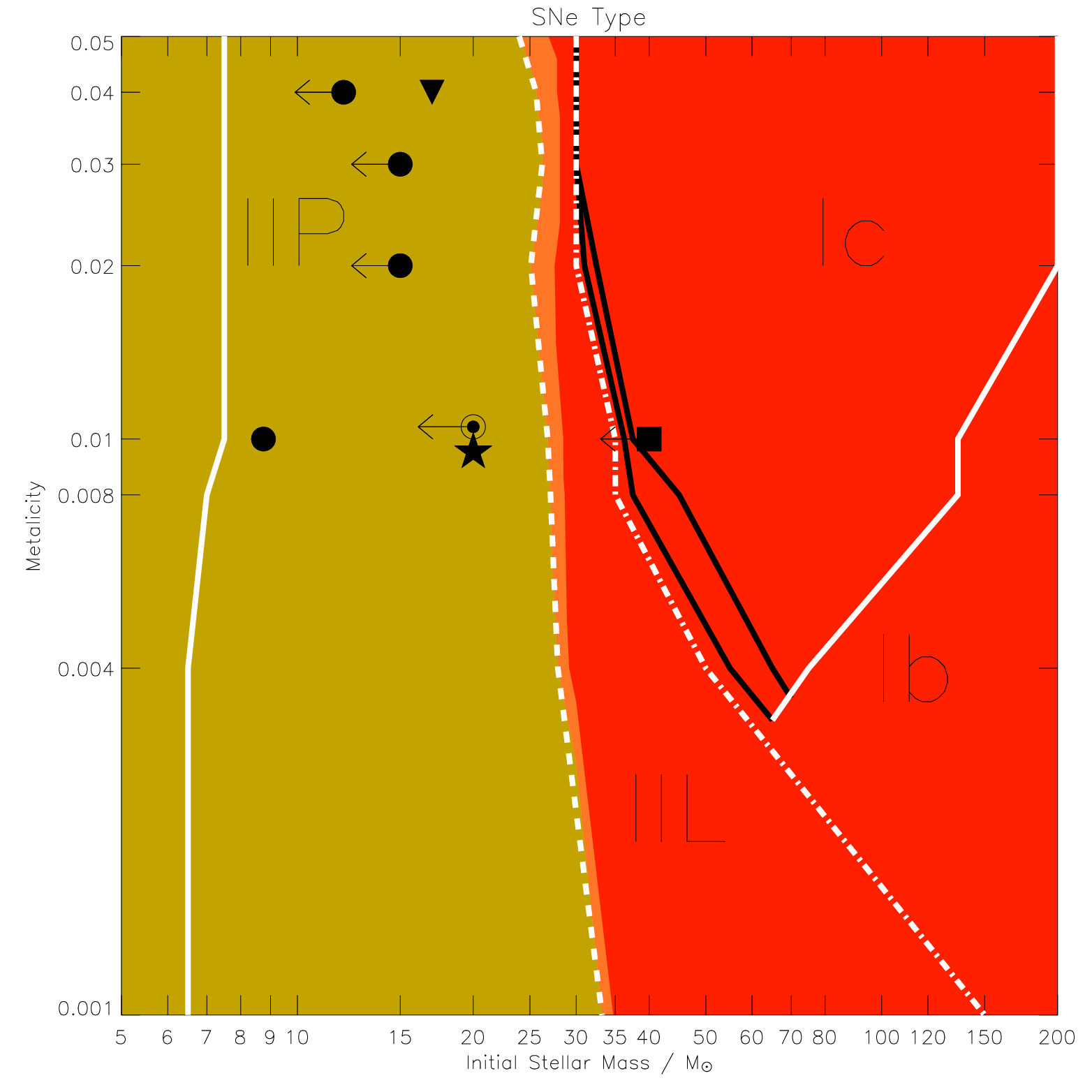}
\includegraphics[height=79mm,angle=0]{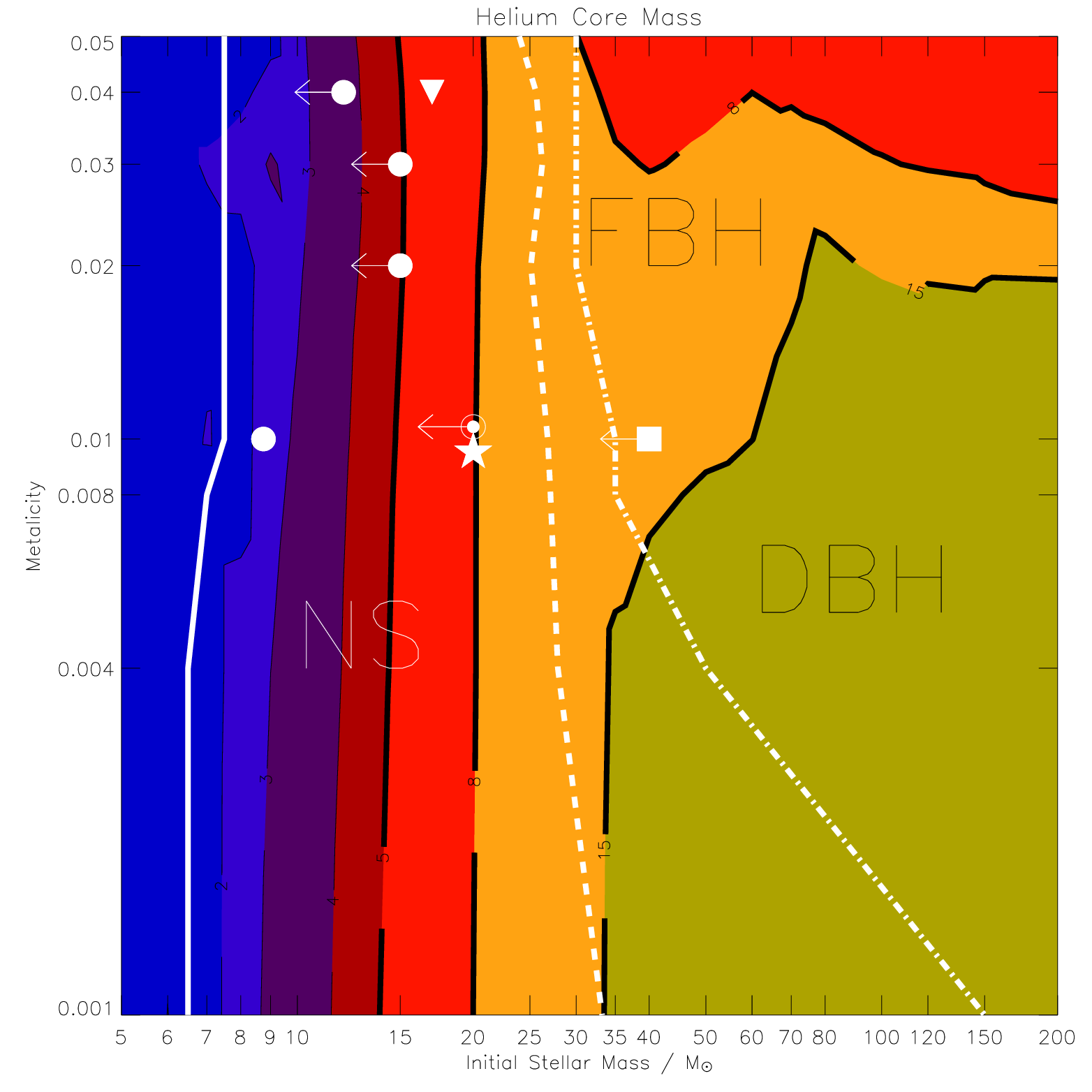}
\includegraphics[height=79mm,angle=0]{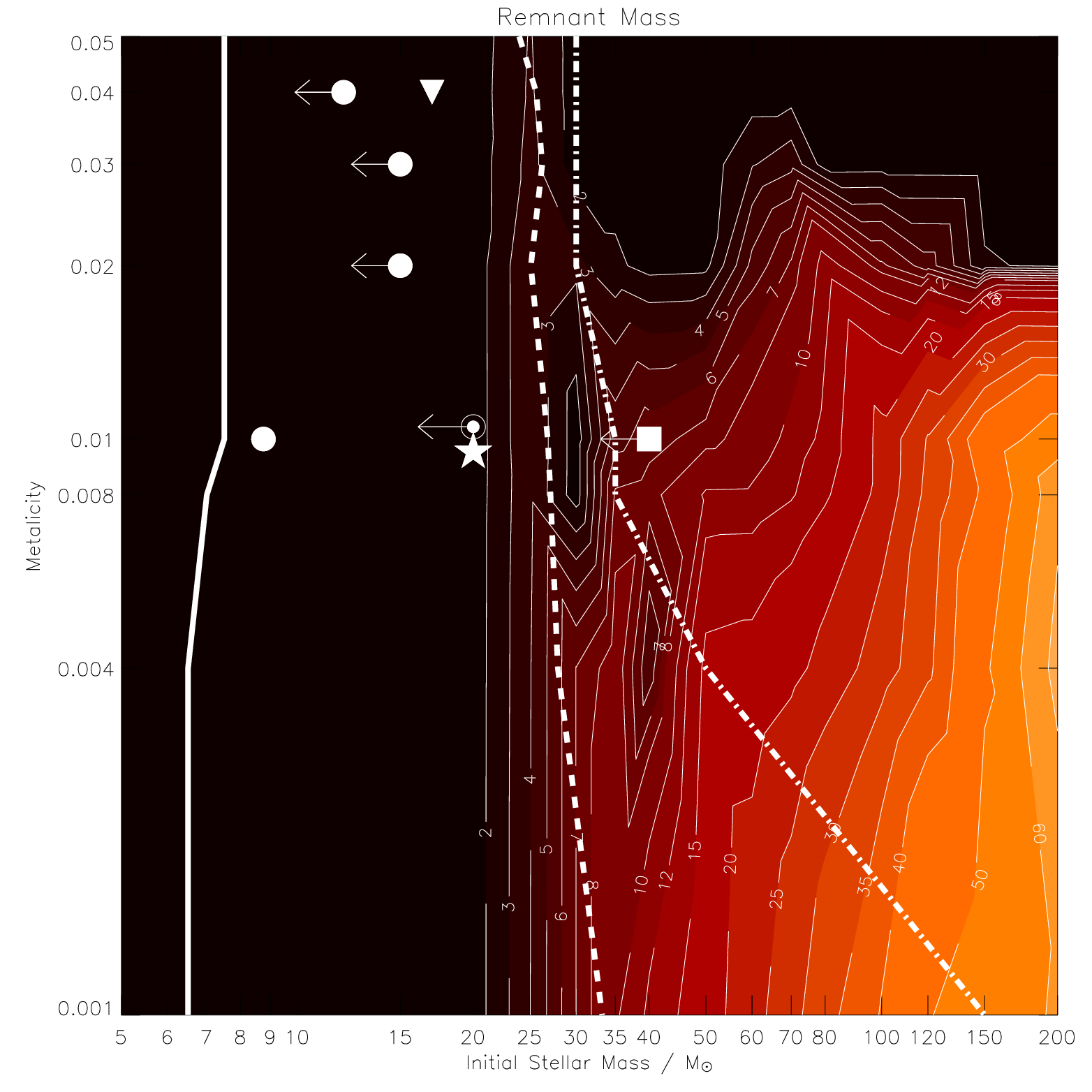}
\caption[Grids for comparison with observations of SN progenitors, plateau length, SNe type, helium core mass and remnant mass.]{Grids for comparison with observations of SN progenitors. For all grids the thick solid white line is the minimum mass for SNe. The dashed line separates type II SNe with $M_{\rm H}>2M_{\odot}$ from those with $M_{\rm H}<2M_{\odot}$. The dashed-dotted line separates type II and type I SNe. Top-left the contours are the length of the plateau phase in days. The top-right panel shows the boundaries between SN types. The thick black lines in this grid represent the change from Ib to Ibc to Ic SNe. Bottom-left the helium core mass. Bottom-right the final remnant mass, contours are in $M_{\odot}$. The filled circles represent IIP SN, the circle with filled centre a IIL, the upright triangle IIn, the inverted triangle IIb, the star symbol IIpec and the square a Ic. The IIP SNe are from top to bottom 1999gi, 1999em, 2001du and 2003gd. The IIL SN is 1980K. The IIb SN is 1993J. The IIpec is 1987A. The Ic SN is 2002ap.}
\label{map6-2}
\end{center}
\end{figure}

We have used the results for S-AGB SN progenitors from chapter 3 to extend our preferred grid from chapter 4. We have not included mass loss in these models and have created models at every integer mass up to $10M_{\odot}$. We assume the maximum mass S-AGB star lies $0.5M_{\odot}$ above the highest mass S-AGB model we have produced, $8.5M_{\odot}$ at $Z=Z_{\odot}$. The minimum mass S-AGB SN progenitor is then $0.5M_{\odot}$ below our highest mass model, $7.5M_{\odot}$ at $Z_{\odot}$. The results are shown in figures \ref{map6-1} and \ref{map6-2}. On the figures we have plotted the location of SN progenitors observed to date. The primary problem is that, in the type IIP region, we have SNe that are other type IIs. There is evidence that two of these, 1987A and 1993J were in binary systems \citep{Maund04,podsi92}. Therefore the remaining two SNe, 1980K and 1997bs, could also be binaries.

SN1980K, as already discussed in chapter 3, indicates that IIL SNe do not necessarily come from massive stars that have been stripped of a large fraction of their hydrogen envelope but they could also come from S-AGB stars that go SN. A binary model with RLF or CEE would also produce a IIL SN. The problem would be fitting a binary to the change in the mass-loss rate over the short period of 10,000 years prior to the SN. However it may be possible to observe the companion of this system in the future to test this possibility. The progenitor would not have been a single massive star in the range of $15 \le M_{\rm initial}/M_{\odot} \le 20$ that has lost a great amount of envelope before exploding because our models in that region have too great a hydrogen envelope remaining for such a SN. Furthermore our IIL single star progenitors might be classified instead as IIn SN as these stars have high mass loss leading up to their eventual SN event. This would make a dense circumstellar environment that could give the IIn appearance. Therefore the most massive single stars, that have only small amounts of hydrogen in their envelope, could be IIn progenitors not type IIL. \citet{cap1} say that type IIn are about $5-10\%$ of all observed type II progenitors. This certainly agrees with our models at solar metallicity. The range of models that fit this star are 23 to 26$M_{\odot}$.

The type IIP SN progenitors are all in the region we expect IIP progenitors from our models and the plateau lengths are in reasonable agreement. However SN2003gd requires special attention as it draws our eye to the minimum mass for S-AGB SN progenitors. From the derived luminosity of the progenitor of 2003gd, $\log(L/L_{\odot})=4.3 \pm 0.3$, we have a problem that our models only just go under the upper limit of $4.6$. We can look at this from two views. The first is that we can place quite stringent limits on the mass of 2003gd since it must be around $8.5M_{\odot}$ otherwise it would have gone through second dredge-up and be a great deal more luminous. We could also say that the host galaxy must be more distant than other observations indicate. 

The more correct view we should take is that this progenitor does agree with our models although there is now a mystery that can only be solved by more observations of progenitors in this region. Second dredge-up before core collapse can be prevented by making convection less efficient at the base of the convective envelope. Therefore if 2003gd was around $8M_{\odot}$ and did not undergo second dredge-up this has important ramifications for stellar structure models and the details of convection.

The only type Ic SN in the field, 2002ap, is in the right place. From the graph we can see that the star's mass is also tightly constrained. This is also the only SN that is definitely in the region where we expect a black hole was formed; the remnant should have been around $6M_{\odot}$. However any position to the right of this is also possible since there is only a limit on it mass from observations. However that would require a binary system to explain a mass lower than the theoretical Ic region.

The super-star SN1987A, the best observed SN since the invention of the telescope, can be seen right on the limit of where it may or may not have formed a black hole by fall back on to a neutron star. A remnant has yet to be observed at 1987A although it might have been ejected from the system and so not been discovered but there is a chance it could have been a black hole. However if it was the result of a binary the primary would have been a lower-mass star to begin with and so would be in the region that would not have formed a black hole, although this relies on our very approximate methods to decide on the remnant left by the SN.

When observers find more SN progenitors we will be able to place them on this diagram and draw more conclusions. This will force us to refine our models over time changing the mass-loss rates, our details of convection and other physical details in the code. Only time will tell the quality of our current models.

\subsection{Remnant masses}
\begin{figure}
\begin{center}
\includegraphics[height=79mm,angle=0]{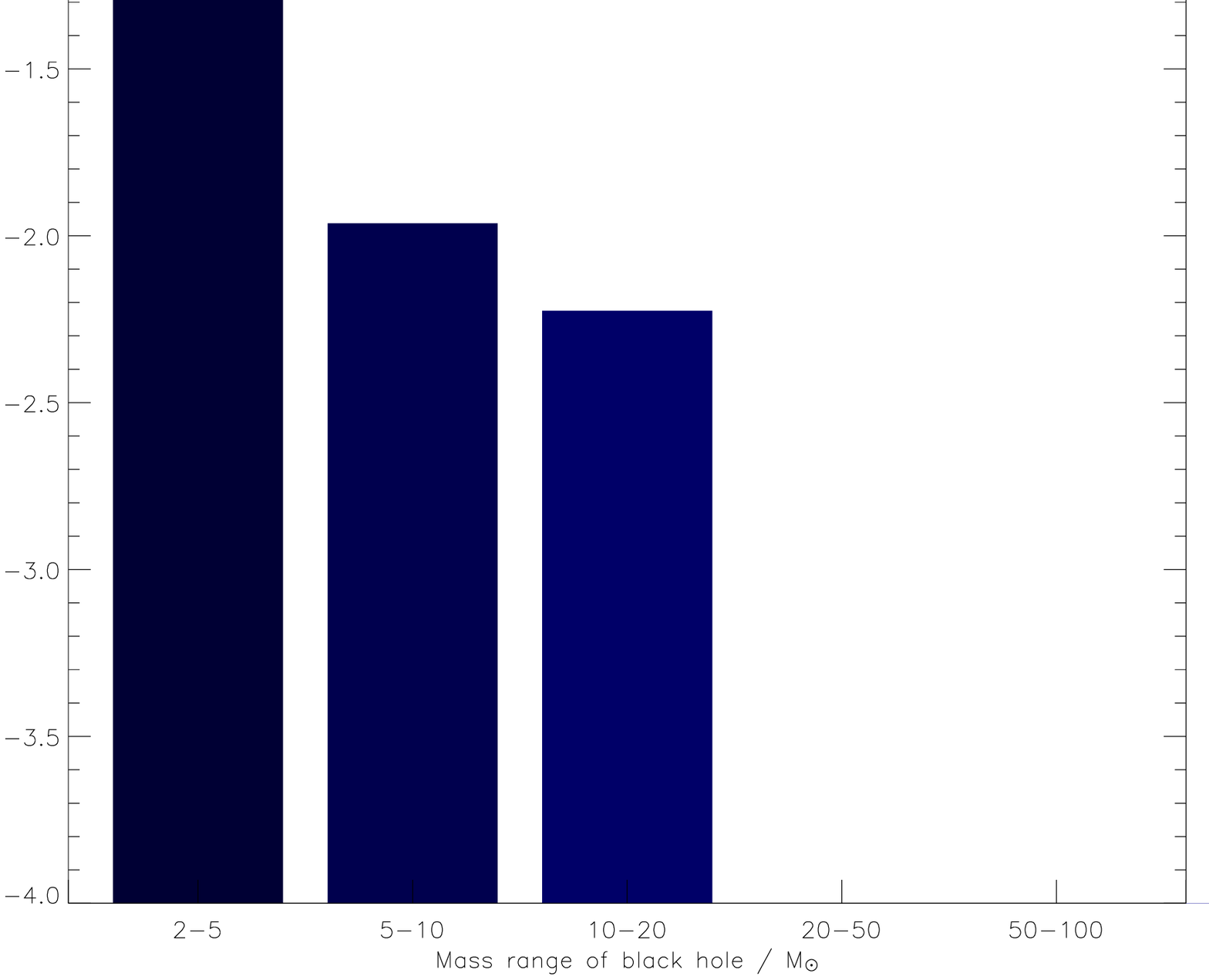}
\includegraphics[height=79mm,angle=0]{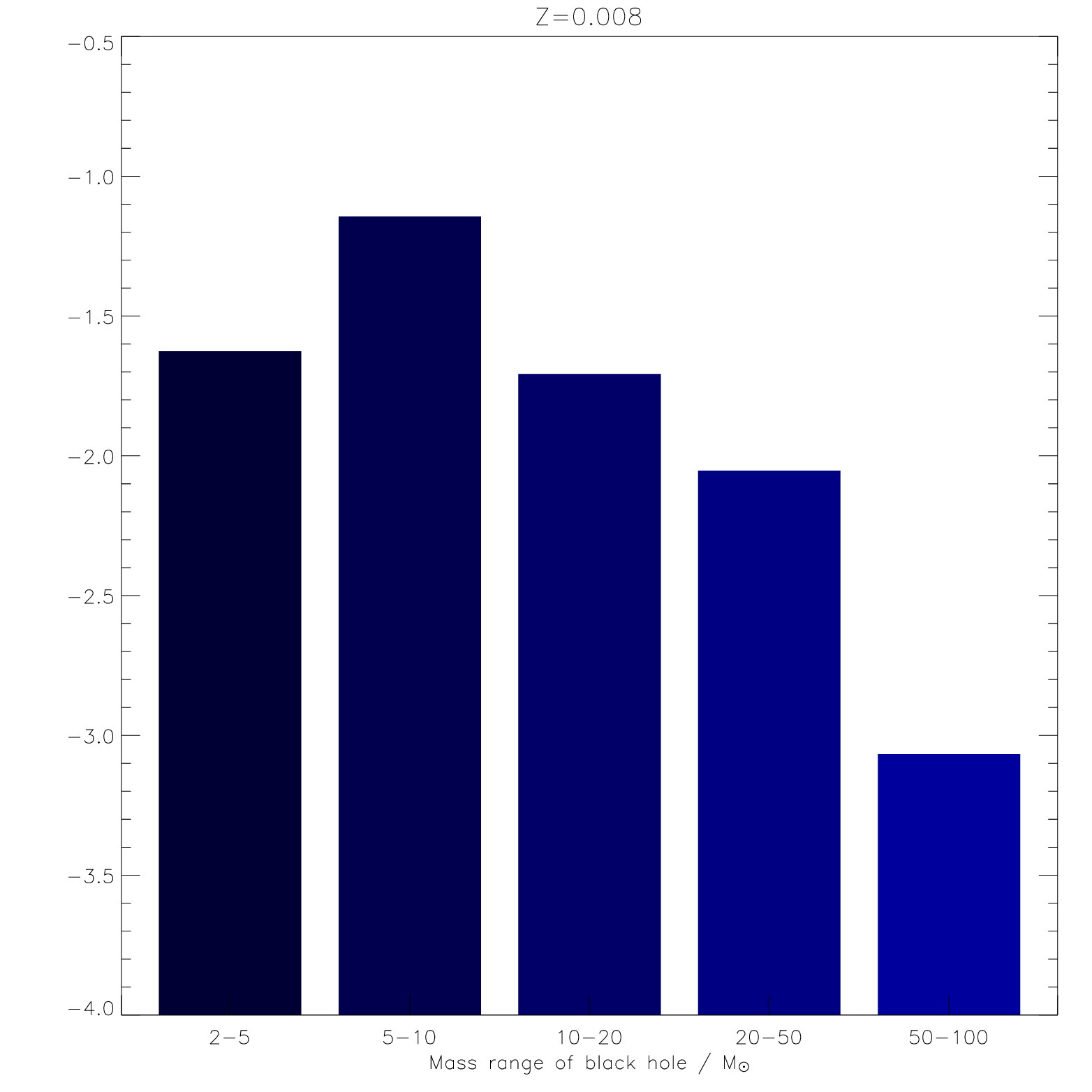}
\includegraphics[height=79mm,angle=0]{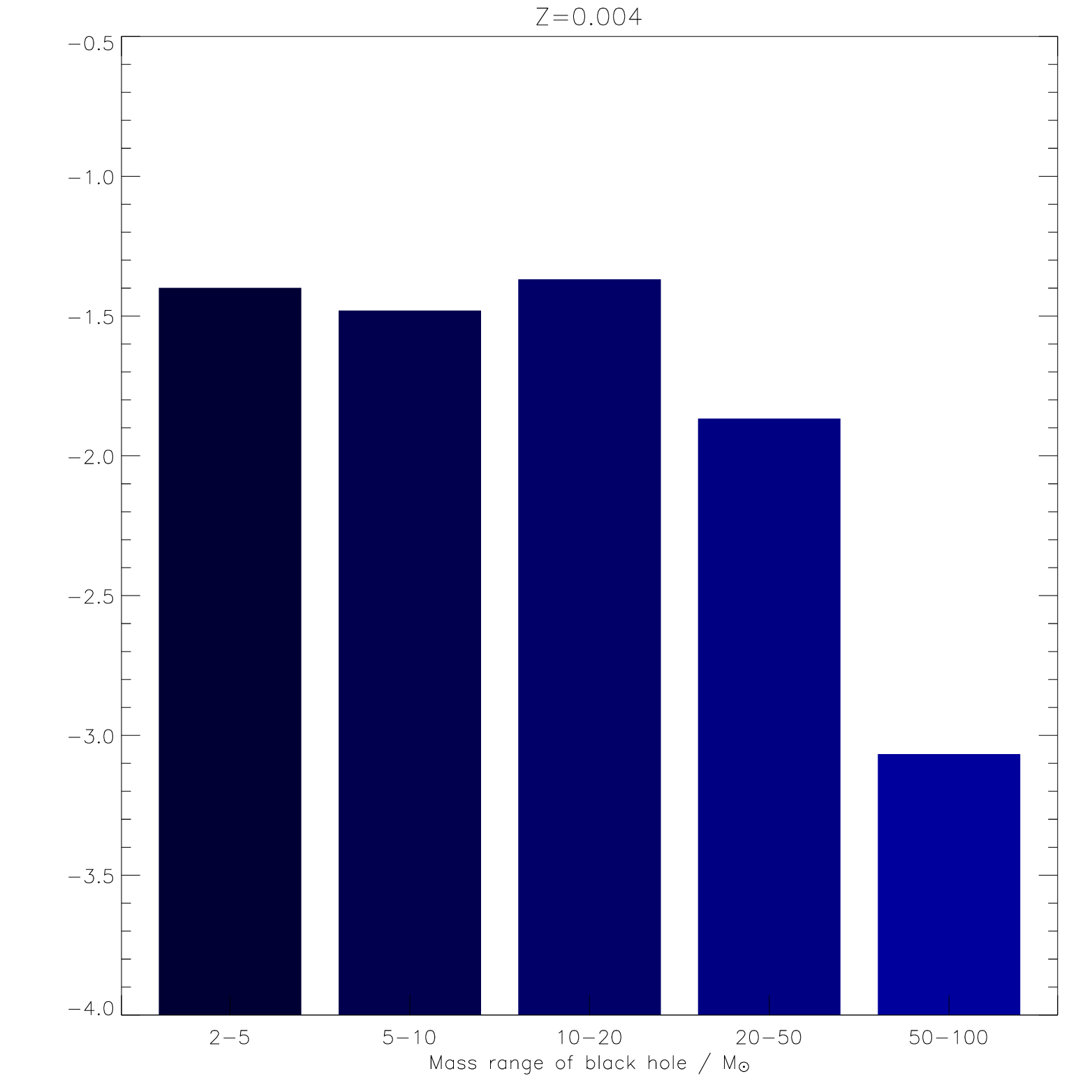}
\caption[Black hole remnant mass distribution at various metallicities.]{Black hole remnant mass distribution at various metallicities. The y-axis is the base ten logarithm of the fraction of remnants in the mass range of remnants listed, out of all remnants from SNe. All remaining remnants are taken to be neutron stars.}
\label{map6-3}
\end{center}
\end{figure}

Figure \ref{map6-3} shows the logarithm of the fraction of remnants with mass $M_{\rm rem}$ in the ranges listed. At the highest metallicities the remnants are all of a few $M_{\odot}$ with neutron stars dominant. This means there is a maximum mass for remnants at lower metallicity. Then as metallicity decreases the maximum remnant mass increases until the production of black holes greater than $100M_{\odot}$ becomes possible. The remnant masses used for these diagrams come from the very simple calculation explained in chapter 3. They are at best estimates of the distribution.

Very massive $(M \ge 100M_{\odot})$ black holes can form in SNe but this can only happen at low metallicity ($Z < 10^{-3}$). The population of such black holes is very small as they are only formed in the most massive stars. At higher metallicities massive black holes are not formed because mass loss limits the size of the SN progenitor and thus the size of the remnant. Any black holes observed to be very massive in high metallicity environments must therefore have grown via accretion from a binary companion or must have formed from an earlier population of stars.

\subsection{HR Density Plots}

\begin{figure}
\begin{center}
\includegraphics[height=79mm,angle=0]{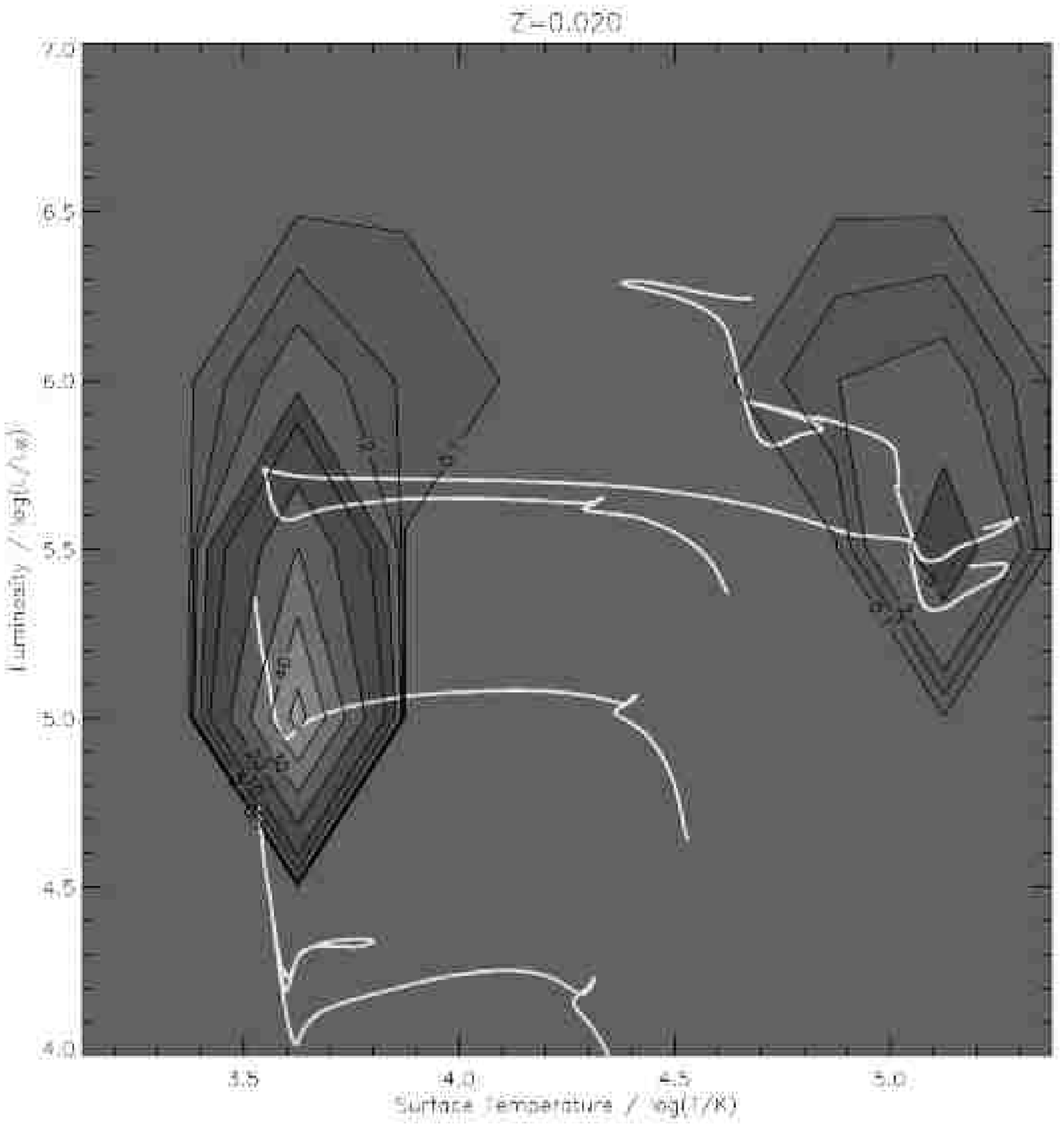}
\includegraphics[height=79mm,angle=0]{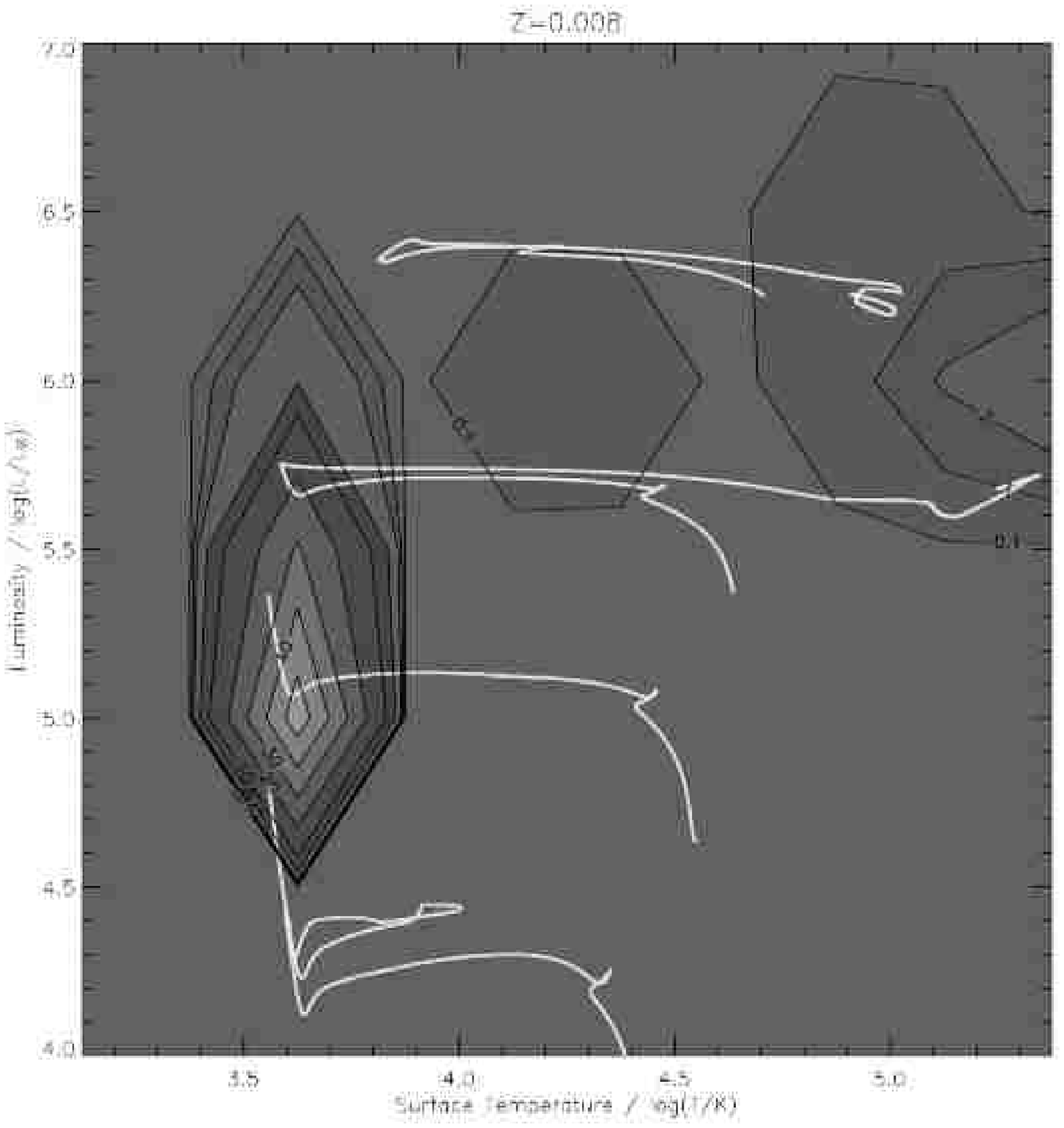}
\includegraphics[height=79mm,angle=0]{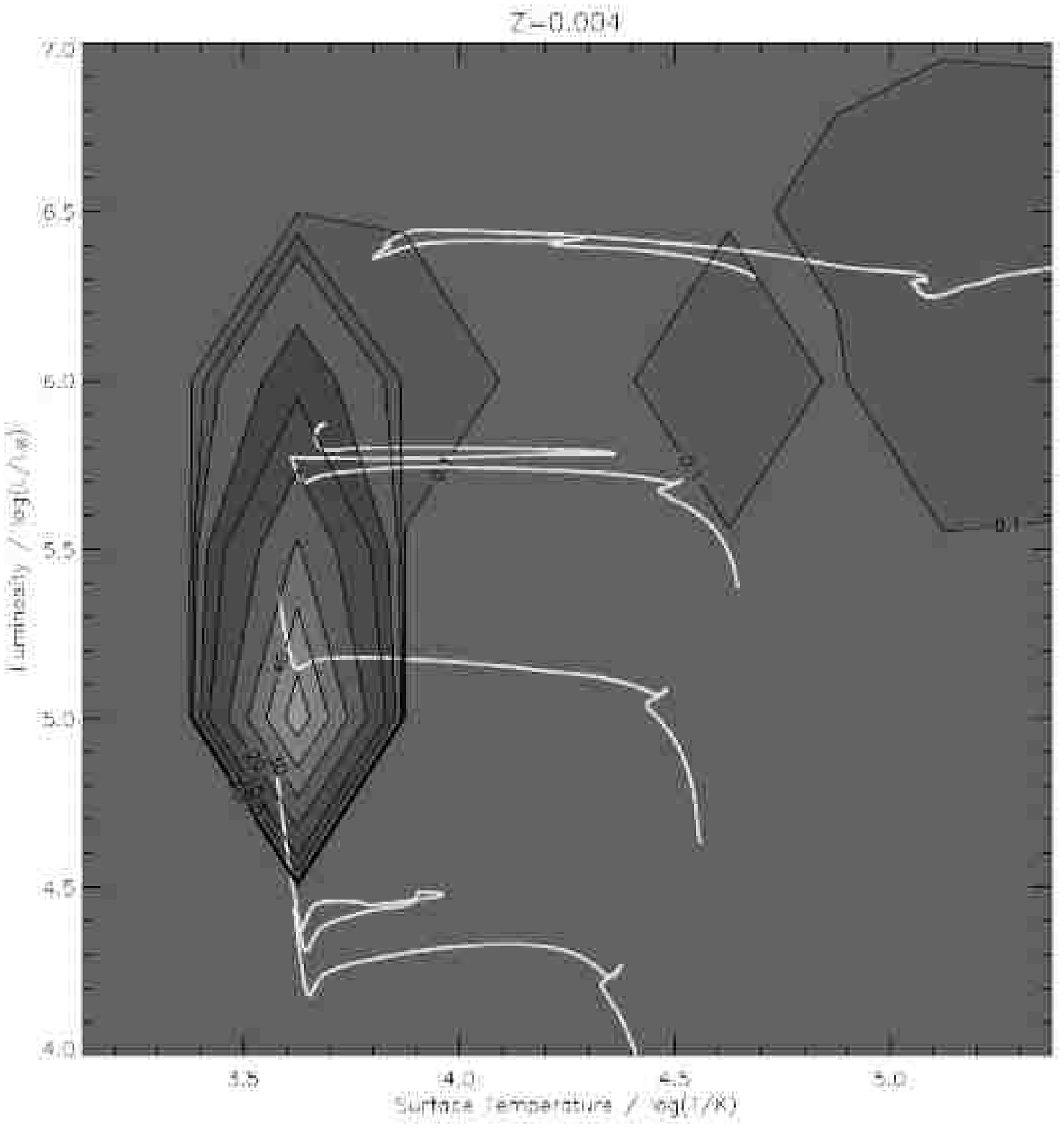}
\caption[HR diagram progenitor population plots.]{HR diagram progenitor population plots. Displaying the population over the HR diagram for single star SN progenitors. Contours are in \% of total population. The yellow lines are the evolution tracks of 10, 20, 40 and 120 $M_{\odot}$ stars at the metallicity of the contour plot.}
\label{map6-4}
\end{center}
\end{figure}

Figure \ref{map6-4} shows where we can expect single star progenitors to lie on the HR diagram. Red giants are the dominant group. If we decrease metallicity there are fewer WR star progenitors and the luminosities increase. The temperatures used here are from hydrostatic models. We know that the temperatures should be slightly lower but we can clearly see WR stars form a separate population from the red giants.

We must ask the question where would binary progenitors lie on this diagram? Type II progenitors can be blue rather than red super giants or have their envelopes removed to become type Ibc progenitors. If the type II progenitors become blue super giants through a merger event or mass-gain near the end of their evolution this would move them over to the right of the diagram but the luminosity would remain the same because it is fixed by the hydrogen burning shell. The situation is quite different if the hydrogen envelope is removed completely because the primary source of a star's luminosity is the hydrogen burning shell. This event would produce a range of low-luminosity progenitors in between the two populations in the diagram.

Table \ref{tab61} gives the details of some example Ib progenitors with equivalent results for single stars that retain their hydrogen envelope. This has important implications when we try to fit the masses of progenitors to type Ib SNe. Stars that are in binaries that are greater than $20M_{\odot}$ are probably massive enough that interactions lead to the helium stars becoming WR stars and therefore increase the WR population. It is not possible to use the standard stellar evolution models from the hydrogen main sequence alone to identify type I progenitors. A sequence of models from the helium main sequence should be constructed and evolved and then these used for comparison to observations of the progenitor. Once the helium star initial mass is known we would be able to then determine scenarios for the source of the star and its initial mass.

\begin{table}
\caption{Type Ib SNe progenitor details compared to the details for the a similar star that does not lose its hydrogen envelope.}
\label{tab61}
\begin{center}
\begin{tabular}{|c|c|c|}
\hline
Initial Mass & With Hydrogen Envelope  & Without Hydrogen Envelope \\
$/M_{\odot}$   &  $\log(T/{\rm K}), \, \log(L/L_{\odot})$&  $\log(T/{\rm K}), \, \log(L/L_{\odot})$\\
\hline 
11			&	3.5, 4.8		&   4.0, 4.5\\
12			&	3.5, 4.9		&   4.3, 4.7\\
13			&	3.5, 4.9		&   4.4, 4.8\\
15			&	3.5, 5.1		&   4.6, 4.9\\
20			&	3.5, 5.3		&   4.8, 5.1\\
\hline
\end{tabular}
\end{center}
\end{table}

\section{Combining Single Stars and Binaries}

Now we take the SN type ratios for single stars from chapter 3 and the ratios for binary stars from chapter 5 and mix them together. This should lead to a better agreement with what we see in observations. We use the same SN type classification scheme for both sets of data. For type II SNe we assume all stars with $M_{\rm H}<2M_{\odot}$ are type IIL SNe and the rest are plain type II. We split the Ic and Ib SNe to those with a surface mass fraction of helium either greater than or less than 0.5.

\begin{table}
\caption{Mixing different populations of single and binary stars. Not including a S-AGB star contribution to IIL SN.}
\label{mixmain}
\begin{center}
\begin{tabular}{|c|c|ccc|}
\hline
  Single &    Binary   &    &     &          \\
  Fraction &    Fraction  &  Ib/Ic   &   II/Ibc  &   II/IIL       \\
\hline
0&1&3.62&0.99&3.43\\
0.1&0.9&3.28&1.70&6.82\\
0.2&0.8&2.94&2.41&10.21\\
0.3&0.7&2.60&3.12&13.61\\
0.4&0.6&2.25&3.84&17.00\\
0.5&0.5&1.91&4.55&20.40\\
0.6&0.4&1.57&5.26&23.78\\
0.7&0.3&1.23&5.97&27.18\\
0.8&0.2&0.89&6.68&30.57\\
0.9&0.1&0.55&7.39&33.96\\
1&0&0.21&8.10&37.35\\
\hline   
\end{tabular}
\end{center}
\caption{Mixing different populations of single and binary stars. Including a S-AGB star contribution to IIL SN.}
\label{mixmain2}
\begin{center}
\begin{tabular}{|c|c|ccc|}
\hline
  Single &    Binary   &    &     &          \\
  Fraction &    Fraction  &  Ib/Ic   &   II/Ibc  &   II/IIL       \\
\hline                            
0&1&3.62&0.99&3.43\\
0.1&0.9&3.28&1.70&4.65\\
0.2&0.8&2.94&2.41&5.87\\
0.3&0.7&2.60&3.12&7.09\\
0.4&0.6&2.25&3.84&8.31\\
0.5&0.5&1.91&4.55&9.53\\
0.6&0.4&1.57&5.26&10.75\\
0.7&0.3&1.23&5.97&11.97\\
0.8&0.2&0.89&6.68&13.19\\
0.9&0.1&0.55&7.39&14.41\\
1&0&0.21&8.10&15.63\\
       
                 \hline       
\end{tabular}
\end{center}
\end{table}

To summarise the observations of \citet{cap1} and \citet{cap2} we know that the II/Ibc ratio must be $5.00 \pm 3.45$  and the II/IIL ratio must be a few. However depending on galaxy type the II/Ibc ratio can take values from $3.82 \pm 2.71$ to $6.14 \pm 3.96$. This makes the situation very difficult to fit with models because the value is so uncertain. In fact ratios calculated from any of our mass-loss schemes when mixed with binaries agree with one of these values and indicate that binaries are very important for models of SNe ratios. They are much more important than mass-loss prescription or inclusion of extra mixing from rotation.

When we look at table \ref{mixmain} and use the ratio of \citet{cap2} to determine the binary fraction we see that it should be around 45\% but, due to the errors in the measurement, the errors in our estimation are $\pm 40\%$. The expected binary fraction is expected to be higher than this and to not be constant \citep{binaryman}. If we assume a fraction of 70\% we find a ratio of $3.12$ which is on the low side of the observations. The binary fraction is likely to vary with initial mass and the distribution of mass ratio and separation. Also we can favour the lower ratios if we assume that some Ibc SNe, that we have determined occur, are not observed because they are too compact to eject much material and remain unseen. Because of this is it likely that we should favour the higher binary fractions.

The type II/IIL ratio is a problem. From single stars we have only a tiny fraction of possible type IIL progenitors and some of these are quite likely to be IIn progenitors. If we include the lowest mass S-AGB, stars between 7.5 and $7.6M_{\odot}$, as in table \ref{mixmain2}, we still find poor agreement. The other option is to lower the mass for massive single star IIL progenitors. However this doesn't produce a large change until it is lowered to below$18M_{\odot}$ and this would produce too many WR stars for the WR ratios.

There are just too few IIL SNe at the binary fraction that agrees with the II/Ibc ratio. We must get a larger population of these models. The main source must be binary stars. In fact the need to produce more IIL SNe has important implications for our models. We require a finer grid to find the exact region where IIL SN could occur. The other possibility is that CEE could be less efficient than we think and that not all hydrogen is removed in such an interaction. It is also possible, that during the CEE, hydrogen is mixed into the helium core by an unknown dynamical process that would leave enough hydrogen for a signal in the SNe spectra. However we must remember that \citet{cap1} are very vague in their estimation of the relative IIL population so the error bars are extremely large. \citet{IILstats} suggest the IIL fraction of all type II SNe could be a little as one tenth. Our results indicate the later however we will need to wait for more observations to tie this ratio down. Also there could be selection effects in the observations that are not understood.

The other solution would be to increase the mass-loss rates to produce more IIL progenitors from single stars but this would then reduce the agreement with the WR ratios and furthermore decrease the II/Ibc ratio that would in turn decrease the binary fraction needed to reproduce observed values. Therefore this problem must be addressed by further observations.

\section{Conclusion}

In this chapter we have combined and analysed our data from the previous sections to draw some conclusions on the total population of SN progenitors and shown that it does fit observations well and that from observations we can provide vital constraints on stellar evolution models.

\chapter{Conclusion}

\begin{center}

``Answers are easy. It's asking the right questions which is hard.''\\
\textit{The 4th Doctor Who, The Face of Evil.}
\end{center}
\section{Summary}
This dissertation is long and a final summary to highlight some of the important points is necessary. Also we should draw some final overall conclusions on the progenitors of SN progenitors.

The preliminary chapters 1 and 2 provided an overview to stellar evolution and supernovae. Chapter 2 described in detail the stellar evolution code used and presented the refinements made to the Cambridge code as part of this study. The most important of these was the construction of new and unique opacity tables that refine the models of stars during the latest stages of evolution.

Later chapters covered the evolution of SN progenitors. Chapter 3 contains a detailed look at the lowest mass stars to go SNe, S-AGB stars undergoing ONe core collapse by electron capture. The exact nature of these stars can only be determined by observations of SN progenitors which can provide tough constraints on stellar physics, most critically on convection. This chapter also contained a high resolution study of single star progenitors at solar, LMC and SMC metallicities using our favoured mass-loss scheme. By using the results we made suggestions on how to relate these models to types of SNe, showing that there is much to study in the region of type I SNe with a diverse range of possible progenitors and that the difference between type Ib and Ic SNe may be more than just the absence or presence of helium. For type II progenitors we discussed that IIL progenitors come not only from binary stars but from the S-AGB stars, while IIn progenitors may come from single WN stars that still have hydrogen envelope but have lost a large fraction via severe stellar winds.

In chapter 4 we have described the multiple data sets created to compare the diverse mass-loss schemes to fix on one that fits best to observations. We showed that the inclusion of metallicity scaling of mass loss for WR stars makes important changes to the population of SNe progenitors and that this scaling should be included. From arguments using observations we then settled on our favoured mass-loss prescription that uses the rates of \citet{dJ}, \citet{VKL2001} and \citet{NL00}.

Next in chapter 5 we moved on to study binary progenitors. After describing our method we discuss the results showing that the type I progenitors differ from those from single stars, adding further evidence to the idea that there are a few different populations of type I SNe that we do not currently distinguish.

Finally in chapter 6 we collated some of our information to draw conclusions over the entire mass range for SNe and compare this directly to observations. We also then mixed together our results from single and binary stars to see that the evidence points toward the fact that the interacting binary fraction of stars is around 45\% to best fit the observations. However the errors are large at $\pm 40\%$ and higher values are required to fit the II/IIL fraction. This shows that understanding of single and binary stars is of equal weight when we consider how the population of SN progenitors will be spread over the HR diagram and the distribution of black hole masses from supernovae. We importantly showed that black holes more massive than $100 \, M_{\odot}$ are only possible at metallicities below $Z<0.001$. We also pointed out that the determination of the type I SN progenitors needs extra consideration to allow for the lower luminosity once the hydrogen envelope has been removed and that this could lead to overestimation of the initial masses.

\section{Conclusion}

Our main conclusions are to define what we can expect the progenitors of SNe to be, all the limits below are for solar metallicity with convective overshooting. For type II SNe they are,

\begin{itemize}
\item IIP, these stars are probably single stars with a mass initially below about $25M_{\odot}$. They have only lost a small fraction of their envelopes and are red giants when the SN occurs. At the low-mass end of this range, around 8 to $8.5M_{\odot}$, they may have undergone second dredge-up shortly before core-collapse to become extreme S-AGB stars. There may also be a low luminosity class of IIPs where $M_{\rm initial}>20 \, M_{\odot}$ and a black hole is formed  rather than a neutron star and affecting the SN evolution.
\item IIb, are mostly likely to be from massive stars in a binary when interactions leave only a tiny fraction of the original hydrogen. It is also probable that this type of progenitor must be in the same mass range as progenitors of type Ib SN. There is probably a smooth progression from IIL to IIb to type Ib SNe.
\item IIL, a large fraction of stars that experience only light mass loss via an interaction with a binary companion to lose some but not their entire envelope. The population of IIL SNe relative to all type II SNe places constraints on binary interactions and the binary mass ratio and separation distributions. A small fraction of the progenitors are massive single stars that have lost their envelope by stellar winds although the models in this class might be more likely to give rise to IIn.  There is also a chance that low-mass stars in the tiny range 7.5 to $7.6M_{\odot}$ give rise to a IIL SNe rather than IIP because they experience second dredge-up a few thousand years before core-collapse. This is enough time for a wind to remove sufficient envelope to prevent a plateau phase in the light curve. The interesting thing is that all these progenitors must produce a similar amount of $^{56}{\rm Ni}\approx 0.1M_{\odot}$ and all of these models do. Observations and further modelling of SNe are required to deal with this group. It is likely there is not a unique progenitor and observations already indicate possible subgroups in this class.
\item IIn, are mostly likely to be from hydrogen rich WR stars about 25-26 $M_{\odot}$ for single stars. Such stars have a dense hydrogen rich circumstellar environment that the SN ejecta collides with to retard its expansion. Binary stars could also provide some of these stars from a RLOF or CEE event close to the occurrence of the SN.
\item IIpec, well everything else really. We should introduce a new group of SN that have undergone mergers and therefore become blue stars such as SN1987A. Most type IIpecs probably suffer from lack of detailed observations to specify what type they are accurately so many in this bin probably belong to one of the other groups. For example there are weak IIP SNe that are thought to form black holes rather than neutron stars that would get misplaced into this group. Only more detailed observations will provide more clues about this region.
\end{itemize}

For type I SNe we should not provide a list as for type II SNe. The observational scheme does not provide enough possible groups to place our progenitors in. The answer to which stars give rise to which type I SN is difficult because we are trying to discover what these stars don't have pre-SN. The answer is probably not just to do with the presence or absence or material but also the size of the object, its structure and the nature of the remnant formed.

However Ib progenitors are probably from the evolution of binary stars. They are low-mass helium stars that become giants, similar to the progenitors of IIb SNe but have been fully stripped of their hydrogen envelopes. Type Ic SNe are probably more compact stars because there is evidence that the ejecta velocities are higher for Ic SNe than for Ib SNe so, if we assume a constant energy from the SN, then less mass is ejected but at a higher velocity. This would be correct if the star is more compact because the material would be more tightly bound.

Therefore our other compact progenitors are probably observationally all classed as type Ic. The more massive progenitors that form a black hole directly in their centre could be the progenitors to hypernovae but these stars would need to be rapidly rotating for this to occur.

There are broadly three classes of progenitors that fit into the Ic group. The first are stars around solar metallicity initially more massive than $30M_{\odot}$ with final masses less than $15 \, M_{\odot}$. They form black holes by fallback or directly. If the black hole forms directly this may lead to a different explosion mechanism and therefore different behaviour. The most massive progenitors in this class may be too tightly bound for enough matter to be ejected to provide a display. They also have less than a tenth of a solar mass of helium remaining so there would be little to affect the spectra of these objects.

The second group are single stars at lower metallicity are more massive and retain more helium than their solar metallicity counterparts. However they are compact. This has an important effect on their SNe. They may only give rise to dim SNe or have no display at all because they are so tightly bound that very little material is ejected. If there is any display it is uncertain whether it would be possible to observe the helium.

The final class are GRBs or hypernovae that fall into the type Ic scheme. These are probably in binary systems to ensure rapid pre-SN rotation. They must be more massive than $15 \, M_{\odot}$ so a black hole is formed directly. The rotation ensures an accretion disc can form to power jets and give rise to an energetic SNe, hypernovae or GRBs.

\section{Future Work}
\begin{center}
``A code is never fully tested until you run it for the last time.''\\
\textit{Sverre Aarseth, 2004.}
\end{center}
The main path forwards from here can be succinctly put; more models and more observations are required. While the more observations point relies on other researchers the more models is easy to do. With 70 years of computer time it is possible to calculate over half a million model stars. That is roughly the total evaluated for this dissertation.

It is important to model the correct thing. Current plans are to produce a vast and detailed grid of model SNe progenitors over the range used in this thesis but with higher resolution in mass and metallicity than used here and to reach zero metallicity. However rather than making a single jump it would be better to lower the metallicity in steps to determine exactly when the behaviour alters. At the current time and for the foreseeable future there will be no observational constraints here.

The binary models also need refinements to produce the remnant and HR population diagrams that we have for single stars. That is the easy part. Waiting for the code to run a large grid is the problem. However time and multiple computers ease the problem. Using the code in parallel with binary population synthesis so we know the interesting places to study will also be necessary.

Comparing binary models to limits for SN progenitors is difficult owing to the large parameter space, especially for Ibc progenitors. Therefore it will be quite useful to produce model helium stars in a similar fashion to the models in chapters 3 and 4. This will be more practical than just using the normal tracks because evolution of low-mass stars, once they have lost their hydrogen envelopes, is significantly different. Once we know the initial helium star mass it will be easier to work out where it came from.

Another step is to take all these models and feed then into a SN simulation code. It doesn't have to be a state of the art 3D magnetohydrodynamic masterpiece of coding either. An existing 1D code will be sufficient. This will replace the need to use analytical approximations and provide further understanding of the difference between SNe.

One important refinement is to take the burning stages of the progenitor models yet further, all the way to the formation of an iron core if possible. However this will require making the code stable at time steps below $10^{-4}$ years by fixing the mesh and calculating the opacity of central regions on the fly to account for the changing composition.

Also with this data looking at how the SN ratios change over time will be another useful analysis of the above data which may provide a better agreement to observations.

Otherwise let's just hope for a few galactic SNe next year....

\begin{center}
``Looks like the end of the line, space-mates!''\\
\textit{Colonel Dan Dare - Eagle 4/7/1952 - Vol.3 No.13.}
\end{center}

{
\clearpage

\thispagestyle{empty}

\vspace*{5cm}
\begin{center}
\end{center}
}
{
\cleardoublepage
\thispagestyle{empty}
\thispagestyle{empty}
\vspace*{5cm}

\begin{center}
``Time is an illusion. Lunchtime doubly so''\\
\textit{Douglas Adams.}\\
 \end{center}
 \begin{center}
``The true delight is in the finding out rather than in the knowing.''\\
\textit{Isaac Asimov.}\\
 \end{center}
 \begin{center}
``I don't pretend we have all the answers. But the questions are certainly worth thinking about.''\\
\textit{Arthur C. Clark.}\\
\end{center}
}

\end{document}